\documentclass[12pt,a4paper]{book}
\usepackage{latexsym}
\usepackage{amssymb,amsfonts,amsmath}
\usepackage{epsfig,epic}
\usepackage{smallcaptions}
\usepackage[english]{babel}
\usepackage{wrapfig}
\usepackage{fancyheadings}
\usepackage{subfigure}
\usepackage{floatfig}
\usepackage{rotating}
\renewcommand{\chaptermark}[1]%
                {\markboth{#1}{}}
\renewcommand{\sectionmark}[1]%
                {\markright{\thesection\ #1}}
\begin{document}

\thispagestyle{empty}
\vspace{-3cm}
\begin{center}
    {\bf Universit\'a degli Studi di Torino - Dipartimento di Matematica}
\end{center}

\begin{center}
    {\bf Dottorato di ricerca in\\
    Matematica\\
    XX ciclo}
\end{center}

\vspace{2 cm}
\begin{center}
    {\Huge Analysis of measurement algorithms and modelling of interferometric signals for infrared astronomy}
\end{center}

\vspace{0.2 cm}
\begin{center}
    {\Large Analisi di algoritmi di misura e modellizzazione di segnali interferometrici per l'astronomia infrarossa}
\end{center}

\vspace{1cm}
\begin{center}
    {\bf Tesi di dottorato\\
    presentata da}
\end{center}
\begin{center}
    {\huge Donata Bonino}
\end{center}

\vspace{1 cm}
\begin{center}
    {\bf Advisors:}
\end{center}
\begin{center}
    {\Large Prof.ssa Laura Sacerdote\\
    Dott. Mario Gai}
\end{center}

\vspace{1 cm}
\begin{center}
    {\bf Coordinatore del Dottorato:}
\end{center}
\begin{center}
    {\large Prof. Luigi Rodino}
\end{center}
\begin{center}
    {\bf Anni accademici:\\
    2004-2008}
\end{center}

\vspace{0.2 cm}
\begin{center}
    {\bf Settore Scientifico-disciplinare di afferenza: MAT/06}
\end{center}

\newpage{\pagestyle{empty}\cleardoublepage}
\pagenumbering{roman}
\addcontentsline{toc}{chapter}{Contents}
\rhead[\fancyplain{}{\bfseries Contents}]%
   {\fancyplain{}{\bfseries\thepage}}
\lhead[\fancyplain{}{\bfseries\thepage}]%
      {\fancyplain{}{\bfseries Contents}}
\tableofcontents
\chapter*{Summary}
\addcontentsline{toc}{chapter}{Summary}
\setcounter{page}{1}
\pagenumbering{arabic}
\lhead[\fancyplain{}{\bfseries\thepage}]%
      {\fancyplain{}{\bfseries Summary}}
\rhead[\fancyplain{}{\bfseries Summary}]%
      {\fancyplain{}{\bfseries\thepage}}
\noindent Interferometry has been widely used for astronomy in the last century at radio wavelengths; in the last decades, it has gained an important place also in the medium and near infrared wavelengths range. There are many differences between the two fields, due especially to the constraints posed by different behaviour of noise sources, flux intensities, instrumental limits at different wavelengths.\\

\noindent However, the potentiality of interferometry with respect to observation with a telescope, especially the higher angular resolution, has encouraged the astronomical community to concentrate big efforts on this subject, in terms of research, money and time. Nowadays, several interferometric arrays, working at infrared wavelengths, have been built all over the world, or are under construction. One of them is the VLTI, an ambitious project of the European Southern Observatory (ESO). From the beginning of this millennium, its first interferometric data in the near infrared have been recorded. \\

\noindent The aim of studying ever fainter sources, with increasing angular resolution, requires a great accuracy in the control of the interferometric process. In particular, if the correlation between the interfering beams is maintained high and stable, the integration time can be increased sensibly, still providing a meaningful integrated flux. Otherwise, the phase information is lost, e.g. due to atmospheric and environmental disturbances. For these reasons, ESO decided to equip the VLTI with fringe trackers, i.e. instruments able to sense the relative position of the interfering beams and to correct it to a nominal position.\\

\noindent In this framework, the Astronomical Observatory of Torino, part of the Italian National Institute of Astrophysics (INAF-OATo), has been involved from the late nineties in the design and development of a first fringe tracker, FINITO and then of a fringe sensor for the PRIMA instrument, i.e. the PRIMA FSU.\\
Designing and building a fringe sensor is a challenging task, with great difficulties. One of them is how to extract information about the fringes parameters from raw data. This is the subject of this thesis.\\

\noindent The two instruments differ for the opto-mechanical layout, for the choice between to temporal vs. spatial modulation, and for the quantity and type of data available. From the point of view of simulation, the fundamental difference is the model adopted for the interferogram pattern. For FINITO, it is very simple and based essentially on theoretical predictions. The algorithms for the optical path difference identification that we present in chap. \ref{chap:FINITO} use extensively this model. They are able to work with good results in the central area of the coherence length, but their principal limit is that they need to process a normalized interferometric signal. \\
In principle they can be modified in order to adapt to the inputs, but this leads to the necessity of changing the underlying model.\\

\noindent In chapter \ref{chap:PRIMA} we describe a more detailed model that is still based on the previous theoretical one, but it contains a number of parameters to be opportunely tuned to easily adapt to the current signal. The availability for the PRIMA FSU of a larger number of interferometric signal samples (twelve instead of two) allows the implementation of a weighted least squares fit of the measured data to the new model. The algorithm works well and fast, thanks to the use of tabulated functions for the reference signal template.\\
Of course, this is true if one assumption we make is true, i.e. that the template model is consistent with the current interferometric measurement conditions. Every discrepancy between the real signal and the tabulated template gives an error on the fringe parameters estimation.\\
Some of the model parameters can be assumed to remain stable or very-slowly changing during the life time of the instrument, or at most to require a check on the time scale of months. Other terms, indeed, needs to be properly determined before any observational night or even more often, such as source-dependent ones.
For this reason, we implement a set of calibration procedures. Their primary goal is to estimate the value of the critical parameters of the model, such as the overall instrumental transmission and phase functions, the visibility and the magnitude of the source. 
These values can then be fed into the template of the least square algorithms for the fringe location.\\
\noindent We tested it with both simulated and laboratory data, and we were able to reconstruct very well the spectral features of the measured signal. There are still some discrepancies between the intensities, especially for channels with lower flux. We can suspect that there is some phenomenon we do not include or properly model.\\

\noindent Such a doubt highlights the particular condition in which we are working. Our model is given by a deterministic equation, describing the optical power of two electromagnetic waves that interfere. In this sense, it is a {\it correlation} between the two waves. However, it does not describe what happens to the other term of the model, i.e. the noise.\\
These considerations have lead to the idea of analyzing interferometric data using classical instruments of the Statistical Sciences, such as analysis in the time and in the frequency domains. We have used VLTI data. Their peculiar nature has required efforts to tailor the statistical methods and to understand their results. For a particular problem, that is, the impact of estimation and subtraction of a signal trend on the estimated spectral density function, we give a mathematical derivation of the bias on the spectral density in appendix \ref{appendixA}.\\
Since the treated signals are not stationary, we analyze their variability, making use of statistical tests, in order to have a significance, and with regression analysis tool, trying to get out as much information as possible. All this statistical part is the subject of chapter \ref{chap:stat}.\\

\noindent This work has been developed in collaboration between the Astronomical Observatory of Torino and the Department of Mathematics of the University of Torino. From my point of view, the collaboration between mathematicians and astronomers, the exchange of knowledge, the needs to find a common statement of the problem, the twofold interpretation of each results have been a challenging opportunity for improvement. \\
Several of the results achieved in this framework deserve further investigation on a more complete set of experimental conditions, and many of the tools proposed could fruitfully be included in either on-line or off-line diagnostics and data analysis software for interferometric instruments. 

\chapter*{Plan of the thesis}
\addcontentsline{toc}{chapter}{Plan of the thesis}
\lhead[\fancyplain{}{\bfseries\thepage}]%
      {\fancyplain{}{\bfseries Plan of the thesis}}
\rhead[\fancyplain{}{\bfseries Plan of the thesis}]%
      {\fancyplain{}{\bfseries\thepage}}
This work can be divided in two parts.\\

\noindent In the first part, the attention is focused on OPD and GD algorithms that we have proposed for FINITO (chapter \ref{chap:FINITO}) and PRIMA FSU (chapter \ref{chap:PRIMA}). We describe the interferometric model on which they are tailored and their theoretical performances. For PRIMA, we also test the model reconstruction from laboratory data. \\
These algorithms reflect, in their increasing complexity, the increasing knowledge we gain on the manipulation of interferometric data, especially on the model. It must be noted that fringe tracking present different data analysis problems with respect to visibility extraction and interpretation. Indeed there are few good estimators for fringe position and their properties depend on instrumental features.
The analysis of this first part showed us that, even if algorithms have good performances, efforts are still needed to deepen the relation between source flux, noises and so on.\\

\noindent The second part (chapter \ref{chap:stat}) is devoted to the identification and test of possible statistical tools able to answer some of our questions. We started from this problem: is it possible to check the presence of a noise due to combination, and in the affermative case, how to estimate it? We use data from different instruments, more suited to our purposes.
There are many other questions, that can be posed, and lot of work has still to be done. The last paragraph of chapter \ref{chap:stat} will point out some of these questions.\\

\noindent In Appendix A we consider the power spectral density function (PSD) of beams characterized by a linear trend, evolving in time. We are interested on the effect on the frequency spectrum determined by removing an estimated linear trend. This problem, which arises while analyzing interferometric data in chapter \ref{chap:stat}, is discussed in the special case of a detrended process that results wide sense stationary process.\\

\noindent Finally, appendix \ref{appendixB} collects all graphics that were not inserted in the corresponding sections to avoid a too heavy presentation.

\chapter{Interferometry: from theory to fringe tracking}
\lhead[\fancyplain{}{\bfseries\thepage}]%
      {\fancyplain{}{\bfseries\rightmark}}
\rhead[\fancyplain{}{\bfseries\leftmark}]%
      {\fancyplain{}{\bfseries\thepage}}
\label{chap:intro}
In this chapter, we present the application of interferometry to astronomical observation: the historical development, the state of art, why it is useful and what are the goals. We describe the physical process, and how it can be modeled, and we justify the need for fringe tracking. Finally, we introduce the working environment of the thesis: the VLTI and its instruments.

\section{Introduction}
\label{sec:intro1}  

The sky and the stars have been the subject of enthusiastic research since the oldest records of human activity. In the last years, several missions have begun to scan the sky from the space, but ground observations are still the dominant means.\\
When looking at the sky from the ground, one of the great limits to accuracy and resolution is certainly the atmospheric turbulence. To face this problem, in the last century one branch of the technical development was devoted to the improvement of large telescope performance, using adaptive optics to flatten the incoming wavefront. Another branch that is becoming important in the last decades is interferometry in the near infrared part of the electromagnetic spectrum, after the success achieved in the second half of the XXth century by radio interferometry. Its most appealing feature is its {\it angular resolution}, i.e. the minimum distance between stellar sources at which the instrument is able to recognize the sources as distinct. \\ When observing with an array of telescopes in interferometric mode, the highest achievable angular resolution $\theta$ is limited by the longest {\it baseline}, i.e. the maximum separation between pairs of telescopes in the array:
\begin{equation}\label{eq:interf_angRes}
\theta \propto \frac{\lambda}{B},
\end{equation}
where $\lambda$ is the observing wavelength and $B$ is the baseline.
For a single telescope, the angular resolution is inversely proportional to the {\it diameter} of the collecting surface:
\begin{equation}\label{eq:tel_angRes}
\theta \propto \frac{\lambda}{D},
\end{equation}
where $\lambda$ is the observing wavelength, and $D$ is the aperture diameter.\\
Since the angular resolution is related to a ratio of lengths, it depends upon the order of magnitude of both terms.\\
When working with radio wavelengths (from millimiters to meters), to achieve a good angular resolution is necessary to have large baselines, up to kilometers, but the sensitivity is high. For short wavelengths (from a fraction to tenths of $\mu m$ for optical and infrared observation) the angular resolution is acceptable also for a single aperture, but not comparable with that achievable with baselines of hundreds of meters. Moreover, having two collecting areas should increase the limiting sensitivity.\\
Actually, bigger telescopes do not guarantee a better sensitivity, because the atmospheric turbulence degrades rapidly their performances. The coherence area $a_C$, i.e. the area where the wavefront can be considered flat, limits the angular resolution; it depends on the wavelength and the Fried parameter $r_0$:
\begin{equation}\label{eq:coh_area}
a_C \propto \frac{\lambda}{r_0} \; rad.
\end{equation}
The Fried parameter is a characteristic of the observing site and of the current observing conditions, and can be measured. For short wavelengths, the coherence area is small, and without correction of the wavefront, the big aperture is useless.\\
These considerations drove both the development of telescopes of increasing aperture, with sophisticated procedures for wavefront flattening (active and adaptive optics), and the construction of interferometers. Given that for technological issues the biggest apertures now achievable are of orders of $10$ m, interferometry has today an important place, and is the subject of a crucial research field.
\\

\noindent As an example, let us consider a large telescope with aperture $D=10 \; m$ observing in the near infrared range ($\lambda = 2 \; \mu m$). Its angular resolution will be\cite[p. 36]{Michelson}:
\begin{equation}\label{eq:ex_angRes}
\theta_{tel} = 1.22 \frac{\lambda}{D} = 2\cdot10^{-7}  \; rad \; = 0.05  \; arcsec
\end{equation}
If we are using an interferometer with baseline $B = 100 \; m$, the angular resolution will be\cite[p. 39]{Michelson}:
\begin{equation}\label{eq:ex_angRes-intIR}
\theta_{int} = \frac{\lambda}{2B} = 10^{-8}  \; rad \; = 0.002  \; arcsec
\end{equation}
To reach the same angular resolution with a radio wavelength, say $\lambda = 1$ mm, we would need a baseline of:
\begin{equation}\label{eq:ex_angRes-intRadio}
0.002 \; arcsec \sim \frac{1 \; mm}{2B} \rightarrow B = 5 \cdot 10^{6} \; m
\end{equation}
that is, 5000 km! This is the best currently achievable by VLBI, i.e. radiointerferometers using the whole Earth as observing baseline.
\\

\noindent There are different ways to produce interference images, and they will be briefly presented in par. \ref{sec:interf_principles}. All interferometers, however, share some common components: two or more telescopes, connected with the combination laboratory through a beam-transport system, a delay line, to compensate the optical path introduced by the observing geometry, a beam combiner and a detector. 
With every solution, however, the ambition of observing fainter and fainter sources imposes strong conditions on the instrument sensitivity and on the control of optical and instrumental variables. In particular, the optical path difference (hereafter, OPD) between the beams before the combination has a crucial role, because if it is maintained near zero, the integration time can be increased from a fraction of second to minutes, or hours, with a great benefits on the sensitivity.\\
This has led to the conception and construction of dedicated interferometer' subsystems, the {\it fringe sensors} and {\it fringe trackers}, able to measure and correct the optical path difference between the beams at nanometer level.

\section{History of interferometry}
\label{sec:interf_history}

The real angular size of stellar objects, compared to the observed one, and the way to measure them, have been central questions for astronomers from centuries. It was in 1801 that Thomas Young isolated the light interference in laboratory experiments: "homogeneous light, at certain equal distances in the direction of its motion, is possessed of opposite qualities, capable of neutralizing or destroying each other, and of extinguishing the light, where they happen to be united" (Young, 1804). It was William Herschel \cite{Herschel} that first observed, in 1805, that unfilled apertures allowed to obtain better angular resolution than the whole aperture, but we have to wait until 1835 for a theoretical explanation (Airy), and 1867 for the proposal of multiple apertures, with Fizeau\cite{Fizeau}. The first practical results came with St\'{e}fane \cite{stephane1874}, at the Observatoire de Marseille in 1874 and Hamy\cite{hamy1893} at the Observatoire de Paris in 1893. Albert Michelson in 1891\cite{Michelson1891} measured the angular diameter of the Jupiter satellites with great precision, and in 1921 the first stellar interferometer was mounted at the Mount Wilson Telescope in California (Michelson \& Pease\cite{Michelson1891}). The technological challenge involved was heavy: to limit mechanical instabilities, the baseline was of about fifteen meters; photometric evaluation on fringes pattern were made by human eye.\\
After the second world war, the higher resolution offered by an interferometer made this technique to become a standard in radio astronomy, that knew a great development, thanks to the relaxed tolerances offered by macroscopic wavelengths. We recall the work of Hambury Brown and Tiss, that in 1956 showed that photons coming from a common source are correlated, and this correlation survives the process of photoelectric emission on which detectors are based \cite{hambury56}. Baselines grew from meters to kilometers, imaging procedures through efficient sampling of sky regions became well established. A detailed exposition of a number of theoretical and practical aspects can be found in \cite{Thompson}.\\

\begin{figure}[!htb]
    \begin{center}
        \epsfig{figure=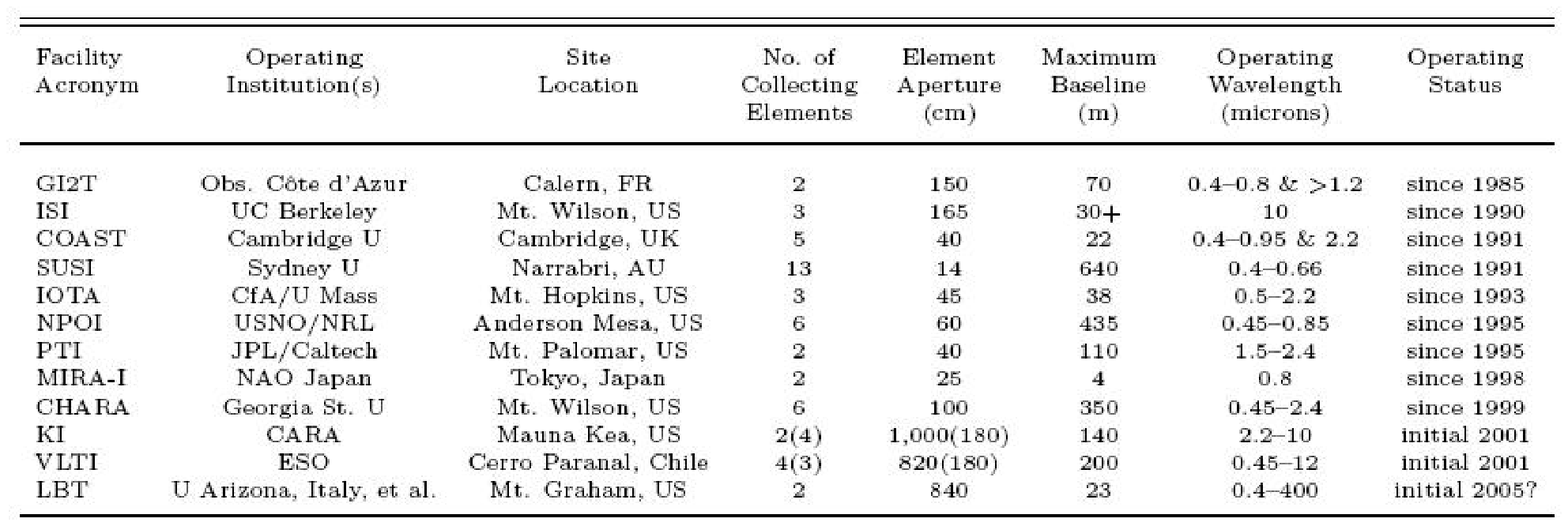,width=15cm}
        \caption{List of available optical/infrared ground telescopes arrays, taken from \cite{Michelson}.}
        \label{fig:interf_list}
    \end{center}
\end{figure}

\noindent When photon-counting detectors became available in the seventies, and laser control of the optical path went to the micrometers level, allowing baselines to grow to tens of meters, the radio techniques could be adapted to short wavelengths: modern optical and infrared interferometry was born.\\
Here we just mention, in chronological order, the pioneering work of Labeyrie, with the speckle interferometry, the phase tracking stellar interferometer from Shao \& Staelin in 1977, the first fringes seen at $2.2 \; \mu m$ by Di Benedetto \& Conti (1983), the first fully automated interferometer (the Mark III at Mount Wilson, Shao \& Colavita, 1988), the introduction of single mode optical fibers (e.g., Coud\'{e} du Foresto in 1992), the optical synthesis imaging from Cambridge, in 1996 with Baldwin. References can be found, e.g., in \cite[page 330]{Michelson}.\\
Figure \ref{fig:interf_list} shows a list of available optical/IR ground facilities, with the number and size of apertures, the maximum baseline, and their state of development. This table has been taken from \cite{Michelson}.
The most ambitious projects are the Keck Interferometer (KI) in Hawaii and the Very Large Telescope Interferometer (VLTI) at Cerro Paranal, in Chile. References describing these arrays can be found in literature; for the KI, see Colavita \& Wizinowich\cite{colavita00}, for the VLTI, see Glindemann et al.\cite{glindemann00}

\noindent The interferometry research is an open field: there are still unexplored issues, especially technological problems, highlighted by the analysis of the first a\-vai\-lable optical and IR interferometric data. An overview of technological matters and scientific goals of optical interferometry can be found in the review of Monnier, 2003 \cite{Monnier03}.\\

\subsubsection{Interferometry potentialities}
\label{subsubsec:astro_potentiality}

The potentialities of interferometry are of course dependent upon instrumental limitations, such as maximum baseline length for the angular resolution, number of combination for a good sky coverage, flux coupling between apertures for the limiting sensitivity. Data with these good properties could assure the validation of theoretical models.
Some of the most appealing goals are, for example, the study of close binary systems, for a precise determination of stellar masses, precise measurements of stellar diameters and their changes, for pulsational models, large surveys of sky portions at extreme magnitudes (toward twenty!). We also mention the possibility of using interferometry for differential measures, thanks to instruments (such as PRIMA or AMBER at the VLTI) able to simultaneously observe in two different directions: for example, for validation of the range of atmospheric models.

\section{Principles of Interferometry}
\label{sec:interf_principles}

To describe the interference process we can consider the classic experiment of Young.\\
The light from a point source passes into two pinholes at a certain distance. The two resulting beams are then combined and imaged on a surface. Due to the wave nature of the light, the electromagnetic fields interfere alternatively constructively and destructively, depending on the difference of the optical path they have covered. The figure of interference shows black area alternate with bright ones.\\

\noindent There are fundamentally two types of beam combination, requiring different mechanical and optical solution for the superposition of beams, but equivalent in ideal conditions: the {\it image-plane} and the {\it pupil-plane} interferometry. The main difference is {\it where} the beam combination takes place. Their properties make them best suited for intermediate resolution and for high resolution, respectively (see, e.g., \cite[ch. 3]{Michelson} and \cite[ch. 4]{Michelson}).\\
In the image-plane method, each beam is focused on the image plane (a detector, for example), and the images are superposed. This is called Fizeau interferometer. \\
In the pupil-plane method, the beams are superposed {\it before} being focused on the image plane, in a beam combiner, that can be a glass, or an optical fiber. The beams are then separated again and focused separately on the detector. This is called Michelson interferometer.\\ 
We are interested in instruments based on the latter method.
\\

\noindent In the following paragraphs, we recall the fundamental principles of interferometry, following the notation of \cite{Michelson}.

\subsection{Monochromatic interference}
\label{subsec:mono_interf}

\noindent The wave nature of stellar beams, written as electromagnetic fields, gives a simple description of the physical process.
\\
\noindent Let $\phi(\bar{x},t)$ be an electromagnetic monochromatic wave at frequency $\nu = \frac{c}{\lambda}$, where $c$ is the light speed, traveling in the space in the direction $\bar{n}$. We can write it as:
\begin{equation}\label{eq:monowave}
\phi(\bar{x},t) \sim A e^{i(\frac{2 \pi \nu}{c}\bar{n}\bar{x} - 2 \pi \nu t)} = A e^{i(\bar{k}\bar{x} - \omega t)}
\end{equation}
where
\begin{eqnarray}\label{eq:freq_descrip}
\nonumber \bar{k} &=& k \bar{n}\\
\nonumber k &=& \frac{2 \pi \nu}{c} = \frac{2 \pi}{\lambda}\\
\omega &=& 2 \pi \nu
\end{eqnarray}
$\omega$ is the angular frequency\footnote{We are neglecting here the velocity change of the beam due to the travel into a medium, as the atmosphere, which modifies the beam wavelength:
\[v = \frac{c}{n} = \frac{\lambda \nu}{n} = {\lambda_n \nu}\]
where $n$ is the refraction index of the medium. Moreover, we are neglecting the atmospheric turbulence too, which adds a random optical path to the beam.}.
\\

\begin{figure}[!htb]
    \begin{center}
        \epsfig{figure=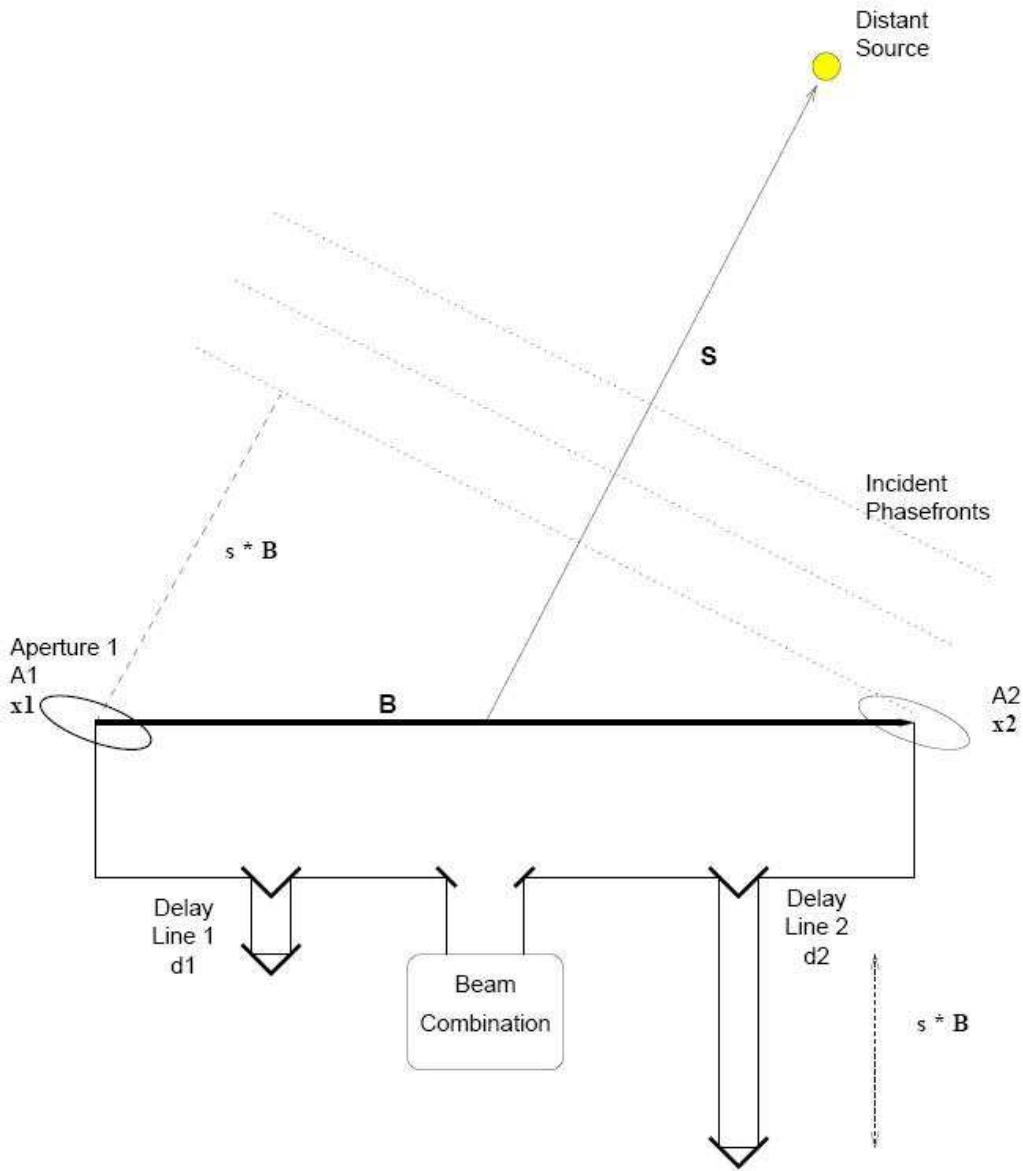,width=14cm}
        \caption{Idealized interferometer scheme.}
        \label{fig:interferometer}
    \end{center}
\end{figure}

\noindent Now let us consider an idealized interferometer, as shown in figure \ref{fig:interferometer}. Two telescopes are given, separated by a baseline $\bar{B}$, and they are both pointing to a distant point source located at a distance $\bar{S}$ from the center of the baseline. Thus, the pointing direction is given by the versor $\frac{\bar{S}}{|\bar{S}|} = \bar{n}$. The source is taken sufficiently distant so that the wavefront can be considered flat.\\

\noindent When the light beam traveling from the distant source arrives at telescope 1, at position $\bar{x}_1$, its equation will be, following eq. \ref{eq:monowave}:
\begin{equation}\label{eq:monowave1}
\phi_1(\bar{x}_1,t) \sim A e^{i(\bar{k}\bar{x_1} - \omega t)} = A e^{-i(k\bar{s}\bar{x_1} + \omega t)}
\end{equation}
and equivalently will be the beam at telescope 2 at position $\bar{x}_2$:
\begin{equation}\label{eq:monowave2}
\phi_2(\bar{x}_2,t) \sim A e^{i(k\bar{s}\bar{x_2} - \omega t)} = A e^{-ik\bar{s}(\bar{x_1} + \bar{B}) - i \omega t} = A e^{-i(k\bar{s}\bar{x_1} + \omega t)}e^{-ik\bar{s}\bar{B}}
\end{equation}
where we have used the fact that the baseline $\bar{B}$ is given by $\bar{B} = \bar{x}_2 - \bar{x}_1$. Comparing eqs. \ref{eq:monowave1} and \ref{eq:monowave2}, we can notice that the only difference between the two waves is given by $e^{-ik\bar{s}\bar{B}}$, a term that depends just on the geometry of the system including the interferometer ($\bar{B}$) and the observed source ($\bar{s}$).
\\

\noindent After the collection at the telescope level, the two beams $\phi_1$ and $\phi_2$ travel into the interferometer arms to reach a common point where they will interfere, covering a distance of $d_1$ and $d_2$, respectively, as shown in fig. \ref{fig:interferometer}. This additional optical path increments the total optical path covered by the beams from the stellar source:
\begin{eqnarray}\label{eq:monowave_bis}
\nonumber \phi_1(\bar{d_1},t) &\sim&  e^{ik d_1 - i\omega t},\\
\phi_2(\bar{B}+d_2,t) &\sim&  e^{-ik \bar{s}\bar{B} + ikd_2 - i\omega t}
\end{eqnarray}
where we have neglected, without loss of generality, the common component $A e^{-ik \bar{s}\bar{x_1}}$. The interference process adds the two waves:
\begin{equation}\label{eq:mono_interf}
\phi(t) = \phi_1(t) + \phi_2(t) \sim e^{- i\omega t} (e^{ik d_1} + e^{-ik \bar{s}\bar{B} + ikd_2})
\end{equation}
\\
\noindent The detector is sensitive to the optical power\footnote{The energy of a beam crossing a unitary area perpendicular to the propagation direction is proportional to the temporal average of the square of the electric field:
\[<\phi^2> = \lim_{T->+\infty} \frac{1}{2T} \int_{-T}^{T} \phi^*(t) \phi(t) = A^2. \]}  $P$ of the resulting beam, defined as $P=\phi^*\cdot\phi$. We finally get the optical power over a unit time integration:
\begin{eqnarray}\label{eq:mono_opticalPower}
\nonumber P \propto \phi(t)^* \phi(t) &=& e^{- i\omega t} e^{i\omega t}(e^{ik d_1} + e^{-ik \bar{s}\bar{B} + ikd_2})(e^{-ik d_1} + e^{ik \bar{s}\bar{B} - ikd_2}) = \\
\nonumber &=& 1 + e^{-ik d_1 - ik \bar{s}\bar{B} + ikd_2} + e^{ik d_1 + ik \bar{s}\bar{B} - ikd_2} + 1 =  \\
&=& 2 [1 + cos k(\bar{s}\bar{B} + d_1 - d_2)] = 2 [1 + cos kD].
\end{eqnarray}
where we have posed $D =  \bar{s}\bar{B} + d_1 - d_2$.\\
Looking at this equation, we can see that the optical power is given by an offset and a sinusoidal wave with frequency $k = 2 \pi / \lambda$ over the spatial variable $D$, called the {\it optical path difference}, or OPD. It will have a crucial importance all over the work of this thesis.
\\

\begin{figure}[!htb]
    \begin{center}
        \epsfig{figure=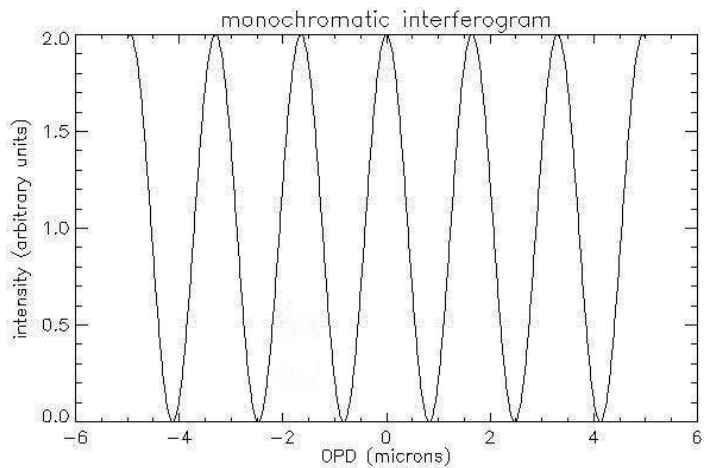,width=10cm}
        \caption{Idealized interferometer scheme.}
        \label{fig:mono_fringes}
    \end{center}
\end{figure}

\noindent In an ideal but concrete example, if each telescope has a collecting area of $A$ $[m^2]$, if the source emits at wavelength $\lambda$ a constant flux $F$ $[{\rm photons} \cdot ({\rm time} \; {\rm unit})^{-1} \cdot m^{-2}]$, in a time unit we get an optical power:
\begin{equation}\label{eq:mono_optPow_bis}
P = 2AF[1 + cos kD]
\end{equation}
A representation of this ideal function is shown in figure \ref{fig:mono_fringes}. The optical power obscillates infinitely over the OPD variable. Each period is called {\it interfe\-ro\-metric fringe}. Two consequent fringes are separated by $\frac{\lambda}{B}$. This has a physical explanation, too. A difference $\Delta \bar{s}$ of the observation direction can be interpreted as an angle in the sky (in radians), so two different fringe peaks projected in the sky are separated by an angle given by \cite[page 12]{Michelson}:
\begin{equation}\label{eq:mono_angleSepar}
\Delta\bar{s} = \frac{\lambda}{B}.
\end{equation}

\subsection{Polychromatic source}
\label{subsec:poly_interf}
Let now allow the source to be polychromatic. We report the detailed computation of the interferogram pattern in this ideal case because we will use it in the future work.\\
We can consider the source flux as the harmonic composition of components at different frequencies, as photons do not interfere with each other.\\
Let us suppose that the source flux is constant over a range of frequencies $\nu \in [\nu_1, \nu_2]$: $F_{\nu} = F_0$. Also the interferometer will have a finite spectral response $\eta(\nu)$. The total optical power will be, modifying eq. \ref{eq:mono_optPow_bis}:
\begin{equation}\label{eq:poly_optPow}
P = 2 \int A F_{\nu} \eta(\nu) [1 + cos kD] d\nu
\end{equation}

\noindent In the ideal case, the interferometer response is a perfect bandpass filter, i.e. a rectangle over the band $[\nu_1, \nu_2]$, centered in $\nu_0 = \frac{\nu_1+\nu_2}{2}$ and with length $\Delta\nu = \nu_2 - \nu_1$, with constant value $\eta_0$:
\begin{equation}\label{eq:poly_optPow_bis}
P = 2A \int_{\nu_0-\Delta\nu / 2}^{\nu_0+\Delta\nu / 2} F_{\nu} \eta_0 [1 + cos kD] d\nu
\end{equation}
Remembering that $k=2\pi\nu / c$, the argument of the cosine becomes:
\begin{equation}\label{eq:cosine_arg}
k D = \frac{2 \pi \nu}{c} D = 2 \pi \nu \frac{D}{c} = 2 \pi \nu \tau
\end{equation}
where $\tau$ has the dimension of time. Substituting in eq. \ref{eq:poly_optPow_bis}, we obtain:
\begin{eqnarray}\label{eq:poly_optPow_ter}
\nonumber P &=& 2A \int_{\nu_0-\Delta\nu / 2}^{\nu_0+\Delta\nu / 2} F_{\nu} \eta_0 [1 + cos (2 \pi \nu \tau)] d\nu = \\
\nonumber &=& 2A F_0 \eta_0 [\nu + \frac{sin (2 \pi \nu \tau)}{2 \pi \tau}]_{\nu_0-\Delta\nu / 2}^{\nu_0+\Delta\nu / 2} = \\
\nonumber &=& 2A F_0 \eta_0 [\Delta\nu + 2 \frac{sin (\pi \tau \Delta\nu)}{2 \pi \tau}cos(2 \pi \tau \nu_0)] = \\
&=& 2A F_0 \eta_0 \Delta\nu [1 + \frac{sin (\pi \tau \Delta\nu)}{\pi \Delta\nu \tau}cos(2 \pi \tau \nu_0)]
\end{eqnarray}

\noindent We have now to consider again the relation between $\tau$ and the OPD $D$ of eq. \ref{eq:cosine_arg}. We obtain:
\begin{eqnarray}\label{eq:poly_sinc_arg}
\nonumber \pi \tau \Delta\nu &=& \pi \frac{D}{c} \Delta\nu = \pi \frac{D}{c} (\nu_2 - \nu_1) = \pi \frac{D}{c} (\frac{c}{\lambda_2} - \frac{c}{\lambda_1}) = \pi \frac{D}{c} c \frac{\lambda_1 - \lambda_2}{\lambda_1 \lambda_2} \sim \\
&\sim& \pi D \frac{\Delta\lambda}{\lambda_0^2} = \pi \frac{D}{L_C}
\end{eqnarray}
where we have used the relation $\nu\lambda = c$, where $\lambda_0 = c / \nu_0$ is the {\it central wavelength} of the wavelength range, and where $L_C = \frac{\Delta\lambda}{\lambda_0^2}$ is the {\it coherence length}, where the fringes are formed. In a similar manner, for the cosine argument we find:
\begin{equation}\label{eq:poly_cos_arg}
2 \pi \tau \nu_0 = 2 \pi \frac{D}{c} \frac{c}{\lambda_0} = \frac{2 \pi D}{\lambda_0} = k_0 D.
\end{equation}

\begin{figure*}[!htb]
    \begin{center}
        \epsfig{figure=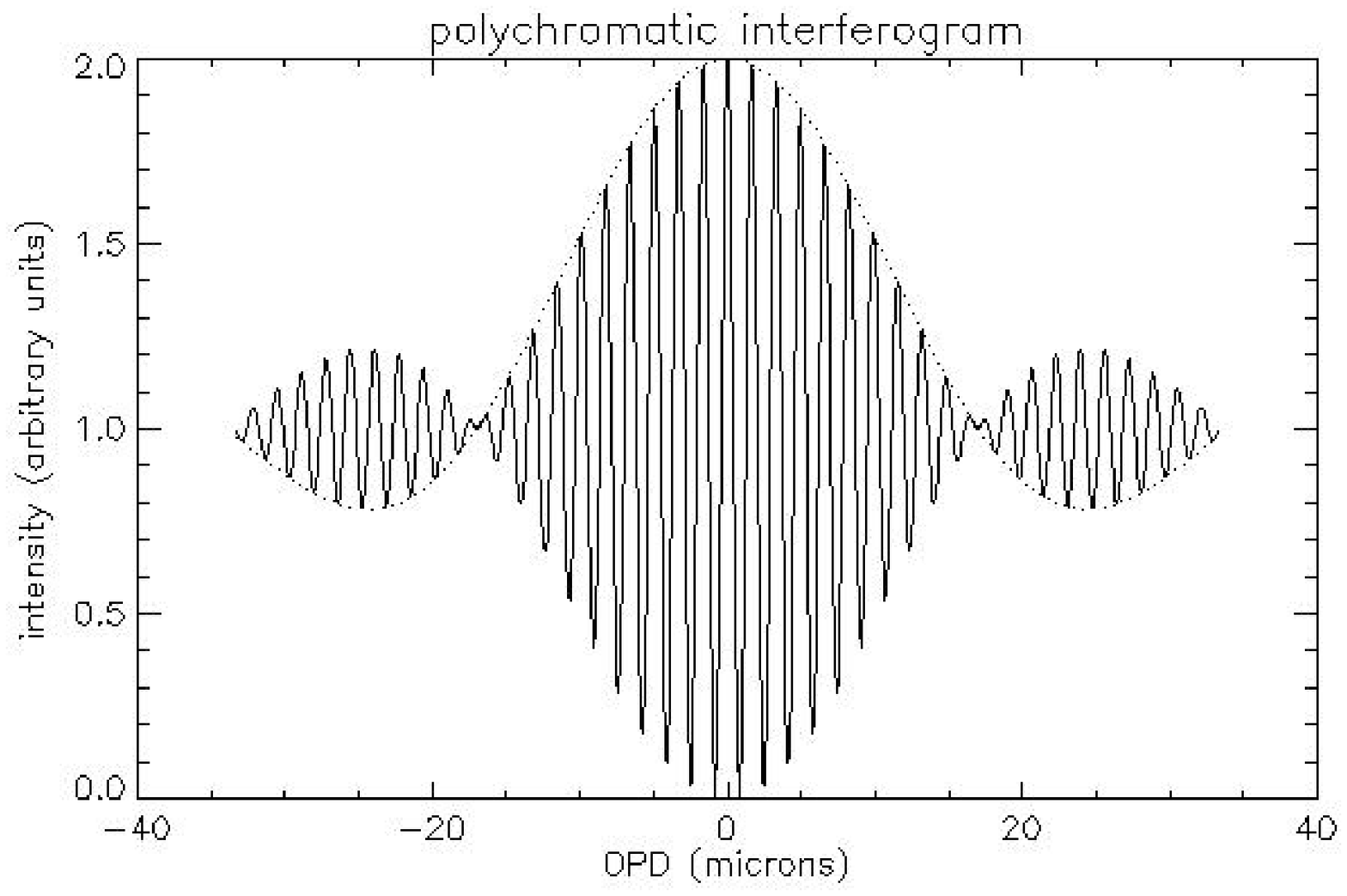,width=7.5cm}
        \epsfig{figure=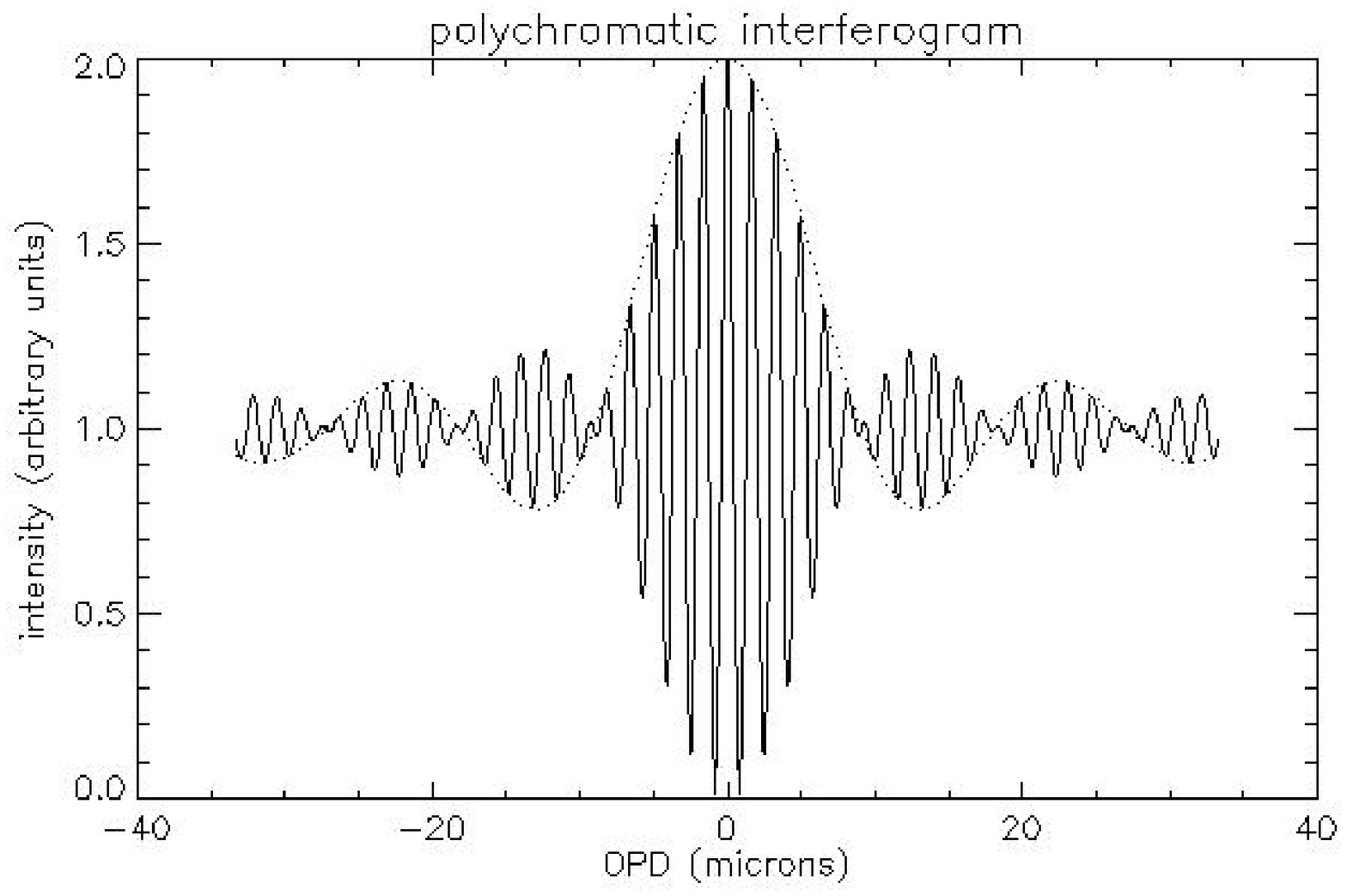,width=7.5cm}
        \caption{Polychromatic interferogram in H band. The dotted line is the envelope, i.e. the sinc function. The central wavelength is $\lambda = 1.65 \mu m$ for both graphics, while the waveband changes from $0.16 \mu m$ (left) to $0.30 \mu m$ (right), giving a different coherence length: $17.016 \mu m$ and $9.075 \mu m$, respectively.}
        \label{fig:poly_interf}
    \end{center}
\end{figure*}

\noindent Substituting into eq. \ref{eq:poly_optPow_ter} we finally get:
\begin{eqnarray}\label{eq:poly_optPow_quater}
\nonumber 2A F_0 \eta_0 \Delta\nu [1 + \frac{sin (\pi \tau \Delta\nu)}{\pi \Delta\nu \tau}cos(2 \pi \tau \nu_0)] = \\
= 2A F_0 \eta_0 \Delta\nu [1 + sinc \frac{\pi D}{L_C} cos(\frac{2 \pi D}{\lambda_0})]
\end{eqnarray}
where the function $sinc(x) = \frac{sin(x)}{x}$ is the `sinus cardinalis', or cardinal sine. Looking at the last equation, we can recognize two different modulation patterns. One is the cardinal sine: it has a modulation frequency that depends on the filter range $\Delta\nu$, here hidden in the coherence length. We remark that it comes from the Fourier transform of the rectangular bandpass; the variables $\nu$ and $\lambda$ form a Fourier pair, with the normalization factor $c$. The other pattern is the cosine, modulated at the frequency $k_0$, correspondent to the central wavelength $\lambda_0$.
Figure \ref{fig:poly_interf} shows a typical interferogram pattern: fringes are modulated over the optical path difference D, and the sinc creates an {\it envelope} that smooths the fringes amplitude as far as D is far from the zero OPD (ZOPD), where we have the maximum of the envelope.

\noindent Geometrically, if the OPD is zero the two beams overlap perfectly even at different wavelengths. If the OPD is not adjusted to be zero, beams at different wavelengths will have their maximum coherence at different position, causing a decrease in the interference amplitude. Figure \ref{fig:poly_interf} can help visualizing this concept.
\\

\noindent If the interferometer bandpass filter was described by a different function, the Fourier Transform of this new filter would modulate the envelope pattern.

\subsection{The complex visibility}
\label{subsec:visibility}
Information on the observed source have to be extracted from the interferometer output. A fundamental Fourier transform relationship holds between the optical power measured by the interferometer and the source brightness function. This relation is known as the van Cittert-Zernike theorem.
To describe intuitively the nature of this relation, let us come back to the monochromatic description, and let us introduce a spatial variable on the sky, say the solid angle $\Omega$, with reference to the vector of observing direction $\bar{s}_0$. The angle $\Omega$ is moving on the star surface S.\\
\begin{figure}[!htb]
    \begin{center}
        \epsfig{figure=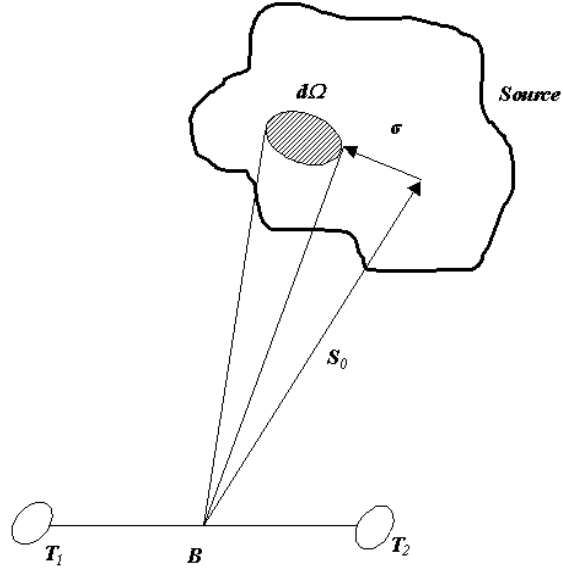,width=7.5cm}
        \caption{Area $d\Omega$ on the source S.}
        \label{fig:ext_source}
    \end{center}
\end{figure}

\noindent Let us assume that the instrument response is described by a function $\eta(\bar{s}_0, \bar{\sigma})$ and the flux intensity by $F(\bar{s}_0,\bar{\sigma})$. As $\bar{\sigma}$ moves on the star surface, it identifies a different region on the star. A portion of source surface with dimension $d\Omega$ (see fig. \ref{fig:ext_source}), related to the observing direction $\bar{s}_0 + \bar{\sigma}$, sufficiently small to ensure that $\eta(\bar{s}_0, \bar{\sigma})$ and $F(\bar{s}_0, \bar{\sigma})$ can be considered constant over $d\Omega$, will generate an optical power $P_{d\Omega}$:
\begin{equation}\label{eq:smallPort}
P_{d\Omega} = \eta(\bar{s}_0, \bar{\sigma})F(\bar{s}_0, \bar{\sigma})[1 + cos k(\bar{B}\cdot (\bar{s}_0 + \bar{\sigma}))] d\Omega
\end{equation}
If we further assume that all surface portions on the source add up incoherently, we can write the interferometer output as an integration, putting together eq. \ref{eq:poly_optPow} with \ref{eq:smallPort} as:
\begin{eqnarray}\label{eq:visib_deriv}
\nonumber P &=& 2 \int_S \eta(\bar{s}_0, \bar{\sigma}) F(\bar{s}_0, \bar{\sigma}) [1 + cos k(\bar{B}\cdot (\bar{s}_0 + \bar{\sigma}))] d\Omega = \\
\nonumber &=& 2 \int_S \eta(\bar{s}_0, \bar{\sigma}) F(\bar{s}_0, \bar{\sigma}) d\Omega + 2 \int_S \eta(\bar{s}_0, \bar{\sigma}) F(\bar{s}_0, \bar{\sigma}) cos k(\bar{B}\cdot (\bar{s}_0 + \bar{\sigma})) d\Omega = \\
\nonumber &=& P_0 + 2 cos (k \bar{B}\cdot \bar{s}_0) \int_S \eta(\bar{s}_0, \bar{\sigma}) F(\bar{s}_0, \bar{\sigma}) cos (k \bar{B}\cdot \bar{\sigma}) d\Omega \\
&-& 2 sin (k \bar{B}\cdot \bar{s}_0) \int_S \eta(\bar{s}_0, \bar{\sigma}) F(\bar{s}_0, \bar{\sigma}) sin (k \bar{B}\cdot \bar{\sigma}) d\Omega =\\
\nonumber &=& P_0 + 2 cos (k \bar{B}\cdot \bar{s}_0) Re\{V\} + sin(k \bar{B}\cdot \bar{s}_0) Im\{V\} = P_0 + Re\{Ve^{ik \bar{B}\cdot \bar{s}_0}\}
\end{eqnarray}
where the function $V$ is the {\it complex visibility} of the brightness distribution $F$, referred to the phase reference $\bar{s}_0$, and is defined by:
\begin{equation}\label{eq:visib}
V = V(k, \bar{B}) = \int_S \eta(\bar{s}_0, \bar{\sigma}) F(\bar{s}_0, \bar{\sigma}) e^{ik\bar{B}\cdot \bar{\sigma}} d\Omega
\end{equation}

\noindent This relation is known as the van Cittert-Zernike theorem. The complex visibility is function of the observing wavelength, through $k$, and of the baseline $\bar{B}$. Remembering that $\frac{\lambda}{\bar{B}}$ can be considered angles in the sky, noting as $B_x$ and $B_y$ the projection of the baseline over the ground coordinate system, we define {\it spatial frequencies} the coordinates $(u, v)$ defined as:
\begin{equation}
u = \frac{B_x}{\lambda}, \;\;\; v = \frac{B_y}{\lambda}
\end{equation}
With a good coverage of the sky in the $(u,v)$ coordinates it is possible to invert the complex visibility to obtain a {\it dirty brightness distribution}
, so called because it is biased by the sampling function of the sky, i.e. the $(u,v)$ coverage.
\\
There is a huge effort in the domain of cleaning the brightness distribution and of reconstructing images from it.

\subsubsection{Visibility properties}
\label{subsubsec:visib_prop}
The complex visibility has several properties, directly derived from its definition. Being $A$ and $F$ real, for the visibility holds the following relation:
\begin{equation}
V(-u, -v) = V^*(u,v)
\end{equation}
If we add a delay $\bar{\sigma}$ to the reference phase $\bar{s}_0$, we have seen in eq. \ref{eq:visib_deriv} that this is equivalent to adding a delay in the phase domain:
\begin{equation}
P_0 + Re\{Ve^{ik \bar{B}\cdot \bar{s}_0}\}
\end{equation}
so the fringes are translated in the optical path difference space.

It is a dimensional quantity. It is common to use a normalized visibility, first introduced by Michelson, that compares the intensities of the dark and the bright areas of the interferogram:
\begin{equation}
V_M = \frac{P_{max} - P_{min}}{P_{max} + P_{min}}
\end{equation}
This is a non dimensional quantity, varying in the interval $[0,1]$. It goes to zero when the intensity does not vary along the OPD ($P_{max} = P_{min}$), and is $1$ when $P_{min} = 0$.
It can be shown that \cite[page 20]{Michelson}
\begin{equation}
V_M = |V|.
\end{equation}

\noindent We can resume saying that the visibility is a complex quantity; its modulus is the fringe contrast, while the phase contains information about the shift of the fringes from the zero OPD, i.e. from the reference position.

\section{Fringe Tracking}
\label{sec:ft}

We have seen in the previous paragraph how it is possible to find a relationship between the brightness distribution of the observed source and the optical power recorded by an interferometer. Moreover, we have mentioned the potentiality of interferometry with respect to traditional monolithic telescopes.\\ 
We have pointed out, however, that a delay in the observational direction introduces a delay in the phase domain. The same applies when a delay is introduced in one arm of the interferometer. In both cases, the result is an unbalance of the optical path of the two beams before the combination.
\\

\noindent We have mentioned in section \ref{sec:interf_history}, that a differential optical path between the beams impedes the fringes at different wavelengths to overlap perfectly: the interference is not maximal. The principal responsible of this phenomenon is the {\it atmospheric turbulence}, that forces the beams to do additional optical path before the collection at the telescopes, in a random way, different at all wavelengths. Also instruments can add an extra path, usually static or slowly varying. The consequences on the visibility depends from several factors: the estimator used for the measurements of the visibility itself, the kind of OPD (a simple shift, or a OPD with a velocity and an acceleration). Detailed studies on the subject can be found in literature (see \cite{Kervella04} and references therein).
\\

\noindent The visibility degradation, added to the possibility of increasing the integration time well beyond the coherence time\footnote{The wavefront can be considered flat over a time interval, called coherence time, which depends on the observing wavelength, and it is characteristic of the site} if the beams are maintained aligned, explains why the astronomic community judged necessary the introduction of dedicated instruments able to measure and correct the OPD in real time, equa\-li\-zing the paths of the beams. These instruments are called the {\it fringe sensors}, used in the fringe tracking closed control loop. Their role is to follow the OPD very quickly, well beyond the atmospheric change rate, and to correct it. The challenge is great: the operations must be very agile and accurate.
\\

\noindent All big interferometers have been equipped with a fringe tracking system. The Keck Interferometer was equipped with FATCAT (see, e.g., Vasisht\cite{Vasisht03} and re\-fe\-rences therein). The VLTI was first equipped with FINITO; the VLTI PRIMA instrument, dedicated to astrometry, has its own fringe sensor, PRIMA FSU.\\

\noindent For the subject of this thesis, we will focus onto VLTI instrumentation and data reduction.

\subsection{Fringe Tracking at the VLTI}
\label{subsec:VLTIft}

The ESO VLTI (European Organisation for Astronomical Research in the Southern Hemisphere Very Large Telescope Interferometer, www.eso.org) has been designed to combine up to four Unit Telescope (UT), with apertures of $8$ meters, and several Auxiliary Telescopes (AT) of 1.8 meters diameter. Each UT can work as a stand-alone conventional telescope, but the array of telescopes can guarantee a maximum baseline of $\sim 200$ m for ATs and $\sim 100$ m for UTs.
There is a first generation of instruments working in the near infrared, and dedicated mainly to the visibility measurement: VINCI was the commissioning instrument\cite{coudedeforesto97}\cite{Kervella00}, responsible for testing the working performance of the VLTI, working in K band ($[2.0 - 2.5] \mu m$) from the beginning of this century; AMBER\cite{petrov07}, the first attempt to combine up to three beams, working in J ($[1.1 - 1.3]\; \mu m$), H ($[1.4 - 1.9]\; \mu m$) or K band, whose fringes were first recorded in 2004, MIDI\cite{leinert98}, dedicated to thermal infrared (N band, $10 \mu m$).
All these instruments have a temporal resolution worse than the average coherence time, so they can not do fringe tracking alone.\\
The first prototype for a dedicated fringe tracker was developed by the Observatoire de la C\^{o}te d'Azur (OCA, Nice). A description can be found in \cite{GayRabbia94}, and the working principle will be resumed in chapter \ref{chap:FINITO}. From this prototype, the Astronomical Observatory of Torino, in collaboration with ESO, developed the first fringe sensor for the VLTI: FINITO (Fringe tracking Instrument of NIce and TOrino). It was constructed in Torino, and delivered to ESO in 2004. During the commissioning the fringe tracking loop could not be closed, meaning that the instrument wasn't able to check the actual OPD value and to correct it. It was not an instrument fault, at the contrary it was useful to identify several problems of the overall VLTI system\cite{Bonnet06}: the delay lines had a residual alignment error, the Adaptive Optics (AO) left an internal turbulence, and the source flux was subject to intensity fluctuations due to saturations of the mirrors of AO, there were vibrations that induced a distortion of the mo\-du\-la\-ted fringe pattern. The fringe-tracking loop was finally closed in 2006, and now FINITO is routinely used in association with other scientific instruments, such as AMBER.
\\

\noindent In the same time frame, the OATo was involved in the implementation of a fringe sensor (FSU) facility for a second generation instrument, PRIMA\cite{DelPlancke00} (Phase Referenced Imaging and Microarcsecond Astrometry). PRIMA aims not only at visibility estimation, i.e. the modulus of the complex visibility as we have seen before, but also to the {\it phase} of the complex visibility, in order to be able to reconstruct images from interferometric data. Moreover, it has two separate FSUs, to simultaneously track the scientific object, potentially a faint source, and the reference bright star.
\\

\noindent The difficulties encountered by FINITO, which suffered the lack of information on the environmental condition and on the received flux features, forced to elaborate a more sophisticated interferometric model and to add a number of instrumental degrees of freedom in order to be able to properly calibrate each FSU.

\subsection{Interferometric working condition at VLTI}
\label{subsec:VLTIworking_cond}

The atmospheric turbulence induces disturbance on OPD and intensity fluctuations. Less important, but not negligible are the perturbations caused by instruments and by the environmental conditions. These effects must be rejected by the interferometric system in order to work properly.\\
A description of the influence of the main subsystem of an interferometer and of the turbulence model can be found in \cite{Gai-SPIE04}. Here we recall some of the principal aspects.

\subsubsection{Atmospheric turbulence}
\label{subsubsec:atm_turbulence}

Lot of researches have been carried on for the description of the atmospheric turbulence, and it still is an open field. For our purposes, we use the model developed for the ESO VLTI, described in \cite {Gennai01} and based on the Kolmogorov model.\\
Above a cut-off frequency $\nu_c = 0.22 v/B$, where $v$ is the wind speed (meteorological conditions) and $B$ is the baseline (observing configuration), the power spectral density of the disturbance on the OPD can be approximated by a power-law formula:
\begin{equation}\label{eq:atm_spectrum}
PSD_{OPD}(\nu) = S_0 \cdot \lambda_0^2 \cdot r_0^{-5/3} \cdot v^{5/3} \cdot \nu^{-8/3},
\end{equation}
where $S_0$ can be measured and is equal to $0.0039$ in standard VLT condition, $\lambda_0$ is the central wavelength and $r_0$ is the Fried parameter. Below the cut-off frequency, the spectrum can be simply approximated by $\nu^{-2/3}$, so it is independent on the baseline and on the wavelength, and for very low frequencies the slope become positive.


\subsubsection{Instrumental issues}
\label{subsubsec:instr_issues}

The atmospheric coherence length depends on the wavelength, it ranges from few milliseconds in the visible range to few ten ms in the near infrared and few hundred ms in the thermal (medium) infrared.\\
Since the collecting area of each single telescope is larger than the atmospheric coherence length, an adaptive optic system is required. Its role is to flatten the incoming wavefront, correcting low frequencies turbulence. It is essential, especially for large apertures. We can say that interferometric performances depend on those of AO.\\
However, the AO correction is done on each single telescope, which is affected by a different wavefront corrugation with respect to the other apertures, cau\-sing a residual OPD to be introduced in the system. For baselines longer than few meters it is necessary to track this OPD disturbance and to correct it.\\

\noindent After the collection, the beams are sent into the delay lines (DL), that have a double role: first, they carry the beams from individual telescopes to the combination laboratory. Second, they correct the OPD sensed by the fringe sensor, thanks to the OPD controller, that receives information from the sensor and sends them to the delay lines, according to a specified control algorithm.\\
In the VLTI, the DL are not evacuated, and this fact leaves a mismatch between the zero OPD, where interference at each wavelength reaches its maximum, and the overall group delay\footnote{The group delay (GD) is a measure of the optical path difference that takes into account the dependence of the refractive index from the wavelength. Beams can travel through different paths in air, with different refraction indexes. See, for example, P. Lawson in \cite[p. 115]{Michelson}} caused by the wavelength range. (longitudinal dispersion).
\\

\noindent The fluctuations caused by instruments (delay lines, residuals from AO, from the electronic boxes of other instruments before the fringe tracker and of the fringe tracker itself) add up to the atmospheric turbulence. A metrology sy\-stem can be foreseen, at the FT level, to check the internal optical path, and its information can be used locally, to stabilize the internal path, or sent back to the OPD controller.
\\

\noindent At the end of the combination chain, a detector records the interferometric and, eventually, photometric intensities. Modern detectors can reach high recording rates with an acceptable noise level. Both FINITO and PRIMA FSU adopted integrating detectors: the integration time can be set by the user, together with the read-out mode. Thanks to the interposition of a dispersing prism, the PRIMA FSU detectors record also different spectral bands in contiguous pixels. After the data saving, the optical path difference can be evaluated, using a proper combination of the interferometer output. \\

\noindent The role of the algorithms responsible of this evaluation is of course very important. They must have good performances both in accuracy and in velocity. In the ideal working condition, the limiting sampling rate should be due to the finite photon flux from sources rather than to instrumental performances.

\chapter{Location algorithms for the VLTI FINITO fringe tracker}
\lhead[\fancyplain{}{\bfseries\thepage}]%
      {\fancyplain{}{\bfseries VLTI FINITO}}
\rhead[\fancyplain{}{\bfseries VLTI FINITO}]%
      {\fancyplain{}{\bfseries\thepage}}
\label{chap:FINITO}
\noindent In this chapter, we will focus on real-time fringe tracking algorithms that are suitable in the framework of temporal modulation. Starting from a general approach due to J. W. Goodman (\cite{Goodman}), based on the Fourier Transform of the interferometric data, we will introduce a temporal version of it, used for a laboratory prototype of fringe tracker, the PFSU (\cite{Rabbia94}), and finally we will describe the location algorithms proposed by the Observatory of Torino for the VLTI ESO FINITO. In this last framework, I worked on the simulation for the adapted demodulation algorithm (par. \ref{subsec:demodulation}) and on iterative techniques for the flux intensity monitoring task (par. \ref{sec:int_Fluct_SPIE04}).

\section{Classical algorithms for fringes location}
\label{sec:Intro2}

We have explained in chapter \ref{chap:intro} that it is important to know as precisely as possible the position along the OPD scan of the fringe packet, in order to reach high sensitivity with longer observational intervals. This is possible thanks to fringe sensors, that evaluate the current OPD using dedicated algorithms, and send the information back to OPD correctors.\\
The fringe sensor provides regular measurements of the fringes intensity. Algorithms require the knowledge of the essential parameters of the fringes, such as intensity, visibility, working wavelength and spectral range. These parameters are matched with those of an interferometric model, to finally obtain the desired differential phase between interfering beams.\\

\noindent There are at least two ways to physically record the fringe packet pattern, and they are known as `spatial' and `temporal' modulation. The former consists in the simultaneous recording on the detector of selected points on the fringe, separated by a known phase. The latter is implemented recording the intensity values throughout a controlled modulation, with known applied phase, of internal OPD.\\ In both situation, the basic measurement scheme is the AC (two points for fringe, separated by a phase of $\pi$ rad, or equivalently by an OPD of $\lambda/2$). However, the phase can be evaluated modulo $\pi$, leaving a position uncertainty inside the fringe. The ABCD scheme (four points, separated each by $\pi/2$ rad) avoids this problem.

\noindent In the ideal noiseless model, the performances of ABCD algorithm can be analytically estimated, and we report them in par. \ref{subsec:ABCD}. These values are taken as reference even with other algorithms, for which the analytical evaluation is less straightforward. \\

\noindent For the FINITO fringe tracking instrument, described in par. \ref{sec:FINITO} and based on temporal modulation of the internal OPD, two algorithms have been proposed, apart a modified ABCD (par. \ref{subsec:FINITO_ABCD}). The first is a classical demodulation scheme, adapted from the PFSU prototype algorithm, developed by the Observatoire de la C\^{o}te d'Azur and described in par. \ref{subsec:lamp}. The latter is based on correlation with a template, and is resumed in par. \ref{subsec:correlation}. For both, we present the interferometric model and the assumptions on which they rely. The simulations are done with the IDL programming environment. Additional algorithms are required to solve the fringe uncertainty, i.e. to remove the periodic degeneration within the modulated envelope.




\subsection{Ideal ABCD}
\label{subsec:ABCD}

\begin{figure}[htb]
      \begin{center}
      \epsfig{figure=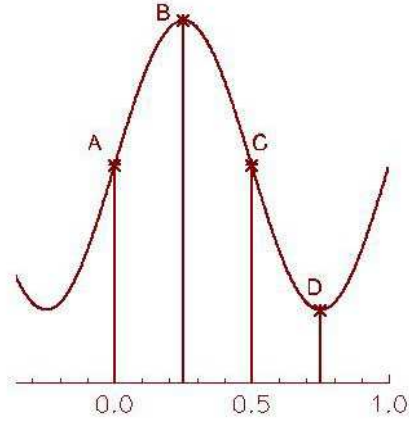,width=7cm}
      \caption{ABCD scheme}
        \label{fig:ABCDscheme}
    \end{center}
\end{figure}

We resume here the classical fringe-tracking ABCD scheme, which is able, in ideal condition, to estimate the essential fringe parameters modulo $2\pi$. The ABCD sampling of the fringe is represented in figure \ref{fig:ABCDscheme}; it consists of four points in quadrature over a single fringe of constant flux intensity $F$ and visibility $V$:
\begin{eqnarray}
\nonumber A &=& F \cdot (1 + V \cdot \sin \phi) \\
\nonumber B &=& F \cdot (1 + V \cdot \sin (\phi + \pi/2)) = F \cdot (1 + V \cdot \cos \phi) \\
\nonumber C &=& F \cdot (1 + V \cdot \sin (\phi + \pi)) = F \cdot (1 - V \cdot \sin \phi) \\
D &=& F \cdot (1 + V \cdot \sin (\phi + 3\pi/2)) = F \cdot (1 - V \cdot \cos \phi)
\end{eqnarray}

\noindent from which we can easily obtain, through trigonometric relations:
\begin{eqnarray}\label{eq:idealABCD}
\nonumber \phi = \frac{2 \pi}{\lambda} OPD = \arctan \frac{A-C}{B-D} \;\; &\rightarrow& \;\; OPD = \frac{\lambda}{2\pi} \phi\\
\nonumber F &=& \frac{1}{4} (A + B + C+ D) \\
V^2 &=& \frac{(A-C)^2 + (B - D)^2}{F^2}
\end{eqnarray}
where $\lambda$ is the working wavelength.\\

\noindent If we suppose that the only uncertainty on the A, B, C and D estimates is given by the photonic noise and by readout noise (see par. \ref{subsubsec:demodul_perf} for a description), whose variances can be approximated with the mean flux and a constant R, respectively, we obtain the following equations for the residual noise on the flux F, the visibility V and the OPD estimates:
\begin{equation}
\nonumber \sigma^2(A) = A + R; \;\; \sigma^2(B) = B + R; \;\; \sigma^2(C) = C + R; \;\; \sigma^2(D) = D + R
\end{equation}

\begin{eqnarray}\label{eq:ABCDperf}
\nonumber \sigma(\phi) = \frac{\lambda}{V \cdot SNR} \;\; \rightarrow \;\; \sigma(OPD) &=& \frac{\lambda}{2 \pi V \cdot SNR} \\
\nonumber \sigma(F) &=& \sqrt{F + R} \\
\sigma(V) &=& \frac{1}{SNR}
\end{eqnarray}
where the Signal to Noise Ratio (SNR) is, in this case, given simply by:
\begin{equation}
SNR = \frac{2F}{\sqrt{F+R}}
\end{equation}


\subsection{Demodulation algorithm at low light level}
\label{subsec:goodman}

In 1973 Walkup and Goodman\cite{WalkupGoodman73} described the limitation of the fringe parameters estimation at low light levels, both for the spatial and the temporal modulation. In this approach, with a good dispersion of the fringes over a sufficient number of pixels of a detecting system, it is possible to extrapolate information about phase and amplitude of the interferogram from the zero component of the Discrete Fourier Transform (DFT) of recorded data:
\begin{eqnarray}
X(m) = \frac{1}{N} \sum_{j=0}^{N-1} n(j) e^{-2\pi ijm/N}, \;\; m=0, \ldots N-1
\end{eqnarray}
where $N$ is the number of pixels, $n(j)$, $j = 0 \ldots N-1$ are the counts, $X(m)$, $m=0, \ldots N-1$ is the DFT of the pixels counts.

\noindent We recall the definition of the coherence length for a polychromatic interferogram over the wavelengths range $[\lambda_1, \lambda_2]$, given in chapter \ref{chap:intro}, as a function of the range width $\Delta \lambda = \lambda_2 - \lambda_1$ and of its central wavelength $\lambda_0$:
\begin{equation}
\mbox{CL} = \frac{\lambda_0^2}{\Delta\lambda}
\end{equation}
Let $f_0$ be the spatial frequency corresponding to $\lambda_0$, that will be the working wavelength of the modulation. If $\mbox{L} \sim 2 \cdot \mbox{CL}$ is the modulated path corresponding to the central lobe of the interferogram, rounded to an integer number $n_0$ of fringes, then the following statements hold:
\begin{eqnarray}
\nonumber L = n_0 \cdot \lambda_0 \;\; &\rightarrow& \;\; \lambda_0 = \frac{L}{n_0}\\
f_0 = \frac{2\pi}{\lambda_0} \;\; &\rightarrow& \;\; f_0 = \frac{2 \pi n_0}{L}
\end{eqnarray}
\noindent The number of fringes $n_0$  will be $\sim 2 \cdot CL / \lambda_0$.
Then the mean number of counts $\bar{n}(j)$ can be written, according to the simple interferometric model introduced in chap. \ref{chap:intro}:
\begin{equation}
\bar{n}(j) = x_t \left[1 + V \cos \left(\frac{2\pi j n_0}{N} + \phi \right)\right], \;\; j= 0 \dots N-1
\end{equation}
with $x_t$ the mean number of the pixel values over a number of recording, comprising both signal and background, and $V$ the fringe visibility.\\
The mean values of $R(n_0)$ and $I(n_0)$, i.e. the real and the imaginary parts of $X(m)$ when $m = n_0$, are given by:
\begin{eqnarray}
\nonumber \bar{R}(n_0) = \frac{1}{N} \sum_{j=0}^{N-1}\bar{n}(j)\cos \left(\frac{2\pi j n_0}{N} \right) = \frac{x_t V}{2}\cos\phi\\
\bar{I}(n_0) = \frac{1}{N} \sum_{j=0}^{N-1}\bar{n}(j)\sin \left(\frac{2\pi j n_0}{N} \right) = \frac{x_t V}{2}\sin \phi
\end{eqnarray}
thanks to summation over complete periods. From this equation the phase $\phi$, the visibility $V$ and the mean flux $x_t$ information can be retrieved, in a similar procedure as ABCD scheme.
\\

\noindent In low flux regime, the counts register can be modeled as Poisson variables, which mean can be approximated with an average of pixel values, recorded subsequently. Assuming fluctuation noise negligible and background noise as independent upon the signal counts, it can be shown that the mean of the real and the imaginary part of the FT follows a circular gaussian distribution, and it is possible to give a measure of the error in the estimation of the fringe parameters. The analytical derivation can be found in \cite{WalkupGoodman73}.

\noindent These estimation, however, are valid for low flux level, and to reach them it is necessary to average over a number of measurements, and so it is not so useful when phase information are needed at high frequency rate, and there is no time to perform a good average.\\

\noindent The basic idea to retain is that the Fourier transform of data contains information on fringe parameters. \\

\subsection{PFSU and LAMP}
\label{subsec:lamp}

The first fringe sensor concept proposed for the VLTI has been LAMP(\cite{Rabbia94}), acronym for Large Amplitude Modulated Path, developed by the Observatoire de la C\^{o}te d'Azur (Nice, France). The fringe sensor is devoted to detection of the error on the optical path and to communicate its value to the dedicated OPD corrector.

\begin{figure}[htb]
      \begin{center}
      \epsfig{figure=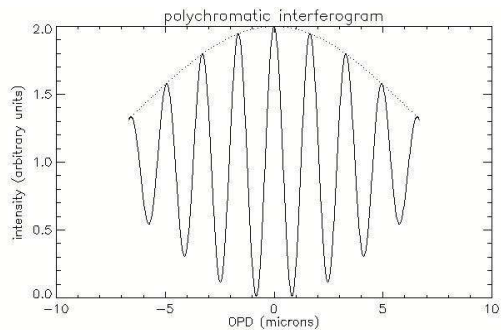,width=7cm}
      \caption{Interferogram over an OPD scan smaller than the coherence length.}
      \label{fig:interf_trunc}
    \end{center}
\end{figure}

\noindent The basic idea of the LAMP algorithm(\cite{GayRabbia94}) is that it is possible to retrieve information on the fringes parameters through the spectral analysis of the flux intensity data. To use this technique, it is essential to modulate the optical path over several wavelengths. In this way, the resulting signal contains both information about the phase of the white fringe (cophasing) and about the absolute position in the envelope of the polychromatic fringes (coherencing).\\
The control range plays an essential role. If the modulation path is larger than the coherence length of the fringes, the interferogram shape does not show truncation effects. In fig. \ref{fig:interf_trunc} a truncated interferogram is shown: the OPD scan is smaller than the coherence length, so the number of fringes is not integer, the total energy of the recorded modulation path is not maximum.

\noindent The monochromatic interferogram produced by the modulation of the optical path will be expressed as:
\begin{equation}
I(\xi) = I_s \left[1 + V_s cos \left(\frac{2 \pi \xi}{\lambda_0} + \phi\right)\right], \;\; 0 \le \xi \le L
\end{equation}
where $\xi$ varies in the modulation range, $I_s$ is the signal intensity (source and background), $V_s$ is the overall visibility (different from the effective visibility of the source), and $\phi$ is the unknown phase of the fringe.
\\

\noindent We assume that the phase varies slowly with respect to the modulation period, otherwise the phase values given by the algorithm would be a sort of mean of the varying phase, and the correction would be useless.
\\

\noindent The procedure to find the phase modulo $2 \pi$ is analogous to that of Goodman, with the obvious difference that we are working with a temporal modulation, and not with a spectral dispersion, which allows to have a spatial distribution at each time instant.\\
The modulated path is a symmetric triangular periodical function of time (sawtooth) with frequency $f_0$ and amplitude $n_0\lambda_0$, chosen approximately equal to twice the coherence length ($L \sim 2 \cdot CL$). Note that $n_0$ is an integer, so the modulation is done over an integer number of fringes. Let $N$ be the total number of samples covering $L$.\\
The signal component at frequency $f_0$ carries the information about the phase $\phi$, so it is detectable with the analysis of the corresponding component of the Fourier transform (\cite{OCA93}):
\begin{equation}
\nonumber r_{\phi} = \int{\cos (2 \pi \xi / \lambda_0 + \phi) \cdot \sin (2 \pi \xi / \lambda_0)} = \sum_{i=1}^N{\cos (2 \pi \xi_i / \lambda_0 + \phi) \cdot \sin (2 \pi \xi_i / \lambda_0)} =
\end{equation}
\begin{equation}
= \frac{1}{2}\sum_{i=1}^N{\sin (2 \pi \xi_i / \lambda_0) – \sin(\phi)} = - \frac{1}{2} \sin(\phi)
\end{equation}
thanks to the summation over an integer number of fringes, i.e. over the complete range $[0, 2\pi]$. In the same way:
\begin{equation}
i_{\phi} = \int{\cos (2 \pi \xi / \lambda_0 + \phi) \cdot \cos (2 \pi \xi / \lambda_0)} =  \frac{1}{2} \cos(\phi)
\end{equation}
Apart the factor $1/2$, these are the real and imaginary part of the complex number $e^{-i \cdot \phi}$. We find the phase $\phi$ and the corresponding OPD simply as:
\begin{equation}
\phi = \arctan (-\frac{r_{\phi}}{i_{\phi}})\;\; \rightarrow \;\; OPD = \frac{\lambda}{2 \pi}\phi
\end{equation}

\noindent Note that this is a measure modulo $2 \pi$ (or $\lambda$), for the properties of the arctangent function. To do {\it coherencing}, i.e. to find the absolute position within the coherence length, a proper combination of the lateral frequencies $f_{\pm 1} = 2 \pi (n_0 \pm 1) / L = f_0 \pm 2 \pi / L$ is needed. In an analogous way as before, the modulated interferogram is filtered to find the phases $\phi \pm \frac{\phi}{n_0}$ modulo $2 \pi$. The differential value is $\frac{\phi}{n_0}$ modulo $2 \pi$, from which we finally obtain $\phi$ modulo $2 n_0 \pi$, i.e. modulo the coherence length.

\noindent This last measure is complementary to the phase modulo $2 \pi$ because it gives the position of the intensity maximum over all the coherence length, while the phase modulo $2 \pi$ does not guarantee that the evaluated phase is a secondary maximum in a lateral fringe, instead of the central peak.
\\

\noindent Two aspects have to be stressed. First of all, the presence of an integer number of fringes is important to filter out the modulation path. Then, these formula are based on the modulated part only, and do not include the offset of the interferogram $I_S$ or the amplitude of the modulation ($I_s \cdot V_s$). This means that the original signals have to be normalized before using them.

\section{Algorithms for FINITO}
\label{sec:FINITO}

After an introduction on the VLTI FINITO fringe tracker, we describe the location algorithms that we proposed for the VLTI FINITO fringe tracker. The first is based on the demodulation algorithm, the second on a correlation with a template.

\subsection{Instrument description}

\begin{figure}[htb]
      \begin{center}
      \epsfig{figure=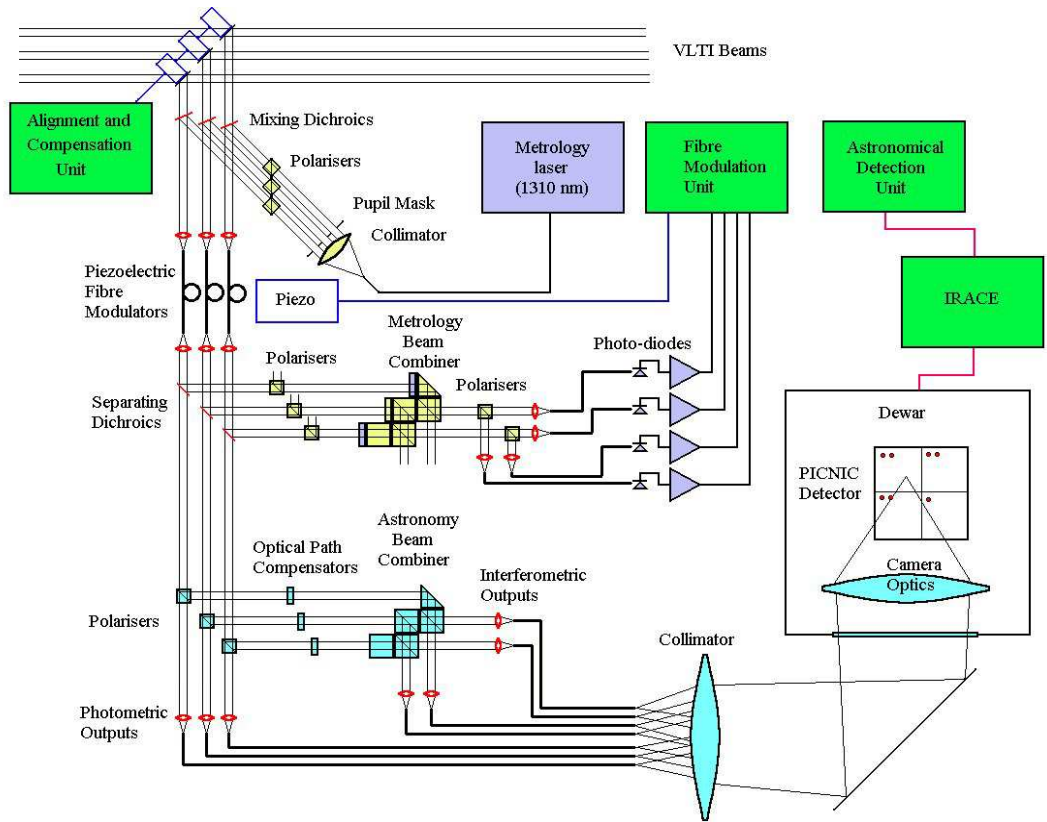,width=14cm}
      \caption{FINITO layout}
        \label{fig:FINITOstructure}
    \end{center}
\end{figure}

FINITO (Fringe tracking Instrument of NIce and TOrino) is a fringe sensor unit developed by the Osservatorio Astronomico di Torino, in collaboration with ESO. It is based on the PFSU laboratory prototype.\\
It is an interferometric instrument based on amplitude combination of either two or three telescopes beams. It operates in H band ($\lambda \in [1.48 - 1.78]\;\mu m$). Figure \ref{fig:FINITOstructure} shows the overall scheme of the instrument. The astronomical beams are injected into monomode and polarisation maintaining optical fibres. The fibres act as spatial filters, because the injection system retains just the central lobe of the incoming wavefront, and reject the most aberrated lateral one. The optical paths of the two beams are modulated by piezoelectric devices, that stretch the fibres in order to modify the paths. The differential path is monitored by a metrology system, based on the superposition of a laser source at $\lambda=1.31 \mu m$. This laser beam shares the same optical path of the astronomical beams during the modulation, then it is removed and separately combined, and the current differential path is achieved and used for correction.\\

\noindent Therefore, the modulation applied to the astronomical beams through this closed control loop can be considered as ideal. In turn, the phase detected on astronomical beam combination is a measurement of the external disturbances which are to be compensated for stable integration on science combiners.
\\

\noindent After the separation of metrology from astronomical beams, the two polarization components of each beam are separated; one is retained for photometry purpose, while the other is sent to a beam combiner. This approach avoids to deal with phase differences between polarization components.\\
\noindent In the beam combiner, one of the three input beam is split and superposed separately with the others. The former acts as a reference beam, and the interferometric signals allow detection of the relative phase of each beam pair. Four interferometric outputs are produced, two destructive and two constructive; they are finally focused on a detector, together with the three photometric outputs.
\\

\noindent The goal of FINITO is the measurement in real time, allowing correction by the OPD controller, of the perturbation to the optical path, caused {\it in primis} by atmospheric turbulence that affects photons by the stars. The instrument adds some disturbance, such as modulation and readout noise (RON); optical elements can produce fluctuations of the phase and the amplitude. Moreover, all these factors limit the performance of the instrument. \\
Its set-up is optimized taking into account the operational conditions: the scan amplitude should be comparable with the coherence length (for the H band, i.e. $[1.5 - 1.8] \; \mu m$, with $\Delta\lambda = 0.3 \; \mu m$ and central wavelength $\lambda_0=1.65 \; \mu m$, the number of fringes is about $2\lambda_0/\Delta\lambda \sim 10$), the fringe scanning rate must be faster than the typical atmospheric turbulence, even if higher rates correspond to shorter integration times, and so to lower sensitivity.
\\

\noindent Different algorithms have been proposed for the evaluation of the current optical path difference to be compared with the modulation one. The mandatory request for all the algorithms is to execute in real time, so they can't make use of too many computations. In the following sections, we review them with their performances.
\subsection{Demodulation algorithm}
\label{subsec:demodulation}

The demodulation algorithm proposed for FINITO is an implementation of the LAMP concept.\\
Let us consider the following simplified description of the two complementary outputs of each metrology beam combination:

\begin{eqnarray}\label{eq:constr_destr}
\nonumber  F(p) &=& \frac{1}{2} \{I_1 + I_2 + 2 \cdot V \cdot \sqrt{I_1 \cdot I_2} m(p) \cos (2 \pi \frac{p-p_1}{\lambda_0})\} \\
G(p) &=& \frac{1}{2} \{I_1 + I_2 + 2 \cdot V \cdot \sqrt{I_1 \cdot I_2} m(p) \cos (2 \pi \frac{p-p_1}{\lambda_0} + \pi)\}
\end{eqnarray}

\noindent where $F(p)$ and $G(p)$ are the signals from the constructive and the destructive outputs, $I_1, I_2$ are the incident beams intensities, $V$ is the overall visibility (composition of the source and of instrumental visibilities), $p$ is the optical path difference, $p_1$ is an additional OPD that comprises the contributes of different kind of noises, $\lambda_0$ is the central wavelength of the working range and $m(p)$ is the envelope, i.e. the contribution to the fringe pattern due to the polychromaticity of the beams. The two signals have a phase difference of $\pi$. We can rewrite them as:
\begin{equation}
F(p);G(p) = \frac{1}{2} \left\{I_1 + I_2 \pm 2 \cdot V \cdot \sqrt{I_1 \cdot I_2} m(p) \cos \left(2 \pi \frac{p-p_1}{\lambda_0}\right)\right\}
\end{equation}
that explains the "constructive" and "destructive" definition of $F(p)$ and $G(p)$.
If we subtract $F(p)$ and $G(p)$, we eliminate the common offset:
\begin{equation}
S(p) = F(p) - G(p) =  2 \cdot V \cdot \sqrt{I_1 \cdot I_2} m(p) \cos \left(2 \pi \frac{p-p_1}{\lambda_0}\right)
\end{equation}

\begin{figure}[ht]
      \begin{center}
      \epsfig{figure=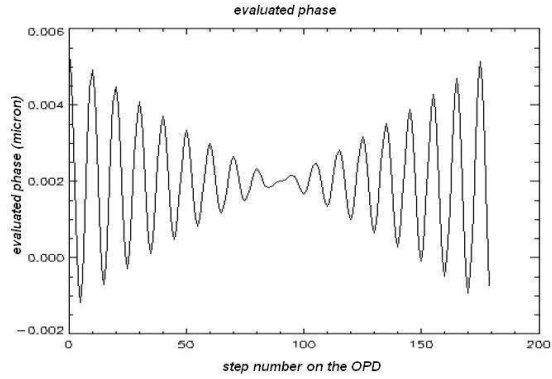,width=7.5cm}
      \caption{Phase error evaluated with the demodulation algorithm, using as templates simple sinusoidal functions (sin and cos) at same frequency than the central fringe. The introduced phase error is constant ($2$ nm). The abscissa axis reports the number of points over the OPD scan, while on the ordinates the evaluated phase amplitude is shown.}
      \label{fig:demod_OPDwith-sin-and-cos}
    \end{center}
\end{figure}

\noindent In order to apply the LAMP concept, we have to make some adjustments. First of all, the identification of the amplitude factors is needed, in order to define a proper filter for the Fourier analysis. If we use a simple sinusoidal wave, as explained in section \ref{subsec:lamp}, the results are worsened by the lack of matching between the sinusoidal wave and the interferogram. Figure \ref{fig:demod_OPDwith-sin-and-cos} shows the results of a simulation of the OPD evaluation using a demodulation over 10 fringes, with a step of a twentieth of fringe, when the modulation law is a ramp and the introduced phase error is a constant atmospheric piston of $2$ nm.

\noindent The bad quality of the phase error evaluation is due to the shape of the interferometric beam, which presents the envelope pattern. In the area near to the zero OPD, this effect is weakened because the interferometric intensity has a maximum, together with the envelope.\\

\noindent We then define a shape for $m(p)$. The polychromatic spectrum is a superposition of monochromatic components $\omega(\lambda,p)$:
\[m(p) = \int_{\lambda_1}^{\lambda_2}  \omega(\lambda, p) d\lambda \]
In the nominal case, wavelengths in the selected range give the same contribute to the polychromatic beam, while all other wavelengths outside the band give no contribution:
\[\omega(\lambda, p) = \left \{ \begin{array} {ll}
                    1    & \mbox{if $\lambda_1 \leq \lambda \leq \lambda_2$} \\
                    0   & \mbox{otherwise}
                  \end{array}
                  \right.\]

\noindent The function $\omega(\lambda, p)$ acts as a perfect rectangular filter:
\[ m(p) = \int_{\lambda_1}^{\lambda_2}  \omega(\lambda, p) d\lambda = \int_{-\infty}^{\infty}  \omega(\lambda, p) d\lambda.\]

\noindent We are in the situation described in par. \ref{subsec:poly_interf} of chapter \ref{chap:intro}. The resulting model for $m(p)$ is a {\it sinc} function:
\begin{equation}
m(p)=\frac{sin \pi x}{\pi x}
\end{equation}
where $x$ is the ratio OPD to coherence length: $x = OPD / CL$.\\

\noindent We define two functions for the demodulation, tailored on the modulating envelope: \begin{eqnarray}
\nonumber t_1 (x) = \sin (2 \pi x / \lambda_0) \cdot \mbox{sinc} (\pi x)\\
t_2 (x) = \cos (2 \pi x / \lambda_0) \cdot \mbox{sinc} (\pi x)
\end{eqnarray}
in order to better suit the shape of our interferogram, and we apply the LAMP concept, filtering the signal $S(p)$ with the modified templates $t_1(x)$ and $t_2(x)$.\\
Note that the subtraction of the destructive and constructive waves cancel out the common offset; if this offset is not equal, it must be evaluated (for example, by photometric measurements) and eliminated.\\

\subsubsection{Expected performances}

As for the ABCD case, also for this algorithm the performance is influenced by the SNR. An estimation for the minimum detectable phase can be found in \cite{GayRabbia94} and is given by:
\begin{equation}
\hat{\sigma}(OPD) = \frac{\lambda}{2 V \pi SNR}
\end{equation}
Piston variation below this limit cannot be recognized by the algorithm, so this is a limiting performance.


\subsubsection{Performance}
\label{subsubsec:demodul_perf}
The performance of this adapted algorithm depends on the type of noise that affects the optical path difference or the interferogram intensity, and so is reflected in the SNR.

\noindent Different kind of noises are considered.
\begin{itemize}
\item{The photonic noise is linked to the particle nature of light and to the fact that the arrival time of the photons is random. Under some conditions (semi-classic theory of detection, see \cite{Goodman} for references), it follows a Poisson distribution with rate proportional to the square root of the intensity of the incident light.} 
\item{The scintillation is defined as a variation in the intensity of the flux collected by a telescope. It can be caused by refraction effects in atmospheric layers, especially in small structures caused by turbulent phenomena. It depends from the position of the source in the sky, and from observational conditions. It can be modeled as a time sequence having a slowly decreasing spectrum $s(\theta)$, depending also from the wavelength: $s(\theta) \propto \lambda_0^2 \cdot \theta^{-8/3}$ in order to simulate also high-frequency turbulence typical of the atmosphere.}
\item{The RON (Read Out Noise) is caused by the detector, and can consist in the uncertainty of the digitalization as well as small charges induced by electronic components. The easiest way to model it is with a constant value representing a statistic of the error on the detector read out, directly proportional to the integration time and the flux intensity
    }
\item{The shotnoise is linked to the quantization of the receiving matter.
    An approximation is with a random variable drawn from a normal distribution of mean zero and standard deviation $\sigma = \frac{f}{SNR}$. This formula takes into account the SNR: the bigger it is, the smaller is the shot noise.}
\item{The atmospheric disturbance spectrum has been studied for long (for VLTI environment, see, e.g., \cite{Daigne99}). 
    It is function of the geographical location and of weather conditions. An approximation of the low frequencies ($f \le 100$ Hz) of such noise is given in figure \ref{fig:noise1}. The power spectrum density of the atmospheric variation is proportional to $f^{-\frac{8}{3}}$ and to atmospheric parameters, such as the wind speed and the Fried parameter (a measure of the coherence length of the atmosphere). The phase is randomly distributed following a uniform distribution in the range $[-\pi, \pi]$ .
\begin{figure}[ht]
      \begin{center}
        \epsfig{figure=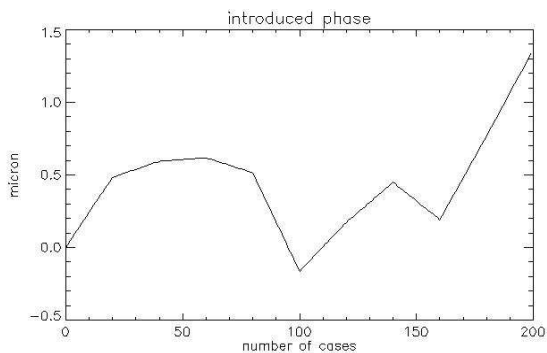, width=7.5cm}
        \caption{Stochastic sequence of atmospheric induced phase.}
        \label{fig:noise1}
    \end{center}
\end{figure}
}
\end{itemize}

\noindent In the following simulations, an interferometric signal is generated using the description of equation \ref{eq:constr_destr} and the following parameters:
\begin{itemize}
	\item{working wavelength: $\lambda = 1.65 \; \mu m$ in H band}
	\item{source magnitude = 13}
	\item{estimated visibility: $0.72$}
    \item{flux at zero magnitude: $4.8e10$}
\end{itemize}

\noindent With the listed parameters, the reference performance for the OPD is given by:
\begin{equation}\label{eq:demod_ref_perf}
\sigma_{OPD} = \frac{1.65}{2 \pi \cdot 0.72 \cdot 10} = 0.036 \mu m
\end{equation}

\begin{figure*}[htb]
     \begin{center}
     \epsfig{figure=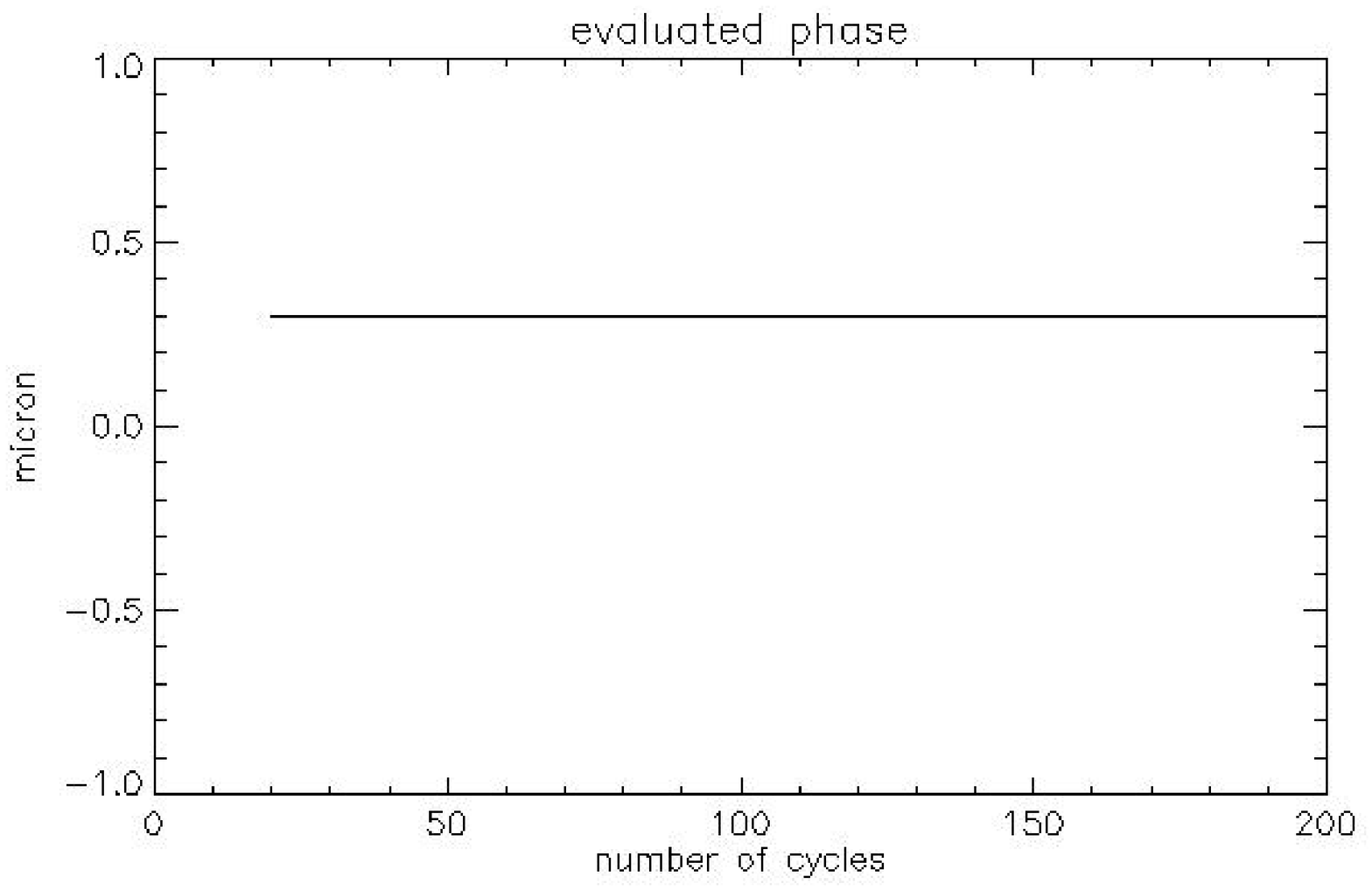, width=5cm}
     \epsfig{figure=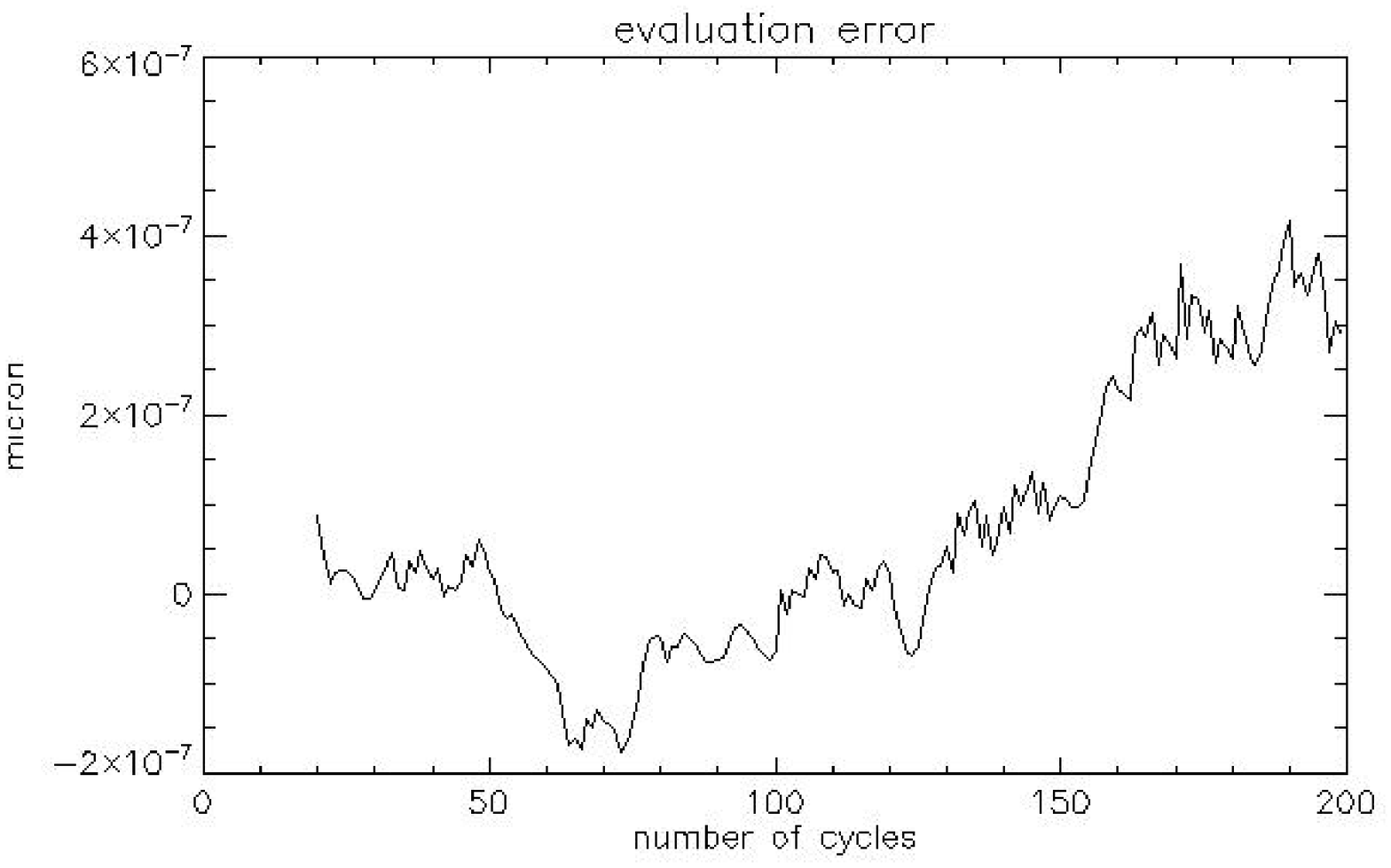, width=5cm}        \caption{Left, the phase evaluated by the algorithm, right, the error between the evaluated and the introduced one. The mean value of this last discrepancy is: $9.6 \cdot 10^{-9} \; \mu m$ and the standard deviation is: $2.8 \cdot 10^{-8} \; \mu m$. The two graphics have different ranges of values for comprehension reasons.}
     \label{fig:modul_nonoise}
    \end{center}
\end{figure*}

\noindent We first do not add any noise on the amplitude, to assess the nominal performances of the algorithm. We can see (figure \ref{fig:modul_nonoise}) that when the introduced phase is a constant ($0.3\; \mu m$), the error between the evaluated phase and the introduced one is of order of $10^{-7} \; \mu m$. \\

\begin{figure*}[htb]
     \begin{center}
     \epsfig{figure=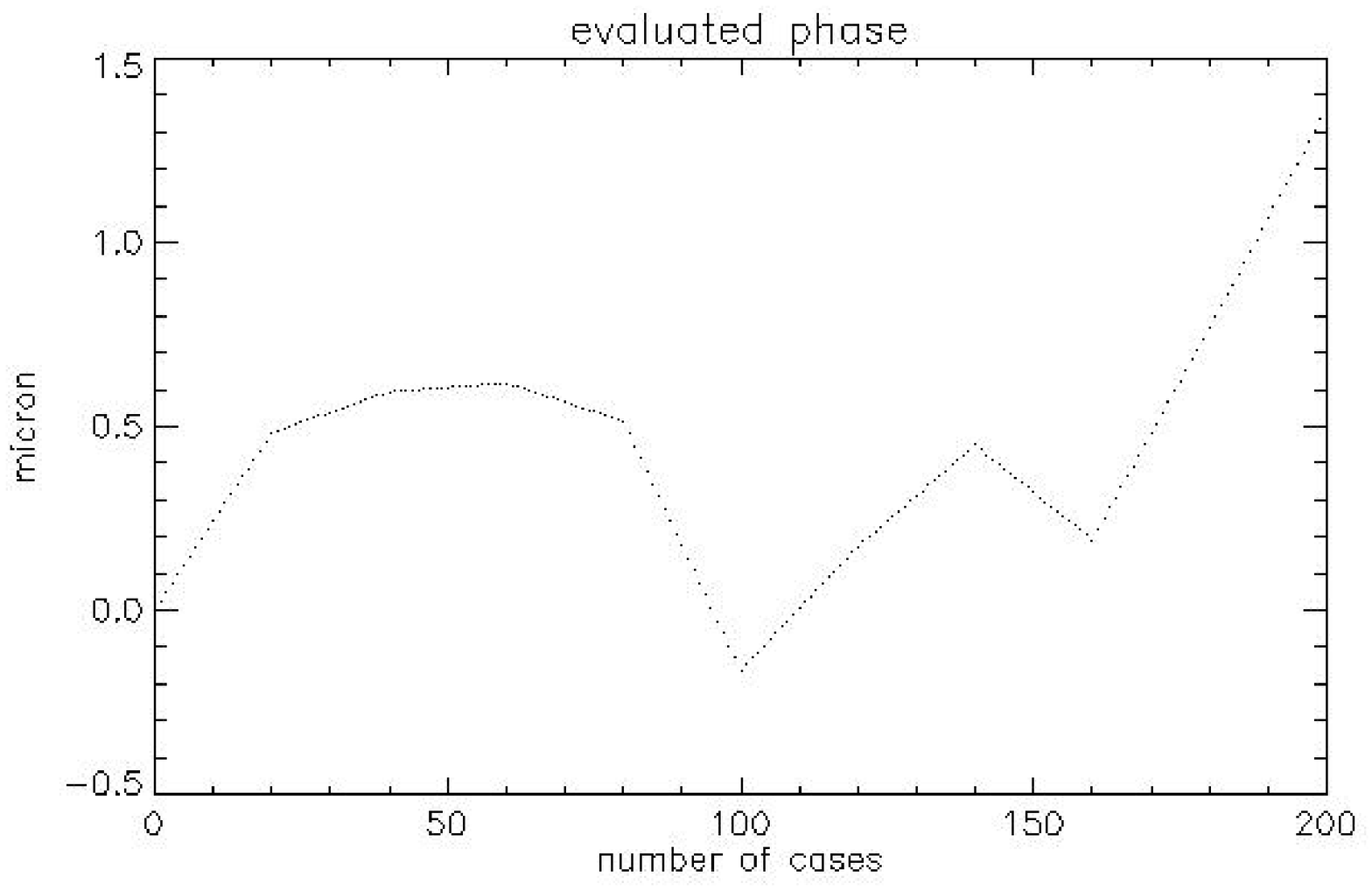, width=5cm}
     \epsfig{figure=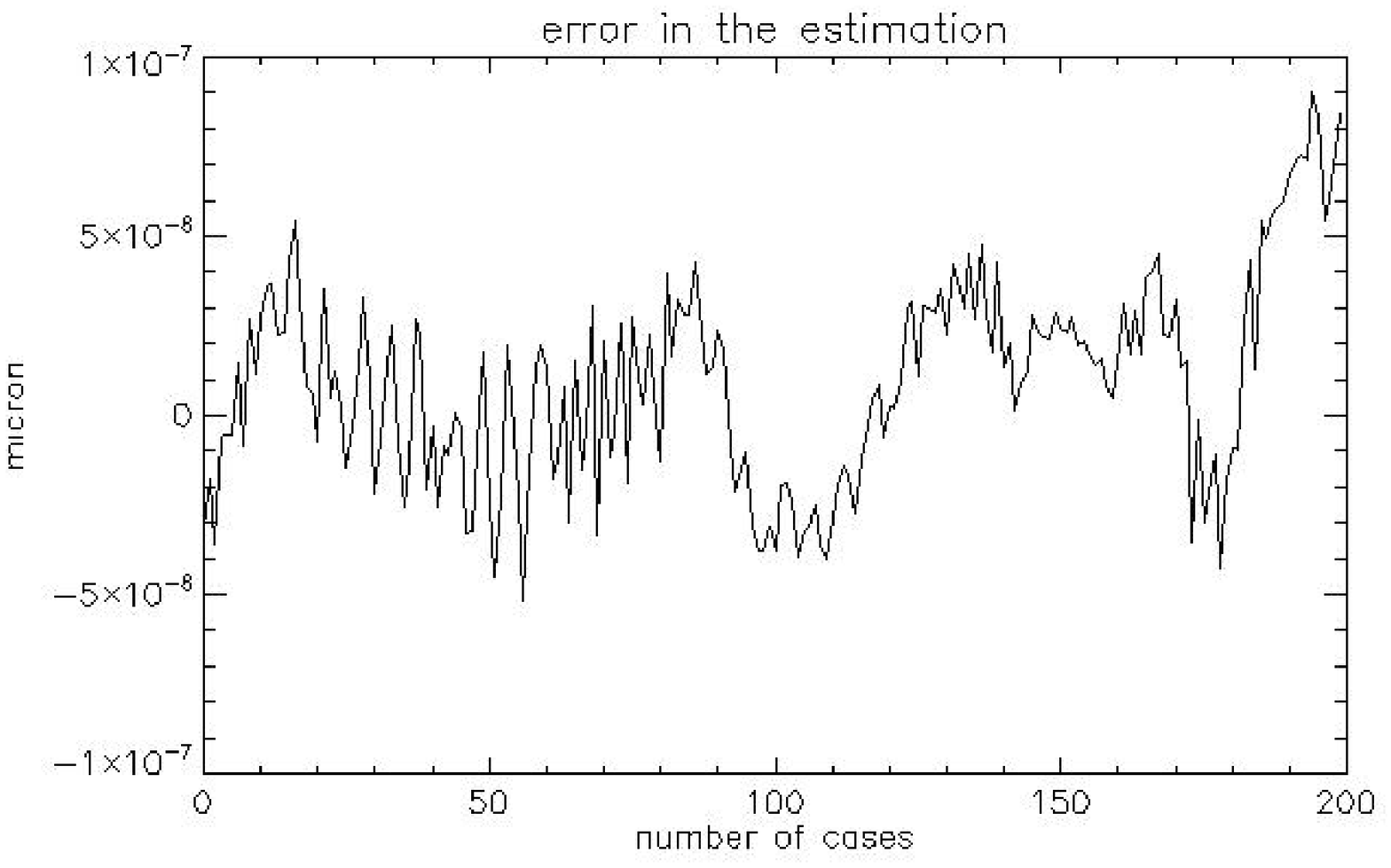, width=5cm}        \caption{Left, the phase evaluated by the algorithm, right, the error between the evaluated and the introduced one, when the introduced noise is of the kind described in figure \ref{fig:noise1}. The mean value of this last discrepancy is: $9.6 \cdot 10^{-9} \; \mu m$ and the standard deviation is: $2.8 \cdot 10^{-8} \; \mu m$.}
     \label{fig:modul_noise1}
    \end{center}
\end{figure*}

\noindent We now add an atmospheric disturbance on the OPD, with the same features of the one shown of figure \ref{fig:noise1}. Of course, the estimation error increases, and also shows a pattern inherited from the introduced OPD (figure \ref{fig:modul_noise1}). However, it is well beyond the reference limit set by eq. \ref{eq:demod_ref_perf}, since its standard deviation is of order of $10^{-8} \; \mu m$.\\
Therefore, the noise induced by model/algorithm error is quite negligible with respect to that associated to physical noise.

\noindent We finally add also noises on fringes amplitude, both observational (photonic noise and scintillation) and instrumental (RON and shot noise):
\begin{itemize}
	\item{RON = 10 photons/sec}
	\item{photonic noise with rate equal to the square root of the flux of each incoming beam}
	\item{shot noise, modelled as a normal variable with $\sigma = \frac{flux}{SNR}$, with SNR = 10}
	\item{scintillation, following the previous description.}
\end{itemize}

\begin{figure*}[ht]
     \begin{center}
       \epsfig{figure=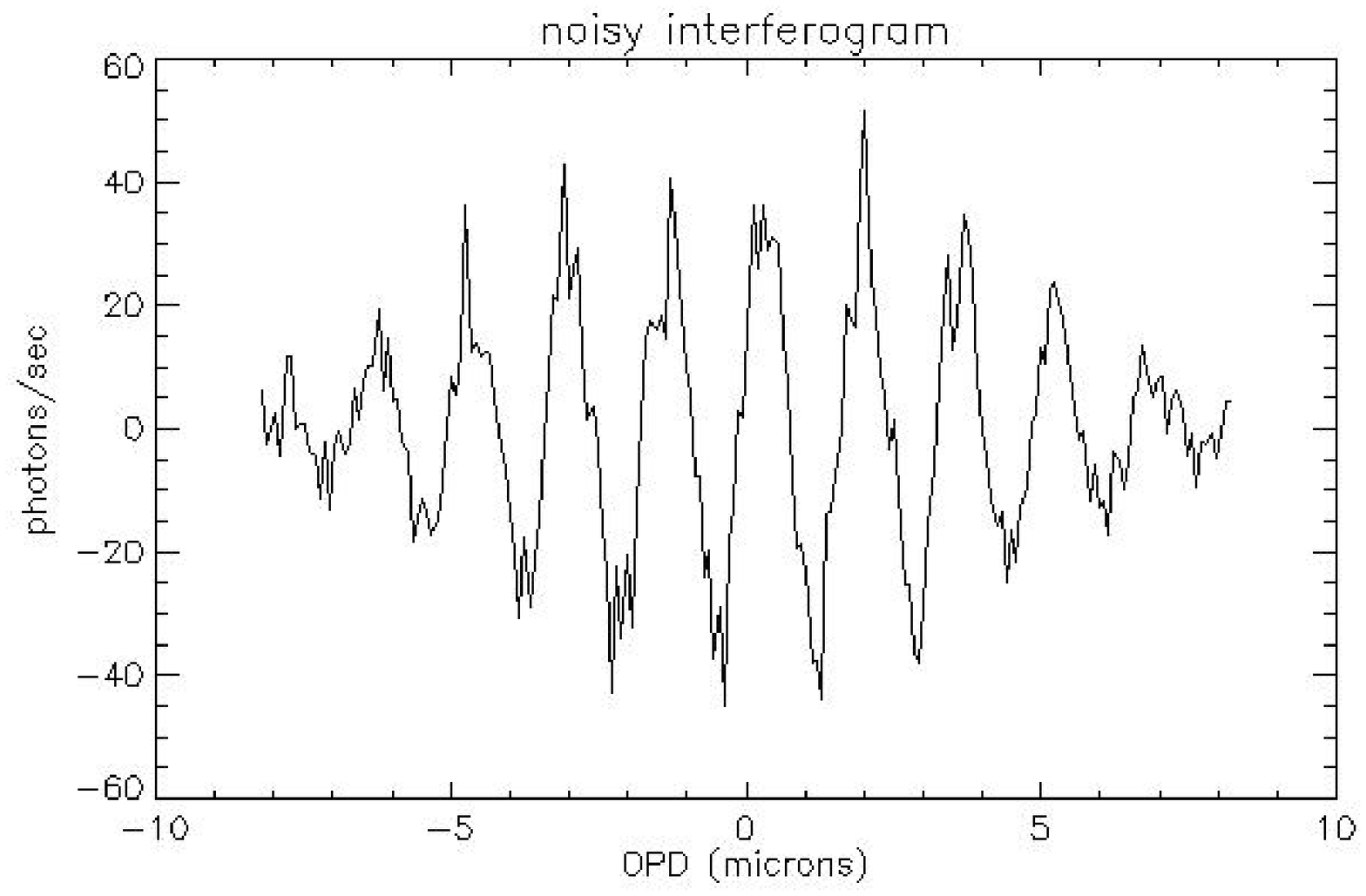, width=4.5cm}
       \epsfig{figure=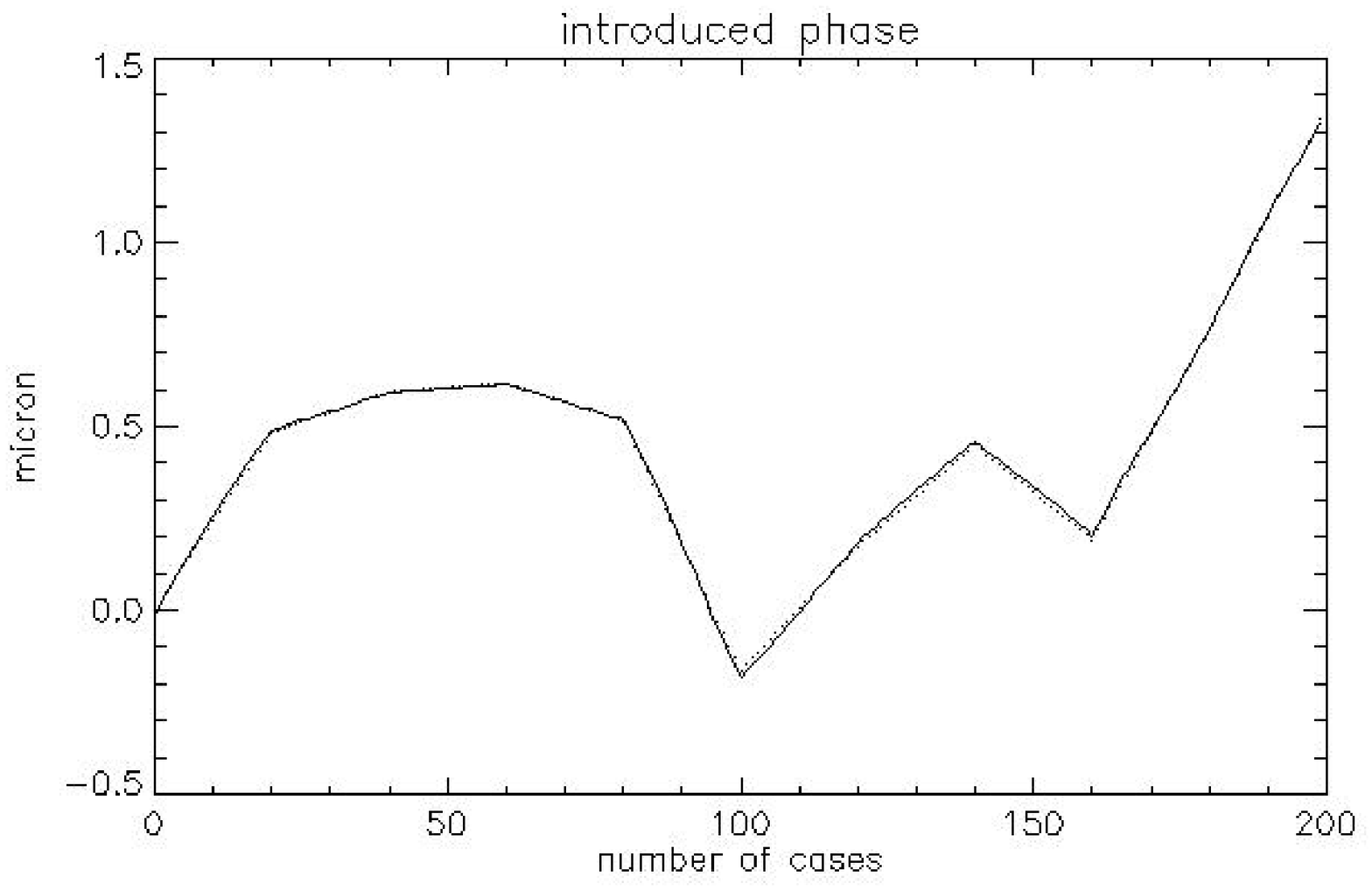, width=4.5cm}
       \epsfig{figure=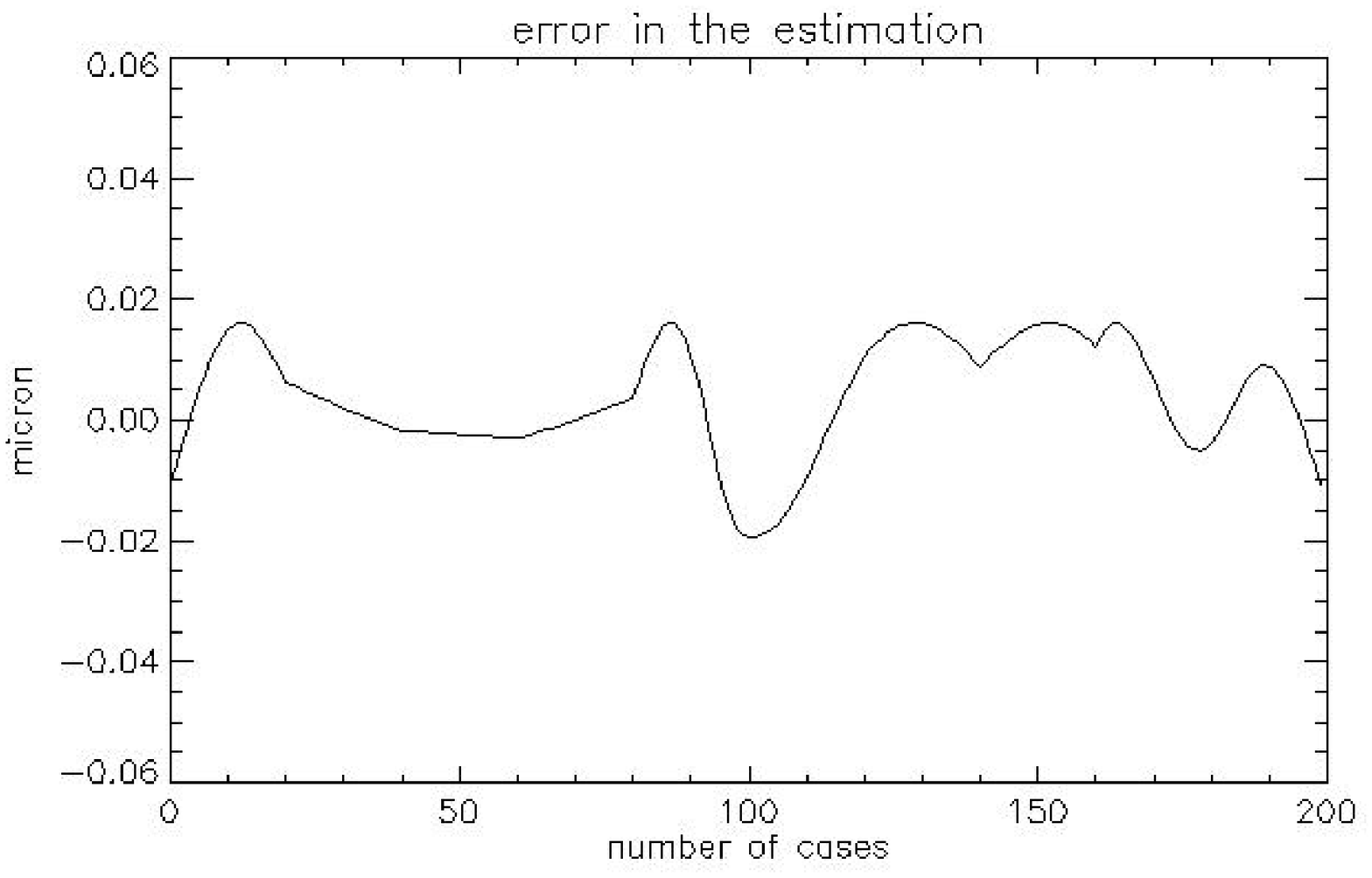, width=4.5cm}
	   \caption{Left, the noisy interferogram; centre, the introduced phase (solid line) and the  evaluated one (dotted), right, the error between the evaluated and the introduced one. The mean value of this last discrepancy is: $3.8 \cdot 10^{-3} \; \mu m$ and the standard deviation is: $9.2 \cdot 10^{-3} \; \mu m$}
        \label{fig:noise1_noisyInterf}
    \end{center}
\end{figure*}

Even if the interferogram shows a noisy pattern, due to shot noise and photonic noise, the performances of the algorithm are still quite good (figure \ref{fig:noise1_noisyInterf}). Again, its standard deviation ($9.2 \; nm$) is in accordance with the reference one.\\

\begin{figure}[htb]
     \begin{center}
       \epsfig{figure=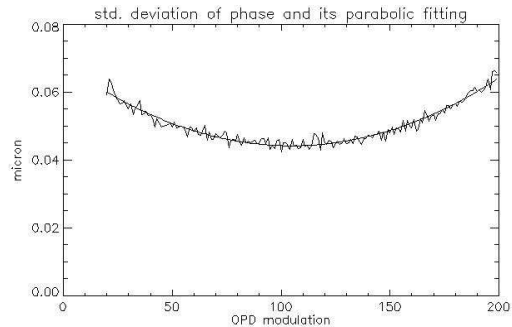, width=7cm}
	   \caption{Standard deviation of the error on the evaluation of phase as a function of the OPD}
       \label{fig:stdev_phase}
    \end{center}
\end{figure}

\noindent In order to figure out features of the algorithm, we generate an high number of intensity noise realizations ($1000$). For each of them, the interferogram is generated with the same constant OPD deviation added to the optical path. This is not a realistic case, but it is easy to understand and to control. As we could expect, the standard deviation of the evaluation error, averaged on the number of realizations, depends upon the current OPD, i.e. from the distance from the zero OPD, where the flux intensity reaches its maximum (figure \ref{fig:stdev_phase}). If the OPD noise is not constant, this feature is covered by error threshold due to the atmospheric OPD.

\subsubsection{Spectral performances}

Our task is now the comparison of the power spectral density\footnote{A description of PSD and its principal features can be found in chapter \ref{chap:stat}} (PSD) of the atmospheric phase with the PSD of the evaluated phase, to check how spectral features of the introduced phase are reproduced in the spectrum of the evaluated one.\\
We will first analyze the spectral characteristic with a noisy interferogram, but with a poor realistic noise, i.e. a linear ramp, in order to check particular behaviour of the algorithm.
We will then use a realistic noise, checking the influence of the noise on intensity comparing the algorithm performance on interferogram with or without intensity perturbance.\\

\begin{figure*}[ht]
     \begin{center}
       \epsfig{figure=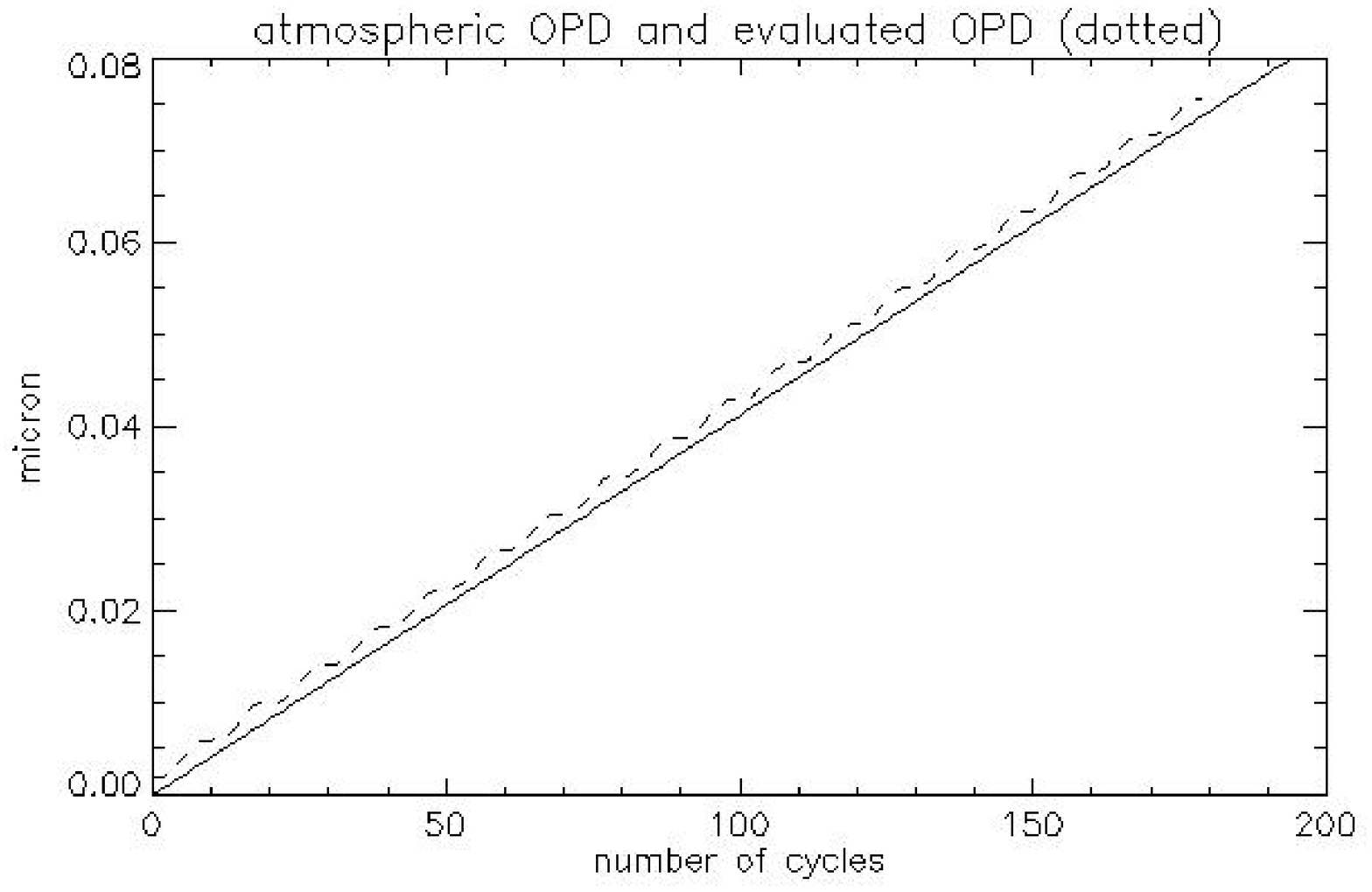, width=6.5cm}
       \epsfig{figure=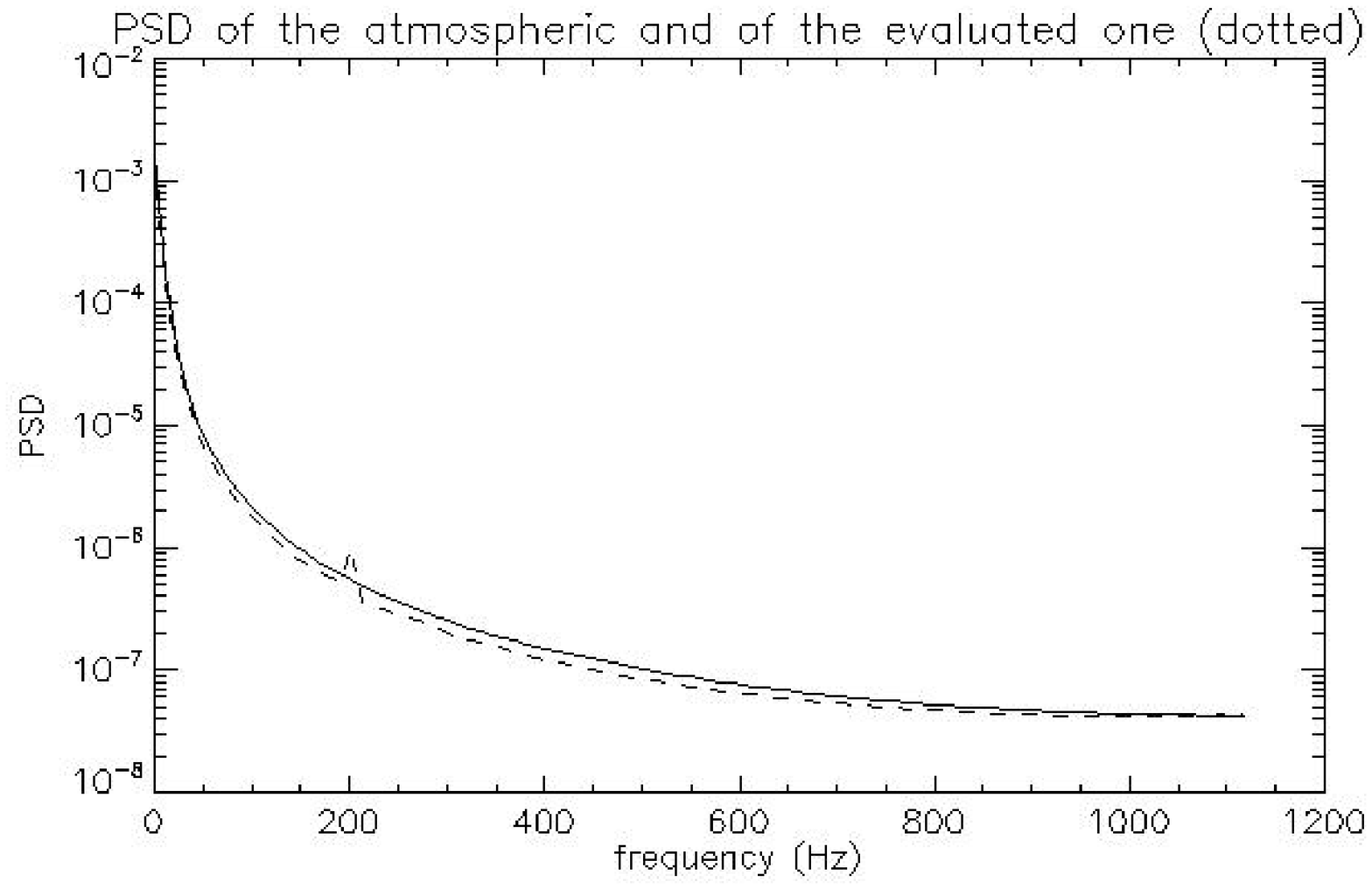, width=6.5cm}
       \caption{Phase evaluation on signal with noisy intensity. Left, introduced versus evaluated (dashed) phase; right, the respective PSD}
       \label{fig:PSD_linearOPD}
    \end{center}
\end{figure*}

\noindent So, let us begin adding to the interferometric signals the intensity noises described in the previous paragraph: realizations of shotnoise with a SNR=10, photonic noise and scintillation, drawn as a noise with a decaying spectrum ($\approx \theta^{-2}$). The sampling is set at 4 kHz, with 20 samples for fringe (a fringe in 5 ms, i.e. 200 Hz). \\
\noindent We first introduce an atmospheric phase drawn as a linear ramp with coefficient $0.5$ of the modulated path.\\
The PSD behaviour remains the same before and after the algorithm application (see fig. \ref{fig:PSD_linearOPD}). We have to notice, however, a peak in correspondence of the 200 Hz frequency, corresponding to the fringe frequency. This is clearly an artefact, but it can be recognized on the evaluated OPD, too, as a superposed sinusoidal pattern over the linear shape. This is a residual of the interferometric modulated component.\\

\begin{figure*}[htb]
     \begin{center}
       \epsfig{figure=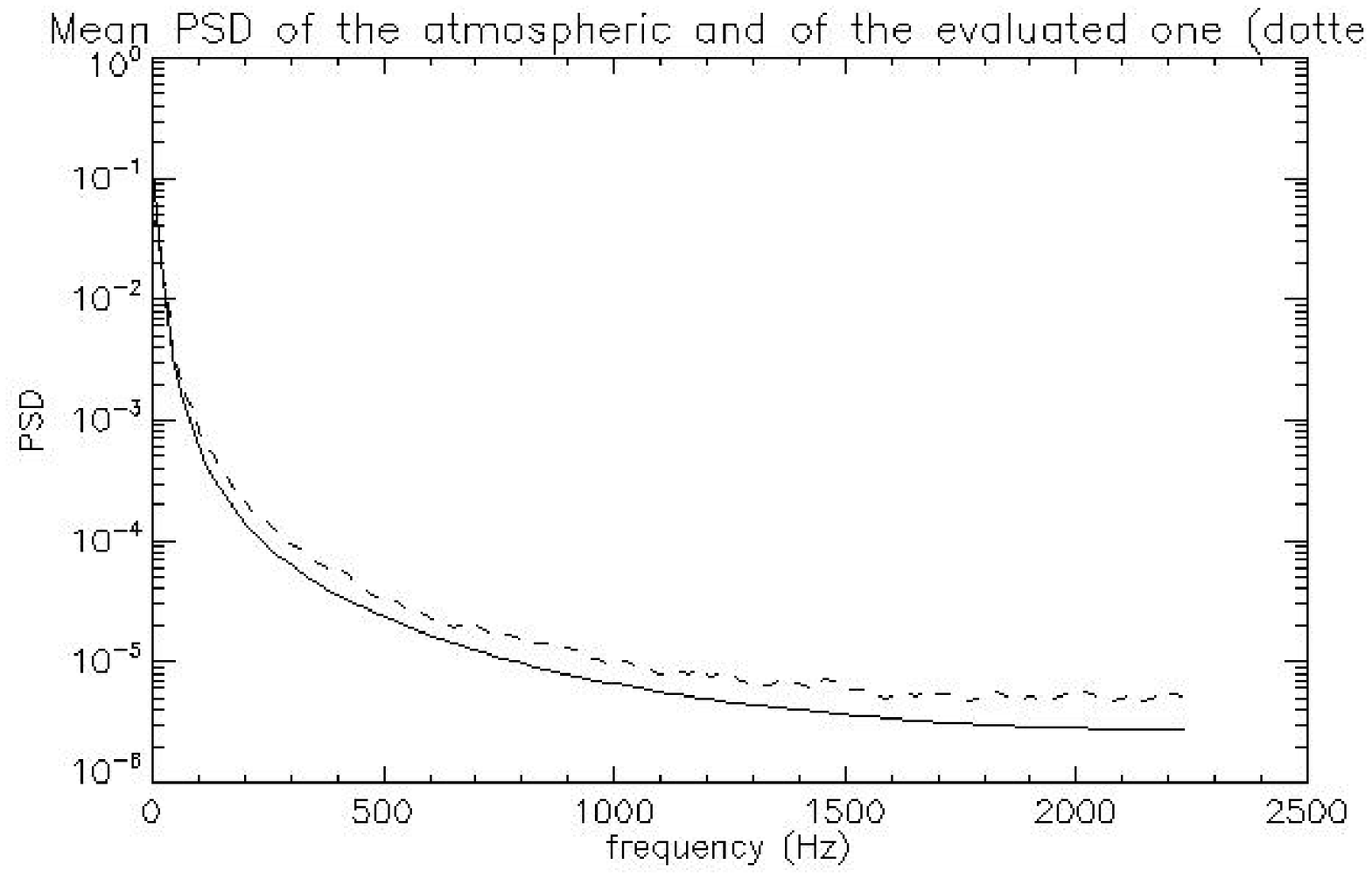, width=6.5cm}
       \epsfig{figure = 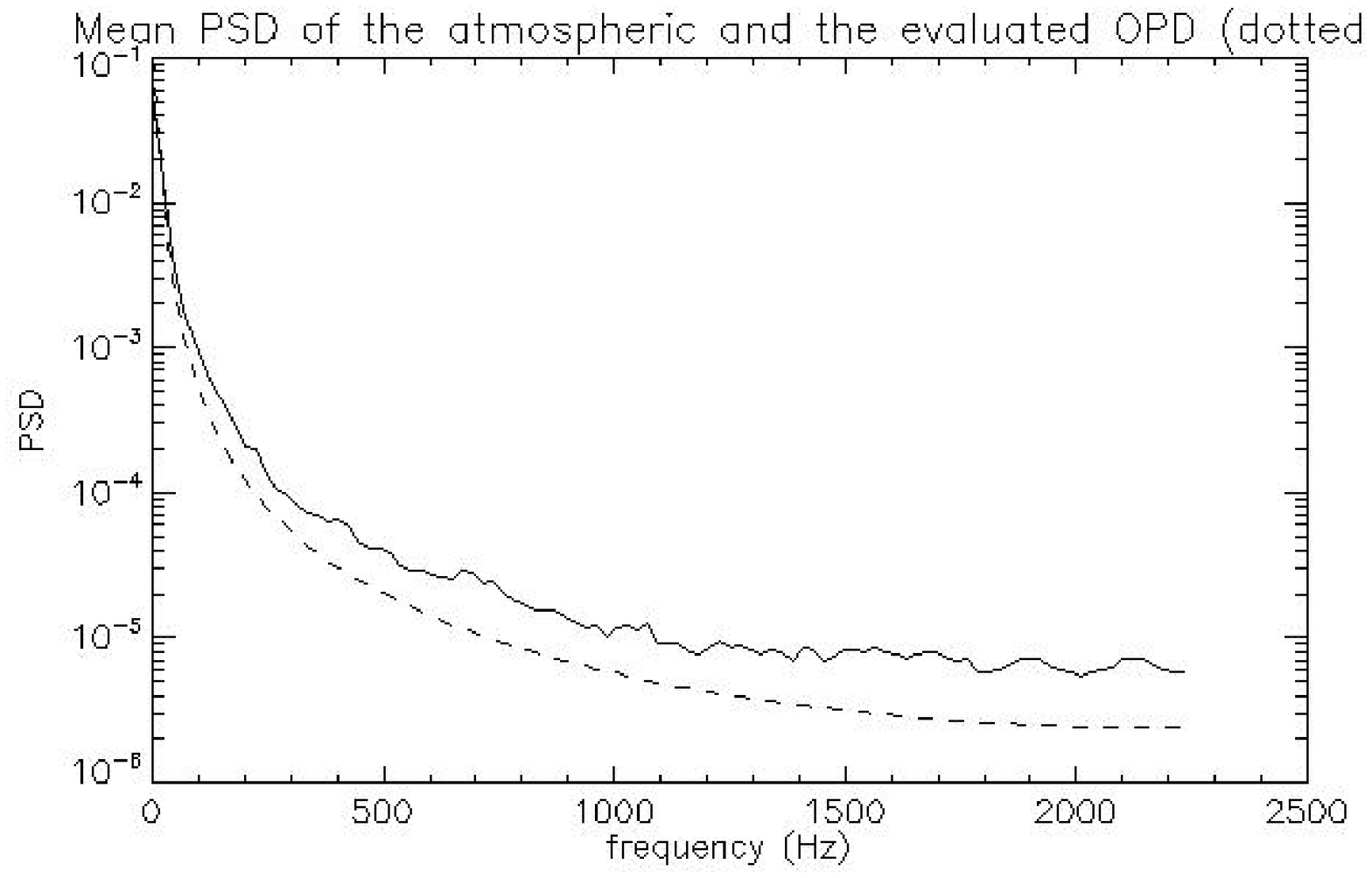, width=6.5cm}
       \caption{PSD of introduced and evaluated (dashed) phase in two different simulations: without noise on interferometric intensity (left) and with (right). For both, the noise on the optical path is a realistic one (figure \ref{fig:noise1})}
       \label{fig:PSD_noise1_200stat}
    \end{center}
\end{figure*}

\noindent We now perform the more realistic simulation adding the atmospheric noise described in figure \ref{fig:noise1} to the optical path.\\
For both cases described above (nominal and noisy intensities), we generate a statistic of $N=200$ noise realizations, we evaluate the phase for each of them and we compute the PSD of the two series. Finally, we average all PSDs averaged over the number of realizations. Fig. \ref{fig:PSD_noise1_200stat} shows this averaged PSD for the nominal (left) and corrupted (right) interferometric intensities.  The left picture reveals that the algorithm augments the noise on the phase, especially at higher frequency. It seems, however, that this added noise source affects all the frequencies over a threshold. A similar situation is the case of interferometric outputs with perturbation on the intensities (figure \ref{fig:PSD_noise1_200stat}, right), but the added noise is greater, i.e. the offset between the PSD of the atmospheric phase and of the evaluated one is greater.\\
This is due to the modeling of disturbance on intensities. In fact, the shotnoise is based on the realization of a normal random variable, the photonic noise is designed as a Poisson variable (even if, at this flux level, it is approximated by a normal variable too), so these components have a flat spectrum. 

\subsubsection{Remarks}

\noindent The magnitude of the source is an important variable. Given its flux intensity at the reference magnitude (zero magnitude) for a wavelength, the flux intensity of the light reaching the detector surface is linked to the magnitude $m$ by the relation:
\[f_m = f_0 \cdot 10^{-\frac{m}{2.5}}. \]

\noindent However, this algorithm is robust against the source flux intensity, because the flux offset must be eliminated before applying it. At the same time, it requires a good knowledge of this offset. 
Another limit is the need of integration over complete fringes, in order to be able to apply the trigonometric relations.\\ 
Therefore, a good knowledge of the source spectrum is also necessary, to define a correct effective wavelength.\\

\noindent We remark, finally, that in practice a suitable effective spectral bandwidth with appropriate intensity can be used. This produces different templates $t_1$ and $t_2$, but allows to adapt this algorithm to a wide range of different instruments.



\subsection{Correlation with a template}
\label{subsec:correlation}

Another algorithm we proposed for the detection of the phase modulo $\lambda$ with FINITO is the `correlation method'. It is based on the correlation between the measured interferogram and a template, in order to find the maximum of the correlation function. The OPD correspondent to that maximum in the template is the current OPD.\\
Its greater advantage with respect to the demodulation algorithm is the fact that it doesn't require the modulation over an integer number of fringes, relaxing the requirements on the knowledge of the working spectral range.\\

\noindent The correlation is limited to a single fringe. At the beginning of the scan, the interferogram intensity is collected for a fringe, then the template is correlated to find the actual phase on the OPD. For the next step, every collected sample can be added to the fringe pattern, discarding the oldest one, and correlated again with the template.  In this way, there is a loss of information just for the time needed for the first fringe collection.\\
The maximum of the correlation function is found searching the maximum of the interpolating polynomial, of second order being the involved functions sinusoidal.\\
Being the correlation limited to a single fringe, the template does not need to be modulated. Besides, the introduction of modifications to take into account envelope modulation, or slight departure from the full fringe assumption (i.e. variations of the effective wavelength), is straightforward.
\\

\noindent We first seek for the behaviour of the algorithm when there is no noise on the OPD modulation, and on the intensity pattern. The input interferogram is modelled following the description of eq. \ref{eq:constr_destr}, and with the following parameters:
\begin{itemize}
    \item{nominal intensities for $I_1$ and $I_2$: $3.085\cdot 10^5$, for magnitude 13}
    \item{visibility V = 0.73}
    \item{waverange: H band, with central wavelength $\lambda_0 = 1.65 \mu m$}
	\item{quantum efficiency: 60\%}
	\item{sampling frequency: 4 kHz}
	\item{fringe per semi-ramp: 10}
	\item{sample per fringe: 20}
    \item{OPD range: $16.5 \mu m$, i.e. 10 fringes}
\end{itemize}

\noindent  and the template is constructed as the sinusoidal wave:
\begin{equation}
t(x) = sin (\frac{2\pi}{\lambda_0} \cdot OPD_t)
\end{equation}
where the subscript $i$ for the OPD is the same stepping of the modulation ramp (a twentieth of $\lambda_0$).

\begin{figure}[htb]
      \begin{center}
        \epsfig{figure=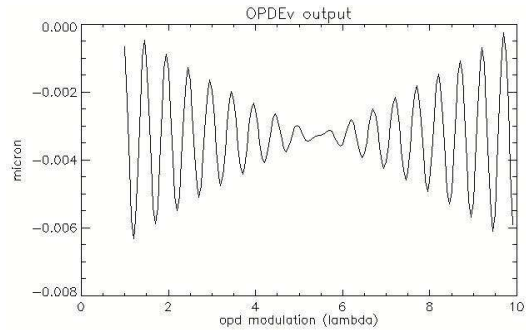,width=7cm}
        \caption{Error on the phase evaluated by the algorithm when no noise is introduced over the modulated OPD. The mean error is $-0.003 \; \mu m$, its standard deviation is $0.003 \; \mu m$.}
        \label{fig:correl_noNoise}
    \end{center}
\end{figure}

\noindent Figure \ref{fig:correl_noNoise} shows the evaluated maximum of the correlation function. The most remarkable result is the superposed oscillatory behaviour. Its mean is of $-3.3 nm$ and its frequency is roughly double the fringe frequency. This effect is probably due to a beating between the interferometric signal and the template.\\ 
This is the best performance of the algorithm in nominal conditions, and its performance is still good, compared to the ideal case. The model induced error is acceptable, in most observing case.\\

\begin{figure*}[htbp]
      \begin{center}
        \epsfig{figure=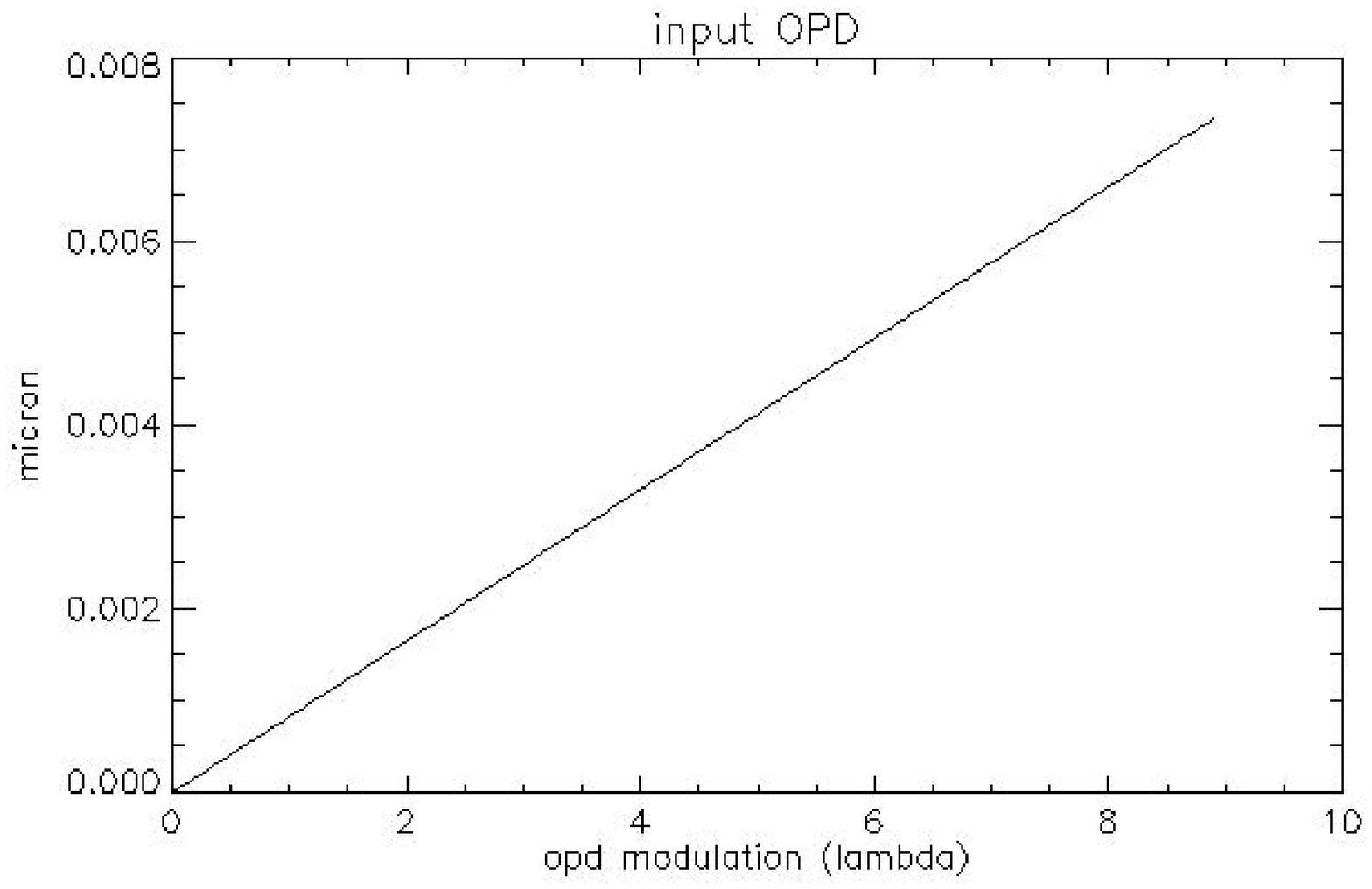,width=4.5cm}
        \epsfig{figure=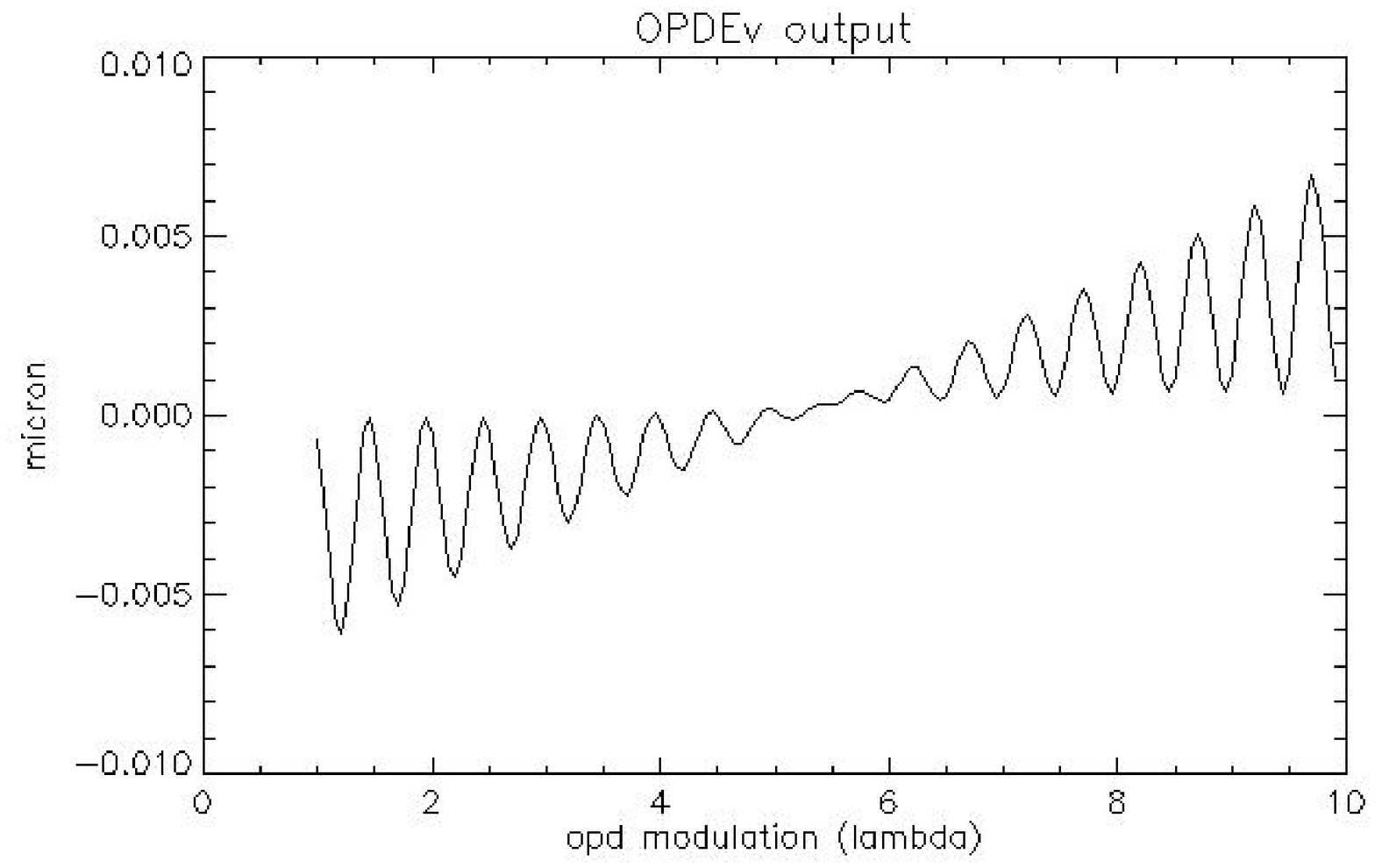,width=4.5cm}
        \epsfig{figure=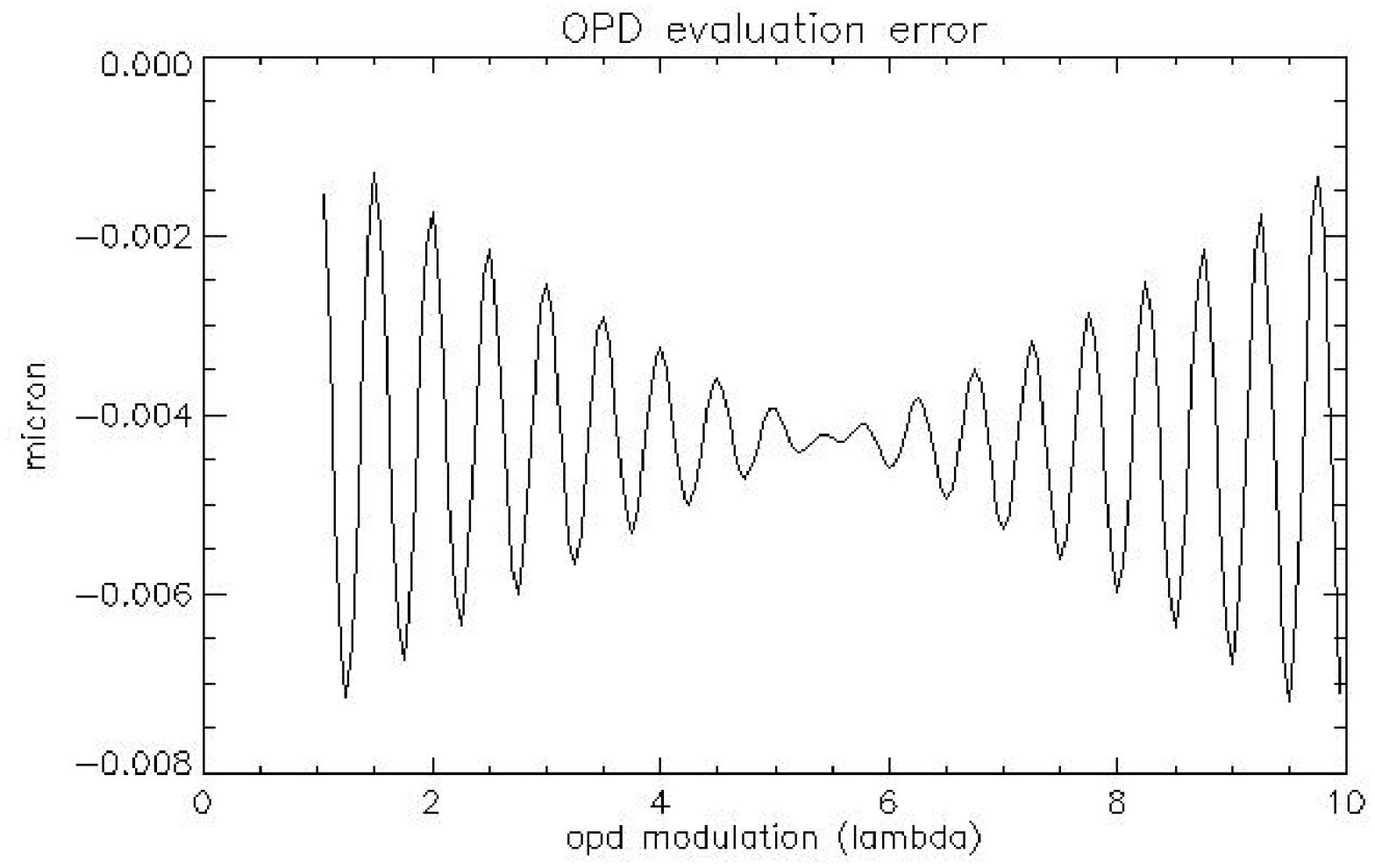,width=4.5cm}
        \caption{Linear noise on the OPD, nominal intensity. Left: input phase on OPD, center: evaluated phase, right: difference between the previous. The mean error is $0.2 \; nm$, its standard deviation is $0.002 \; \mu m$.}
        \label{fig:correl_noiseOnOPD}
    \end{center}
\end{figure*}

\begin{figure*}[htbp]
      \begin{center}
        \epsfig{figure=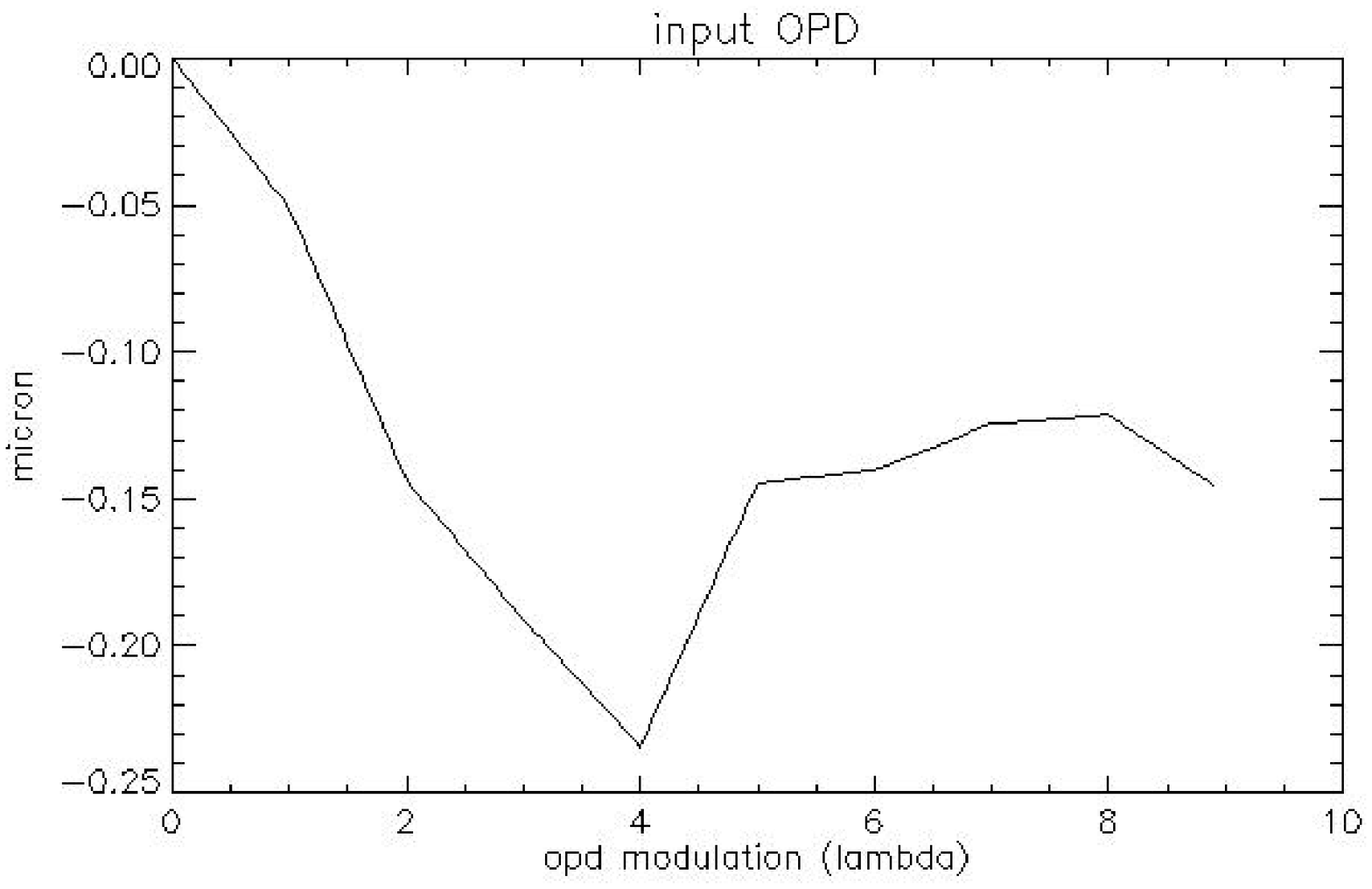,width=4.5cm}
        \epsfig{figure=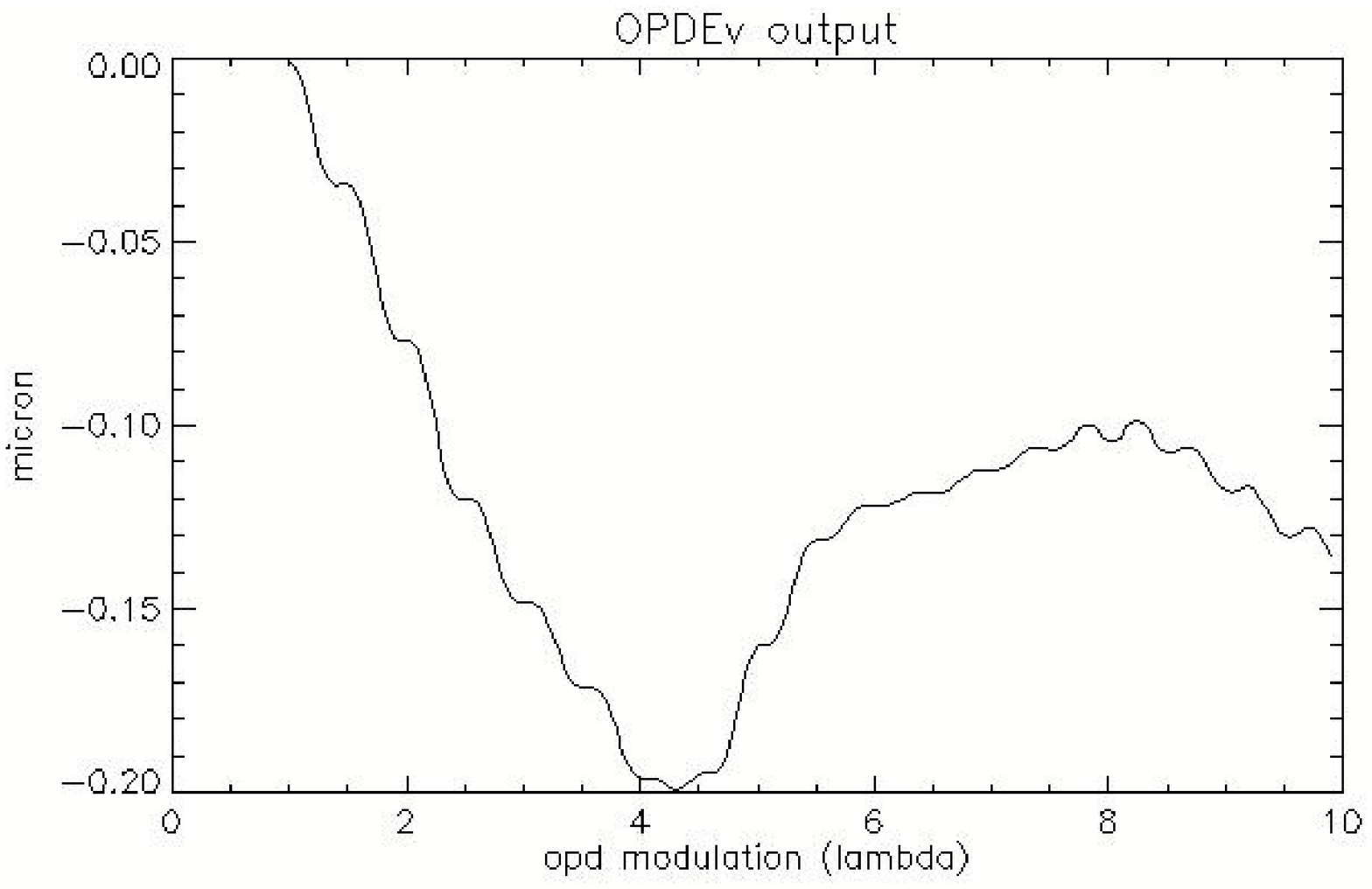,width=4.5cm}
        \epsfig{figure=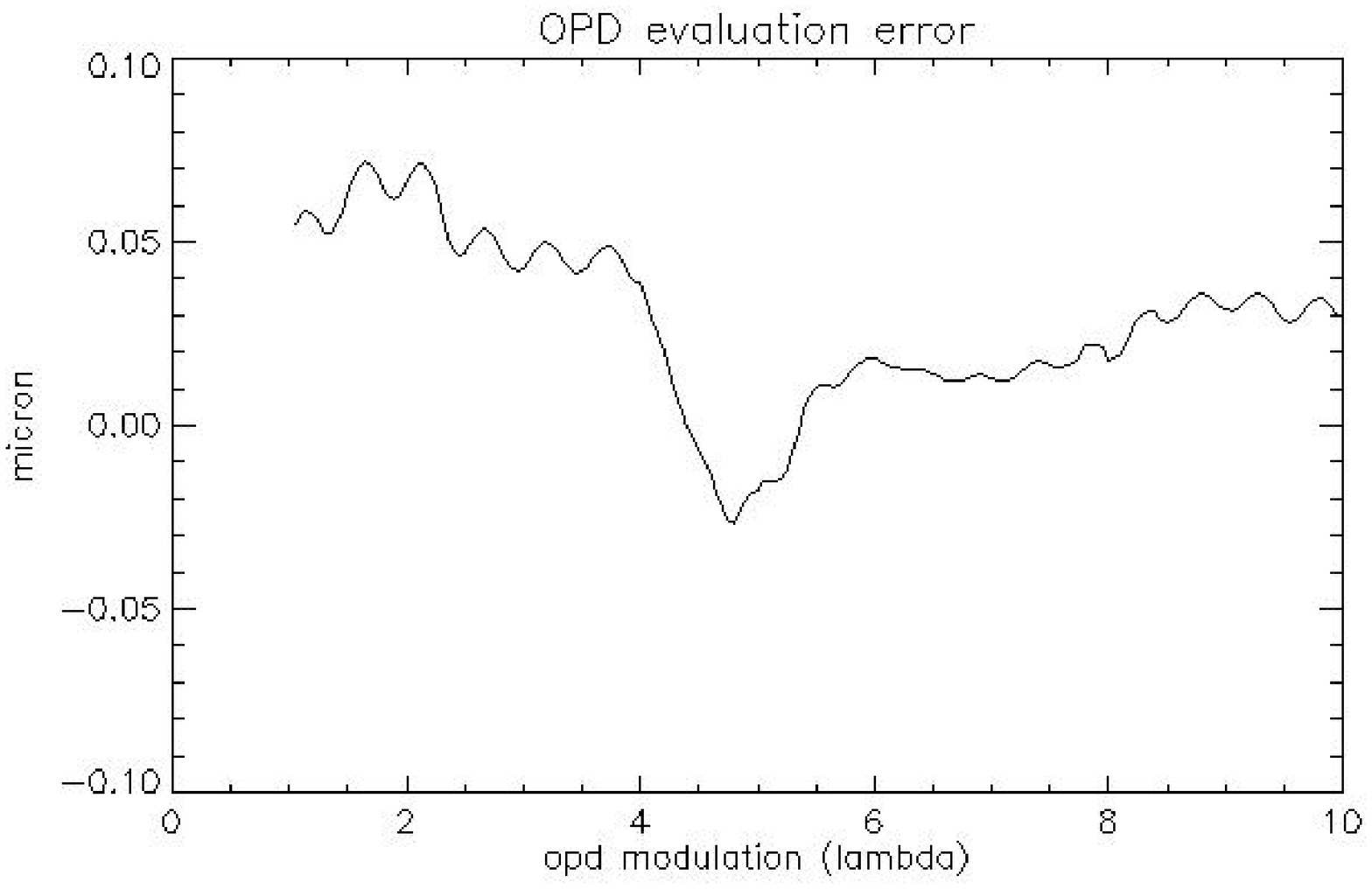,width=4.5cm}
        \caption{Atmospheric noise on the OPD, nominal intensity. Left: input phase on OPD, center: evaluated phase, right: difference between the previous. The mean error is $-0.122 \; \mu m$, its standard deviation is $0.130 \; \mu m$.}
        \label{fig:correl_noiseOnOPD_fringeJump}
    \end{center}
\end{figure*}

\noindent When the introduced phase on the OPD is no longer zero, we find again the oscillatory phenomena, even if the algorithm is able to follow the OPD pattern. Different cases are shown below (see figure \ref{fig:correl_noiseOnOPD}), with a linear phase on the modulated OPD, and then with a more realistic noise (see fig. \ref{fig:noise1}).\\
Figure \ref{fig:correl_noiseOnOPD_fringeJump} shows what happens when the introduced phase induces the overall OPD to be greater than $\lambda$, i.e. when the intensity jumps in a lateral fringe instead of remaining into the central one. This phenomenon is called {\it fringe jump}. To simulate it, we generate the same atmospheric noise than before, but with a greater amplitude (factor 10), so that the noise induces big shift of the whole interferogram.\\

\begin{figure*}[htbp]
      \begin{center}
        \epsfig{figure=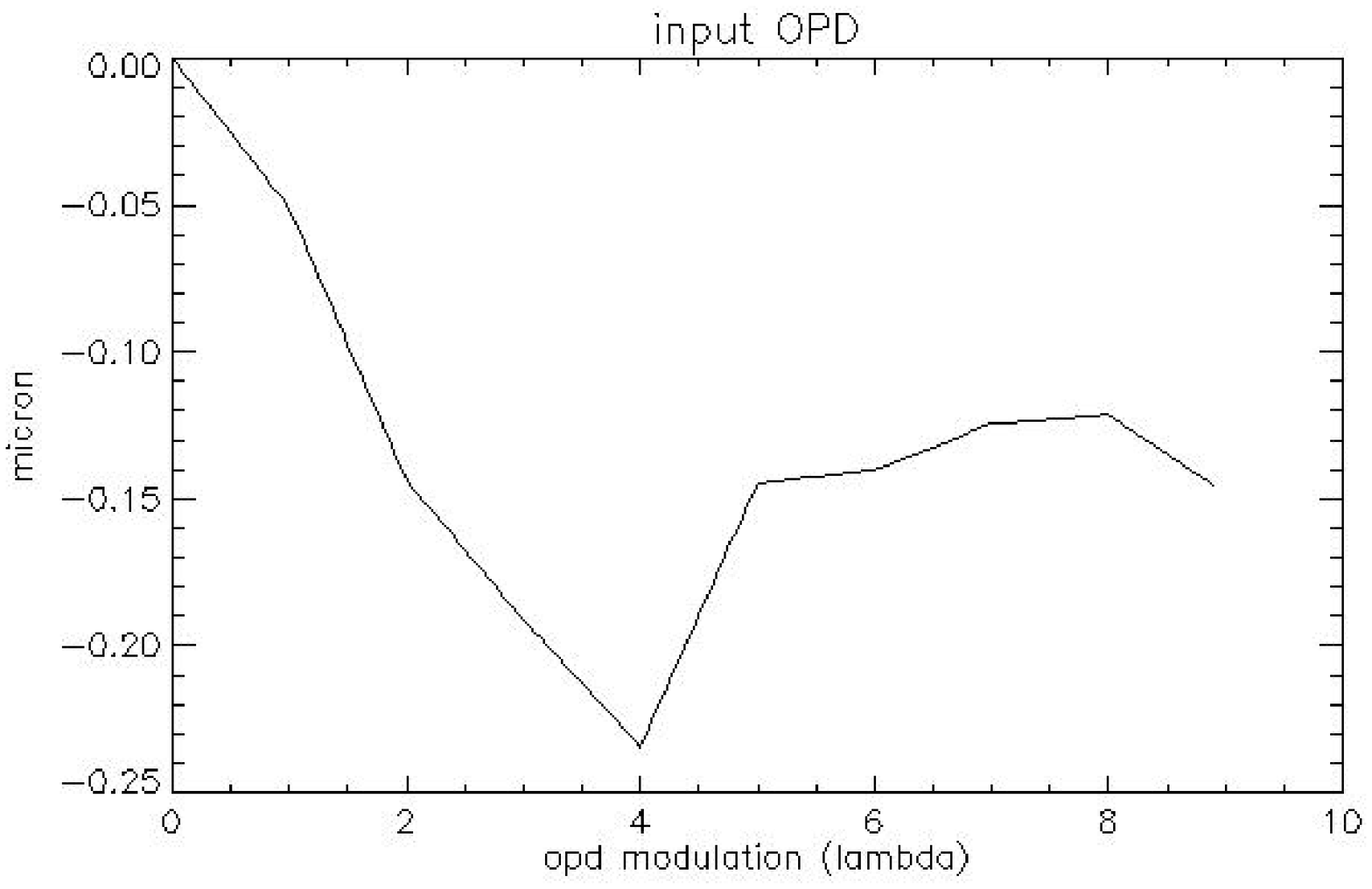,width=4.5cm}
        \epsfig{figure=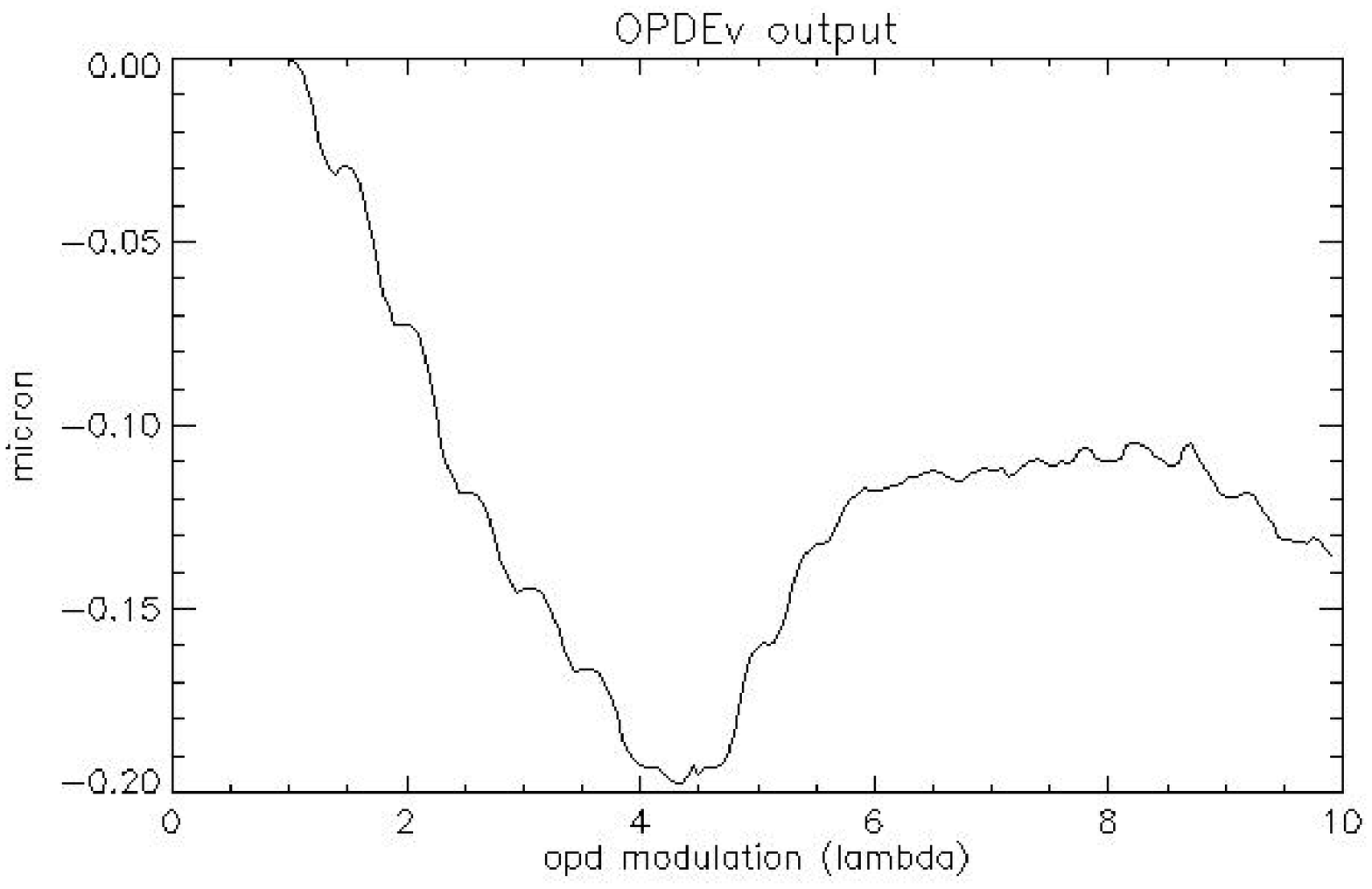,width=4.5cm}
        \epsfig{figure=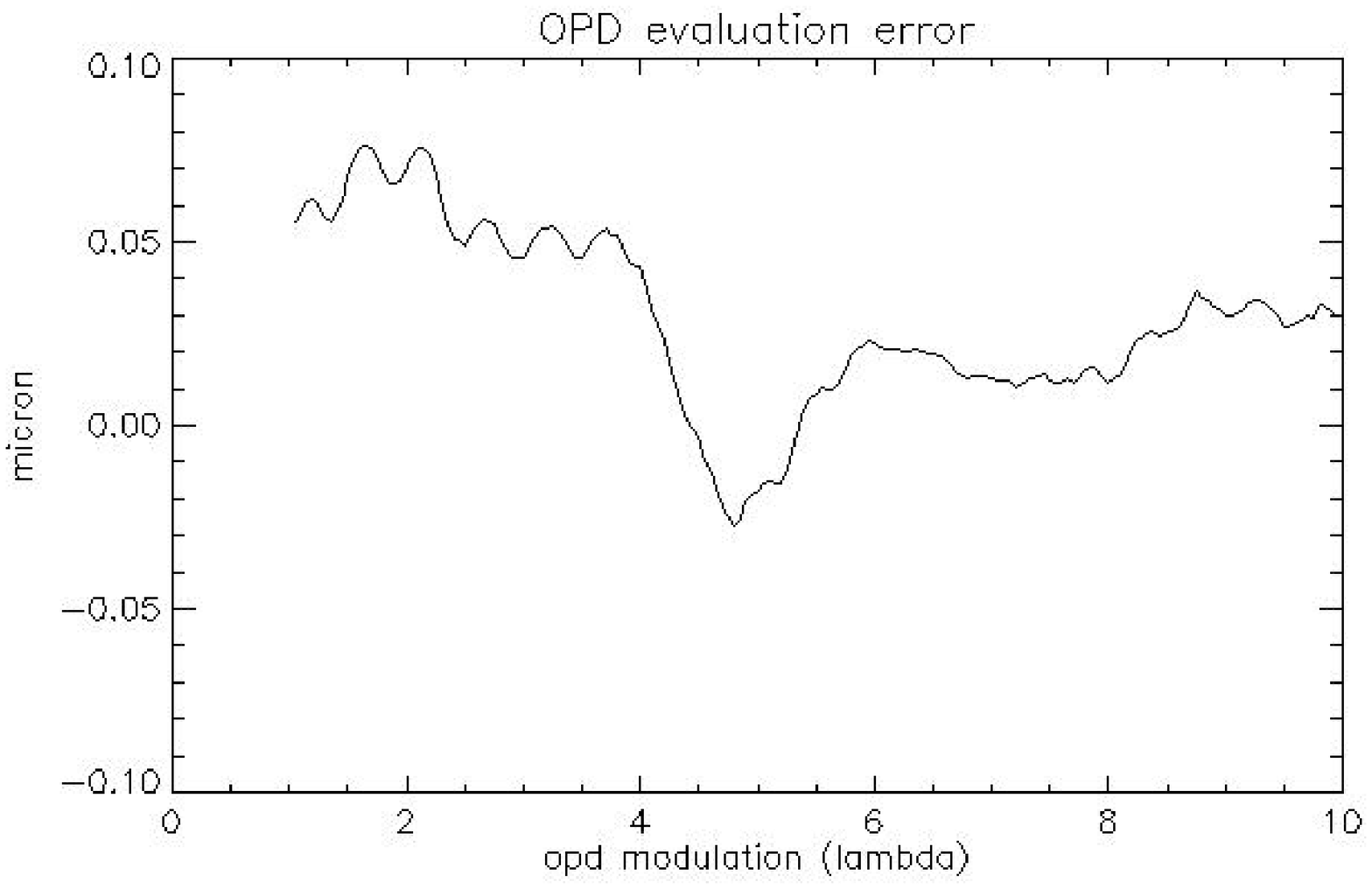,width=4.5cm}
        \caption{Atmospheric noise on the OPD, noisy intensity. Left: input phase on OPD, center: evaluated phase, right: difference between the previous. The OPD noise does not cause fringe jumps. The mean error is $-0.121 \; \mu m$, its standard deviation is $0.129 \; \mu m$}
        \label{fig:correl_noise1+noisyFringe}
    \end{center}
\end{figure*}

\noindent Finally, we perform a realistic simulation adding both atmospheric turbulence on the OPD and noise on the interferogram intensities. The results are shown in figure \ref{fig:correl_noise1+noisyFringe}.\\

\noindent We have to notice, however, that for atmospheric noise the performance is worsening, even with nominal intensity.\\

\begin{figure*}[htbp]
      \begin{center}
        \epsfig{figure=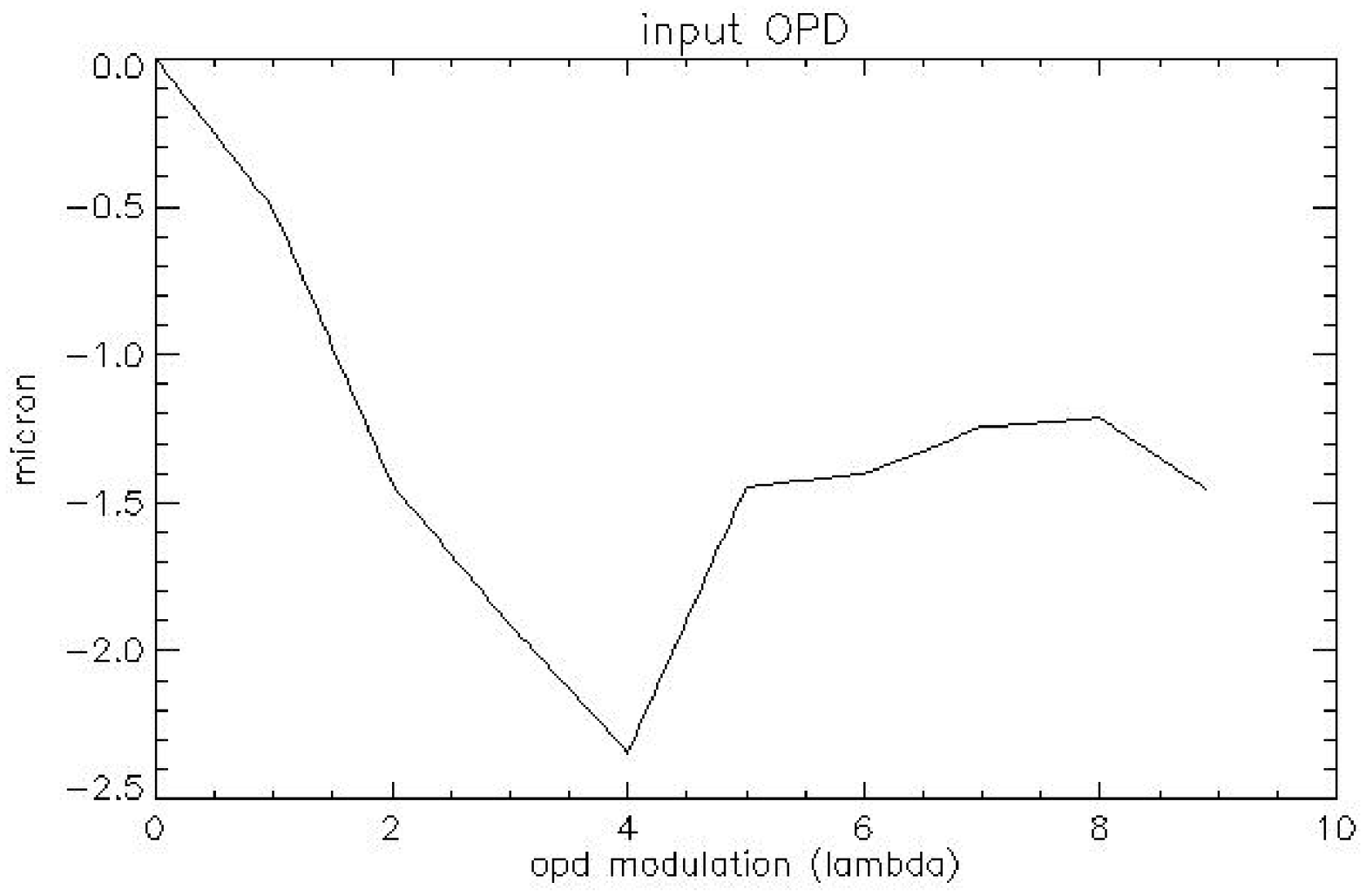,width=4.5cm}
        \epsfig{figure=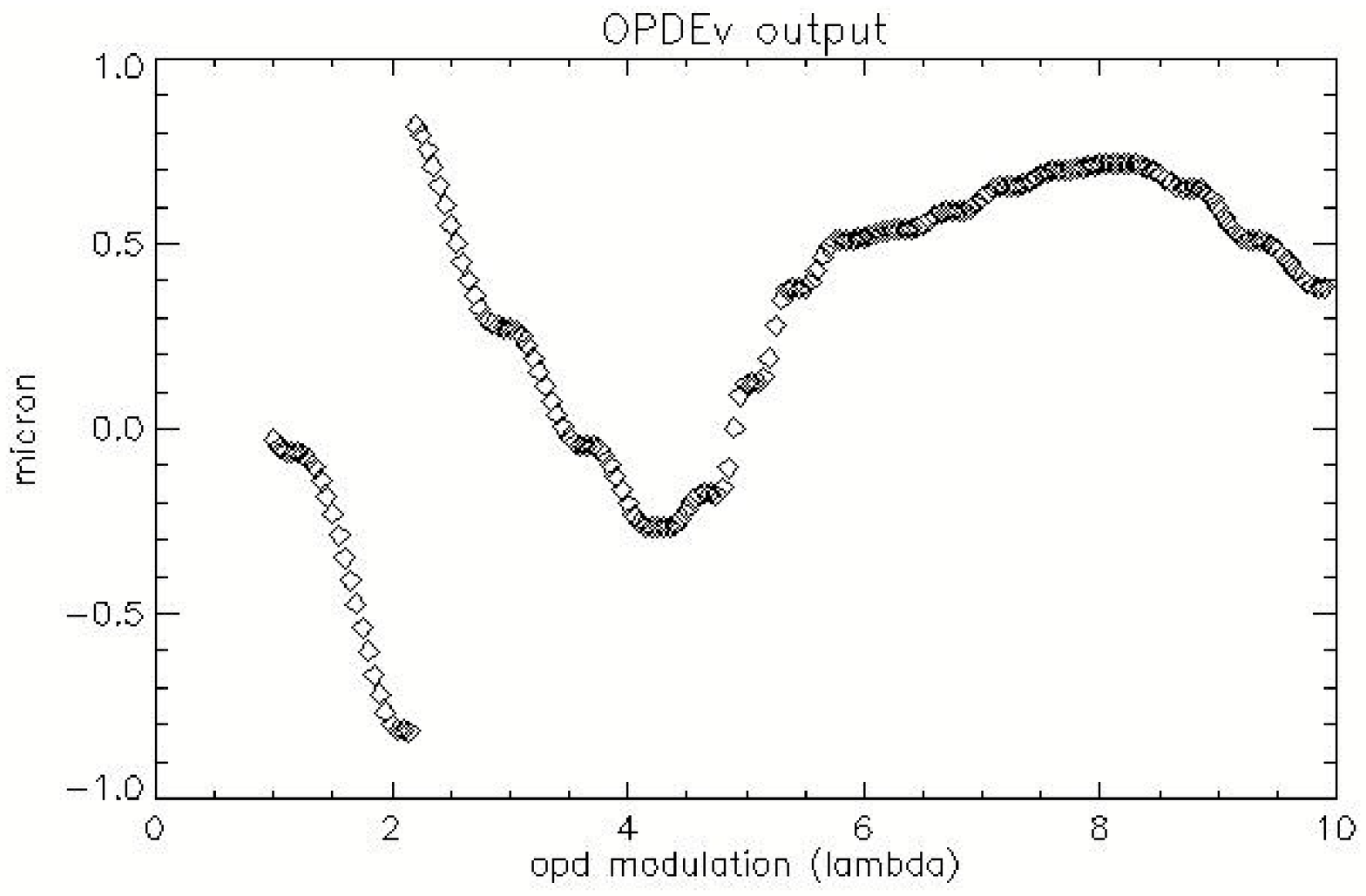,width=4.5cm}
        \epsfig{figure=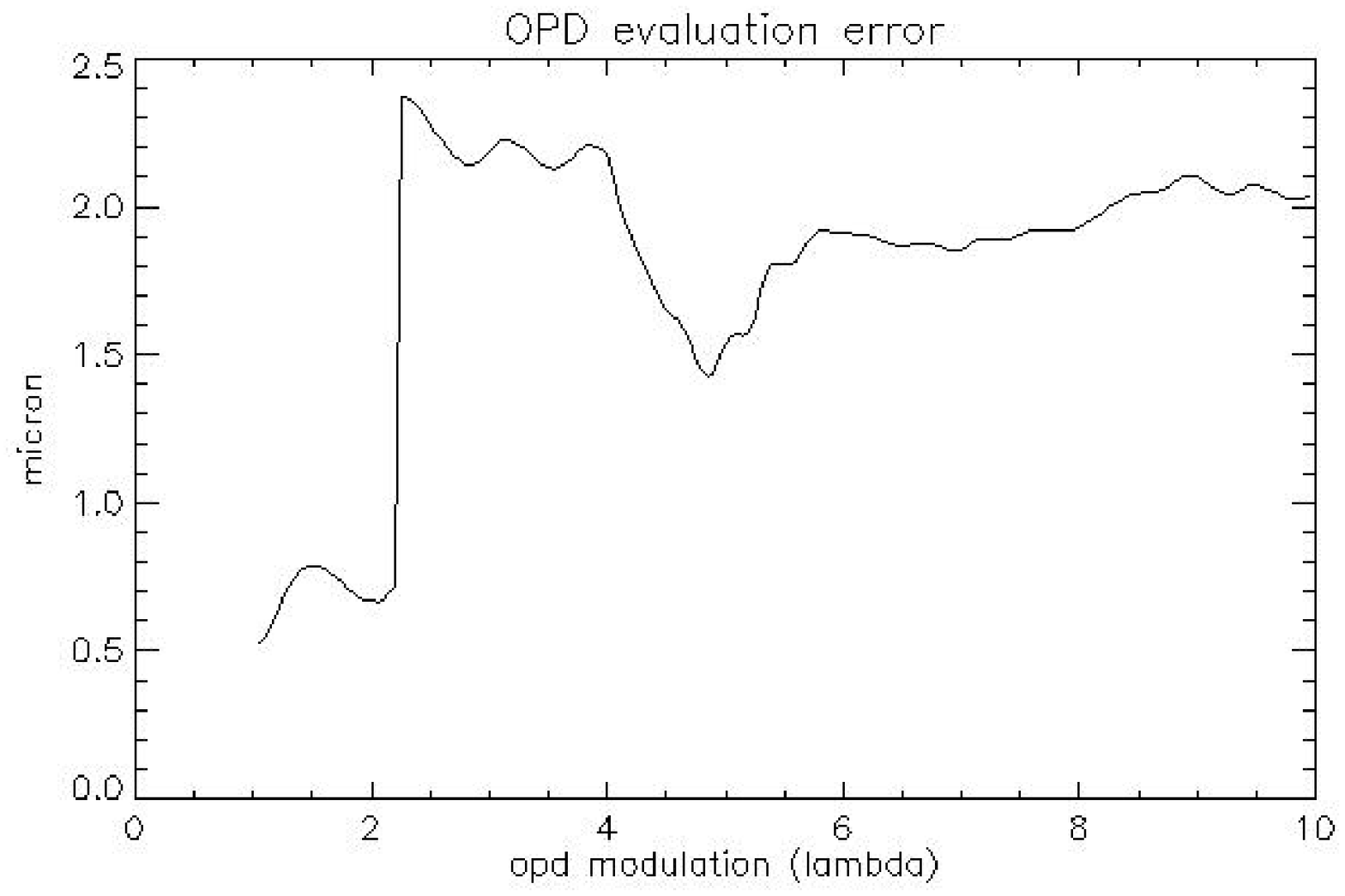,width=4.5cm}
        \caption{Atmospheric noise on the OPD, nominal intensity. Left: input phase on OPD, center: evaluated phase, right: difference between the previous. The graphic range is different from before, since the amplitude of the noise is greater than \ref{fig:correl_noiseOnOPD_fringeJump} of a factor 10. The mean error is $0.283 \; \mu m$, its standard deviation is $0.495 \; \mu m$.}
    \end{center}
\end{figure*}

\noindent Analitically, it can be convenient to model the interferometric beam and the template as sinusoidal waves at different frequencies\cite{Gardiol}:
\begin{eqnarray}
\nonumber s_i(t) &=& \cos ([\omega_m + \omega_{atm}] \cdot t)\\
s_t(t) &=& \cos (\omega_m \cdot t)
\end{eqnarray}
where $\omega_m = \frac{2 \pi}{N \lambda}$ is the frequency of the modulation, and N is such that $t$ covers a wavelength in a period T. At the same way, $\omega_{atm}$ is the unknown atmospheric frequency.
\\
\noindent The correlation function between the two beams is:
\begin{eqnarray}
\nonumber C(\phi) &=& \int_\tau^{\tau + T} {s_i(t) \cdot s_t(t+\phi) dt} = \int_\tau^{\tau + T} {cos ([\omega_m + \omega_{atm}] * t) cos (\omega_m \cdot t + \phi) dt} = \\
\nonumber &=& cos (\phi) \int_\tau^{\tau + T} {cos ([\omega_m + \omega_{atm}] * t) cos (\omega_m \cdot t) dt} + \\
&-& sin (\phi) \int_\tau^{\tau + T} {cos ([\omega_m + \omega_{atm}] * t) sin (\omega_m \cdot t) dt}
\end{eqnarray}
\noindent Maximum of the correlation function are among the zero of the derivative function $\frac{dC(\phi)}{d\phi}$:
\begin{eqnarray}
\nonumber sin (\phi) \int_\tau^{\tau + T} {cos ([\omega_m + \omega_{atm}] * t) cos (\omega_m \cdot t) dt} + \\
- cos (\phi) \int_\tau^{\tau + T} {cos ([\omega_m + \omega_{atm}] * t) sin (\omega_m \cdot t) dt} = 0
\end{eqnarray}
from which we obtain:
\begin{equation}
tan (\phi) = \frac{\int_\tau^{\tau + T} {cos ([\omega_m + \omega_{atm}] * t) sin (\omega_m \cdot t) dt}}{\int_\tau^{\tau + T} {cos ([\omega_m + \omega_{atm}] * t) cos (\omega_m \cdot t) dt}}
\end{equation}
which leads to: 
\begin{equation}
tan (\phi) = \frac{sin(\omega_{atm}\tau +  \pi \alpha) - \gamma sin[(\omega_{atm} + 2 \omega_m)\tau + \pi \alpha]}{cos(\omega_{atm} \tau + \pi \alpha) + \gamma cos[(\omega_{atm} + 2 \omega_m)\tau + \pi \alpha]}
\end{equation}
where we have defined:
\begin{equation}
\alpha = \frac{\omega_{atm}}{\omega_m}, \;\; \gamma = \frac{\alpha}{\alpha+2}
\end{equation}

\noindent A closer look to that formula leads us to a few considerations. \\
First of all, we recognize the doubled modulation frequency that we have discovered on the evaluated phase. This factor is generated by the algorithm itself, since the other terms depend just on the introduced phase, referred to the initial condition ($\omega_{atm} \tau$), and can be weakened by taking as reference the measurement of the OPD at time $\tau$.
\\
\noindent If we assume the atmospheric frequency to be low with respect to the modulation one, so $\omega_{atm} \ll \omega_m$, then $\alpha\sim0$, $\gamma\sim0$, and we are left with:
\[tan(\phi) \sim tan (\omega_{atm}\tau).\]
\\
\noindent However, in our first simulation no external phase was introduced. In that case the error had a constant threshold and a modulation. This effect can be caused by the fact that the template is calibrated on the central fringe, while the interferogram is polychromatic, and the effect of polychromaticity is stronger at increasing distance from the zero OPD.

\subsubsection{Remarks}
This algorithm provides quite good results, but has a threshold error that can't be avoided even in nominal situation. This is due to the shape of the template for the correlation, which relies only on the introduced modulation path and the working wavelength, that should be known and sufficiently stable. However, a more detailed template should require a proper calibration of fringes parameters such as intensity, spectral range, and so on. These enhancements depends, however, on the knowledge of the instrument.

\noindent Oscillations are often present, and in many cases sufficiently small. They are due basically to the beating induced by the mismatch between the real fringe frequency and the modulation speed or the external OPD rate of variation.

\subsection{Modified ABCD}
\label{subsec:FINITO_ABCD}

The AC and ABCD algorithms are modifications of the simple trigonometric equations that apply in the ideal case (eq. \ref{eq:idealABCD}). With these algorithms there is no need to detail the model of the incoming beam, but its normalization is mandatory, otherwise the trigonometric relations are lost.\\


\noindent In fact, the aim of the normalization is to bring the signal to a sinusoidal wave with unitary amplitude and zero mean. There are two scenarios: if photometry is available (as the case of FINITO) and if not.\\
In the first case, the photometry values can be used in real time for the correction of the interferometric outputs. This method is used, e.g., by the VINCI interferometer (\cite{coudedeforesto97}), and by FINITO.\\
In the latter case, information about the offset and the current amplitude of the interferometric signal must be deduced independently, e.g. with a slow calibration off-line. This is the case of the PRIMA FSU, and will be discussed later.\\

\noindent If the photometry is available, the interferometric outputs can be described in terms of the photometric ones. A detailed derivation of a calibrated interferometric pattern when working off-line can be found in \cite{coudedeforesto97}. In this case, the working conditions are relaxed, and data can be carefully denoised before using them.
\\
\noindent When working in real time, however, the operations must be reduced to the minimum. This is the case of FINITO, and we illustrate it.\\
If $P_A$ and $P_B$ are the photometric inputs, and $I_1$ and $I_2$ the interferometric signals after the combination, then a simple normalization is, derived from \cite{FINITOTestParanal}:
\begin{eqnarray}\label{eq:FINITO_ESO_Inorm}
\nonumber I_1^{norm} = \frac{I_1 - \beta_{1,A }P_A - \beta_{1,B}P_B}{\sqrt{P_A \cdot P_B}}\\
I_2^{norm} = \frac{I_2 - \beta_{2,A }P_A - \beta_{2,B}P_B}{\sqrt{P_A \cdot P_B}}
\end{eqnarray}

\noindent The $\beta$ coefficients have to be evaluated before the observation, monitored and periodically updated. They intrinsically contain information about the source, e.g. the wavelength distribution, but also instrumental ones, such as the coupling ratio of the photometric beam splitters, the efficiency of the transmission system from the combination to the detection. Their values can be influenced by all these factors.
\\


\noindent After the normalization, eq. \ref{eq:idealABCD} can be used, acquiring four samples for fringe, separated by a $\pi/4$ offset, through a modulation of each interferometric outputs, for example.

\noindent With this approach, the expected performance is that of ideal ABCD (see eq. \ref{eq:ABCDperf}). In that formula, information on the incoming flux and its noise are contained in the SNR term. If we suppose that the astronomical beam is subject to photon noise, background and detector noise, the SNR can be expressed as\cite{VLTScience}:
\begin{equation}
\mbox{SNR}_V = \frac{S\cdot N_p\cdot V}{\sqrt{1 + N_b + N_p + N_d^2}}
\end{equation}
where $N_p$, $N_b$ and $N_d$ are the beam flux, the background flux and the standard deviation of the detector noise, respectively, all expressed in terms of number of detected photons, $S$ is the Strehl ratio, i.e. the ratio of the reference intensity to the measured one, and $V$ is the visibility.

\subsubsection{Application of ABCD method for FINITO at VLTI}
\label{subsubsec:FINITO_Normalization}

\begin{figure}[htb]
      \begin{center}
        \epsfig{figure=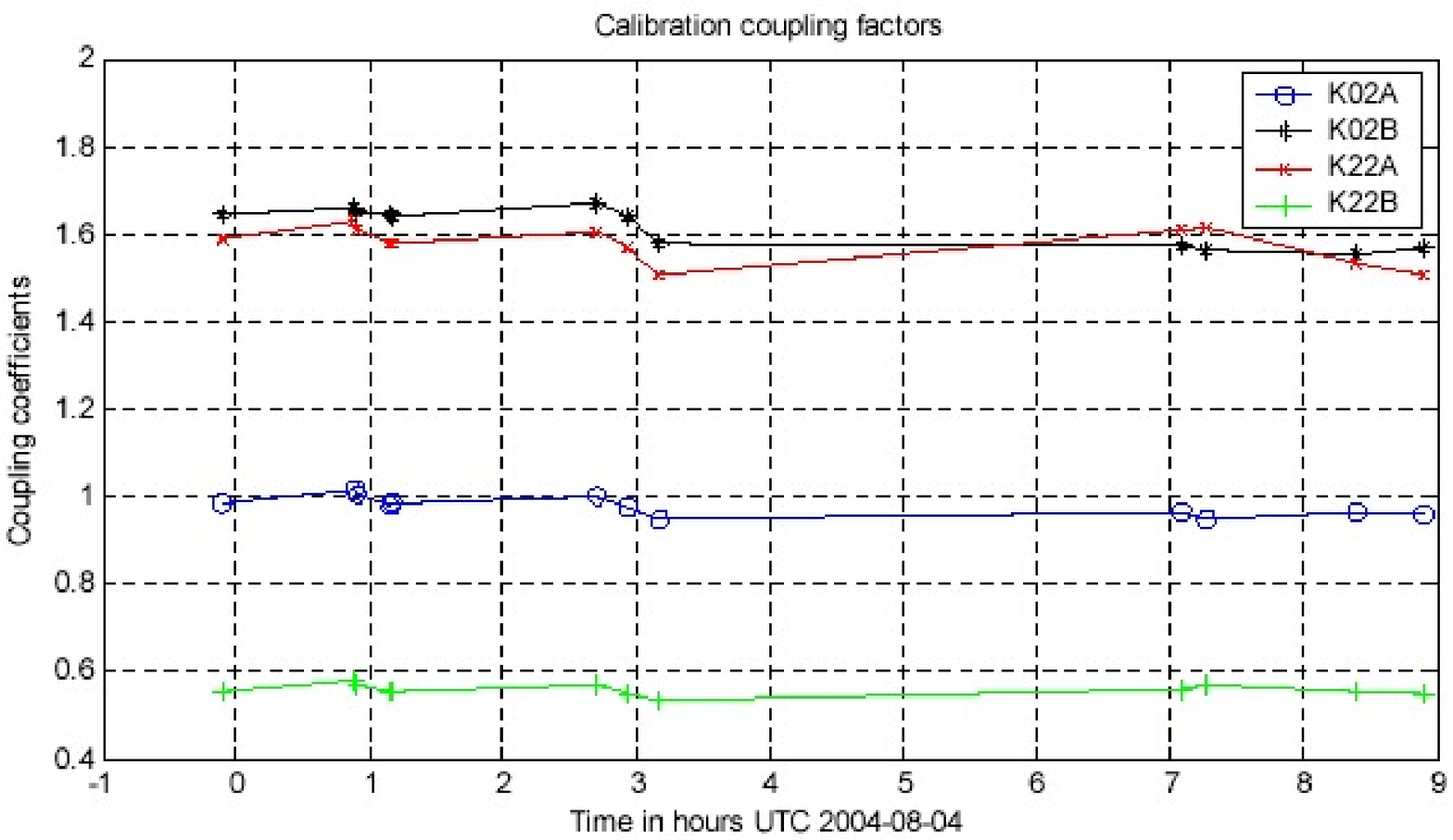,width=6.5cm,height=5cm}
        \epsfig{figure=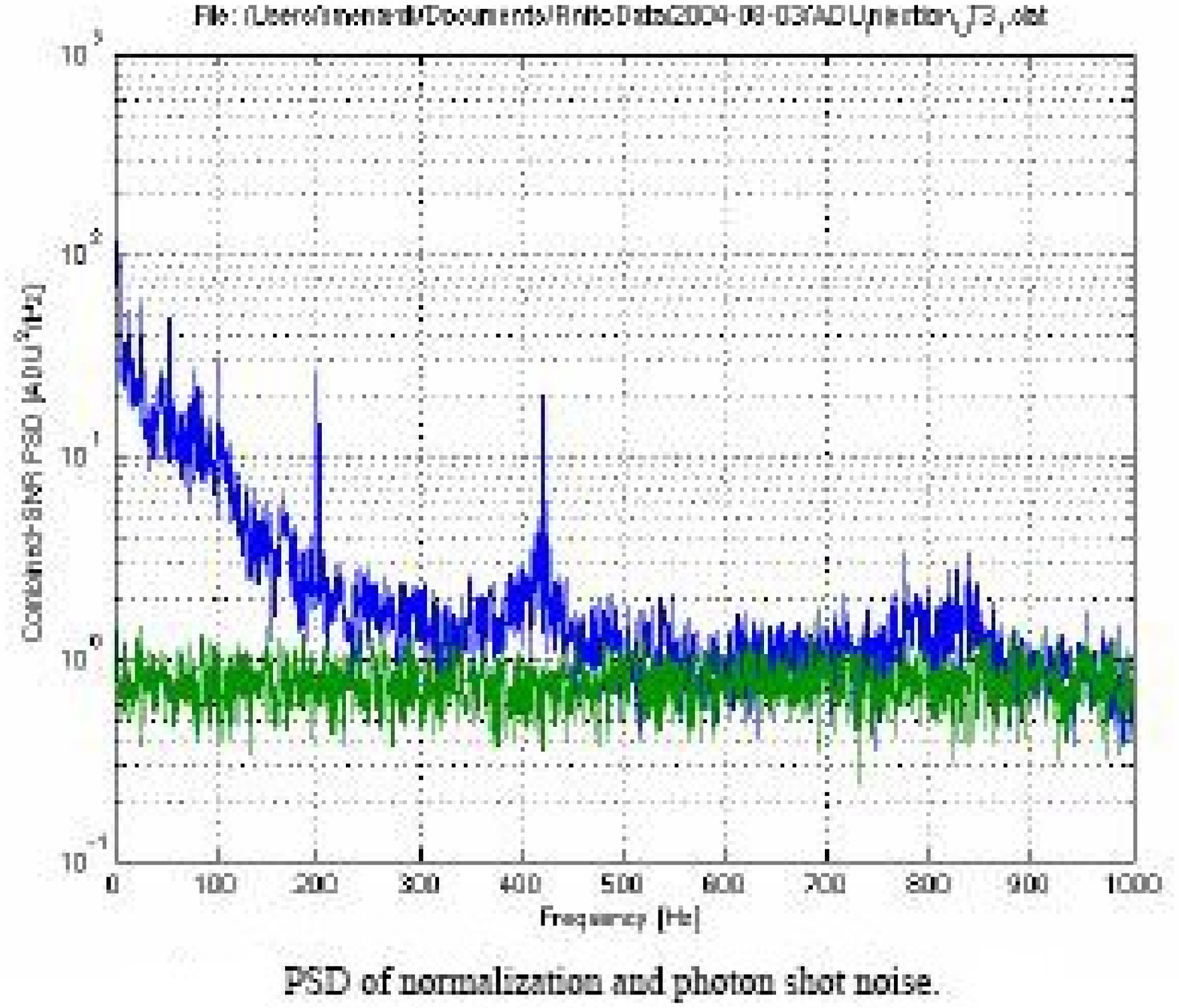,width=8cm}
        \caption{First row, calibration coefficients monitored for one night; second row, comparison between the spectral behaviour of the noise due to normalization and the standard deviation of photometric channel. These figures are taken from \cite{FINITOTestParanal}.}
        \label{fig:FINITO_calibESO}
    \end{center}
\end{figure}

As mentioned before, the FINITO performance when installed at VLTI suffered for the bad working conditions. For this reason, the ABCD method has been chosen, among all proposed, for its robustness.\\
However, this algorithm is sensitive to residual effects due to normalization, in particular to the interaction between the photometric signals $PA$ and $PB$. In presence of such correlations, it is difficult to theoretically estimate their effects on the performance. This was the case of FINITO, subject to unexpected flux fluctuations\cite{FINITOTestParanal}. Figure \ref{fig:FINITO_calibESO}, first row, shows the calibration coefficients evaluation during one observational night: it is evident that they are changing.
The residual noise from calibration, computed as the standard deviation of the normalized signals of eq. \ref{eq:FINITO_ESO_Inorm}, compared with the standard deviation of the photometric inputs reveals that new features where added with the normalization. Figure \ref{fig:FINITO_calibESO}, second row, shows the spectral behaviour of these two noises.

\section{Monitoring of intensity fluctuations}
\label{sec:int_Fluct_SPIE04}

\noindent For the reasons mentioned in the previous paragraphs, the need of intensity fluctuations control arises. In fact, if it is possible to rapidly get aware of a significant flux variation, the coefficients for the normalization can be updated, in order to better fit the injected beams features.

\noindent In this section, we propose\cite{Bonino04} two different intensity estimation algorithms based on the estimated phase. We analyze the effects of a discrepancy of the intensity on the estimation of both phase on the optical path and on intensity itself, in a ideal situation. In fact, for every algorithm seen till now, the knowledge of the current flux intensity is crucial, and we have seen how noise on intensity causes in turn an error on the OPD estimation.

\noindent Let us describe the two complementary outputs on an interferometric system as:
\begin{eqnarray}
\nonumber s_1 (x) &=& I \cdot \left[1 + V \sin \left(\frac{2\pi}{\lambda}x\right)\right],\\
s_2 (x) &=& I \cdot \left[1 + V \cos \left(\frac{2\pi}{\lambda}x\right)\right]
\end{eqnarray}
In the equation, $I$ is the flux intensity, $V$ is the visibility, $\lambda$ is the working wavelength and $x$ is the optical path difference. In ideal case, $I$, $V$ and $\lambda$ are constant. So we can easily find the phase $\phi = \frac{2\pi}{\lambda}x$ as:
\begin{equation}\label{eq:phi}
\phi = \arctan \frac{m_1(x)}{m_2(x)} \; \rightarrow \; x = \frac{\lambda}{2\pi} \phi
\end{equation}
where we have defined:
\begin{eqnarray}
\nonumber m_1 (x) = s_1(x) - I &=& I \cdot V \sin \left(\frac{2\pi}{\lambda}x\right)\\
m_2(x) = s_2 (x) - I &=& I \cdot V \cos \left(\frac{2\pi}{\lambda}x\right)
\end{eqnarray}

\noindent Let now allow the intensity to vary: $I = I_0 + \delta I_0$, where $I_0$ is the known nominal intensity, while $\delta I_0$ is unknown. We find an estimation of the intensity $I$ first minimizing an error function, then with trigonometric elaboration of the original signals.\\

\noindent For the first approach, if we subtract from $s_1(x)$ and $s_2(x)$ the $I_0$ nominal intensity instead of the true $I$, we obtain the following signals:
\begin{eqnarray}
\nonumber \tilde{m_1} (x) = s_1(x) - I_0 &=& \delta I_0 + (I_0 + \delta I_0) V \sin \left(\frac{2\pi}{\lambda}x\right)\\
\tilde{m_2}(x) =  s_2 (x) - I_0 &=& \delta I_0 + (I_0 + \delta I_0) V \cos \left(\frac{2\pi}{\lambda}x\right)
\end{eqnarray}
Substituting them into eq. \ref{eq:phi}, we obtain a perturbed estimation $\tilde{\phi}$ of the phase $\phi$:
\begin{equation}
\tilde{\phi} = \arctan \frac{\tilde{m_1}(x)}{\tilde{m_2}(x)}
\end{equation}

\noindent With this estimate of the phase, we construct two sinusoidal templates:
\begin{equation}
t_1(x) = I_0 \cdot V \sin \tilde{\phi}, \;\; t_2(x) = I_0 \cdot V \cos \tilde{\phi}
\end{equation}
that we use to find the intensity value that minimize the squared error
\[S^2 (\phi) = \sum_{i=1,2} (\tilde{m_1} (x) - t_i(x))^2\]
The minimum is found searching the zero of the derivative function $\frac{dS^2}{dI_0}$. We find:
\begin{equation}
\tilde{I}(x) = \frac{\tilde{m}_1(x)\cdot \sin \tilde{\phi} + \tilde{m}_2(x)\cdot \cos \tilde{\phi}}{V}
\end{equation}
This evaluation suffers for the low number of measurements for each estimate. It is nevertheless interesting because it allows an iterative process for estimates of both phase and intensity:
\begin{eqnarray}\label{eq:int_iterative}
\nonumber \tilde{m}_{1,i}(x) = s_1(x) &-& \tilde{I}_{i-1}(x), \;\;\;\; \tilde{m}_{2,i}(x) = s_2(x) - \tilde{I}_{i-1}(x)\\
\nonumber \tilde{\phi_i} &=& \arctan \frac{\tilde{m}_{1,i}(x)}{\tilde{m}_{2,i}(x)}\\
\tilde{I}_i(x) &=& \frac{\tilde{m}_{1,i}(x)\cdot \sin \tilde{\phi}_i + \tilde{m}_{2,i}(x)\cdot \cos \tilde{\phi_i}}{V}
\end{eqnarray}

\noindent The second iterative method we analyze is based on the manipulation of the equation $s_1^2(x) + s_2^2(x)$, which leads to:
\begin{equation}
\frac{\delta I_0^2}{I_0^2} + 2 \frac{\delta I_0}{I_0} - \frac{s_1^2(x) + s_2^2(x)}{I_0^2[2 + V^2+ 2 V (\sin \phi + \cos \phi)]} +1 = 0
\end{equation}
which has two solutions:
\begin{equation}
\frac{\delta I_0}{I_0} = - 1 \pm \sqrt{ \frac{s_1^2(x) + s_2^2(x)}{I_0^2[2 + V^2+ 2 V (\sin \phi + \cos \phi)]}}
\end{equation}
For $0 \leq V \leq 1$ and $\forall \phi$, we have $2 + V^2+ 2 V (\sin \phi + \cos \phi) > 0$, so the two solutions exist and they are real. The factors of the ratio in the square root are comparable, the only difference is the intrinsic influence of the perturbed $I$ instead of $I_0$ in $s_1^2(x) + s_2^2(x)$: the ratio is near zero. So the two solutions are:
\begin{equation}
\frac{\delta I_0}{I_0} \sim 0, \;\;\;\; \frac{\delta I_0}{I_0} \sim -2.
\end{equation}
The first one is the searched solution.\\

\begin{figure*}[htb]
      \begin{center}
        \epsfig{figure=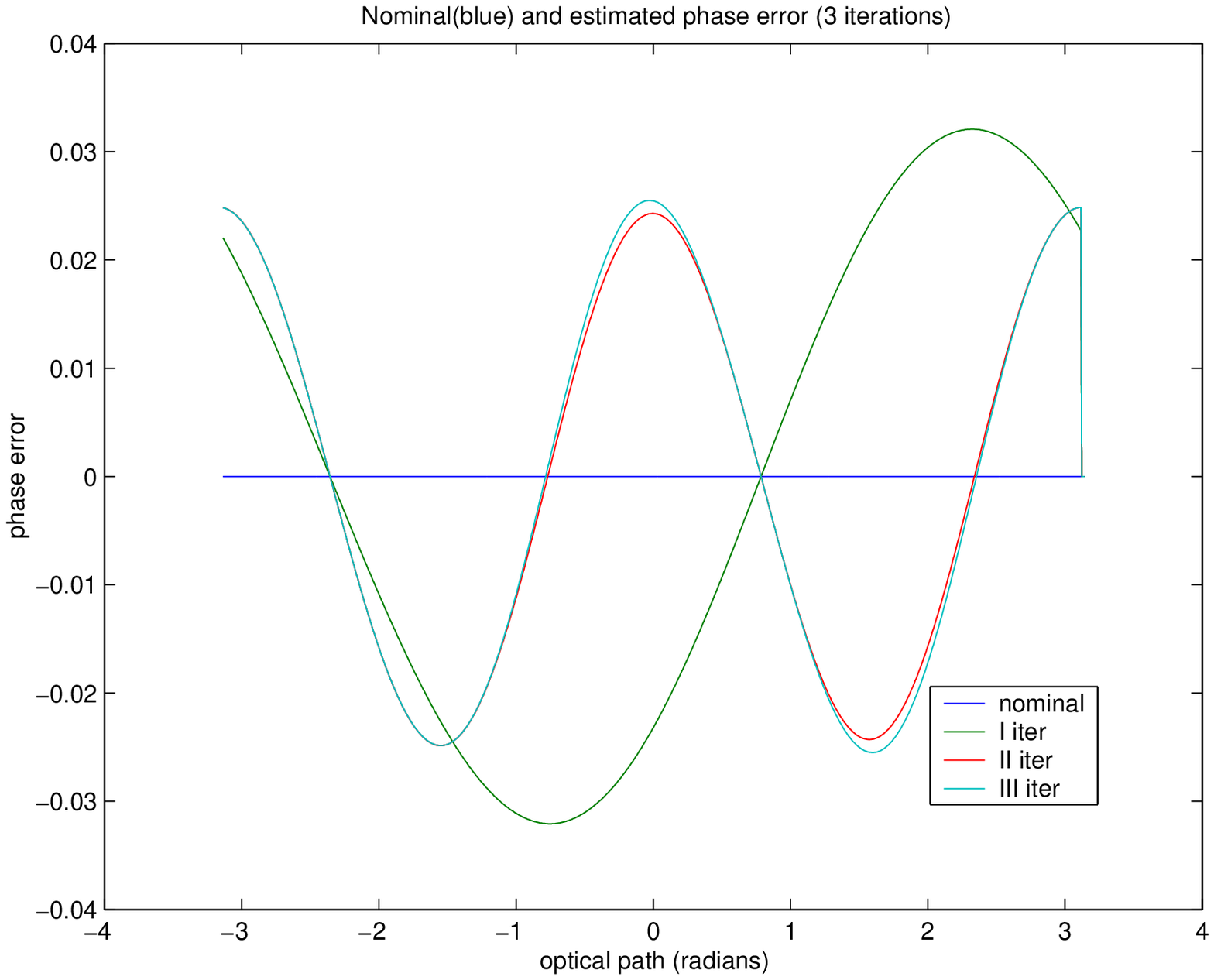,width=6.5cm}
        \epsfig{figure=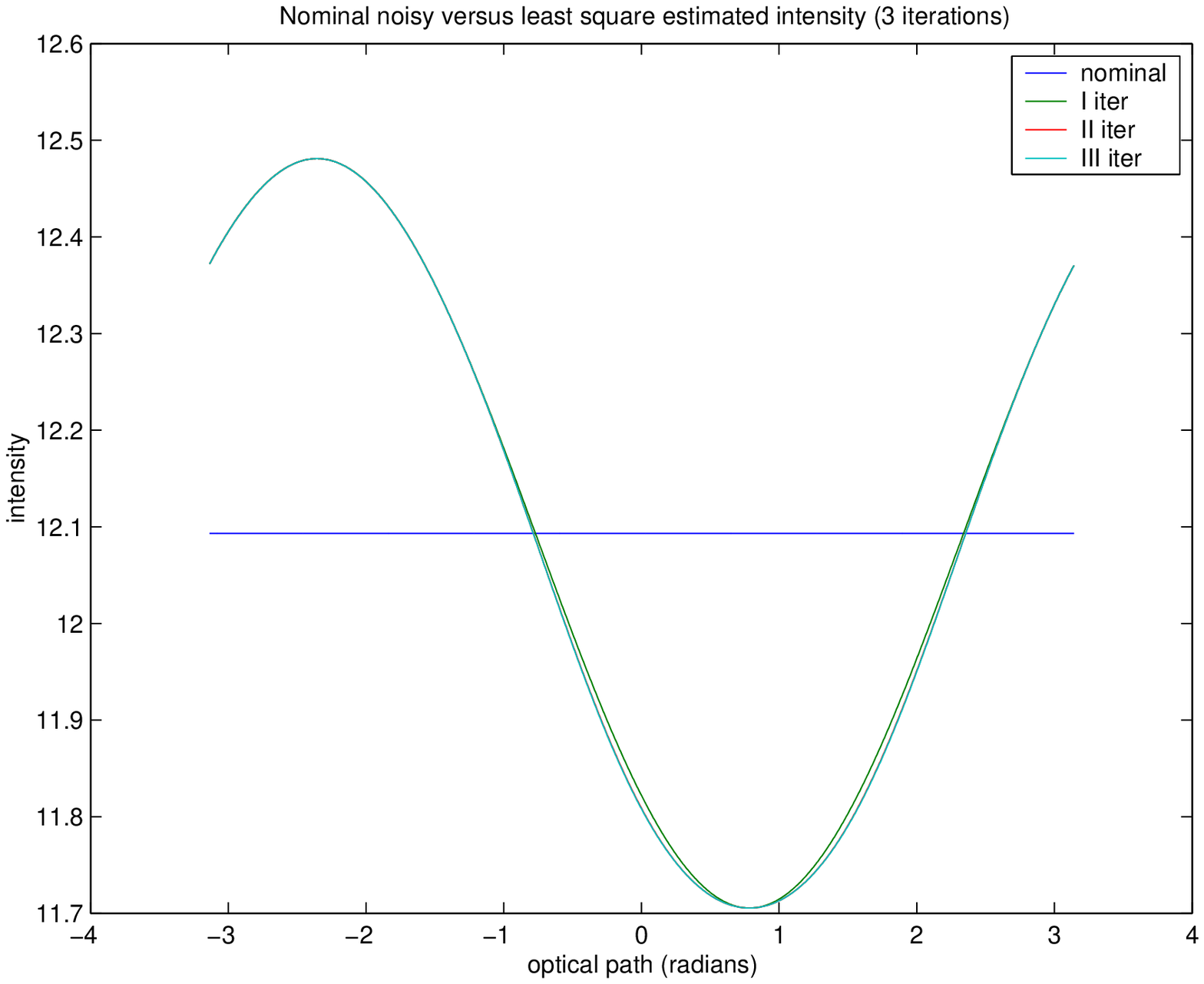,width=6.5cm}
        \epsfig{figure=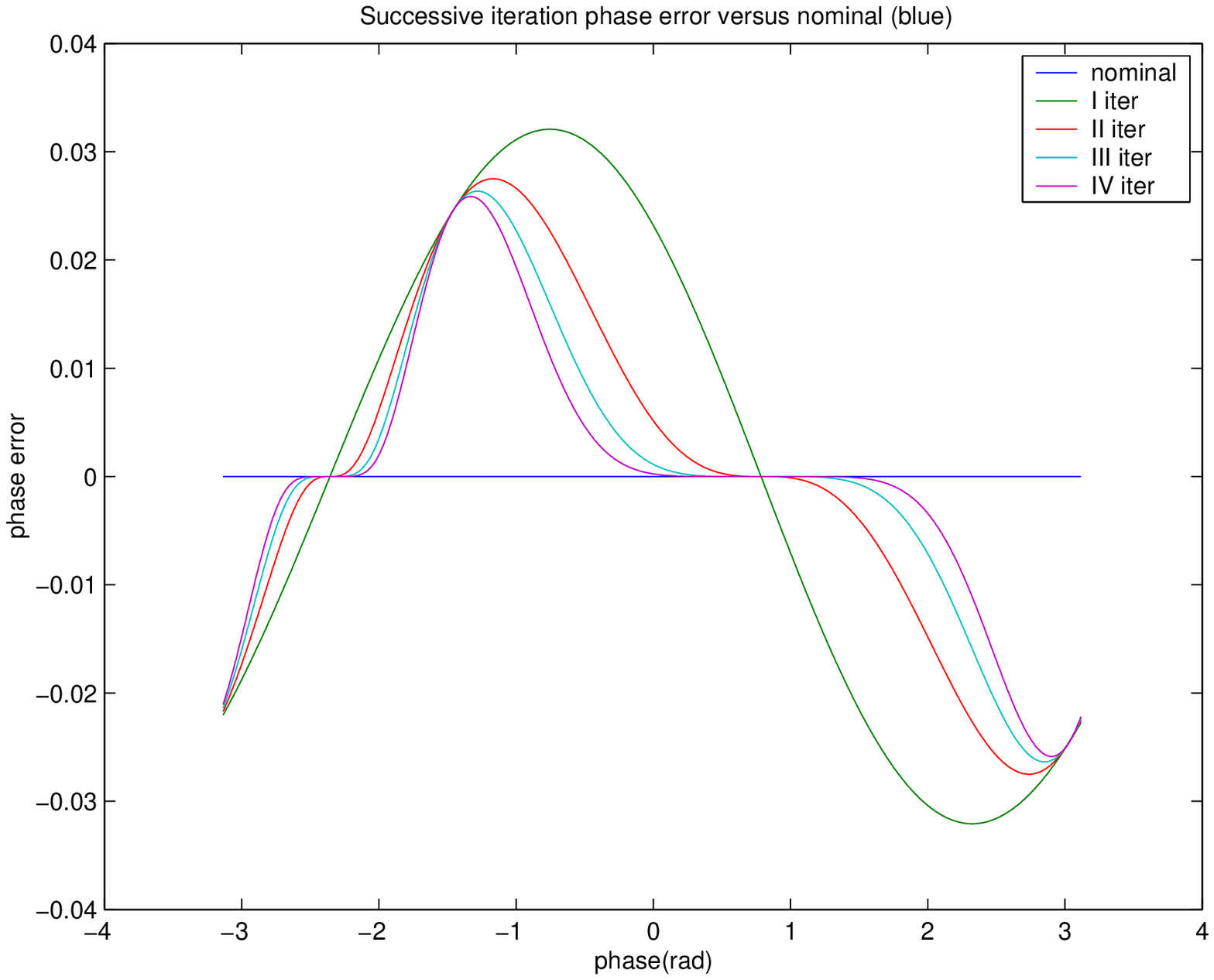,width=6.5cm}
        \epsfig{figure=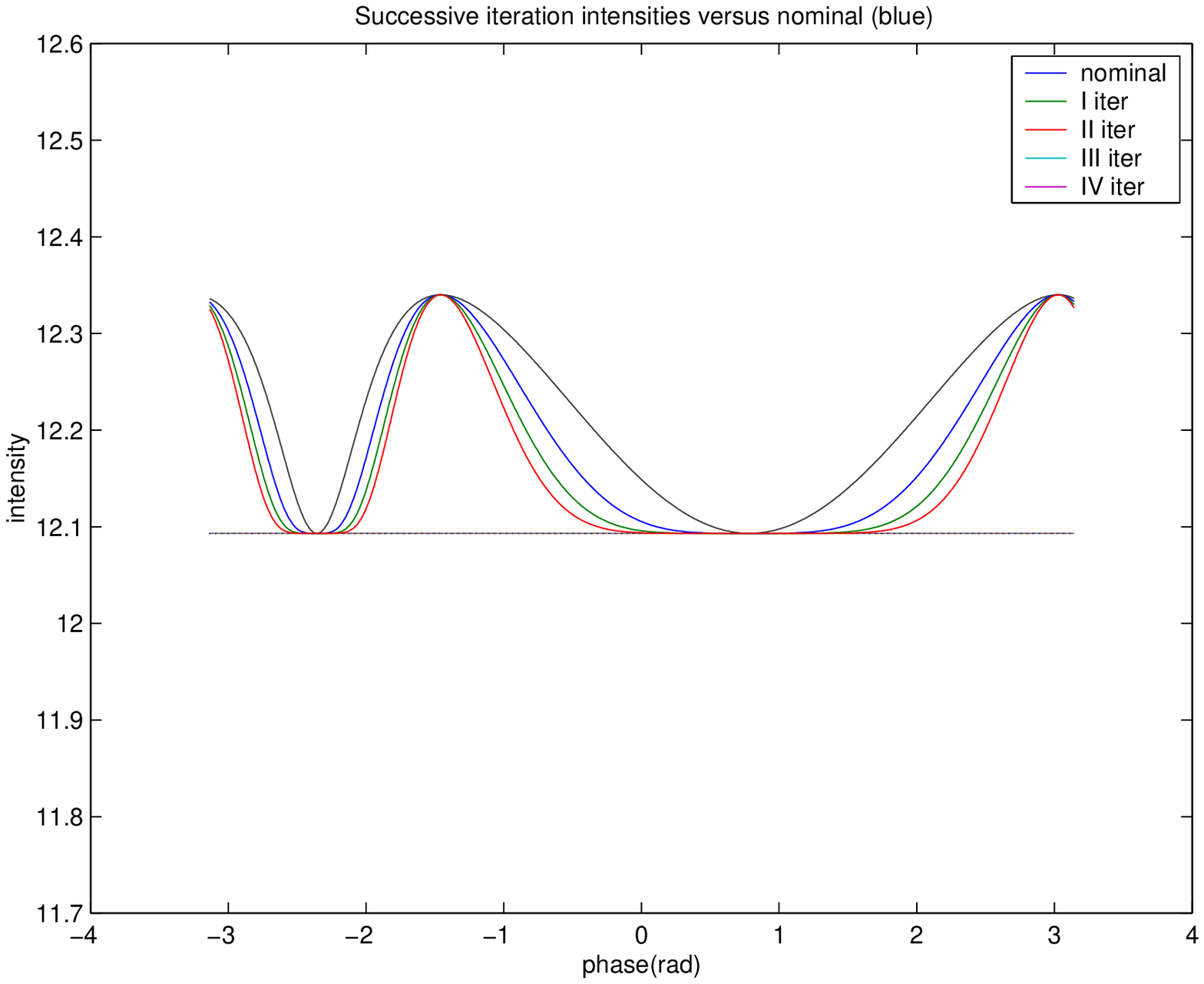,width=6.5cm}
        \caption{Phase (left) and intensity (right) iterated estimates for the error minimization method (first row) and for the analytical one (second row).}
        \label{fig:intens_2004}
    \end{center}
\end{figure*}

\noindent Also this estimate of the intensity $I$ can be used in an iterative process similar to that described in eq. \ref{eq:int_iterative}.\\
Figure \ref{fig:intens_2004} shows some iterative steps in the evaluation of both phase and intensity, for the two described methods. In the simulation, the model parameters take the values listed in table \ref{table:param}.
\begin{table}[htb]
\begin{center}
    \begin{tabular}{lc|lc}
     \hline
      visibility              & 0.9          & nominal intensity $I_0$ & 12.34 (arbitrary units)\\
      optical path diff (rad) & $[-\pi,\pi]$ & noise on intensity & constant ($-2\%$ of $I_=0$) \\
      noise on OPD            & null         & number of samples  & 1000 \\
      \hline
    \end{tabular}
    \caption{Parameters of fringes in the simulations.}
    \label{table:param}
\end{center}
\end{table}

\noindent The optical path difference is given in terms of angles (radians). From the a\-na\-lysis of the first row of the graphs (first method), we can notice a superposed sinusoidal error, both for phase and for intensity. It is at the same frequency of the fringe, it is probably due to the use of an estimated initial phase. From the second iteration the frequency doubles, but then it remains fixed. It can be due to the fact that the first method foresees the comparison with a sinusoidal template at frequency close to that of the signal. The iteration saturates immediately, without any further improvements.\\
The second method shows a better convergence of phase toward the nominal value. However, after some iterations the convergence decreases and then stops. We can notice that the intensity, apart from being modulated with more than a frequency, it is always overestimated. Tables \ref{table:Imeth_iterative} and \ref{table:IImeth_iterative} give a resume of mean and standard deviation of the estimates of both phase and intensity for the two methods. We can see that the standard deviation of the intensity is better (by a factor 10) with the second method.\\

\noindent Finally, we remark that both these methods are based on a simple model, that does not foresee a superposed envelope, so they are applicable in the central fringe. The model extension to a more complex function to get rid of the beat, and to cope with envelope modulation, is conceptually simple but was not yet carried on; it remains thus as part of the possible future developments.

\begin{table}
\begin{center}
  \begin{tabular}{c|c|c|c|c}
  \hline
  Iteration & Mean($\tilde{\phi}_i$) [rad] & $\sigma_{\tilde{\phi}_i}$ [rad] & Mean($I$) & $\sigma_I$ \\
  I   & $-8.95e-5$ & 0.0226 & 12.096 & 0.2743 \\
  II  & $-8.94e-5$ & 0.0173 & 12.091 & 0.2777 \\
  III & $-8.93e-5$ & 0.0177 & 12.091 & 0.2778 \\
  \hline
  \end{tabular}
  \caption{First method: error function minimization}
  \label{table:Imeth_iterative}
\end{center}
\end{table}

\begin{table}
\begin{center}
\begin{tabular}{c|c|c|c|c}
  \hline
  Iteration & Mean($\tilde{\phi}_i$) [rad] & $\sigma_{\tilde{\phi}_i}$ [rad] & Mean($I$) & $\sigma_I$ \\
  I   & $8.95e-5$ & 0.0226 & 12.218 & 0.0873 \\
  II  & $8.95e-5$ & 0.016  & 12.186 & 0.0902 \\
  III & $8.8e-5$  & 0.0136 & 12.171 & 0.0886 \\
  VI  & $8.7e-5$  & 0.0122 & 12.161 & 0.0864 \\
  \hline
  \end{tabular}
  \caption{Second method: analytical solution}
  \label{table:IImeth_iterative}
\end{center}
\end{table}

\chapter{Location and calibration algorithms for the VLTI PRIMA Fringe Sensor Unit}
\lhead[\fancyplain{}{\bfseries\thepage}]%
      {\fancyplain{}{\bfseries VLTI PRIMA FSU}}
\rhead[\fancyplain{}{\bfseries VLTI PRIMA FSU}]%
      {\fancyplain{}{\bfseries\thepage}}
\label{chap:PRIMA}
The Observatory of Turin has been involved in the design and development of a fringe sensor unit for the VLTI PRIMA instrument, the PRIMA FSU. We will describe it in Sec. \ref{sec:PRIMA FSU}.
The location algorithm proposed for this fringe sensor is based on the comparison of the current interferometric pattern with a tabulated template. The difference with the algorithms described in chapter \ref{chap:FINITO} is the use of a highly detailed interferometric model.
Of course, being based on a significant number of variables, the needs of their calibration and monitoring are much more demanding. However, good results can be obtained, in terms of model accuracy and variables estimations. In this chapter, we will introduce the interferometric environment used for PRIMA FSU: model, algorithms and calibration tools.
My work has focused especially on the last task, i.e. development and validation of diagnostics and calibration tools, both with simulated and laboratory data.

\section{Introduction of a detailed interferometric model}
\label{sec:intro3}

In the previous chapter we have analyzed some algorithms for the location of the fringe position with respect to the OPD scan. We have noticed that a relevant problem was the need of a good description of the interferometric output; otherwise, the results of all algorithms suffer of additional error caused by the lack of similarity between the real signal and the template (correlation method) or the filter output (demodulation).

\noindent Methods like the classical AC or ABCD are robust in this sense, but they impose strong requirements in terms of normalization, requiring a deep insight into the calibration.
\\

\noindent The model we have used till now is based on strong assumptions on the source and on the instrument features, such as constant flux over the spectral range for the star, constant response function over a rectangular spectral band for the latter. If we introduce in the model the possibility of tuning as much as possible of these characteristics, we can reach a better accuracy.

\noindent So, before proceeding, we have to introduce a more realistic and detailed model of the interferometric pattern, using more parameters than those used till now, separating the contribution of the source, of the atmosphere and of the instruments.\\

\noindent The instrument is characterized by a spectral transmission factor $\rho(\lambda)$, a visibility $V_I(\lambda)$, a detector quantum efficiency $QE(\lambda)$ and integration time $\tau$ that influence the flux intensity, and it introduces its own phase to the optical path $\phi_I(\lambda)$.
The source has its own spectral distribution of the intensity $I(\lambda)$, a visibility $V(\lambda)$ and a magnitude $m$.
The atmosphere effect can be described by a factor representing the wavefront degradation $\eta_A(\lambda)$ and by reflective indexes $n_i(\lambda)$ that depends from the path in air $p$ of the astronomical beams in the delay line, compensating the distance from the zenith of the observing direction, and so from the relative position of source and telescope.
All these parameters, except the source magnitude, are wavelength-dependent, and so they influence the effective working wavelength $\lambda_0$, which is no longer a fixed nominal value in the middle of the wavelength range.
\\

\noindent We are now able to describe the monochromatic signal for an interferometric channel by a combination of all these parameters:

\begin{eqnarray}\label{eq:signalModel}
f(\lambda,x) &=& I_S(\lambda) \cdot \tau \cdot \rho(\lambda) \cdot QE(\lambda) \cdot \\
\nonumber &\cdot& \left[1 + \eta_A(\lambda) \cdot V_I(\lambda) \cdot V(\lambda) \cdot \sin \left(\phi_I(\lambda) + \frac{2 \pi}{\lambda_0} [n(\lambda) \cdot x + n_i(\lambda) \cdot p]\right)\right]
\end{eqnarray}
where $I_S$ is the effective source flux intensity given by the combination of the emitted flux scaled by the star magnitude and the telescope collecting area, and resulting from the beam interference.\\
The polychromatic beam is the integration over the spectral range $\Lambda = [\lambda_1, \lambda_2]$ of each monochromatic component:
\begin{equation}
S(x) = \int_{\Lambda} s(\lambda,x) d\lambda
\end{equation}


\noindent In the following paragraphs, we will review the algorithm concept (Sec. \ref{subsec:algoritm}), giving an estimate of the expected error, and then we will briefly assess OPD and GD performance (Sec. \ref{subsec:OPD_GDperf}). A detailed analysis of GD and OPD performance, in different observational and atmospherical situations, and including perturbation on instrumental parameters, can be found in \cite{errorbudget}. In Sec. \ref{sec:sensitivity} we will first discuss the importance of sensitivity analysis, with a working example, and then (Sec. \ref{subsec:calibration}) we will concentrate on the calibration issue, i.e. the derivation of the essential features of the model directly from data, and the reconstruction of the interferometric signal.

\section{Algoritm for PRIMA FSU}
\label{sec:PRIMA FSU}

In this Section, we describe the VLTI PRIMA Fringe Sensor Unit instrument and the location algorithm we proposed, based on the interferometric model introduced with eq. \ref{eq:signalModel}.

\subsection{Instrument description}

A description of the PRIMA (Phase Referenced Imaging and Micro-arcsecond Astrometry) instrument can be found in \cite{DelPlancke00}, or at the web site: \\ http://www.eso.org/projects/vlti/instru/prima/index\_prima.html.\\
Its primary goals are the imaging of faint sources with an high angular resolution and a high precision astrometry, at the $\mu$as level. Finally, it also aims at increasing the sensitivity in order to be able to reach prohibitive magnitudes, till 19.\\

\begin{figure}[htbp]
   \begin{center}
      \epsfig{figure=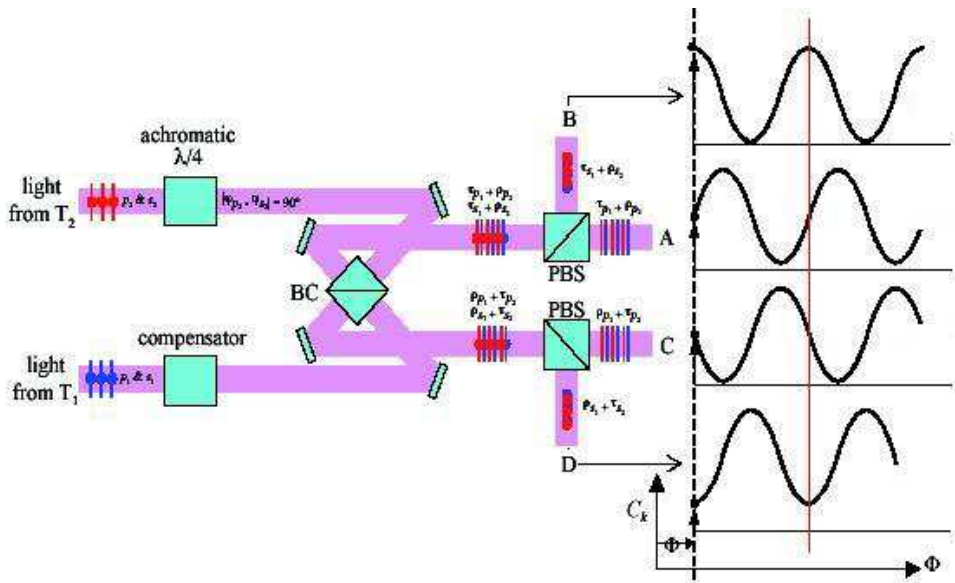,width=13cm}
      \caption{Schematic of PRIMA FSU design.}
        \label{fig:PRIMAdesign}
    \end{center}
\end{figure}

\noindent For such reasons, a fringe tracking system is mandatory. PRIMA requires two identical dedicated fringe sensors, called FSU A and B, tailored for its needs.\\
The PRIMA FSU (Fringe Sensor Unit) concept is schematized in figure \ref{fig:PRIMAdesign}. The working wavelength range is the K band ($[1.9-2.6] \; \mu m$). Differently from FINITO, it does not foresee an active modulation of the internal optical path, since the estimate of OPD is based on the phase distribution among the different polarization and wavelength components.

\noindent Each FSU is fed by two telescope beams, that pass through an alignment system with five degrees of freedom per beam, aimed to correction of different aberration effects: two lateral (decenter) for pupil alignment, two angular (tilt) for image alignment, and a longitudinal term for OPD alignment. Their control is in charge of a dedicated software module, which convert mechanical to optical degrees of freedom through a kynematics matrix.
The fringes are spatially sampled following an ABCD scheme. To implement it, before combination one beam is retarded of $\pi/4$, by an achromatic phase shifter implemented by a K-prism, then the two beams enter the combiner, a splitting cube with nominal transmission and reflection of $50\%$, then the two combined outputs are further split according to the polarization components with two independent polarising beam splitters, finally obtaining four beams with a relative $\pi/4$ phase shift: the A, B, C and D beams. Each of them is then injected in individual optical fibres, that also act as spatial filters, as we mentioned in the case of FINITO. The fibres carry the four beams into the dewar, where they are mounted onto a mechanical reference, aligned thanks to four degrees of freedom: two lateral, one longitudinal (for image magnification) and one rotational. Each beam is collimated and spectrally dispersed by a prism, before being focused on a PICNIC detector: most of the flux, about $70\%$, is retained in the central spot, while the remaining flux is split into two side bands and focused in neighbour pixels. In total, after each integration period the detector gives 12 values, corresponding to three sub-bands for each beam. In another way, it gives three sets of ABCD points, one for each spectral band.\\

\noindent Also PRIMA FSU is monitored by a metrology system, that shares the same optical components with astronomical beams, giving an alternative monitoring of the optical paths followed by stellar photons. The metrology monitors the whole optical path from the FSU combiner up to the telescopes, thus providing sensitivity to environmental disturbances which degraded the performance of previous instruments.\\

\noindent We remark some differences with respect to the FINITO design, apart the obvious mechanical and optical approach, based on the symmetry between the two arms. First of all, all photons are retained for the location algorithms, and there are no photometric beams such as in FINITO.
Then, the fibres enter directly into the cryostat, in order to minimize the effect of thermal background, which is relevant in K band. This means that their position must remain stable. After delivery, this part of the instrument has been upgraded, allowing motorized control of their positions, to further improve stability. Finally the spectral dispersion allows to simultaneously sample the interferogram in three contiguous spectral bands. In this way, it is possible to check the OPD position with the central pixel values, but also to directly assess the differential phase shift among the spectral channels. This allows to have a chromatic phase information, which in turn may be used to provide an estimate of the absolute OPD, removing the fringe period degeneracy, i.e. the group delay (GD).\\

\noindent The location algorithms proposed for both OPD and GD take advantage from the availability of a discrete number of values for each measurement step. They are based on the comparison between the measured signals and the global signal model, described in eq. \ref{eq:signalModel}, using a weighted least square fit. Note that the variables of eq. \ref{eq:signalModel} are allowed to change between each channel, in order to properly model the overall system, so eq. \ref{eq:signalModel} becomes:
\begin{eqnarray}\label{eq:signalModel_channeled}
&\left.\right.&f_i(\lambda,x) = I_{S,i}(\lambda) \cdot \tau_i \cdot \rho_i(\lambda) \cdot QE(\lambda) \cdot \\
\nonumber &\cdot& \left[1 + \eta_{A,i}(\lambda) \cdot V_{I,i}(\lambda) \cdot V_i(\lambda) \cdot \sin \left(\phi_{I,i}(\lambda) + \frac{2 \pi}{\lambda_{0,i}} [n(\lambda) \cdot x + n_i(\lambda) \cdot p]\right)\right]
\end{eqnarray}
with $i = 1, \ldots, 12$. The choice of the cutoff wavelengths between the three sub-bands of K band, that we will hereafter indicate as K1, K2 and K3, is of crucial importance, because all instrumental parameters depend on it.\\

\subsection{Algorithm concept}
\label{subsec:algoritm}

The choice of spatial modulation of the fringe pattern over the detector through the combination of beams separated by known phase offsets makes available four samples in approximate quadrature for each K sub-band.
The OPD/GD evaluation is based on the comparison of the measured interferometric signal with  tabulated FSU output, computed for a set of OPD, respectively GD, and source, atmosphere and FSU parameters computed by a calibration before integrating. The most likely OPD or GD value is identified in the least squares sense, by an iterative technique.

\noindent Instead of minimizing the quadratic error between measured and tabulated measurements, the algorithm searches the zero of the derivative function\linebreak (Newton-Raphson method of zero-crossing). An initial approximation is needed, and from it depends the descent of the error gradient. The minimum search is done over the three sub-bands simultaneously.

\noindent The error function {\it e}, called hereafter {\it discrepancy}, between the measured $s_n(z)$ and the tabulated $f_n(x)$ values depends on $x$ and $z$, i.e. the template and the unknown OPDs, respectively. Each measurement is affected by an error, represented by the random variable $\epsilon_n$. We assume that the $\{\epsilon_n\}_{n\leq0}$ are mutually uncorrelated, and such that
\begin{equation}
E[\epsilon_n]= 0,  \;\;\;\; E[\epsilon_n^2] = \sigma^2_n,  \;\;\;\; \forall n
\end{equation}
So the discrepancy has to be weighted to take into account the variability $\sigma^2_n$ of each measure:
\begin{equation}\label{eq:leastSquares}
e(x, z) = \sum_n [s_n(z) – f_n(x)]^2  / \sigma^2_n
\end{equation}
The subscript $n$ varies over the three sub-bands and over the number of channels in each sub-band ($4$, i.e. the ABCD points).\\

\noindent If we seek an estimator for the $z$ unknown, the natural choice is $x$. The best estimate of the current OPD/GD value is reached when $x = z$. Thanks to the assumption on the measurement errors, provided that the signal model is adequate, it can be derived (\cite[p.853]{Gai-PASP1998}) that the estimation obtained minimizing eq. \ref{eq:leastSquares} is unbiased and optimal in the least square sense:\\
\begin{eqnarray}
\nonumber E[(x-z)] &=& 0 \\
E[(x-z)^2] = \sigma^2(x) &=& \left(\sum_n{\frac{(f^\prime_n)^2}{\sigma^2_n}}\right)^{-1}
\end{eqnarray}
when $ f^\prime_n $ is the template signal derivative.

\noindent The derivative function in the $x$ variable of the discrepancy is given by:
\begin{equation}
h(x,z) = - \sum_n {\frac{[s_n(z) - f_n(x)] \cdot f^\prime_n(x)}{\sigma_n^2}} = \sum_n {[s_n(z) - f_n(x)] \cdot g_n(x)}
\end{equation}
where $g_n(x) = - f^\prime_n(x)/\sigma_n^2$ is the weight function. The minimum discrepancy is reached when $h = 0$. 
We separate the template factors of the preceding formula from the one containing the measured signals, and we obtain:
\begin{equation}h(x,z) = \sum_n {s_n(z) \cdot g_n(x)} - \sum_n {f_n(x) \cdot g_n(x)} = \sum_n {s_n(z) \cdot g_n(x)} - l(x)
\end{equation}
having defined the {\it bias} function $l(x)$ as $\sum_n {f_n(x) \cdot g_n(x)} = - \sum_n {\frac{f_n(x) \cdot f^\prime_n(x)}{\sigma^2_n}}$.
The first order approximation with Taylor series in the $x$ point for the discrepancy is given by:
\begin{equation}
h(z,z) = h(x,z) + h^\prime(x,z) \cdot (z-x) + o((z-x)^2)
\end{equation}

\noindent This formulation gives an iterative procedure for the approximation of the estimate $x$ of $z$; given the $z_{j-1}$ estimate, the following $z_j$ is:
\begin{equation}
z_j = z_{j-1} - \frac{\sum_n {s_n(z) \cdot g_n(x)} - l(x)}{h^\prime(z_{j-1},z)}
\end{equation}
with
\begin{equation}\label{eq:dh_in_dx}
h^\prime(x,z) = \sum_n \frac{[f^\prime_n(x)]^2}{\sigma^2_n} + \sum_n{[s_n(z) - f_n(x)] \frac{f^{\prime\prime}_n(x)}{\sigma^2_n}}
\end{equation}

\noindent The useful thing to notice is that the quantities $h^\prime(x,y)$, $g_n(x)$ and $l(x)$ are known once the template $f_n(x)$ has been defined in terms of the source and environmental conditions shown in section \ref{sec:intro3} before starting the observation, so they can be tabulated off-line, with a significant saving of computational load for real-time operation. This defines an iterative method for fast on-line computation of OPD and GD.

\subsubsection{Derivation of the formula and error estimation}
Let us consider the equation $t(x)=0$, and the $\mbox{j}^{\mbox{th}}$ approximation $x_j$ of the zero crossing abscissa. The Taylor expansion of $f(x)$ in the neighborhood of $x_j$ is given by:
\begin{equation}
t(x) = t(x_j) + (x - x_j)t^\prime (x_j) + R_2(x_j)
\end{equation}
where $R_2(x_j)= o((x-x_j)^2)$ is the error term. We easily obtain:
\begin{equation}
t(x) = 0 \;\; \leftrightarrow \;\; x = x_j -\frac{t(x_j) - R_2(x_j)}{t^\prime (x_j)}
\end{equation}

\noindent The error done with this truncation is $-\frac{t(x_j) - R_2(x_j)}{t^\prime (x_j)}$.

\noindent If we substitute $t(x)$ with $h(x,z)$, we can write the error term as:
\begin{equation}
e(x_j) = - \frac{h(x_j,z) - R_2(x_j)}{h^\prime (x_j,z)}
\end{equation}
where $R_2(x_j)$ is of order $o((x-x_j)^2)$. So, the convergence to zero of $e(x_j)$ depends on the distance $z-x_j$ and from the value of $h^\prime (x_j,z)$. We have seen in eq. \ref{eq:dh_in_dx} that the latter depends on the first and the second derivative of the signal model $f_n(x)$, we check when this term is zero, to avoid discontinuities.
From eq. \ref{eq:dh_in_dx}, we have that $h^\prime (x_j,z) = 0$ if:
\begin{equation}f^\prime_n(x_j)=0,  \;\; \forall n \;\;\; \wedge \;\;\;s_n(z) - f_n(x_j) = 0,  \;\; \forall n
\end{equation}
or if:
\begin{equation}
f^\prime_n(x_j)=0, \;\; \forall n \;\;\; \wedge \;\;\; f^{\prime\prime}_n(x_j) = 0, \; \; \forall n
\end{equation}
In the first case, the formula is exact, so the error term is zero. For the second case, we schematically write $f_n(x)$, following eq. \ref{eq:signalModel}, as a sinusoidal function $f_n(x) = A_n \sin(2\pi x + \phi)$, which leads to the following statements:
\begin{eqnarray}
\nonumber f^\prime(x) &=& 2 \pi A_n \cos (2\pi x + \phi)\\
f^{\prime\prime}(x) &=& -4 \pi^2 A_n \sin (2\pi x + \phi)
\end{eqnarray}
and they are equal to zero if:
\begin{eqnarray}
\nonumber f^\prime(x) = 0 \;\; &\leftrightarrow& \;\; x = \frac{2k+1}{2} - \frac{\phi}{2 \pi}, \;\; k = 0, \pm 1, \pm 2, \ldots\\
f^{\prime\prime}(x) = 0 \;\; &\leftrightarrow& \;\; x = k - \frac{\phi}{2 \pi}, \;\; k = 0, \pm 1, \pm 2, \ldots
\end{eqnarray}
so they can not be both zero. So, we are sure that the error function never diverges to $\infty$.\\
Of course, the magnitude of the error on the estimation depends on the goodness of the approximation $x-x_j$.

\subsubsection{GD estimation}
For the GD estimation, applying this procedure on the coherence length would require a large amount of time, incompatible with the desired high OPD control rate. A useful approach is to adapt the described algorithm to a large path, i.e. over three or five fringes around the central one. The OPD estimate is a good first approximation, since it is a minimum; the search in nearby fringes assures that the result is a global minimum, and not a local one, as the OPD could be. The estimation error can increase, because $x_{OPD}-z$ at the first iteration could be greater than $2\pi$.

\subsection{Algorithm performance on OPD and GD}
\label{subsec:OPD_GDperf}
The algorithm description highlights the need of a signal model, from which all the tabulated functions can be derived. In this section, we describe the software implementation of this model, and the principal parameters.

\subsubsection{Numerical description of the FSU}
We have already said, in Sec. \ref{sec:intro3}, that the parameters that appear in eq. \ref{eq:signalModel_channeled} are all wavelength dependent. This fact has suggested us an implementation procedure, based on the description of the interferometric model parameters as a function of the wavelength. Moreover, a version of each function can be tailored on the characteristic of the single channels.\\
From this description, it is possible to write the monochromatic interferogram at wavelength $\lambda$ for a selected channel, over an assigned OPD scan, just selecting the corresponding values, and substituting them in the eq. \ref{eq:signalModel}.\\
The polychromatic interferogram of each channel is now simply the sum of the corresponding monochromatic ones.\\

\noindent It is evident that this approach allows to manage the spectral description of the channel very easily, because it construct the polychromatic interferogram simply as a sum.

\noindent Instrumental parameters can be calibrated with laboratory test, starting from the reference description given by the constructors of each component. We now describe how we modeled the source, the background noise and the atmospheric turbulence.

\subsubsection{Source spectral description}
\begin{figure}[htb]
   \begin{center}
      \epsfig{figure=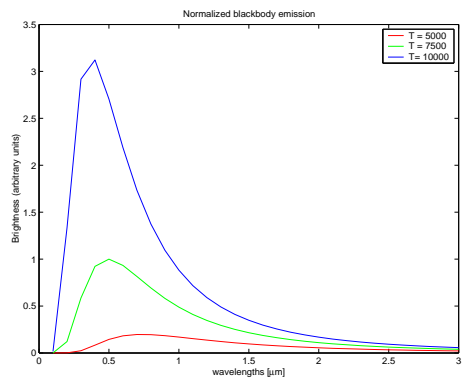,width=6cm}
      \caption{Normalized blackbody emission for three different temperatures [K].}
        \label{fig:blackbodies}
    \end{center}
\end{figure}

The model we have chosen for the source is the black body at a given temperature. The black body is a theoretical object that absorbs all the radiation that hits it, without any reflections. It is also a perfect emitter of radiation, at all wavelengths because it has to be able to absorb at every wavelength.\\
Even if in practice no material has been found to be a perfect blackbody, it is a convenient model because it is defined at all wavelengths. As the temperature decreases, the peak of the radiation curve moves to lower intensities and longer wavelength. Figure \ref{fig:blackbodies} shows the normalized blackbody intensity for three temperatures.

\subsubsection{Background noise}
\begin{figure}[htb]
   \begin{center}
      \epsfig{figure=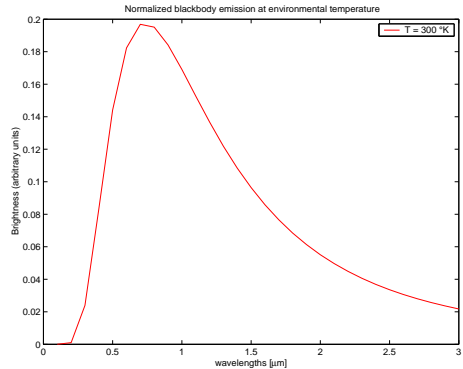,width=6cm}
      \caption{Normalized blackbody radiation for background noise [300 °K].}
        \label{fig:ambient_bb}
    \end{center}
\end{figure}

The background noise, i.e. the radiation emitted by the environment on which the instrument is immersed, can be modeled as a blackbody at environmental temperature (300 °K). It emits in the near infrared, and its spectral distribution is depicted in figure \ref{fig:ambient_bb}.

\subsubsection{Atmospheric noise}
The atmospheric noise can be modeled as in chapter \ref{chap:FINITO}, Section \ref{subsubsec:demodul_perf}. We recall here that it can be described with its power spectral density, proportional to $f^{-\frac{8}{3}}$, where $f$ is the frequency, and it depends on atmospheric parameters, such as the wind speed and the Fried parameter; its phase is assumed to vary randomly with a uniform distribution in the range $[-\pi, \pi]$. The parameters depends on the geographical location, and on weather conditions.


\subsubsection{OPD performance}
The nominal OPD performance is limited by the expected noise on the signal with respect to the template one. Additionally, we can expect systematic errors due to model limitations and non-linearity of the interferometric process.
To assess the OPD performance, we look at different parameters: the discrepancy between the input OPD and the evaluated OPD, the linearity of the correlation of the two OPDs, and the number of fringe jumps, i.e. the percentage of choice of a lateral maximum instead of the correct maximum in the central fringe.

\begin{figure}[htbp]
   \begin{center}
      \epsfig{figure=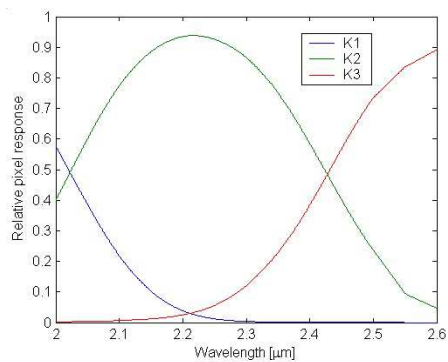,width=6cm}
      \caption{Source spectral division in the three K bands.}
        \label{fig:source_int}
    \end{center}
\end{figure}

\begin{figure}[htbp]
   \begin{center}
      \epsfig{figure=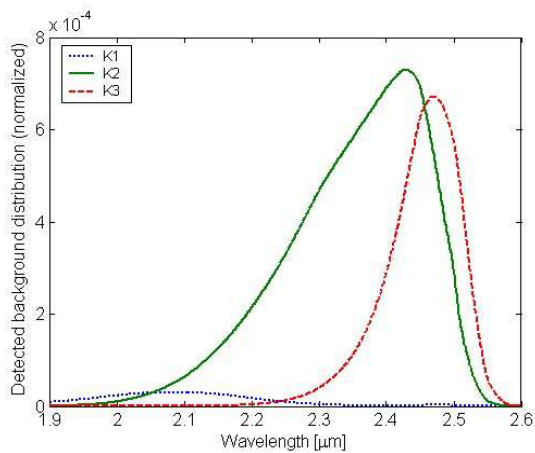,width=8cm}
      \caption{Background radiation in the three K sub-bands.}
        \label{fig:bck_int}
    \end{center}
\end{figure}

\noindent We model the celestial source as a black body with effective temperature of 3500°K, that is, a quite faint source. Fig. \ref{fig:source_int}, shows the spectral distribution of the source spectrum, split into the three sub-bands K1, K2 and K3. We set the model parameters to their nominal values; they will be described in more detail in Sec. \ref{subsec:calibration}. In particular, the relative fluxes in the three sub-bands are scaled accordingly to realistic values for the transmission and phase of the FSU, of the VLTI, to take into account the fluctuations sources outside the fringe sensor, and for the quantum efficiency of the detector.
The background noise is distributed as a blackbody at 300°K, but it is also split following the spectral division between K subbands (\ref{fig:bck_int}, right).

\noindent We express the signal conditions in terms of the Signal to Noise Ratio (hereafter SNR).
We define it approximately as:
\begin{equation}
SNR = \frac{N_s * QE}{\sqrt{N_s * QE + RON^2}}
\end{equation}
where $N_s$ is the total received flux, comprehensive of the visibility and of instrumental factors, while $QE$ is the quantum efficiency. For the noise term, we consider the photonic and the readout (RON) noises.\\

\noindent We consider a set of observational situations, with sources at different magnitudes and integration time, which leads to different SNR. The described situation are quite critical, for short integration time and limiting magnitudes. They are listed in table \ref{table:perf_OPD_workcase} for decreasing SNR values.

\begin{table}[htbp]
\begin{center}
    \begin{tabular}{cccc}
    \hline
    mag & integration time (ms)  & RON (electron) & SNR \\
    7   & 0.25                   &   11           & 352.5201\\
    8   & 0.25                   &   11           & 192.8585\\
    10  & 0.25                   &   11           & 45.7332 \\
    11  & 0.25                   &   11           & 19.9307\\
    13  & 2                      &   11           & 23.4404  \\
    19  & 10000                  &   4            & 23.0549   \\
    14  & 2                      &   4            & 18.1428  \\
    19  & 2000                   &   4            & 10.3002   \\
    19  & 1000                   &   4            &  7.2743    \\
    16  & 4                      &   4            &  5.1856   \\
    17  & 10                     &   4            &  3.8227    \\
    18  & 20                     &   4            &  2.3320    \\
    19  & 100                    &   4            &  2.2520    \\
    19  & 20                     &   4            &  0.9327    \\
    \hline
    \end{tabular}
    \caption{SNR vs. magnitude and exposure time.}
    \label{table:perf_OPD_workcase}
\end{center}
\end{table}

\begin{figure}[htb]
   \begin{center}
      \epsfig{figure=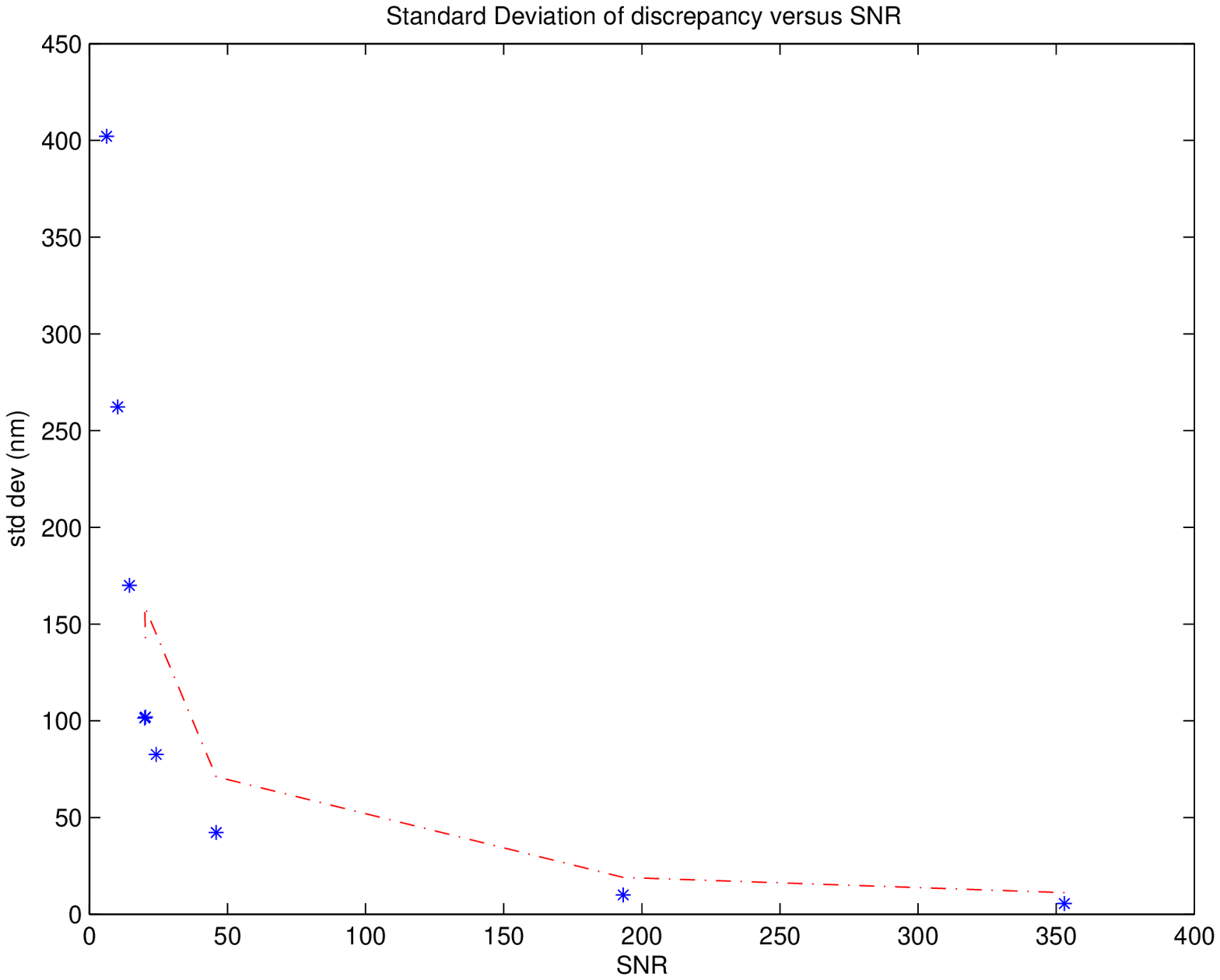,width=6.5cm}
      \epsfig{figure=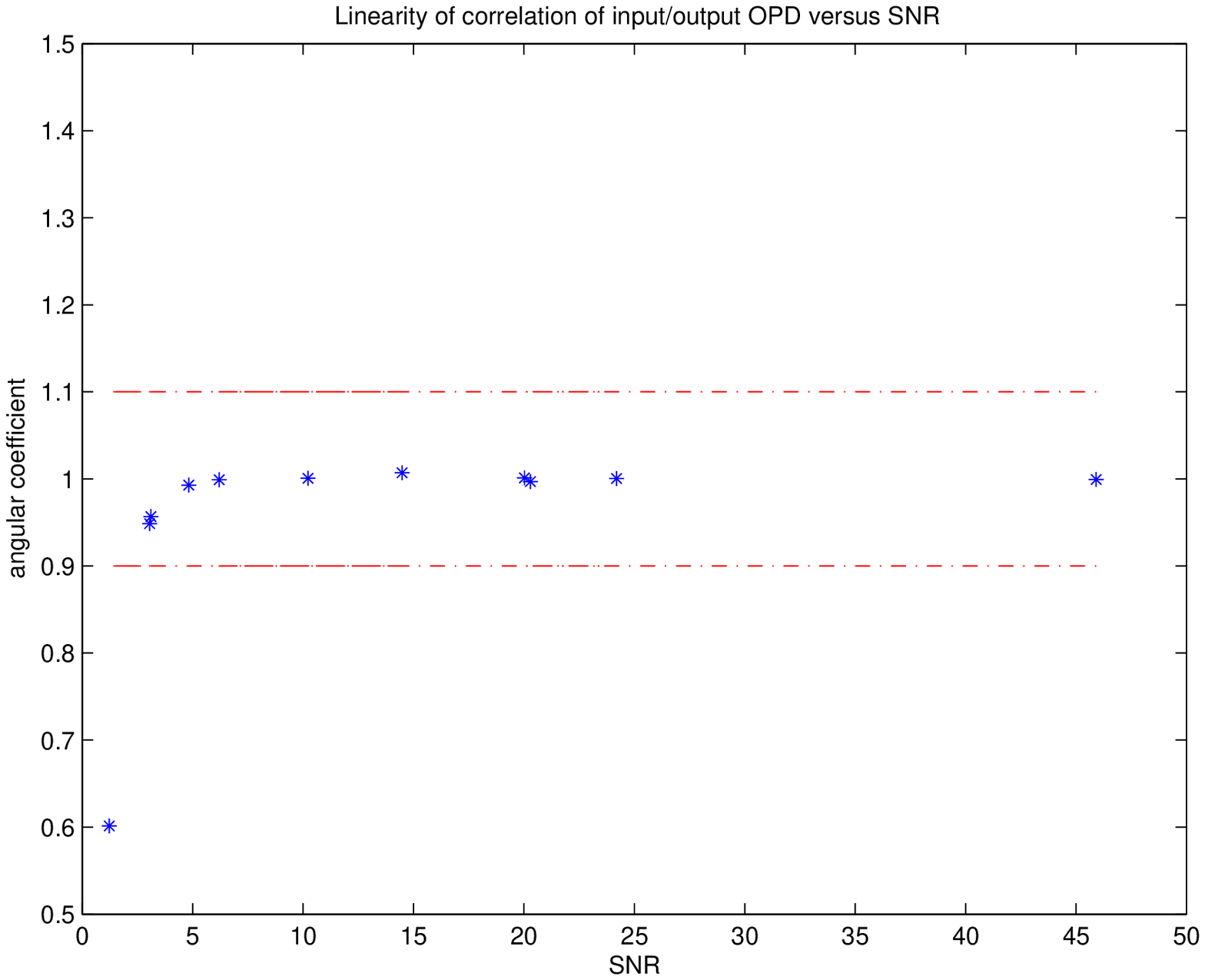,width=6.5cm}
      \caption{Standard deviation of the discrepancy between nominal and evaluated OPD (left) and linearity between the two measures (right).}
        \label{fig:OPD_RMSdisc}
    \end{center}
\end{figure}

\begin{figure}[htb]
   \begin{center}
      \epsfig{figure=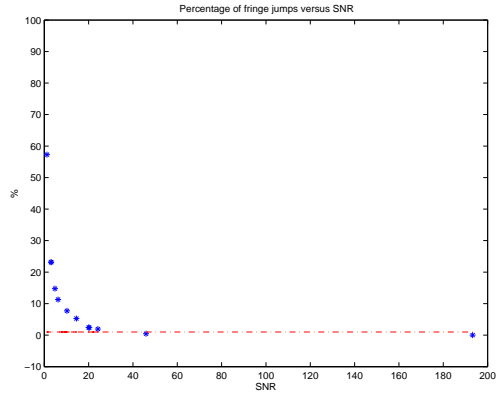,width=6.5cm}
      \caption{Percentage of fringe jumps at different SNR.}
        \label{fig:OPD_fj}
    \end{center}
\end{figure}

\noindent The standard deviation of the discrepancy between the nominal and the evaluated OPD and the linearity between the two are reported in figure \ref{fig:OPD_RMSdisc}, while fig. \ref{fig:OPD_fj} shows the percentage of fringe jumps. In all picture, the red dotted line corresponds to limiting performance requirements given by ESO. We can see that the algorithm gives good results in terms of both evaluation error and linearity, while the fringe jumps are a more critical issue.

\subsubsection{GD performance}
In this section, we report the GD performance in different observational situations, with a range of values for integration time, magnitude and readout noise. The simulated source is again a blackbody with temperature $3500$ °K, so a rather red source. Results are reported in table \ref{table:GD_perf}. Its analysis reveals that the algorithm performance is good. However, the integration time is an important variable: for faint magnitudes, it can affect the performance dramatically (fourth row in the table). This imposes some constraints on the GD  evaluation: while the OPD can be estimate at high rate, for a good GD estimation a longer integration time is required. This fact have impact on the software development of the detector readout. Thus, the side spectral band pixels can be read at the lower GD rate, whilst the central band (``white light'') pixels are read at the faster OPD rate.

\begin{table}
    \begin{center}
    \begin{tabular}{c|c|c|c}
      \hline
      case & magnitude & Integration time & RON  \\
      a & 10 & 5    & 5.1  \\
      b & 12 & 5    & 5.1  \\
      c & 14 & 200  & 0.8  \\
      d & 16 & 200  & 0.8  \\
      f & 16 & 2000 & 0.3   \\
      \hline
    \end{tabular}
    \caption{Description of GD observational situations.}
    \label{table:GD_perf_descript}
    \end{center}
\end{table}

\begin{table}
    \begin{center}
    \begin{tabular}{c|c|c|c|c|c|c}
      \hline
      case & GD noise sd[nm] -req & GD noise sd [nm] & fj (\%) req. & fj (\%) & l req. & l \\
      a & 900  &  5.394   & 1\% & 0\%  & 1 $\pm$ 0.1 & 1.0001 \\
      b & 3300 &  138.38  & 1\% & 1\%  & 1 $\pm$ 0.1 & 1.0011 \\
      c & 800  &  7.302   & 1\% & 0\%  & 1 $\pm$ 0.1 & 1 \\
      d & 1900 &  1071.79 & 1\% & 59\% & 1 $\pm$ 0.1 & 1.019  \\
      e & 600  &  11.39   & 1\% & 0\%  & 1 $\pm$ 0.1 & 1.004  \\
      \hline
    \end{tabular}
    \caption{GD performance, in terms of standard deviation of GD noise (GD noise sd), percentage of fringe jumps (fj) and linearity coefficient (l). For each of them, the limit requirements are reported (req.)}
    \label{table:GD_perf}
    \end{center}
\end{table}

\begin{figure}[htb]
   \begin{center}
      \epsfig{figure=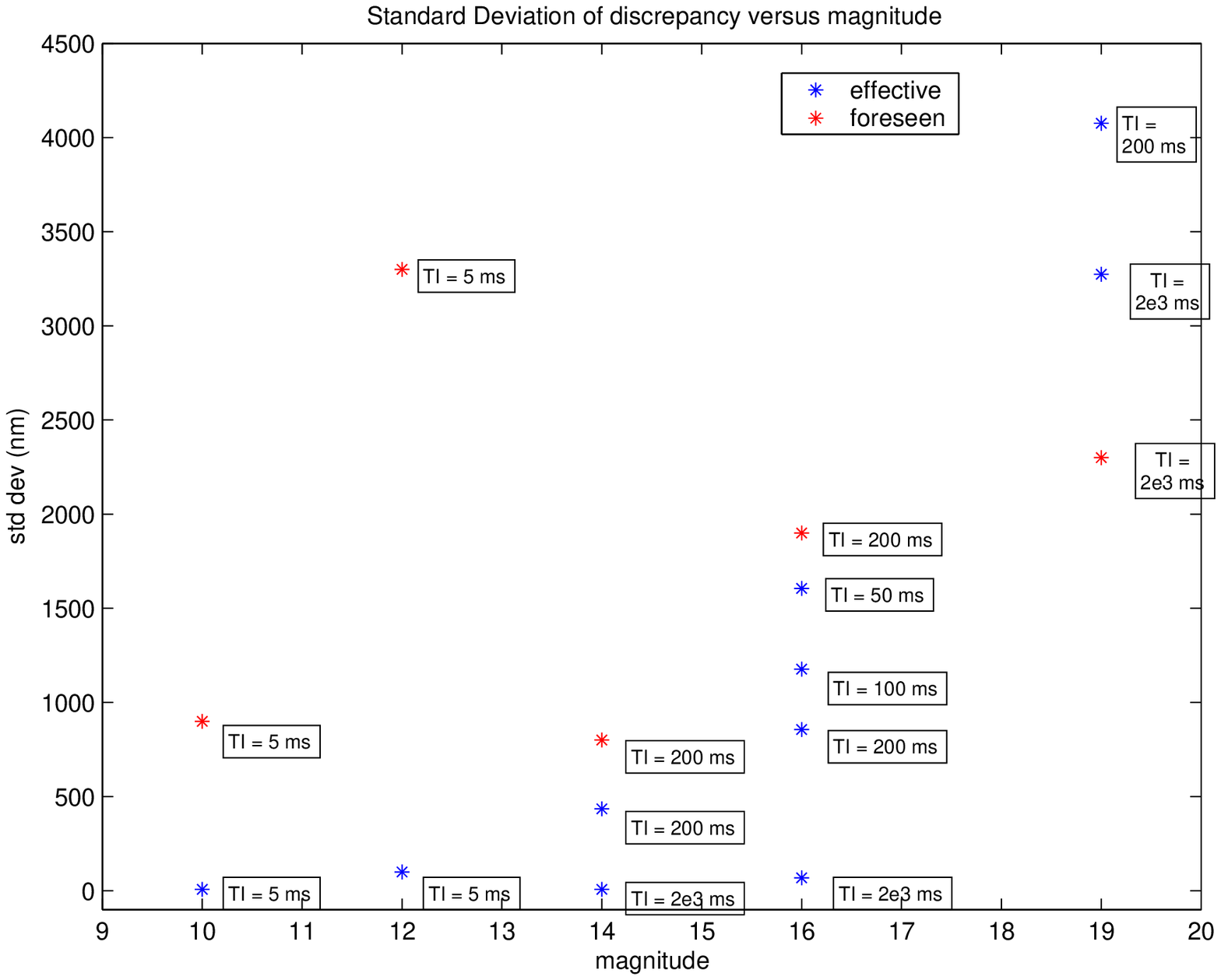,width=6.5cm}
      \epsfig{figure=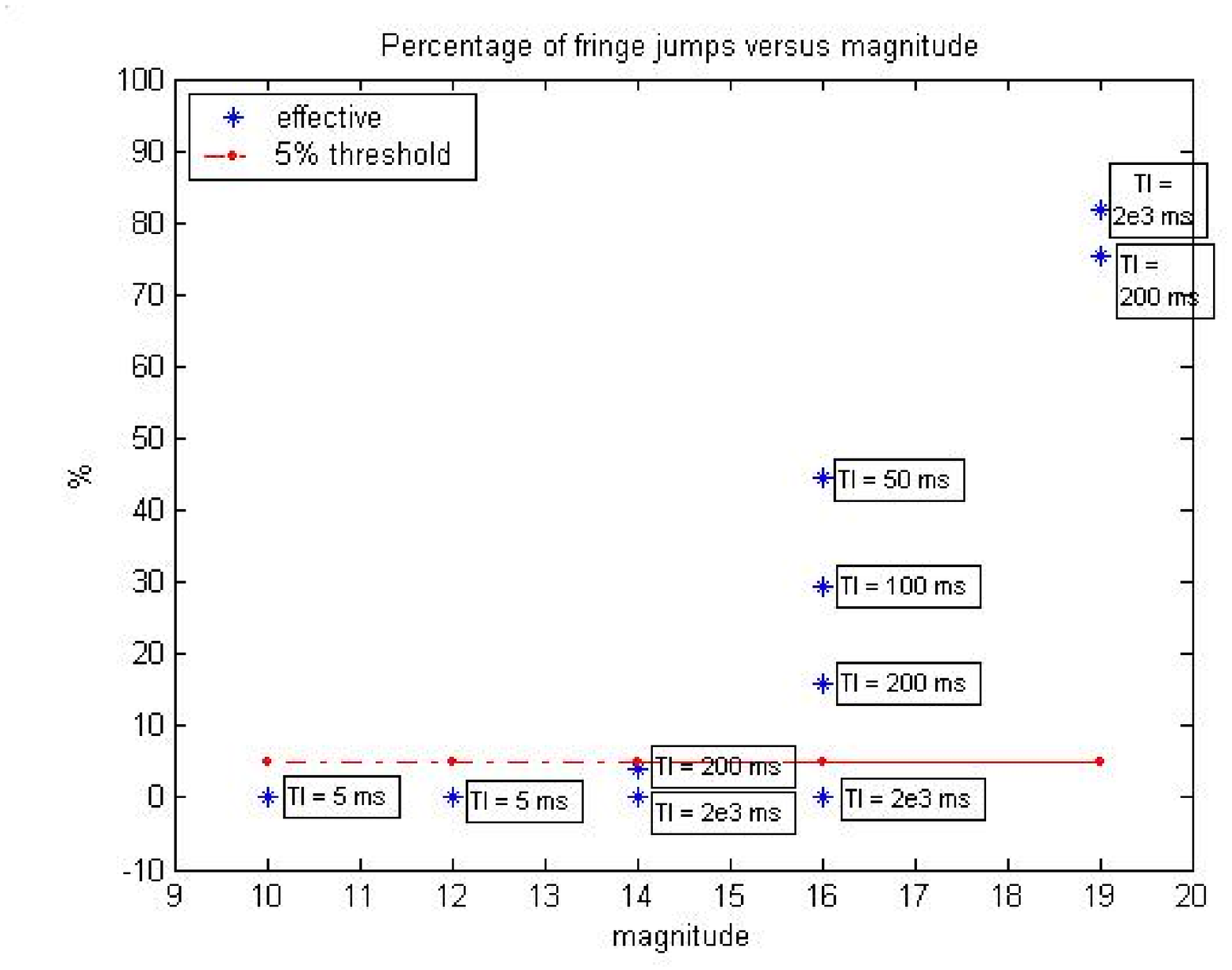,width=7.3cm}
      \caption{Standard deviation of the discrepancy between nominal and evaluated GD (left) and number of fringe jumps (right).}
        \label{fig:GD_RMSdisc}
    \end{center}
\end{figure}

\begin{figure}[htb]
   \begin{center}
      \epsfig{figure=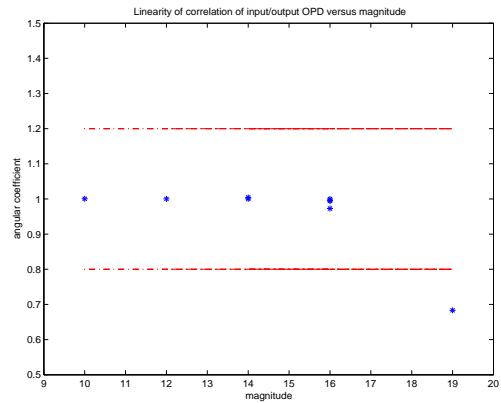,width=6.5cm}
      \caption{Linearity between nominal and evaluated GD.}
        \label{fig:GD_lin}
    \end{center}
\end{figure}

\noindent As for the OPD case, we resume the GD algorithm performance in terms of the standard deviation of the discrepancy between the nominal and the evaluated GD, and of the percentage of fringe jumps (reported in figure \ref{fig:GD_RMSdisc}). All quantities are given as a function of the magnitude of the stellar source and of the integration time. The limiting performance required by ESO is represented as red dots. We can notice that the integration time is a crucial variable for the algorithm performance, especially for high magnitudes.\\
Finally, we show in fig. \ref{fig:GD_lin} the linearity between the nominal and the evaluated GD. We find again that the performance is good till the limiting magnitude $m = 19$. 

\section{Calibration and Sensitivity analysis}

The proposed description of the interferometric signal requires a good knowledge of a lot of parameters. The uncertainty on them causes the error on the OPD and GD to grow.\\

\noindent In the described implementation of the FSU model, the needed parameters can be divided in four categories, resumed in the following list.
\\
Source parameters:
\begin{itemize}
	\item{effective temperature}
	\item{flux at zero magnitude}
	\item{effective magnitude}
\end{itemize}

\noindent Atmospheric parameters:
\begin{itemize}
	\item{refractive index of air (in laboratory or in site)}
	\item{unbalanced air path (in laboratory or in site)}
\end{itemize}

\noindent Instrumental parameters:
\begin{itemize}
	\item{phase and transmission of the VLTI before the FSU}
	\item{visibility of the VLTI before the FSU}
	\item{phase and transmission of the FSU}
	\item{visibility of the FSU}
	\item{thermal background at the input of the FSU}
	\item{detector quantum efficiency, and conversion factor between photons and photo-electrons at the output of the detector}
	\item{cutoff wavelengths between FSU bands}
	\item{wavelength array for K band: $[1.9-2.6] \; \mu m$}
\end{itemize}

\noindent Observation-depending parameters:
\begin{itemize}
    \item{pointing parameters, such as air path}
	\item{integration time}
	\item{detector read-out noise (changing as a function of the readout mode)}
\end{itemize}

\noindent The knowledge of all these parameters with a sufficient accuracy is of crucial importance. A discrepancy from the nominal value can cause perturbations over other model parameters. The study of the possible effects of parameters variation is called sensitivity analysis. We give an example of it, investigating the effect on the working wavelength of a misalignment of the fibres with respect to their nominal position.

\subsubsection{Fibers displacement}
\label{sec:sensitivity}
\begin{figure}[htbp]
   \begin{center}
      \epsfig{figure=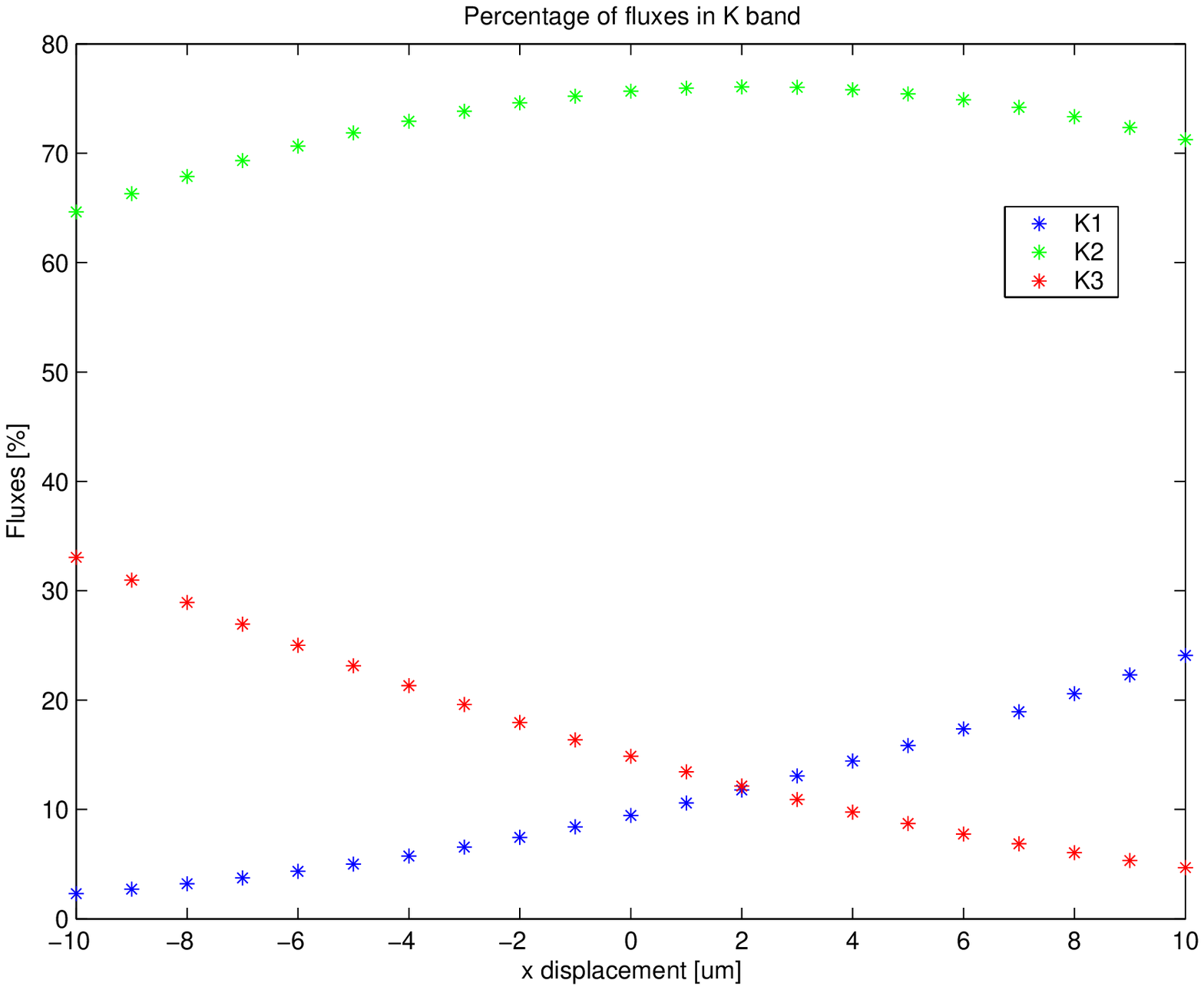,width=6.5cm}
      \epsfig{figure=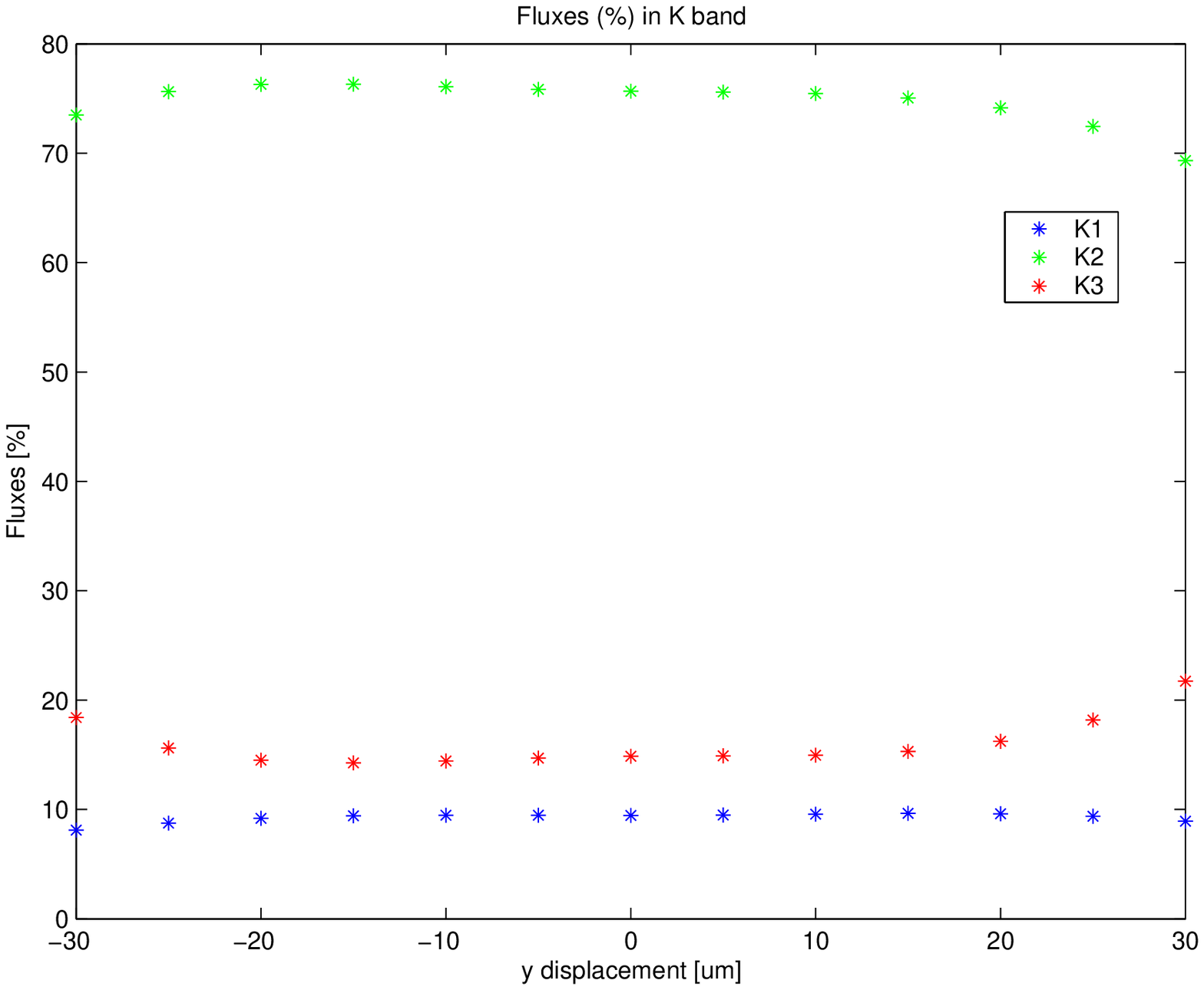,width=6.5cm}
      \caption{Flux intensities variations when a displacement of the fibres is applied on the $x$ direction (left) and on the $y$ direction (right).}
        \label{fig:perf_fluxVsFibDispl}
    \end{center}
\end{figure}

\begin{figure}[htbp]
   \begin{center}
      \epsfig{figure=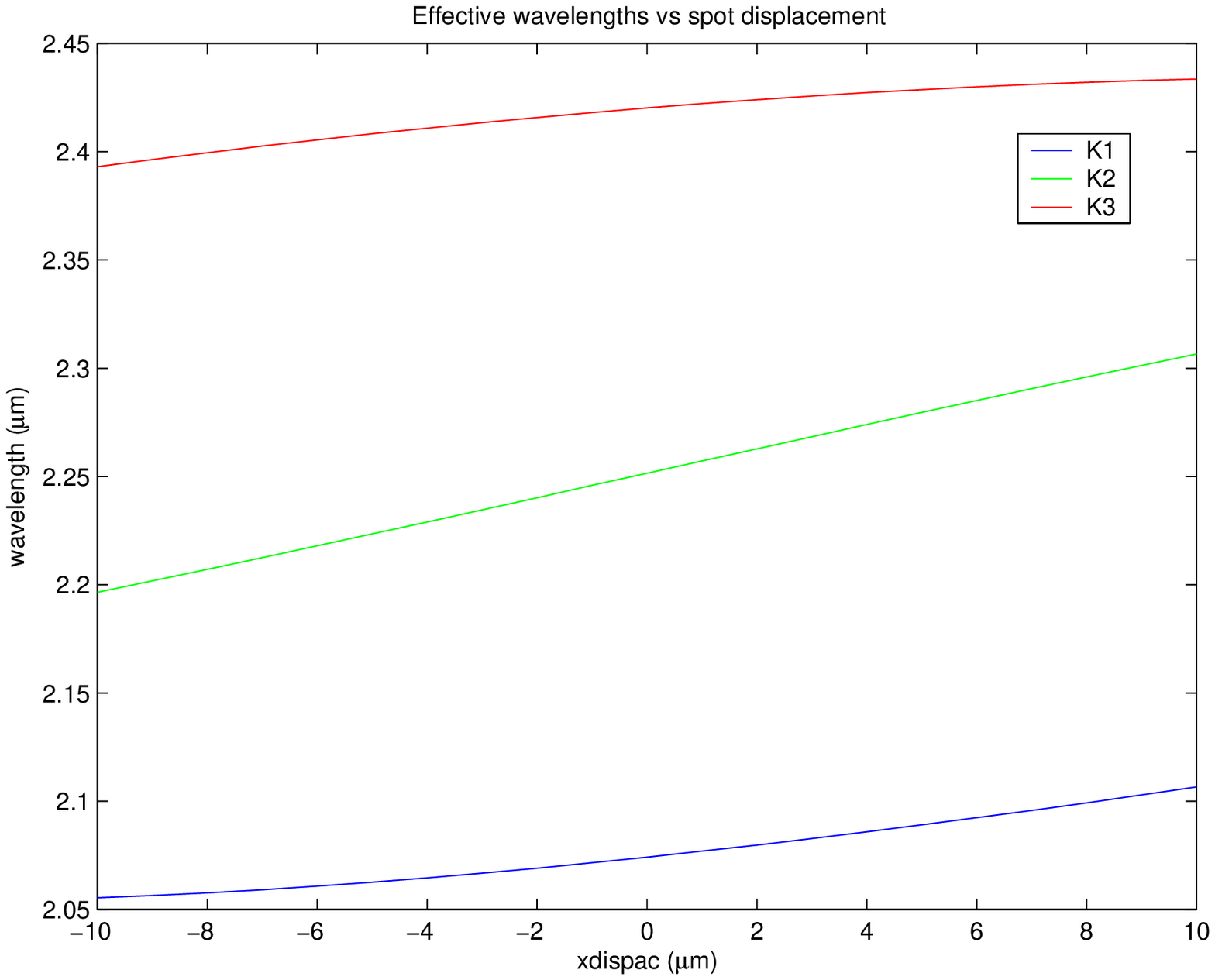,width=5cm}
      \epsfig{figure=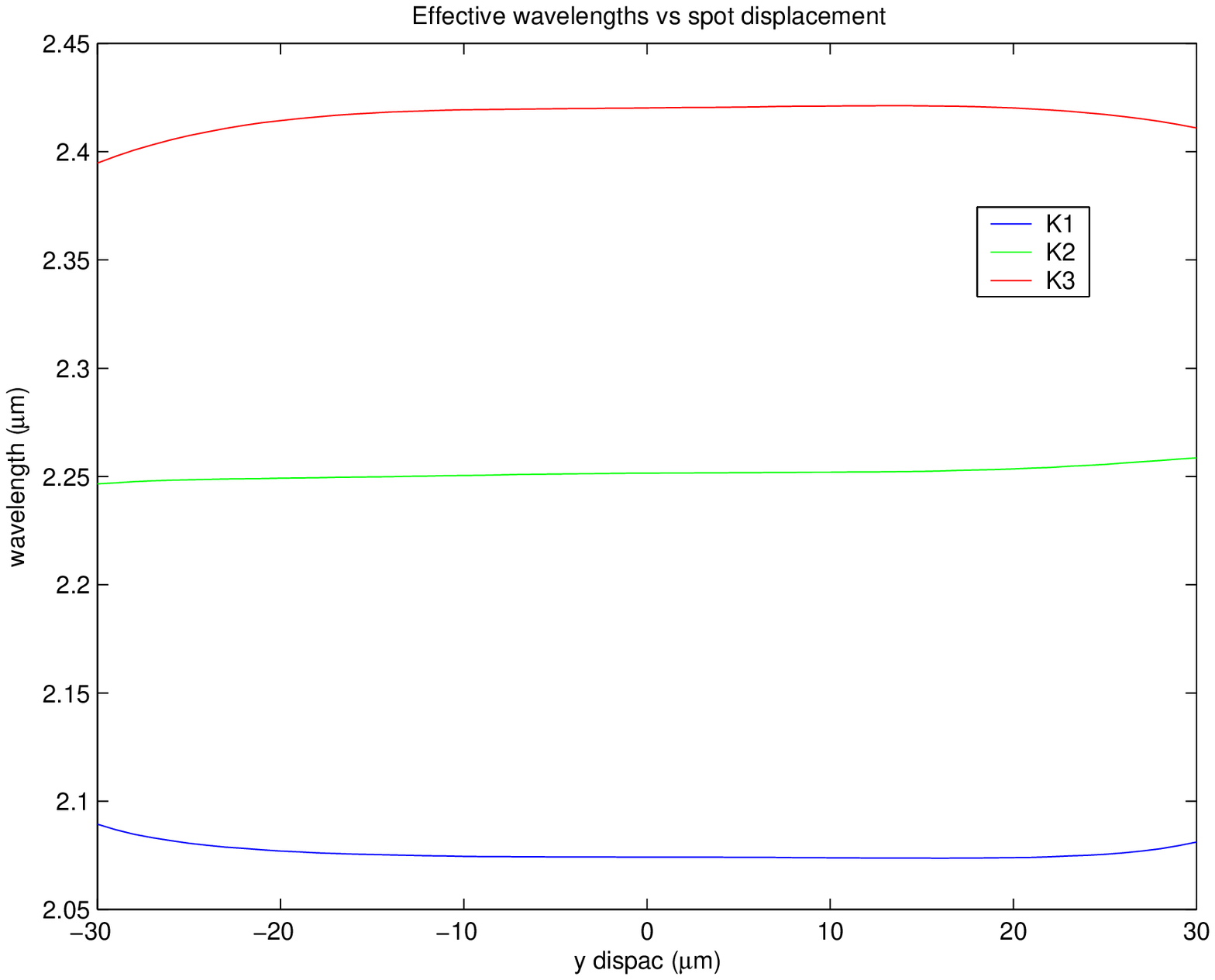,width=5cm}
      \caption{Effective wavelength perturbation when a displacement of the fibres is applied on the $x$ direction (first row) and on the $y$ direction (second row).}
        \label{fig:perf_effwlVsFibDispl}
    \end{center}
\end{figure}

The nominal position of the fibres carrying the FSU outputs to the detector are supposed known. A perturbation of their alignment modifies the energy distribution on the detector, modifying the intensity on each spectral channel, but also the effective wavelength. This effect could induce an apparent phase shift. Due to spectral dispersion over at least three pixels, a displacement along the dispersion direction (say $x$) has more impact than a perpendicular perturbation (say $y$), modifying the effective wavelengths in each band. The simulation of the imaging quality on the detector (summarized in the point spread function) is done by Code V, a ray-tracing software. The model can be tailored on the real optical system. To obtain the overall PSF, the point spread functions at each wavelength are summed together. The charge diffusion on each pixel is not uniform, but is modeled by a Gaussian distribution, whose parameters are determined by the physical instrument. Also, the pixel response distribution is modeled, from literature data, to account for the lower sensitivity close to the pixel edges.\\
Figures \ref{fig:perf_fluxVsFibDispl} and \ref{fig:perf_effwlVsFibDispl} show the modification on the total fluxes in each K subband and the variation of their effective wavelengths, respectively, when we simulate a fibre movement in the $x$ and $y$ directions. For the $y$ direction, we can notice that the effects are not negligible only for large displacement, comparable with half the pixel size, i.e. the sensitivity to this perturbation is low.

The estimated effective wavelength ranges on an interval around the nominal one, i.e. at zero displacement. We evaluate the width of this interval, in terms of percentage of the nominal wavelength, and we list the results in table \ref{table:perf_effWlRange}.
\begin{table}[htbp]
\begin{center}
    \begin{tabular}{lcc}
    \hline
       & $x$ displac                                 & $y$ displac \\
    K1 & $[2.07 \mu m - 0.91\%, 2.07\mu m + 1.56\%]$ & $[2.07\mu m + 0.73\%, 2.07\mu m + 0.34\%]$ \\
    K2 & $[2.25 \mu m - 2.44\%, 2.25\mu m + 2.45\%]$ & $[2.25\mu m - 0.22\%, 2.25\mu m + 0.32\%]$ \\
    K3 & $[2.42 \mu m - 1.12\%, 2.42\mu m + 0.55\%]$ & $[2.42\mu m - 1.05\%, 2.42\mu m -0.38\%]$ \\
    \hline
    \end{tabular}
    \caption{Effective wavelengths variations.}
    \label{table:perf_effWlRange}
\end{center}
\end{table}




\subsection{Calibration}
\label{subsec:calibration}

The example we have illustrate in the previous paragraph shows the importance of the calibration. Moreover, due to the properties of the PRIMA FSU location algorithm, the calibration is a fundamental issue for OPD and GD performance, because it provides the estimates of the current value of all the source and environmental parameters needed for the definition of the tabulated templates.

\subsubsection{Calibration strategy}
\noindent The calibration strategy proposed for PRIMA FSU must be implemented before an observation, is performed with laboratory sources (such as blackbodies and lasers) and can be described by the following steps:
\begin{enumerate}
	\item{evaluation of the values of the previous parameters, with the exception of the phase and the transmission of the FSU}
	\item{definition of the parameters related to the OPD scan}
	\item{Fourier transform of the FSU signal in the wavelength space, for the estimation of the overall phase and transmission}
	\item{estimation of the effective magnitude}
	\item{FSU template signal build-up}
\end{enumerate}

\noindent For the first step, dedicated measurements have to be performed, as for the instrumental parameters, whereas source and atmospheric parameters depend on the chosen observational target, i.e. its magnitude and position in the sky, as well as observation- dependent ones.
\\
The second step defines the operational range of the FSU template in terms of OPD scan and its conjugated quantity, the wavenumber range.
The range and sampling of the OPD scan must take into account the different coherence lengths of the three sub-bands, characterized by different spectral range and central wavelength, and the recovery of the FSU phase and transmission through the Fourier Transform at step 3.
\\
From data effectively acquired in a calibration run it is possible to retrieve the current FSU transmission and phase, as the modulus and the phase of the Fourier Transform of measured data in the wavenumber space.
These measures are directly linked with the complex visibility. The interferogram is the real part of the complex visibility, and being polychromatic is the sum over all working wavelengths. With a FT of the interferogram, we can separate the different components.

\noindent Although in simulation the OPD scan is performed with equally spaced step, when working with real data this is no longer true. The metrology system provides the actual value of the instrumental OPD, with an accuracy of order of nanometers, but the different step size have to be taken into account. This is done using the Discrete Fourier Transform instead of the FFT over the desired wavelength range.\\ 
It must be considered that these values implicitly depend on all the parameters we listed before. Laboratory tests can isolate the FSU phase and transmission values, then they must be scaled with terms depending by external factors, such as the phase and transmission of the VLTI at the input of the FSU, the environmental conditions such as background noise, and so on.
\\

\noindent The estimation of the magnitude of the emitting source is a difficult task, because it is corrupted by several other terms, such as visibility, noise sources, atmospheric conditions and so on. The effective magnitude, hovewer, can be evaluated with a comparison with a set of sources at different magnitudes, but all in the same operational conditions, and searching the magnitude that best fit the measured one.
\\

\noindent The FSU signal at step 5 is constructed starting from the flux emitted by an ideal source, modeled as a blackbody at the given temperature, scaled for the source magnitude and the instrumental parameters. The interferometric output is designed as in equation \ref{eq:signalModel_channeled}, with the phase contribution of the FSU, the VLTI and the known offset between A, B, C and D channels, and the estimated contribution of the air path and refractive indexes.
Finally, noise sources are introduced. The thermal background noise is modeled as a blackbody source, too, at a temperature of  300 K, see figure \ref{fig:ambient_bb}. Indeed, detector and photonic effects are modelled as random uncorrelated variables.
\\

\subsubsection{Simulation parameters}
We now list the relevant parameters used in the simulations, with their nominal values:


\begin{itemize}
	\item{source effective temperature: 3500°K}
	\item{flux at zero magnitude: 3.78e7 ph/msec, effective magnitude: 8}
	\item{The refractive index of air is evaluated following Daigne $\&$ Lestrade (\cite{Daigne99} and references herein) as:
\[n(\sigma) = 1 + \alpha + \beta \sigma^2 + \gamma \sigma^4 \] 

where $\alpha = 199.329e-6$, $\beta = 1.129e-6$ and $\gamma = 9e-9$ are air index parameters, and their values have been measured at VLTI 
}
	\item{atmospheric transmission: modelled  as in figure \ref{fig:atm_transm}
    \begin{figure}[htb]
        \begin{center}
        \epsfig{figure=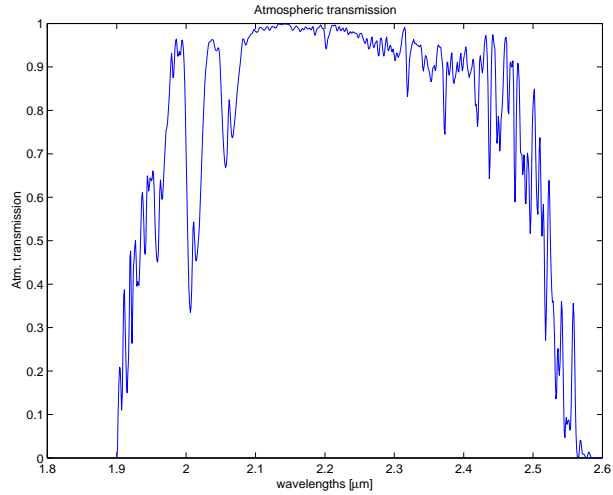,width=8cm}
        \caption{Atmospheric transmission as a function of the wavelengths.}
        \label{fig:atm_transm}
    \end{center}
\end{figure}
    }
	\item{unbalanced air path: 0 m}
	\item{phase and transmission of the VLTI: nominal values (0 and 1, respectively, for all wavelengths)}
	\item{average visibility of the VLTI: 0.75}
	\item{phase and transmission of the FSU: the nominal values are reported in table \ref{table:FSUtransm-phase}
    \begin{table}[htb]
    \begin{center}
       \begin{tabular}{lcccccccc}
       \hline
       $\lambda$    & 1.90 & 2.00 & 2.10 & 2.20 & 2.30 & 2.40 & 2.50 & 2.60 \\
       FSU tr A-C & 48.8 & 48.8 & 51.3 & 50.2 & 45.5 & 42.9 & 33.8 & 33.8 \\
       FSU tr B-D & 45.4 & 45.4 & 47.4 & 47.6 & 43.9 & 40.6 & 34.0 & 34.0 \\
       FSU ph A & 2.35 & 2.35 & -5.63 & 2.62 & 0.23 & 7.69 & -2.98 & -2.98\\
       FSU ph B & 89.26 & 89.26 & 89.52 &	89.80 & 90.09 & 90.40 & 90.72 & 90.72 \\
       FSU ph C & 175.35 & 175.35 & 184.17 & 176.82 & 180.15 & 173.69 & 185.43 & 185.43\\
       FSU ph D & 269.26 & 269.26 & 269.52 & 269.80 & 270.09 & 270.40 & 270.72 & 270.72\\
       \hline
       \end{tabular}
       \caption{Nominal transmission and phase values for the four channels A, B, C and D of the FSU. Channels A and C, and B and D share the same transmission, respectively, because they are separated before the detection. The nominal phase is around the phase angle 0, $\pi/2$, $\pi$ and $3\pi/2$, respectively.}
       \label{table:FSUtransm-phase}
    \end{center}
    \end{table}
    }
	\item{visibility of the FSU: function of the wavelengths. Its nominal values are reported in table \ref{table:FSUvisib}.
    \begin{table}[htb]
    \begin{center}
       \begin{tabular}{lcccccccc}
       \hline
       $\lambda$     & 1.90 & 2.00 & 2.10 & 2.20 & 2.30 & 2.40 & 2.50 & 2.60 \\
       FSU visibility & 0.934 & 0.938 & 0.942 & 0.946 & 0.950 & 0.953 & 0.955 & 0.955\\
       \hline
       \end{tabular}
       \caption{Nominal instrumental visibility, near 1.}
       \label{table:FSUvisib}
    \end{center}
    \end{table}
}
	\item{thermal background at the input of the FSU: modelled as a black body at temperature: 300°K (27°C) and with an amplitude of 1.22e5 ph/msec}
	\item{detector quantum efficiency: a decreasing function of the wavelength $\lambda$, tabulated in table \ref{table:template_det_QE}.
\begin{table}[htb]
\begin{center}
    \begin{tabular}{lcccccccc}
    \hline
    $\lambda$ & 1.90 & 2.00 & 2.10 & 2.20 & 2.30 & 2.40 & 2.50 & 2.60 \\
    QE & 0.65 & 0.65 & 0.62 & 0.62 & 0.61 & 0.58 & 0.45 & 0.03 \\
    \hline
    \end{tabular}
    \caption{Detector quantum efficiency spectral distribution.}
    \label{table:template_det_QE}
\end{center}
\end{table}
}
	\item{wavelength array for K band: $[1.9-2.6] \mu m$, with sampling at $0.05 \mu m$}
	\item{integration time: 1 msec}

	\item {OPD range: $[-50, 50] \mu m$, with a step of $0.2 \mu m$, for a total of $501$ points}
\end{itemize}

\subsubsection{Results of simulation}
\begin{figure*}[htb]
   \begin{center}
      \epsfig{figure=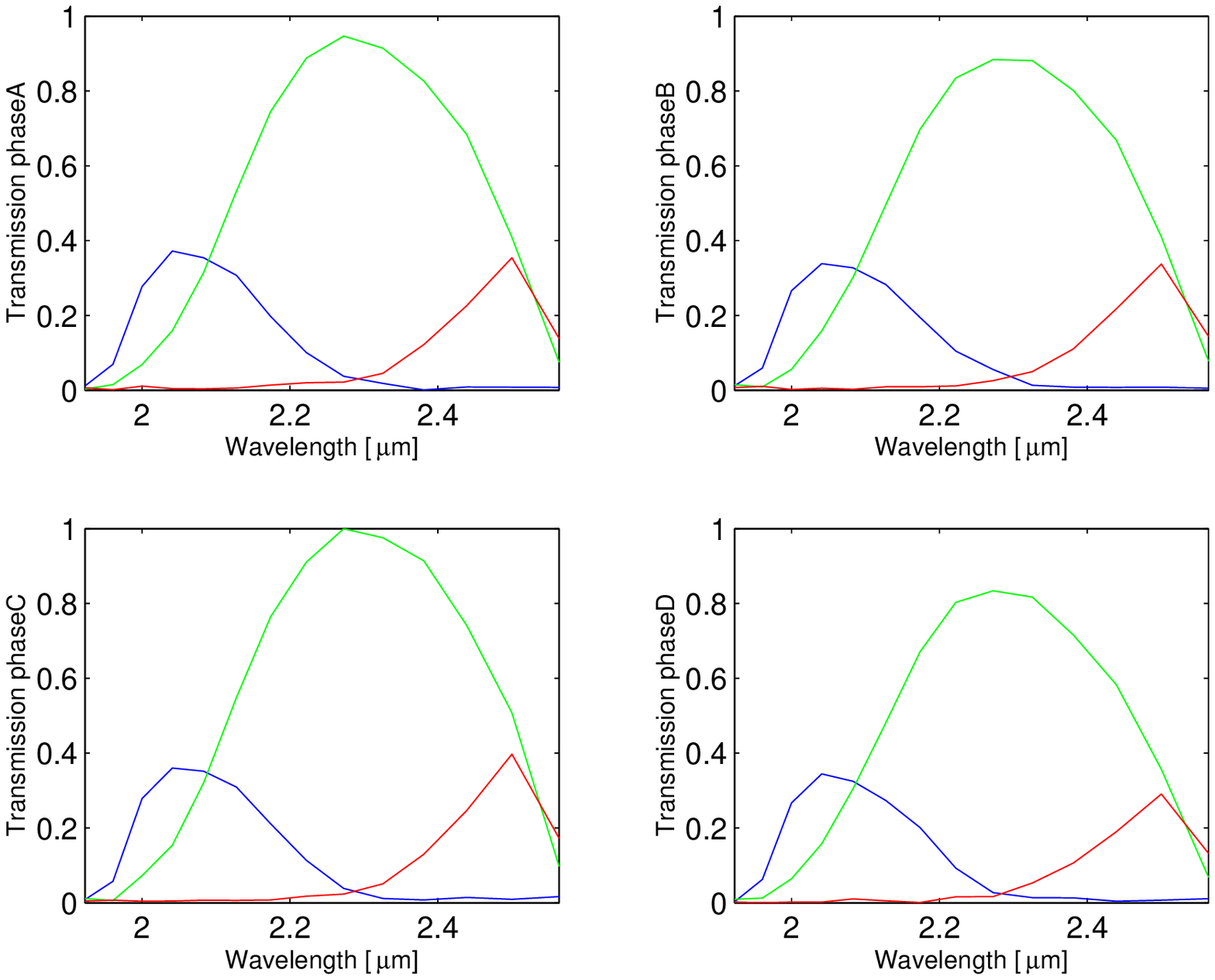,width=10cm}
      \epsfig{figure=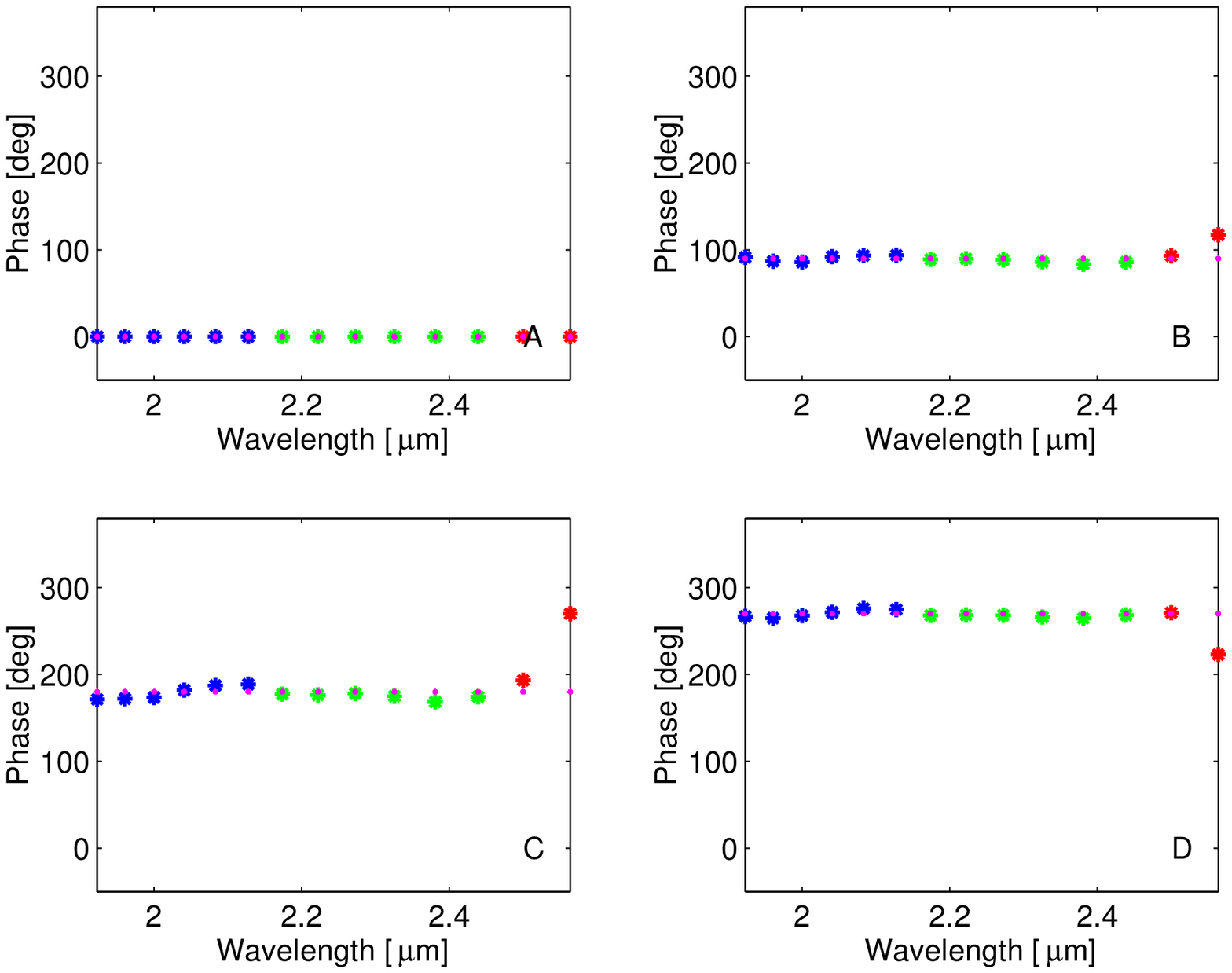,width=10cm}
      \caption{Transmission (four top) and phase (four bottom) evaluated by means of the calibration.}
        \label{fig:calib_simul}
    \end{center}
\end{figure*}

We simulate a noisy signal, adding a perturbation on the interferogram intensity, due to photonic noise and to detector read-out noise, with mean $\mu_R=20$ ph/msec and variance $\sigma^2_R = 20^2$. Both noise sources are modelled as normal random variables, thanks to the approximation of the Poisson distribution with the normal one at these flux levels. Figure \ref{fig:calib_simul} shows the results, while table \ref{table:noisytemplate_eff_wl} displays the effective wavelengths, evaluated with a weighted mean of the transmission functions. The differences between phase channels are small, but not zero.\\


\begin{table}[htb]
\begin{center}
    \begin{tabular}{lcccc}
    \hline
   & phase A & phase B & phase C & phase D \\
  K1 & 2.099 & 2.096 & 2.100 & 2.097 \\
  K2 & 2.279 & 2.281 & 2.283 & 2.277 \\
  K3 & 2.440 & 2.442 & 2.452 & 2.450 \\
    \hline
    \end{tabular}
    \caption{Effective wavelengths for the three sub-bands and the three phase channels, when there is noise on the interferogram.}
    \label{table:noisytemplate_eff_wl}
\end{center}
\end{table}




\subsection{Calibration of laboratory data}
\label{subsec:calibr_labData}

\begin{figure*}[htb]
   \begin{center}
      \epsfig{figure=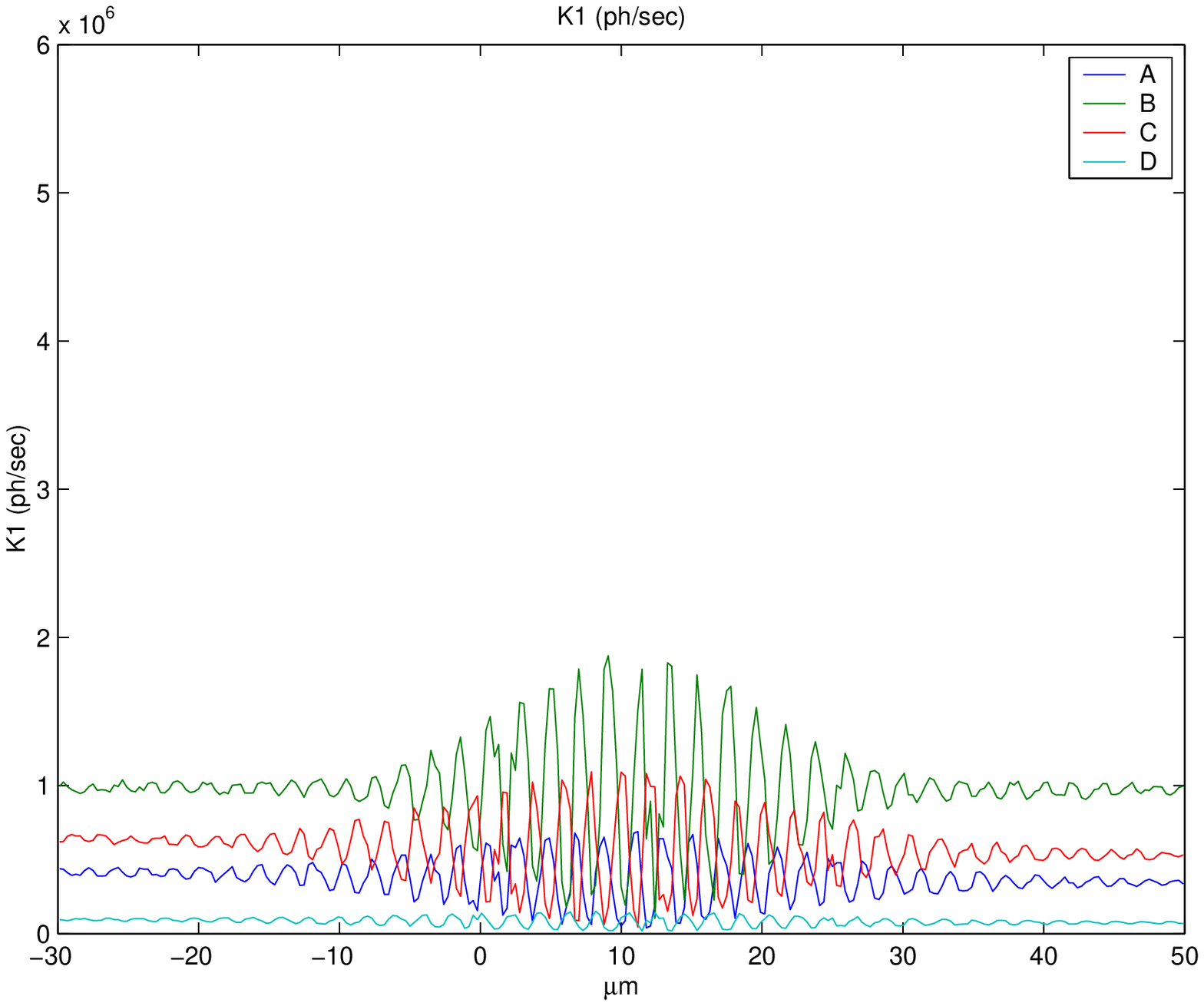,width=5cm}
      \epsfig{figure=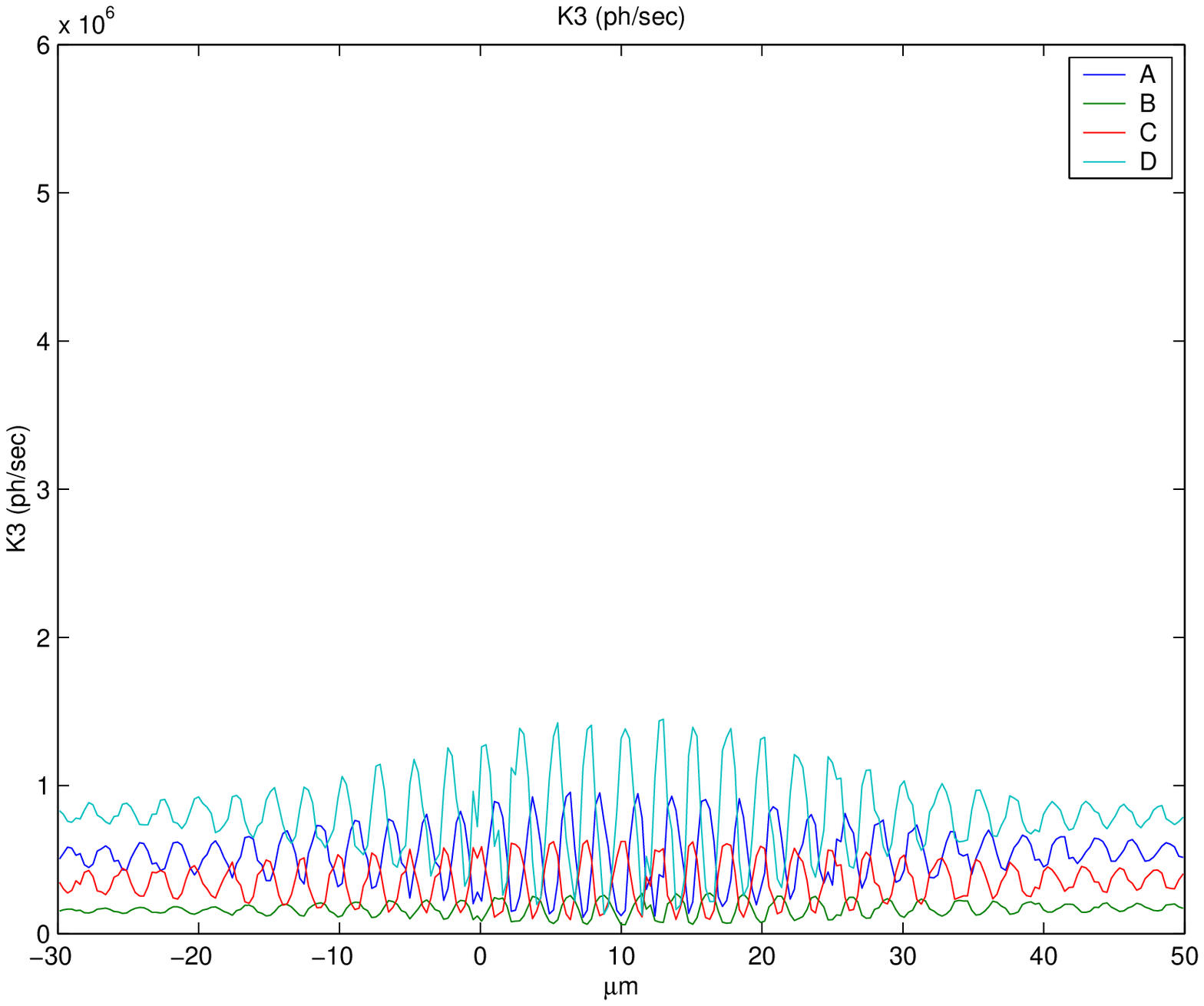,width=5cm}
      \epsfig{figure=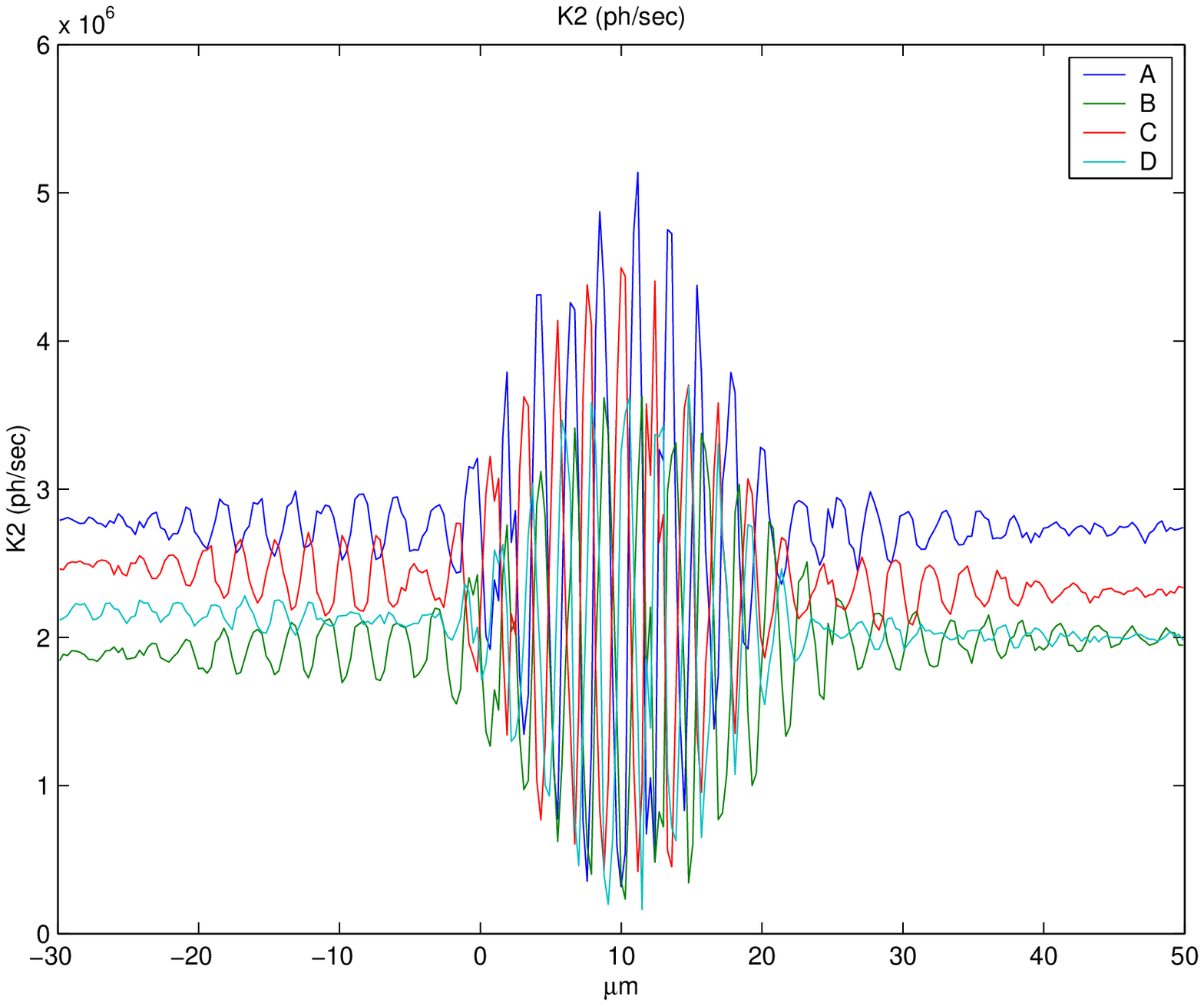,width=5cm}
      \epsfig{figure=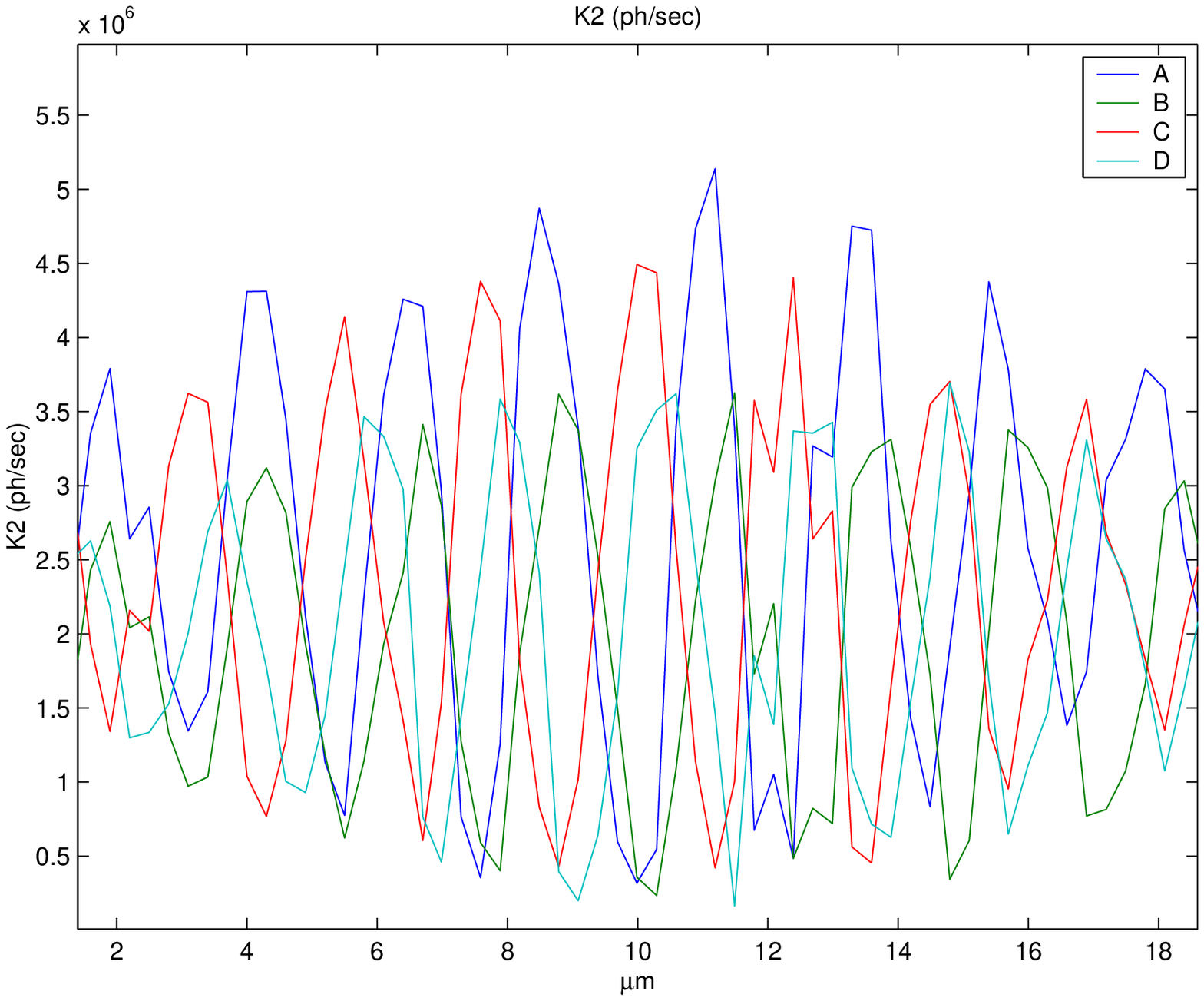,width=5cm}
      \caption{First row, raw signals in lateral bands K1 (left) and K3 (right); second row, raw signals in K2, the wider band, and a zoom of the area near to the maximum of intensity. Note that there is a good phase opposition between A and C, and B and D, and that the maximum intensity is not at the zero OPD, meaning that there is an offset of about $10 \mu m$ between arms of both combiners (A and C and B and D)}
        \label{fig:calib_rawdata}
    \end{center}
\end{figure*}

During the development of PRIMA FSU at the Alenia-Alcatel, now Thales, laboratories, several tests were performed with laser or white light input sources at different temperature and so with different spectra. The calibration procedure has been tested with these data sets.
We choose a data set as representative of the process. We recorded it on December 2005, the 14th. Data are organized as a text file with a matrix containing the recorded values for the twelve channels (A, B, C and D phase for K1, K2 and K3 bands) over an OPD scan of about $80 \;\mu m$, the record time and the OPD position sent by the software. 
\\
The OPD is sampled over the range $[-29.878, 49.896] \; \mu m$ for a total of $N = 267$ points. The mean step is $0.2999 \; \mu m$, with a variance of $5.12 \cdot 10^{-6} \; \mu m$. 
Fig. \ref{fig:calib_rawdata} shows the outputs of the channels. Each figure contains four signals, corresponding to the four phases A, B, C and D for a single band. The phase offset between the signals is nominally of $90$ degrees; the zoom of K2 signals shows a quite good phase opposition between corresponding outputs (A - C and B – D).

\noindent The laboratory temperature of the source is estimated in $800$°C, the integration time is set to $1$ second. The flux at the reference magnitude is $3.78 \cdot 10^{7}$ photons/sec.

\subsubsection{Evaluation of the overall phase and transmission functions}
\label{subsubsec:ph_transm}
\begin{figure}[htb]
   \begin{center}
      \epsfig{figure=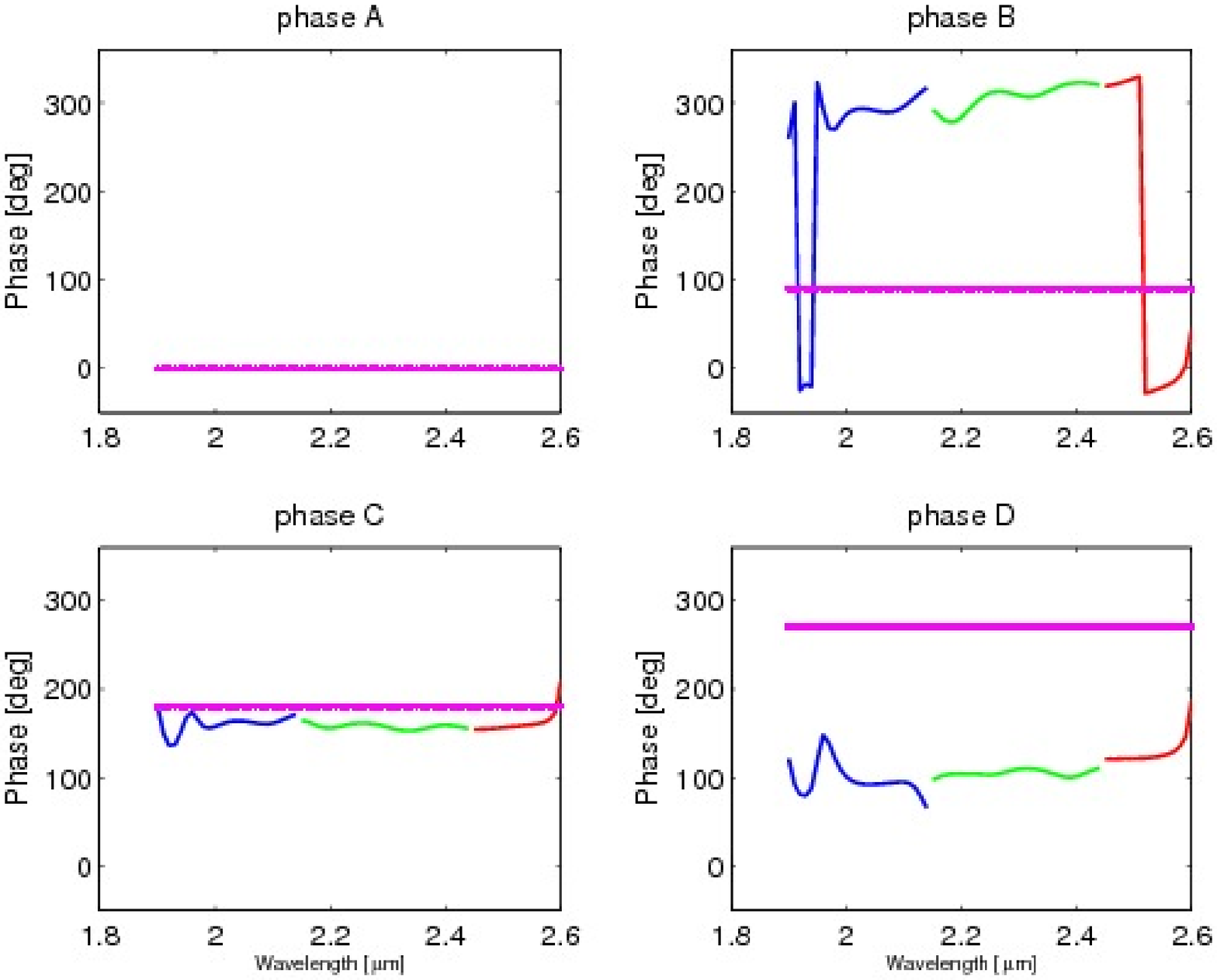,width=10cm}
      \caption{Instrumental differential phase of the four phase channels A, B, C and D; the pink line is the nominal value. The blue line is the evaluated phase for K1, the green one for K2 and the red one for K3.}
        \label{fig:calib_originalPhase}
    \end{center}
\end{figure}

The transmission and phase of the instrument are evaluated through the Discrete Fourier Transform. The wavelength range is $[1.9,2.6]\; \mu m$, with a step of $0.01 \; \mu m$. The differential phase is of interest, i.e. the relative offset between channels. In the simulation, phase A is chosen as reference one. The analysis of the phase results, plotted in figure \ref{fig:calib_originalPhase}, shows immediately that phase D and phase B are exchanged. This fact was due to a wrong pixel indexing during the reading of the detector, and it was corrected.\\

\begin{figure}[htbp]
   \begin{center}
      \epsfig{figure=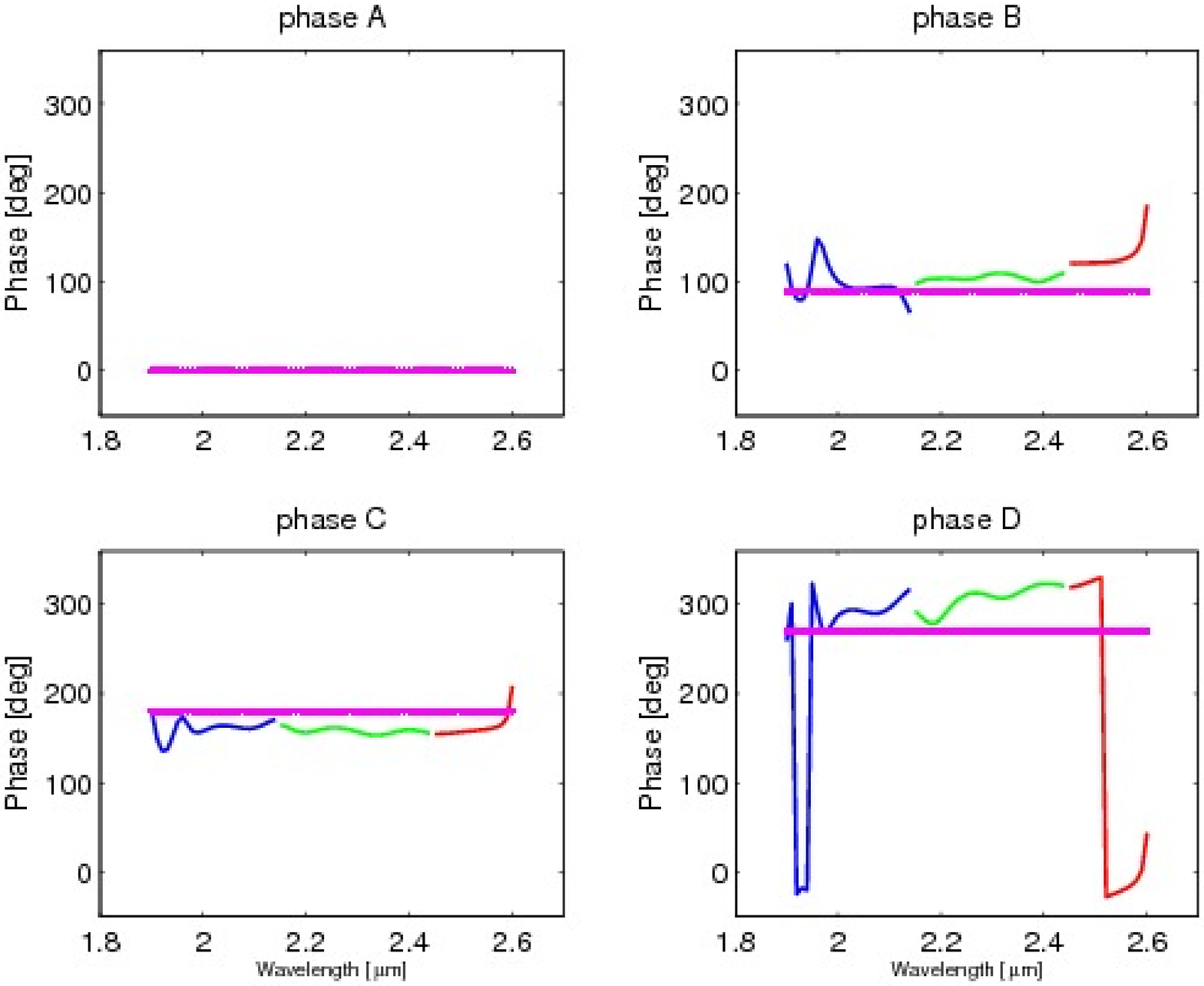,width=10cm}
      \vspace{1cm}
      \epsfig{figure=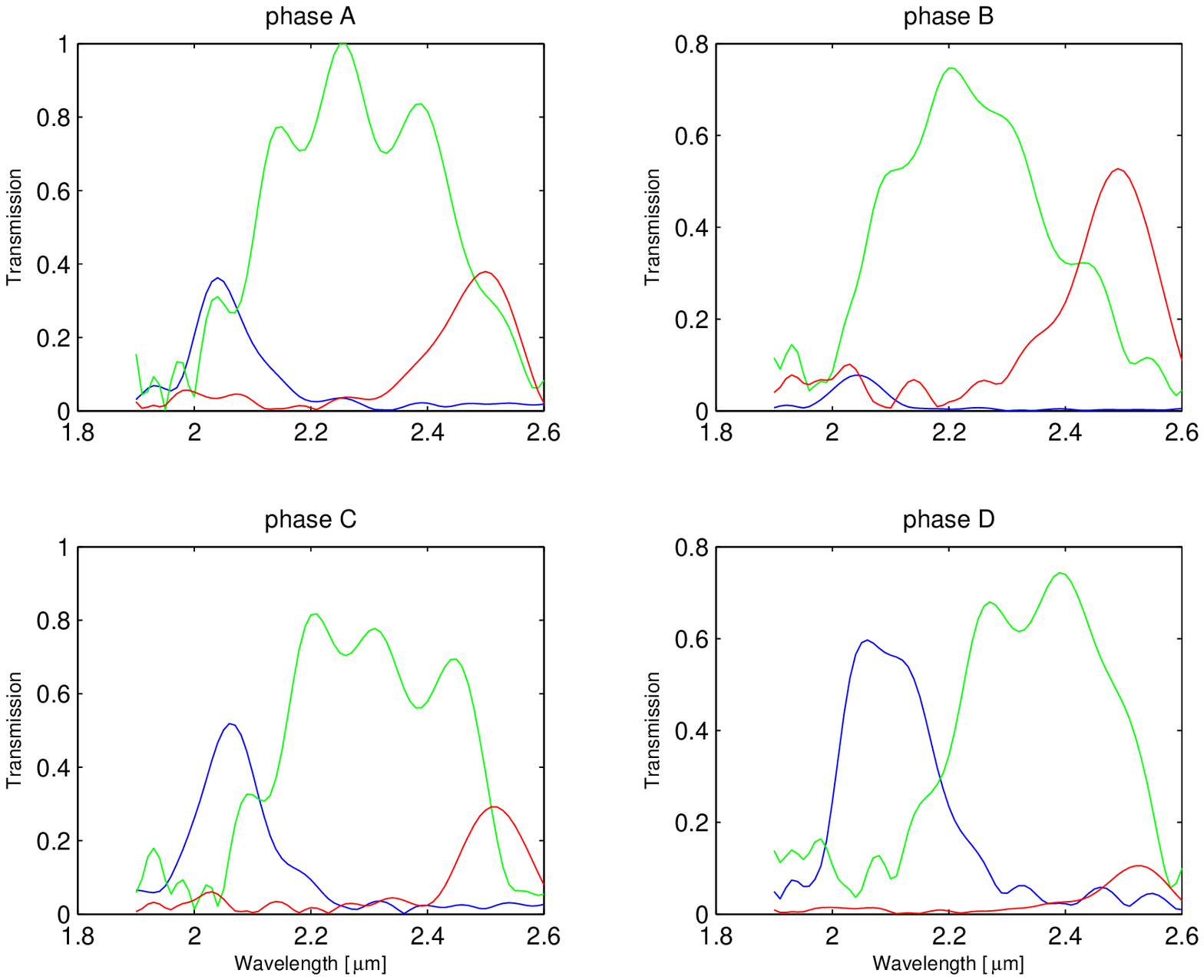,width=10cm}
      \caption{At the top, corrected instrumental differential phase of the four phase channels A, B, C and D; the pink line is the nominal value. The blue line is the evaluated phase for K1, the green one for K2 and the red one for K3; at the bottom, transmission functions. Each picture groups together the transmission curves of each phase channel.}
        \label{fig:calib_ph_transm}
    \end{center}
\end{figure}

\noindent After the right indexing of the phases, the transmission and the phase curves have the pattern shown in figure \ref{fig:calib_ph_transm}, to be compared with the nominal example of figure \ref{fig:calib_simul}. It can be noted that the flux is not equally separated between different sub-bands, but there is a redistribution of it. The pattern of the transmission function is also different from the nominal one, especially for the side bands, that are more sensible to small misalignment of the spot on the detector. This has a weight on the effective wavelengths of the channels, reported in table \ref{table:calib_eff_wl}. There are some differences depending on the phase channel, and in particular we can notice that in general band K3 has a lower wavelength than the nominal situation. This can mean that part of the flux of K2 has migrated into K3 pixels, or that flux in K3 was lower than expected from the instrument transmission, lowering the weight of longer wavelengths.\\
The phase functions shows a stability over the phase channels, even if there are some discrepancies between the nominal values (0, 90, 180, 270 degrees).

\begin{table}[htb]
\begin{center}
    \begin{tabular}{lcccc}
    \hline
   & phase A & phase B & phase C & phase D \\
  K1 & 2.099 & 2.088 & 2.101 & 2.128 \\
  K2 & 2.279 & 2.240 & 2.296 & 2.322 \\
  K3 & 2.412 & 2.402 & 2.426 & 2.422 \\
    \hline
    \end{tabular}
    \caption{Effective wavelengths for the three sub-bands and the three phase channels.}
    \label{table:calib_eff_wl}
\end{center}
\end{table}

\subsubsection{Evaluation of the visibility}
\label{subsubsec:visib}
\begin{figure*}[htb]
   \begin{center}
      \epsfig{figure=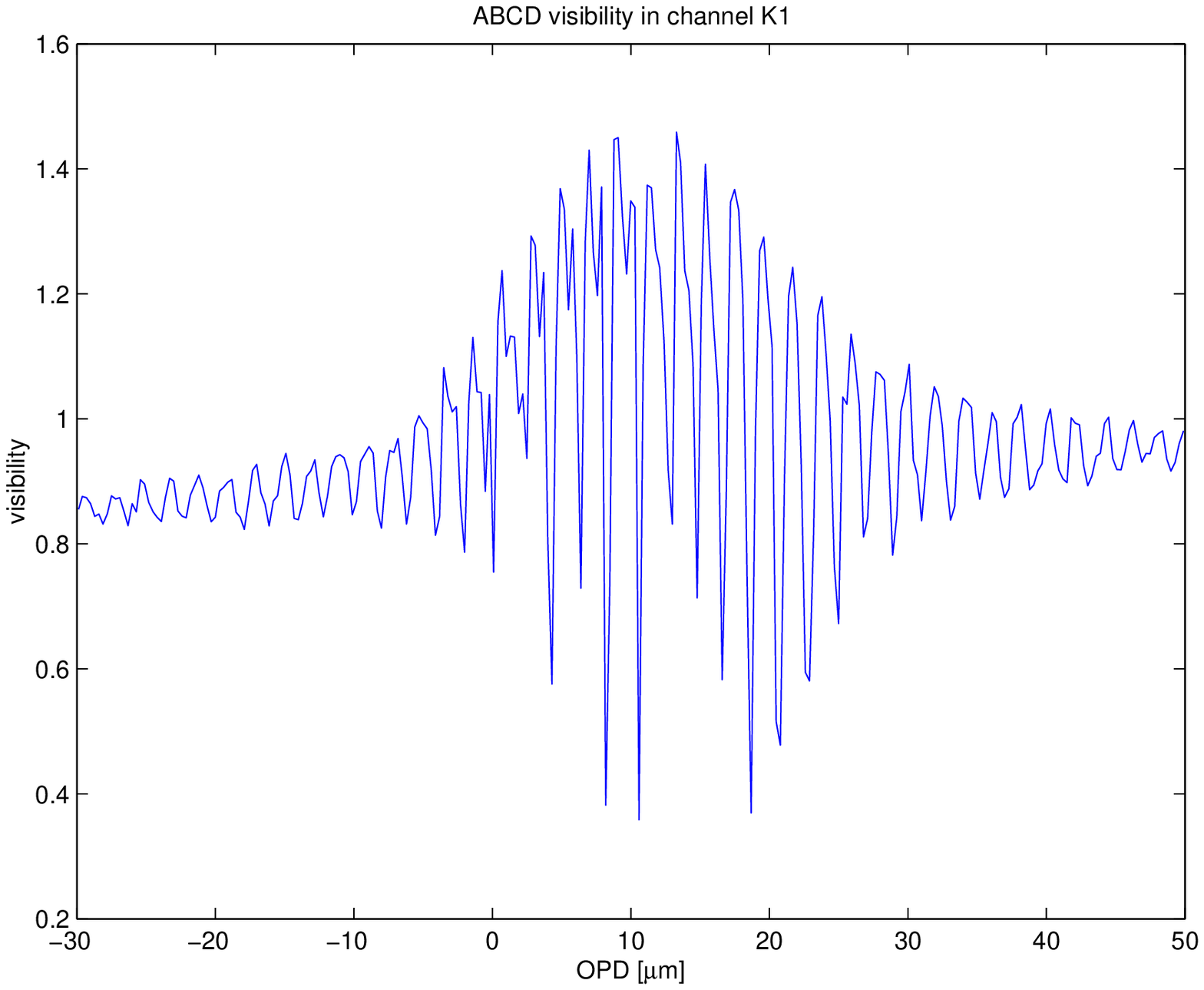,width=6.5cm}
      \epsfig{figure=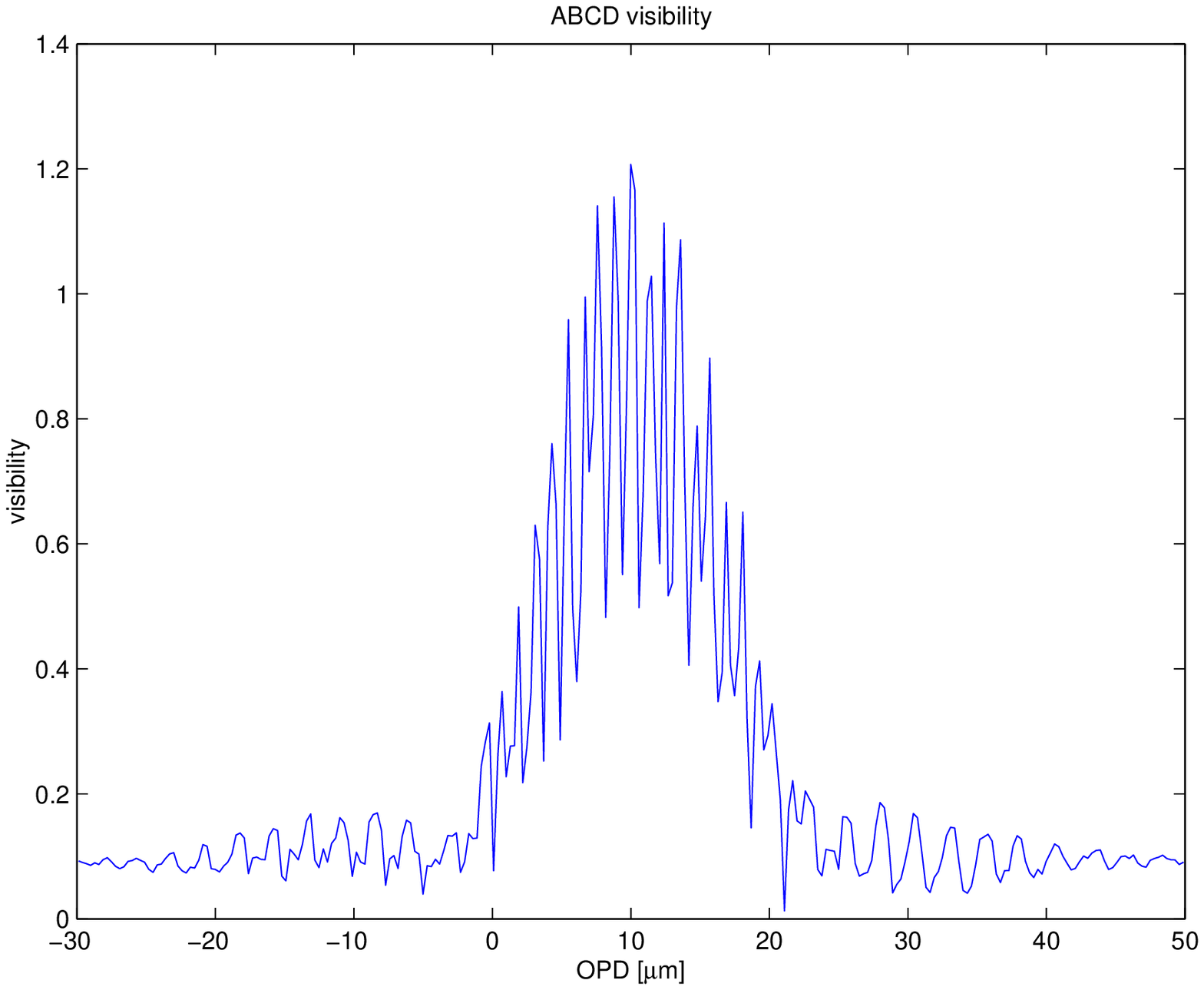,width=6.5cm}
      \caption{ABCD visibility curves for K1 (left) and K2 (right). It is evident the effect of the envelope and of the modulation.}
        \label{fig:ABCDvis}
    \end{center}
\end{figure*}

The evaluation of the overall visibility can be performed in several ways. Being in the quadrature case, so with the A, B, C and D outputs, we can use the standard ABCD method, seen in chapter \ref{chap:FINITO}, but here the lack of normalization of the beams can not be resolved through photometry, and it causes the visibility to be corrupted by the envelope shape. The ABCD visibilities in one side subband (K1) and in the central K2 are reported in figure \ref{fig:ABCDvis}. One way is to search the maximum of the visibility function.

\noindent Another way is to evaluate the modulo of the complex visibility function, and from theory we know that it is equal to the Michelson visibility \cite{Michelson}:
\begin{equation}
V_M = \frac{I_{max} - I_{min}} {I_{max} + I_{min}}
\end{equation}
where I is the intensity in a channel. With this approach, we find a visibility value for each sub-band and each phase channel, and we can interpolate them over the wavelength range.
The values found for each channel are reported in table \ref{table:calib_michel_visib}.

\begin{table}[htb]
    \begin{center}
        \begin{tabular}{lcccc}
        \hline
        & phase A & phase B & phase C & phase D \\
        K1 & 0.891 & 0.789 & 0.920 & 0.895 \\
        K2 & 0.884 & 0.916 & 0.829 & 0.879 \\
        K3 & 0.794 & 0.840 & 0.740 & 0.640 \\
        \hline
        \end{tabular}
        \caption{Michelson visibilities from calibration data}
        \label{table:calib_michel_visib}
    \end{center}
\end{table}

\subsubsection{Evaluation of the magnitude}
\begin{figure}[htb]
   \begin{center}
      \epsfig{figure=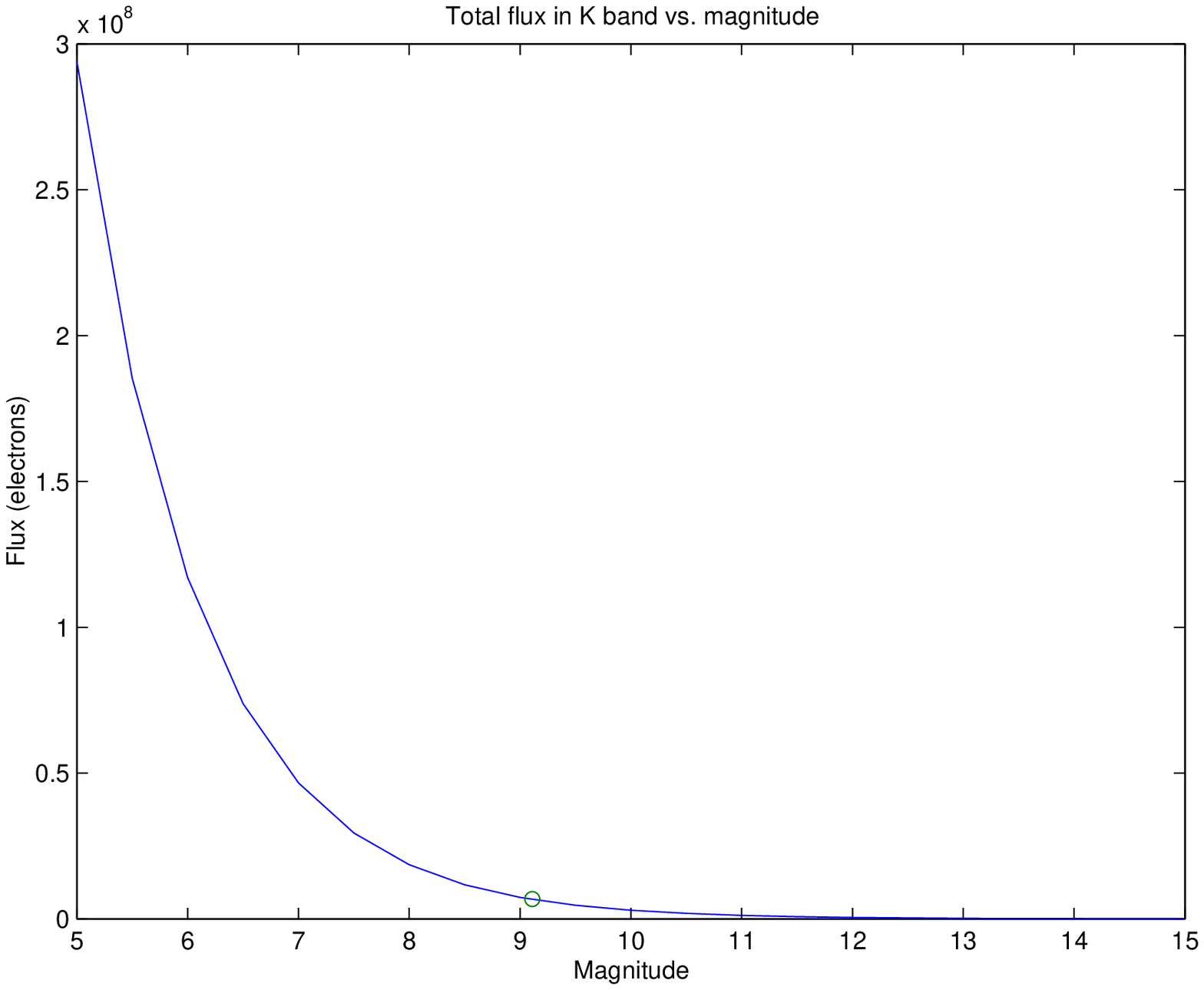,width=8cm}
      \caption{Total flux in K band and the four phases for different magnitudes. The current magnitude is evaluated interpolating the total sum of the data flux over this curve.}
        \label{fig:calib_mag}
    \end{center}
\end{figure}

With the values of the transmission function, of the phase (section \ref{subsubsec:ph_transm}) and of the overall visibility (section \ref{subsubsec:visib}), it is possible to evaluate the magnitude of the source. The followed method foresees the generation of a set of template with different magnitudes, but with the evaluated values of transmission, phase, visibility, and leaving all other parameters unchanged. The flux level of the real data is then compared with the template ones, and the estimated magnitude is interpolated from the flux curve (see figure \ref{fig:calib_mag}). The value is $9.11$.

\subsubsection{Signals reconstruction}
\begin{figure}[htbp]
   \begin{center}
      \epsfig{figure=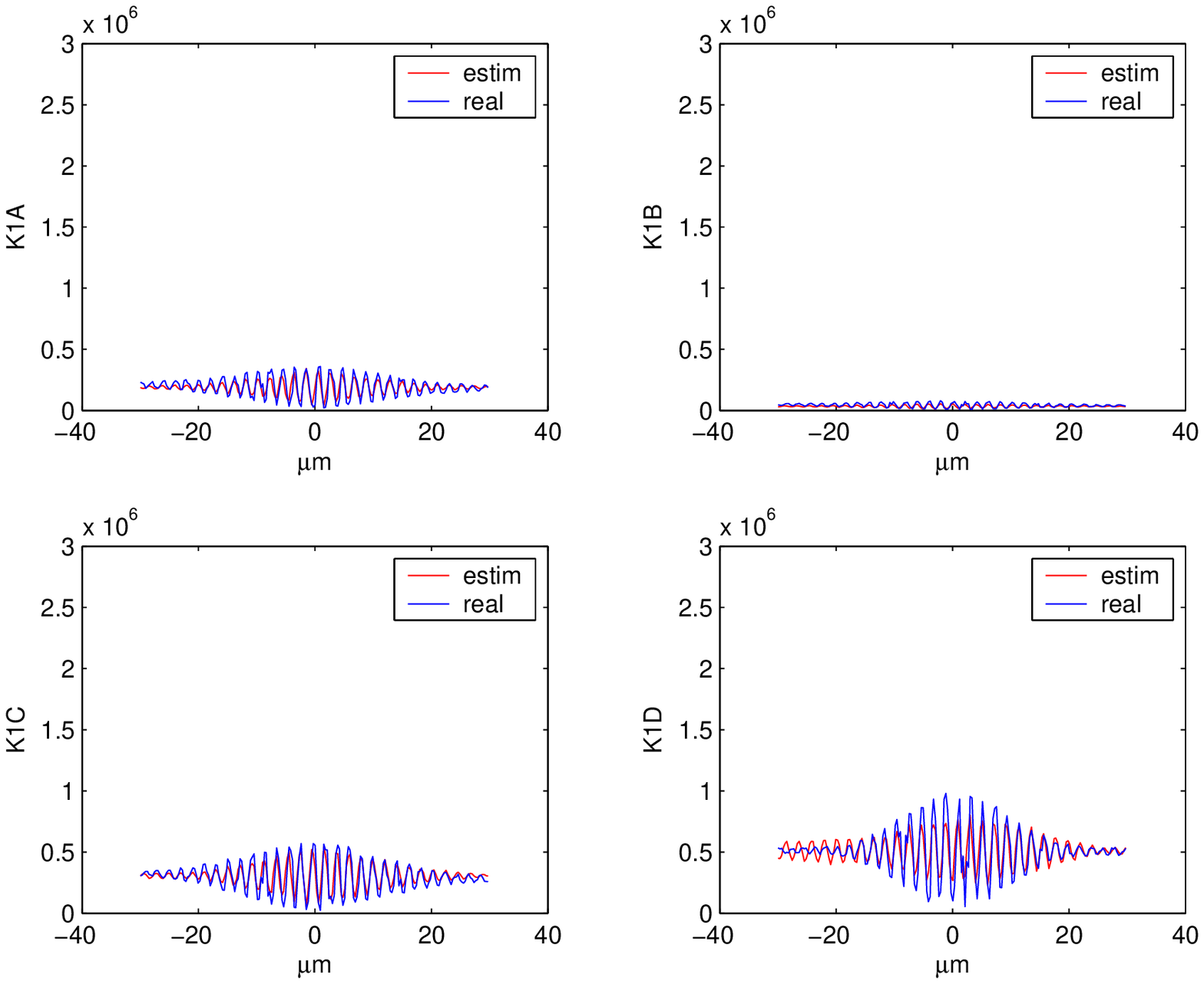,width=8cm}
      \vspace{1cm}
      \epsfig{figure=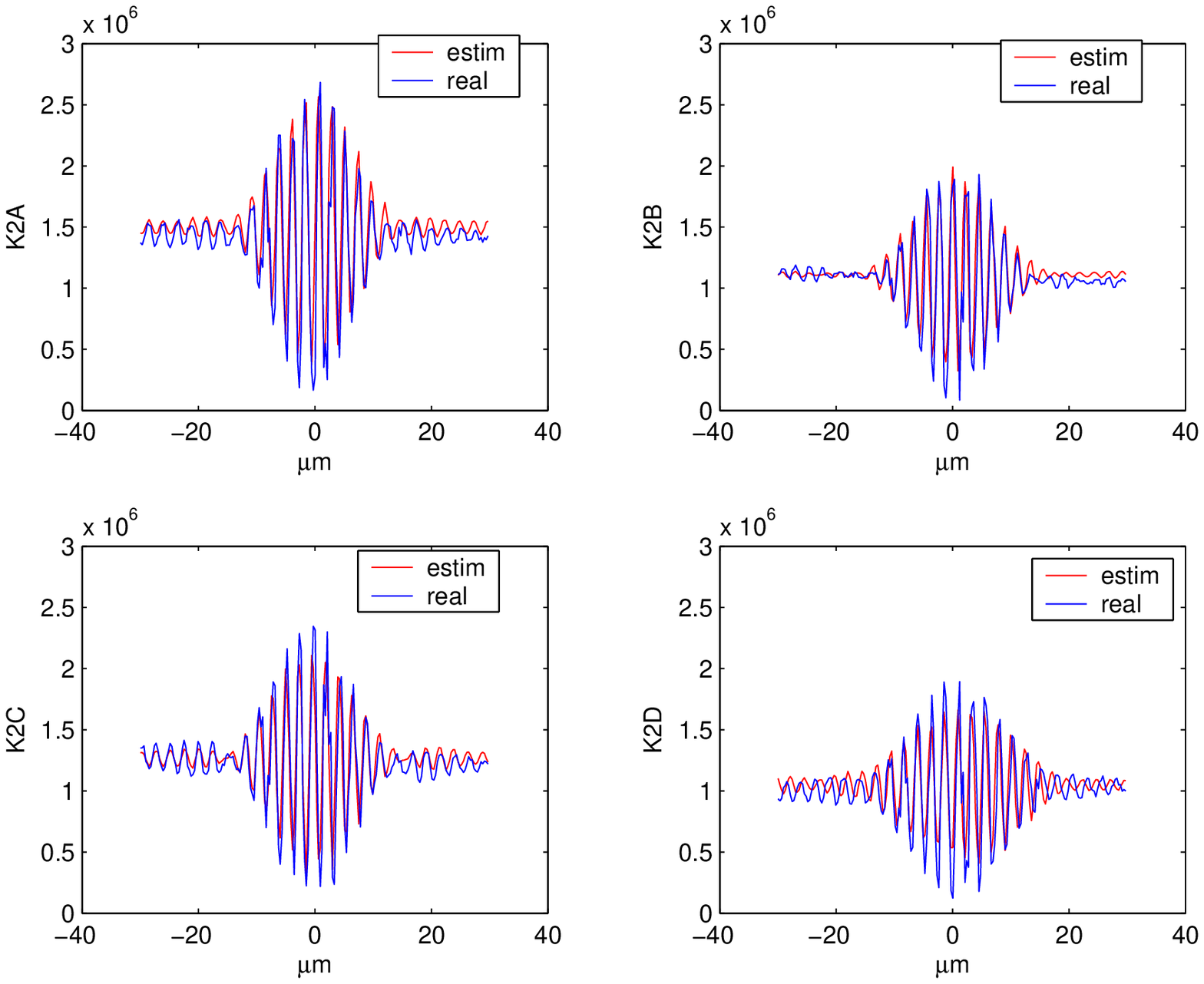,width=8cm}
      \vspace{1cm}
      \epsfig{figure=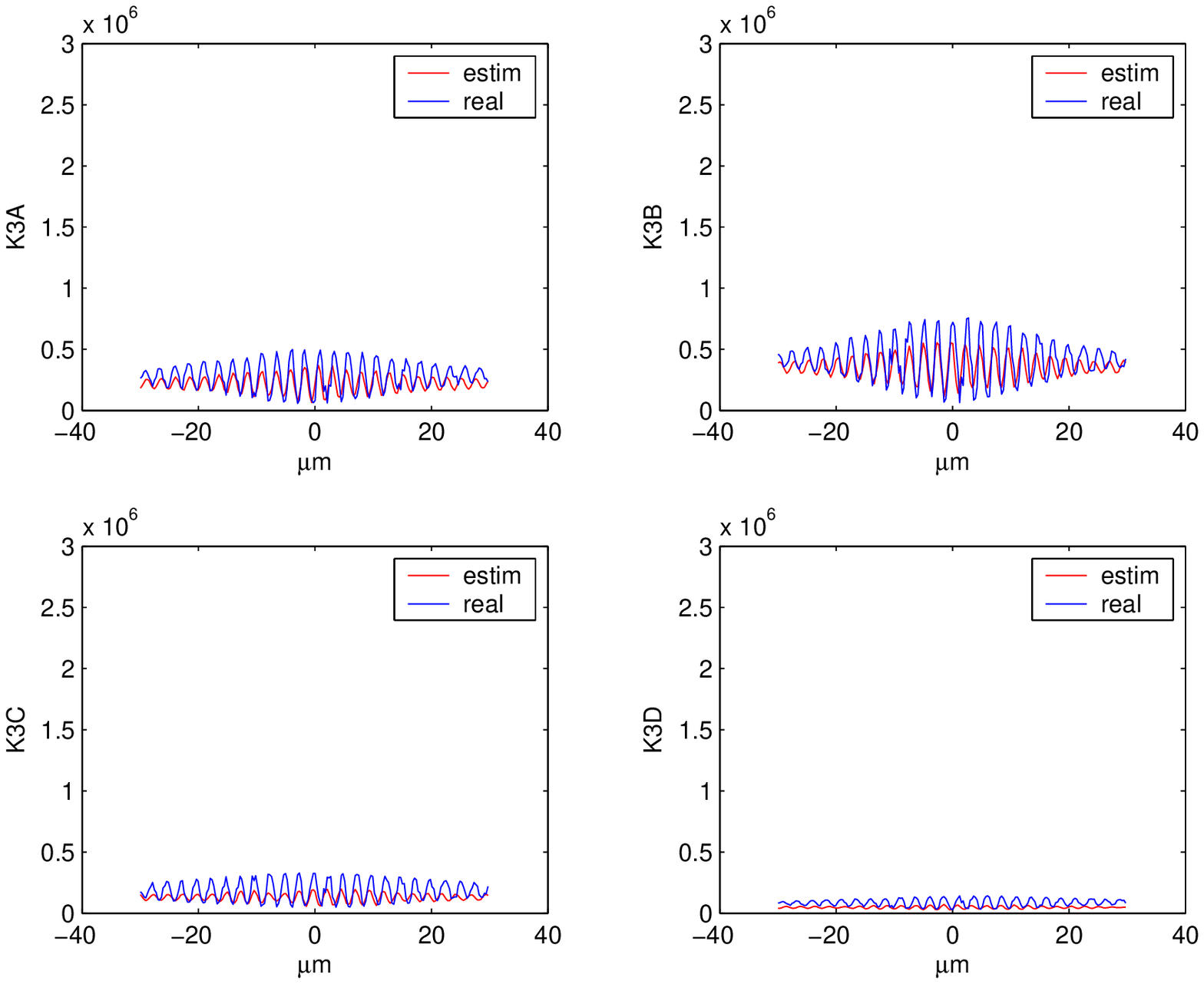,width=8cm}
      \caption{From top to bottom, measured (blue) and reconstructed (red) interferogram for K1, K2 and K3. }
      \label{fig:calib_reconstr}
    \end{center}
\end{figure}

\begin{figure}[htbp]
   \begin{center}
      \epsfig{figure=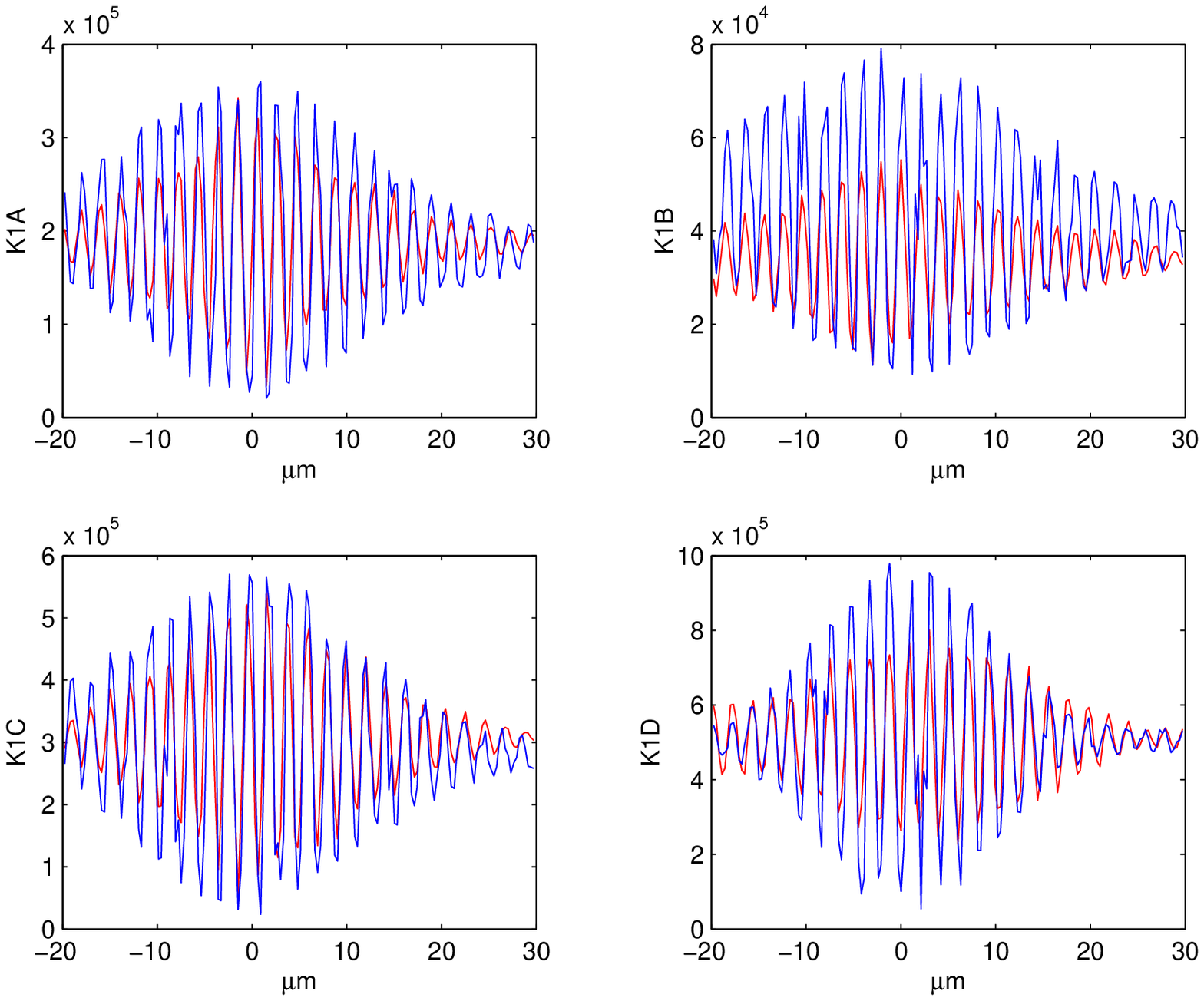,width=8cm}
      \vspace{2cm}
      \epsfig{figure=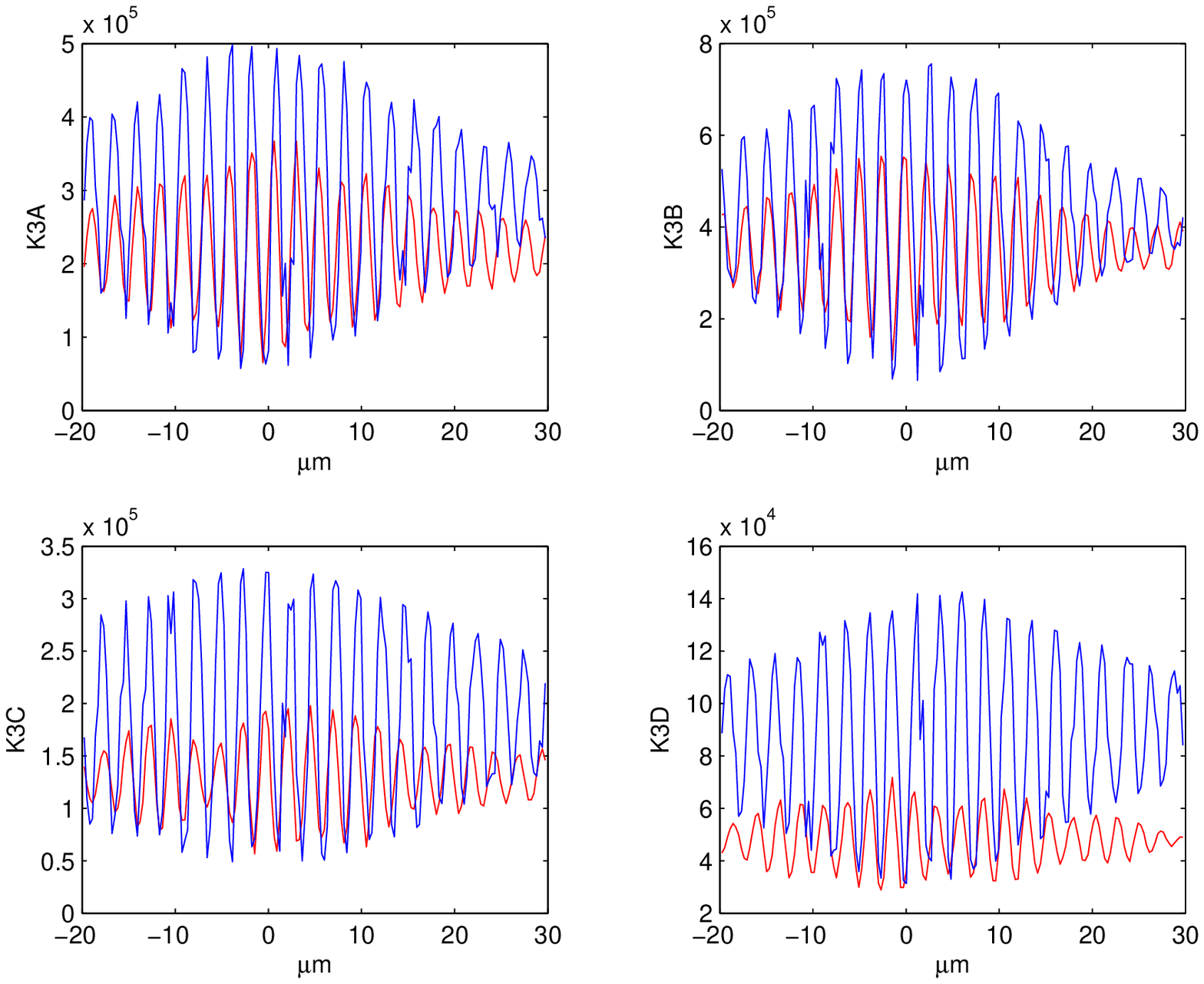,width=8cm}
      \caption{Zooms of K1 (first row) and K3 (last row) signals. }
      \label{fig:calib_reconstr_zoom}
    \end{center}
\end{figure}

With the information collected before, it is possible to reconstruct the signals, following equation \ref{eq:signalModel}. The monochromatic interferograms are computed, and then added together to form the final polychromatic signal. Figure \ref{fig:calib_reconstr} shows both the measured (blue) and the reconstructed signals. It must be noted the perfect similarity of fringe separation, while the flux level needs further adjustments. This is true especially for band K1 and K3 (see figure \ref{fig:calib_reconstr_zoom} for a zoom), for which the flux is weaker, while in K2 the reconstruction is very faithful.

\subsection{Conclusions}
In this chapter, we have analyzed a new approach to the problem of estimating the current position of the fringe packet with respect to the optical path difference scan, based on an accurate modeling of the interferometric signal. We have found good simulated results, but we have also highlighted the need of a precise calibration procedure. We could reproduce very well the spectral behaviour of the measured signal, and this is an important goal. However, we pointed out that there are still some uncertainty on the evaluation of the signal intensities magnitude.

\chapter{Statistical analysis of interferometric data}
\lhead[\fancyplain{}{\bfseries\thepage}]%
      {\fancyplain{}{\bfseries Statistical analysis of interferometric data}}
\rhead[\fancyplain{}{\bfseries Statistical analysis of interferometric data}]%
      {\fancyplain{}{\bfseries\thepage}}
\label{chap:stat}
\noindent In this chapter, we will analyze typical interferometric data with classical statistical techniques, both in the time and in the frequency domain. Then we will select some other statistical instruments, such as tests on variance and multiple regression analysis, to properly identify features of interferometric signals, in particular the variability.

\section{Introduction}
\label{sec:intro4}

In the first part of this thesis, we have described the physical and mathematical framework of interferometry, and used different models in the algorithms for estimating the fringe location with respect to the optical path. We have used the ideal model (see chap. \ref{chap:intro}, par. \ref{subsec:poly_interf}):
\begin{eqnarray}\label{eq:FINITO_model}
\nonumber I(p,\lambda) &=& (I_1 + I_2) \left[1 + \frac{\sqrt{I_1 \cdot I_2}}{I_1 + I_2} \cdot sin ( \frac{2 \pi}{\lambda} p) \right] \\
I(p) = \int_{\lambda_1}^{\lambda_2} I(p, \lambda) &=& (I_1 + I_2) \eta_0 \Delta\nu \left[1 + sinc \frac{\pi p}{L_C} cos\left(\frac{2 \pi p}{\lambda_0}\right)\right]
\end{eqnarray}
where $I(p,\lambda)$ and $I(p)$ are the monochromatic and the polychromatic interferogram, respectively, $I1$ and $I2$ are the intensities of the signal from each telescope, $p$ is the differential optical path (OPD) between the two incoming beams, $\Delta\nu$ is the spatial frequency range, linked to the wavelength range $[\lambda_1,\lambda_2]$, $\eta_0$ is the constant instrument response, and $L_C$ is the coherence length.
We have seen that, when working with real data, in real context, this model is not sufficient (chap. \ref{chap:FINITO}). Many sources of noise add to the real signal, such as atmospherical turbulence, that we have described from theoretical model (see chap. \ref{chap:intro}, par. \ref{subsubsec:atm_turbulence}). Their residuals fluctuations after the adaptive optics correction, plus instrumental noises, can be described in their simpler form by gaussian processes. Also instrumental characteristics, such as transmission and phase, can be properly described as spectral distributions, as they are not constant over the wavelength range.\\
To include such contributions in the signal model, it is necessary to add new variables to the model:
\begin{equation}\label{eq:PRIMAmodel}
I(p) \simeq \rho I \left[1+\eta^p \cdot \eta^C \cdot \eta^I \cdot V sin\left(\varphi + \frac{2 \pi} {\lambda_0} [n \cdot p + (n - n_0) \cdot p_A]\right)\right]
\end{equation}
where $I$ the intensity of the incoming signal, $\rho$ is the transmission factor, $\eta^p$ the instrumental visibility, associated to the photometric unbalance of channels, $\eta^C$ the source spectral distribution, $\eta^I$ the wavefront degradation, $V$ the source visibility, $\varphi$ the instrumental phase, $n=n(\lambda)$ the air refraction index and $p_A$ the optical path in air. \\

\noindent All these parameters have to be measured, and their fluctuations have a consequence on the system stability, as we said in chap. \ref{chap:PRIMA}. Even if this model can give good results, there are still some discrepancies between measured and reconstructed signals, or degeneracy among parameters increasing the difficulty of correct estimates.
We can expect some residual fluctuations after the wavefront correction done by adaptive optics, or higher order interactions between the two beams. There are still some efforts to do to describe correctly the incoming beams, and to understand if some features (flux intensities variations, or spectral characteristics, and so on) are systematic or random, in order to properly model them in \ref{eq:PRIMAmodel}.\\

\noindent Is it possible to define a more manageable equivalent signal model, together with a set of diagnostics and estimate algorithms? And if it is the case, is it possible to use them for several different data, in order to compare the results?\\

\noindent If we read the values given by a detector after each integration time as a time series, the ideal approach would be to have a mathematical model containing the signal static features, as eqs. \ref{eq:FINITO_model} and \ref{eq:PRIMAmodel}, the noise sources suggested by experimental and theoretical evidence, their correlations and temporal evolution: in a word, a stochastic equation. The difficulties, however, are great. In literature it is possible to find statistical distributions for photometric signals in different conditions (see, for example, the treatment of Goodman\cite[ch. 4]{Goodman}). But when the beams get combined, it is not clear which is the most convenient description of what happens inside the instrument. Even in the case we could assign a distribution to some variables (for example, noise), it is not easy to identify their evolution in the final interferogram, because we just have a deterministic description of the physical phenomenon, and we have no simple way of propagating the random process distributions.\\

\noindent An alternative solution is to use statistical methods. First of all, we have to restrict the field of research, i.e. the questions that could benefit from an analysis of this type, then to properly model the physics involved and to find a related sufficient amount of data to have statistical significance, and finally to select the appropriate statistical tools.\\

\noindent In this work, we will focus on the analysis of astronomical beams before and after the combination, in order to determine features maintained or changed in the interference. For this, we have selected data coming from the VLTI commissioning instrument, VINCI, for several reasons, such as the fact that both photometric and interferometric outputs are available, for large data sets. The description of VINCI and of its data is the subject of par. \ref{sec:data_descr}. We use the classical instruments of the statistical analysis, both in the time (par. \ref{sec:time-stat}) and in the frequency (par. \ref{sec:spectral_analysis}) domain.\\
We then focus on the variability of all these signals, and we use more sophisticated tools to follow the time evolution of this variability (par. \ref{sec:var_analysis}). We also try to identify the contribution of the combination system to the variability of the output beams. For this scope, we retain the simpler model of eq. \ref{eq:FINITO_model}, and we use the regression analysis and its tests (par. \ref{sec:GLM}).\\

\noindent Finally, we list in par. \ref{sec:future} some questions that arose throughout this analysis, and that could also benefit from statistical analysis.\\

\noindent We have encountered several difficulties while applying the described tools to our data. First of all, sampled data values are integers, and this poses some problems while using tests based on normal distribution, which is continuous. Then, features of the signals, such as time-varying trends, required some care even in applying estimators of functions, like covariances and correlations, that are well known and deeply studied in literature.
Finally, stationarity of time series is an important property to validate the results of statistical and probabilistic approach, but we gathered evidence that we are handling time series that are not stationary, not even weakly.\\



\section{Instrument and data description}
\label{sec:data_descr}

The analysis has been performed using sets of measurements acquired by the VLTI VINCI (VLT INterferometer Commissioning Instrument) instrument, working in K band ($[2.0-2.5] \; \mu m$). This choice was justified by the characteristic of VINCI and of its data.
The principal request on the data is to have both photometric and interferometric signals, recorded synchronously, in order to be able to link photometric level at a given time to the corresponding interferometric signal.\\
This is possible also with other instruments, such as FINITO in combination with scientific instruments like AMBER, that are now working with real data (see par. \ref{subsec:VLTIft} of chapter \ref{chap:intro}). The VINCI data were preferable for the availability of a large set of homogeneous data, collected in a comparably long period, since the instrument was used since the beginning of this century. \\
The amount of data allows a statistical analysis with some confidence in the validity of the results.
\\
\noindent A detailed description of the VINCI instrument can be found in literature \cite{Kervella00}. Here we give a summary of its principal characteristics. The stellar beams collected by two telescopes are set in nominal phase by the delay lines, then they enter the instrument and are injected into optical fibres, that bring them into the core box, called MONA. The beams first enter two beam separators that send half of each beam directly to the detector for photometry, while the other half are sent to a common coupler, where they can interfere thanks to the electric fields superposition within the coupler. The OPD scan is modulated by a mirror mounted on a piezoelectric translator. The coupler provides two complementary outputs, containing the interferometric modulation and that are sent to the detector. Only the four illuminated pixels on the detector are read, to increase the readout rate.\\
While the OPD is modulated on a complete scan of the coherence length, the detector is read at a frequency up to few kHz. This procedure allows to have time series of modulated interferometric pattern together with the correspondent photometric time series.
\\
\noindent We have to highlight a peculiarity: the sampling of the photometry and the interferometric outputs is synchronous, but the photometry is just one half of the incoming beam, and it is {\it not} the half that contributes to the interferometric output. In the following analysis, we will assume that the two halfs (the one sent to the detector and the one sent to the interferometric coupler) are equal and are subject to the same amount of noise. It is a reasonable hypothesis, because the beams are traveling into optical fibres, and not in air.\\
\\
\noindent Each OPD scan provides a record of data, read from the detector, composed by four time series, two for the photometry and two for the interferometry. The flux intensities are given in ADU (Analog Digital Unit) and they are integers. The scans are repeated, and a number of records is stored.
\\
\noindent Each observation provides four different sets of data. The first three sets are for calibration purposes. The fourth contains the results of the interferometric observation. We will hereafter refer to these different sets as case $1$ to $4$:

\begin{itemize}
	\item [case 1.] {Off source. The two arms of the beam combiner are opened, but not fed by source flux. It is a record of the noise level and noise propagation inside the instrument.}
	\item [case 2.] {One arm of the combiner (arm A) is fed with stellar source, while the other is closed; it still contains background noise. It is useful to check feature of the single arm.}
	\item [case 3.] {The same as case 2, but specular with arm B.}
	\item [case 4.] {On source: both arms are fed with stellar source, so the interferometric combination is possible.}
\end{itemize}

\begin{figure*}[ht]
    \epsfig{figure=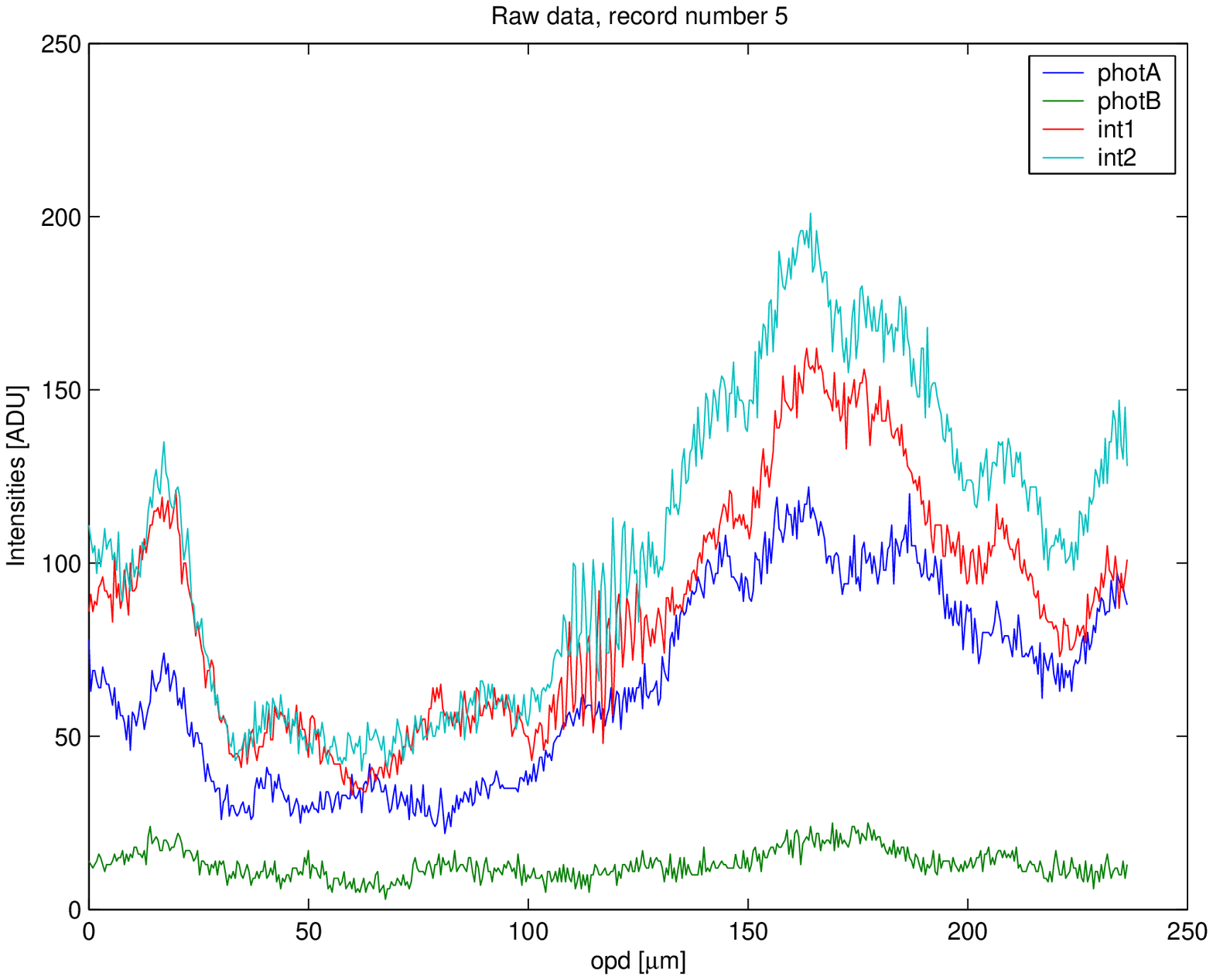,width=7.5cm}
    \epsfig{figure=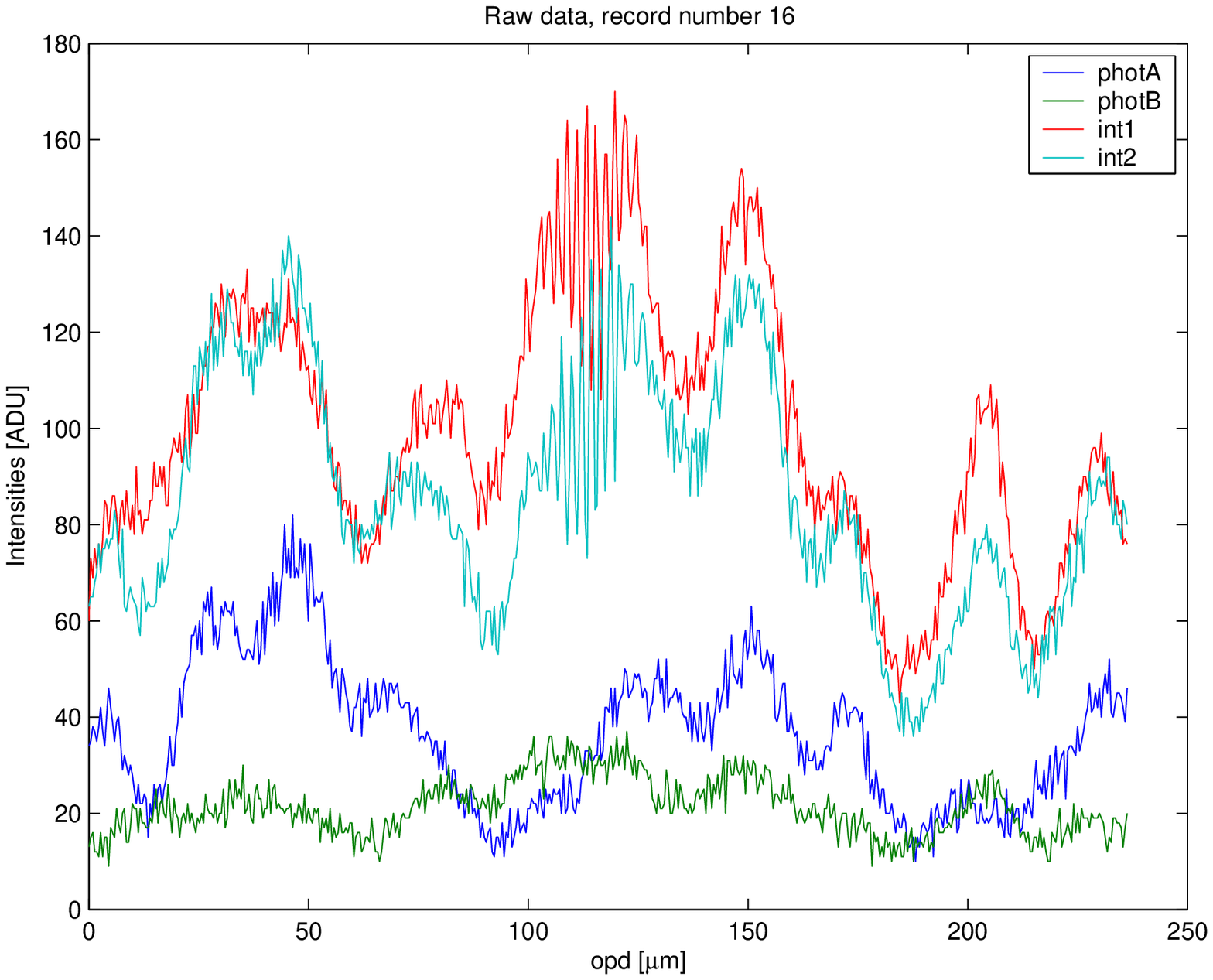,width=7.5cm}
    \begin{center}
        \caption{Raw data, record 5 and 16}
        \label{fig:rec5-16alldata}
    \end{center}
\end{figure*}

\noindent Each case provides the recording of the four pixels (two for the photometry and two for the interferometry), for a number of complete OPD scans ($100$ for case from 1 to 3, and $500$ for case 4).
\\
\noindent The data set analyzed is based on an observation done on July 15, 2002. The reference target was $\Theta$ Centauri, while the scientific star was $\alpha$ Centauri A. 
Both stars are bright, but $\Theta$ Centauri is smaller than $\alpha$ Centauri A, so its visibility is higher. The data set, used to describe the VINCI data processing \cite{Kervella04}, is part of a series of observations used for the determination of the angular diameter of $\alpha$ Centauri A \cite{Kervella03}.
We choose the reference star, and we retrieved the data from the ESO archive (http://archive.eso.org).
\\
\noindent The integration time was set at $0.364$ seconds, at a piezo frequency of $650$ $\mu m/sec$, which gives a scan of $236.6 \; \mu m$. Each scan contains $526$ points, so the step is $0.45 \; \mu m$. The observation is done in the K spectral band ($[2.0-2.5] \; \mu m$), this means that each fringe contains $\sim 5$ points.\\ 
Data are counts of photons occurrences (ADU). This means that, as said before, they are integer numbers.
\\
Picture \ref{fig:rec5-16alldata} shows two typical records of VINCI data. It is possible to recognize the two photometric inputs, in pink and green, and the two interferometric outputs, in blue and red. Hereafter, we will refer at them as $PA$, $PB$ for the photometric signals and $I1$ and $I2$ for the interferometric ones.

\noindent Data immediately show a specific feature, i.e. the presence of a trend changing over time. This trend makes clear the correlation between photometric and interferometric channels, but it can cause features on the autocorrelation function and on the spectrum.

\section{Statistical analysis in the time domain}
\label{sec:time-stat}

Even if we could have some theoretical information about the behaviour of photometric signals in very controlled conditions\cite[ch. 4]{Goodman}, we still need techniques of statistical inference to properly characterize the time series we are dealing with. We have to face the problem of estimating unknown quantities, such as correlation functions, directly from the data. In the next paragraphs, we will adapt classic statistical instruments to the particular features of VLTI signals.

\subsection{Statistical methods description}
\label{subsec:timedom-method}

Since we do not have any {\it a priori} information on the relevant data features, such as mean values or variances, our work is based on the data analysis. The `unit' samples set is the record, i.e. $526$ points corresponding to a single OPD scan.
\\
\noindent The presence of a non negligible trend varying over time imposes some considerations on the sample mean definition. It is useless to consider a global mean over the record, because it can be very different from `local' mean.
Instead of subtracting a mean, we subtract the linear trend, evaluated over sub-intervals on the record. This approach permits to obtain a zero mean signal. The subtracted signal is an essential feature of the beams; being a time-variable trend, it can not be considered a seasonal trend, but a characteristic of the time series. The effect of the subtraction of these two different means (the global and the local one) on the correlation between beams are analyzed in the next paragraphs. Moreover, the evaluation of the mean over different subintervals changes the number of degrees of freedom in the estimators.\\
The detrend operation is done through the Matlab {\it detrend} operation (see help page at http://www.mathworks.com/support/functions/alpha\_list.html). A continuous, piecewise linear trend is subtracted, using set of user-defined breakpoints. The coefficients of the piecewise polynomial are computed with a least squares fit.
\\

\noindent With these considerations in mind, we estimate autocovariances and autocorrelations of single detrended signals and covariances and correlations between channels using as estimators the sample version of these functions, i.e. the autocorrelogram and the correlogram, respectively. These functions are estimated over each record of interest, and then averaged over all records.
\\

\noindent Following \cite[page 321]{Priestley}, we define the sample covariance $\hat{\gamma}_{12}(l)$ and the sample correlation $\hat{\rho}_{12}(l)$ between the signals $s_1(t)$ and $s_2(t)$, where $t$ is a discrete variable, as:
\begin{equation}\label{eq:covariance}
	\hat{\gamma}_{12}(|l|) = \frac{1}{N-|l|} \sum_{i = 1}^{N - |l|} (s_1(i) - \hat{s_1})(s_2(i+t)- \hat{s_2}),  \;\; 0 \leq |l| \leq (N-1)
\end{equation}

\begin{equation}\label{eq:correlation}
	\hat{\rho}_{12}(|l|) = \frac{\hat{\gamma}_{12}(|l|)}{ \sqrt{\hat{\sigma_1}^2\hat{\sigma_2}^2}},  \;\; 0 \leq |l| \leq (N-1)
\end{equation}

\noindent and the sample autocovariance $\hat{\gamma}(l)$ and autocorrelation $\hat{\rho}(l)$ of the signal $s(t)$ as:
\begin{equation}\label{eq:auto-covariance}
	\hat{\gamma}(|l|) = \frac{1}{N-|l|} \sum_{i = 1}^{N - |l|} (s(i) - \hat{s})(s(i+t)- \hat{s}),  \;\; 0 \leq |l| \leq (N-1)
\end{equation}

\begin{equation}\label{eq:auto-correlation}
	\hat{\rho}(|l|) = \frac{\hat{\gamma}(|l|)}{ \hat{\sigma}^2},  \;\; 0 \leq |l| \leq (N-1)
\end{equation}

\noindent where $\hat{s}$ is the sample mean and $\hat{\sigma}^2$ is the sample variance. If not differently specified, the chosen mean will be the linear trend.\\

\noindent For the properties of these estimators, we refer to the section \ref{sec:estimator_properties} of appendix \ref{appendixA}, and to \cite{Priestley} and references herein.
Here we recall that these functions are unbiased estimates of the true functions if the mean is the expectation, while if the mean is estimated from data, these estimators are asymptotically unbiased.
\\

\noindent To take into account the subtraction of different mean values over different subintervals, we propose to change the coefficient $\frac{1}{N-|l|}$ to $\frac{1}{N-|l|-k}$, where $k$ is the number of subintervals used to evaluate the changing mean, meaning that the degrees of freedom of this estimation are reduced by the multiple evaluation of the local mean. Unfortunately this correction changes the properties of the previous estimators, that becomes biased even if the mean coincides with the expectation. However, it is asymptotically unbiased. The proof of this result is given in appendix \ref{appendixA}, par. \ref{subsubsec:autocov_prop}.\\

\noindent Hence, we choose the following estimators, proposed by e.g. Parzen (see \cite{Priestley} for references):
\begin{eqnarray}\label{eq:covarianceN}
    \nonumber \hat{\gamma}(|l|) &=& \frac{1}{N} \sum_{i = 1}^{N - |l|} (s_1(i) - \hat{s_1})(s_2(i+t)- \hat{s_2}), \;\; 0 \leq |l| \leq (N-1) \\
    \hat{\gamma}_{12}(|l|) &=& \frac{1}{N} \sum_{i = 1}^{N - |l|} (s(i) - \hat{s})(s(i+t)- \hat{s}),  \;\; 0 \leq |l| \leq (N-1)
\end{eqnarray}

\noindent It is biased even if the mean is not estimated from the data, but in general has a smaller mean square error than the previous one. In particular, in our case we want to smooth the lobe effect of the $1/(N-|l|-k)$ factors for great values of $|l|$, caused by the average over a decreasing number of factors.
\\

\subsection{Void channels}
\label{subsec:void_ch}

\begin{figure*}[htb]
    \begin{center}
    \epsfig{figure=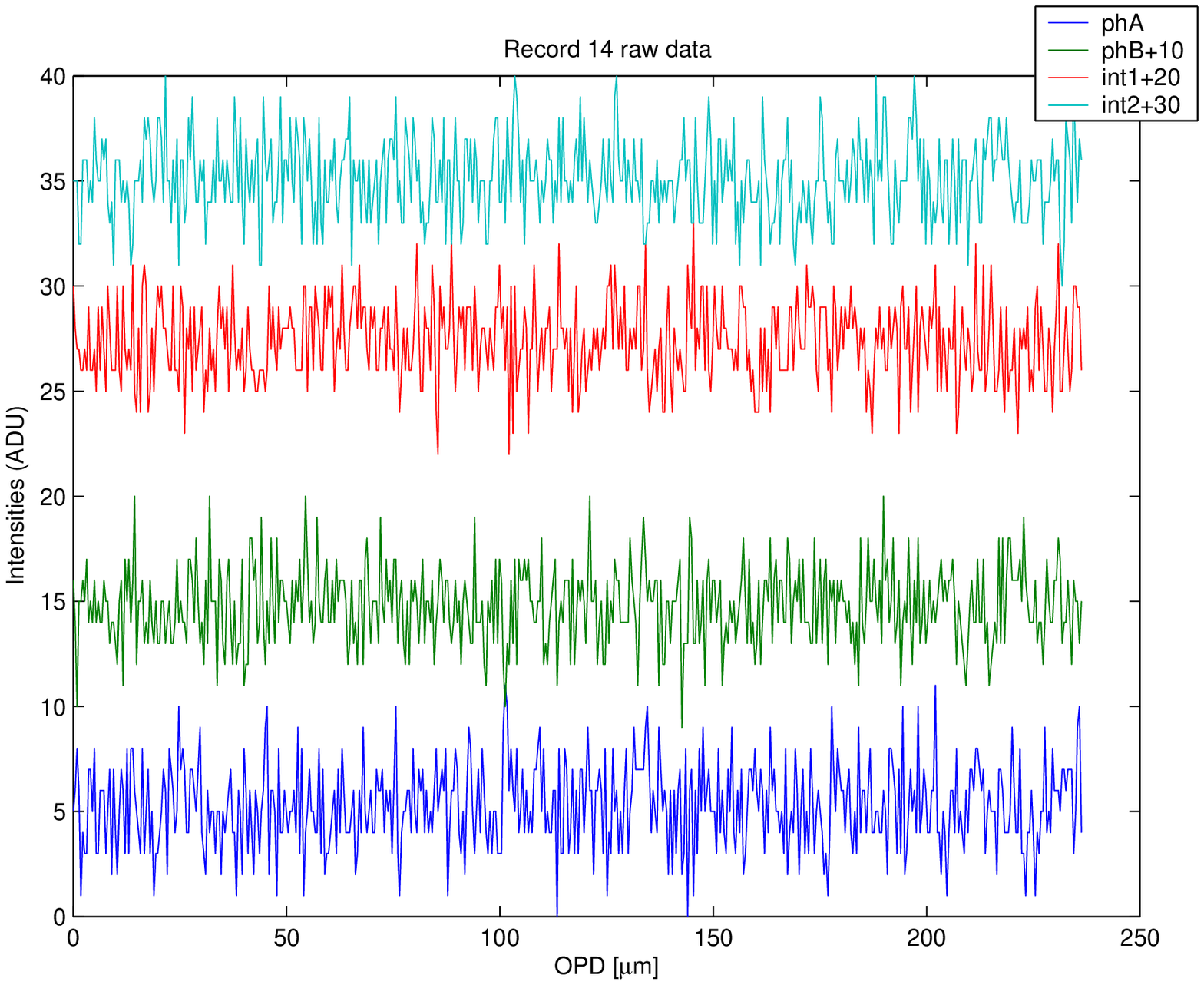,width=6.5cm}
    \epsfig{figure=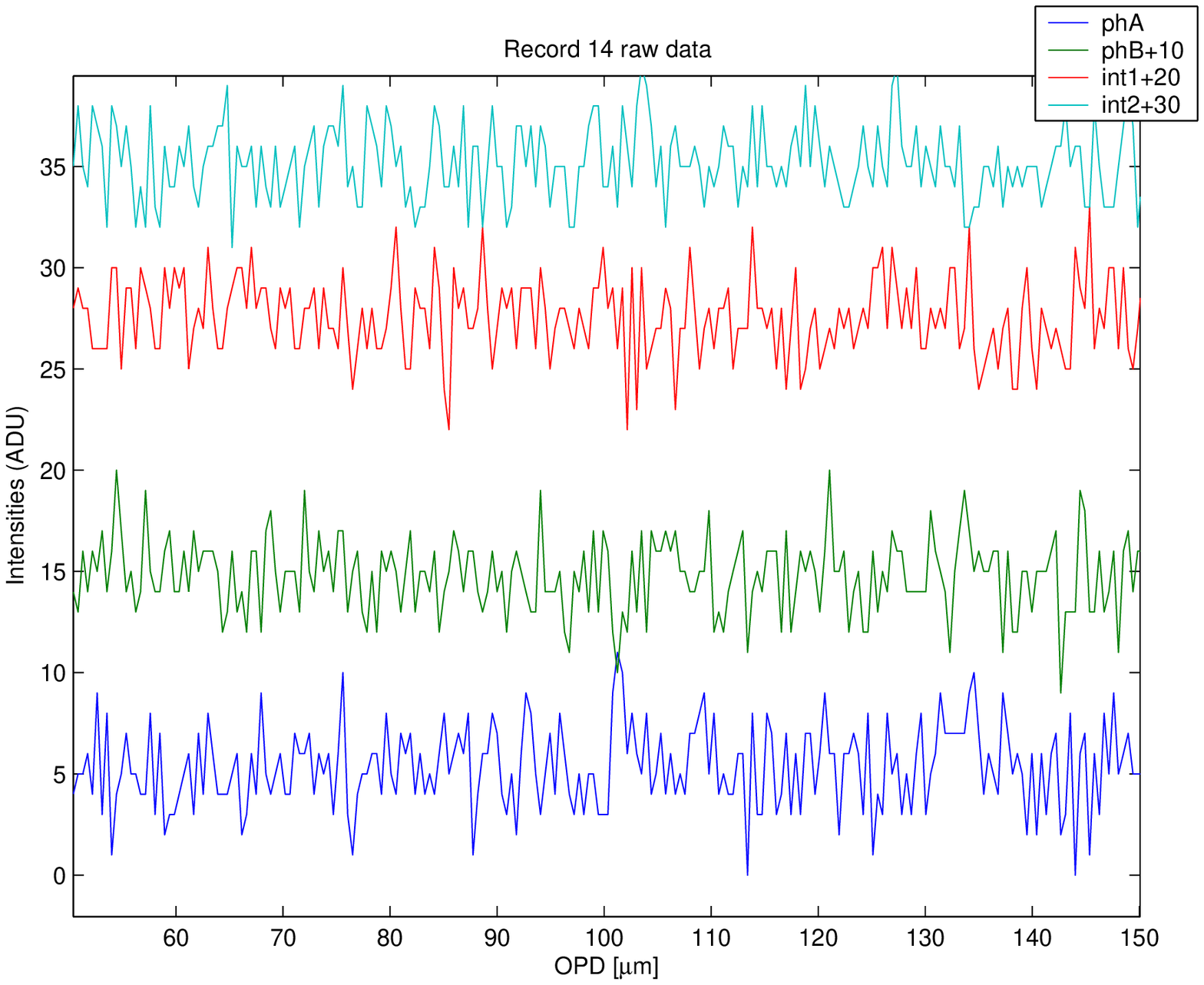,width=6.5cm}
    \caption{The four void channels representation with a zoom (right); the channels are vertically shifted for better understanding}
    \label{fig:void_noise}
    \end{center}
\end{figure*}

\noindent The analysis of void channels, i.e. channels in which the stellar beams are not injected, is useful to identify the environmental working condition. The presence of some noise can be expected, due to the laboratory thermal background and scattering. Figure \ref{fig:void_noise} shows a record of the four outputs with a zoom.

\begin{figure*}[htb]
    \begin{center}
    \epsfig{figure=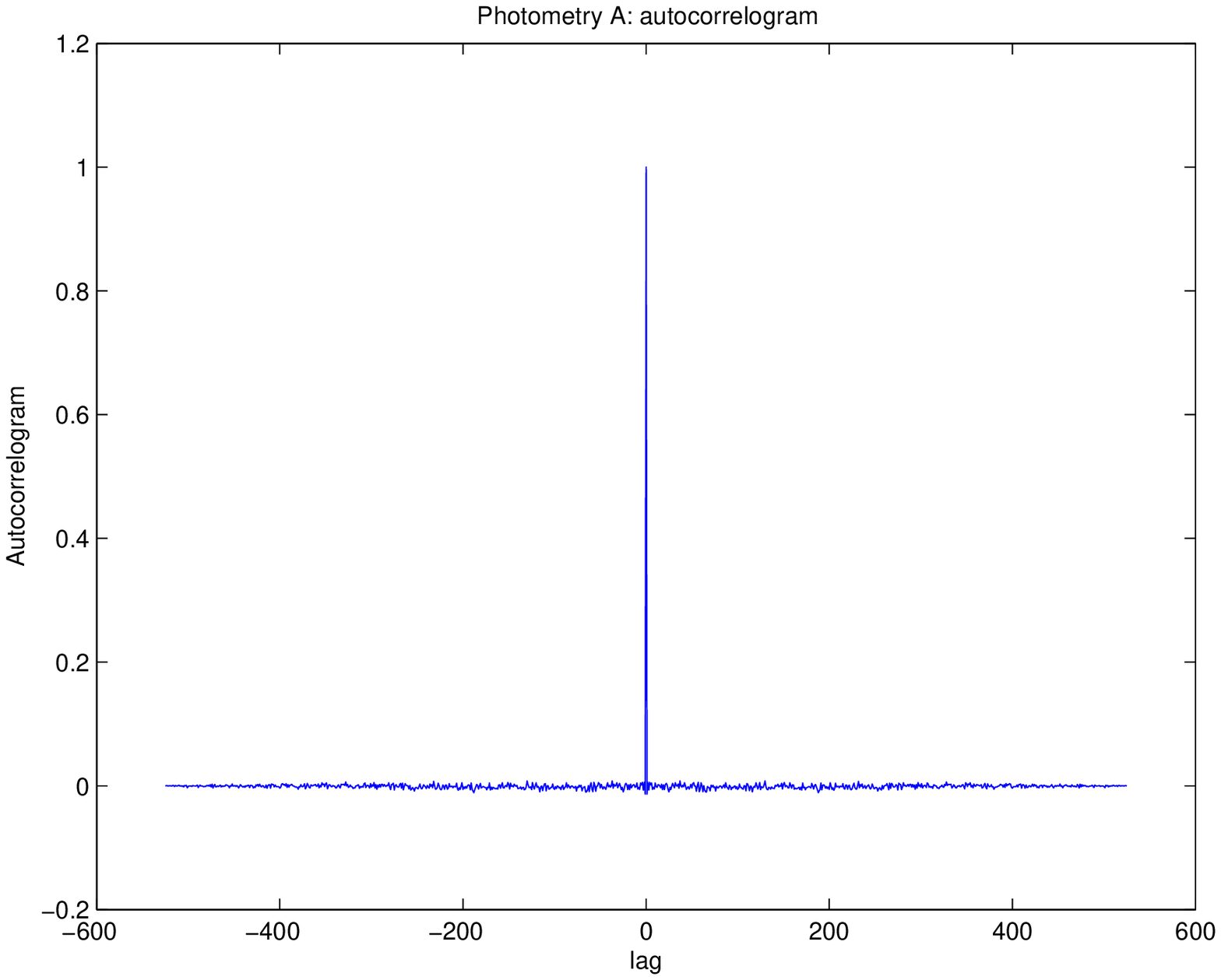,width=6.5cm}
    \epsfig{figure=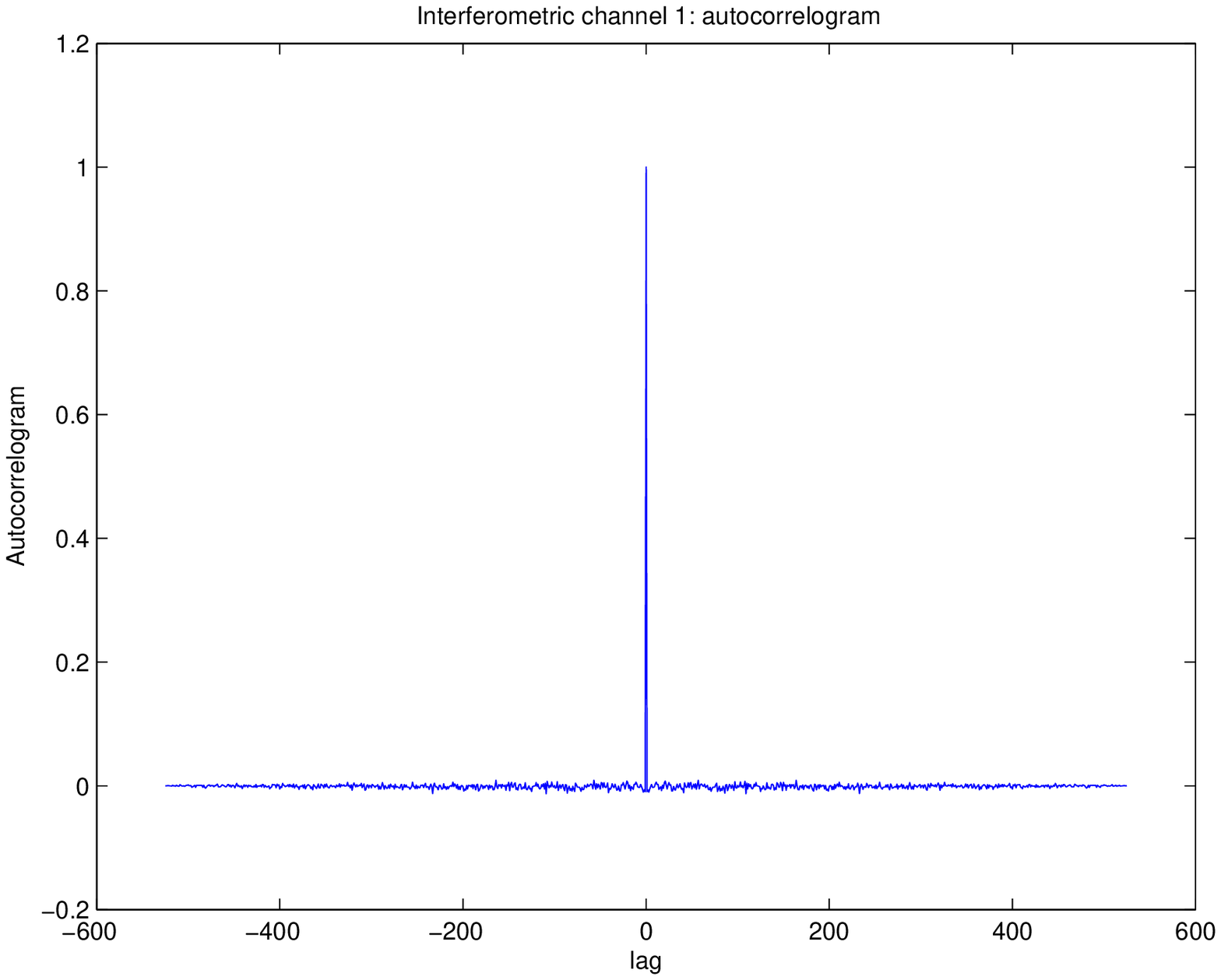,width=6.5cm}
    \caption{Case 1: Autocorrelation function for photometric $PA$ (left) and interferometric $I1$ (right) channels when no flux is injected in}
    \label{fig:void-autocorr}
    \end{center}
\end{figure*}

\begin{figure*}[htb]
    \begin{center}
    \epsfig{figure=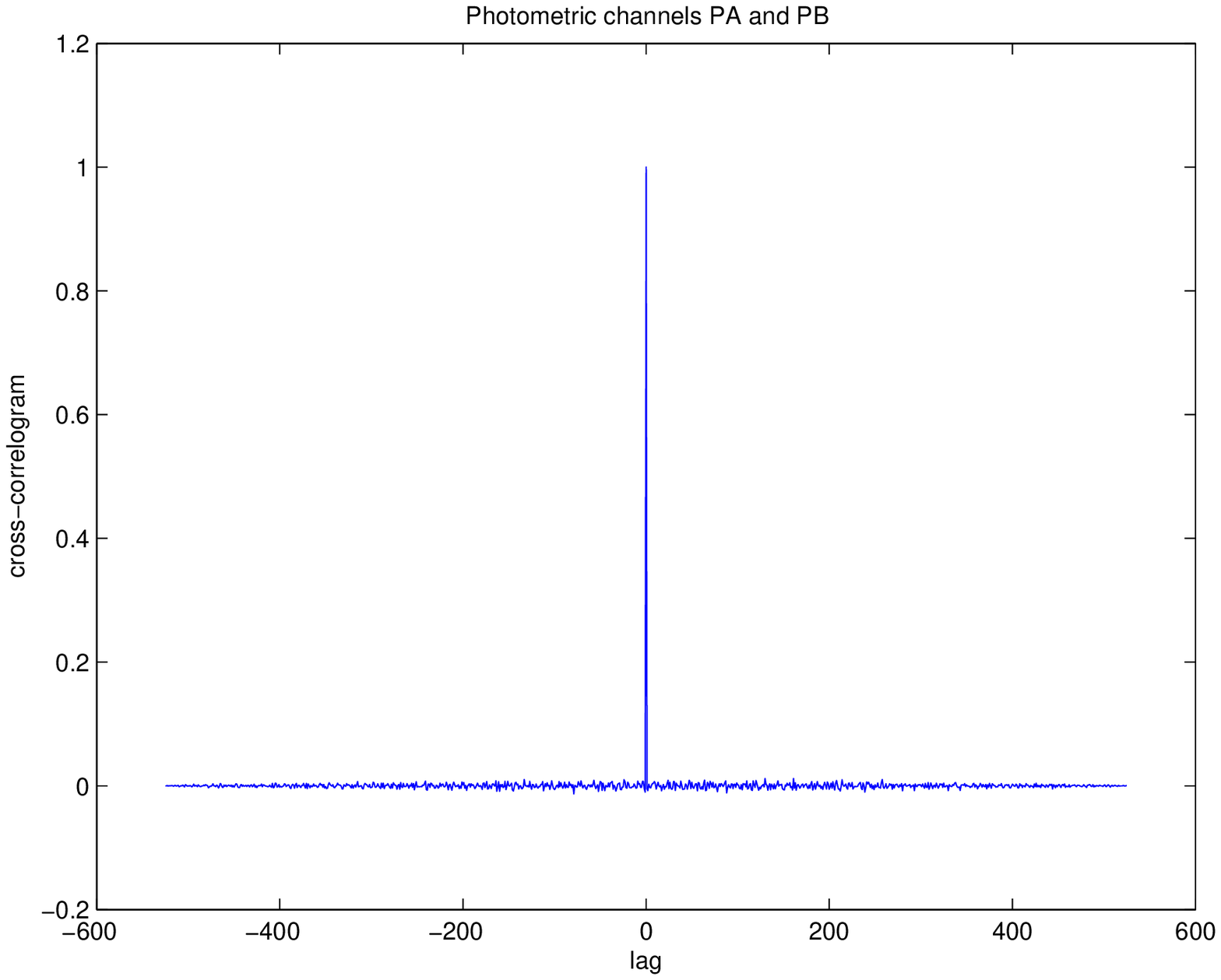,width=6.5cm}
    \epsfig{figure=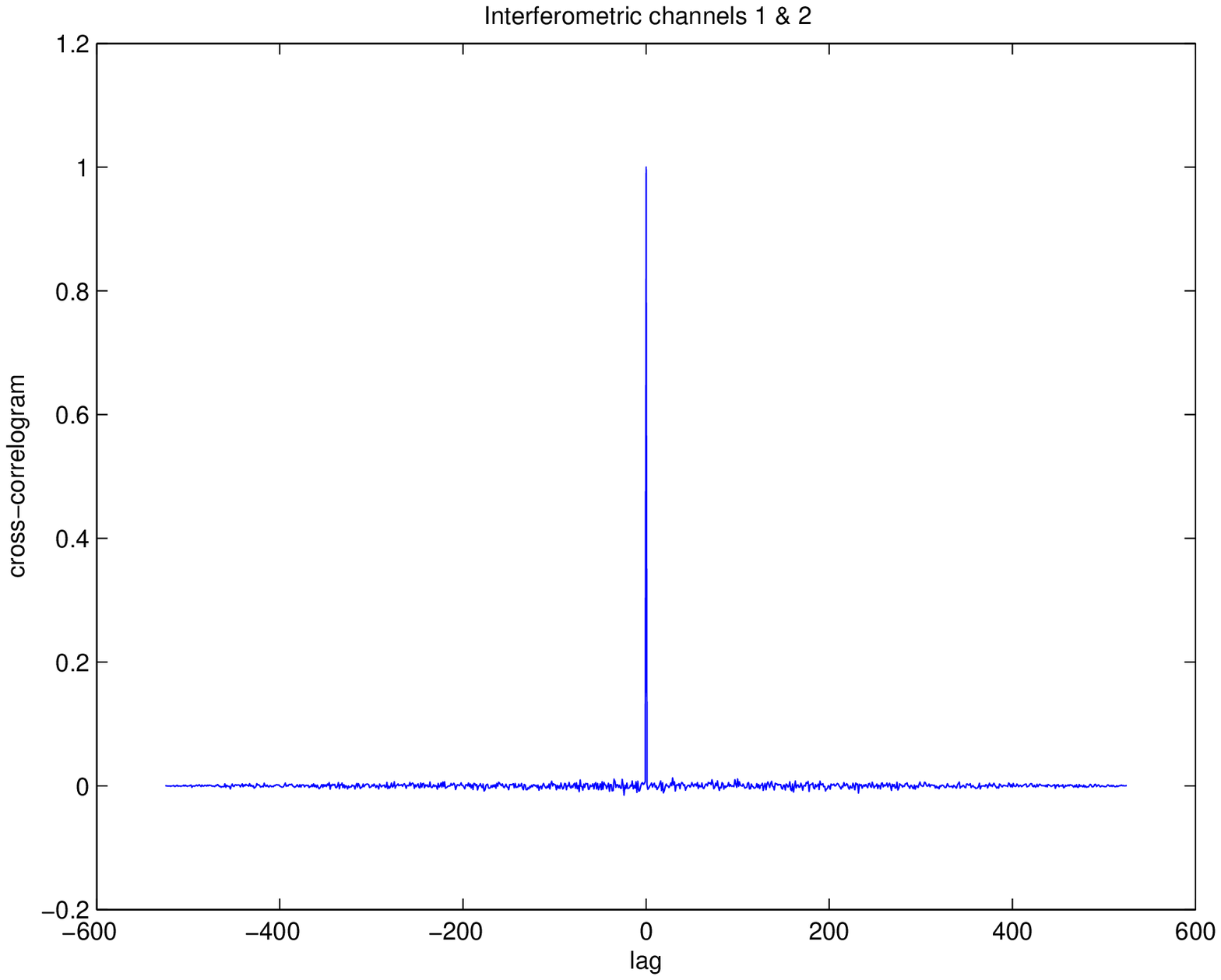,width=6.5cm}
    \epsfig{figure=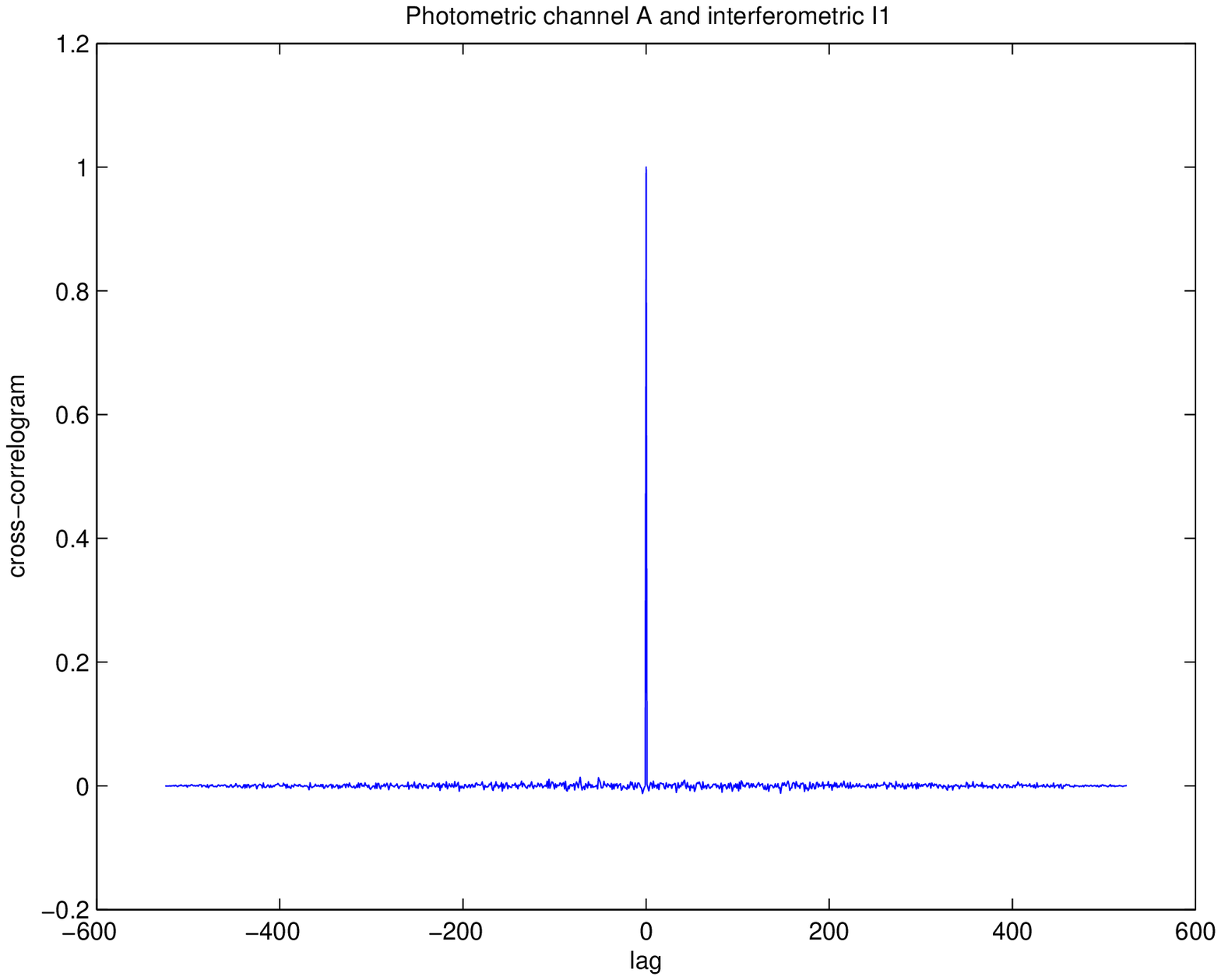,width=6.5cm}
    \epsfig{figure=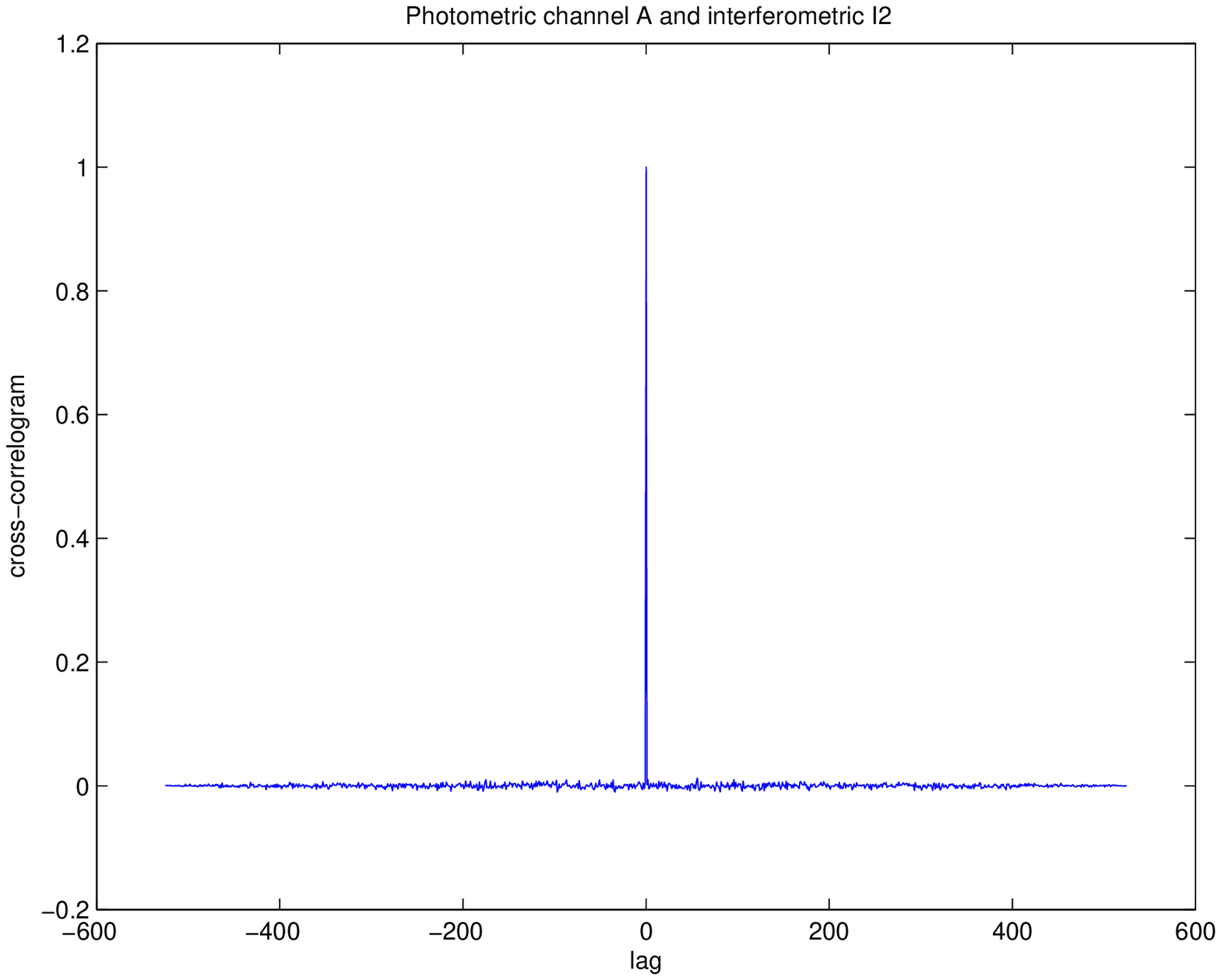,width=6.5cm}
    \caption{Case 1: First row: cross-correlogram functions for photometric $PA$ and $PB$ (left) and for interferometric $I1$ and $I2$ (right); second row: cross-correlogram between inputs and outputs - $PA$ and $I1$ (left) and $PA$ and $I2$ (right)}
    \label{fig:void-crosscorr}
    \end{center}
\end{figure*}

\noindent The analysis of the autocorrelograms reveals that noise channels are self-uncorrelated (see fig. \ref{fig:void-autocorr} where the $PA$ and the $I1$ channels are taken as representative), while the cross-correlograms of fig. \ref{fig:void-crosscorr} show that they are also cross-uncorrelated, as we could expect and hope.

\begin{figure*}[htb]
    \begin{center}
    \epsfig{figure=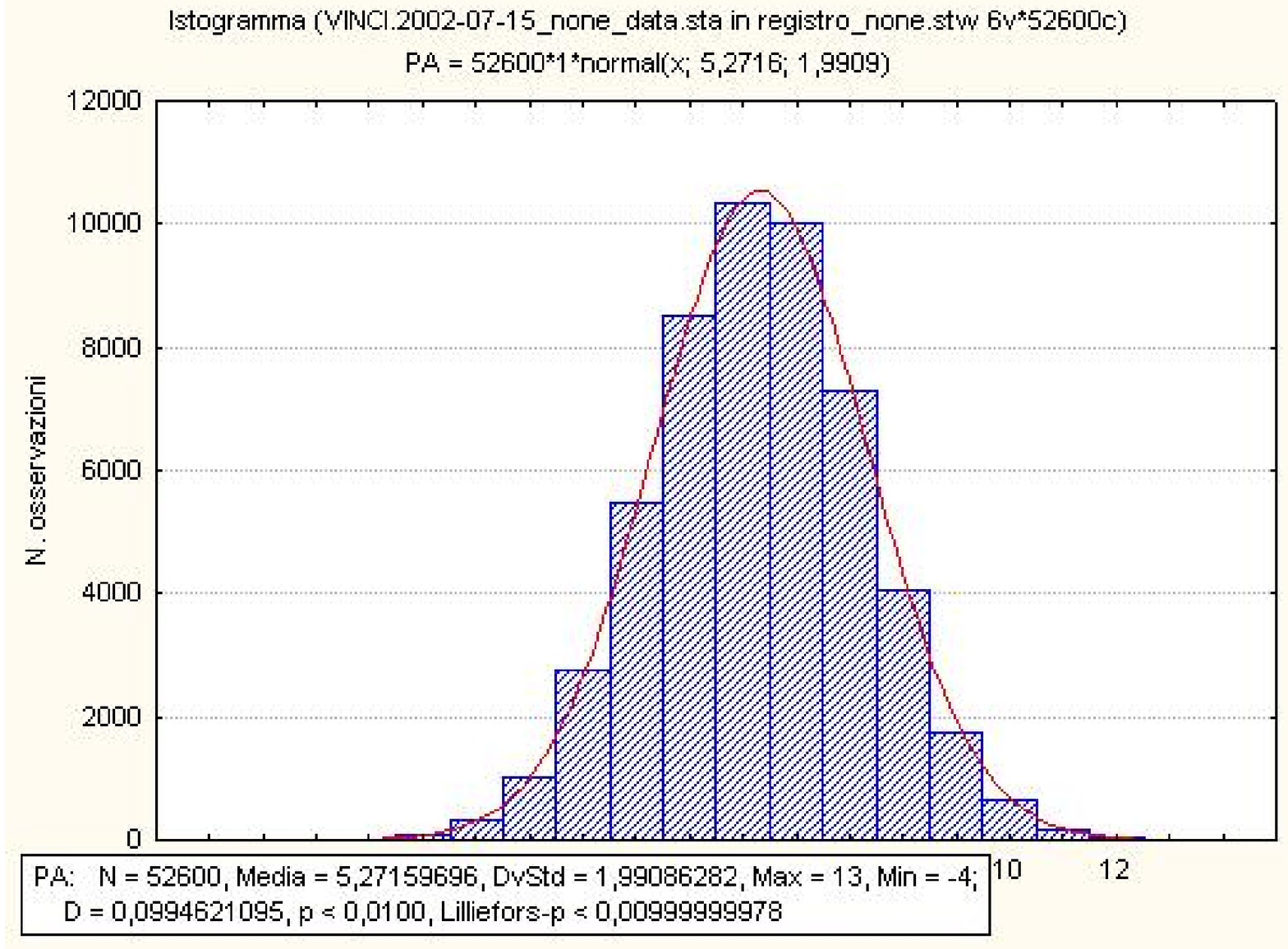,width=6.5cm}
    \epsfig{figure=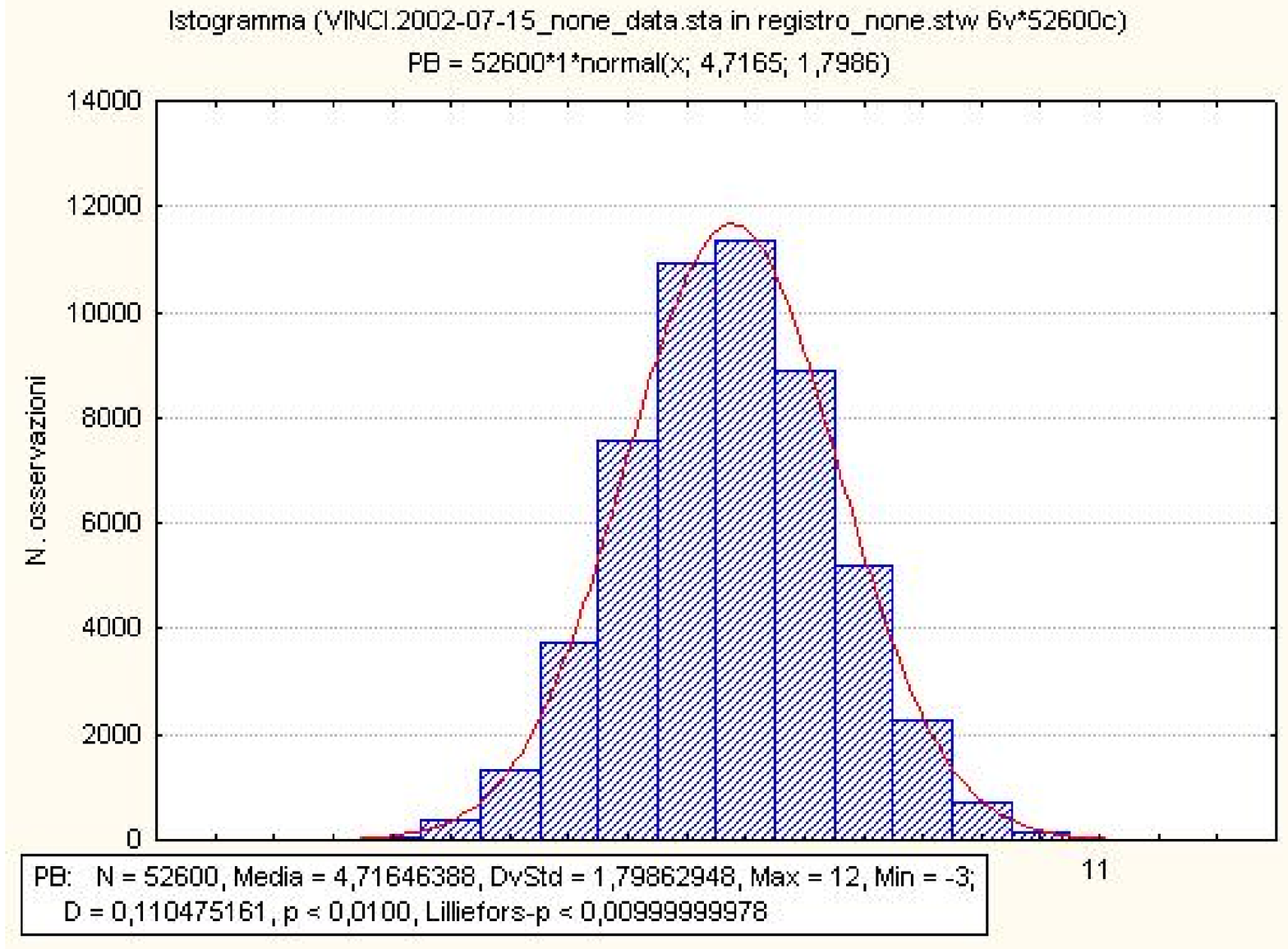,width=6.5cm}
    \epsfig{figure=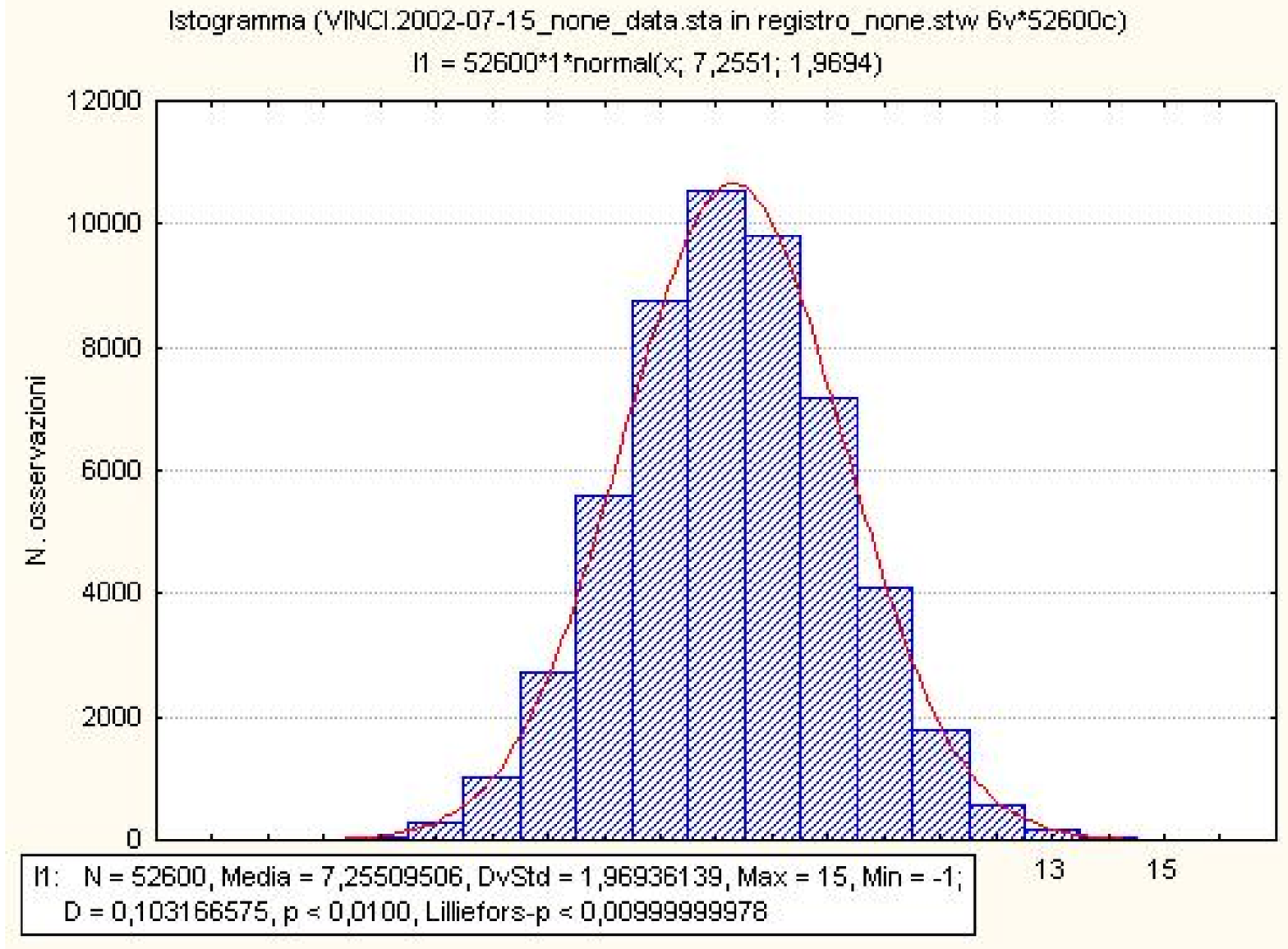,width=6.5cm}
    \epsfig{figure=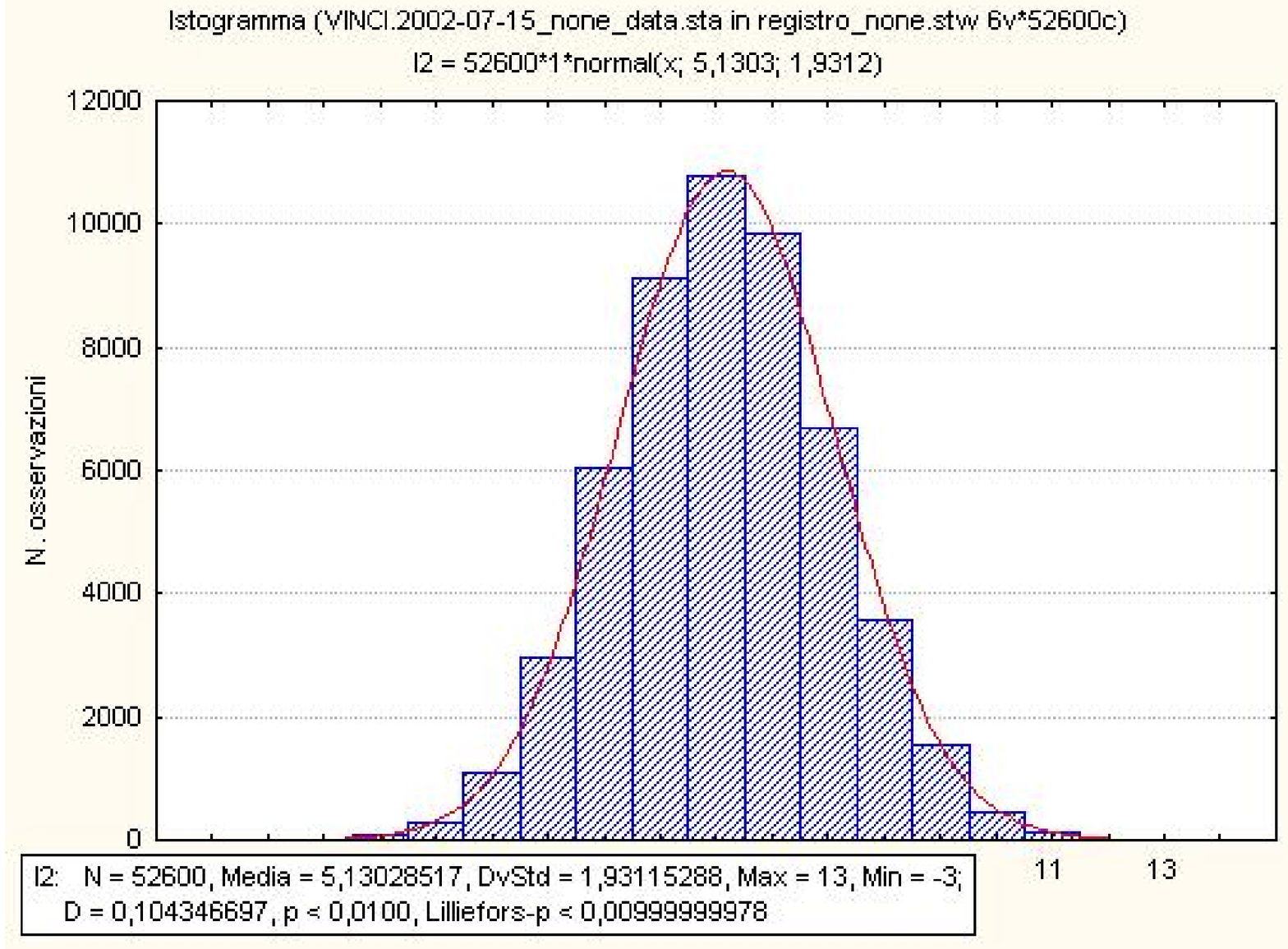,width=6.5cm}
    \caption{Case 1: Histograms for photometric $PA$ and $PB$ (first row) and for interferometric $I1$ and $I2$ (second row). The red line is the density of the normal distribution with parameters estimated from the data. See the Lilliefors p-level for test results.}
    \label{fig:void-hist}
    \end{center}
\end{figure*}

\noindent It is useful to verify the statistical properties of these signals, with our previous considerations in mind. First of all, we take a look at the histograms (fig. \ref{fig:void-hist}). The number of classes is limited by the finite range of possible values taken by data. The red line is the probability density function of a normal distribution with same mean and variance values as the data. The Lilliefors test for normality is evaluated; its p-value is $<0.01$, so there is statistical evidence of normality distribution of data.\\
We recall that we are working with integer values, hence the test could not be reliable, since it is based on continuous distribution. However the sample size set is huge ($52600$ samples - $100$ records of $526$ samples each). This allows a rescaling of the data that determines a continuous limit. The results of the test seems then to confirm the normality hypothesis.

\subsection{Input: photometric signals }
\label{subsec:photometry}

\begin{figure*}[htbp]
    \epsfig{figure=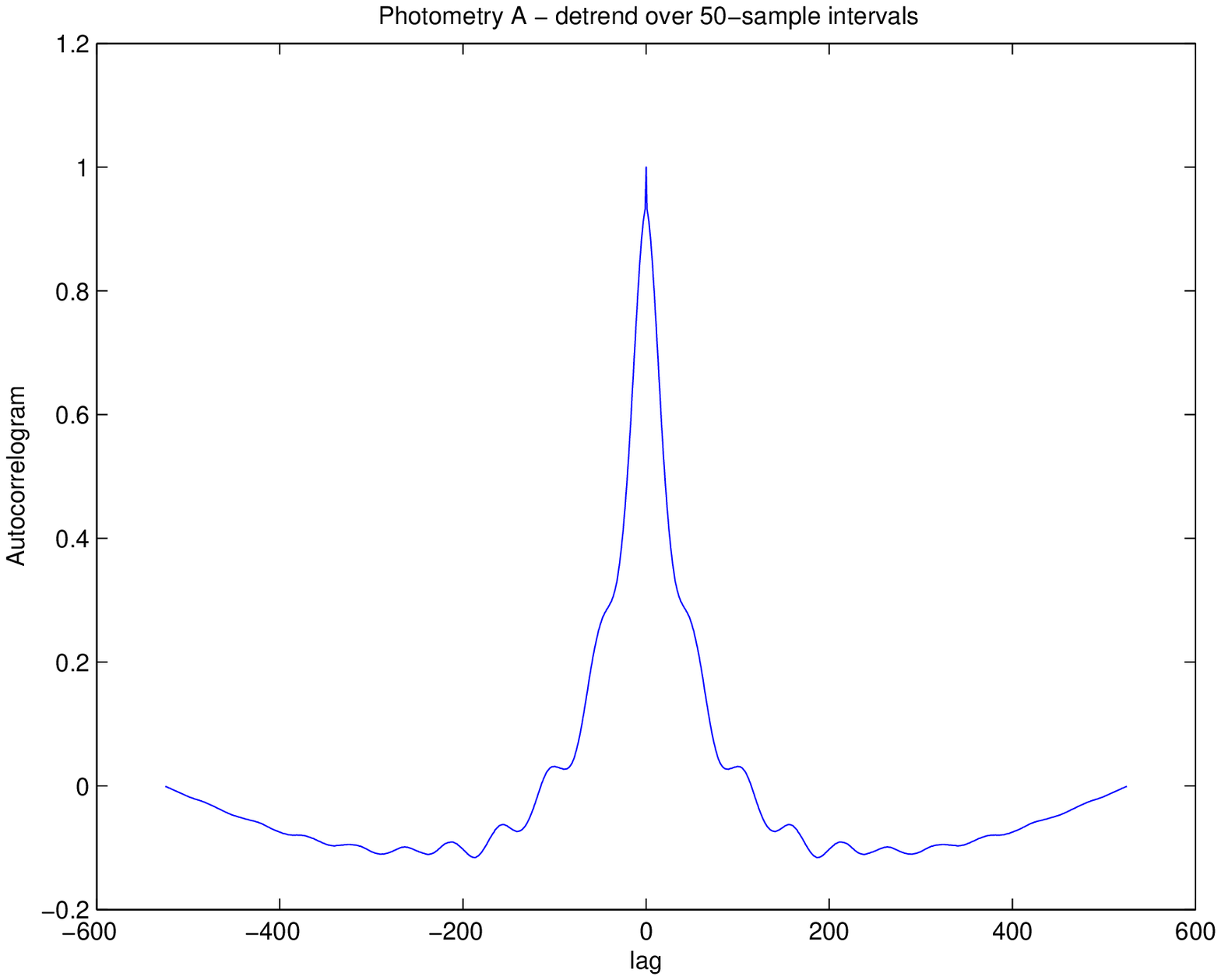,width=6.5cm}
    \epsfig{figure=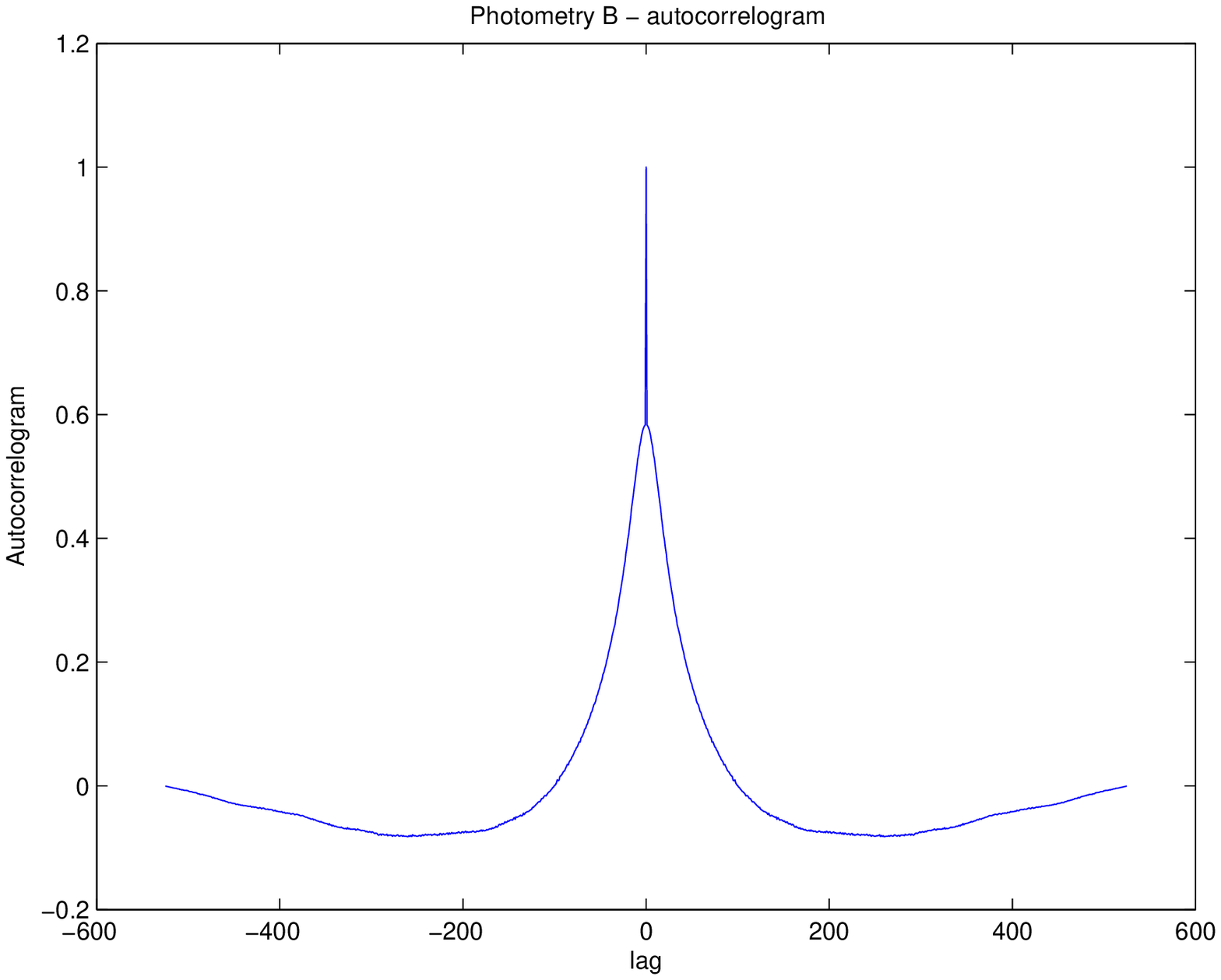,width=6.5cm}
    \begin{center}
        \caption{Case 4: Raw data, autocorrelation functions for photometric channels}
        \label{fig:phot-autocorr}
    \end{center}
\end{figure*}

\noindent We now investigate the sample autocorrelation and sample cross-correlation of photometric channels, following eq. \ref{eq:covarianceN}, when flux from stellar source is injected in both arms of the combiner. We recall that photometric flux corresponds to half the intensity of the input beam fed to the instrument. The mean evaluation problem now arises. We first evaluate the mean, used in the estimation of the autocorrelogram function of the photometry channels, as a global value over each record. Then, all the autocorrelation functions are averaged over the records.  In fig. \ref{fig:phot-autocorr}, the results are shown.

\begin{figure*}[htbp]
    \epsfig{figure=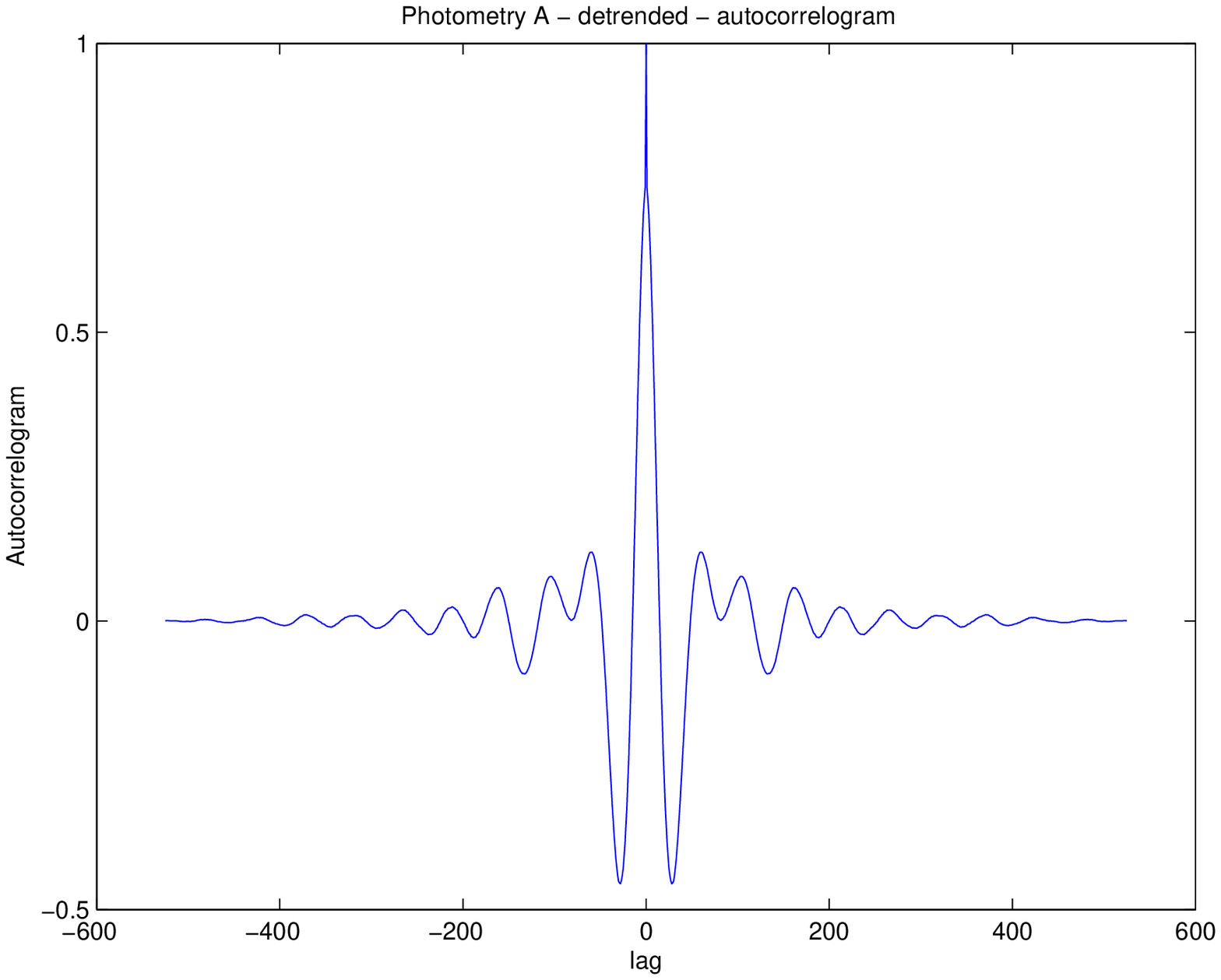,width=6.5cm}
    \epsfig{figure=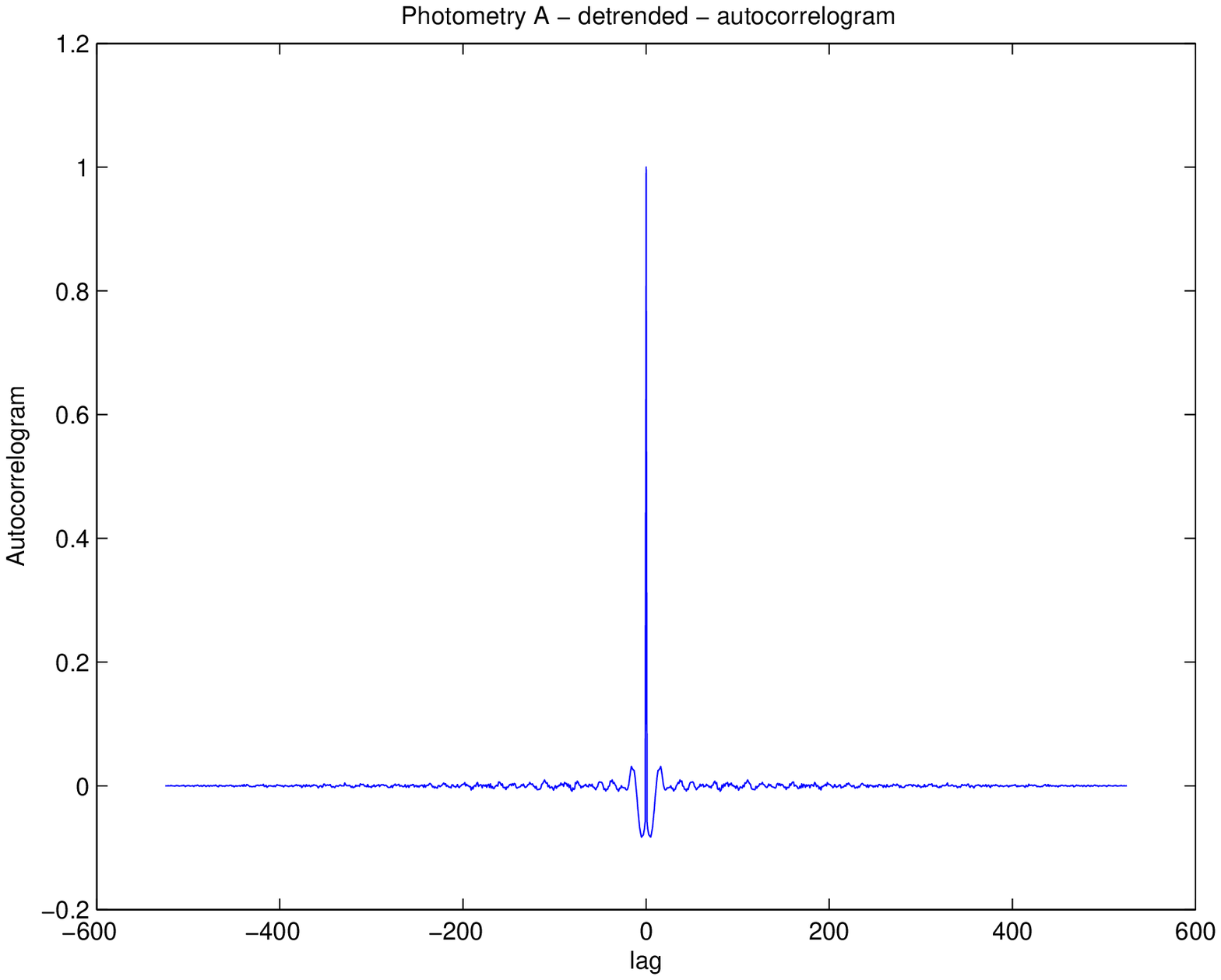,width=6.5cm}
    \begin{center}
        \caption{Case 4: raw data, autocorrelation functions for photometric channel A. Left, the linear trend to subtract is evaluated as a piecewise polynomial with breakpoints every $50$ samples; right, breakpoints are every $10$ samples.}
        \label{fig:photA-autocorr-detr}
    \end{center}
\end{figure*}

\noindent The autocorrelograms tend to zero very slowly. This is a consequence of the presence of the linear trend, that is not affected by the offset elimination. A linear trend is a strong correlation between consequent time samples.
\\
\noindent We then perform a detrend operation, using the {\it detrend} function of Matlab described in par. \ref{subsec:timedom-method}. We compute again the autocorrelogram functions, and we compare two different situations: for the former, the detrend operation is performed over fifty-sample subintervals, for the latter, over ten-samples subintervals. The results are shown in figure \ref{fig:photA-autocorr-detr} for the photometric channel A; the photometric channel B is similar, and can be found in appendix \ref{appendixB}.

\noindent The correlation between distant lags ($l > |20|$) drops to zero, but there is a residual correlation for smaller lags that cannot be explained by the trend. 
\\


\noindent We now investigate the possible cross-correlation between the photometric channels. We know that they come from a common stellar source, that their paths from the collection at the telescope till the detection in the laboratory are similar, but they can be subject to different sources of noise with different amplitudes.

\begin{figure*}[htbp]
    \epsfig{figure=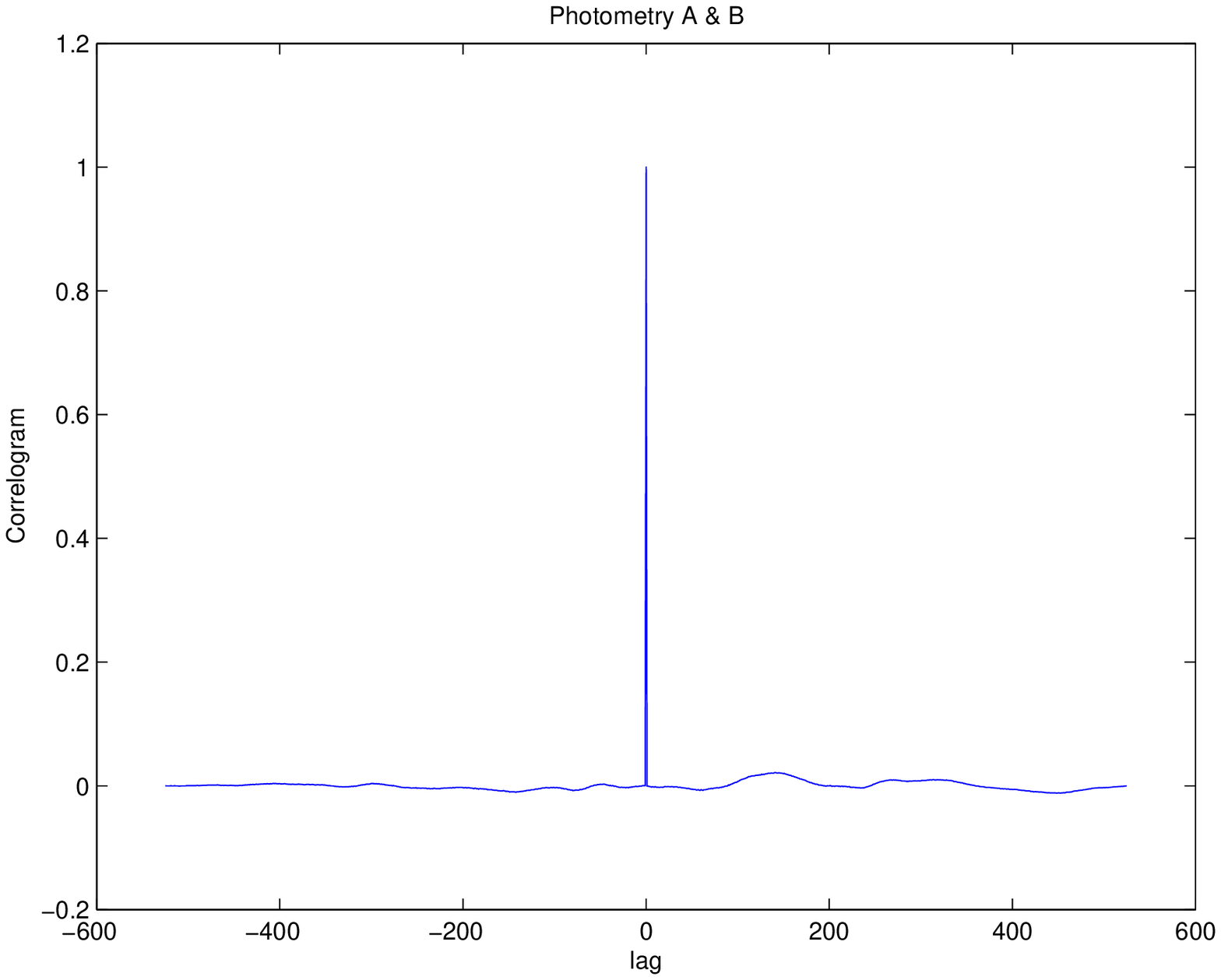,width=6.5cm}
    \epsfig{figure=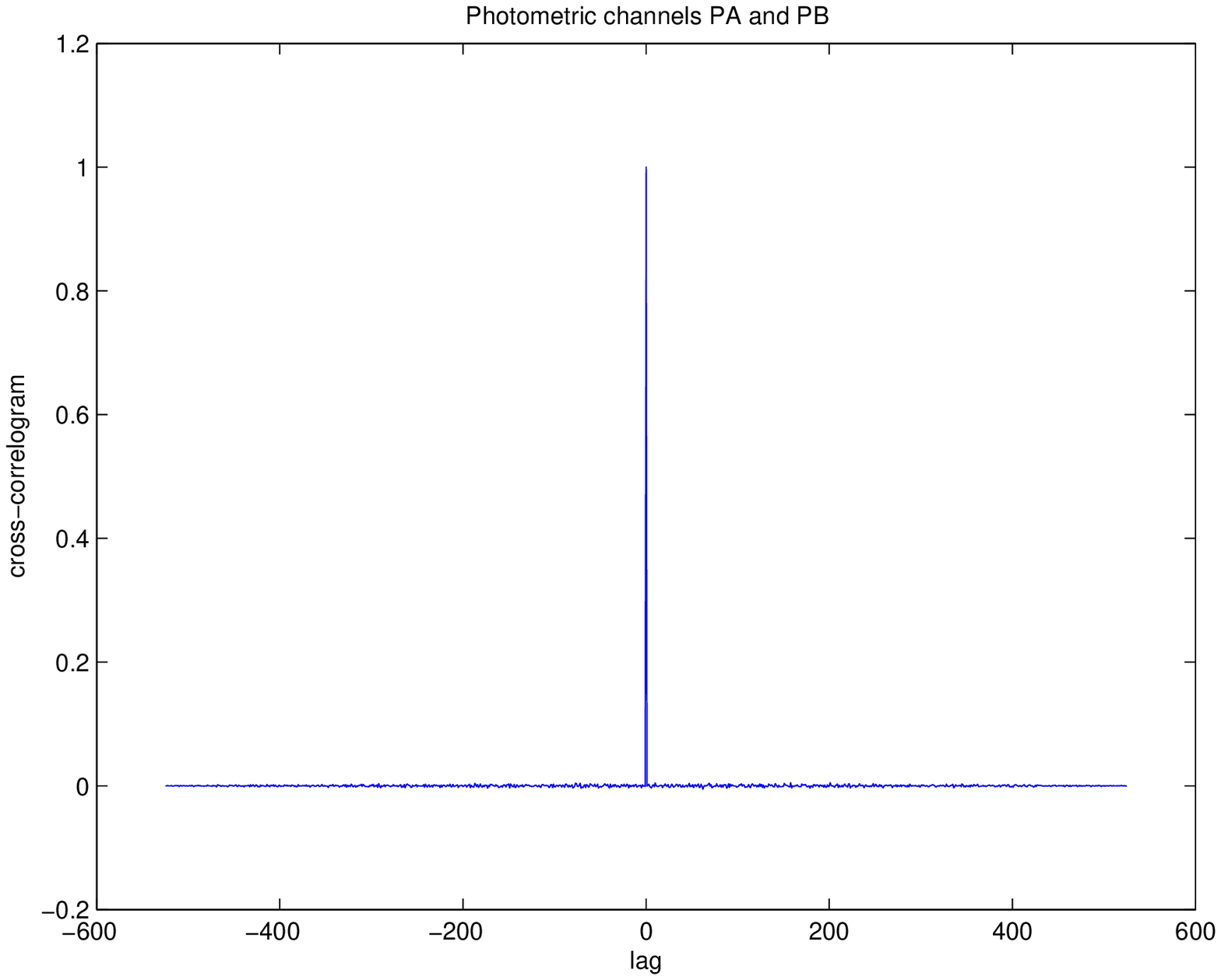,width=6.5cm}
    \begin{center}
        \caption{Case 4: Cross-correlation functions for photometric channel A and B. Left, raw data; right, linear trend subtracted.}
        \label{fig:photAandB-crosscorr}
    \end{center}
\end{figure*}

\noindent Again, we find a difference if signals are detrended or not, as fig. \ref{fig:photAandB-crosscorr} shows.

\noindent However, the differences are smaller than for the auto-correlations.\\

\begin{figure*}[htbp]
    \epsfig{figure=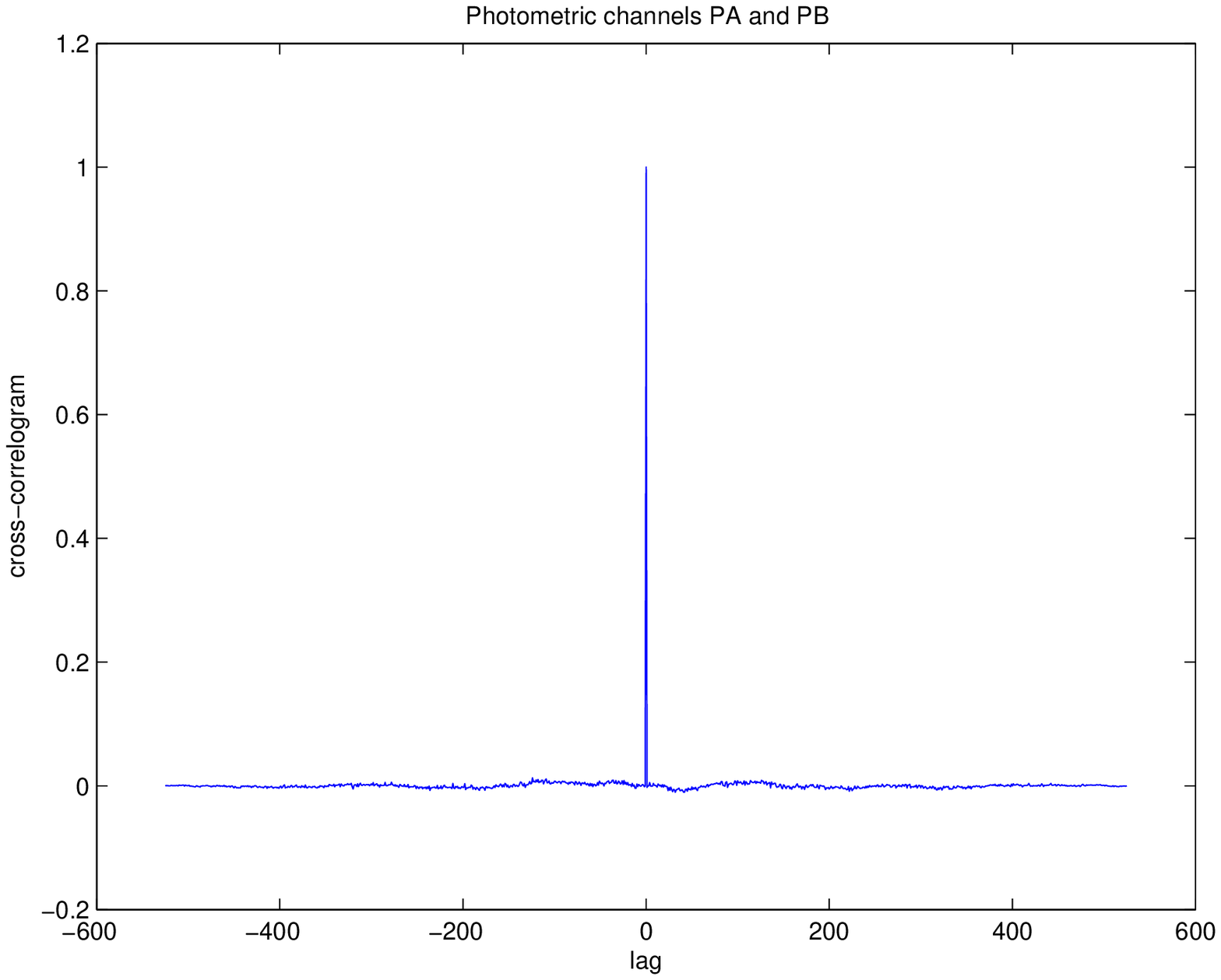,width=6.5cm}
    \epsfig{figure=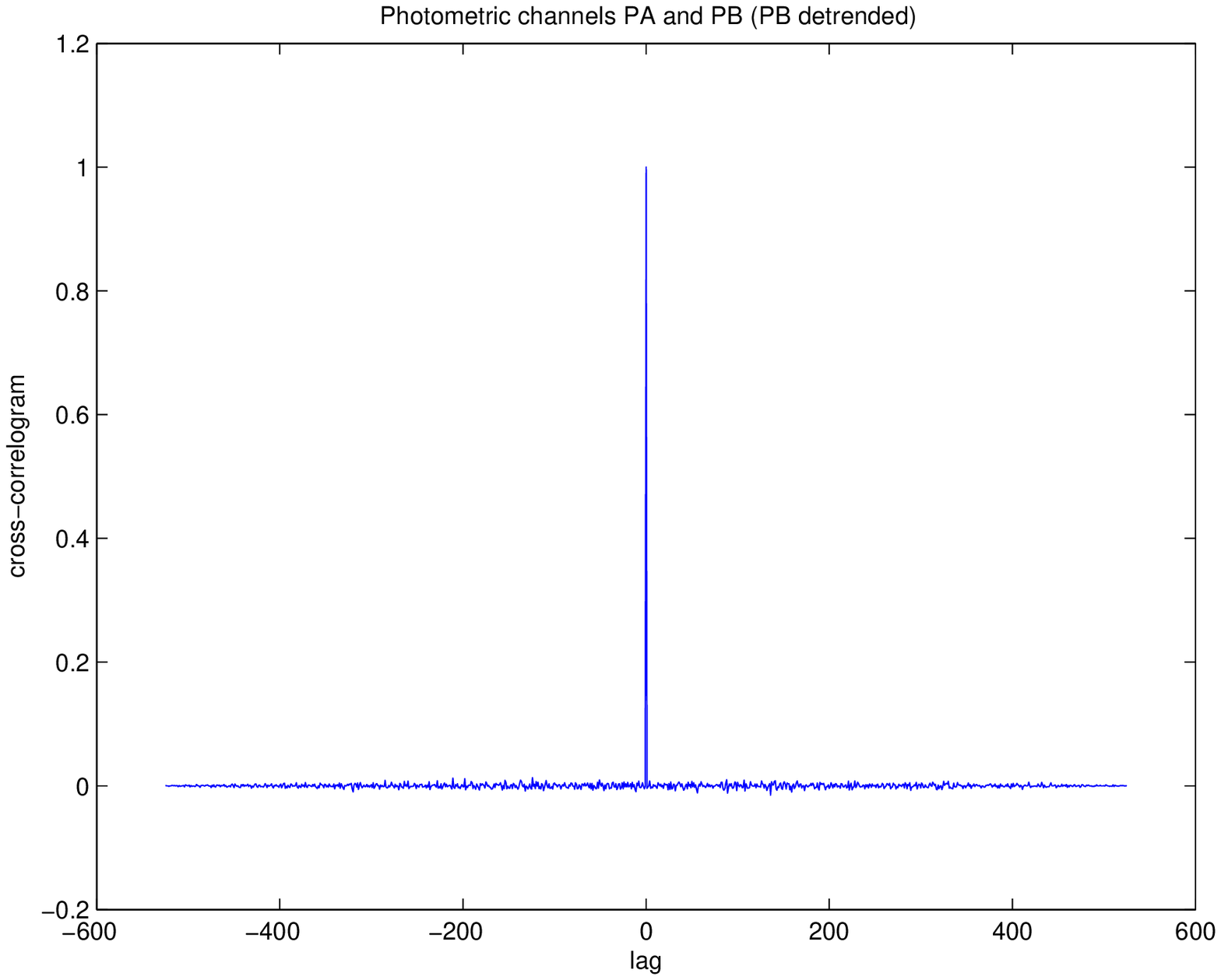,width=6.5cm}
    \begin{center}
        \caption{Case 3: Cross-correlation functions for photometric channel A and B. Left, raw data; right, linear trend subtracted.}
        \label{fig:chB_phA_B-crosscorr}
    \end{center}
\end{figure*}

\noindent We investigate on the cross-correlation estimation for the input channels in case 2 and 3. We remember that in these cases one channel is fed, while the other is void. We have seen in the previous paragraph that void channels contains self-uncorrelated `noise', while in the fed channel the signal is self-correlated. We check if there is a correlation between these different channels. Fig. \ref{fig:chB_phA_B-crosscorr} shows the results for case 3 (channel $PA$ void, channel $PB$ with flux), with raw data (left) and after a detrend of $PB$ (right). We can see that in both cases the cross-correlation drops immediately to zero, as we can expect from the features of the `pure noise' of $PA$. Case 2 is similar, and it is reported in appendix.




\subsection{Output in calibration mode}
\label{subsec:out_calibr}

\begin{figure*}[ht]
    \begin{center}
    \epsfig{figure=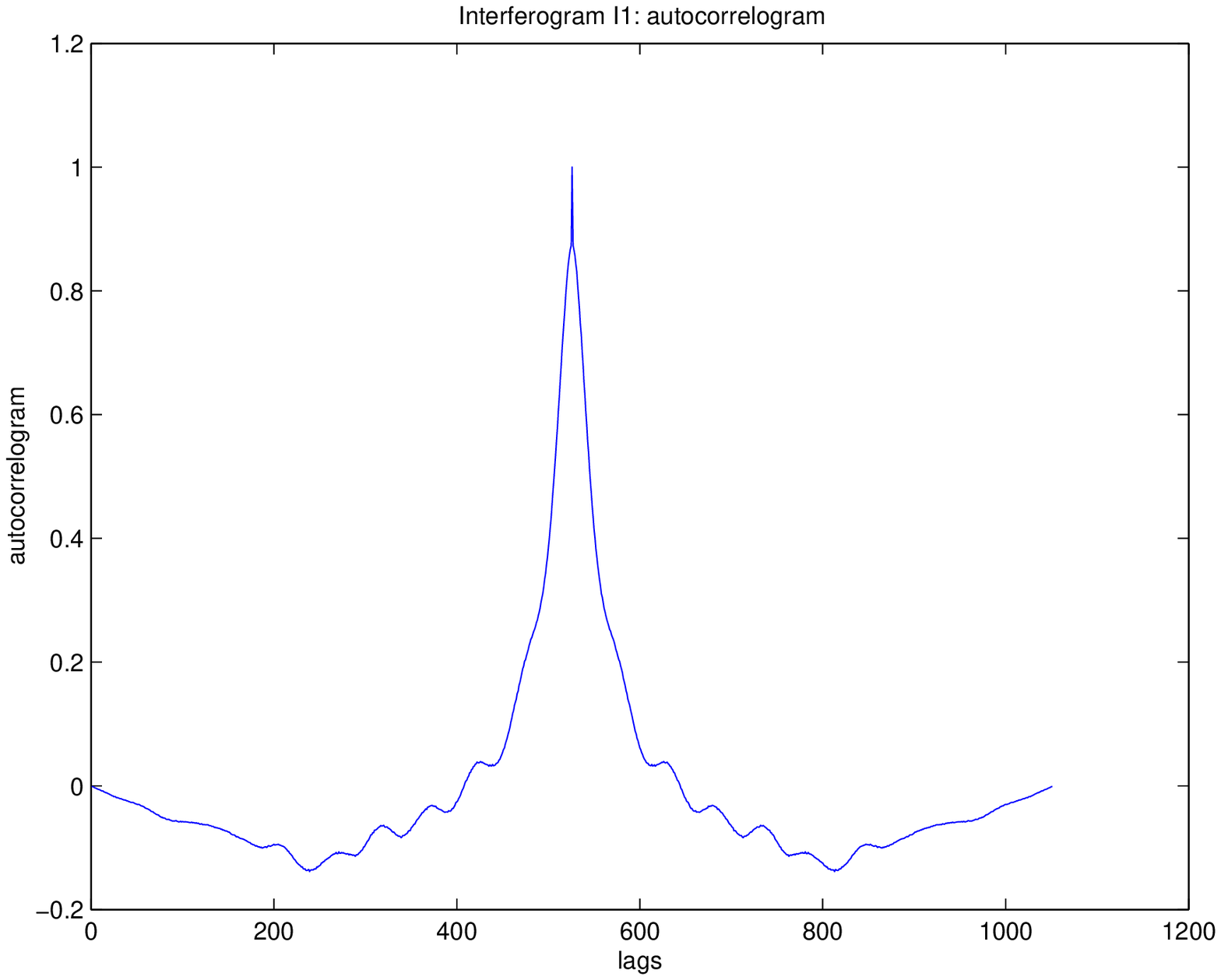,width=6.5cm}
    \epsfig{figure=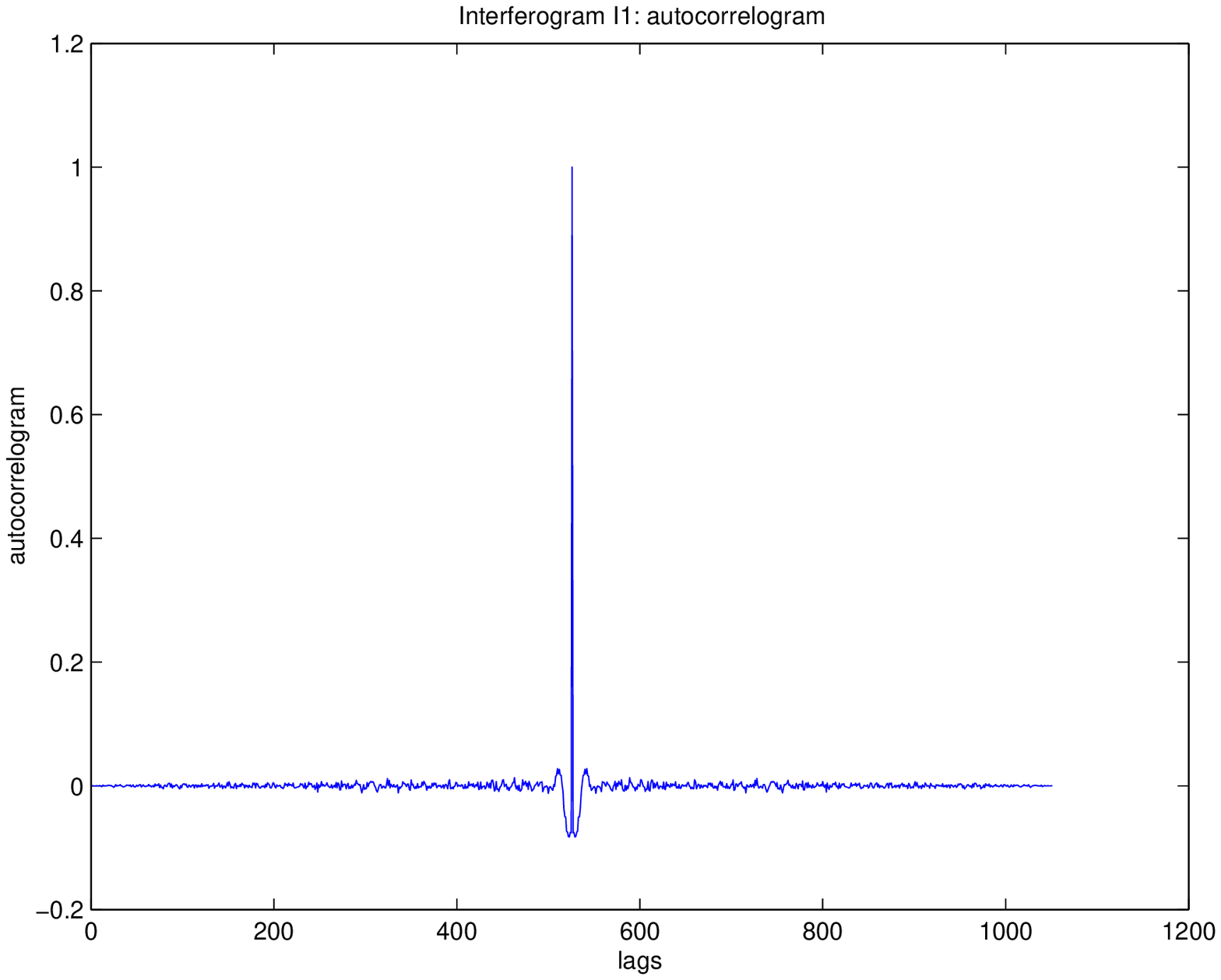,width=6.5cm}
    \epsfig{figure=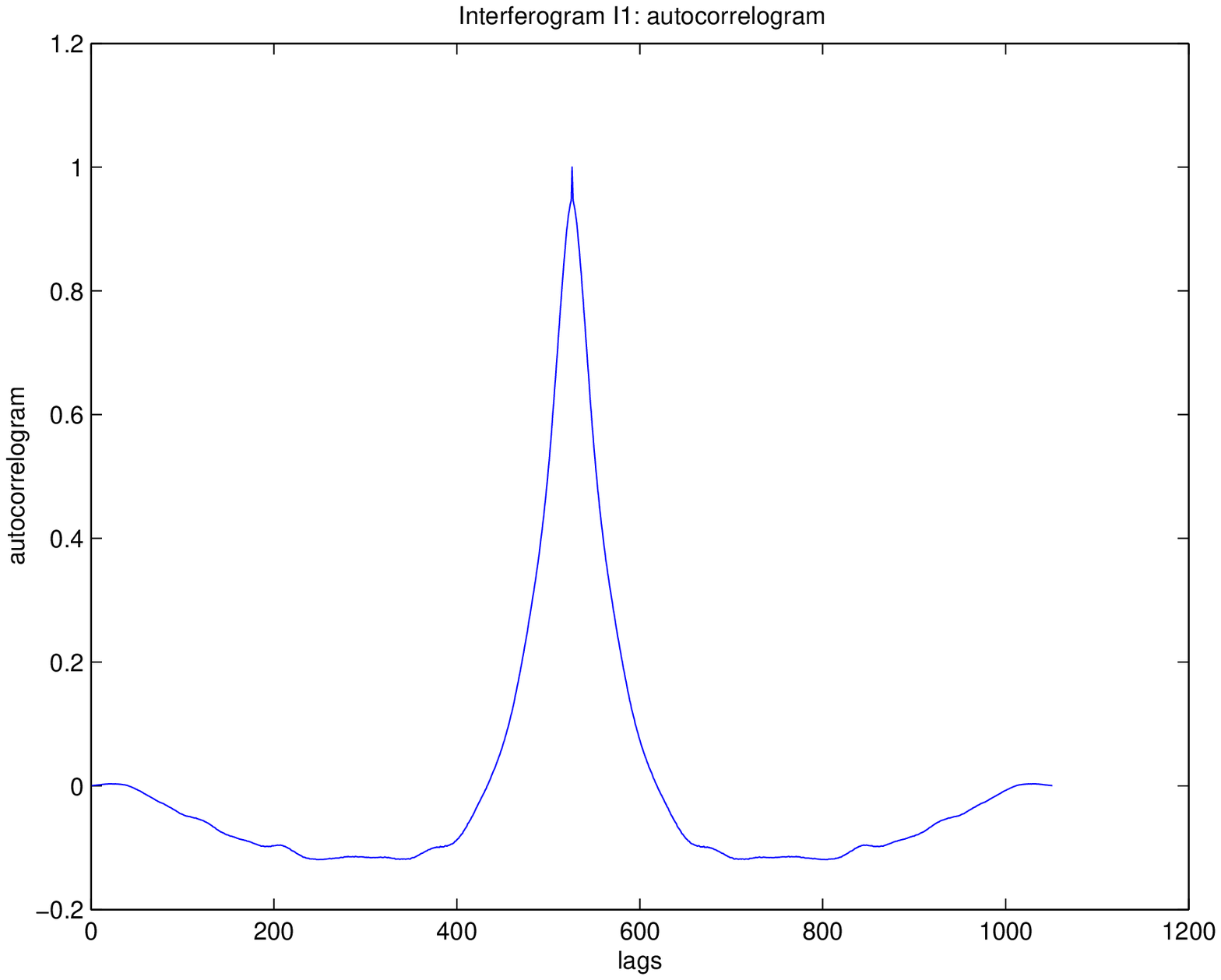,width=6.5cm}
    \epsfig{figure=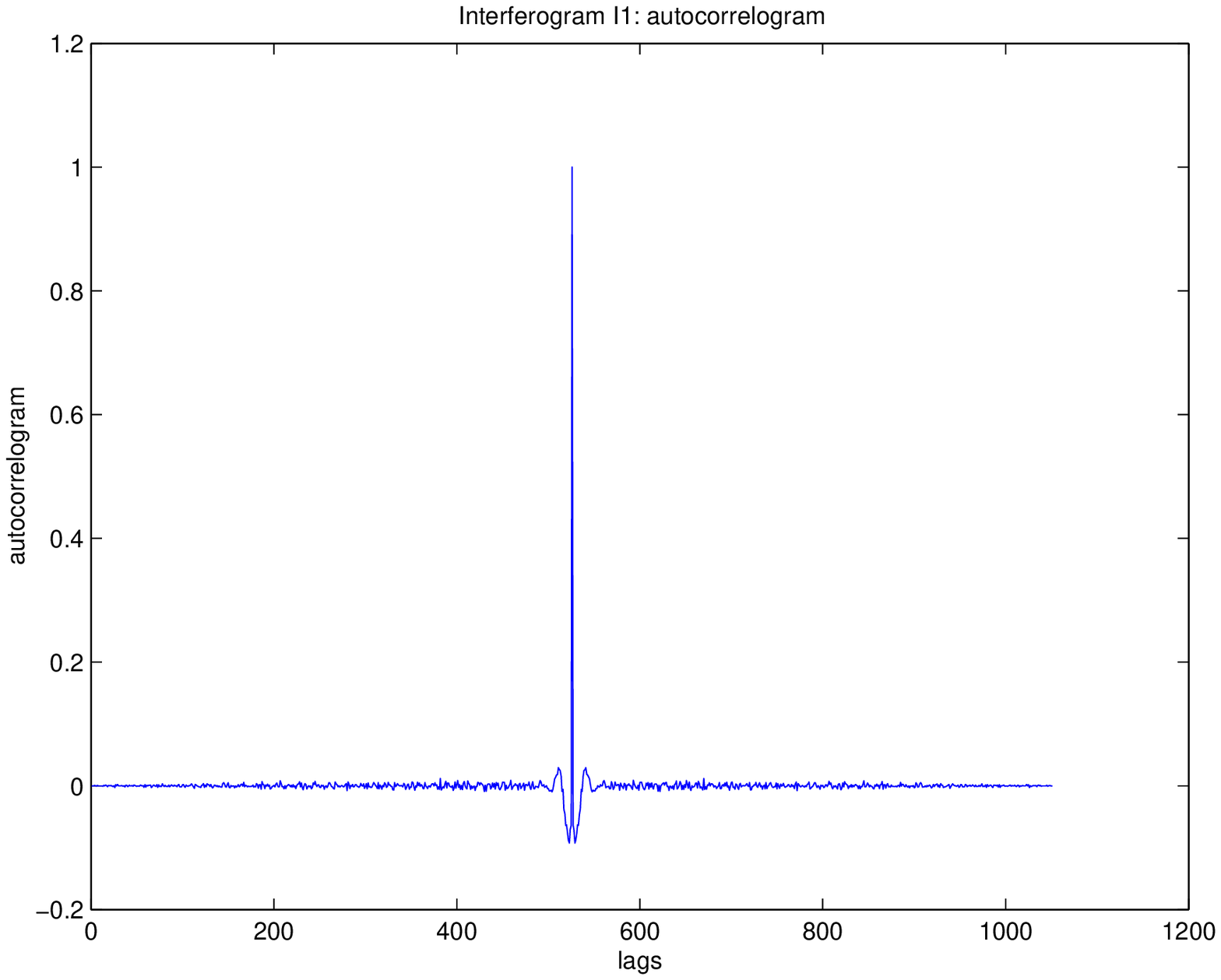,width=6.5cm}
    \caption{Case 2-3: Autocorrelation functions for interferometric channel $I1$ for case 2 (first row) and case 3 (second row): raw data (left) and detrended (right).}
    \label{fig:calOutput-autocorr}
    \end{center}
\end{figure*}

In calibration mode, the output channels do not contain fringes. We analyze their performances, however, to characterize their behaviour. In figure \ref{fig:calOutput-autocorr} the autocorrelation estimates are shown for channel $I1$ for case 2 (first row) and case 3 (second row). There are no big difference from the photometric input case, i.e. raw outputs have a slowly decreasing autocorrelation function, while the detrended signals shows no correlation, apart from lags near zero. So the trend induces self-correlations on signals.

\begin{figure*}[ht]
    \begin{center}
    \epsfig{figure=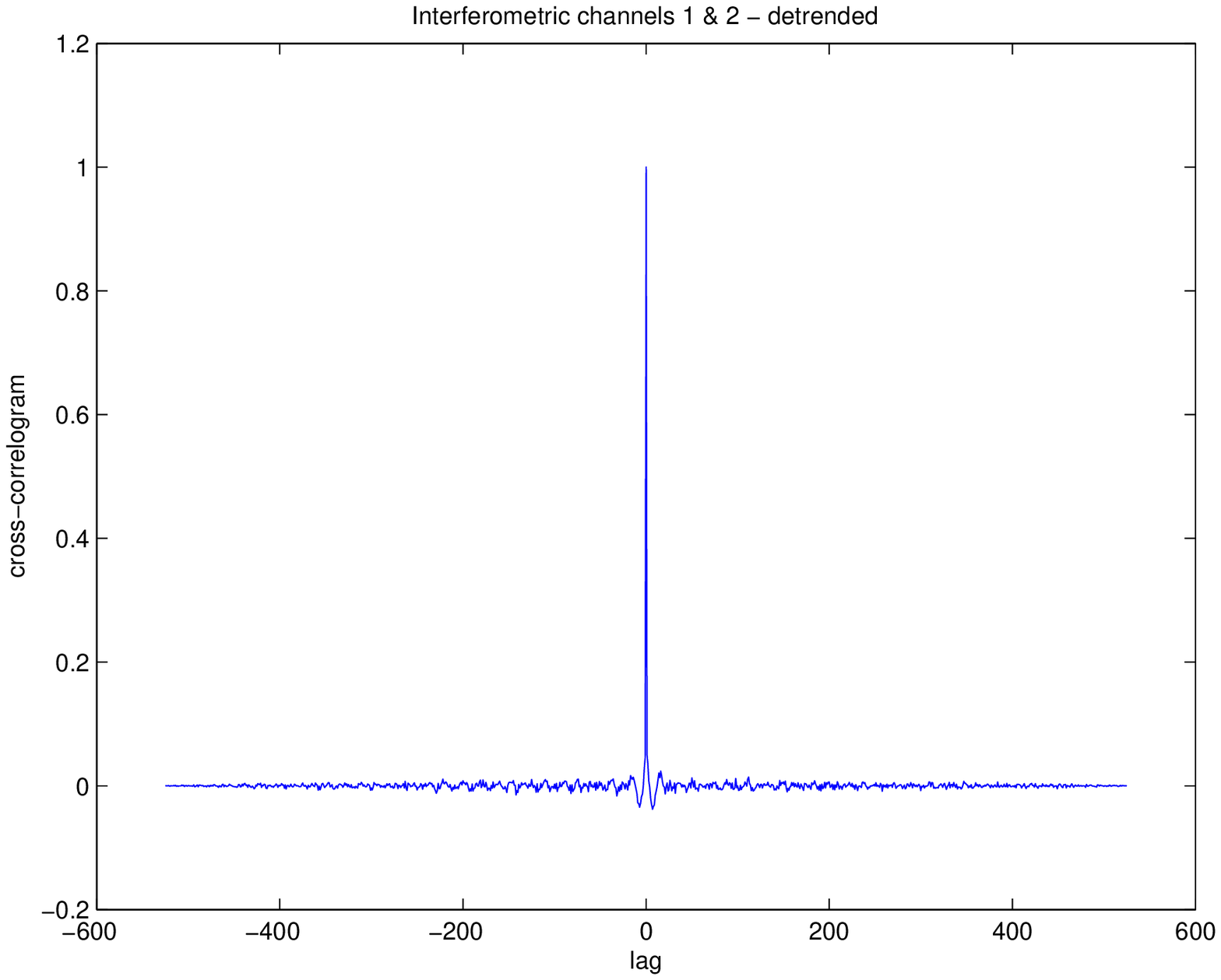,width=6.5cm}
    \epsfig{figure=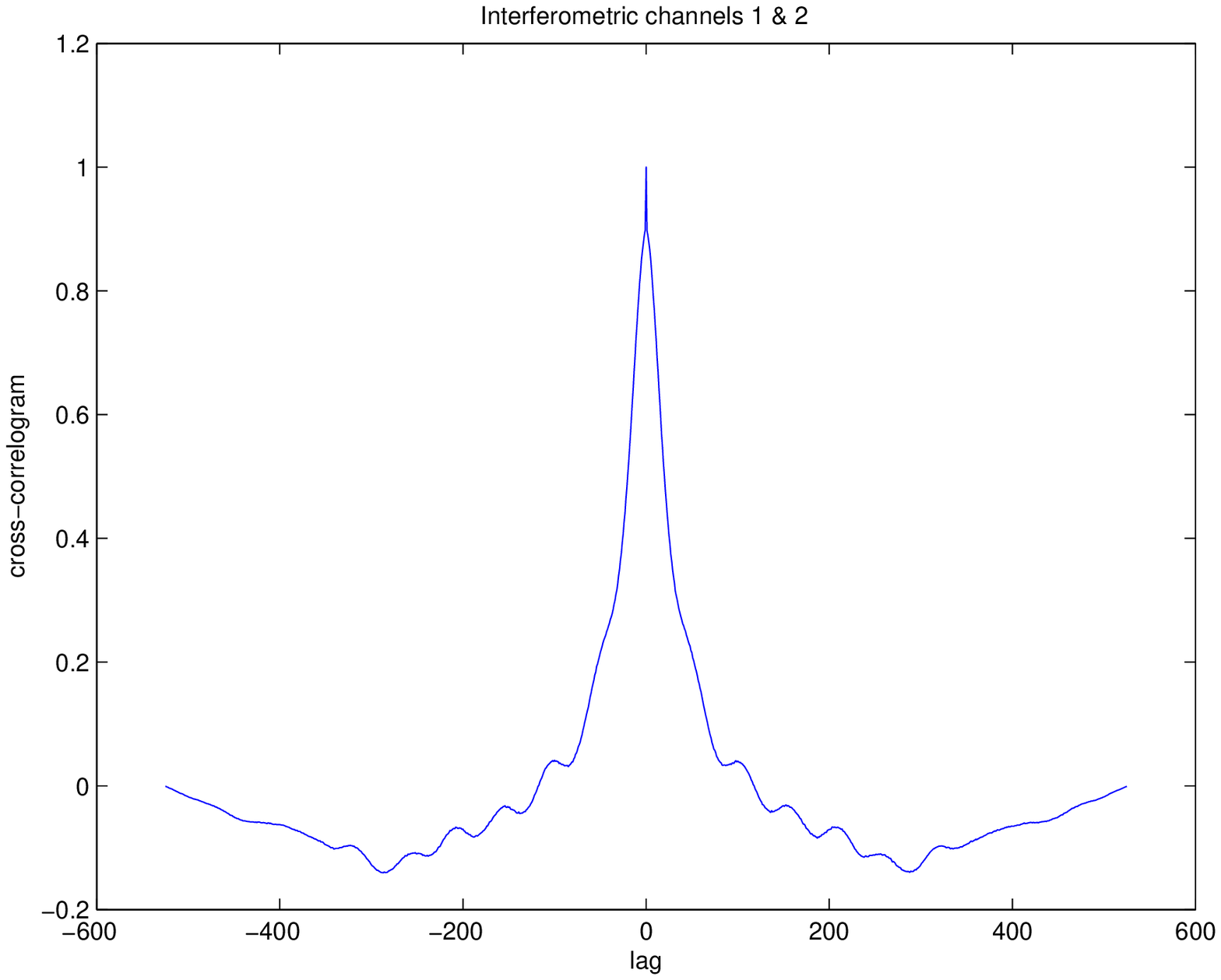,width=6.5cm}
    \caption{Case 2: Cross-correlation functions for interferometric channel $I1$: raw data (left) and detrended (right).}
    \label{fig:calOutput-crosscorr}
    \end{center}
\end{figure*}

\noindent The estimation of the cross-correlation does not show any unexpected pattern. Due to the fact that there are not fringes, the output channels are very similar to the photometric inputs. These functions (see fig. \ref{fig:calOutput-crosscorr} for case 2) do not reveal any particular residual effect due to the combination process that adds up the inputs and then split the sum into the output beams. Case 3 is in appendix \ref{appendixB}.

\subsection{Output in observational mode: Interferometric signals }
\label{subsec:out_interf}

\begin{figure*}[htbp]
    \begin{center}
    \epsfig{figure=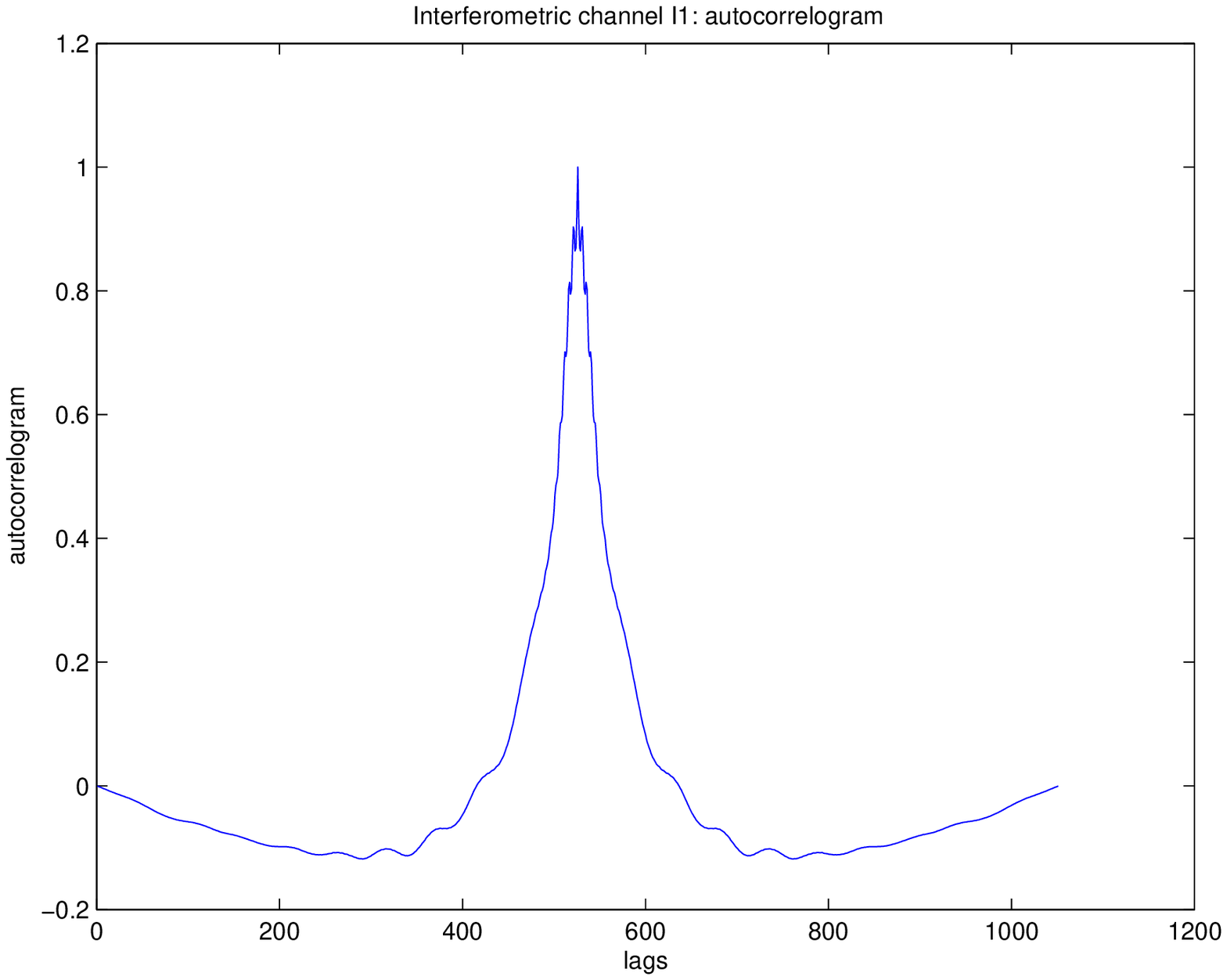,width=6.5cm}
    \epsfig{figure=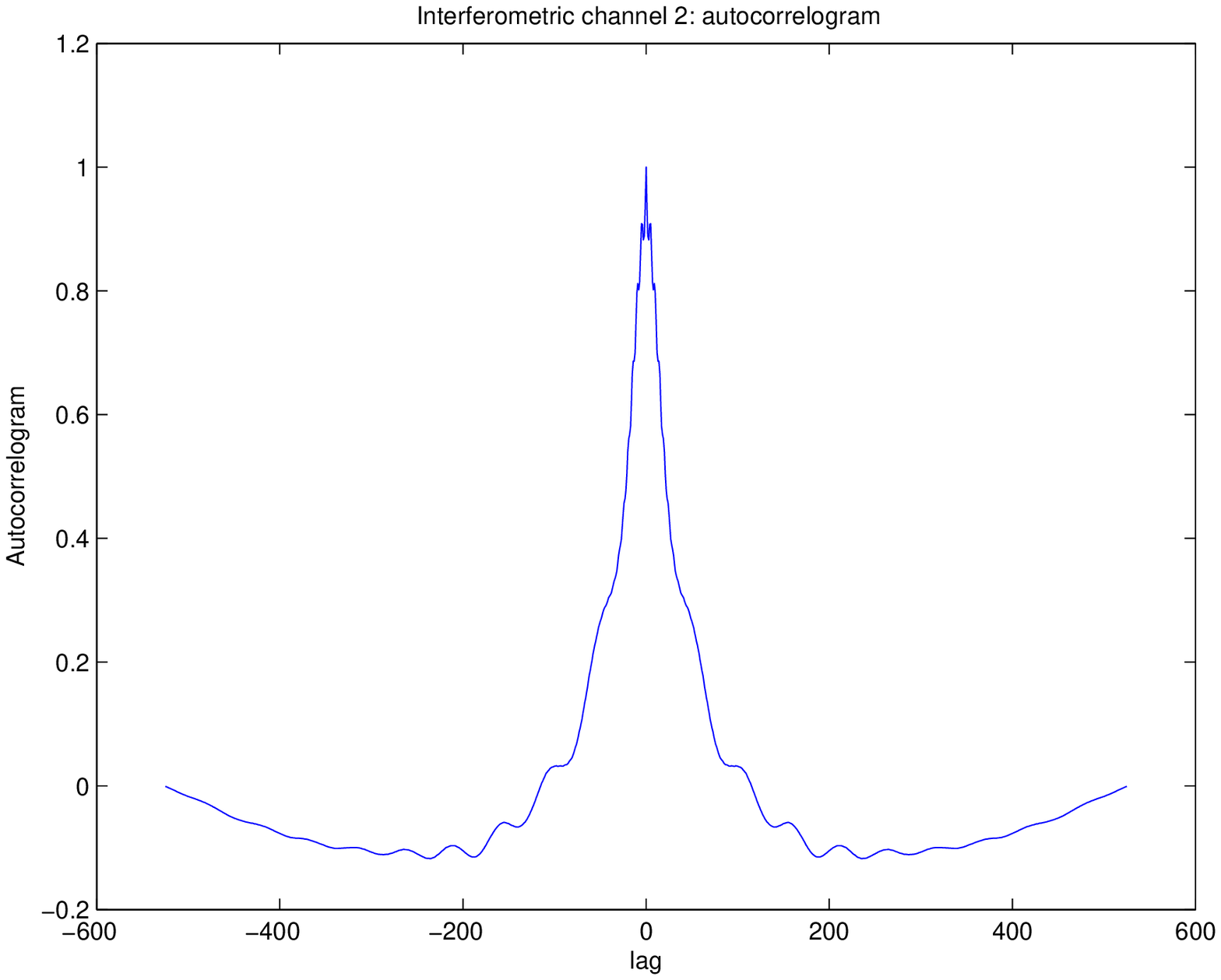,width=6.5cm}
    \vspace{1cm}
    \epsfig{figure=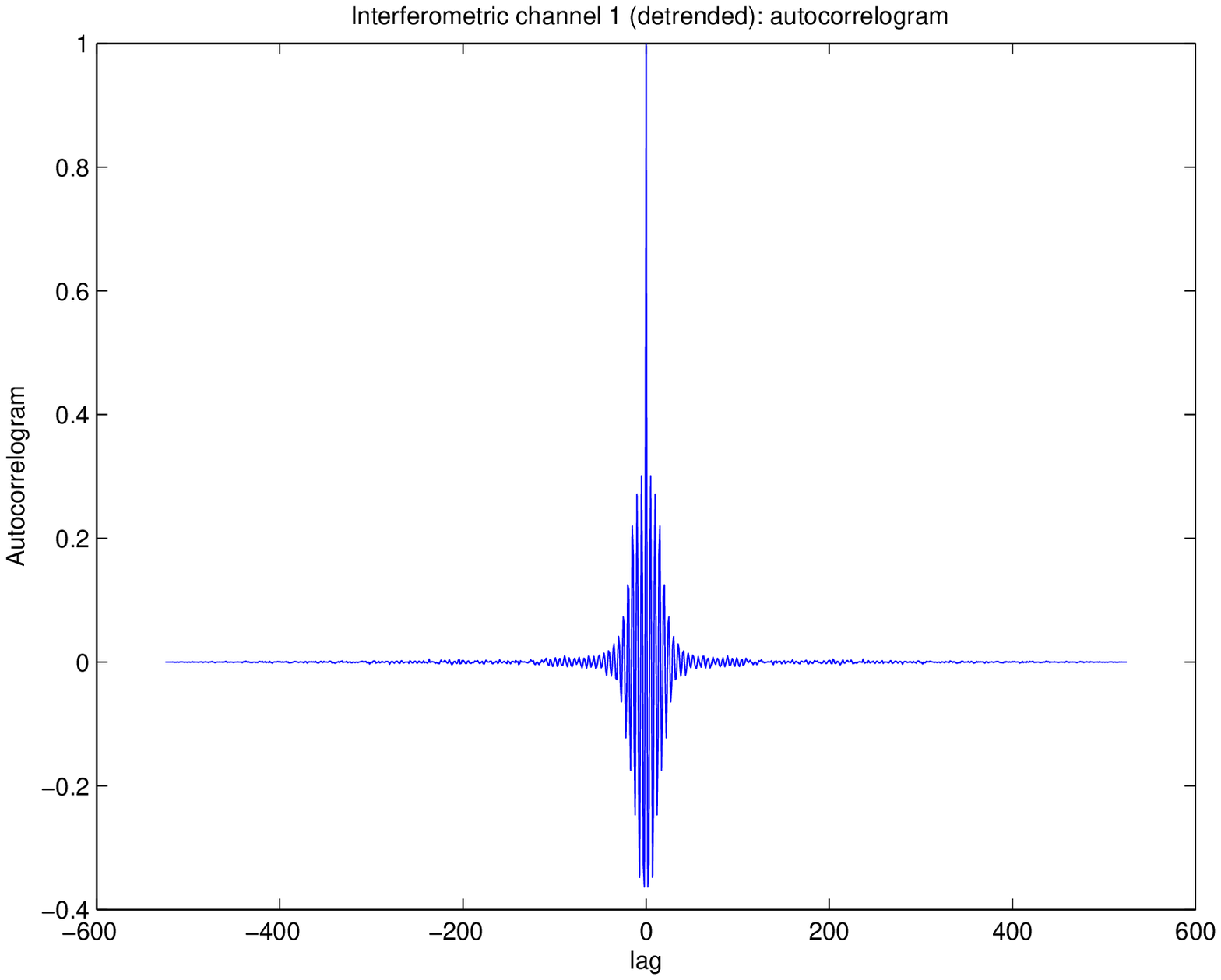,width=6.5cm}
    \epsfig{figure=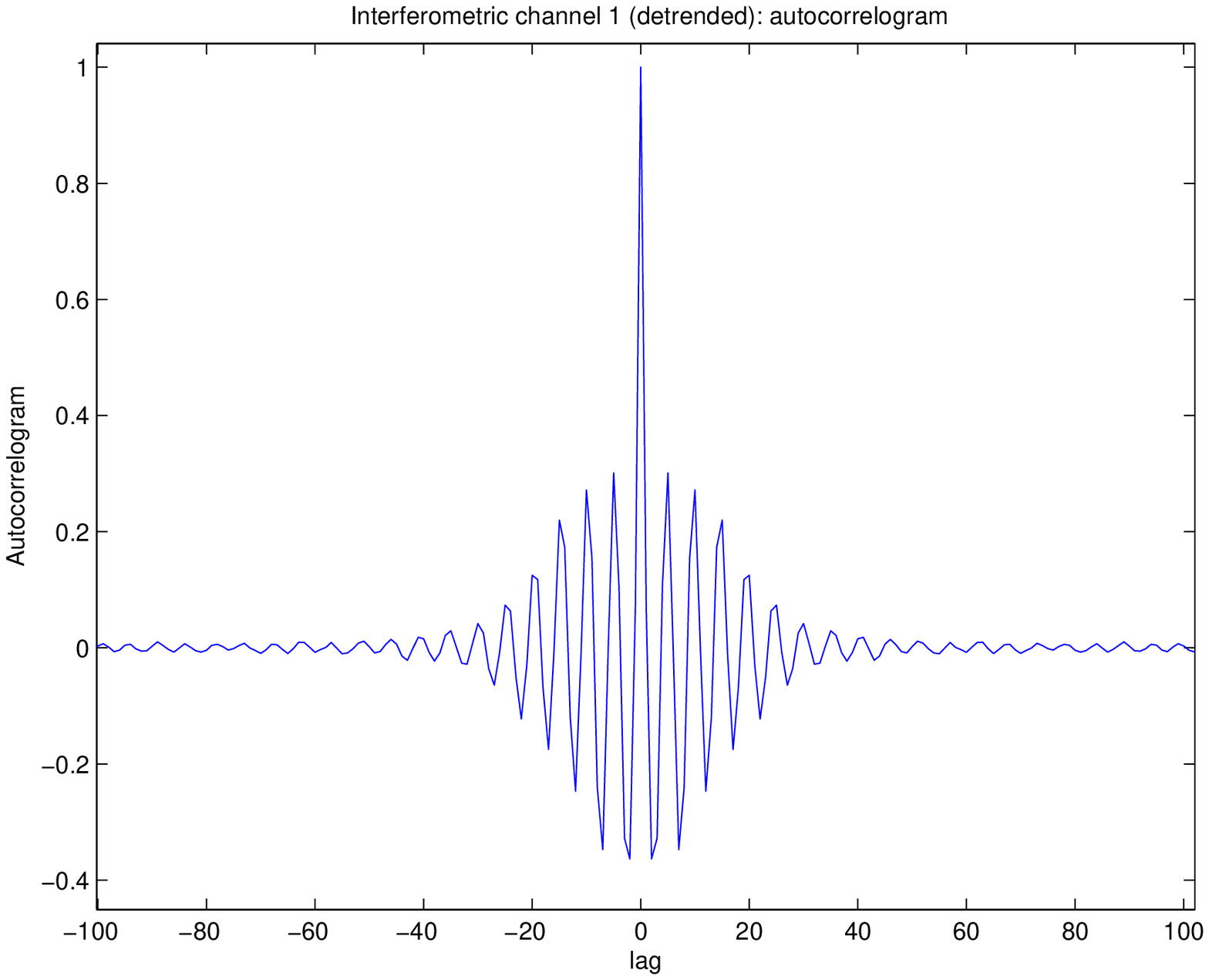,width=6.5cm}
    \caption{Case 4. First row: Raw data, autocorrelation functions for interferometric channels $I1$ (left) and $I2$ (right). The function shape is similar to photometric channels (fig. \ref{fig:phot-autocorr}).
    Second row: autocorrelation function after a detrend for interferometric channel $I1$ (left) and a zoom in the central lags area (right).}
    \label{fig:int12-autocorr}
    \end{center}
\end{figure*}

\noindent The interferometric channels, without the trend subtraction, show a behaviour similar to the photometric ones. They need a careful treatment because they contain fringes, i.e. the modulation part which contains scientific information. Moreover, being the result of an interference between coherent beams, we would like to find this characteristic into the correlation functions.
However, in these sets of measurements the amplitude of beam variations is comparable with the fringe amplitude. The autocorrelation function reflects this feature, i.e. we can't recognize the presence of the fringes, that are lost. Fig. \ref{fig:int12-autocorr}, first row, shows the autocorrelation function for the output beams $I1$ and $I2$.
\noindent Again, we perform the detrend operation. The subtraction of the linear trend does not cancel the fringe patterns, at the contrary, the modulation part is evidenced. Now the autocorrelation function shows very peculiar features, as shown for the $I1$ channel in the second row of fig. \ref{fig:int12-autocorr}. The companion $I2$ is very similar.

\begin{figure*}[htbp]
    \epsfig{figure=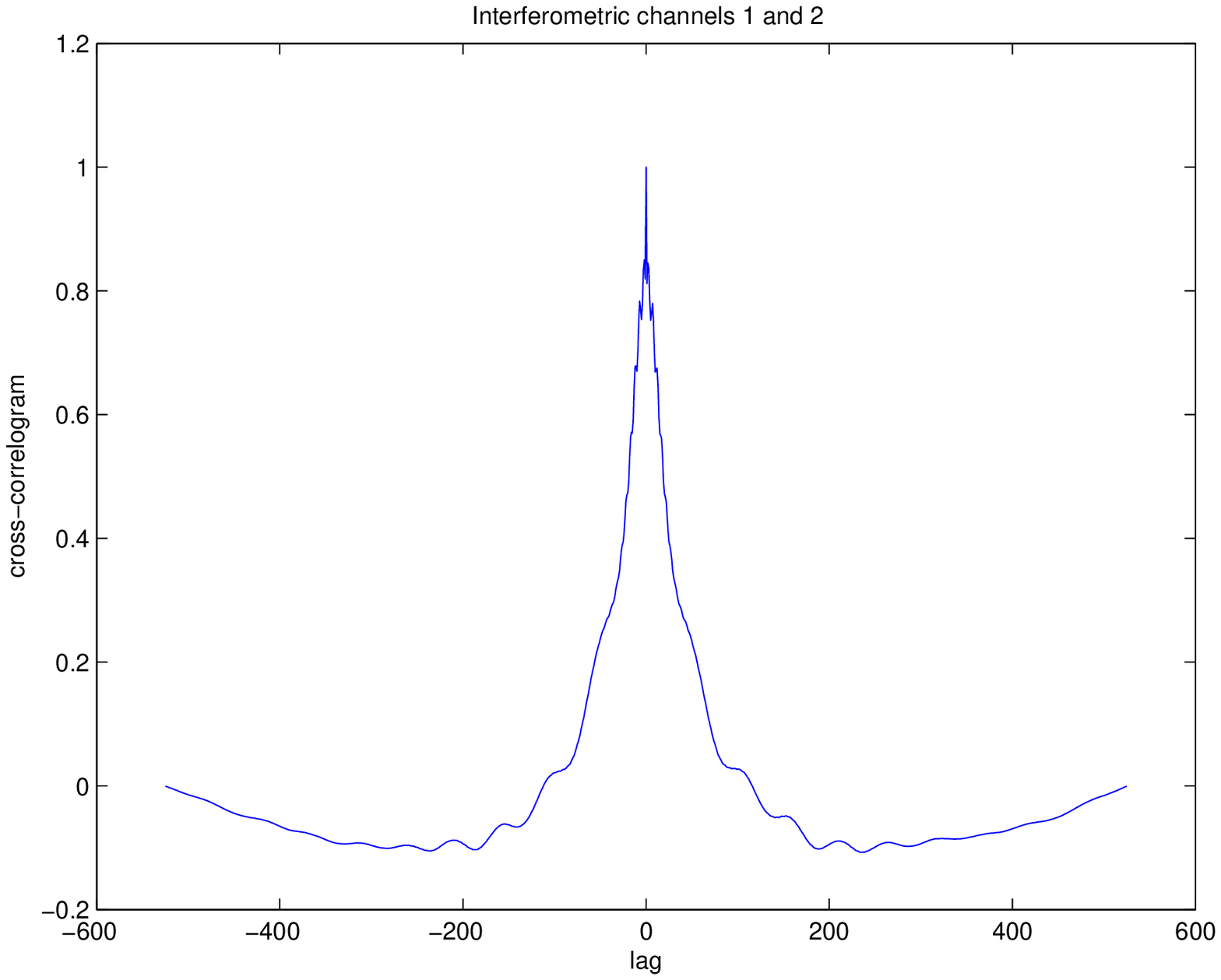,width=6.5cm}
    \epsfig{figure=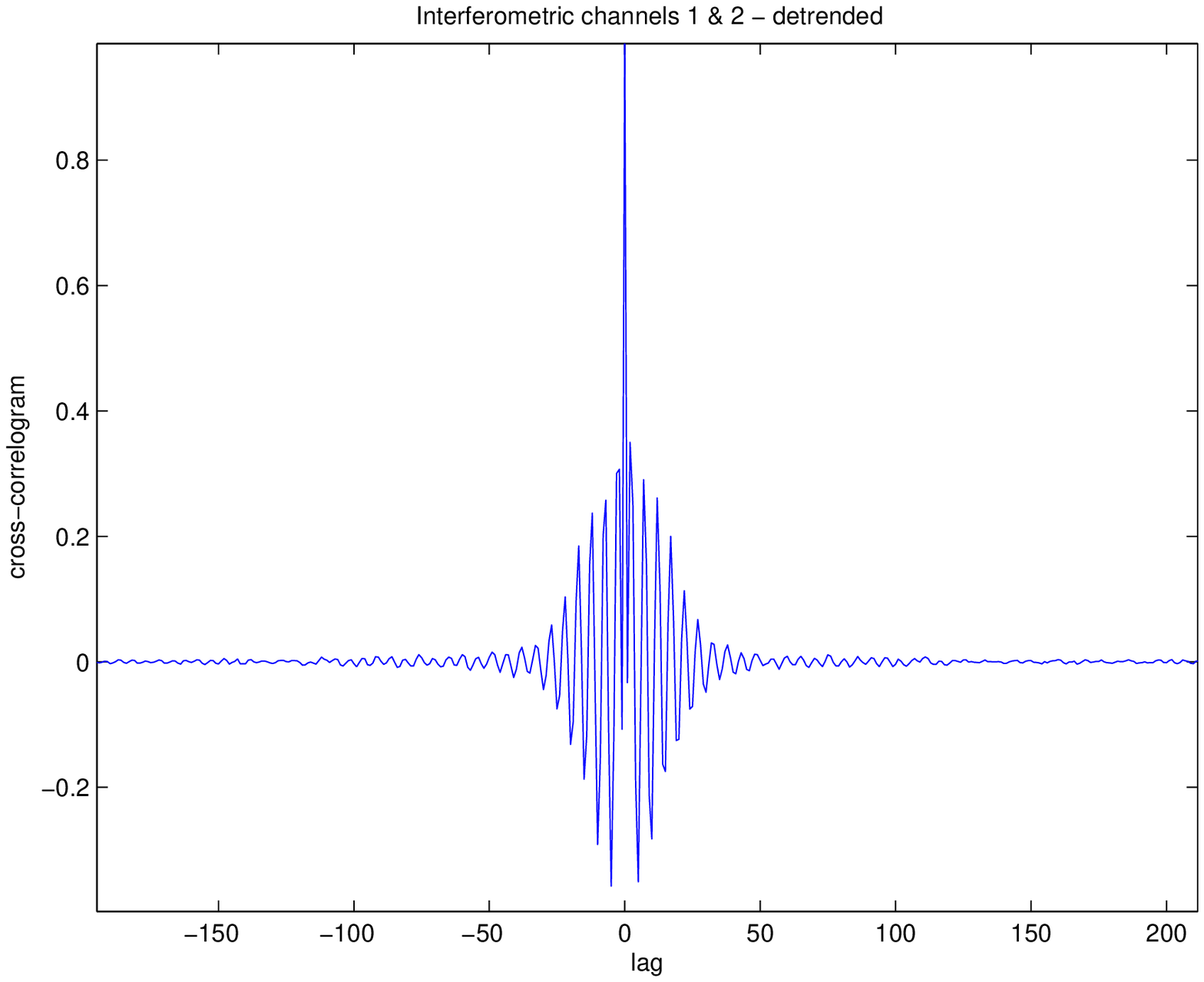,width=6.5cm}
    \begin{center}
        \caption{Case 4.Cross-correlation functions for interferometric channels $I1$ and $I2$. Left, raw data; right, linear trend subtracted.}
        \label{fig:int1and2-crosscorr}
    \end{center}
\end{figure*}

\noindent We can easily recognize two components, associated to the interferometric signal components. The modulation part is the sum of sinusoidal waves at different wavelengths, so we can expect the behaviour of a harmonic process, while the offset is dominated by the photometric fluctuations, with the presence of a long term correlation.
\\
\noindent The same holds for the cross-correlation function, too, as fig. \ref{fig:int1and2-crosscorr} shows.

\subsection{Cross-correlation between photometric inputs and interferometric outputs}
\label{subsec:cross-corrOutIn}
\begin{figure*}[htb]
    \epsfig{figure=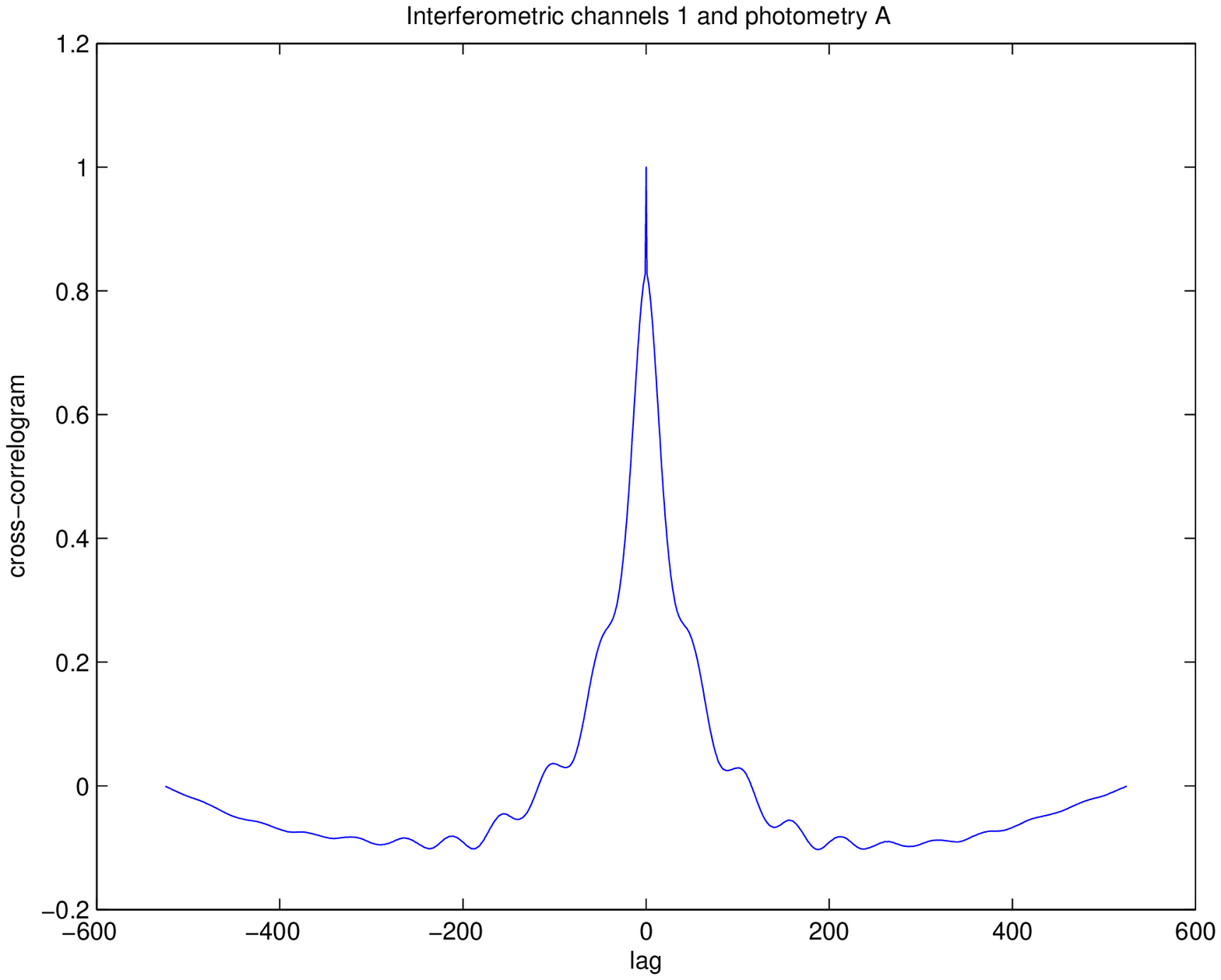,width=6.5cm}
    \epsfig{figure=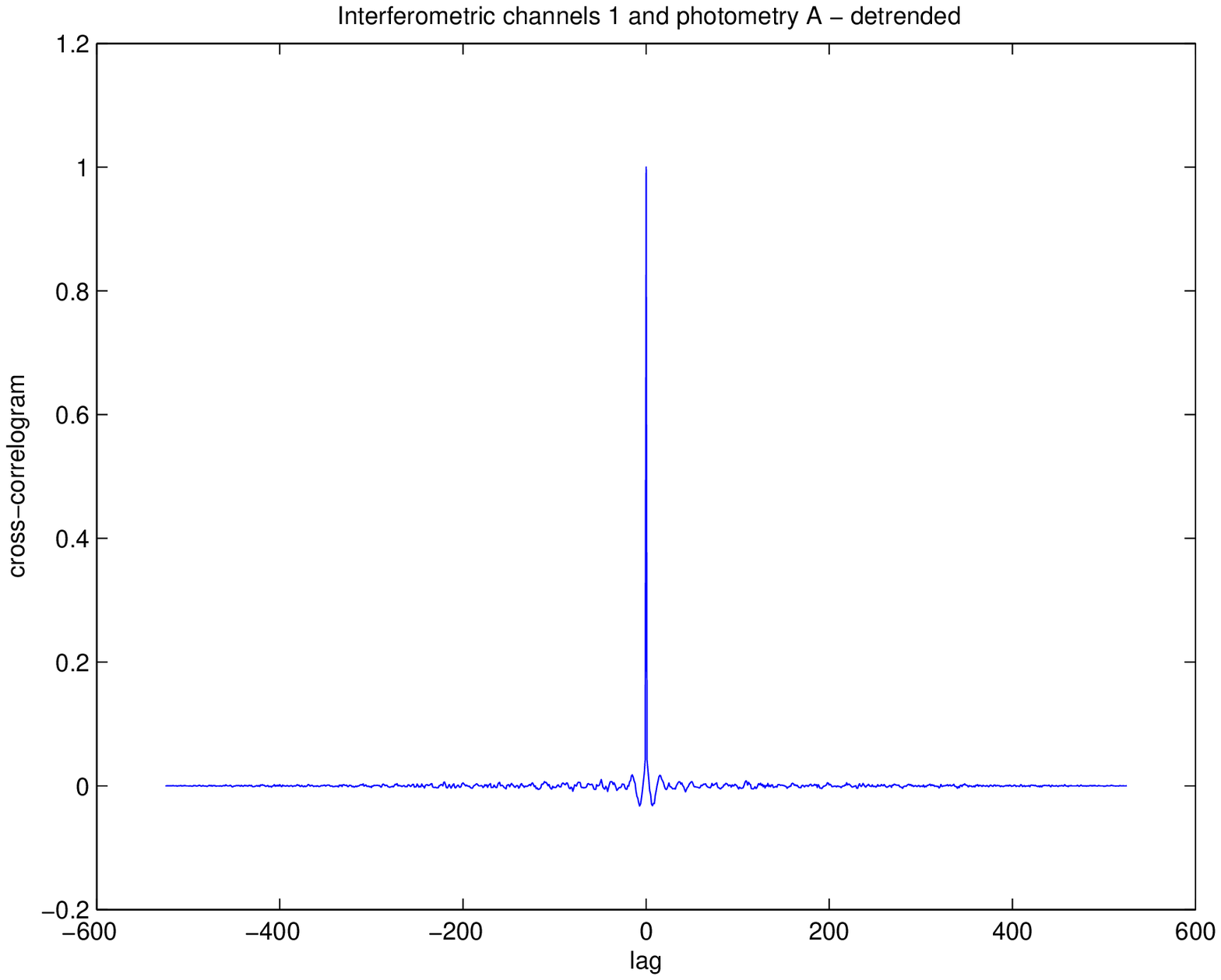,width=6.5cm}
    \begin{center}
        \caption{Case 4.Cross-correlation functions for interferometric $I1$ and photometric $PA$ channels. Left, raw data; right, linear trend subtracted.}
        \label{fig:I1andPA-crosscorr}
    \end{center}
\end{figure*}

The estimate of the cross-correlation function between the photometric and the interferometric signals reveals that the correlation is caused by the trend, as in previous cases. We shows in figure \ref{fig:I1andPA-crosscorr}, as an example, the cross-correlation between the interferometric channel $I1$ and the photometric one $PA$ before and after the trend subtraction to each beam. It has to be noticed that each beam is detrended independently. 

\section{Statistical analysis in the frequency domain}
\label{sec:spectral_analysis}

Spectral analysis can help in understanding the nature of noise on signals under study. Of course, we would expect noise sources to be white processes, and the interference pattern to be well distinguishable from noise.\\
In particular, we investigate the possibility that the trend might mask other higher frequencies features.


\subsection{Spectral methods description}
\label{subsec:freqdom-method}

The analysis in the spectral domain is performed using both the power spectral density function and the Allan variance. There is a relation between these quantities, well established in the case of wide-sense stationary time series. For references, see \cite{Allan}.

\subsubsection{Power Spectral Density}
\label{subsubsec:PSD-method}

The Power Spectral Density (hereafter, PSD) can be defined in several ways. The need of the PSD function estimation instead of the energy spectral density arises when signals are such that their total energy is not finite. For a detailed discussion on this topic, see, e.g., \cite{Priestley} or \cite{Manolakis}.\\

\noindent We estimate the PSD with the periodogram $\hat{\Phi}(k), \; 0 \leq k \leq N-1$, i.e. the squared modulus of the discrete Fourier transform of the signal $s$ time series  $\{s_i\}_{i=0 \ldots N-1}$, with the appropriate normalization:

\begin{eqnarray}\label{eq:fft_PSD}
    \nonumber F(k) &=& \frac{1}{N} \sum_{j=0}^{N-1} s(j) e^{- 2 \pi i j k / N} , \;\; 0 \leq k \leq (N-1)\\
    \hat{\Phi}(k) &=& \frac{1}{N} |F(k)|^2,  \;\; 0 \leq k \leq (N-1)
\end{eqnarray}
where the relation of $k$ with the frequency effectively recoverable from the data can be found with the information on the data scan of Sec. \ref{sec:data_descr}.
From the literature \cite[pg. 209]{Manolakis}, we know that this estimation of the PSD is affected by bias problems, and it is not consistent. 

\noindent To reduce the measurement noise variance, we apply a smoothing operation on $\hat{\Phi}(k)$ averaging over a five-samples window, moving the window one sample at a time.
Moreover, we evaluate the periodogram over all available records for each case, say $M$, and we average all periodograms to get a final estimation of the PSD of the time series.  This procedure is equivalent to a welch smoothing without overlapping of the window, and it leads to a decrease in the standard deviation of the estimation as $\frac{1}{\sqrt{M}}$. 

\begin{figure}[htb]
    \begin{center}
    \epsfig{figure=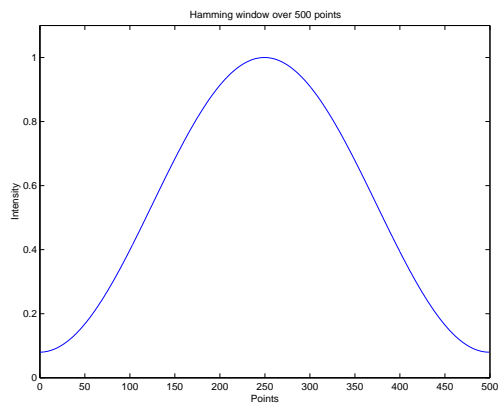,width=6.5cm}
    \caption{Hamming window with $N=500$.}
    \label{fig:hamming_window}
    \end{center}
\end{figure}

\noindent The spectral bias problem can arise from a sharp truncation of the sequence, and can be reduced by first multiplying the finite sequence by a window function which truncates the sequence gracefully rather than abruptly. In our calculation, we chose the Hamming window, defined as:
\begin{equation}
w(k+1) = 0.54 - 0.46 \cdot cos (2  \pi  \frac {k}{N - 1}), \;\;  k = 0, \ldots, N-1
\end{equation}
If n is odd, this window is symmetric around the median point; if n is even, it does not have a central point. Figure \ref{fig:hamming_window} shows the Hamming window for $N = 500$ points.\\

\noindent The presence of a non zero mean has influences especially at low frequencies. Even if the mean was constant over all records, the leakage\footnote{Leakage: contribution of the sinusoidal components with frequencies $\omega \neq \omega_0$ to the periodogram value $\hat{\Phi}(\omega_0)$} phenomenon would spread the frequency peak of the mean on the neightbour frequencies, possibly obscuring low frequencies components of the spectrum, if present.


\subsubsection{Allan variance}
\label{subsubsec:allan-method}

\noindent Another technique we have applied for the understanding of the spectral behaviour is the Allan variance, or two samples variance.
\noindent It was introduced in the sixties \cite{Allan}\cite{Allan87}, in the field of frequency stability measurements for metrology signals, and is now widely used. It was originally conceived to avoid the lack of convergence of the usual variance with some orders of power-shaped spectra. In astronomic science fields, it was first used in radio astronomy, for the phase difference time series \cite[pag. 272]{Thompson}, but also for a general time series \cite{Schieder}. Some authors, such as Colavita \cite{Colavita}, used it for estimating phase difference spectral features in the field of infrared interferometry, with a slightly different formulation (Allan modified).
\\

\noindent We use the original Allan variance, evaluated over a set of time lags, $\tau$, derived from a reference lag $\tau_0$. If
\begin{equation}\label{eq:allan_delta}
\Delta_i^{\tau_0} = \frac{1}{\tau_0}[x_{i+\tau_0} - x_i]
\end{equation}
is the average of the time sequence $x(t)$ over the time interval $[t, t+\tau]$, the Allan variance at lag $\tau = k\tau_0$ is defined as\cite{Allan87}:
\begin{eqnarray}\label{eq:allan}
\nonumber \sigma^2_A(\tau) &=& \frac{1}{2(N-2k+1)} \sum_{n=1}^{N-2k+1} (\Delta_{n+k}^{\tau}-\Delta_n^{\tau})^2 = \\
&=& \frac{1}{2\tau^2(N-2k+1)} \sum_{n=1}^{N-2k+1} (x_{n+2k}-2x_{n+k}+x_{n})^2.
\end{eqnarray}
Since the summation goes from $1$ to $N-2k+1$, the last is the maximum number of independent factors. High lag terms are affected by errors due to the average over a small number of values.
\\
\noindent The exponent of the variance, as a function of the lag $\tau$, can be related directly to a range of power-shape spectra thanks to the following relation:
\begin{equation}\label{eq:allan-spectra}
S_y(f)=a_{\alpha}\cdot f^{\alpha} \;\; \rightarrow \;\; \sigma^2_A(\tau) = \tilde{a}_{\beta} \cdot \tau^{\beta}
\end{equation}
valid for $-2 \leq \alpha \leq 2$, and where $\alpha$ and $\beta$ are linked by:
\begin{equation}\label{eq:allan-coeff}
\beta = - \alpha - 1.
\end{equation}

\noindent In particular, for a flat ($\alpha = 0$) and a `flicker' noise $\alpha = -1$, the Allan variance exponents are $\beta= -1$ and $\beta = 0$, respectively, with coefficients $\tilde{a}_{-1} = 1 / 2 \cdot a_0$ and $\tilde{a}_{0} = 2 ln(2) a_{-1}$.




\subsection{Spectral analysis of raw data}
\label{subsec:spectra}

\begin{figure*}[htb]
    \begin{center}
    \epsfig{figure=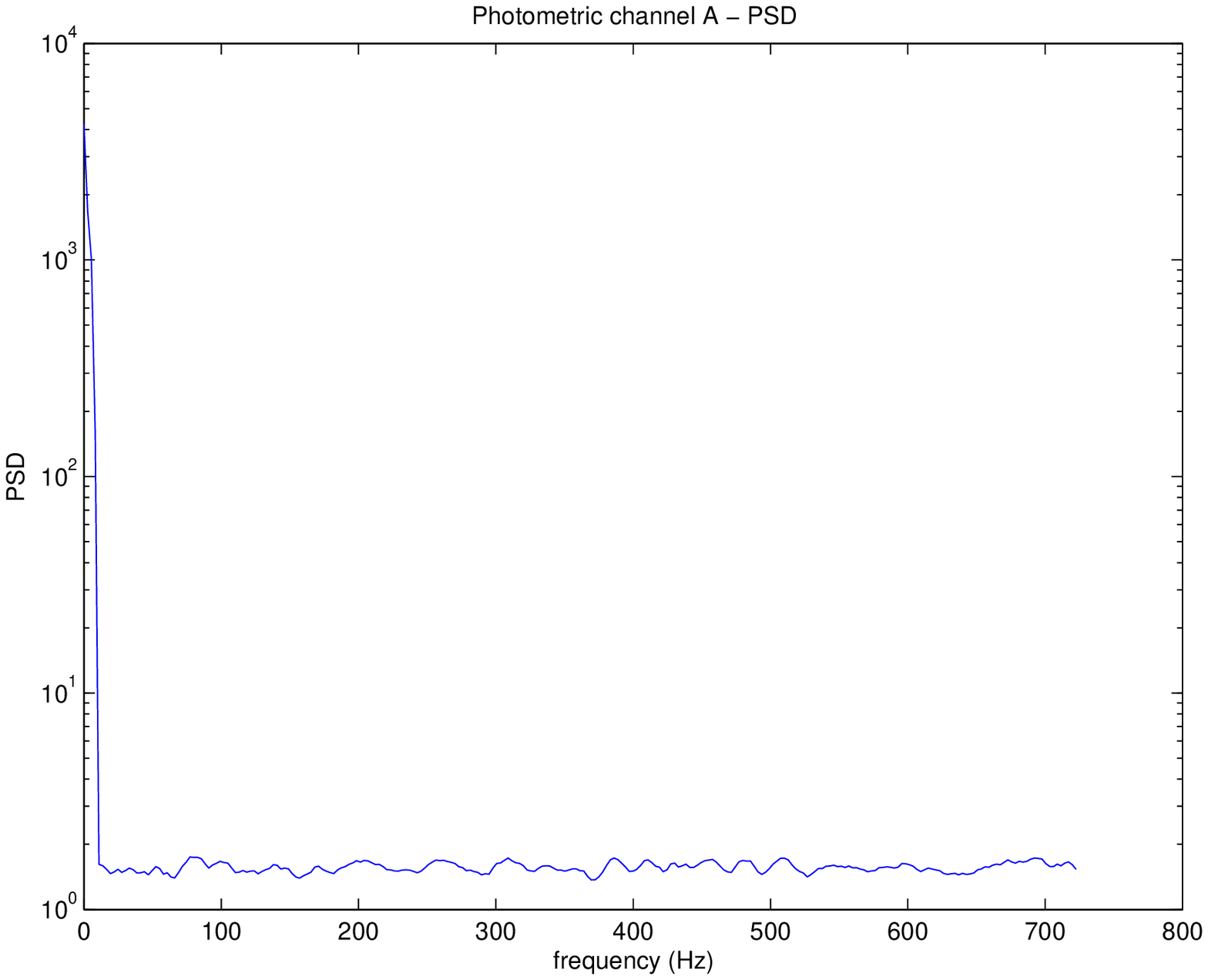,width=6.5cm}
    \epsfig{figure=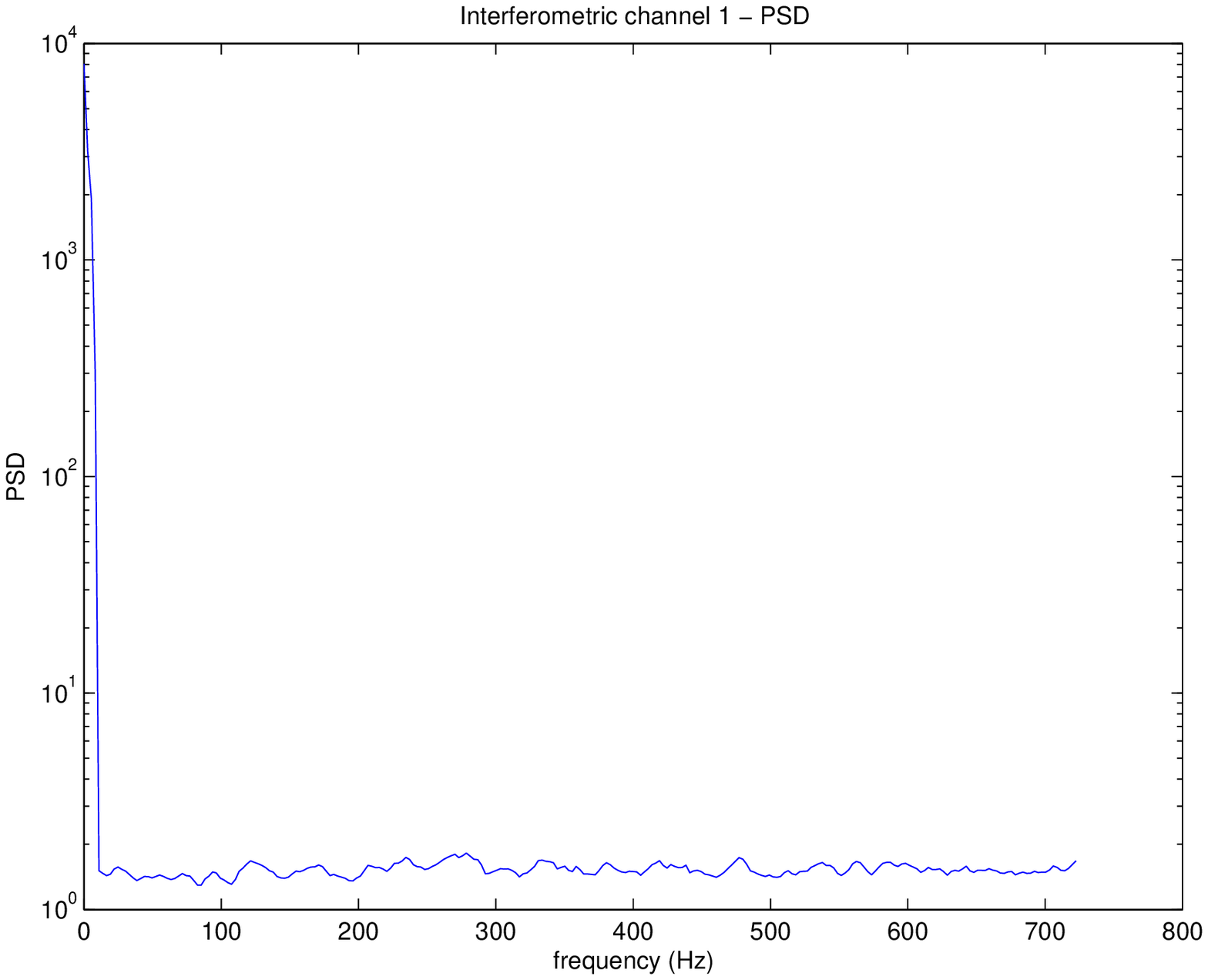,width=6.5cm}
    \epsfig{figure=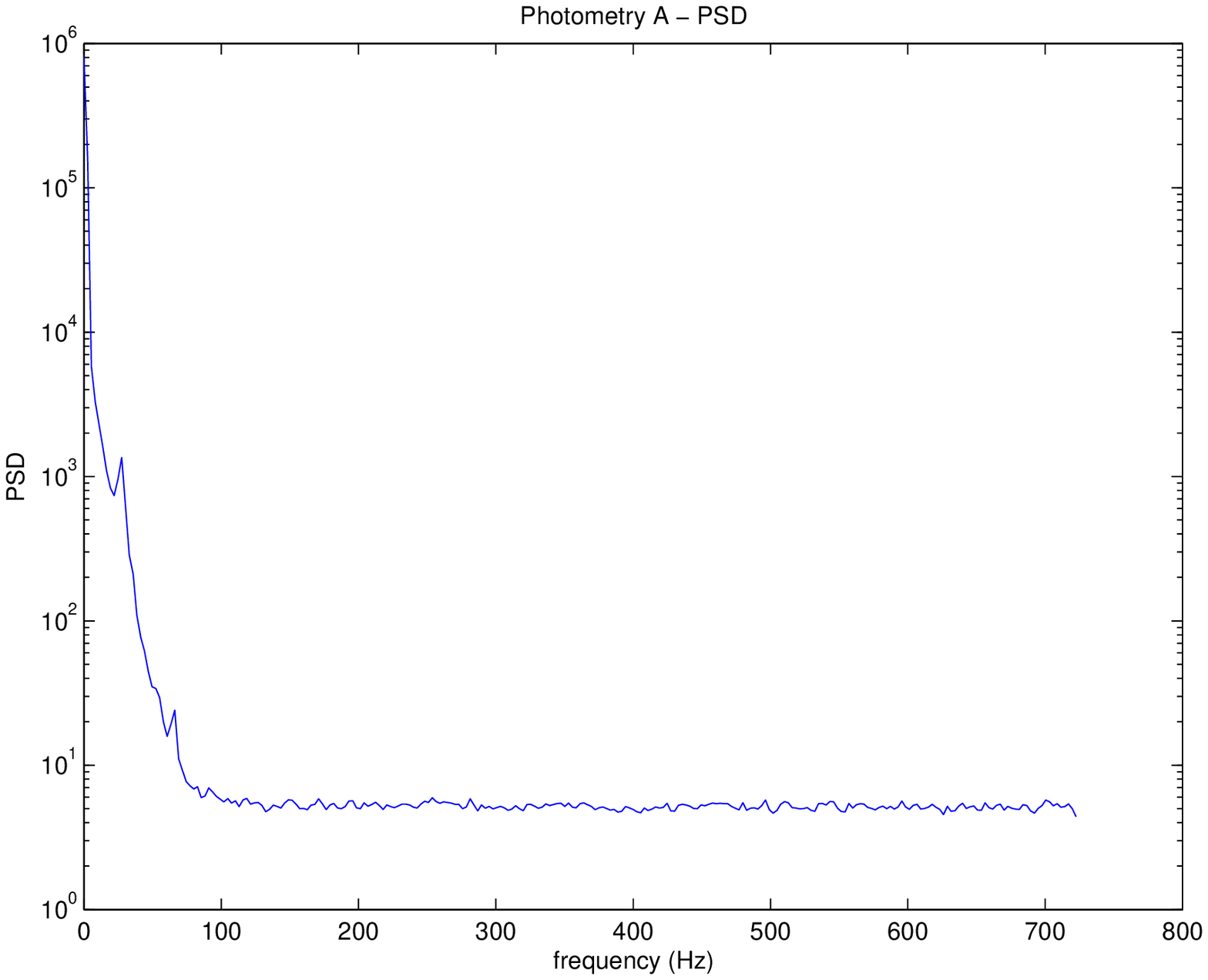,width=6.5cm}
    \epsfig{figure=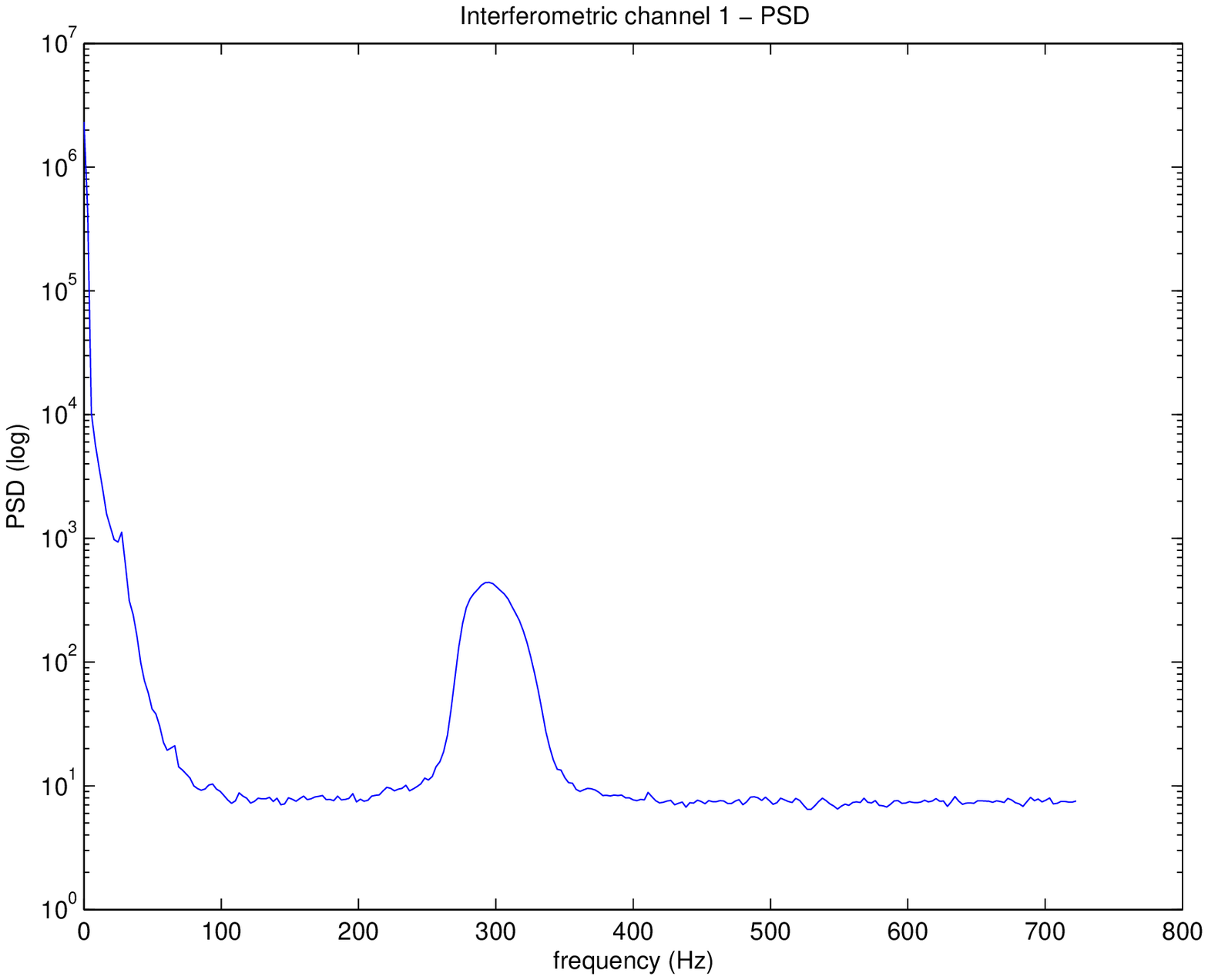,width=6.5cm}
    \caption{PSD estimation for the photometric channel A (left) and for the interferometric channel 1 (right). 
    First row: case 1 - no flux injected - note the low-frequency components.
    Second row: case 4 - flux injected - it is evident the frequency range of the fringes, well distinguished from noise.}
    \label{fig:all_none_PSD}
    \end{center}
\end{figure*}

\noindent We perform a spectral analysis both on calibration channels and on channels with injected fluxes. The former are useful to assess the features of the environmental noise, features that are covered by flux patterns in the latter case. Figure \ref{fig:all_none_PSD}, first row, shows the power spectral density for a photometric and an interferometric channel, when both arms are not fed with stellar source. We process each record according to the described method of section \ref{subsec:freqdom-method}, with the final PSD average over $M = 100$ records.
\\
\noindent We apply the same technique when flux is injected in both arms of the interferometer, with the difference that the final average is performed over $M = 500$ records. Comparing the photometric performances, we can notice that when in the interferometer there is just noise, the PSD drops very quickly to the high frequencies intensity offset, while the presence of flux induces a smoother decrease of the PSD intensity. The interferometric performances, on the contrary, are dominated by the presence of the modulation frequency, clearly identifiable (fig. \ref{fig:all_none_PSD}, second row).

\noindent We can see, in both photometric and interferometric channels, the presence of two low-frequency peaks that have highest magnitude than the surrounding noise.\\

\begin{figure*}[htb]
    \begin{center}
    \epsfig{figure=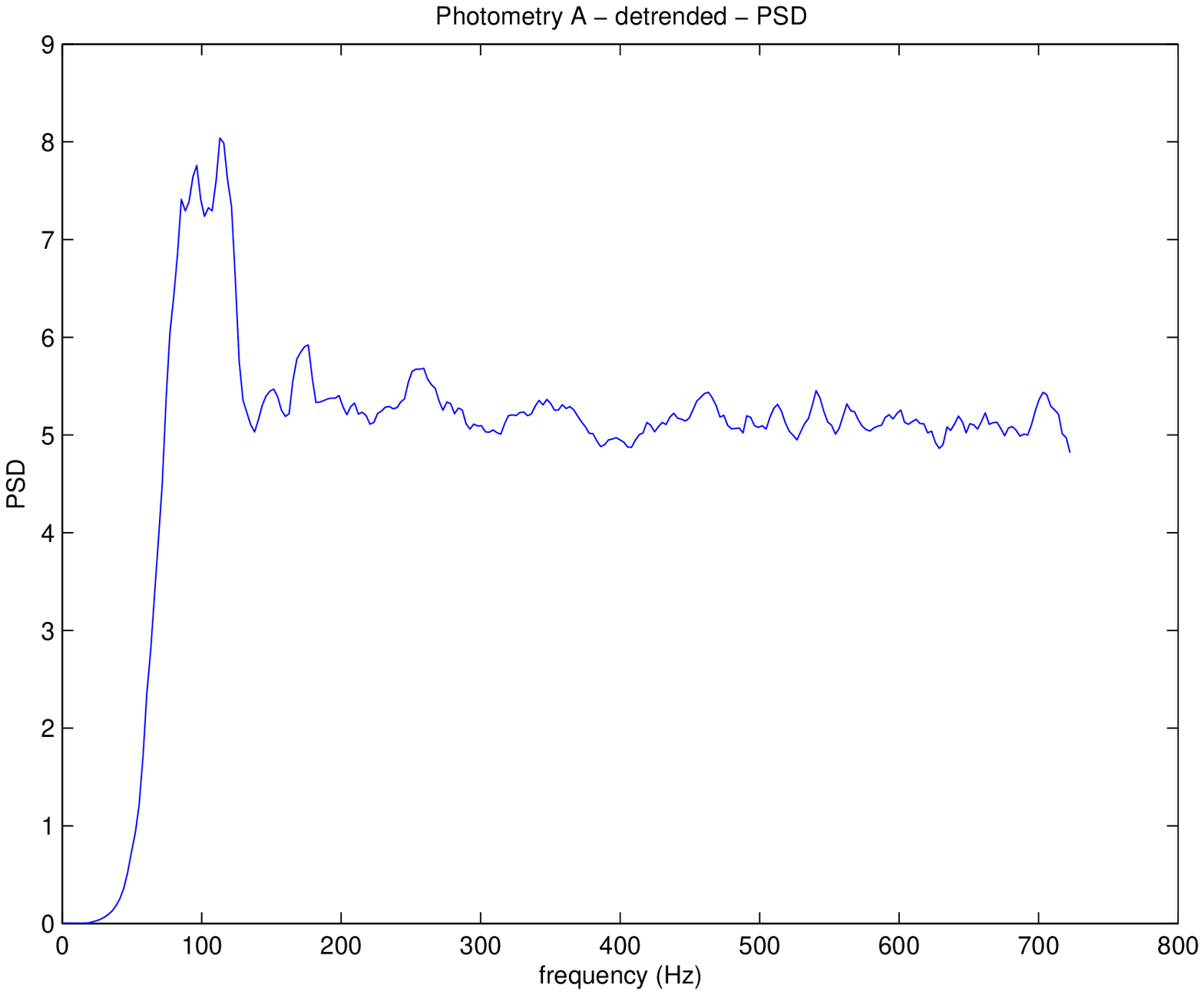,width=6.5cm}
    \epsfig{figure=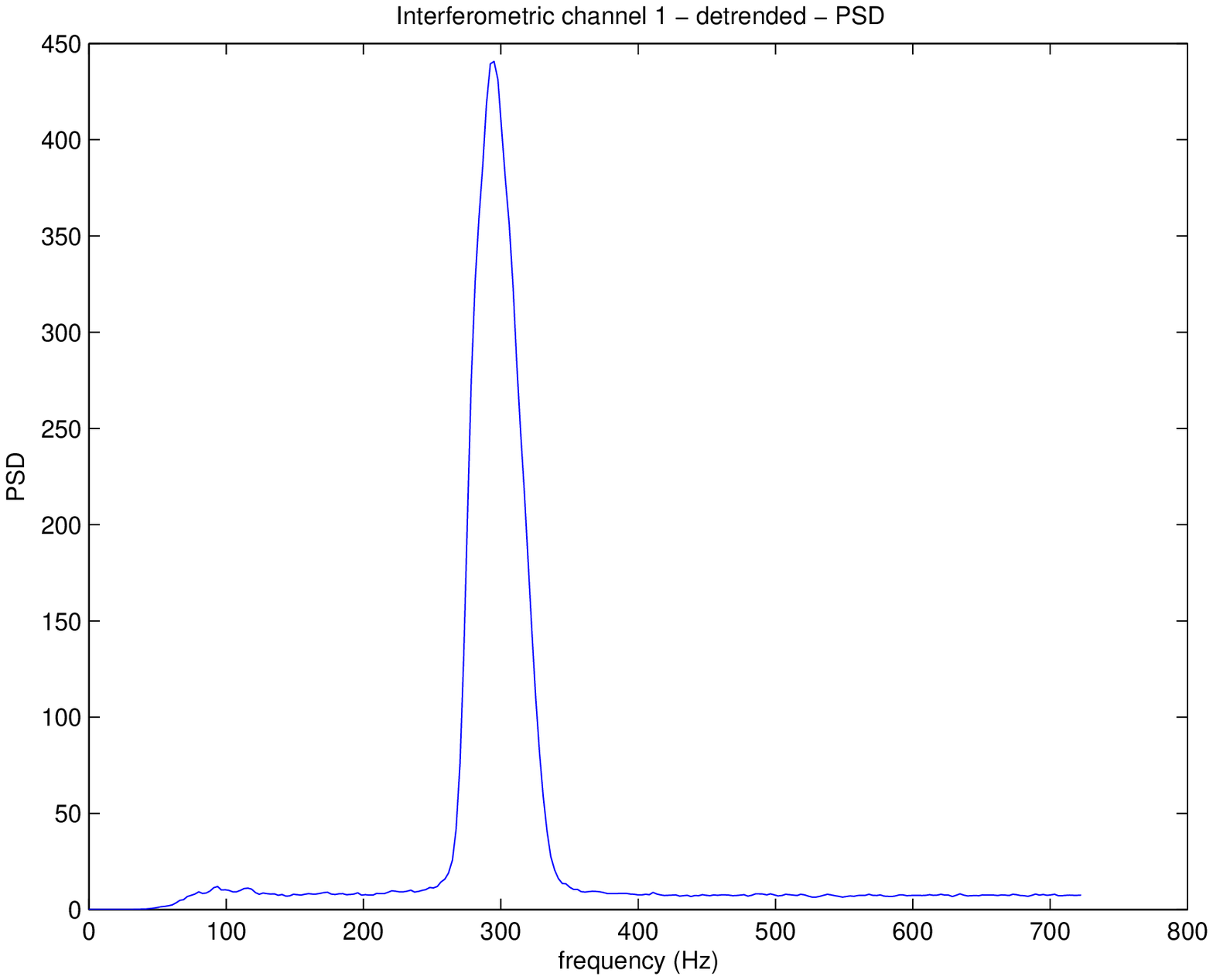,width=6.5cm}
    \caption{Case 4. Power Spectral Density estimation for the photometric channel A (left) and for the interferometric channel 1 (right). All the power area is the frequency range of the fringes. Note the different scale of figures, used to highlight the low-magnitude patterns of the PSD in the photometric case.}
    \label{fig:all_detr_PSD}
    \end{center}
\end{figure*}

\noindent If we apply a detrend operation to the observational data (figure \ref{fig:all_detr_PSD}), low frequencies are suppressed, because of the slow motion of the photometric intensities. The leakage phenomenon is reduced or eliminated. Figure \ref{fig:all_detr_PSD} illustrate this behaviour for the photometric channel $PA$ and for the interferometric one $I1$.

\begin{figure*}[htb]
    \begin{center}
    \epsfig{figure=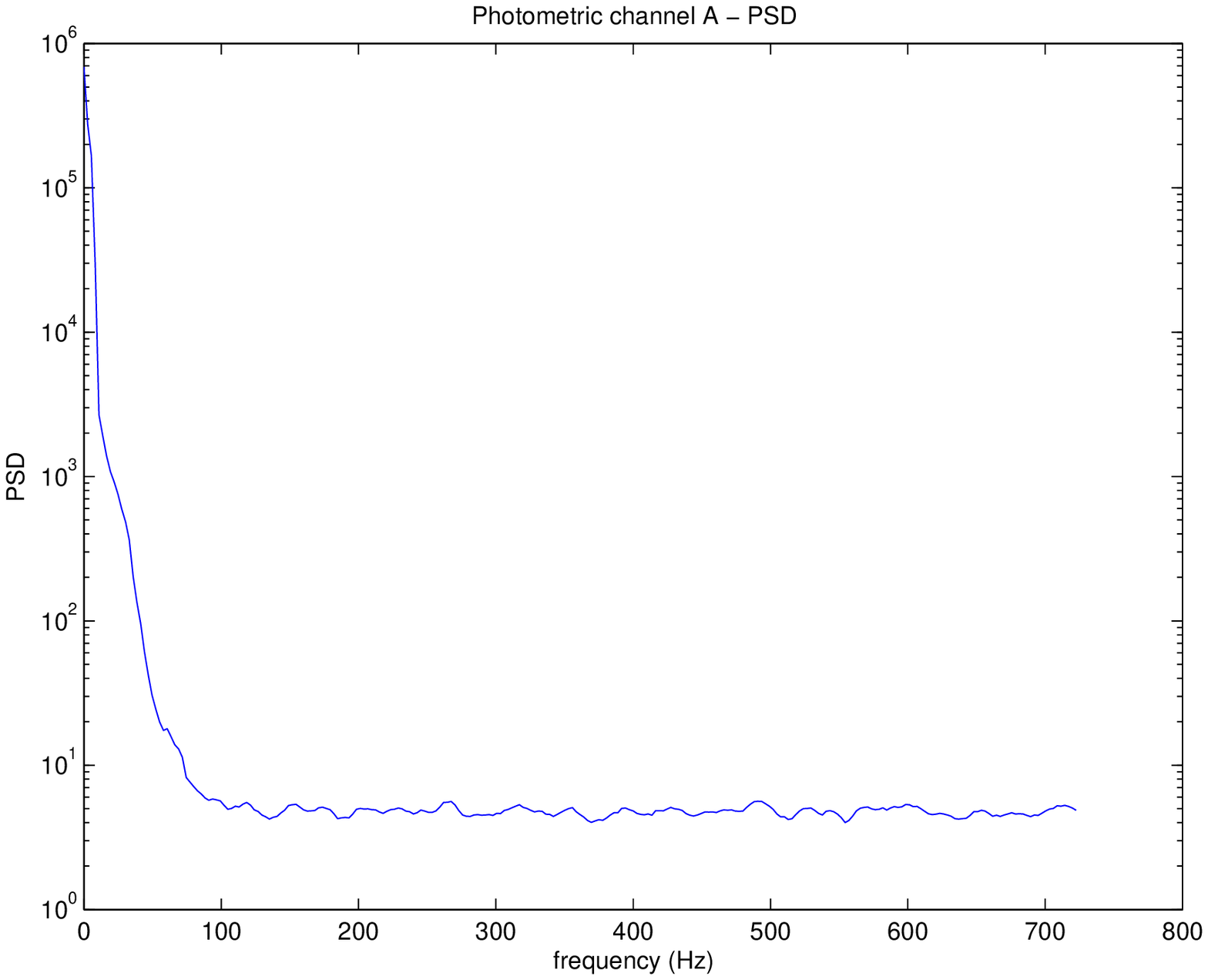,width=6.5cm}
    \epsfig{figure=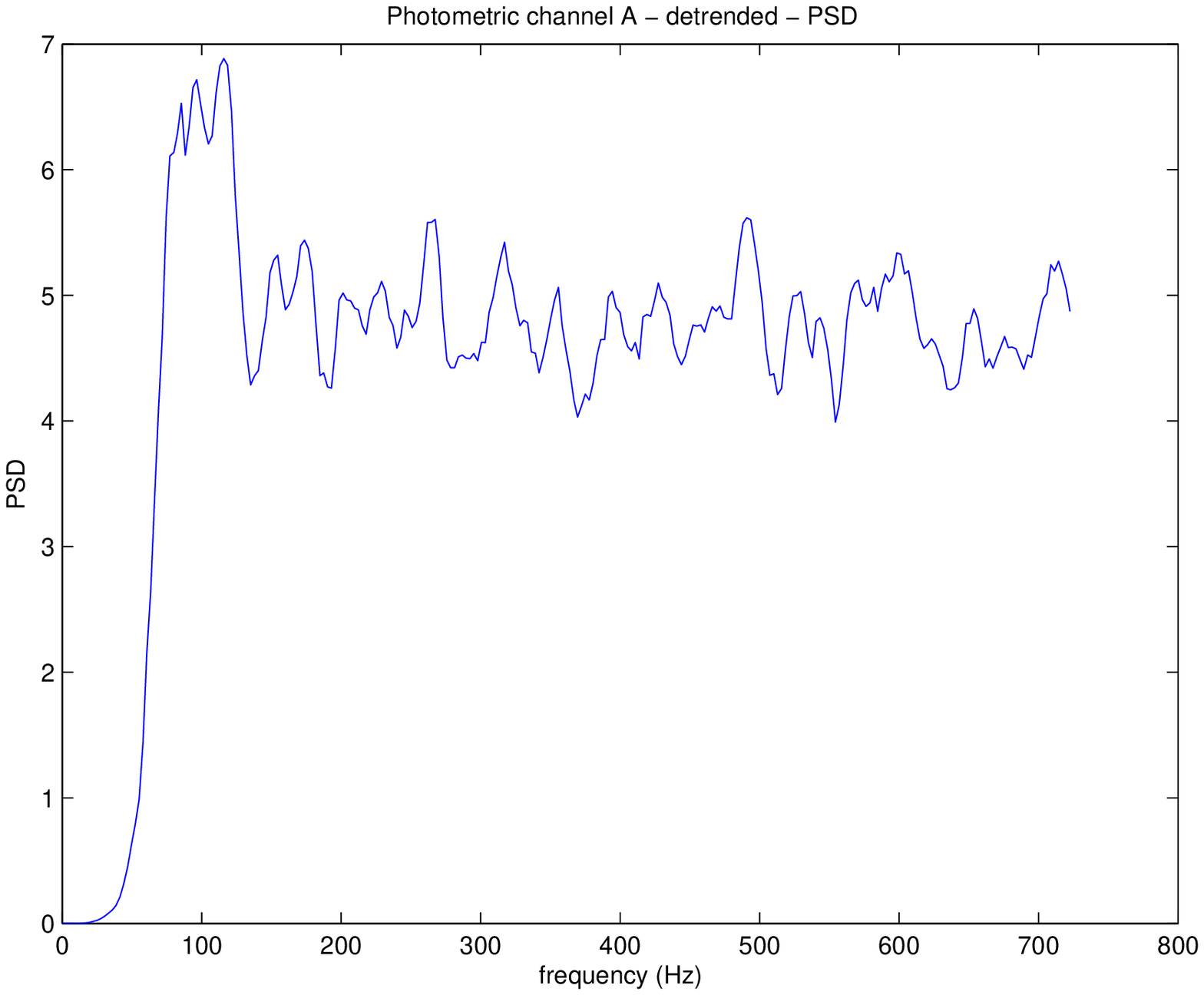,width=6.5cm}
    \epsfig{figure=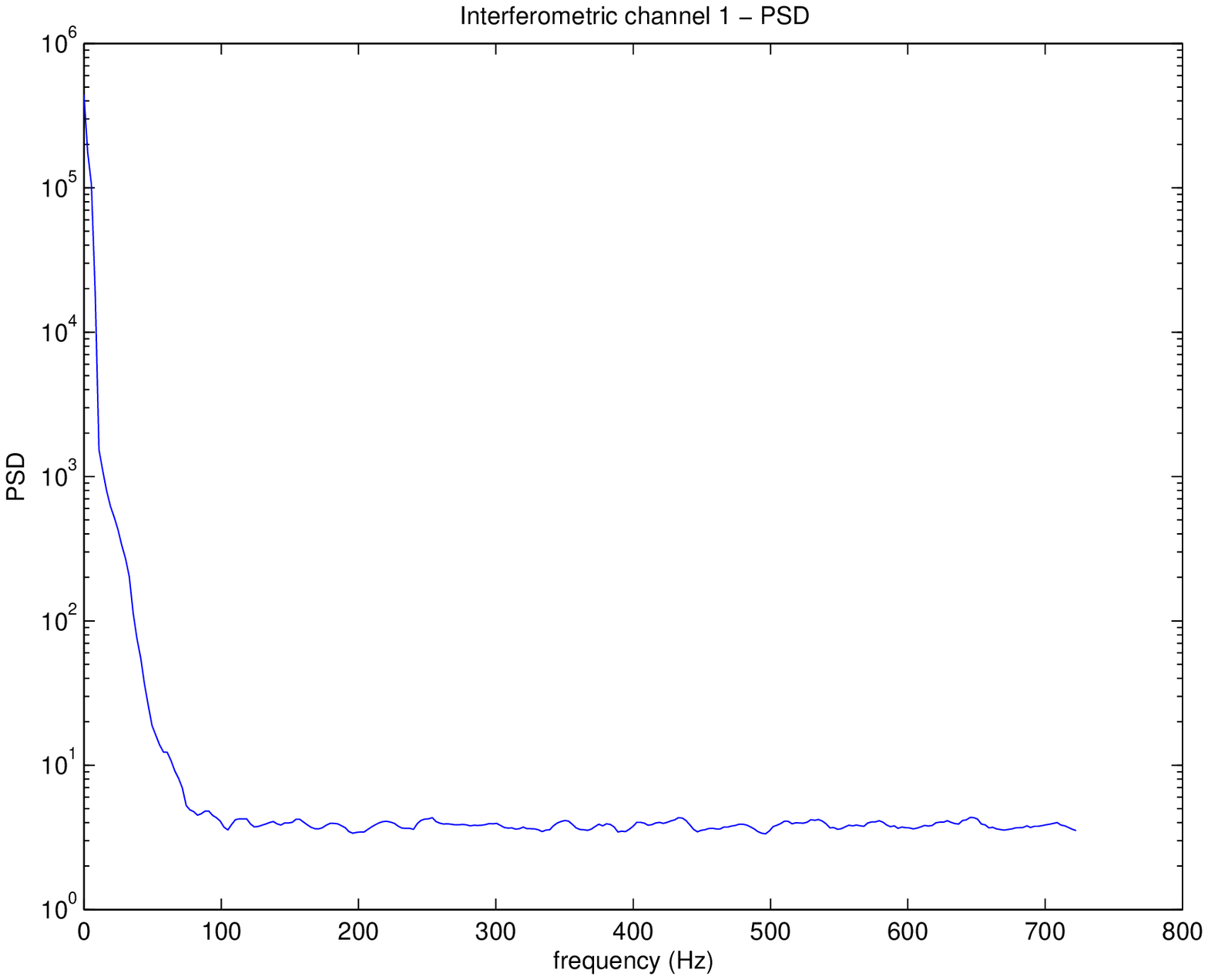,width=6.5cm}
    \epsfig{figure=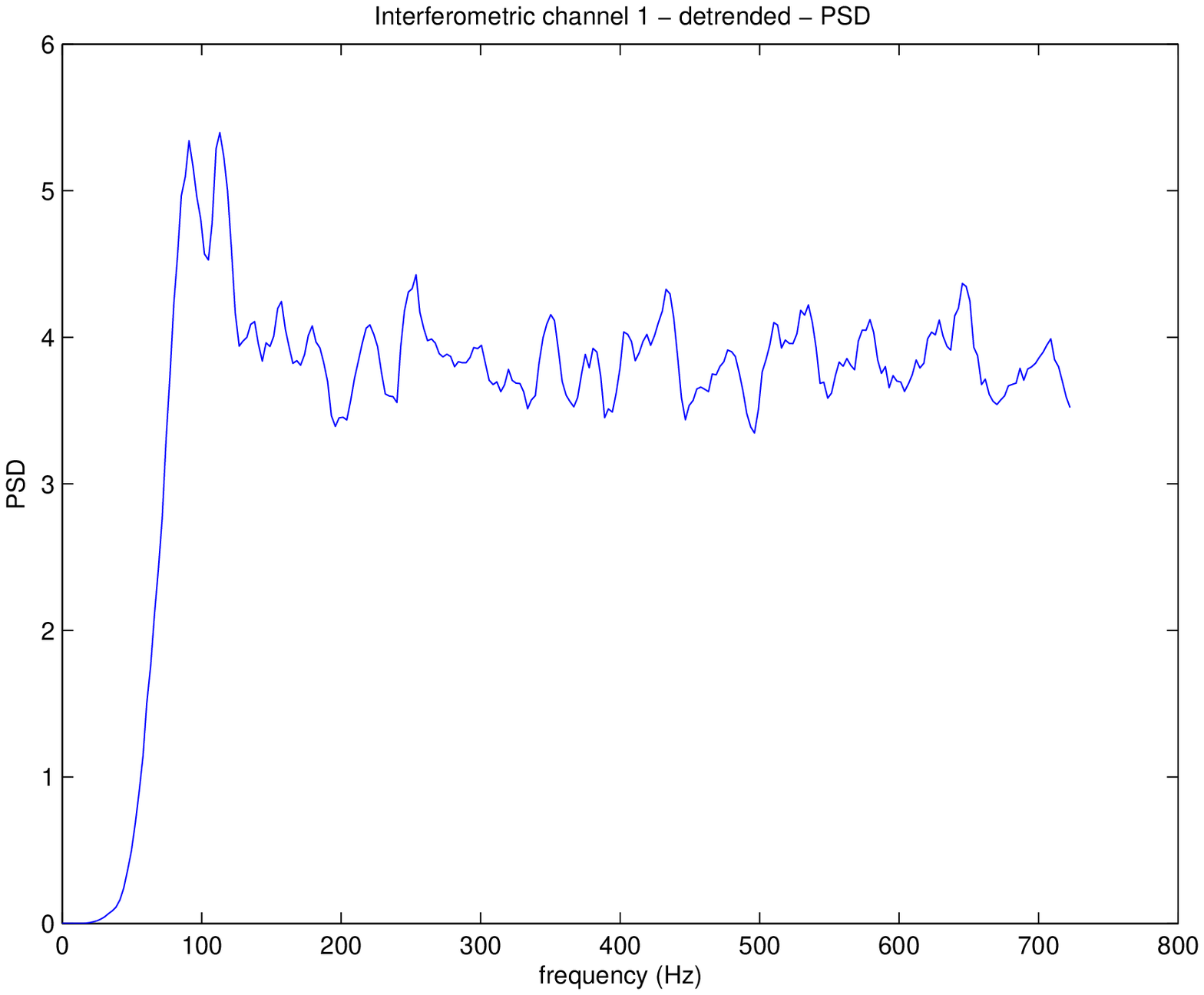,width=6.5cm}
    \caption{Case 2: Power Spectral Density functions for photometric input $PA$ (first row) and interferometric output $I1$ (second row): raw data (left) and detrended (right).}
    \label{fig:calOutput-PSD}
    \end{center}
\end{figure*}

\noindent We compare with the PSD evaluated for case 2 and 3. Figure \ref{fig:calOutput-PSD} shows photometric $PA$ and interferometric $I1$ for case 2. We can notice that the two low-frequency peaks are no longer present, not even in the photometric channel. This fact can be interpreted in two ways: it is due to a cross-effect between the two beams when injected into the instrument, and {\it before} being separated in two parts, or it was merely an observational noise, feature of that set of data, and not repeatable.

\subsubsection{Effects of the application of the spectral window}
\label{subsubsec:hamming}

\begin{figure*}[htb]
    \begin{center}
    \epsfig{figure=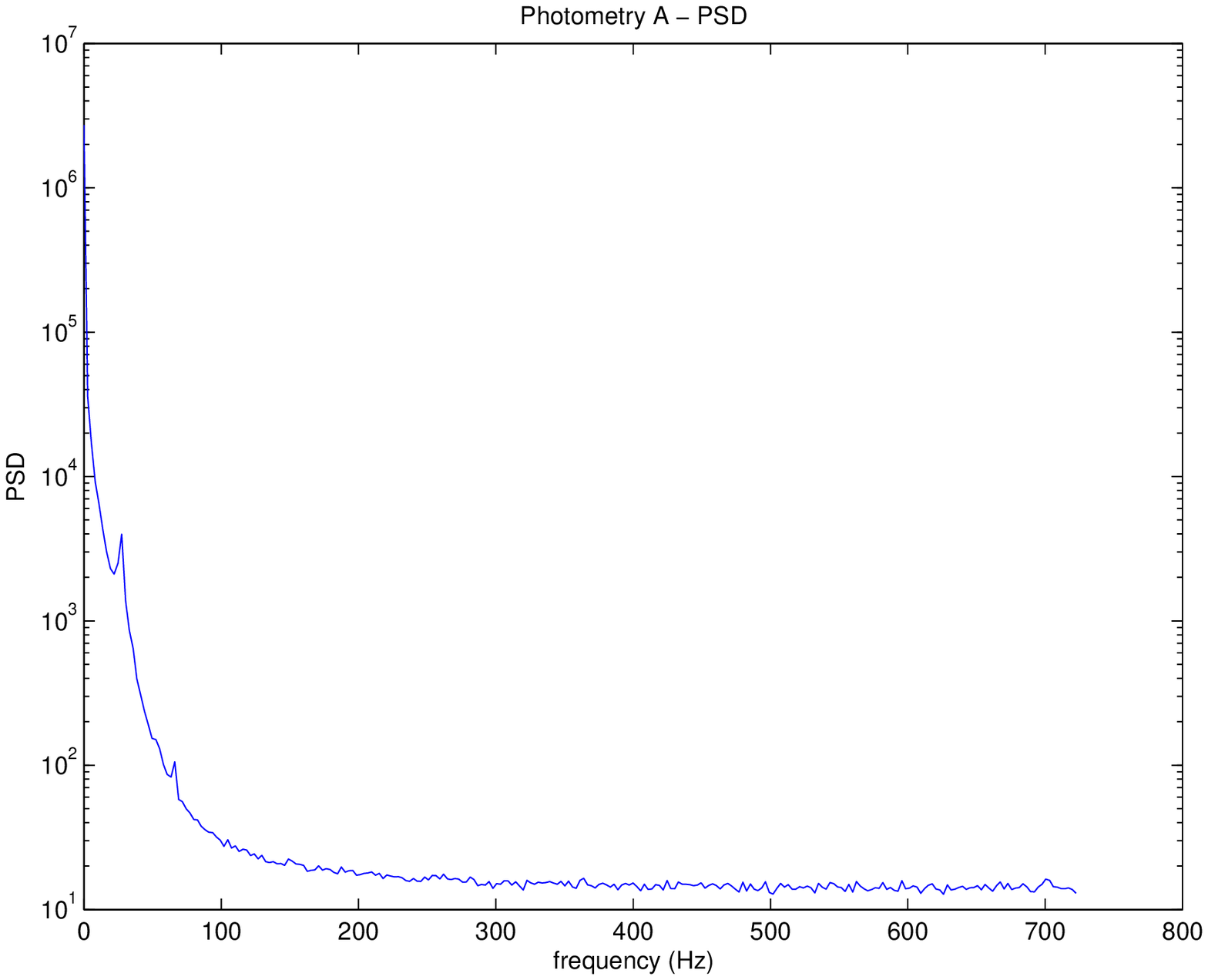,width=6.5cm}
    \epsfig{figure=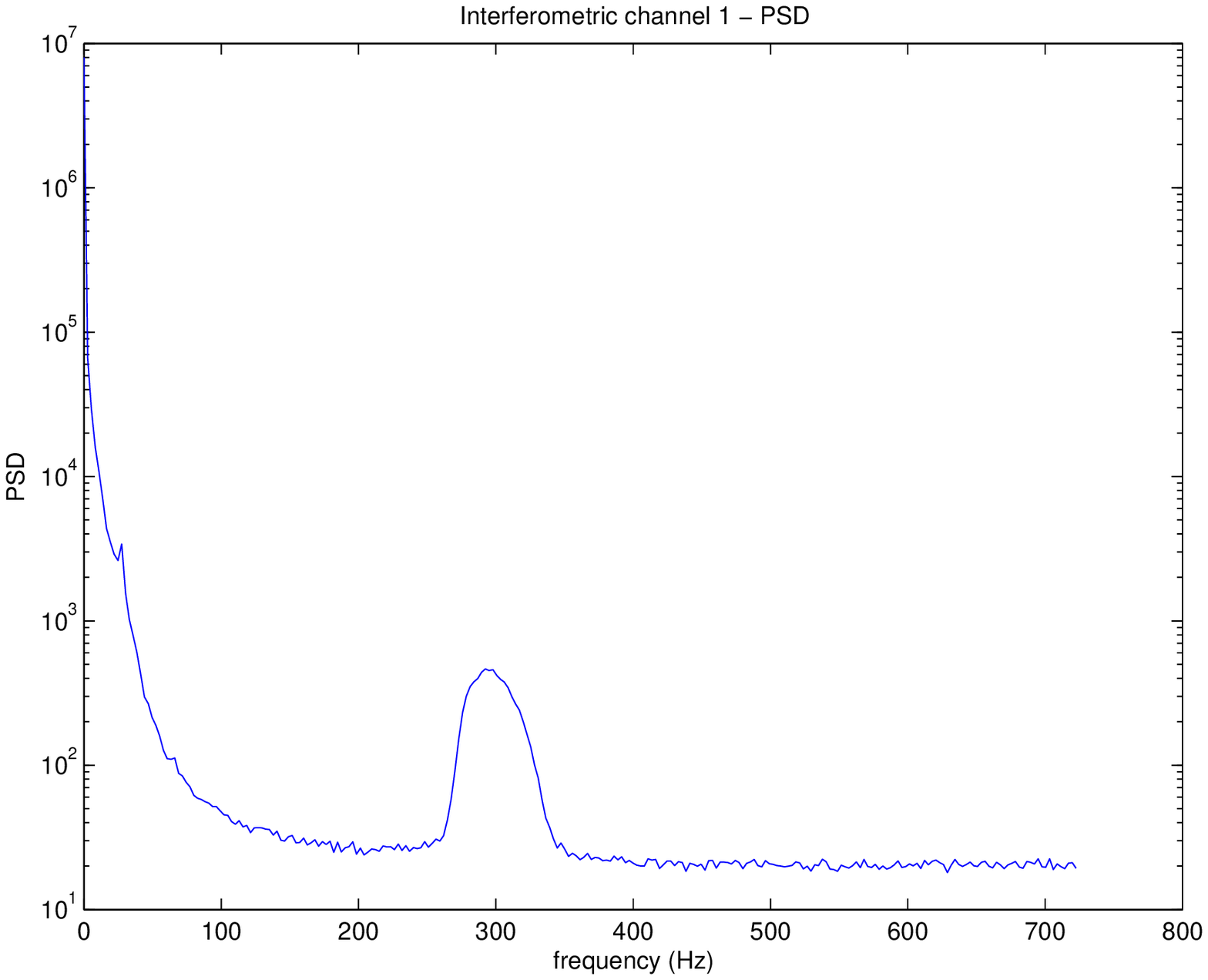,width=6.5cm}
    \caption{Case 4. Power Spectral Density estimation for the photometric channel A (left) and for the interferometric channel 1 (right) without the Hamming window.}
    \label{fig:all_PSD_noHam}
    \end{center}
\end{figure*}

To understand the role of the spectral window, we explore the behaviour of the averaged periodogram without the application of the Hamming window. We notice, comparing the second row of figure \ref{fig:all_none_PSD} with figure \ref{fig:all_PSD_noHam}, that the windowing has the effect of whitening the estimated PSD, sharpening its drop toward the flat offset, and to reduce the offset of the high-frequency white noise, as expected.

\noindent It is interesting to notice that the window effect on the PSD estimation depends on the frequency, since its effect is greater at low frequencies, where the power magnitude is higher. This is probably due to the characteristic of the spectrum of the window, since the multiplication of two functions in the time domain corresponds to the convolution of their Fourier transforms.


\subsubsection{Effects of the subtraction of an estimated trend on the PSD evaluation}
\label{subsubsec:trendSubtraction}

We have seen, in the previous paragraphs, how the slow evolution of the flux observed on the photometric channels has a strong impact on the features of the signals, both in the temporal and in the frequency domain. We have remarked how the estimation of this trend had to be carefully handled, because it is changing in time. \\
We must underline that what we subtract from the raw signal is just an {\it estimation} of the trend of the data, and we can expect that the features of this estimate reflect on all other estimated functions. We have analyzed in details this problem when the trend adds up to a wide sense stationary process, i.e. a stochastic process whose first and second order moments do not depend on time. We have given the error on the expectation of the PSD in the simple case of a constant trend subtraction, and an asymptotical value for this expectation in a general case (see appendix \ref{appendixA}).\\
Here we just mention the result of interest in our situation. Let us suppose that the signal can be expressed as the sum of a trend and a residual process. Let the residual process be such that it possesses a continuous spectral description with density $f(\omega)$, where $\omega$ is the frequency variable. If the trend functional form $t=g(x)$ is such that the $x$ variable has suitable properties (i.e. the regressors have no upper bounds, they increase slowly, they have a correlation matrix which is non singular for the zero lag), the regression estimate of the trend coefficient is the best linear one. Moreover, the asymptotical behaviour of the bias can be formulated and depends on the window size and on $f(\omega)$. Of course, it is necessary to have prior information on the residual process.\\

\subsection{Allan variance}
\label{subsec:allan}

\begin{figure*}[htb]
    \begin{center}
    \epsfig{figure=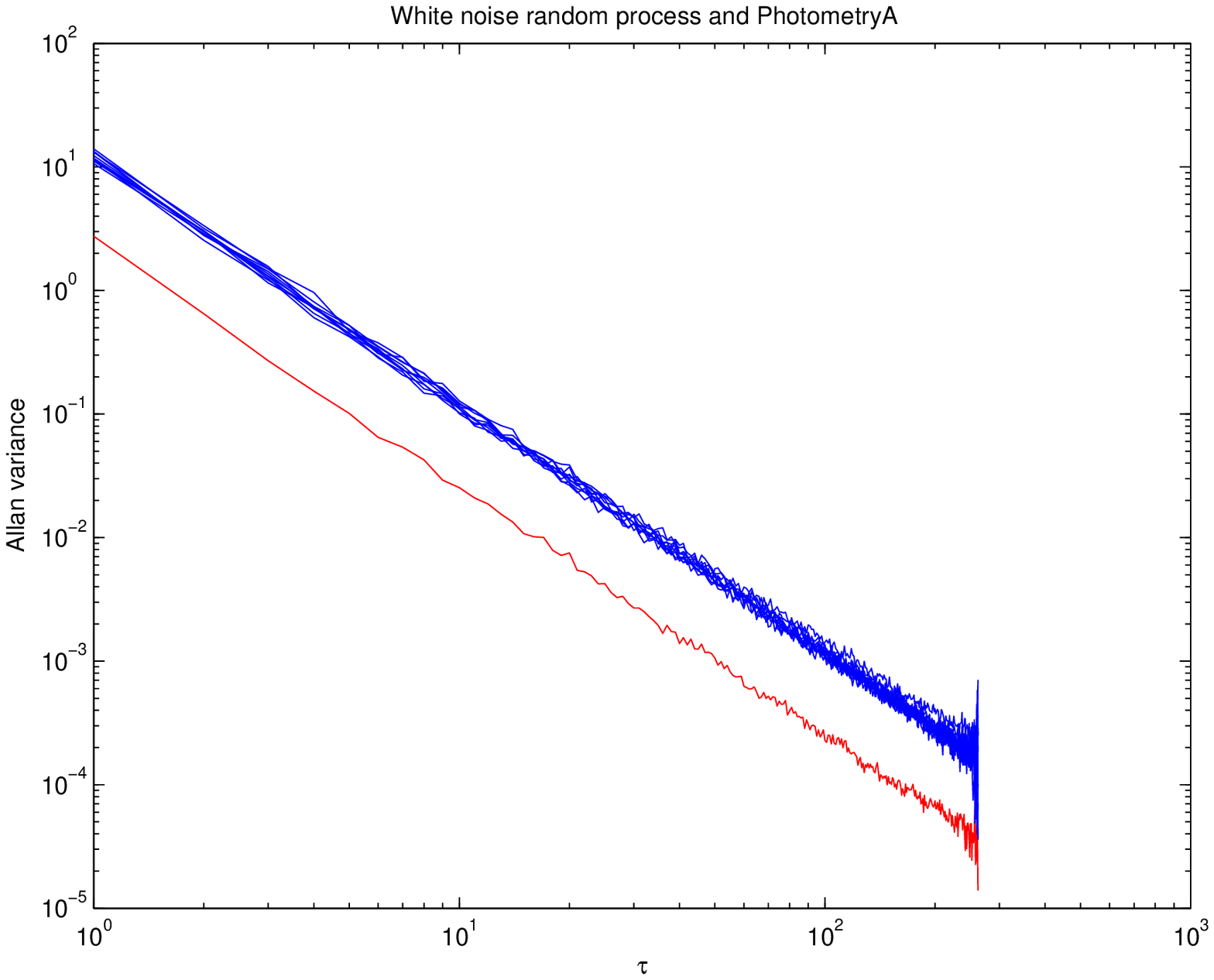,width=6.5cm}
    \epsfig{figure=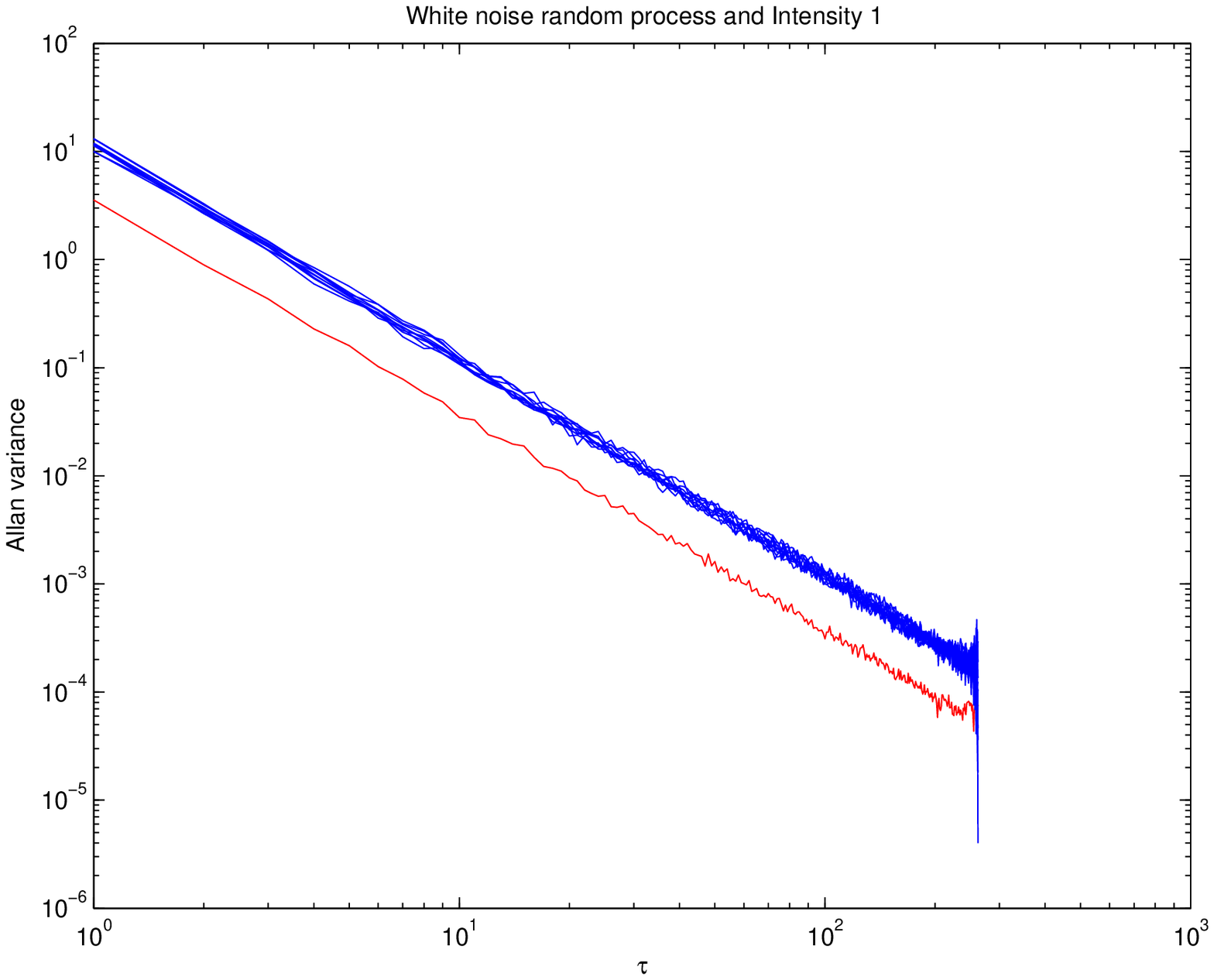,width=6.5cm}
    \caption{Case 1. Allan variance comparison between a realization of a gaussian white noise (red) and ten records (blue) of the photometric channel $PA$ (left) and the interferometric channel $I1$ (right).}
    \label{fig:allan_none}
    \end{center}
\end{figure*}

\begin{figure*}[htb]
    \begin{center}
    \epsfig{figure=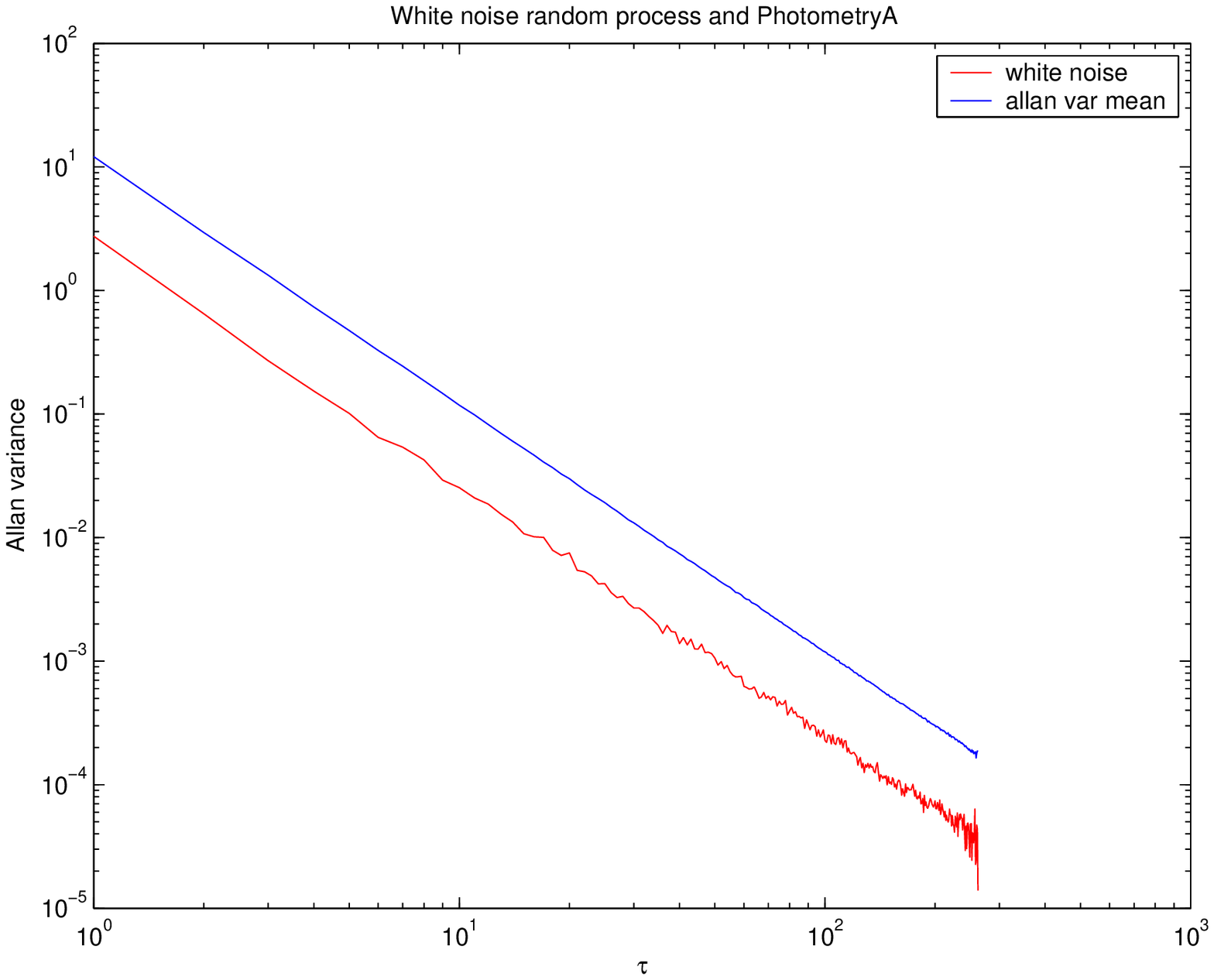,width=6.5cm}
    \epsfig{figure=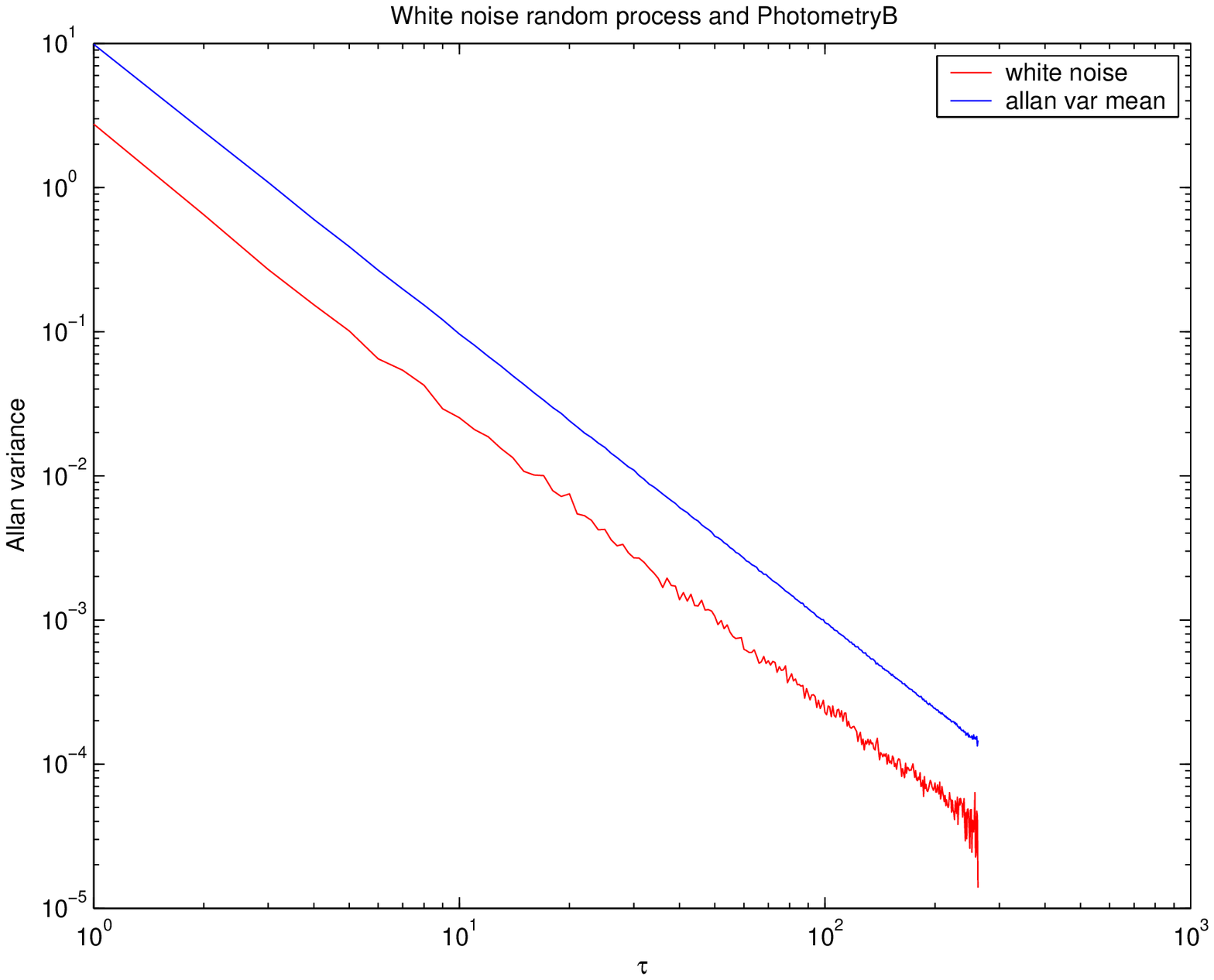,width=6.5cm}
    \epsfig{figure=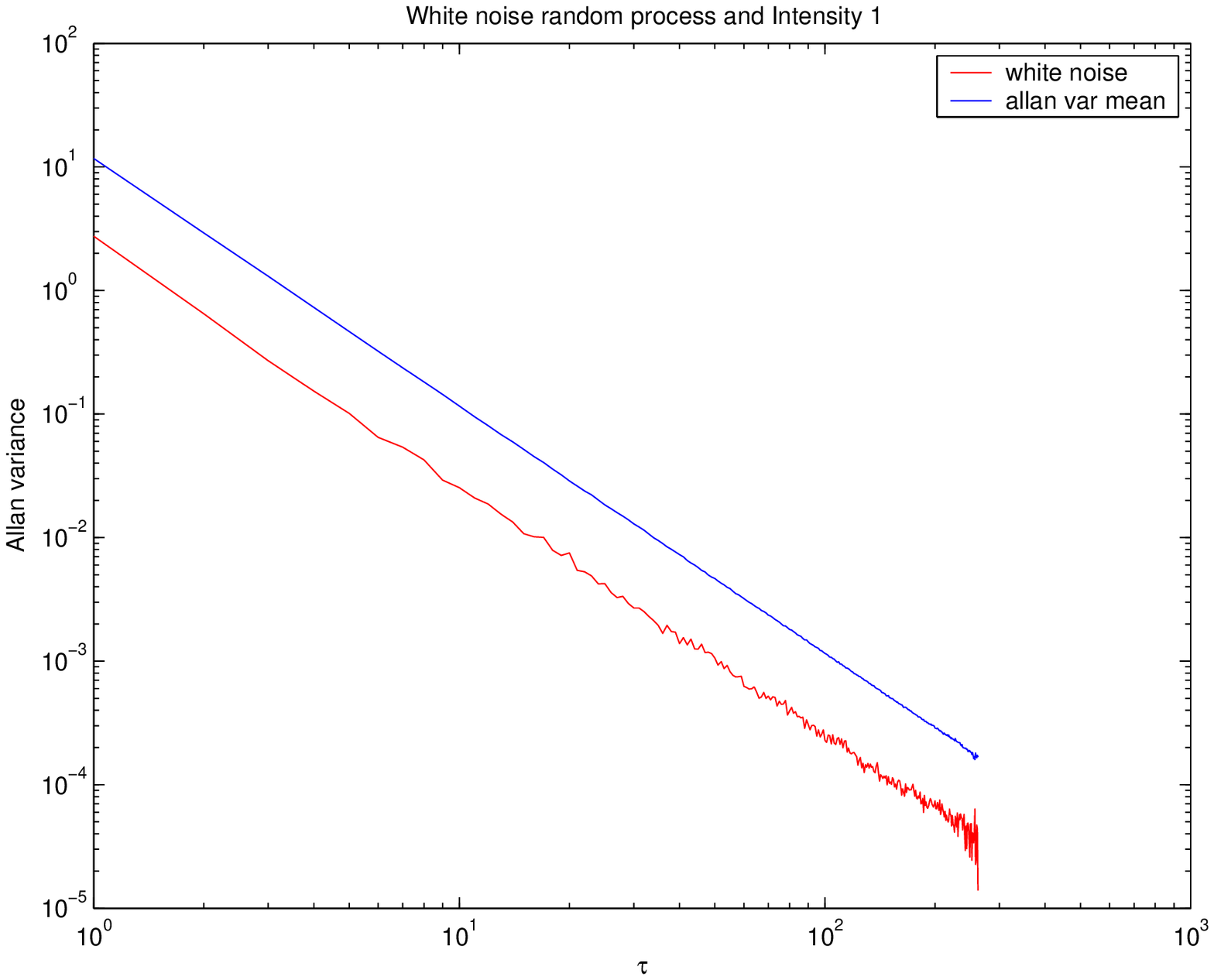,width=6.5cm}
    \epsfig{figure=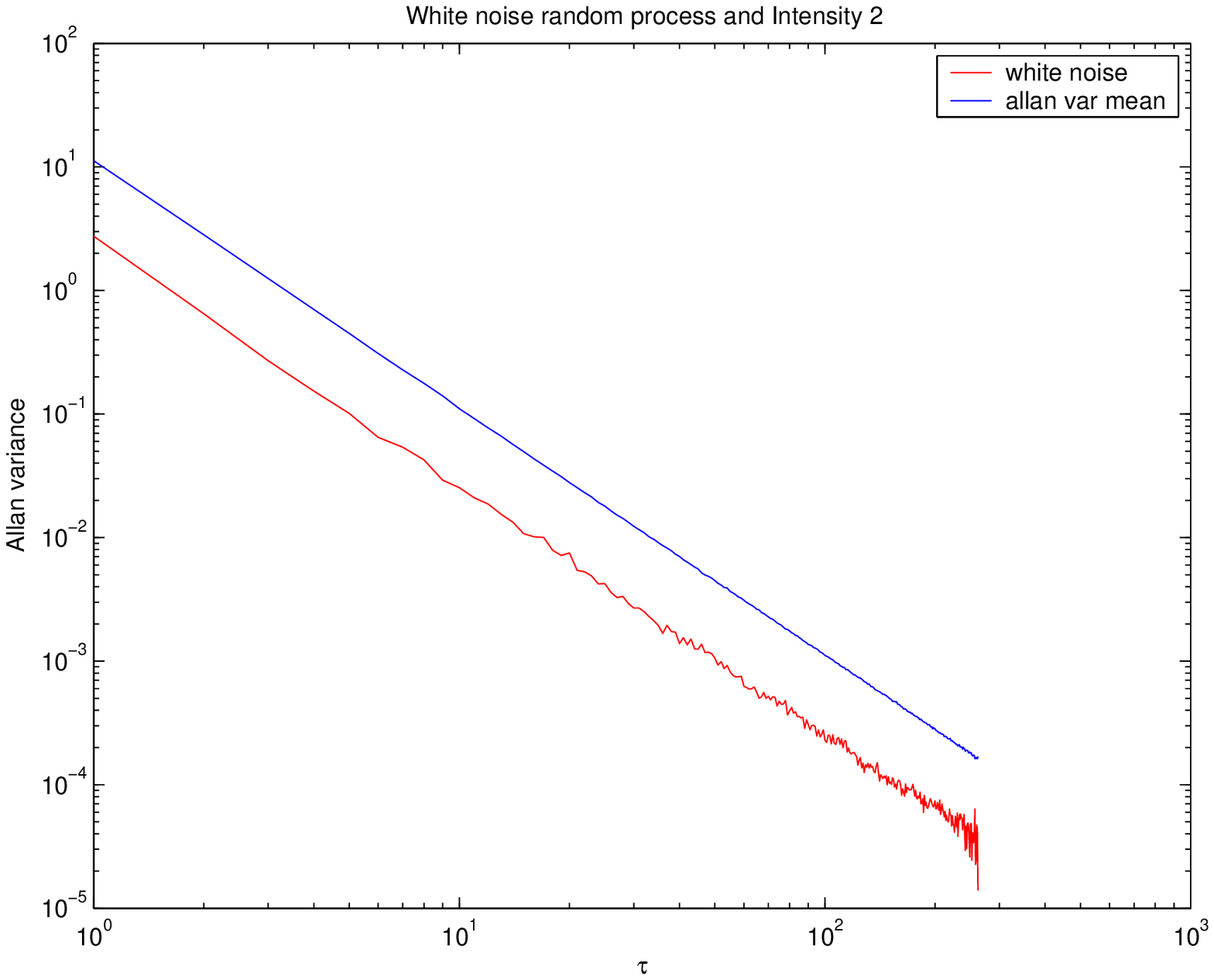,width=6.5cm}
    \caption{Case 1. Allan variance comparison between a realization of a gaussian white noise (red) and ten records (blue) of the photometric channels (first row) and the interferometric ones (second row).}
    \label{fig:allan_none_mean}
    \end{center}
\end{figure*}

We would like to confirm our spectral results using the Allan variance tool.\\
Let us analyze data referred to case 1.
We first generate a realization of a family of random variables, each distributed as standard gaussian r.v. $N(0,1)$. The family dimension is $526$ samples, in analogy with each record analyzed.
We then compare the Allan variance of this family with each record of the inputs channel $PA$ and $PB$ and of the output channels $I1$ and $I2$.
As we could expect from the spectral analysis, we find strong similarities between the gaussian family and the astronomical data. Figure \ref{fig:allan_none} shows this behaviour of the void channels for some records ($10$) for a photometric channel, $PA$, and an interferometric one, $I1$. Due to the regularity of the records pattern, it is useful to consider the mean of all records for the different channels, and this averaged Allan variance is shown in fig. \ref{fig:allan_none_mean}, always compared with the reference white noise.

\begin{figure*}[htbp]
    \begin{center}
    \epsfig{figure=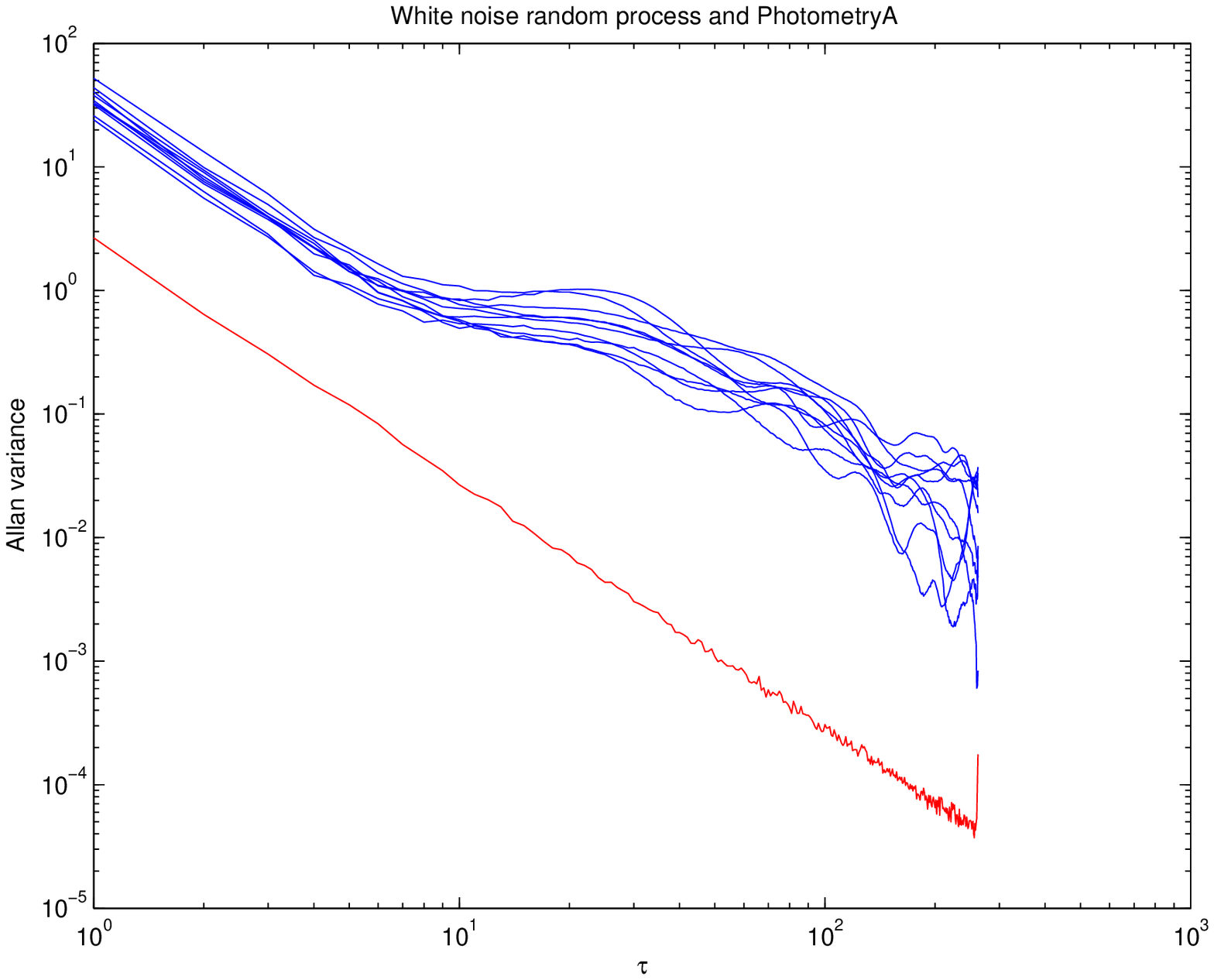,width=6.5cm}
    \epsfig{figure=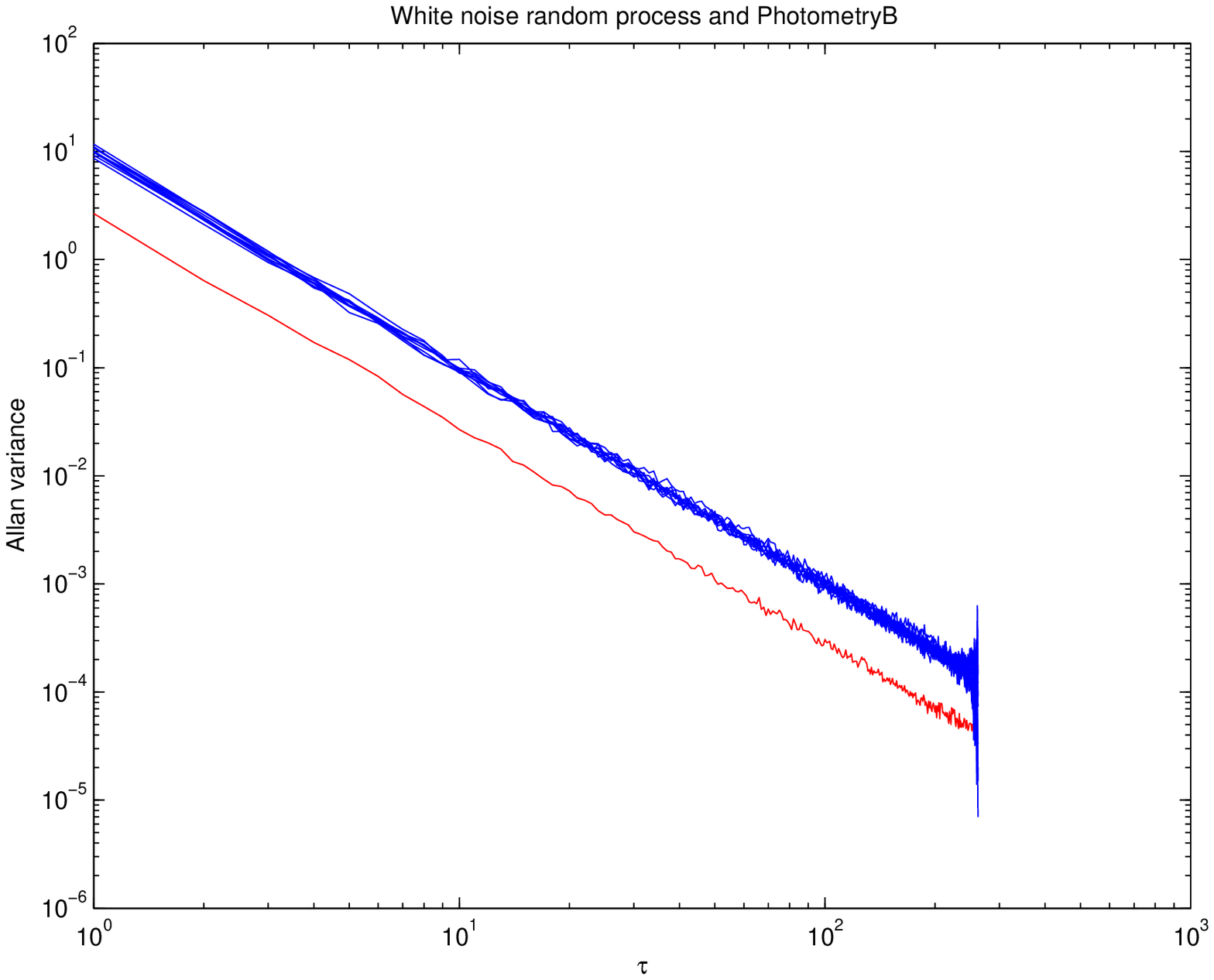,width=6.5cm}
    \epsfig{figure=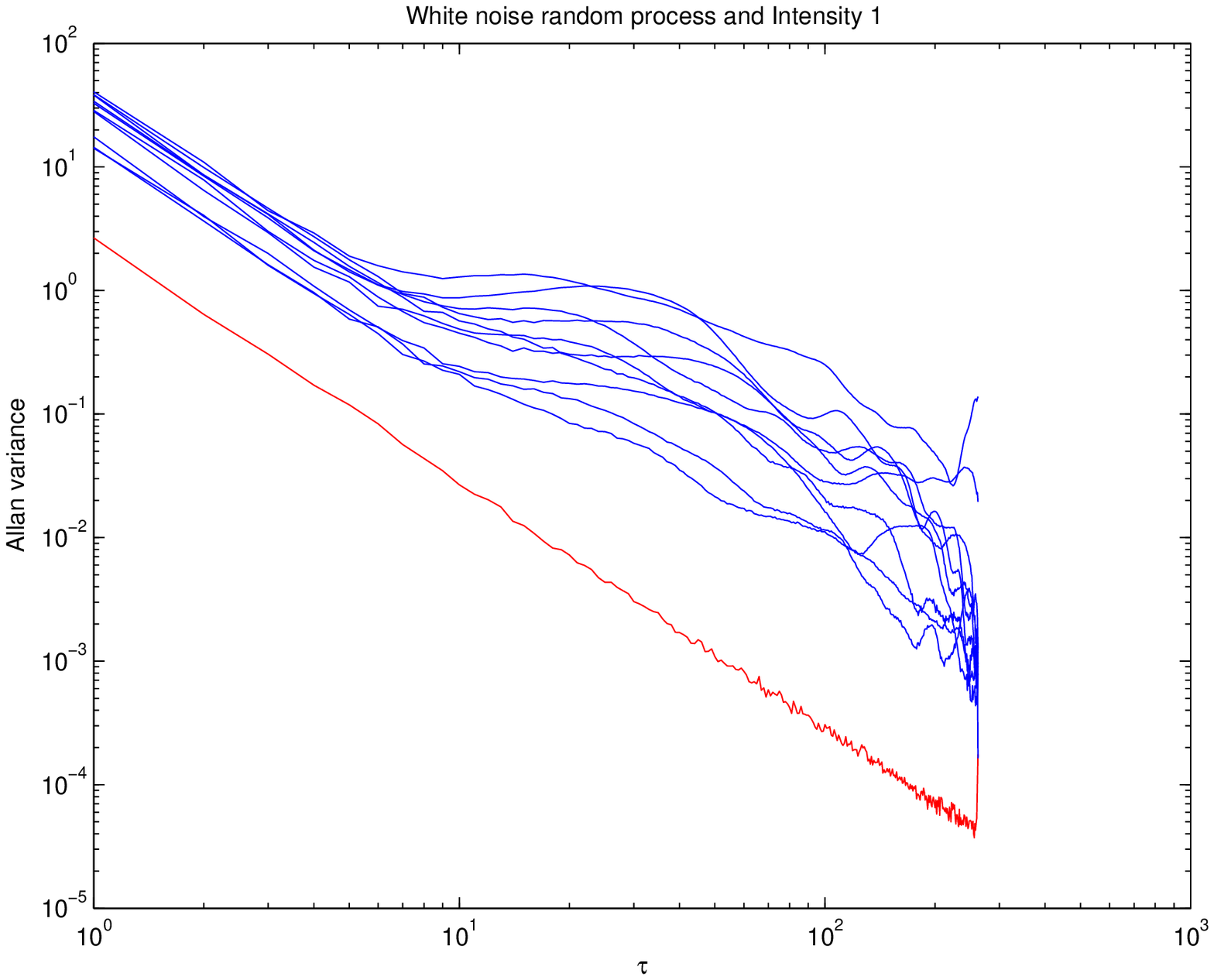,width=6.5cm}
    \epsfig{figure=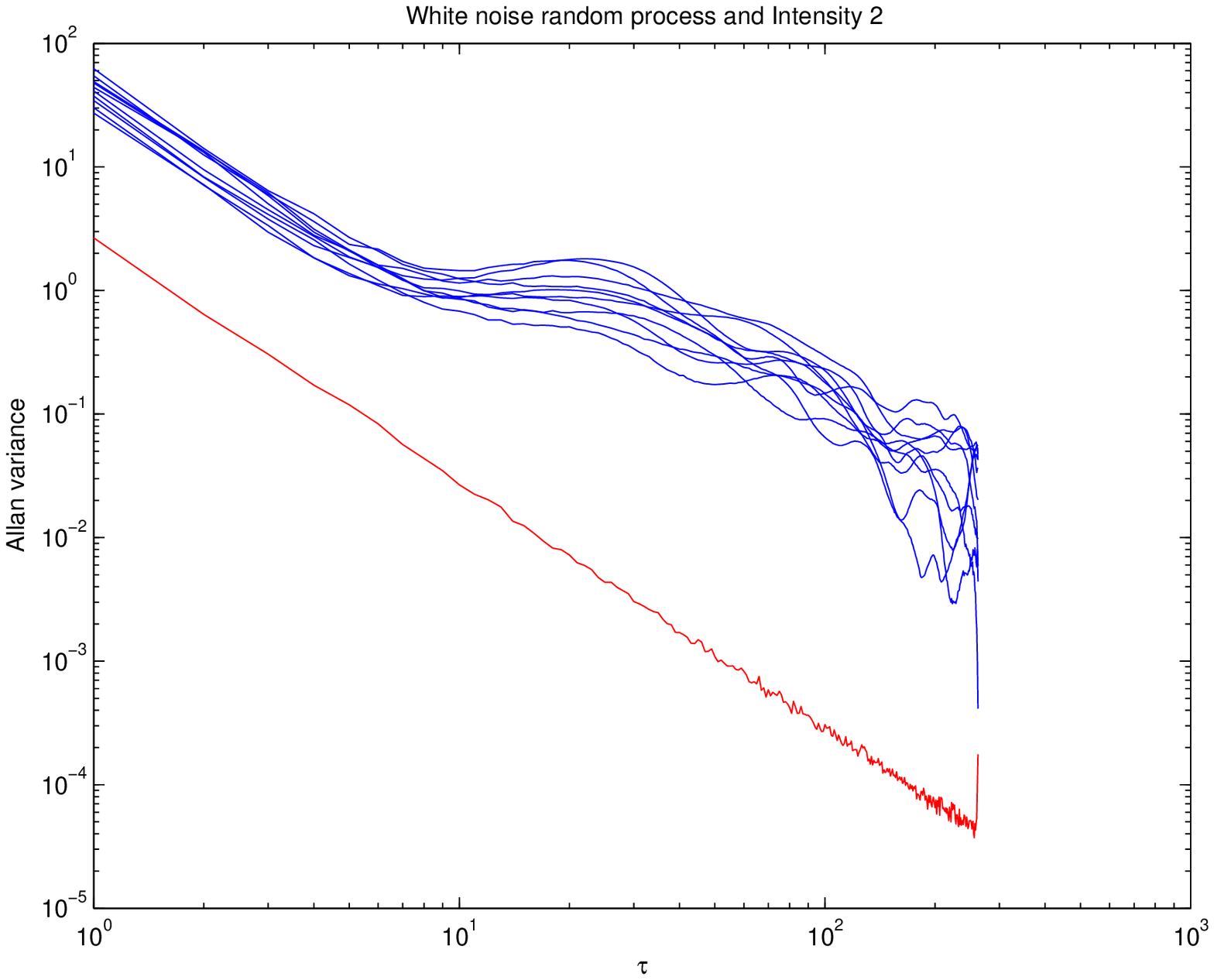,width=6.5cm}
    \caption{Case 2. Allan variance comparison between a realization of a gaussian white noise (red) and ten records (blue) of the photometric channels (first row) and the interferometric ones (second row) for case 2. Flux is injected in channel $PA$ (first row, left), while $PB$ is void (first row, right). Interferometric channels are both interested. We remember that there is no interference pattern.}
    \label{fig:allan_chA}
    \end{center}
\end{figure*}

\noindent The situation changes when we consider data from case 2 and 3, i.e. when just one photometric channel is fed with flux, while the other is left void. Figure \ref{fig:allan_chA} shows, for each picture, ten records of a channel compared again with the gaussian white noise described before. Since the patterns are not regular between records, we do not average. We can however recognize a common pattern: the first part of each variance function is very similar to the reference white noise, and then the pattern changes. Moreover, the intercept of the variance line changes from record to record, and this is caused by a changing variance value (just remember that this value is an average value over all intervals of a certain dimension).

\noindent We can then conclude that locally the signal acts like a gaussian white noise, over moving intervals of about $10$ samples, than other features appears.
Samples separated by a lag $\tau < 10$ can be considered uncorrelated, while for higher lags ($\tau > 10$) the $\tau$ exponent changes dramatically. We can recognize, for certain records, a straight line for intermediate lags ($10<\tau<40$), that is in relation with a noise with spectrum $\propto 1/f$. Some patterns show a curve that could indicate an underlying sinusoidal process, quite strange in photometric signals.\\
For higher lags ($\tau>40$), there is not a clear indication of a known pattern. The final lags are dominated by fluctuations caused by the variance estimation over a poor number of terms.
\\

\begin{figure*}[htbp]
    \begin{center}
    \epsfig{figure=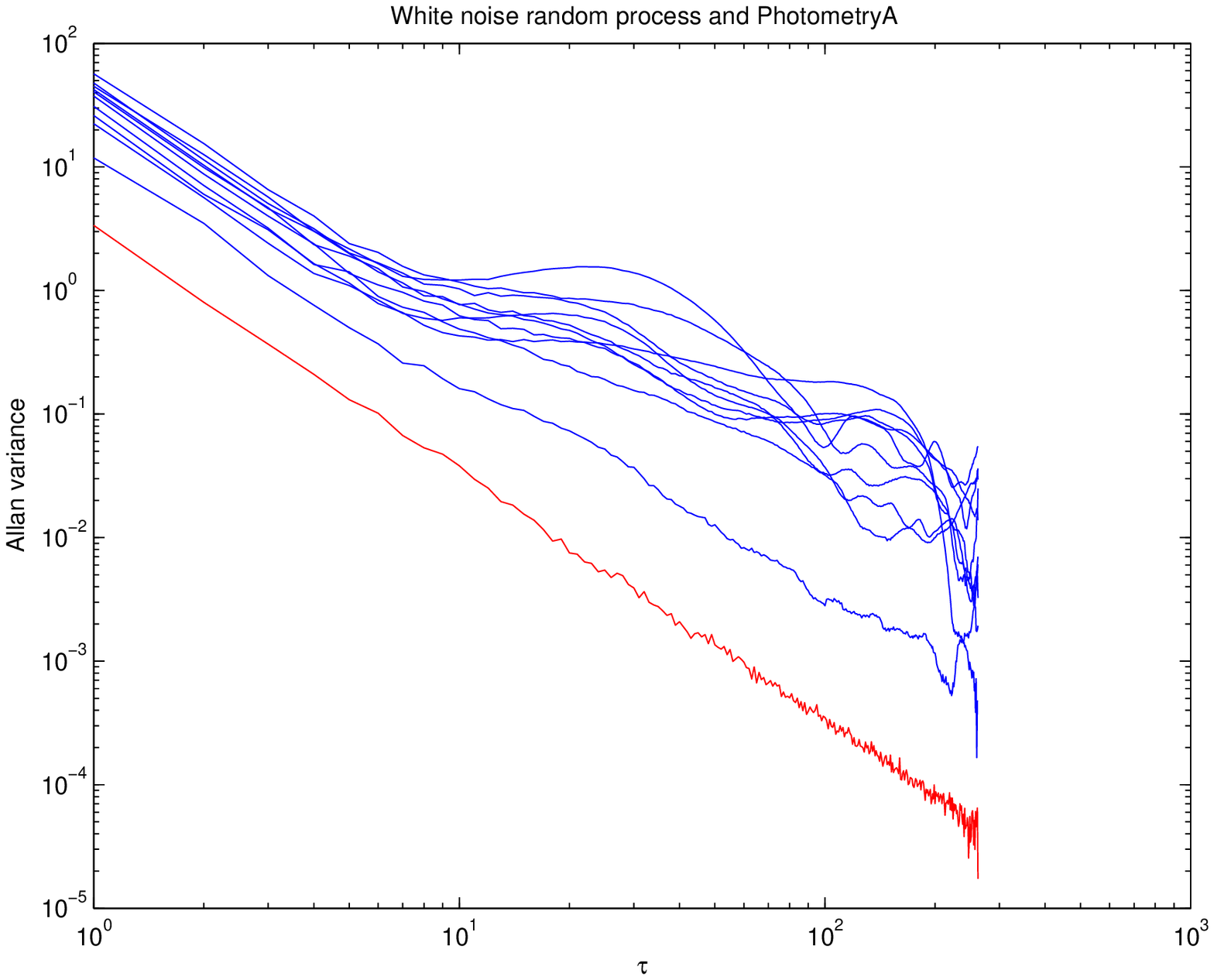,width=6.5cm}
    \epsfig{figure=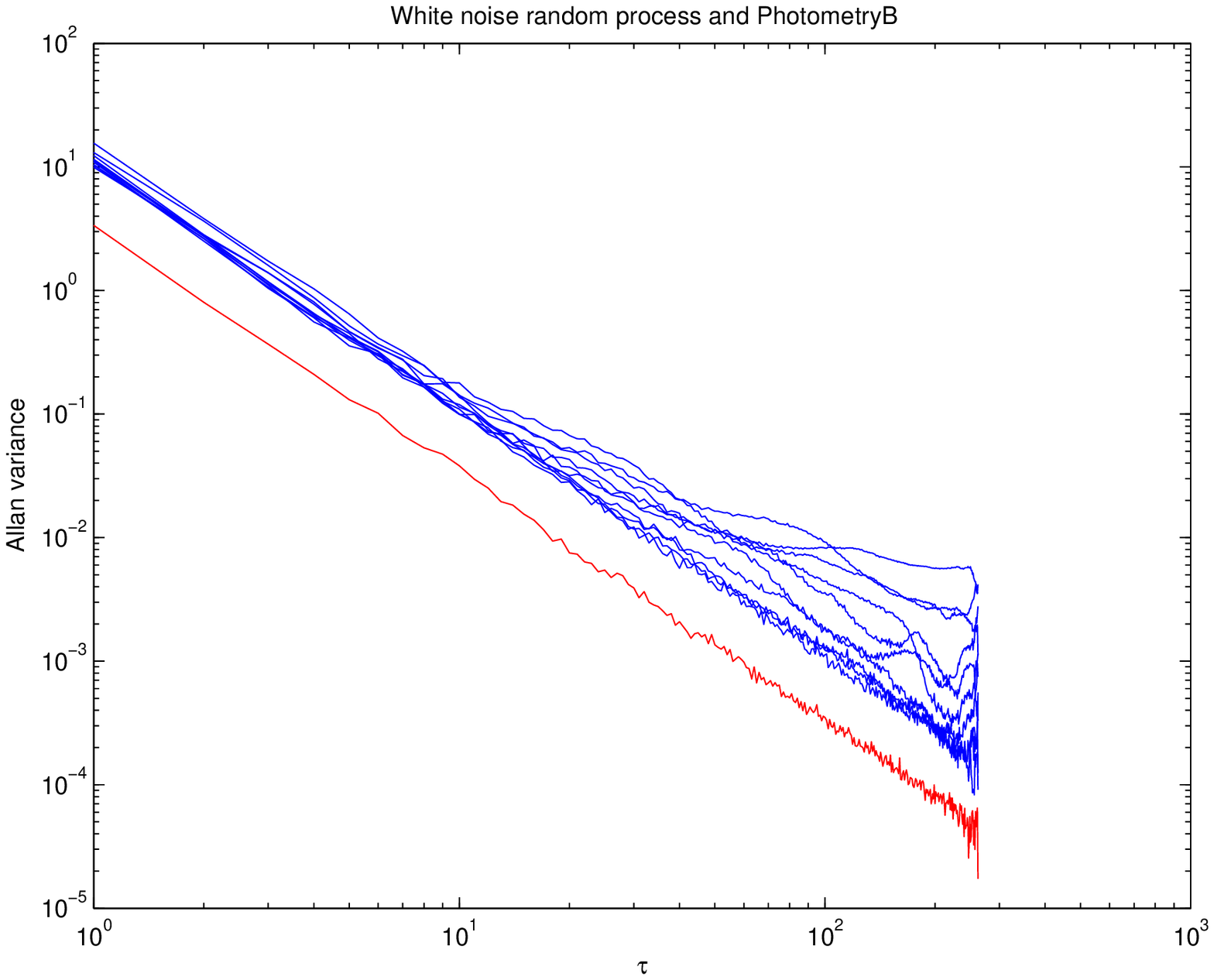,width=6.5cm}
    \epsfig{figure=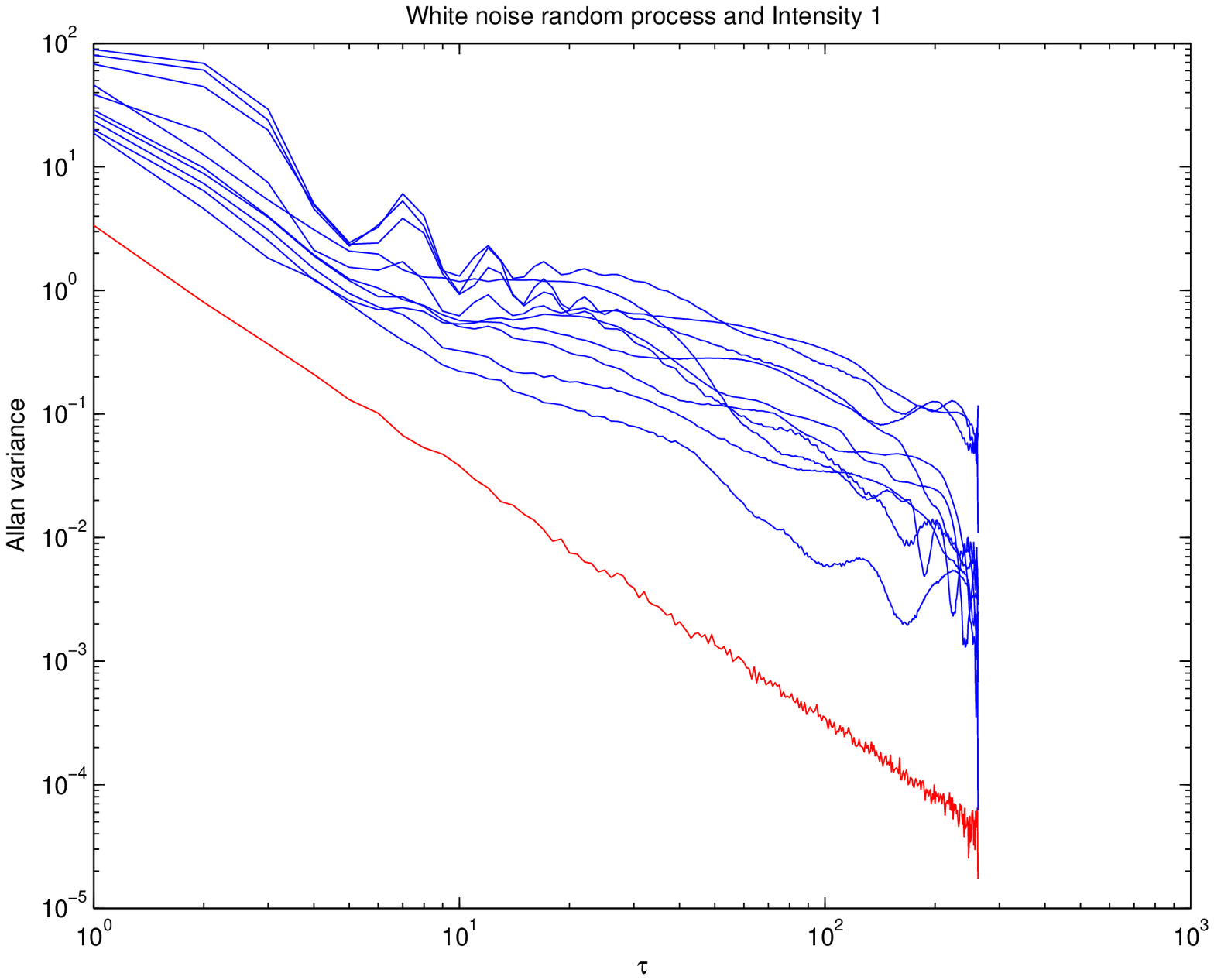,width=6.5cm}
    \epsfig{figure=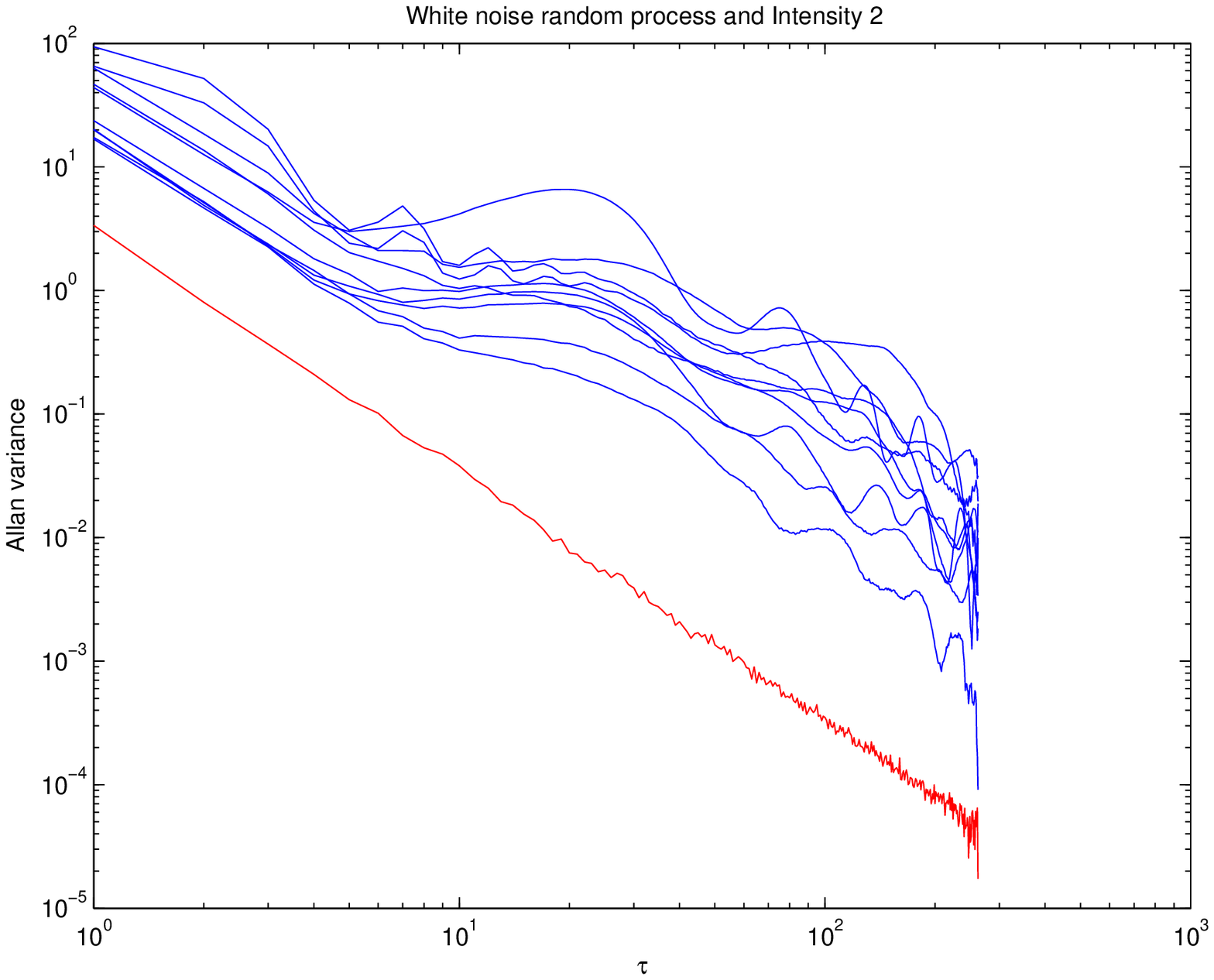,width=6.5cm}
    \caption{Case 4. Allan variance comparison between a realization of a gaussian white noise (red) and ten records (blue) of the photometric channels (first row) and the interferometric ones (second row).}
    \label{fig:allan_all}
    \end{center}
\end{figure*}

\noindent Finally, case 4 of observational data is interesting. It is shown in figure \ref{fig:allan_all}. For the photometric channels (first row), the same considerations as before apply. Moreover, we can say that the photometric channel $PB$ behaves differently from $PA$, its samples seems to be less correlated. This could mean that the different travel of each beam before the combination induces perturbations that are different not only for the magnitude, but also for their statistical properties, such as correlation.

\noindent For the interferometric channels, however, we can notice several things. First of all, for some records it is not possible to find a white-noise-like behaviour, even at low lags. The presence of the modulation is recognizable thanks to the oscillations of the Allan variance. These records are the ones in which fringes effectively formed. \\
Other records, in which fringes are not present, behave like interferometric channels of case 2 and 3. We made the correspondence between variance patterns and fringes formation records by visually comparing them.
\\

\noindent Optimization of this basic evaluation of the Allan variance, such as the Dynamic Allan Variance\cite{Galleani}, should help in identifying also {\it when} different patterns appears or when they are covered by other effects.



\noindent We can finally conclude that the analysis of the Allan variance for these data is useful, because it allows to recognize the scale at which the different types of noise appear.\\
An extensive study, based on a large amount of data in different working conditions, should help to identify these different noise sources, how often they appear, and to search their influence on instrument performances, such as the OPD/GD estimator algorithms we have seen in the previous chapters.


\section{Conclusions for statistical analysis}
\label{sec:stat_concl}


\noindent In the previous paragraphs, we have analyzed in details typical interferometric data with statistical classical techniques, first in time then in frequency domain. We are now able to resume the most important features of our data. \\

\noindent First of all, if we consider the calibration data, we can see that the peculiarity of signals before combination are maintained even after.\\
\noindent We can recognize two different components on the signals: a linear `trend', that contains the macroscopic fluctuations, and a residual variability, that we can define `microscopic'. The former is the responsible of the presence of a slowly decreasing autocorrelation (and cross-correlation between different channels). In fact, if we subtract it from the raw data, all auto/cross-correlations drops immediately to zero, apart lag zero.
The latter component is an uncorrelated process.\\

\noindent This results are confirmed by the spectral analysis. The power spectral densities are affected by low-frequency components that can be linked to the slow-moving trend on data. Once this trend is subtracted, the PSDs confirm that the residuals are uncorrelated signals, apart from frequencies around the zero, which have a pattern that could be a leakage of the zero-frequency components.

\noindent These considerations can appear in contrast with the results of the Allan variance, especially for the photometric inputs. In fact, this technique shows how photometric signals can be considered locally uncorrelated over $\approx$ 10-samples sized intervals.
On the contrary, since the Allan variance is based on the difference of samples separated by a certain time lag, for sufficiently close samples the trend can be considered constant, and it is eliminated by the subtraction. If the time lag is larger than the length of `stationarity' of the trend, it is no longer subtracted, and it induces fluctuations on the variance.\\

\noindent We are able to conclude that the the linear trend can be considered locally constant over an interval of 10 samples, which corresponds to 6.9 msec, or equivalently to $4.5 \; \mu m$. At the working wavelength of VINCI, i.e. $\sim 2.0 \; \mu m$, the linear trend is stable over two fringes.\\

\noindent These considerations imposes some constraints for the use of these data in a fringe sensor, such the need of preprocessing raw data to subtract the linear trend before applying location algorithms.\\

\noindent An Allan variance tool would be very helpful, either on-line or off-line, in the diagnostics of operating conditions of interferometric instruments and for data calibration.


\section{Variance analysis}
\label{sec:var_analysis}

\noindent In previous Sections we have shown that the signal is not stationary, and it is possible to isolate a trend. However, the detrended signal may still not be stationary. For this reason, we study here the evolution of the signal variance. We are no longer interested in signal mean, that we know to be characterized by the trend, so we use detrended data.


\noindent In particular, we are looking for two different tests:
\begin{enumerate}
    \item {given a selected channel, we search for changes of the variance as a function of time}
    \item {given a specific time record, we want to see if the properties of the variance of the input channels are the same of the variance of the output channels, in terms of homogeneity}
\end{enumerate}


\noindent Given the huge size of the data to be analyzed and its organization in a number of signals divided in hundreds of records, the use of a statistical software, such as Statistica, has required an effort to manage data, in order to suit the software requirements (organization of variables in groups, levels, repeated measures and so on).

\subsection{Statistical methods for variance analysis}
\label{subsec:variance_statMethod}

We study the variance of the signals using statistical tests, in particular we test variance patterns synchronously on different channels.
\noindent For the test of variance homogeneity, we use the Levene test, usually contained in the ANOVA analysis tool. Given a group of data sets, also called a level, the test distinguish between the variability of samples in the sets with the variability between different sets, to explain the total variability of all the samples.
In formula, given $k$ sets of samples, each with $N_i$ samples $z_{ij}, \; i = 1 \ldots k, \; j = 1 \ldots N_i$, and marginal mean $z_{i.}$, the test compares the variability between the marginal means and the overall mean $z_{..}$ with the sum of the variability in each set. If the variances are equal, the ratio $F$ approximates $1$:
\begin{equation}\label{eq:levene_def}
     F = \frac{(N-k) \sum_{i=1}^k N_i (z_{i.}-z_{..})^2 }{(k-1) \sum_{i=1}^k \sum_{j=1}^{N_i}(z_{ij}-z_{i.})^2}
\end{equation}

\noindent The numerator is the variability between the sets of samples, while the denominator is the sum of the variability inside each set. The test evaluate the ratio {\it F}, and give a statistical significativity {\it p}.
If the p-value {\it p} is under a specified threshold, the difference between the variances cannot be attributed to a random effect, but it is likely to have been generated by a true variability with a confidence level $(1-$ threshold$)$, so we should reject the hypothesis of homogeneity of variances.
\\

\noindent With our data, the level is composed by $k = 100, 500$ set (calibration sets and observational sets, respectively), while the number $N_i$ is the same for each $i$: $N_i = 526,376$ (the difference for the observational sets is to avoid the presence of fringe, that can disturb the variance evaluation). The two tests proposed foresee a different data organization, that will be described in the relative paragraphs.
\\

\noindent Some authors (Glass and Hopkins \cite{hopkins}) have pointed out that the Levene test and its modification (such as Brown-Forsythe) are based on the variance homogeneity requirement; a lack of symmetry in the distribution of the deviation from the means, for example, can cause a violation of the normality required for the F test. They highlighted that it is not clear if these tests are robust against a great heterogeneity of variances and sets with a different dimension.\\
In our tests, however, the significant number of samples in each set and the fact that sets are equally dimensioned should prevent us from misinterpretation of the results of Levene test.
\\

\noindent In our tests, we choose a confidence level of $95\%$.

\subsection{Analysis of homogeneity of variance}
\label{subsec:levene}

\subsubsection{Levene test 1}
\label{subsubsec:levene-test1}

\noindent For this test, given a selected channel, we want to know if the variance at a time $t_1$ is equal to the variance at time $t_2$.
So we have the following null and alternative hypothesis:
\begin{eqnarray}\label{eq:hyp_testLevene1}
    \nonumber H_0 \;&:&\; \sigma_1(t_1)) = \sigma_1(t_2), \;\;\; 1 \leq t_1, t_2  \leq 100 (500) \\
    H_1 &:& \exists t_1, t_2 \;\;\; s.t. \;\;\; \sigma_1(t_1)) \neq \sigma_1(t_2)
\end{eqnarray}

\noindent The dependent variable is the channel, while the group variable able to distinguish between different sets is the record number, ranging in $[1,100]$ for calibration sets and $[1,500]$ for observational ones.
The record variable creates $100$ or $500$ groups at the same level. In each group, the variance is evaluated and compared with the variance of all other groups. We have just one result in each channel, saying if the variance is changing. The p-value explains the significance of the result: since we have chosen a confidence level of 95\%, if the p-value is $<0.05$, we can't accept the hypothesis of variances equality.

\begin{enumerate}
\item[Case 1.]
{If the two arms are not fed with flux, both photometric and combined channels are stable with respect to the variance, since all p-values are above the critical value $0.05$, as table \ref{table:LeveneA_case1} shows.

\begin{table}[h]
\centering
\begin{tabular}{lcccc}
  \hline
  Channel & MSEffect & MSError & F & p \\
  I1 & 1,372893 & 1,267551 & 1,083107 & 0,269069 \\
  I2 & 1,332706 & 1,221557 & 1,090990 &	0,252061 \\
  PA & 1,394501 & 1,292858 & 1,078618 &	0,279041 \\
  PB & 1,222665 & 1,059943 & 1,153519 &	0,141204 \\
  \hline
\end{tabular}
\caption{Levene test for Homogeneity of Variances - case 1, without flux}
\label{table:LeveneA_case1}
\end{table}
}

\item[Case 2 and 3.]
{If just one interferometer's arm is fed with flux from a source, we find that the only channel that maintains the variance homogeneity property is the void channel (p-value $>0.05$). Table \ref{table:LeveneA_case2} shows this result for the case 2 (channel A fed, channel B empty), while table \ref{table:LeveneA_case3} is similar, but for case 3.

\begin{table}[h]
\centering
\begin{tabular}{lcccc}
  \hline
  Channel & MSEffect & MSError & F & p \\
  I1 & 123,1234 &	3,399613 & 36,21689 & 0,000000 \\
  I2 & 306,9551 & 5,208085 & 58,93818 &	0,000000 \\
  PA & 197,0255 & 4,244176 & 46,42255 &	0,000000 \\
  PB & 0,9752 &	1,067226 & 0,91379 & 0,717938 \\
  \hline
\end{tabular}
\caption{Levene test for Homogeneity of Variances - case 2, channel A fed, channel B without flux}
\label{table:LeveneA_case2}
\end{table}

\begin{table}[h]
\centering
\begin{tabular}{lcccc}
  \hline
  Channel & MSEffect & MSError & F & p \\
  I1 & 90,19290 & 3,182288 & 28,34215 &	0,000000 \\
  I2 & 46,04302 & 2,364202 & 19,47508 &	0,000000 \\
  PA & 1,43348 & 1,284517 &	1,11597 & 0,202605 \\
  PB & 9,92143 & 1,470697 &	6,74608 & 0,000000 \\
  \hline
\end{tabular}
\caption{Levene test for Homogeneity of Variances - case 3, channel A without flux, channel B fed}
\label{table:LeveneA_case3}
\end{table}
}

\item[Case 4.]
{Finally, we considered the case 4 for which both input channels are fed with stellar flux. In this set of data, we have $500$ records instead of $100$; for homogeneity with the other cases the test is repeated over $100$ records at a time (tables from \ref{table:LeveneA_case4a} to \ref{table:LeveneA_case4e}). The results are consistent with those found before, i.e. all channels have inhomogeneous variance, since p-values are smaller than $0.05$.

\begin{table}[h]
\centering
\begin{tabular}{lcccc}
  \hline
  Channel & MSEffect & MSError & F & p \\
  I1 & 327,3119 &	8,94604 & 36,58735 & 0,00 \\
  I2 & 397,9197 & 10,00692 & 39,76444 &	0,00 \\
  PA & 157,4086 & 4,00455 &	39,30744 & 0,00 \\
  PB &	12,4603 & 1,36514 &	9,12750 &	0,00 \\
  \hline
\end{tabular}
\caption{Levene test for Homogeneity of Variances - case 4, channel A and B with flux, record from 1 to 100}
\label{table:LeveneA_case4a}
\end{table}

\begin{table}[h]
\centering
\begin{tabular}{lcccc}
  \hline
  Channel & MSEffect & MSError & F & p \\
  I1 & 143,8233 &  4,893480 & 29,39079 & 0,00 \\
  I2 & 313,9400 & 6,308566 & 49,76409 &	0,00 \\
  PA & 168,0298 & 4,04593 &	41,53058 & 0,00 \\
  PB &	1,7106 & 1,153894 &	1,48249 &	0,001 \\
  \hline
\end{tabular}
\caption{Levene test for Homogeneity of Variances - case 4, channel A and B with flux, record from 101 to 200}
\label{table:LeveneA_case4b}
\end{table}

\begin{table}[h]
\centering
\begin{tabular}{lcccc}
  \hline
  Channel & MSEffect & MSError & F & p \\
  I1 & 999,3393 & 22,03128 & 45,36003 &	0,00 \\
  I2 & 995,6715 & 22,337121 & 44,57475 & 0,00 \\
  PA & 211,2722 & 4,81973 &	43,83484 &	0,00 \\
  PB & 35,1254 & 1,64634 &	21,33541 &	0,00 \\
  \hline
\end{tabular}
\caption{Levene test for Homogeneity of Variances - case 4, channel A and B with flux, record from 201 to 300}
\label{table:LeveneA_case4c}
\end{table}

\begin{table}[h]
\centering
\begin{tabular}{lcccc}
  \hline
  Channel & MSEffect & MSError & F & p \\
  I1 & 668,3774 & 38,57011 & 17,32890 &	0,00 \\
  I2 & 639,9936 & 38,75130 & 16,51541 &	0,00 \\
  PA & 119,9992 & 5,25826 &	22,82107 &	0,00 \\
  PB & 25,0272 & 2,14108 &	11,68906 &	0,00 \\
  \hline
\end{tabular}
\caption{Levene test for Homogeneity of Variances - case 4, channel A and B with flux, record from 301 to 400}
\label{table:LeveneA_case4d}
\end{table}

\begin{table}[h]
\centering
\begin{tabular}{lcccc}
  \hline
  Channel & MSEffect & MSError & F & p \\
  I1 & 630,2230 & 44,64749 & 14,11553 &	0,00 \\
  I2 & 690,7584 & 44,31748 & 15,58659 &	0,00 \\
  PA & 135,0856 & 5,30639 &	25,45718 &	0,00 \\
  PB & 18,8933 & 2,22661 &	8,48522 &	0,00 \\
  \hline
\end{tabular}
\caption{Levene test for Homogeneity of Variances - case 4, channel A and B with flux, record from 401 to 500}
\label{table:LeveneA_case4e}
\end{table}
}
\end{enumerate}

\noindent We can conclude that the beam flux has a real variability along the time scale. This is of course reflected on the combined channels. The Levene test ensures us that this is a true inhomogeneity because it takes care of the changing mean value over different records.

\subsubsection{Levene test 2}
\label{subsubsec:levene-test2}

\noindent With this test, we want to assess if the variance in two channels changes over a fixed time interval $\Delta t$. So we consider the simultaneous records of two different channels, and we evaluate each variance, and then we compare them. We can repeat this procedure for all the records ($100$ or $500$). We remember that we are working, for these tests, with sample variance.
Now, the hypothesis are:
\begin{eqnarray}\label{eq:hyp_testLevene2}
    \nonumber H_0 &:& \sigma_1(\Delta t) = \sigma_2(\Delta t) \\
    H_1 &:& \sigma_1(\Delta t) \neq \sigma_2(\Delta t)
\end{eqnarray}
where $\Delta t$ varies along each record.

\noindent Data has been organized in order to have $100$ ($500$, respectively) variables, representing the repeated measures, each containing one record of the two channels under testing. The channel variable creates 2 groups at the same level, and the variances of these two groups are evaluated and compared.
This test can be repeated on both channel pairs, i.e. input and output.

\begin{table}[h]
\centering
\begin{tabular}{lcc}
  \hline
  Case & PA and PB & I1 and I2 \\
  1 & 40\% & 90\% \\
  2 &  2\% & 16\% \\
  3 & 66\% & 28\% \\
  4 & 5.2\% (26/500) & 65.8\% (329/500) \\
  \hline
\end{tabular}
\caption{Levene test for Homogeneity of Variances in two synchronous channels}
\label{table:LeveneB}
\end{table}

\noindent This test shows (table \ref{table:LeveneB}) an interesting feature of the combination system. In general signals after the combination are more balanced than before, in terms of flux intensity. This means that the combination/splitting system is able to sum up factors with different flux intensities and to split the sum into balanced part. Moreover, it says that when flux from a source is injected in both channels (case 4), beams coming in front of the combining system have very different amplitude.
Case 2 and 3, in which only one arm of the combiner is fed, are less interesting, even if we can say that when there is a strong unbalance between the two input channels, the combined outputs are less stable.

\section{Conclusions for variance analysis}
\label{sec:variance_concl}

The analysis of the variance of the VINCI data has evidenced the following features of the handled signals.
The first test has given us a statistical evidence of the fact that the variance of channels fed with stellar flux changes in subsequent records. This means that the flux is subject to variation in a single observation run (composed of different records), so the parameters of the interferometric models have to be updated at a high rate.\\

\noindent However, the second test has shown that the combination system is able to handle properly even unbalanced inputs, and to split them correctly in two equal part, not only in terms of mean, but also in terms of variance.\\

\noindent A test of this kind can be used to check instrument reliability and repeatability, for both on-line and off-line analysis. In fact, off-line analysis must include instrument parameter estimate at low level, and a real-time instrument like a fringe tracker must foresee an on-line update of operating parameters on a comparably short time scale, e.g. faster than 1 Hz. Suitable diagnostics modules have to be included to ensure that data quality is preserved throughout the observation.

\section{Analysis of interferometric output variability sources}
\label{sec:GLM}

Our further task is the analysis of the variability of the interferometric outputs. Some information can be retrieved from the interferogram itself, with statistical moments or autocorrelations, as we did before.\\
However, the availability of the system inputs, i.e. the photometric channels, allows to study the variability of the output beams as a function of that of the input beams. In this way, it is possible to investigate if the input variability is sufficient to explain the output one, or if we can suspect another variability source, perhaps from instrumental contribution.\\
This subject is addressed in this section.

\noindent To focus the problem, we have to make some assumption:
\begin{enumerate}
    \item{the photon noise, the shot noise and the detection noise have the same properties over each channel;}
    \item{the noise level is comparable inside and outside the coherence length.}
\end{enumerate}
The first assumption is needed because each signal, photometric or interferometric, is subject to the detection process independently from all others, and there is no way to check differences. The second is due to the fact that the modulated part of the signal has a strong variability that can not be considered as `variance'. So it is difficult to analyze the variance in the coherence length, and we must trust the results outside this area as applicable inside, in terms of noise estimate and characterization.
\\

\noindent The model we want to use is a simplification of that represented by eq. \ref{eq:PRIMAmodel}.
The two photometric channels, $PA$ and $PB$, can be factorized in the sum of the `true' values, $\tilde{PA}(x)$ and $\tilde{PB}(x)$ respectively, and of variability sources, $\epsilon_{PA}(x)$ and $\epsilon_{PB}(x)$:
\begin{equation}\label{eq:GLM_phot_model}
PA (x) = \tilde{PA}(x) + \epsilon_{PA}(x), \;\;\; PB(x) = \tilde{PB}(x) + \epsilon_{PB}(x),
\end{equation}
where $x$ is the spatial variable for the optical path difference. Since the OPD is modulated to produce the fringes, we have that $x = x(t)$. With the notation of eq. \ref{eq:GLM_phot_model}, we hide the dependence of the OPD on the time $t$.\\
When $PA$ and $PB$ are physically combined in order to produce the interferometric channels $I1$ and $I2$, also the noises $\epsilon_{PA}(x)$ and $\epsilon_{PB}(x)$ enter the combination, causing a variability on the interferometric outputs.\\ {\em We wonder if the introduction of  $\epsilon_{PA}(x)$ and $\epsilon_{PB}(x)$ in the combination process is sufficient to explicate all the variability on $I1$ and $I2$.}\\ In other words, we want to quantify the weight of $\epsilon_{m_1} (x)$ and $\epsilon_{m_2} (x)$:
\begin{eqnarray}\label{eq:GLM_interf_model}
  \nonumber I1(x) &=& (\beta_{1,A} PA(x) + \beta_{1,B} PB(x)) \cdot [1 + m_1(x)] + \epsilon_{m_1} (x)\\
  I2(x) &=& (\beta_{2,A} PA(x) + \beta_{2,B} PB(x)) \cdot [1 + m_2(x)] + \epsilon_{m_2} (x)
\end{eqnarray}

\noindent where $m_i(x), i=1,2$ is the modulation function containing fringes.\\
The ideal combination is noiseless, i.e. $\epsilon_{m_i} (x) \equiv 0, \; i=1,2$. However, we can expect some kind of superposed noises, caused, e.g., by the physical instruments dedicated to the composition/separation of beams (fibres in this case, or optical combiners). If it is the case, we further assume that this noise due to the combination process is uniformly present on the data.\\ {\em Is it possible to quantify its weight?}
\\

\noindent In our region of interest, outside the coherence length, we can suppose that the linear model is predominant with respect to the modulation function $m(x)$. The residual modulation is covered by photometric fluctuations and noise.\\
{\em Is it possible to quantify also the weight of the modulation outside the coherence length?}
\\

\noindent The statistical tool that can answer these questions is the regression analysis. It is suggested by the interferometric model itself (eq. \ref{eq:GLM_interf_model}), that also address the use of a linear model. In particular, the comparison of a linear model and a `mixed' linear model, i.e. with a higher order factor to describe the non-linearity of the interferometric combination, can tell us something on the third question.

\subsection{Review of the multiple regression analysis}
\label{subsec:GLM_Intro}

Let consider a general regression model (in matrix form):
\begin{equation}\label{eq:GLM_model}
Y = \beta X + \epsilon
\end{equation}
where $Y$ is the vector of the observed data, $X$ the regressors vector, $\beta$ are the coefficients, and $\epsilon$ the model error vector.\\
A core principle of the least squares regression method is the fact that the variability of a dependent variable can be partitioned over the sources of variability, i.e. the predicted 
variables and the residual error.
A fundamental identity of the least squares states the following relation between square sums:
\begin{equation}\label{eq:SS}
\sum_{i=1}^N (y_i - \bar{y})^2 = \sum_{i=1}^N (\tilde{y_i} - \bar{y})^2 + \sum_{i=1}^N (y_i - \tilde{y_i})^2 \sim SST = SSM + SSE
\end{equation}
where {\it y} are the observed values of the dependent variable, with mean $\bar{y}$, and $\tilde{y}$ are its estimated values through regression analysis. The quantity \linebreak $\sum_{i=1}^N  (y_i - \tilde{y}_i)^2$ is the square sum of the residuals (SSError, SSE), while SST is the total squares sum and SSM is the model squares sum. SSM and SSE depend from the model adopted, and they can be evaluated as\cite[pg. 4]{rawlings}:
\begin{eqnarray}
    \nonumber SSM &=& b'X'Y \\
    SSE &=& SS Total - SS Model = Y'Y - b'X'Y
\end{eqnarray}

\noindent The residuals are defined as the difference between the foreseen and the measured values:
\begin{equation}\label{eq:residuals}
\varepsilon_i = y_i - \tilde{y}_i, \;\;\; i = 1 \ldots N
\end{equation}

\noindent If we assume the residuals to be uncorrelated random variables, with zero mean and constant variance, and the regressors to be measured without error, than the estimation through the least squares approach is optimal, in the sense that the variance of all other linear estimators is greater than that of the least squares estimator. Note that the residuals are not required to belong to the same distribution, or to be independent.\\
For tests of significance, the random errors $\varepsilon_i$ are often assumed to be normally distributed. In this case, the least square estimators are also the maximum likelihood estimators.

\noindent The ratio of the sum of squares of the regression model to the total sum of squares ($R^2 = \frac{SSM}{SST} = 1 - \frac{SSE}{SST}$) explains the proportion of variance accounted for the dependent variable (y) by the model. This ratio varies between $0$ and $1$. If $R^2=1$, the variance is perfectly explained by the model and there are no residuals; if $R^2=0$, the model could explain nothing of the observed data. $R^2$ can then be used as an estimator of the goodness-of-fit of the model. However, the analysis of residuals is important to validate or not the test on the $R$ ratio, checking if the assumption of the normal distribution of the residual of the least squares model is valid.

\noindent The correlation matrix of the parameters can give an indication wether the parameters are redundant or not. In fact, if two parameters are highly correlated, we can infer that they carry/bring the same amount of `information'.
\\

\noindent We have remarked that some assumptions have to be done to trust the regression results. But what happens if one of them is violated? 
A treatment of the subject can be found in Rawlings et al \cite{rawlings}.

\noindent The non normality of the residuals does not affect the coefficients estimation, as it is not required in the partition of the squares sums, but can cause troubles on the tests for their significance and for the confidence intervals, that are all based on the normal distribution. Moreover, the estimators are still the best between all linear estimators, but are no longer the Maximum Likelihood estimators. The same apply if there is a correlation between the residuals. Techniques exist to face this problem (generalized least squares), but if the residuals covariance matrix has to be estimated by the data, the results could be worse than before.\\

\noindent The property of the minimum variance of the estimators depends directly also from the hypothesis of homogeneity of the variance of the residuals. If it is not the case, it is necessary to introduce weights on the regression analysis.\\

\noindent A different treatment is necessary if the independent variables and/or the regressors have measurement errors. References can be found in \cite[p. 91]{Ryan}, \cite[p. 123]{DraperSmith}. We can resume saying that, if the dependent variables have measurement errors, these errors increases the residuals, reducing the $R^2$ coefficient, and so leaving more unexplained variability on the model.\\
The situation becomes worst if the regressors are measured with errors.
If the regressors are fixed and the measurement errors are normally distributed with same variance and zero mean, the coefficients estimators will still be unbiased. On the contrary, they will be biased if the regressors are random variables, and the bias will be function of the correlation between the true unknown random regressor and the measurement error variance.
\\

\noindent Finally, we remark that the described method applies also to multiple regression, in which there are two or more regressors, as is our case.

\subsection{Method description}
\label{subsec:GLM_method}

\noindent In this paragraph, we will describe the procedure adopted for the regression analysis, the outputs produced and how we evaluated them. The analysis has been carried out with the software STATISTICA (produced by StatSoft: www.statsoft.com).\\

\noindent Our data, as said before, are integer numbers, they can be considered as counts. It seems that in this case the normality distribution required for the model error are not valid any longer. However, the mean level of counts is high (order of hundreds counts).

\noindent With this preliminary remark in mind, we have prepared the data according with the introductive discussion.\\
First of all, for each record, we've eliminated the coherence length area.\\
In order to have information about the possible lack of homogeneity of the variance on the channels, we have performed a Levene test for homogeneity of variances on the residual data (for description of the Levene test, see section \ref{subsec:variance_statMethod}).
We have divided data into subintervals, to check if the variability of data from one subinterval to another was due to mean variation or effectively to variance. We have repeated this test on both photometric and interferometric pairs of channels, and we have retained records for which the variability on the input channels $PA$ and $PB$ is homogeneous. We have further distinguished between records with homogeneous and inhomogeneous variance on the output channels $I1$ and $I2$, retaining the first.
\\

\noindent The chosen models do not foresee an intercept coefficient. The reason is that in the ideal denoised case of photometric channels set to zero, we want to have interferometric outputs set to zero.
\\

\noindent For each considered regression model, we have performed test on the coefficients of the regression model, to see if they were null, on the residual unexplained variance, and we have studied the residuals.\\

\begin{figure}[!htb]
    \begin{center}
        \epsfig{figure=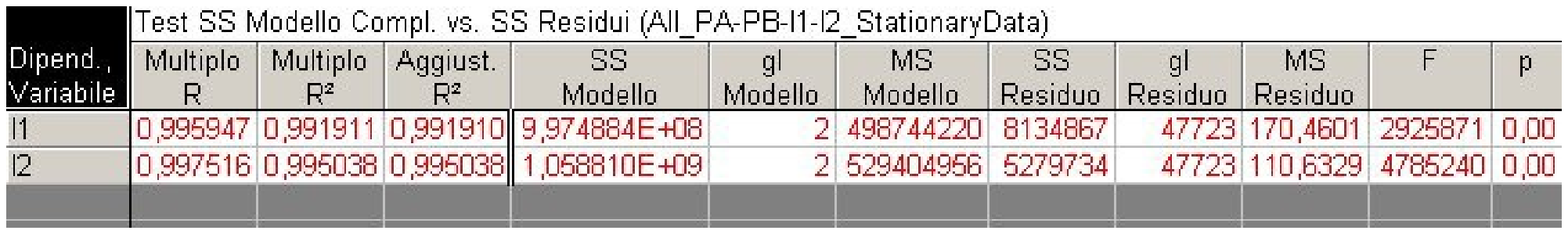,width=15cm}
        \caption{Table of tests on the model. This example is taken from section \ref{subsec:GLM_linearModel} and is referred to the linear model without factors of  higher order.}
       \label{fig:example_SScompleto}
    \end{center}
\end{figure}

\noindent The table summarizing the model description looks like the one in figure \ref{fig:example_SScompleto}.

\noindent Here after we explain the meaning of each column in the tables:
\begin{itemize}
	\item {{\it R multiplo (Multiple R)}: it is the positive square root of $R^2$. }
	\item {$R^2$: is the coefficient of the multiple correlation. It measures the component of the total variability due to the independent variables. It is useful because it takes care of the presence of multiple regressors. We recall its definition as the ratio of the squares sums of the model and total sums squares:
\[ R^2 = 1 - \frac{SSE}{SST}\]}
	\item {{\it $R^2$ aggiustato (Corrected $R^2$)}: it is obtained from the $R^2$ definition dividing the error squares sums and the total squares sums by their degrees of freedom	($n-k$ and $n$ respectively, where $k$ is the number of independent variables and $n$ is the number of cases used in the regression)
}	
	\item {{\it SS Modello and SS Residuo}: squares sum of the regression model and of the residuals, respectively}
	\item {{\it gl Modello and gl Residuo}: degree of freedom (df) of the regression model ($k$, where $k$ is the number of non correlated independent variables) and of the residuals ($n-k$), respectively}
	\item {{\it MS Modello and MS Residuo}: mean square sum ($\frac{SS}{df}$) of the model and of the residuals, respectively}
	\item {{\it F, p}: test to verify the statistical significance of the $R^2$ measures. It is computed:
	\[ F = \frac{MSModel}{MSResidual} \sim F_{(k, n-k)}\]}
\end{itemize}

\noindent Note that Statistica automatically marks in red the results that have a positive significance, for a quicker understanding.\\
Another table of interest is the one shown in figure \ref{fig:example_collinearita}, reporting different statistics for each regressors, and we resume them:

\begin{figure}[htb]
    \begin{center}
        \epsfig{figure=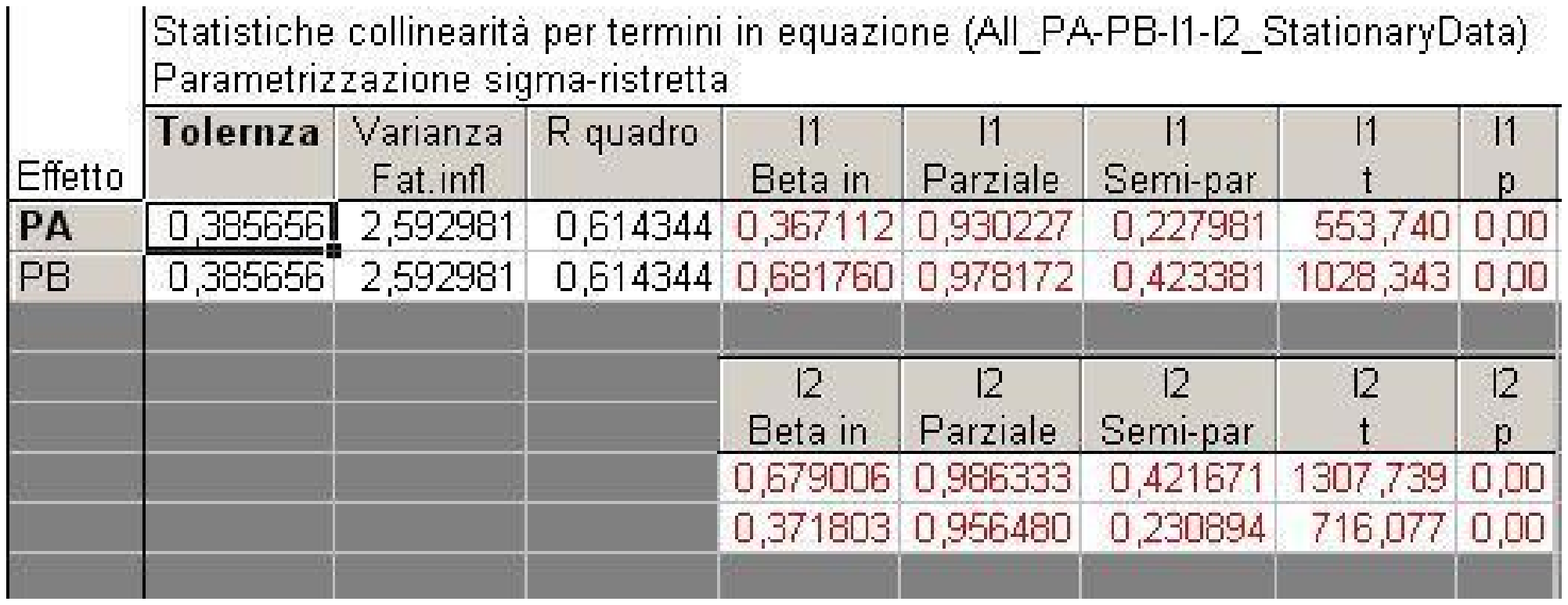,width=15cm,height=5cm}
        \caption{Example of correlation analysis for PA and PB}
        \label{fig:example_collinearita}
    \end{center}
\end{figure}

\begin{itemize}
	\item {{\it Tolerance}: it is defined as $1 - R^2$. If the tolerance is small, the variable is highly correlated with the others variables, and this increases its redundancy. In the example table, we can see that the tolerance is small for every regressors, so there is no redundancy}
	\item {{\it Variance inflation factors (VIF)}: the elements on the diagonal of the inverse of the correlation matrix, used in the model computation. It is another measure of the redundancy of the variables: if the VIF is 1, the predictor variables are uncorrelated. In our case, they are not exactly 1, as the tolerance was not exactly 0

}
	\item {$R^2$: as before, the multiple correlation coefficient}
	\item {{\it Beta inserted ($\beta$) }: the standardized regression coefficient, i.e. the coefficients obtained if the variables were standardized with zero mean and unitary standard deviation before being used in the model. They differ from the `B' coefficients, that could be affected by errors due to different behaviour of the related independent variables.}
	\item {{\it Partial correlation}: the correlation between the dependent variable and the independent ones, taking into account the presence of other correlated variables. It can be interpreted as the percentage of \b{non-explicated} variability of $I_j$, $j = 1,2$ due to a regressor after having `subtracted' the contribution of the other regressors.}
  \item {{\it Semi-partial correlation}: as the partial, but related to the {\underline total} variance of $I_j$.}
  \item {{\it t, p}: test {\it t} for these statistics and related p-value.
}
\end{itemize}

\noindent Finally, the model coefficients come with the B and $\beta$ values, with their standard errors, the test t associated and the confidence intervals.
\\

\begin{figure}[!htb]
    \begin{center}
    \epsfig{figure=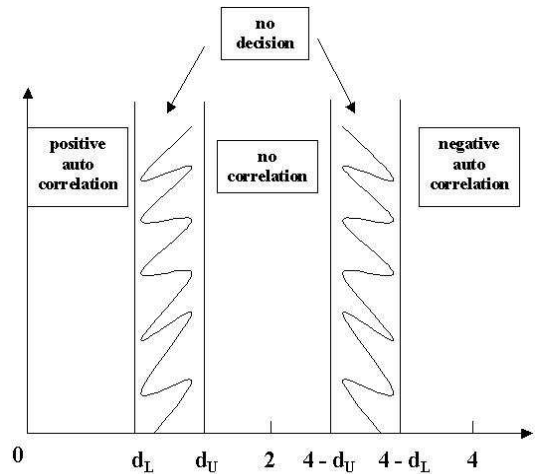,width=7cm}
    \caption{Decisional areas for the Durbin-Watson test. In the figure, $d_L$ and $d_U$ are the lower and the upper bound, respectively.}
    \label{fig:DW-test}
    \end{center}
\end{figure}

\noindent When working with time series, raw residuals, i.e. the differences between the observed and the predicted values, are usually correlated and have a variance that changes, (see, e.g. \cite[page 342]{rawlings}). To test the serial auto-correlation of the raw residuals we have implemented the Durbin-Watson test\cite{durbin-watson51}. The test statistic is
\begin{equation}\label{eq:DWtest}
d = \frac{\sum_{i=2}^{N} (\epsilon_i - \epsilon_{i-1})^2}{\sum_{i=1}^{N} \epsilon_i^2} \approx 2(1-\hat{\rho})
\end{equation}
where $\hat{\rho}$ is the residual sample-autocorrelation at lag $-1$. The statistic $d$ can assume values in the range $[0,4]$ and becomes smaller as the correlation increases. The statistic distribution is not known, but has been tabulated with experimental texts by Durbin \& Watson. Two bounds, a lower and a upper, dependent from the number of residual samples $N$, the number of regressors, without the intercept, and the confidence level, define two doubtful regions, where the test does not permit a decision, a central area of no autocorrelation, and two lateral areas where there is a statistical evidence of positive and negative serial correlation, respectively. Figure \ref{fig:DW-test} shows these areas.

\noindent The presence of a serial correlation modifies the properties of the coefficient estimators: they are still unbiased, but they are no longer the best estimators.\\
For our analysis purposes, since tabulated values foresee up to $N=100$, we have further divided the residuals series in subsets of $100$ samples, and we have set the confidence level at $95\%$.
\\

\noindent To control the assumption of the regression model, several modification of the raw residuals have been proposed. We choose the Standardized residuals, corrected to equalize their variances. The standard residuals are evaluated by Statistica using the following formulation:
\begin{equation}\label{eq:std-residuals}
\epsilon^S_i = \frac{y_i-\tilde{y}_i}{\sqrt{\sum_{i=1}^{N}{(y_i-\tilde{y}_i)^2}} / N}, \;\;\; i = 1 \ldots N
\end{equation}

\noindent In the following paragraphs, we proceed to the description and comparison of regression analysis using both a multiple linear model and a multiple linear model with a factor of higher order, respectively.

\subsection{Estimation of photometric coefficients through calibration analysis}
\label{subsec:GLM_calibr-coeff}

First of all, we use the least squares regression to compute the coefficients $\beta_{i,j}$ of the photometric channels in the eq. \ref{eq:GLM_interf_model}, performing the analysis on calibration records.
These coefficients will be useful for comparison with {\bf successive} analysis.\\
The calibration data consists, as described in par. \ref{sec:data_descr}, in sets recorded with just one photometric channel fed with source flux, while the other is void, and contains just background or environmental noise. The level of the interferometric channels gives immediately the coefficients of the interested photometric channel:
\begin{eqnarray}\label{eq:phot-coeff}
\nonumber I_1 = \beta_{1,A} PA(x); \;\;\; I_2 = \beta_{2,A} PA(x)\\
I_1 = \beta_{1,B} PB(x); \;\;\; I_2 = \beta_{2,B} PB(x)
\end{eqnarray}
using respectively data of case 2 and case 3.\\

\noindent It is clear that in this case we need the regression analysis with just one regressor, $PA$ for case 2 and $PB$ for case 3. In the next figures (\ref{fig:coeff_PA} and \ref{fig:coeff_PB}), we can find a summary of the model properties, the coefficients values with their tests, and the residuals normal probability plots for case 2 (channel $PA$ fed) and case 3, respectively. We have to remark that the analysis has been done selecting only records with variance homogeneity properties on all channels for $PB$ and on regressors for $PA$, since for the latter no ideal record was in the calibration set.

\begin{figure*}[!htbp]
    \begin{center}
    \epsfig{figure=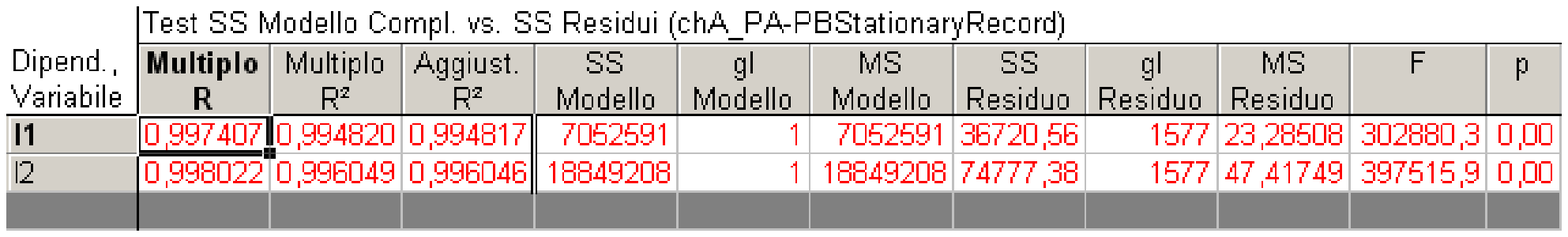,width=15cm}
    \vspace{0.5cm}
    \epsfig{figure=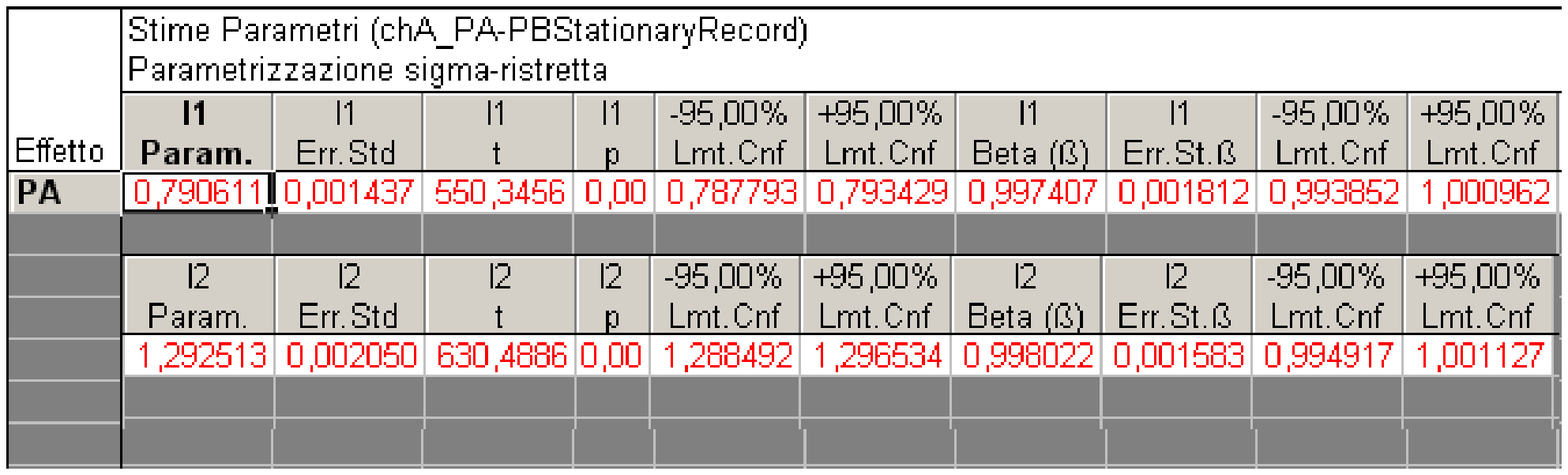,width=15cm}\\
    \vspace{0.5cm}
    \epsfig{figure=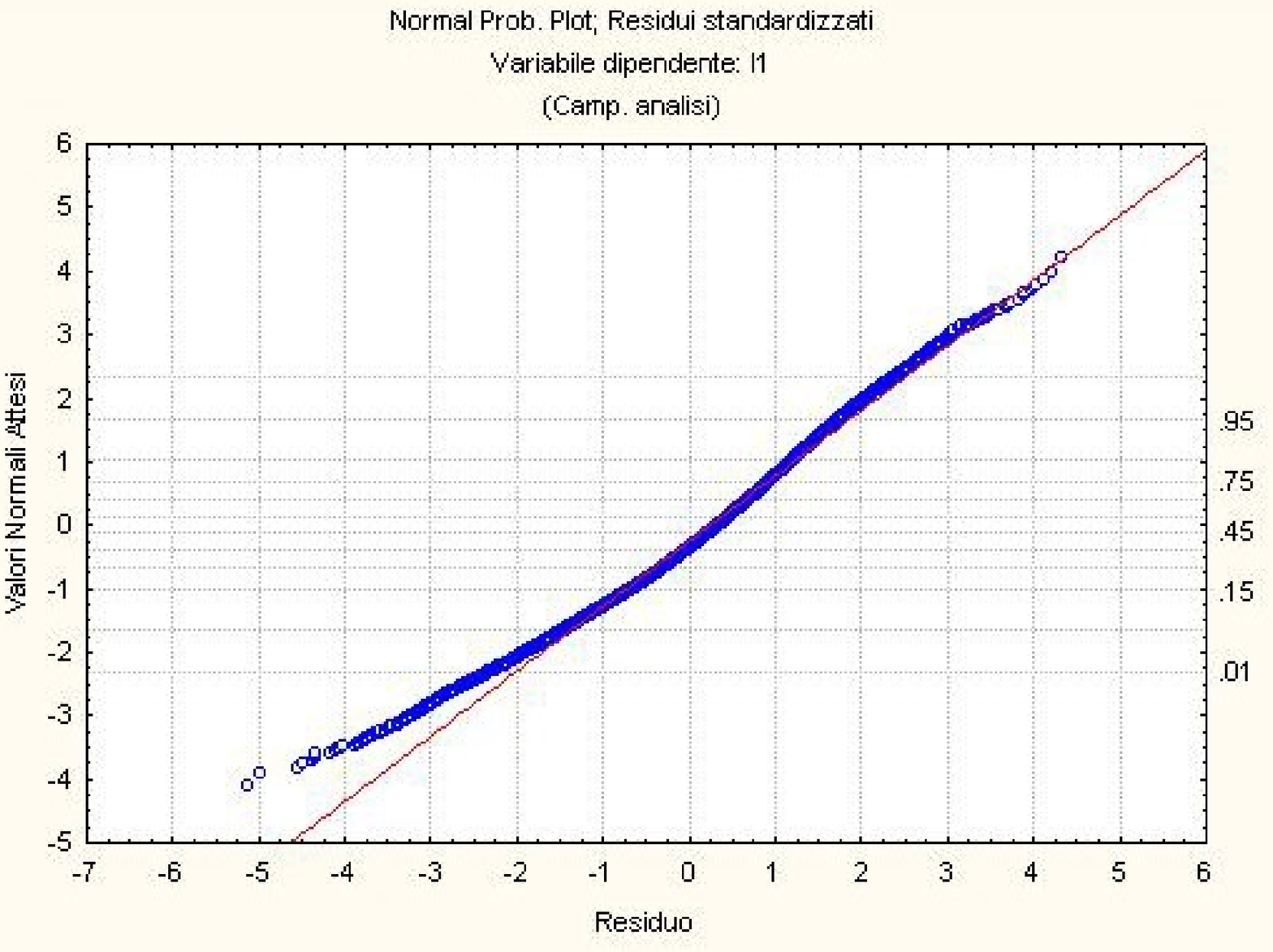,width=6.5cm}
    \epsfig{figure=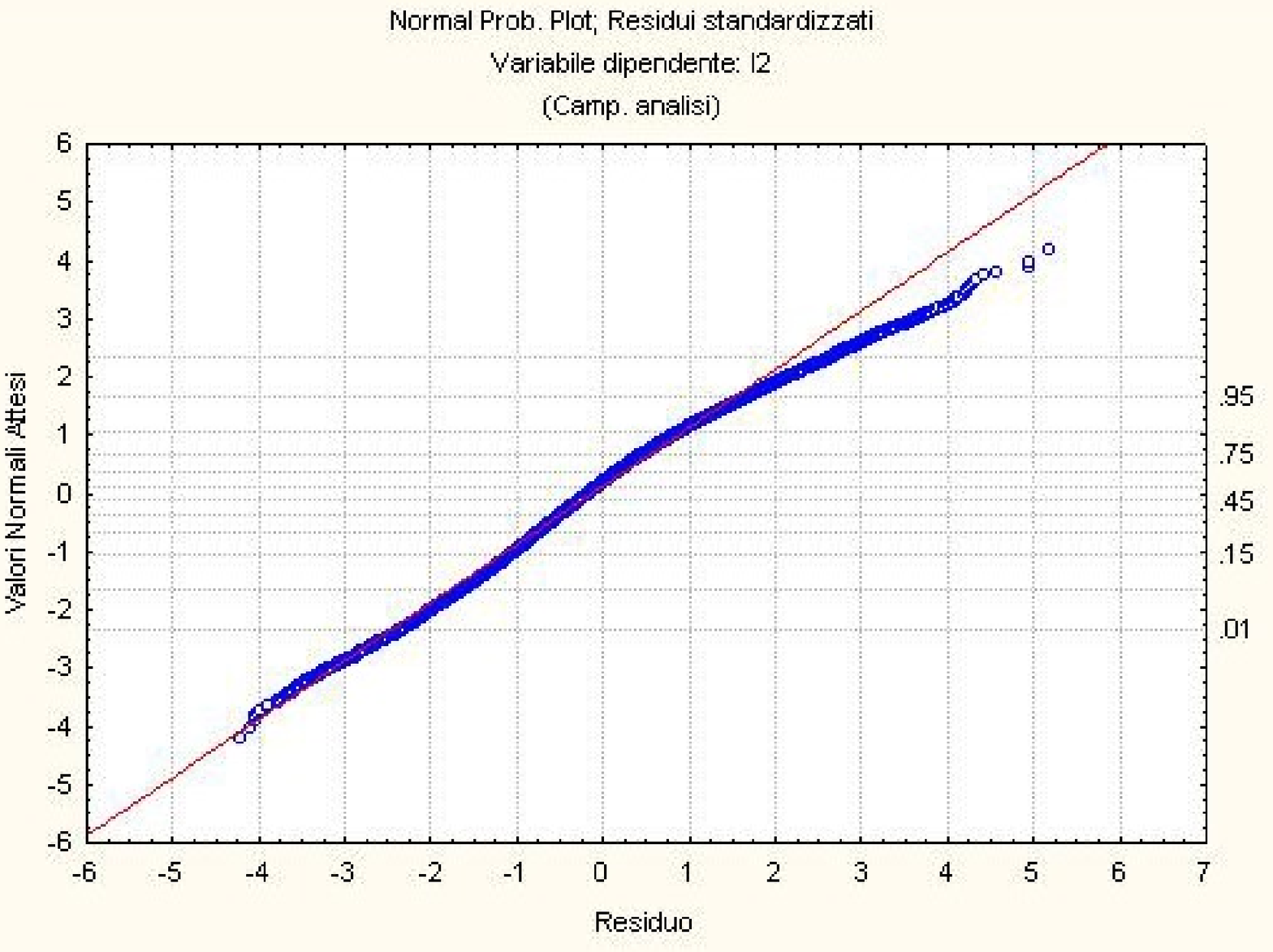,width=6.5cm}
    \caption{Case 2: channel A fed, channel $PB$ void. First row, summary of the model; second row, the regression parameters with tests; third row, normal probability plots of the residuals for the dependent variables $I1$ (left) and $I2$ (right).}
    \label{fig:coeff_PA}
    \end{center}
\end{figure*}

\begin{figure*}[!htbp]
    \begin{center}
    \epsfig{figure=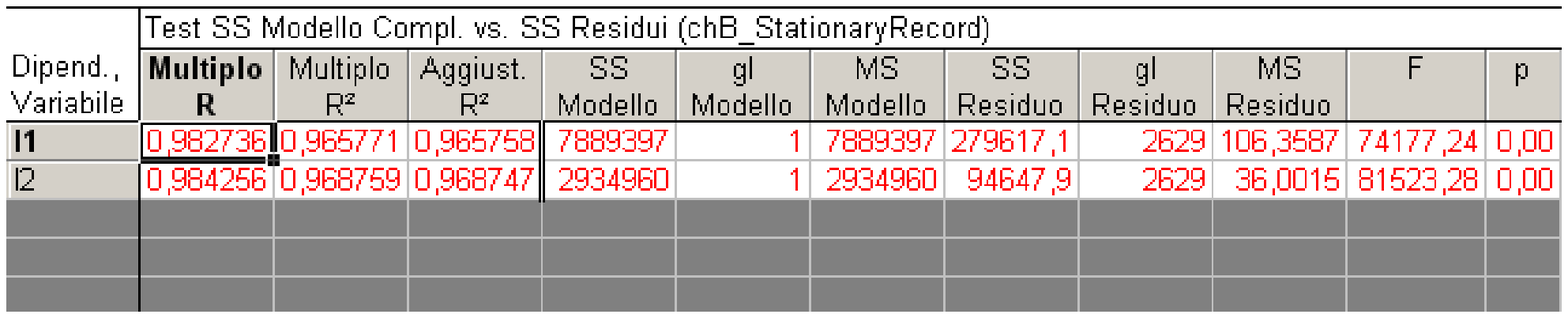,width=15cm}
    \vspace{0.5cm}
    \epsfig{figure=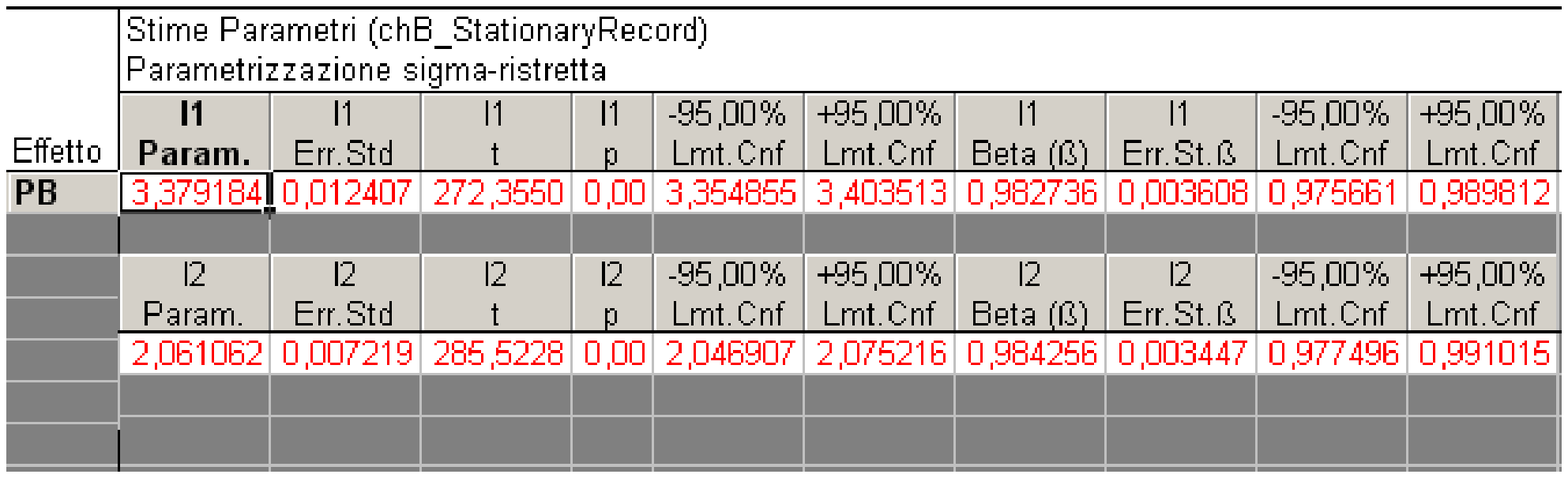,width=15cm}\\
    \vspace{0.5cm}
    \epsfig{figure=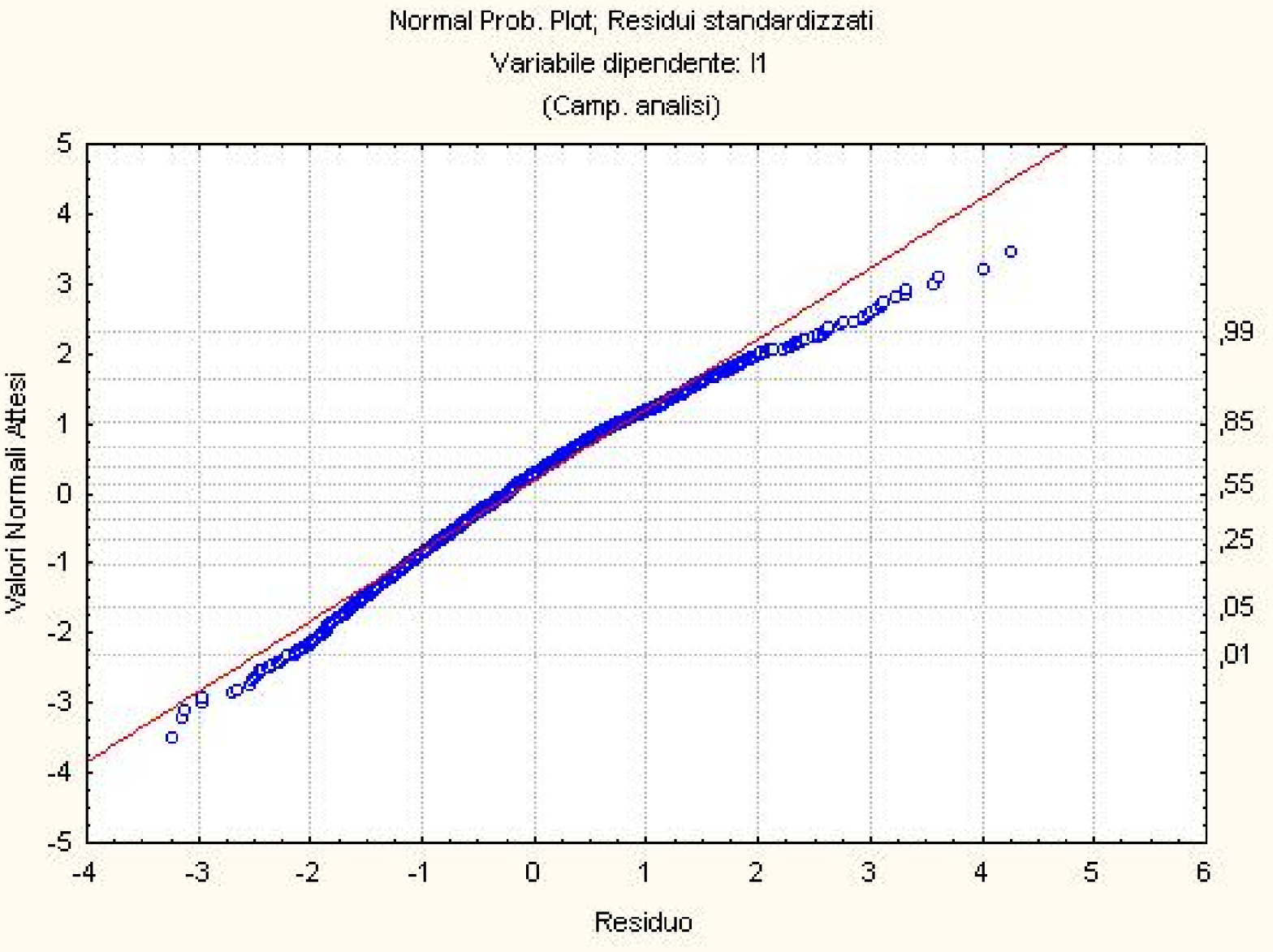,width=6.5cm}
    \epsfig{figure=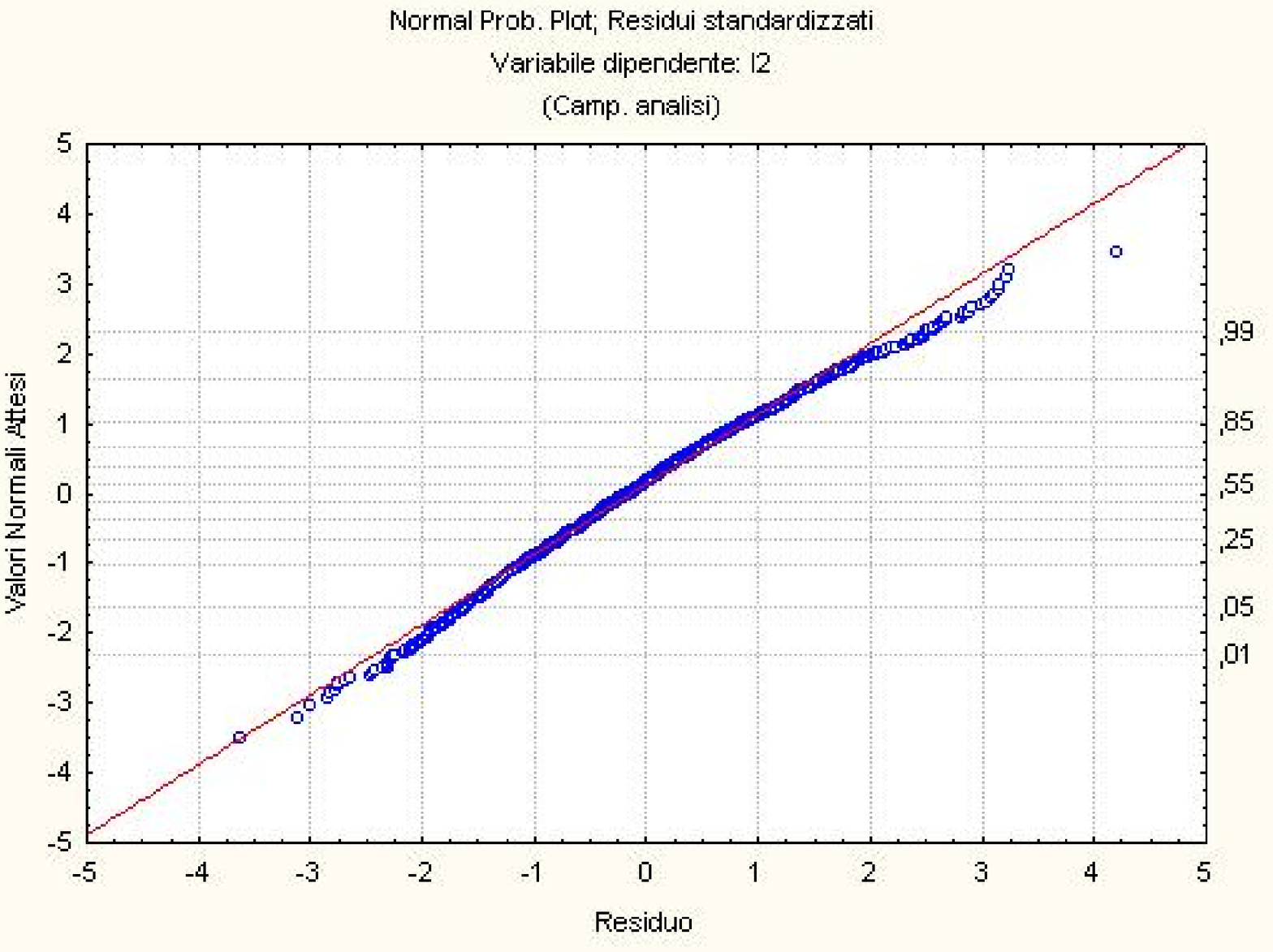,width=6.5cm}
    \caption{Same as fig. \ref{fig:coeff_PA}, but for case 3: channel A void, channel $PB$ fed.}
    \label{fig:coeff_PB}
    \end{center}
\end{figure*}

\noindent The regression model explains very well the variability of the dependent variables $I1$ and $I2$ for the regressor $PA$, for $PB$ there is some more uncertainty, but it is still good. The analysis of the residuals shows their normal distribution for channel $PA$, whereas for $PB$ the residuals have tails that do not respect normality. If we use all the records, the coefficients do not change much, their standard errors decrease, due to the higher number of samples, but the residuals are worse.\\

\begin{figure*}[!htb]
    \begin{center}
        \epsfig{figure=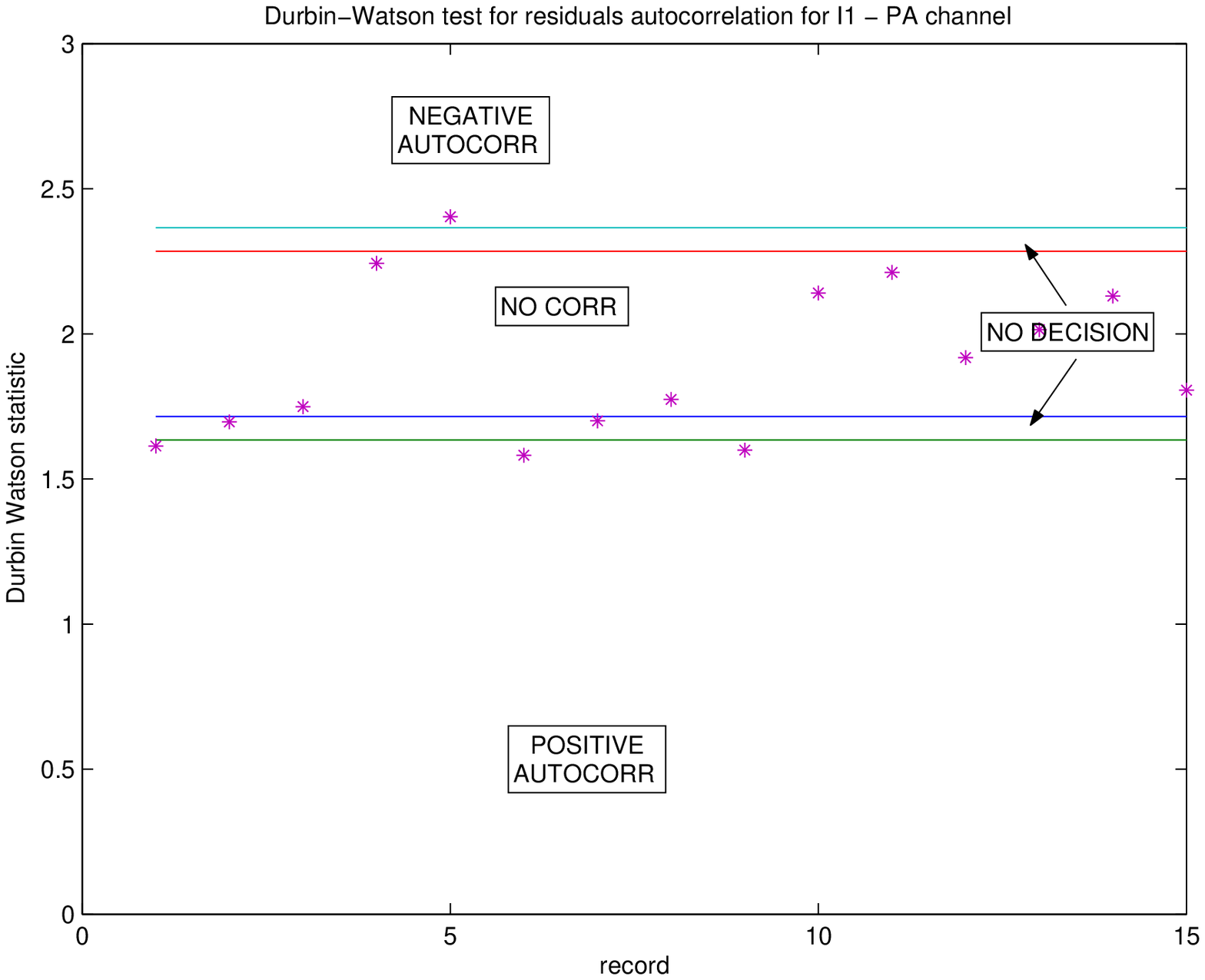,width=6.5cm}
        \epsfig{figure=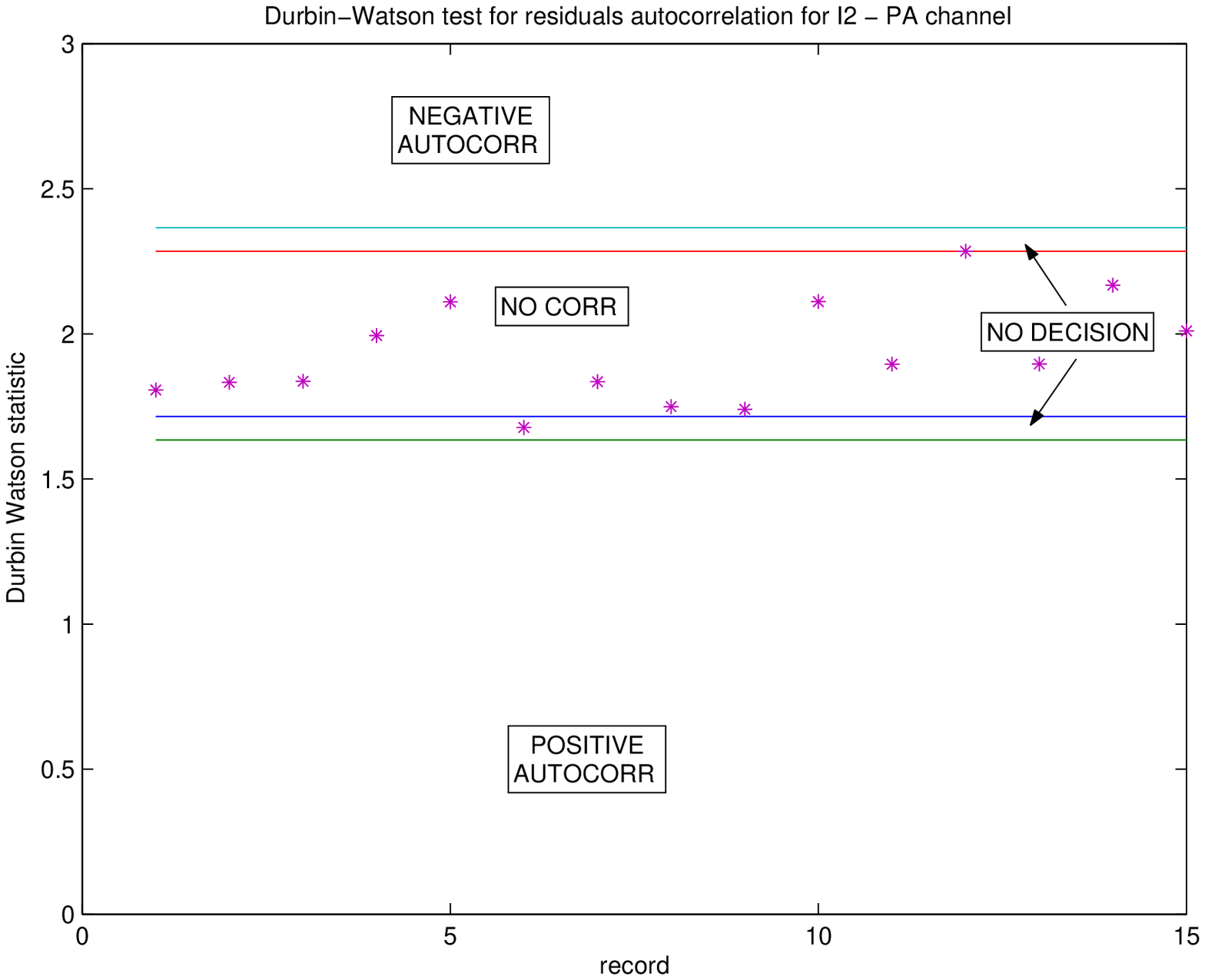,width=6.5cm}
        \epsfig{figure=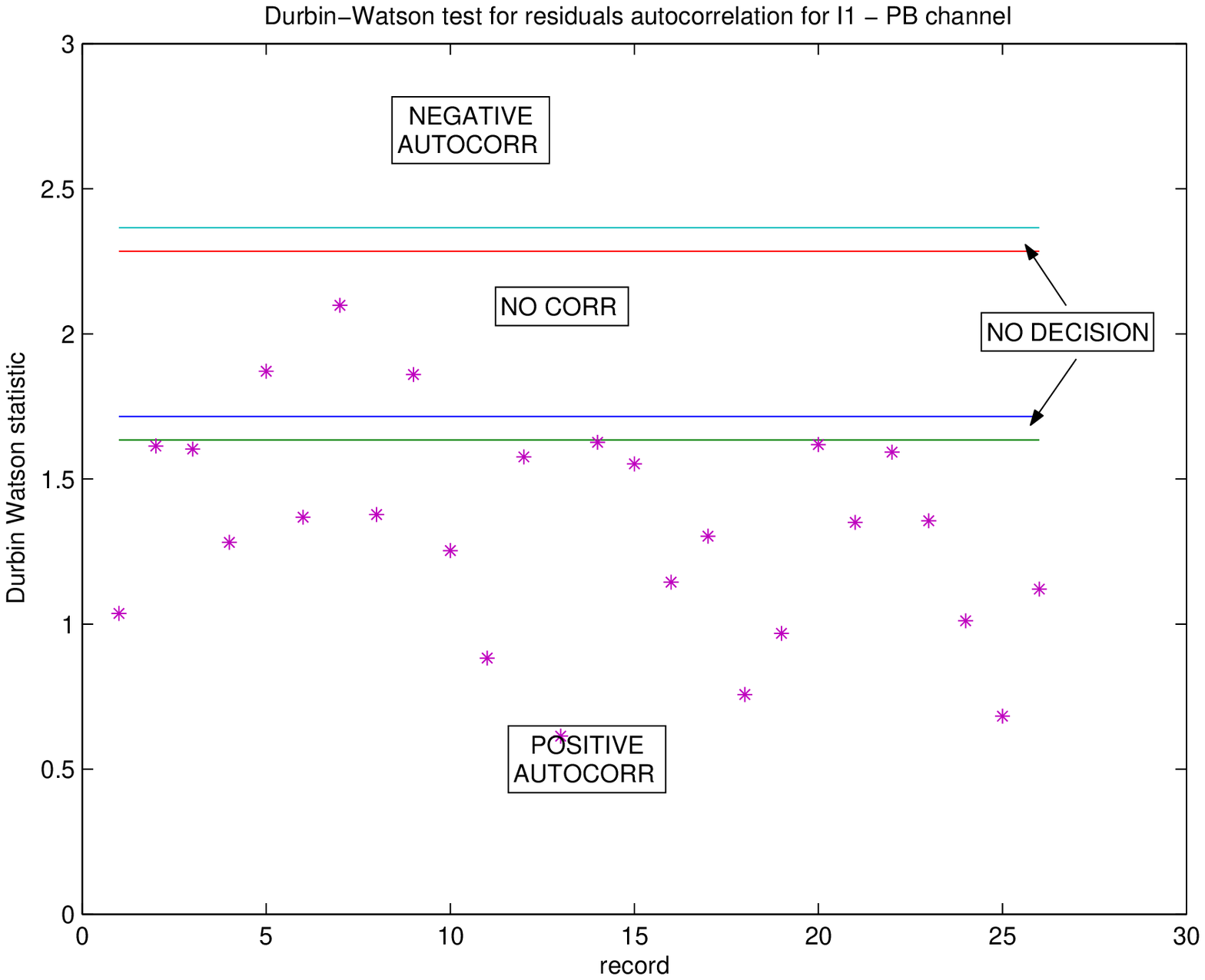,width=6.5cm}
        \epsfig{figure=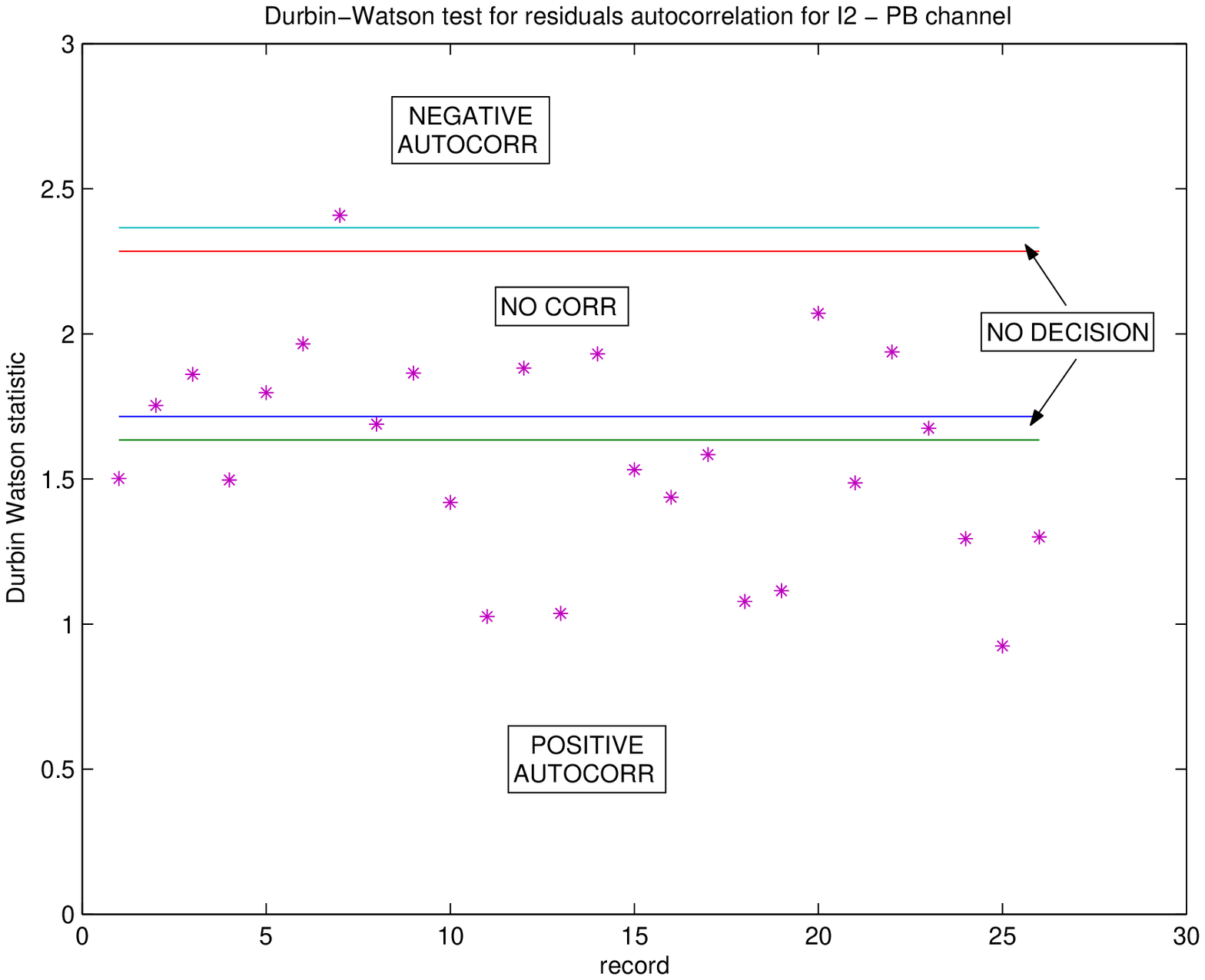,width=6.5cm}
        \caption{Durbin-Watson statistics for the residuals of $I1$ (left) and $I2$ (right): first row, for the regressor $PA$, second row, for $PB$.}
        \label{fig:DW-coeff}
    \end{center}
\end{figure*}

\noindent To better understand the behaviour of the residuals, we perform the Durbin-Watson test for the search of autocorrelation in the time series of the raw residuals. The results are shown in figure \ref{fig:DW-coeff}. The residuals have been divided in 100-samples sized intervals, and the test has been performed over each interval.

\noindent We notice the curious feature of the residuals of the regression with the channel $PA$ as regressor: they are uncorrelated. The only particular difference between the two channels is that the flux in $PB$ is lower than $PA$, and is less subject to fluctuations. So we can say that the regression can easily track the fluctuation. On channel $PB$ there is some different contribution: however, from both time and frequency statistical analysis we could not find anything particular.


\subsection{Regression with linear model}
\label{subsec:GLM_linearModel}

\begin{figure*}[!h]
    \begin{center}
    \epsfig{figure=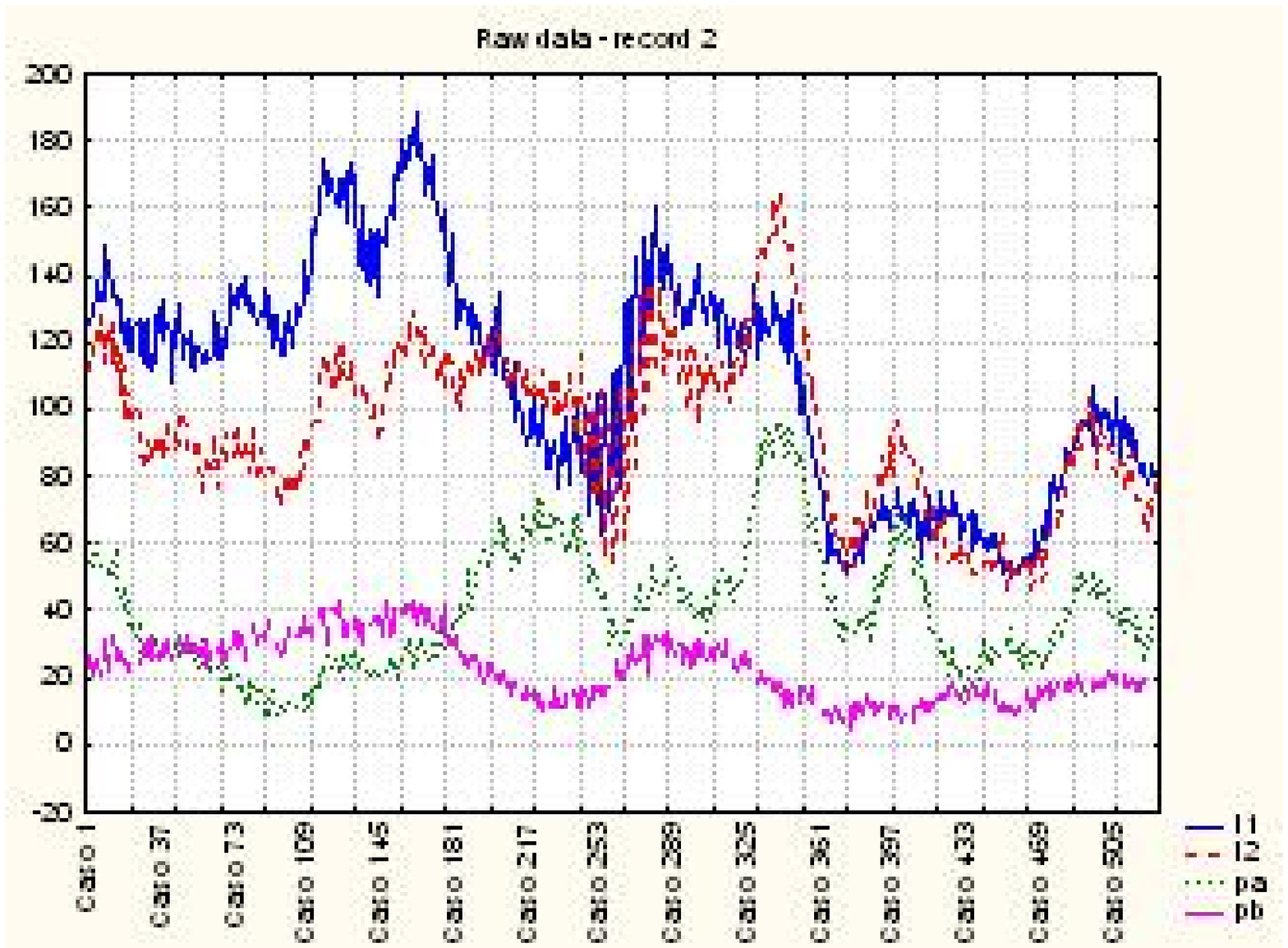,width=6.5cm}
    \epsfig{figure=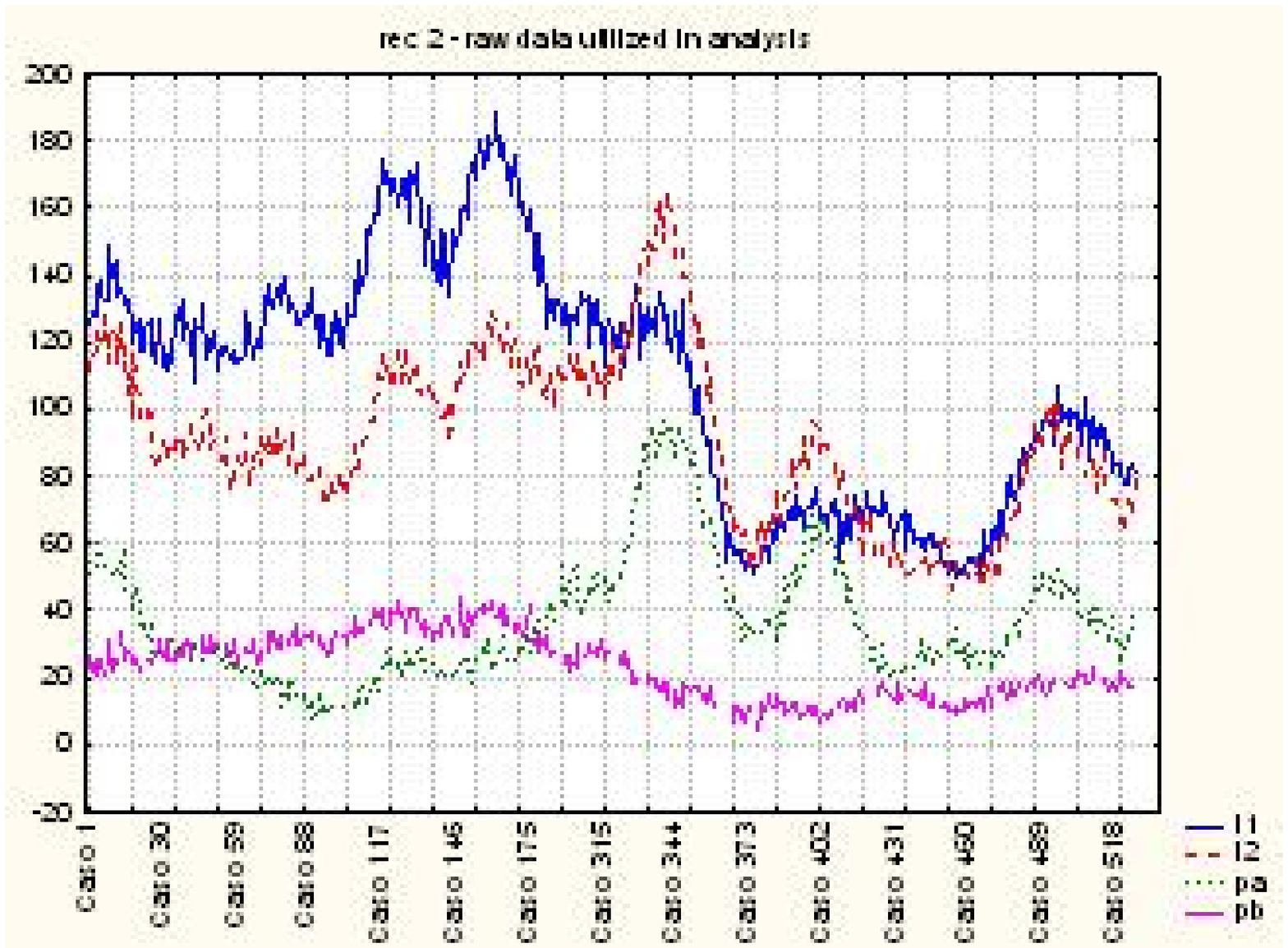,width=6.5cm}
    \caption{Record 2 raw data (left) and utilized data (right). The coherence length has been eliminated from the record to avoid variance variation caused by interferometric fringes.}
    \label{fig:rec2-raw_data}
    \end{center}
\end{figure*}

We now use the observational data to test the model of eq. \ref{eq:GLM_interf_model}. As said before, we limit the analysis over all records for which the variances of the photometric inputs and of the interferometric outputs are not varying along time over the record length. A typical example of the raw data is shown in fig. \ref{fig:rec2-raw_data}.

\begin{figure*}[htb]
   \begin{center}
	 	\epsfig{figure=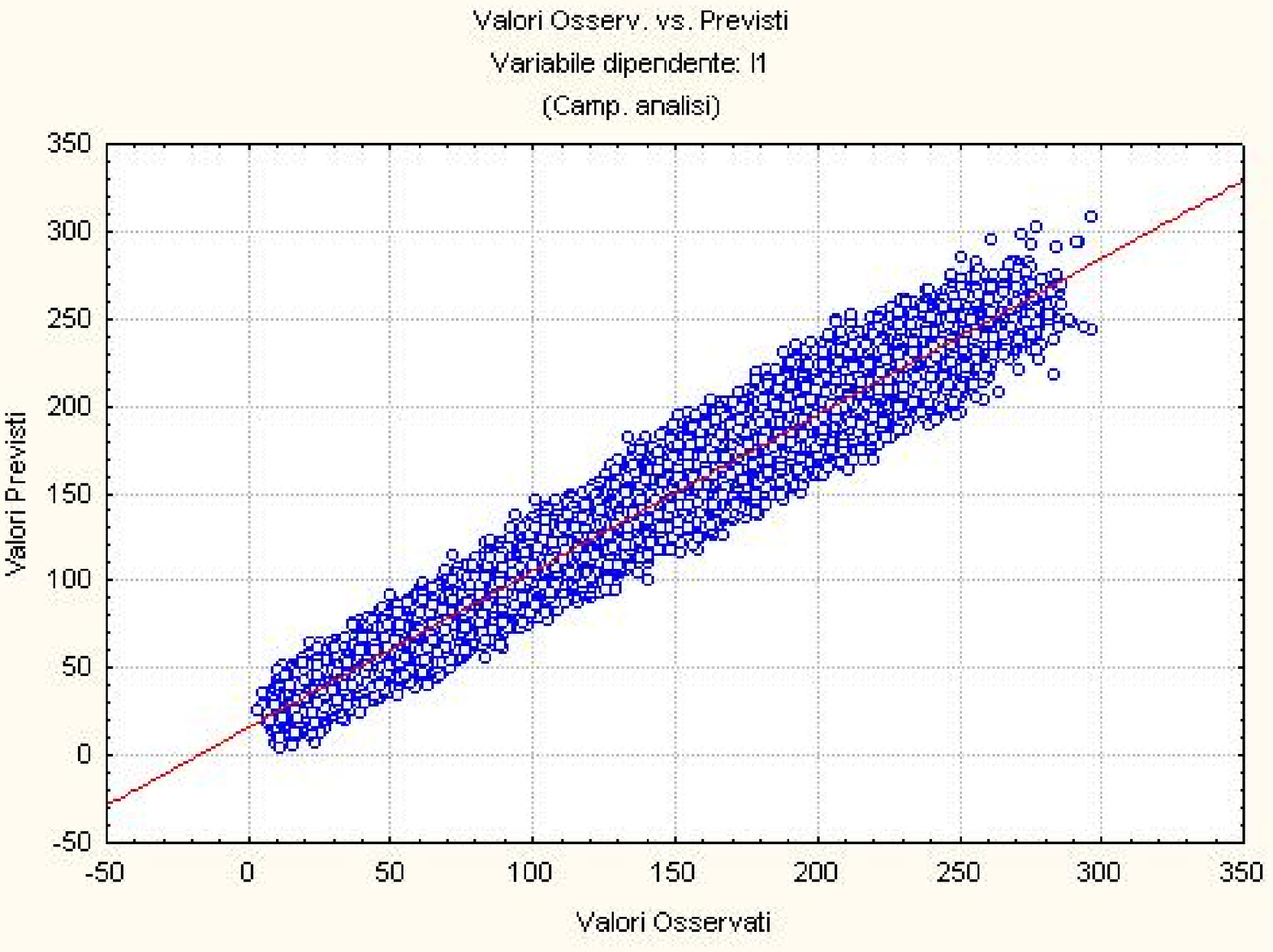,width=6.5cm}
	 	\epsfig{figure=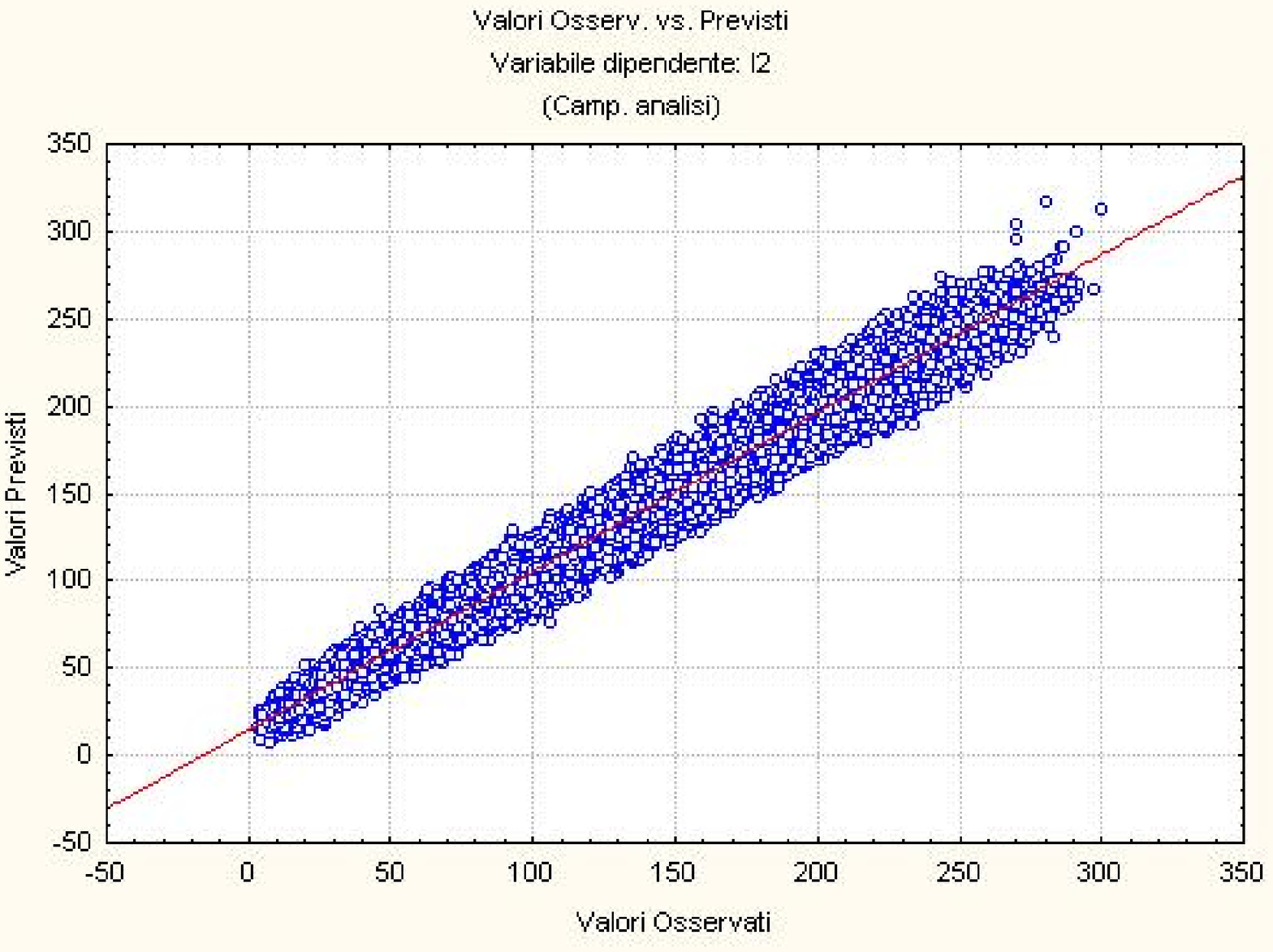,width=6.5cm}
	 	\caption{Scatterplot of observed versus predicted values for I1 and I2}
    \label{fig:lm_obsVsForeseen}
    \end{center}
\end{figure*}

\noindent We use the following linear regression model without intercept:
\begin{equation}\label{eq:GLM_linearModel}
I_i = c_{Ai} PA + c_{Bi} PB,  \;\;\; i=1,2
\end{equation}
where the photometric channels $PA$ and $PB$ are the regression variables, and the interferometric outputs $I1$ and $I2$ are the dependent variables.
This linear model is a good fit of the observed data; in figure \ref{fig:lm_obsVsForeseen}, scatterplots of the predicted vs. observed values are shown. The points follow roughly a straight line; there are no evident outliers.

\begin{figure}[!htb]
    \begin{center}
        \epsfig{figure=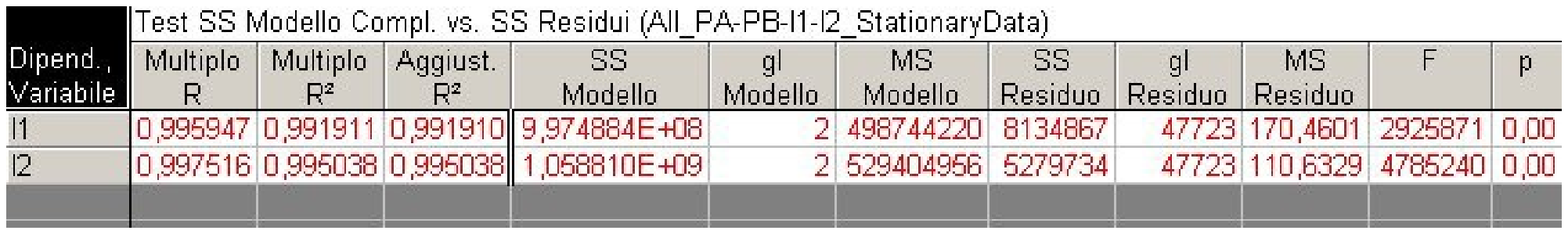,width=15cm}
        \vspace{0.5cm}
        \epsfig{figure=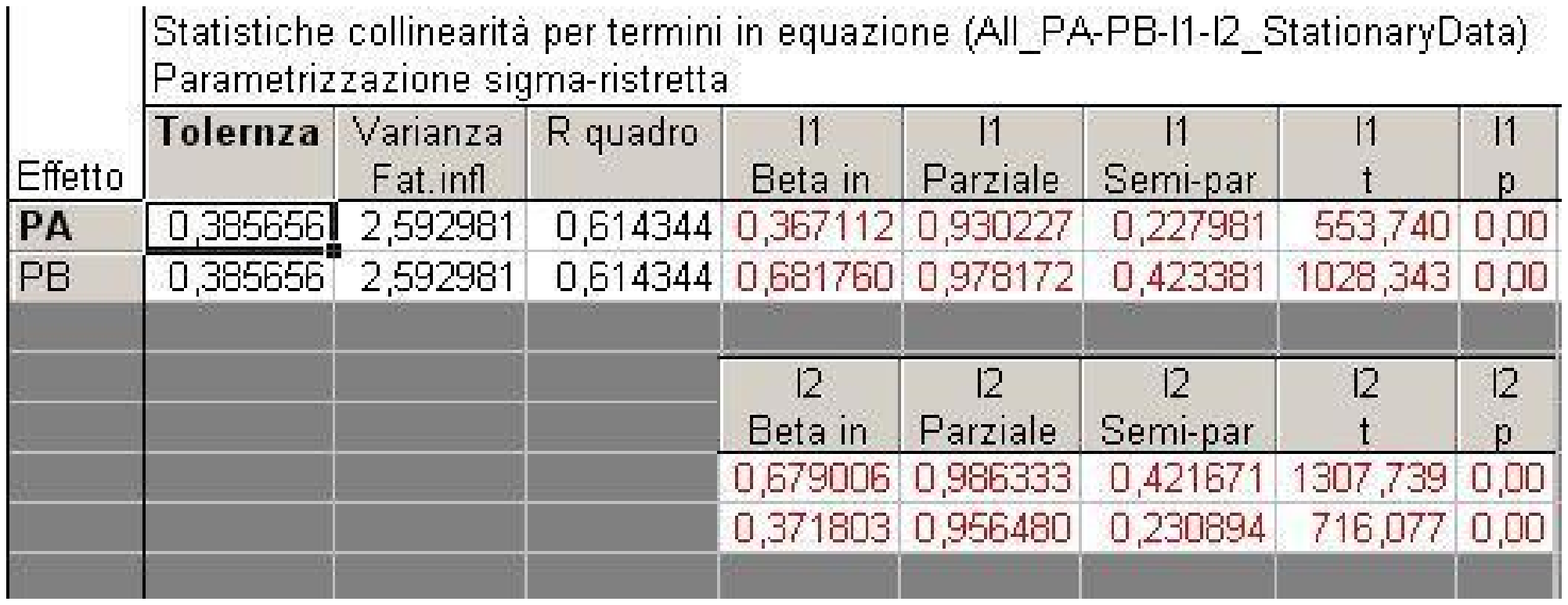,width=15cm,height=5cm}
        \caption{Table of tests on the model (first row) and statistics on the regressors (second row).}
       \label{fig:lm_SScompleto}
    \end{center}
\end{figure}

\begin{figure}[htbp]
    \begin{center}
    \epsfig{figure=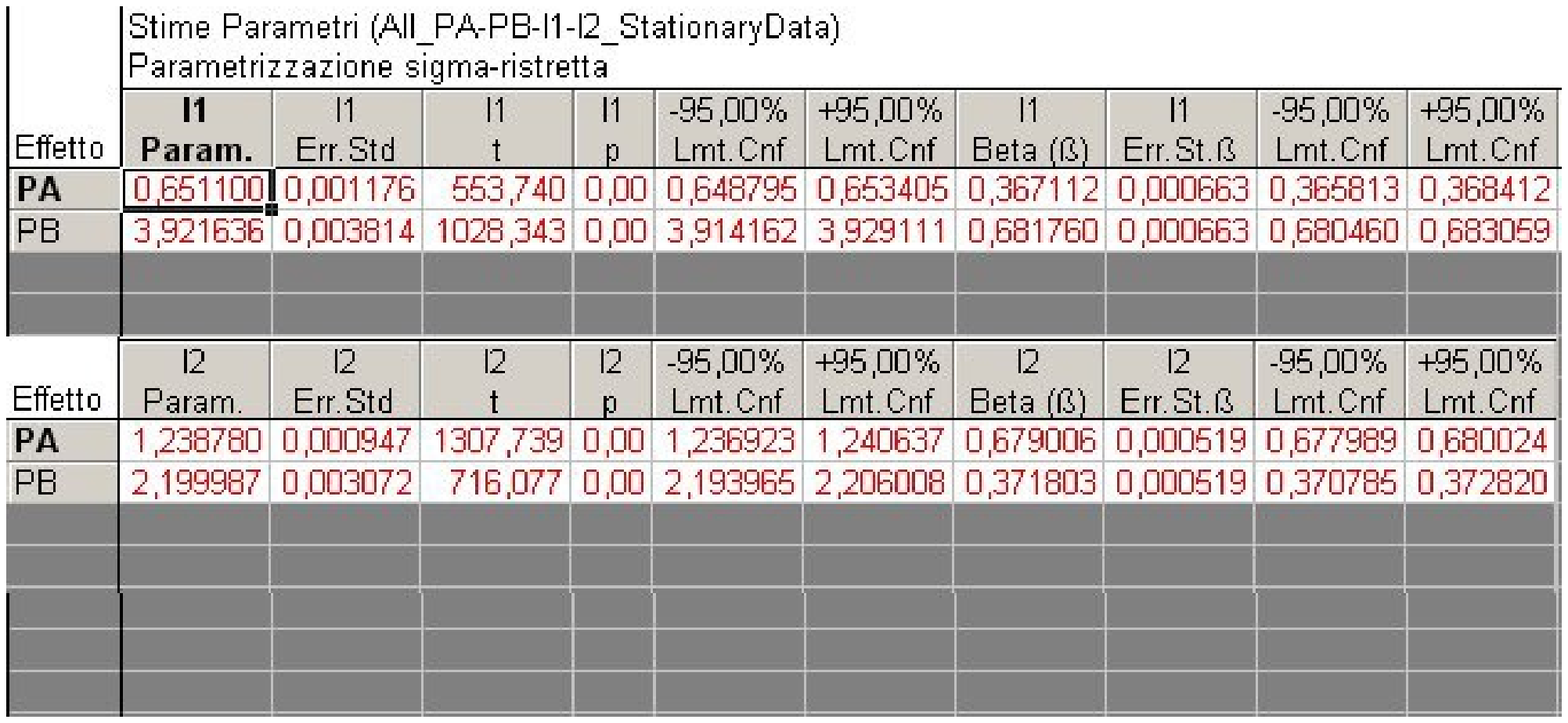,width=16cm,height=8cm}
    \caption{Estimation and statistical tests of regression coefficients}
    \label{fig:lm_coeff}
    \end{center}
\end{figure}

\noindent The model gives a good explanation of the variance of the outputs $I1$ and $I2$, too. In figures \ref{fig:lm_SScompleto} and \ref{fig:lm_coeff} the summary table of the model and the coefficients values and tests are reported. We notice that the $R^2$ values are really high, close to $1$, for both the dependent variables $I1$ and $I2$. They are marked in red, and the p-value is less than $0.01$, so we can accept the results.

\noindent The $\beta$ coefficients values suggest that the division of the incoming beams $PA$ and $PB$ on the outputs $I1$ and $I2$ is not balanced, but has a proportion of about $33\%$ against $65\%$. The regression coefficients are slightly different from those resulting from the regression of par. \ref{subsec:GLM_calibr-coeff}: in this case, the channel $PB$ is reduced with advantage of $PA$. This could be due to the interaction of the two channels, that in the simple regression with just one channel wasn't present.\\

\noindent Even if the model utilized seems to fit very well the data, before validating our results we analyze the residuals, in order to check the linear regression assumption of normal distribution of the residuals, with zero mean and constant variance.
\\

\begin{figure*}[htb]
    \begin{center}
        \epsfig{figure=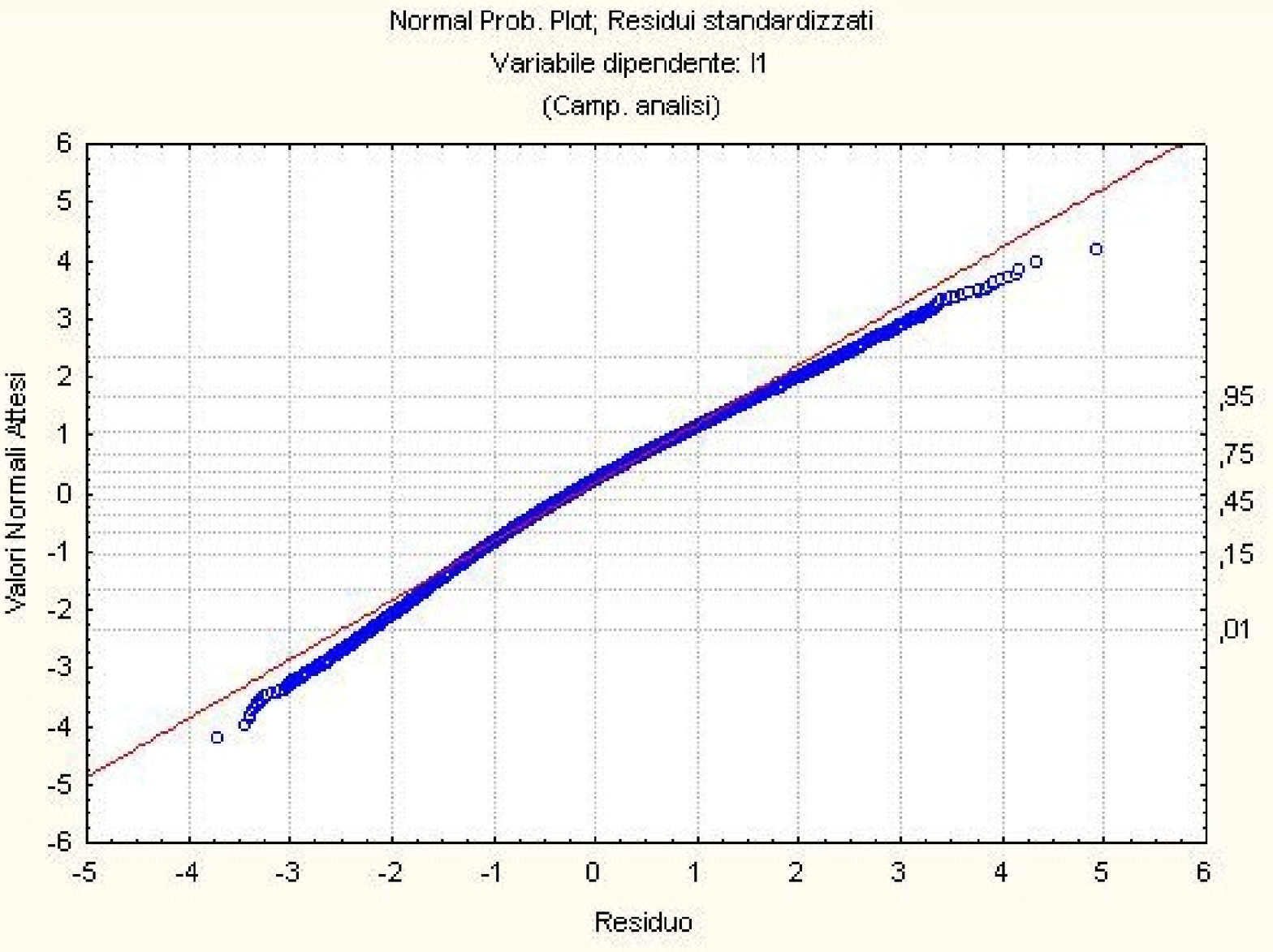,width=6.5cm}
        \epsfig{figure=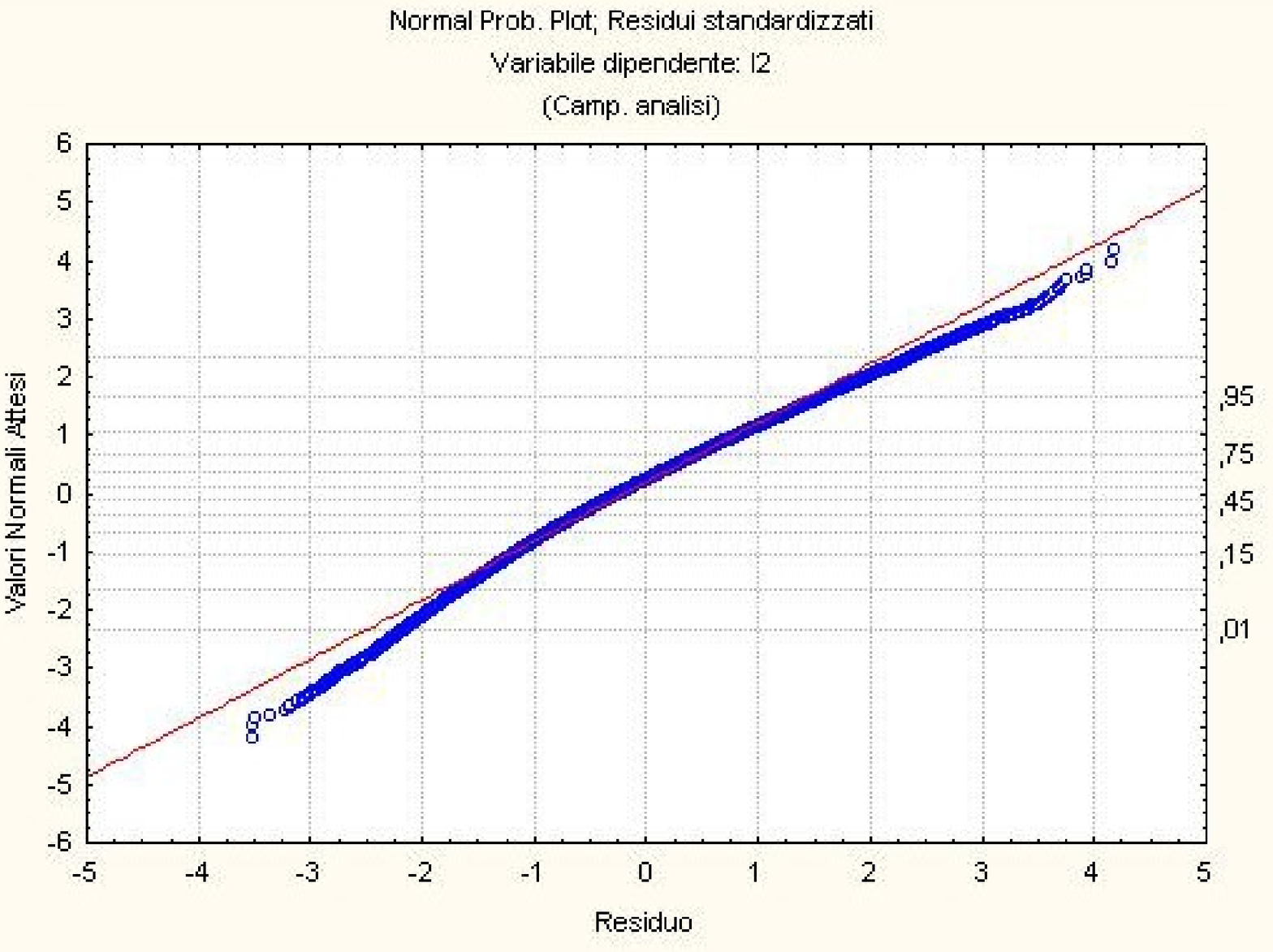,width=6.5cm}
        \epsfig{figure=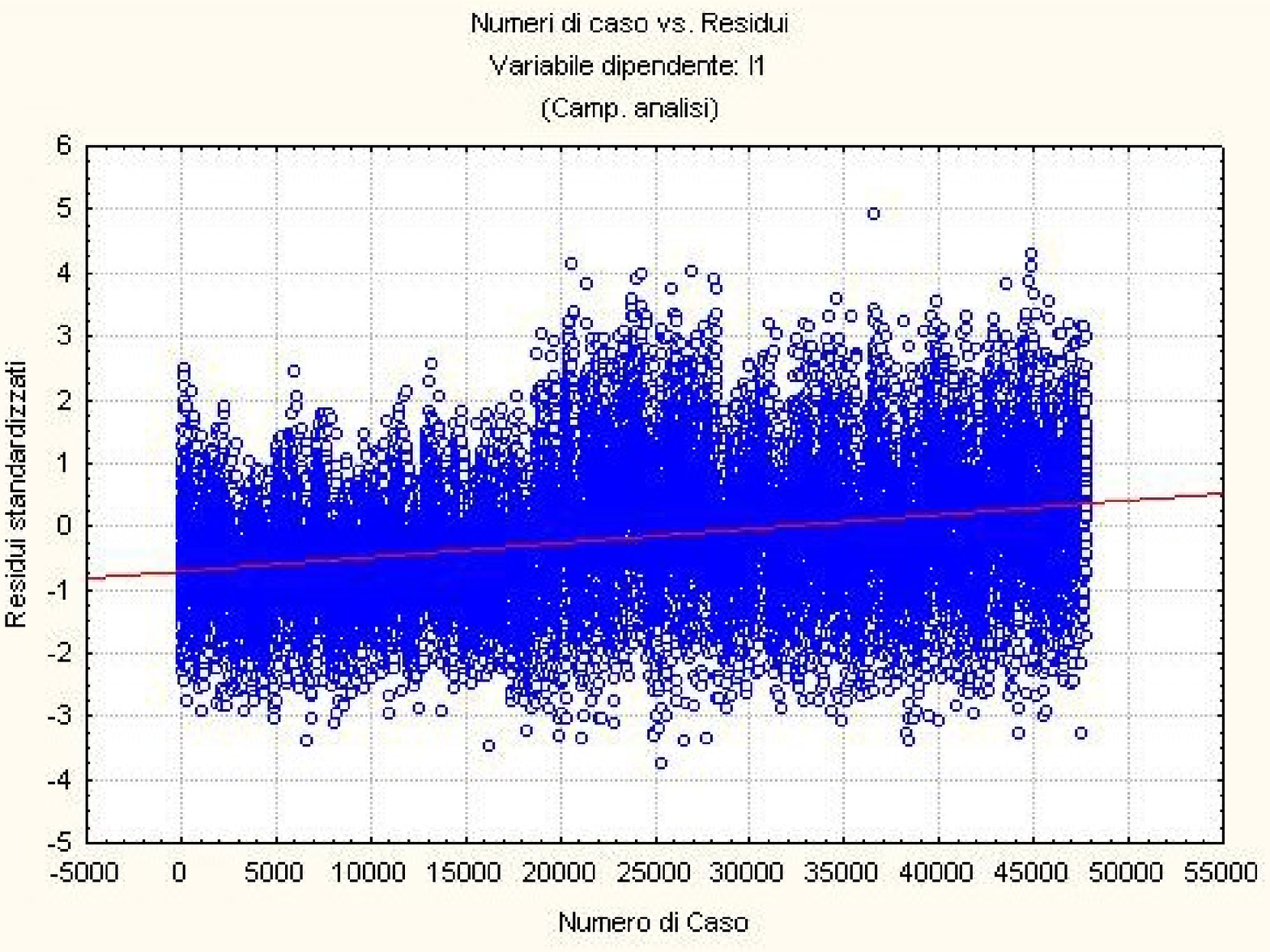,width=6.5cm}
        \epsfig{figure=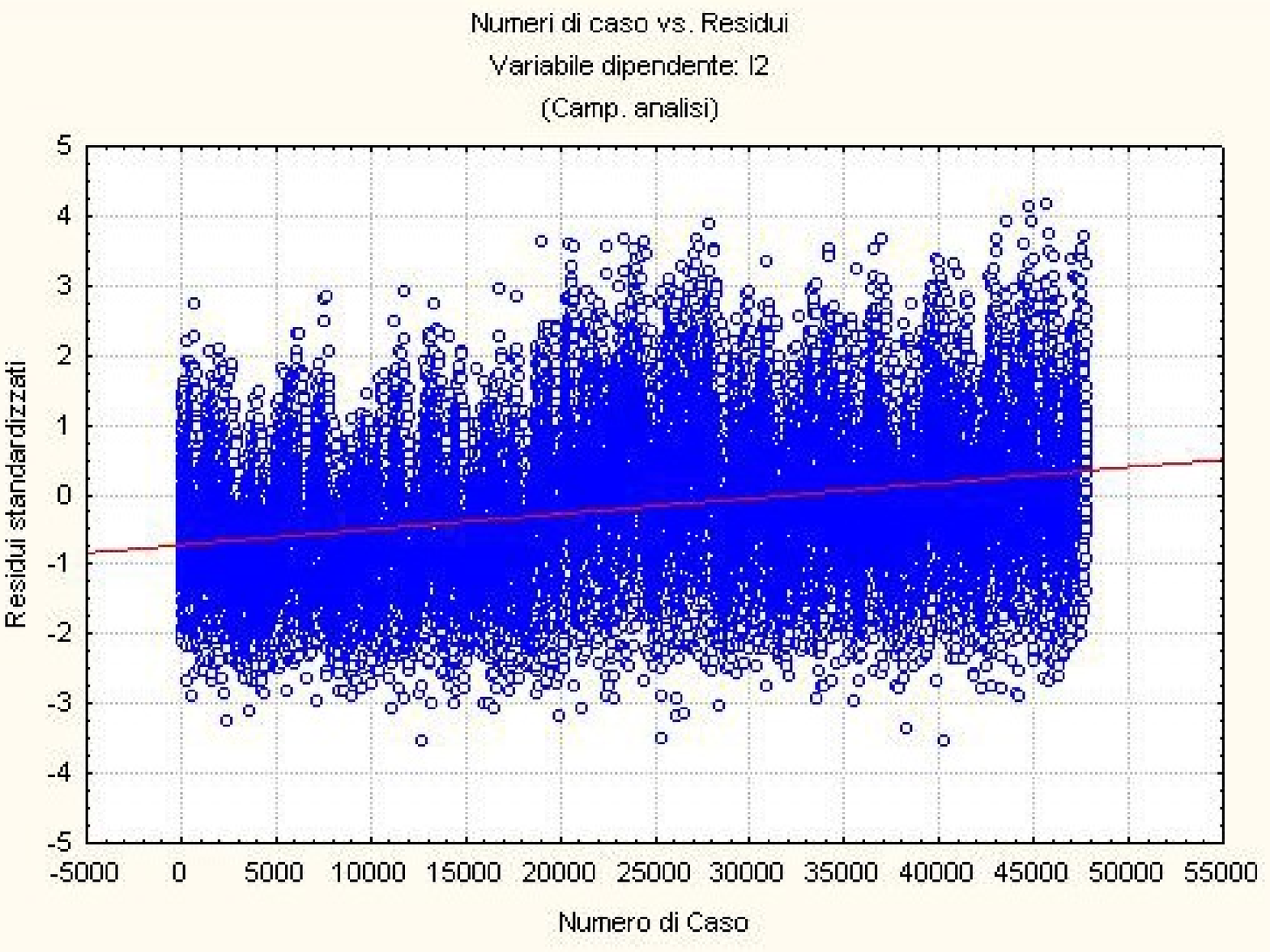,width=6.5cm}
        \caption{First row: normal probability plots of the standardized residuals for channels $I1$ and $I2$. Second row: scatterplot of residuals versus number of cases for $I1$ (left) and $I2$ (right).}
        \label{fig:lm_residuals}
    \end{center}
\end{figure*}

\noindent However, the analysis of the residuals shows that the residuals are not perfectly normally distributed, but they have long tails. The first row of figure \ref{fig:lm_residuals} reports the normal probability plots of the residuals for both $I1$ and $I2$.

\noindent The magnitude of the residuals, plotted against time (i.e. case number) in the second row of figure \ref{fig:lm_residuals}, changes with time, first increasing and then decreasing. We can suspect an inhomogeneous variance of the residuals.
\\

\begin{figure*}[!htb]
    \begin{center}
        \epsfig{figure=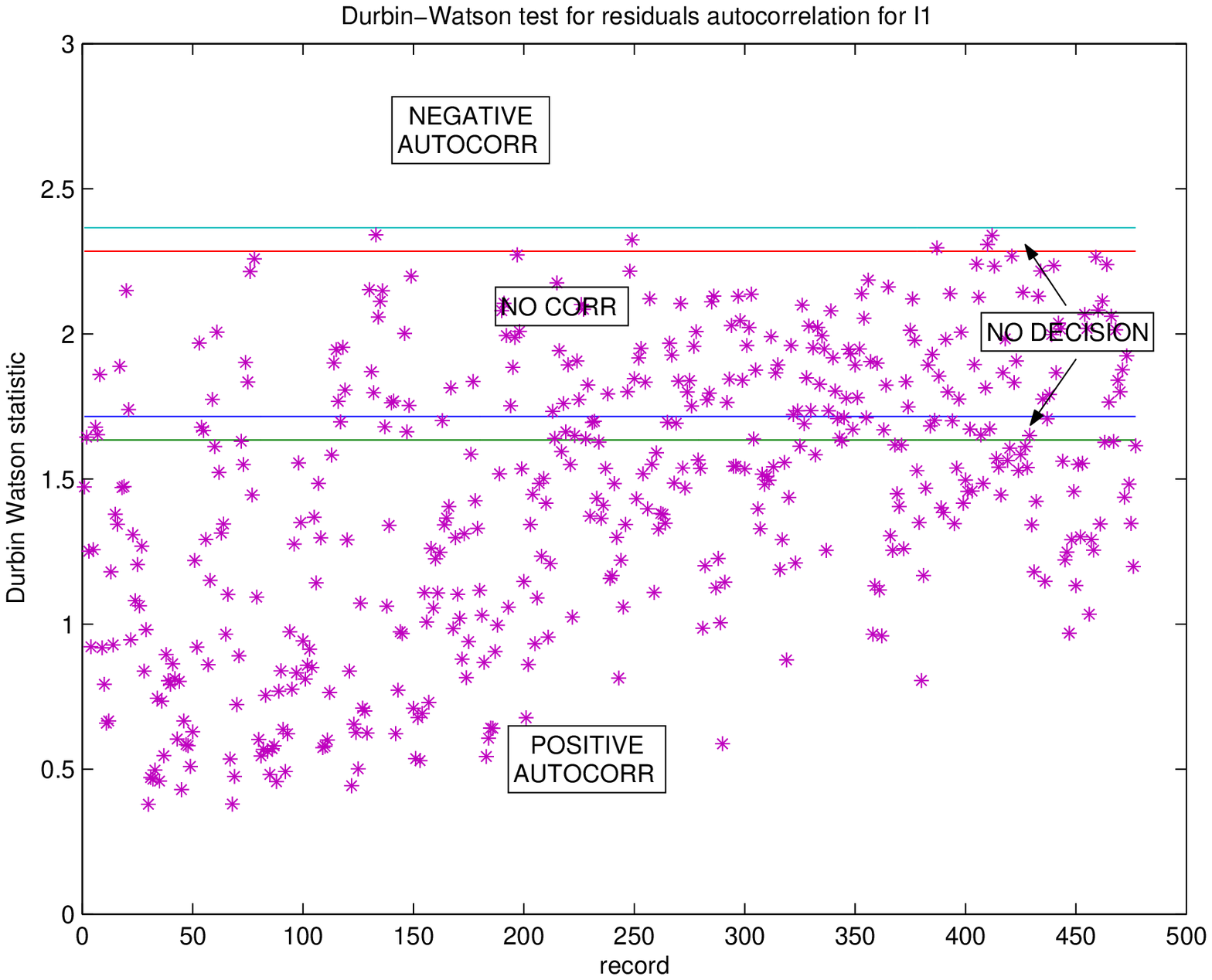,width=6.5cm}
        \epsfig{figure=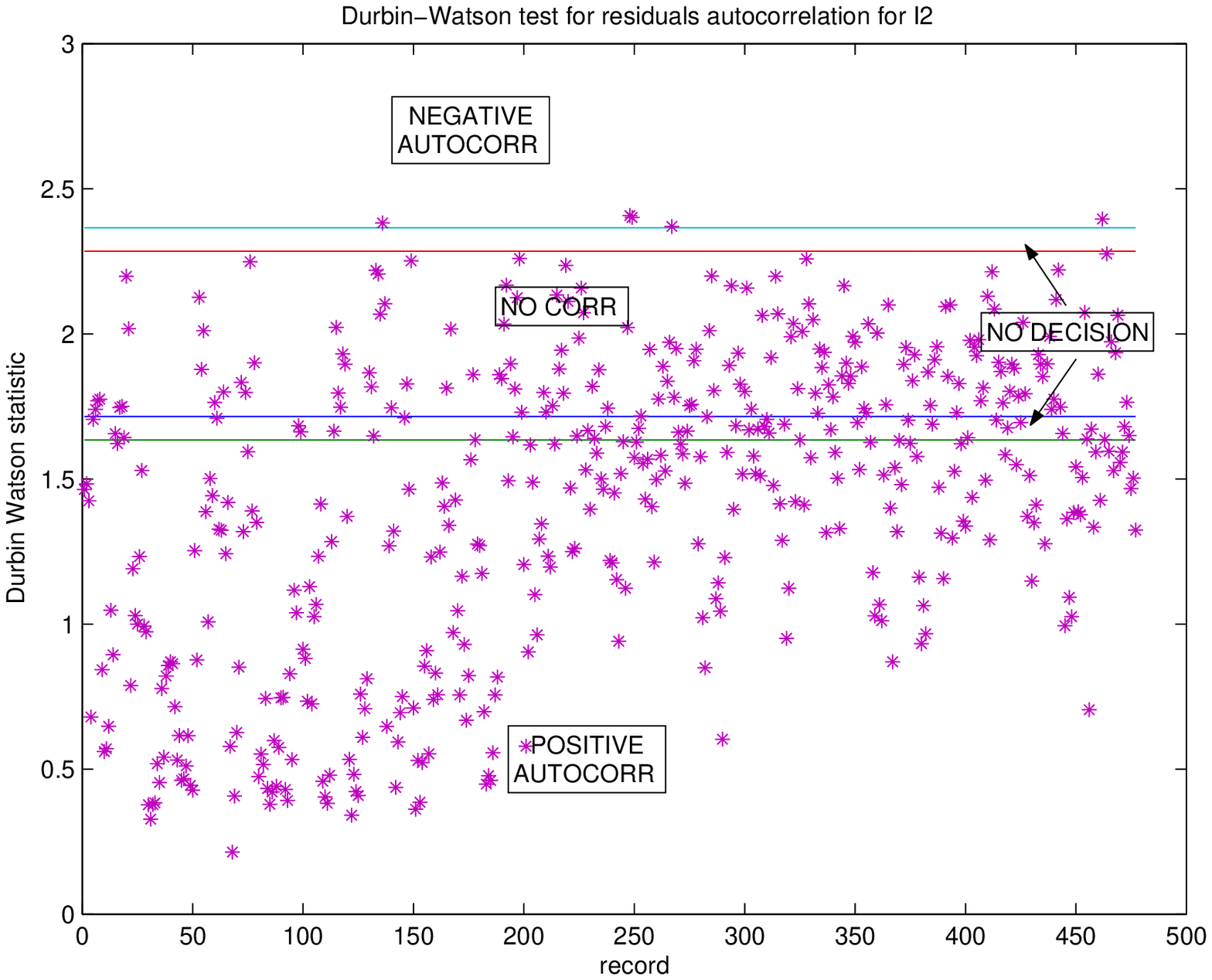,width=6.5cm}
        \caption{Durbin-Watson statistics for the residuals of $I1$ (left) and $I2$ (right).}
        \label{fig:DW-lm}
    \end{center}
\end{figure*}

\noindent The results of the Durbin-Watson test are shown in figure \ref{fig:DW-lm}. It is evident the presence of a positive correlation for a large number of sets.

\noindent Even if the model seems very good, caution is in order, due to the residual distribution and by their changing magnitude.

\subsection{Regression with mixed model}
\label{subsec:GLM_mixedModel}

\begin{figure*}[hbt]
   \begin{center}
	 	\epsfig{figure=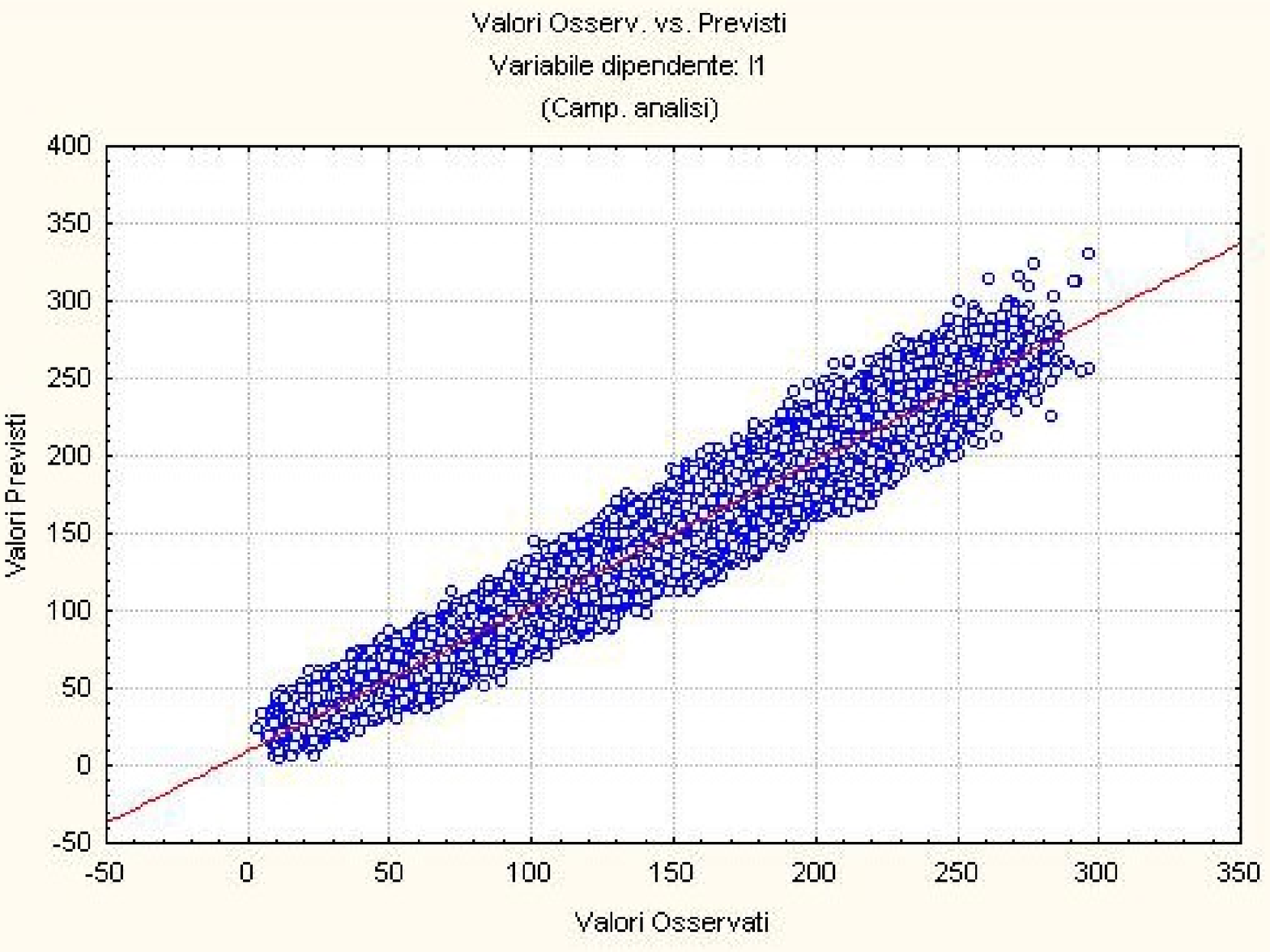,width=6.5cm}
	 	\epsfig{figure=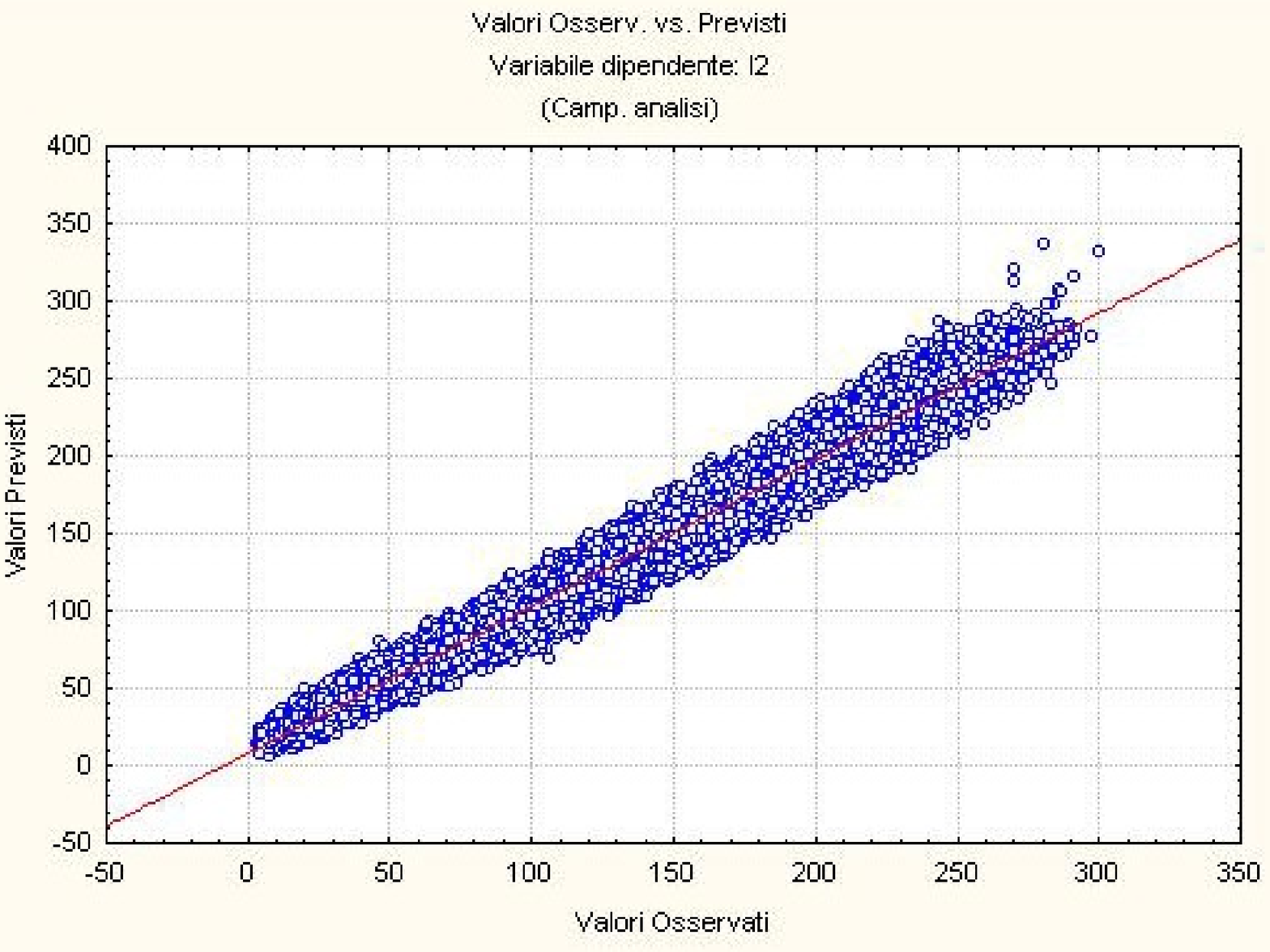,width=6.5cm}
	 	\caption{Scatterplot of observed versus predicted values for I1 and I2}
    \label{fig:mixlm_obsVsForeseen}
    \end{center}
\end{figure*}

\noindent We repeat the same analysis than in the previous paragraph, using the same records of data, but with a different regression model, without intercept but with higher order effects:
\begin{equation}\label{eq:GLM_mixedModel}
I_i = c_{Ai} PA + c_{Bi} PB + c_{ABi} PA*PB, \;\;\; i = 1,2
\end{equation}
where, again, the regressors are the photometric channels $PA$ and $PB$ and the dependent variables are the interferometric outputs $I1$ and $I2$.
\\
The fit of the model to the data is very good, as shown in figure \ref{fig:mixlm_obsVsForeseen} by scatterplots of the predicted vs. observed values. Again, there are no evident outliers.

\noindent

\begin{figure*}[!htbp]
    \begin{center}
        \epsfig{figure=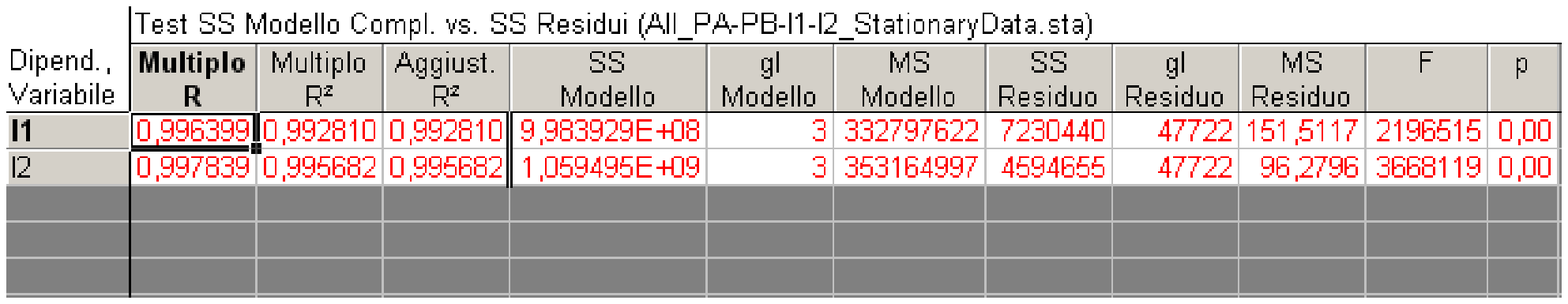,width=15cm}
        \vspace{0.5cm}
        \epsfig{figure=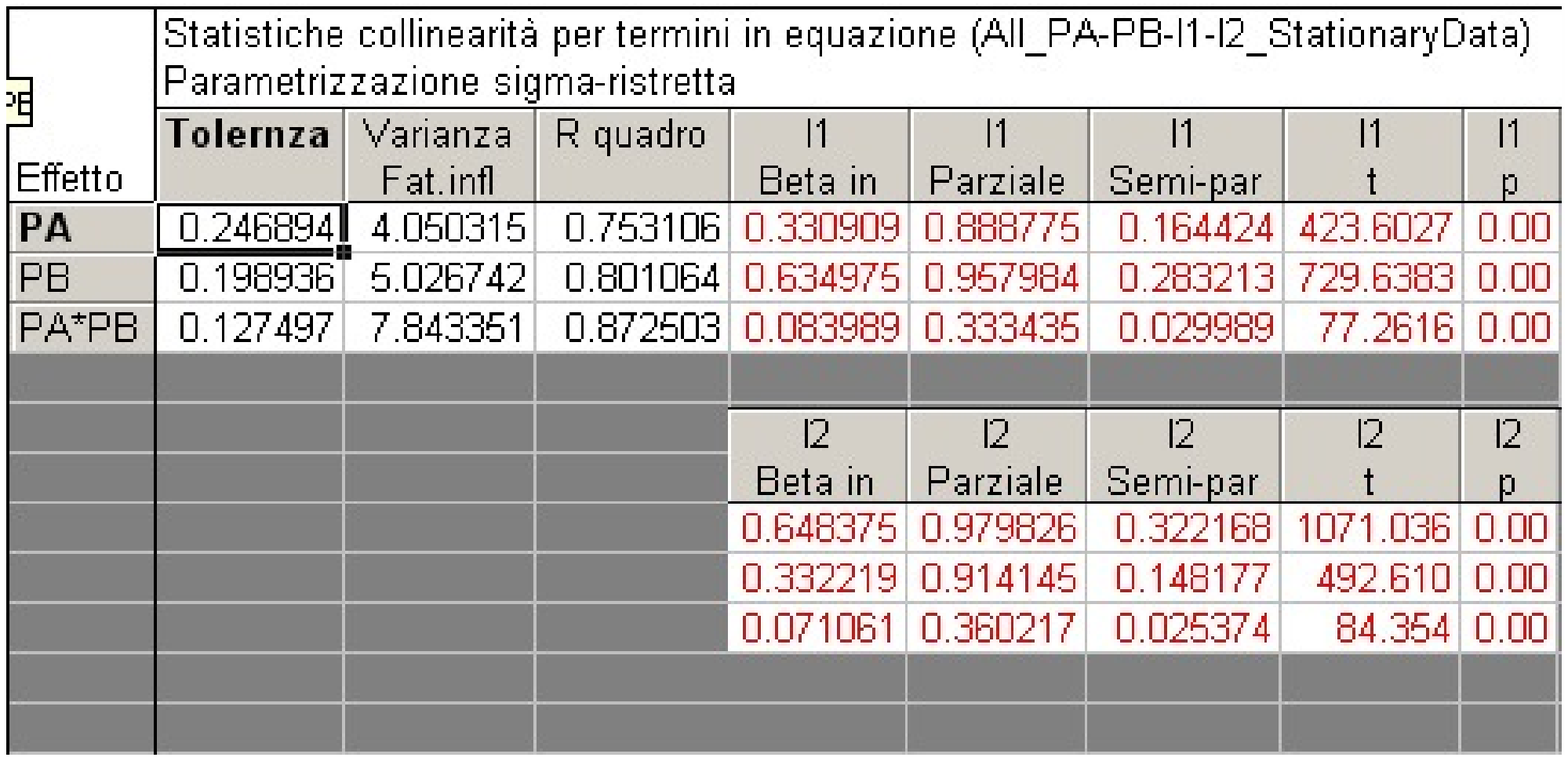,width=15cm,height=5cm}
        \vspace{0.5cm}
        \epsfig{figure=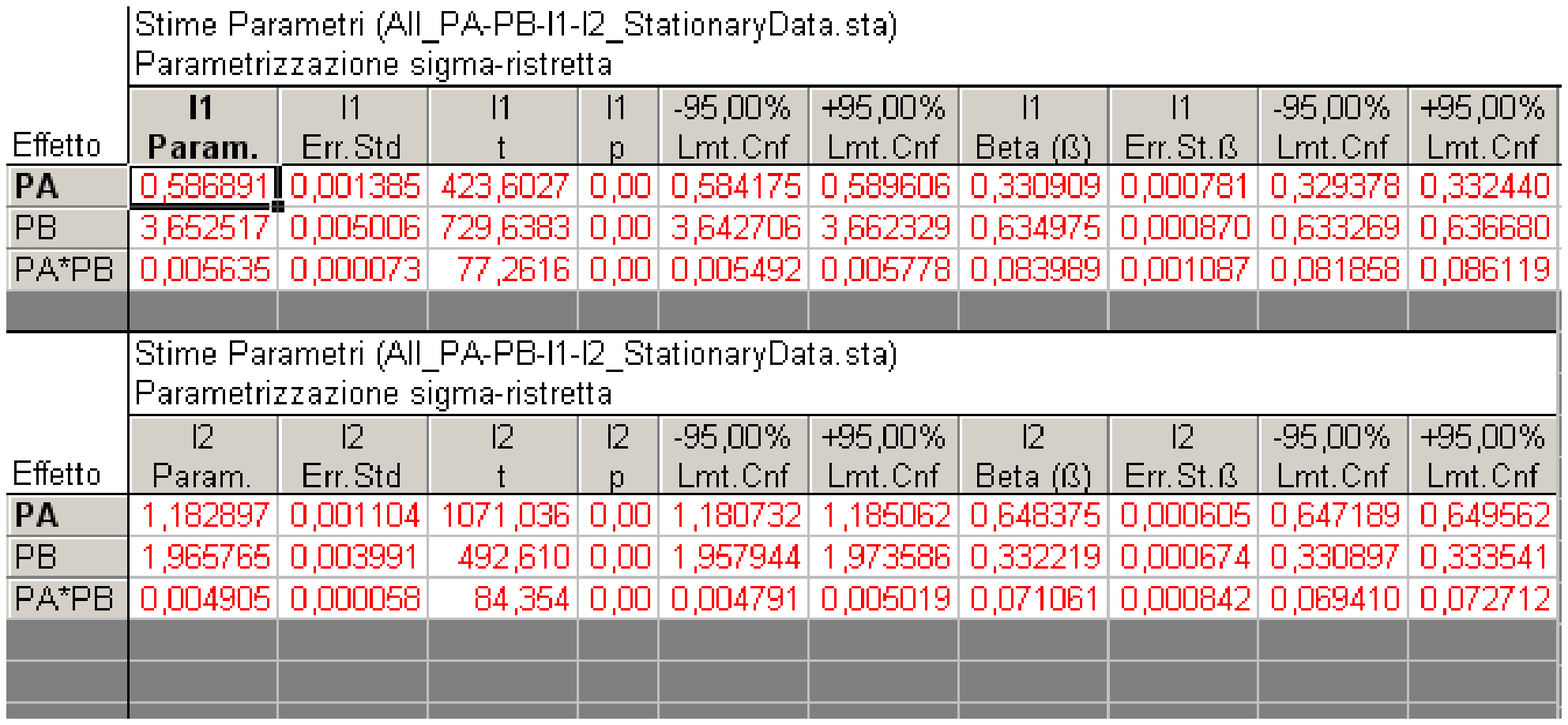,width=16cm,height=8cm}
        \caption{First row, table of tests on the model; second row, correlation analysis for $PA$, $PB$ and $PA*PB$; third row, estimation and statistical tests of regression model coefficients}
        \label{fig:mixlm_tabelle}
    \end{center}
\end{figure*}

\noindent Also this model gives a good explanation of the variance of the outputs $I1$ and $I2$. We can see in figure \ref{fig:mixlm_tabelle}, first row, that the $R^2$ values are the same, just a little better for $I1$.

\noindent Finally, in figure \ref{fig:lm_coeff} the B and $\beta$ coefficients, with their standard errors, the test t associated and the confidence intervals are plotted.

\noindent The correlation analysis of the independent variables (second row of figure \ref{fig:mixlm_tabelle}) shows that, for this kind of analysis, the mixed term can not be excluded, because the test of nullity has a p-value $< 0.01$. Of course its weight is reduced, being outside the coherence length, with respect to $PA$ and $PB$, as we could expect.\\
This term can be easily explained with the presence of the modulation function.
\\

\begin{figure*}[!htb]
    \begin{center}
        \epsfig{figure=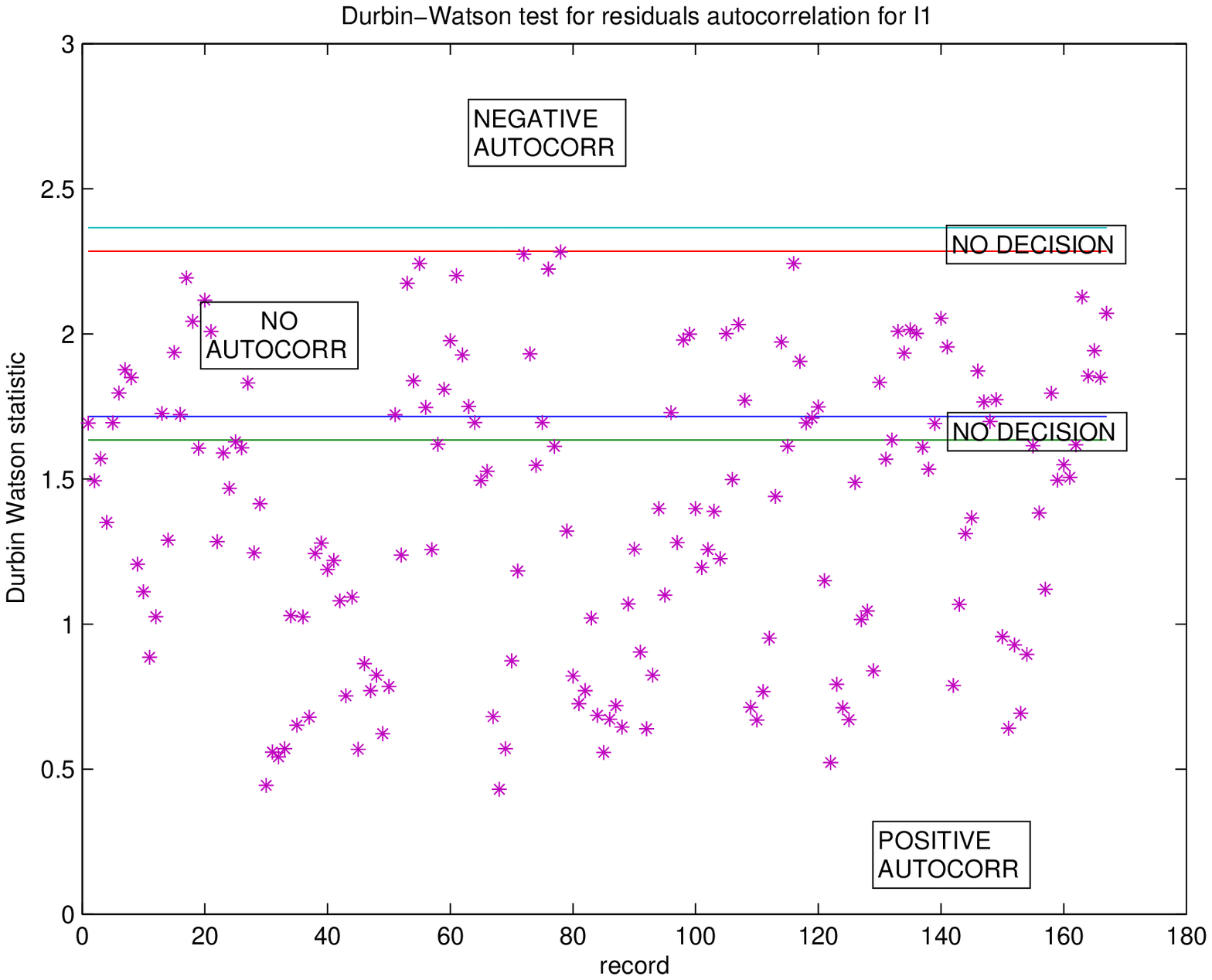,width=6.5cm}
        \epsfig{figure=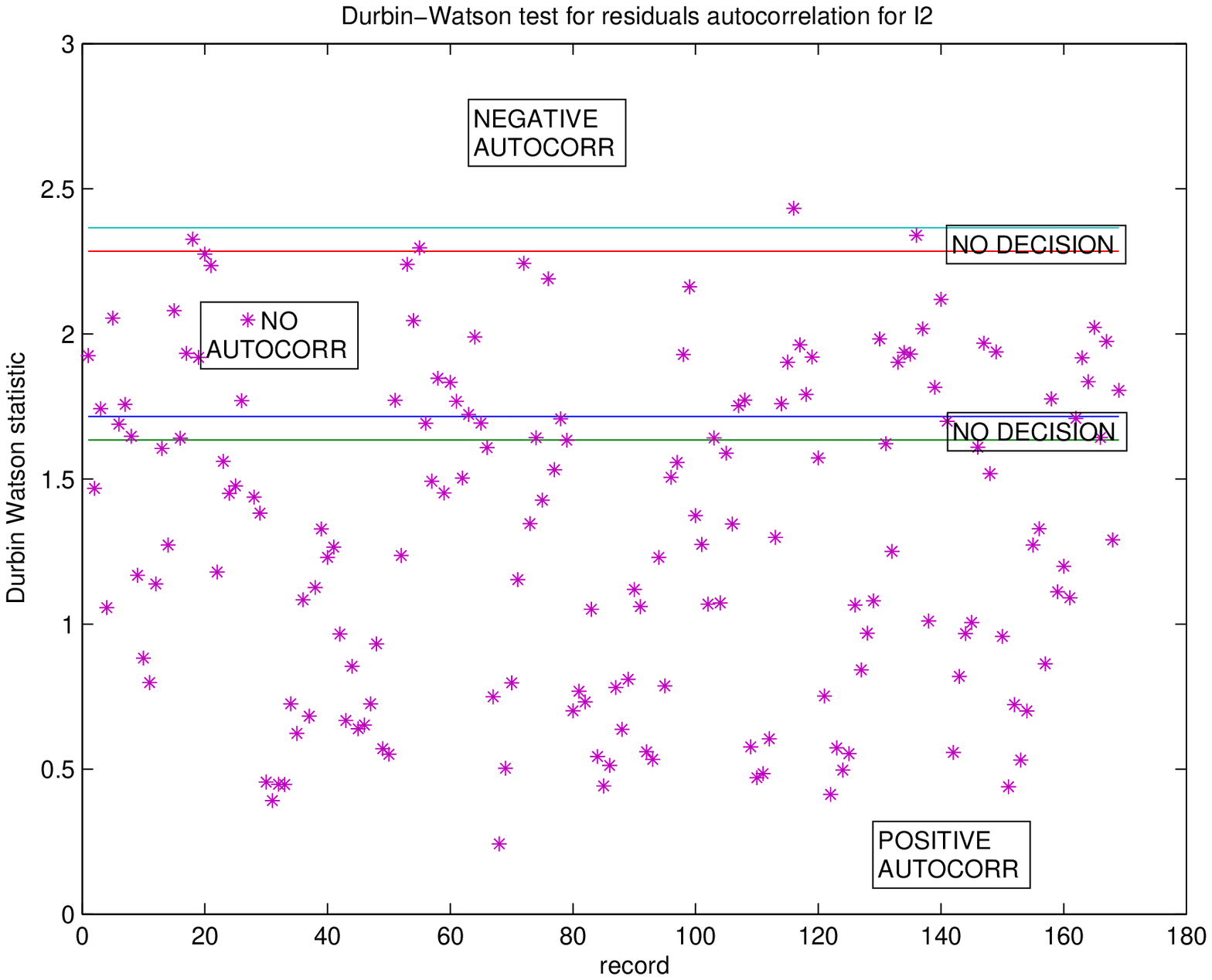,width=6.5cm}
        \caption{Durbin-Watson statistics for the residuals of $I1$ (left) and $I2$ (right).}
        \label{fig:DW-mixlm}
    \end{center}
\end{figure*}

\noindent We now execute the Durbin-Watson test to check for an autocorrelation of the raw residuals. Figure \ref{fig:DW-mixlm} shows the test results for the residuals of the dependent variables for a number of subsequent sets, each of them 100-samples sized. It is evident the presence of a positive correlation for a large number of sets.

\begin{figure*}[!htb]
    \begin{center}
        \epsfig{figure=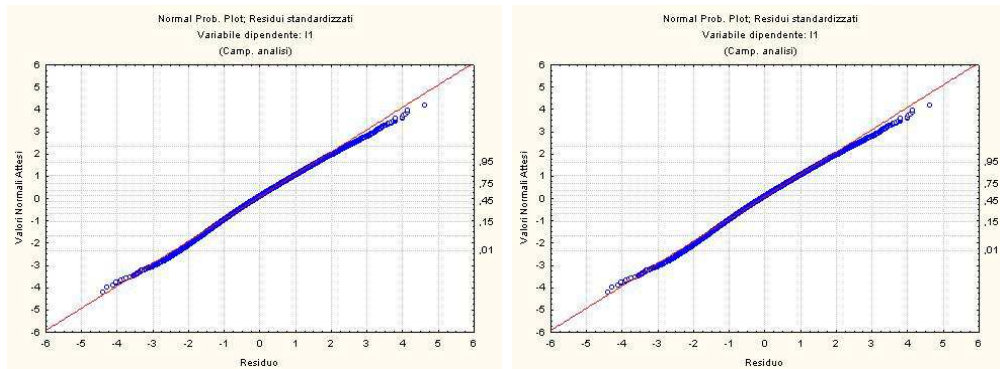,width=6.5cm}
        \epsfig{figure=graphics_cap4/mixlm_npp_resStd_I1.eps,width=6.5cm}
        \caption{Normal probability plots for the standardized residuals for $I1$ (left) and for $I2$ (right) in the mixed linear model.}
        \label{fig:mixlm_residuals}
    \end{center}
\end{figure*}

\noindent We have to look again at the standardized residuals in figure \ref{fig:mixlm_residuals}. We can notice, comparing with figure \ref{fig:lm_residuals}, that the distribution of the residuals of the mixed model is closer to a normal one (first row) and that the magnitude of the residuals is more uniform (second row). We can conclude that in the previous case the residuals contained the variability caused by the factor of higher order.

\noindent This fact is in some way surprising because we have chosen data with homogenous variance. If this higher order term had a strong impact, data should not be homogenous, as shown by simulations. Hence, we can say that presence of noise makes data to be homogeneous! The faint coefficient of the mixed term gives a sort of ratio of modulation / noise.

\section{Conclusions for regression analysis}
\label{sec:GLM_mixedModel}

We are now able to give an answer to the questions we posed at the beginning of this section.\\

\noindent First of all, from the analysis of the $R^2$ multiple coefficients of both models we can say that the variability on the inputs was able to explicate almost all the variability on outputs. There is of course a small part of variability unexplained. Physically, we can identify this quantity with instrumental contribution to the noise of the system. Its low magnitude means that the system does not add strong perturbations to the outputs.\\
We could also give a quantitative estimation of this variability: less than $1\%$. Moreover, the characteristics of the residuals give us some information on this contribution. We can be confident that it is normally distributed, and its variance contains the non-linearity of the combination process. To this variance we have to add the contribution given by the difference of the coefficients of the calibration analysis with respect to the analysis with both channels: we can think of it as a component due to interaction of the two input channels.\\

\noindent However, there are some negative considerations to do. In fact, we could just use a part of our initial data (about 25\% of the total data), to restrict the analysis to the records with homogeneous data. To enlarge the available data, it is necessary to tailor the regression, introducing weights.\\ 
To use data with inhomogeneous variance on the regressors, so on $PA$ and $PB$, we should have more information about the data nature, to be able to properly describe the distribution of the regressor random variable, and to identify measurement errors.
\\

\noindent Finally, the answer to the third question comes from the comparison of the linear and the mixed linear model. The residuals of the first one have some inhomogeneity on the variance that is explained by the latter. The presence of the higher order mixed factor means that the low-magnitude modulation outside the coherence length is not negligible. However, it is very small. The Levene test for the homogeneity of variance, applied to simulations of an ideal interferogram (see eq. \ref{eq:FINITO_model}) without noise, has given evidence of non homogeneous variance for almost all cases (side lobes of the interferometric pattern). This means that the noise covers this patterns, at least in the considered records, but it is still identifiable thanks to the higher order factor. 

\noindent Finally, we have to remark that the presence of serial correlation between residuals does not influence the bias of the estimators, but their variance: they are no longer the best estimators. This fact affects especially the estimation of the photometric coefficients, since they are used in the normalization of interferometric signals. Some authors (see, e.g., Rawlings) have proposed a prior transformation of the regression variables before performing the analysis; but Rawlings 
also says that it is always better to retain a good simple model, even in presence of inhomogeneity of variance or non-normality.

\section{Future improvements}
\label{sec:future}  

First of all, the validity of the analysis described in this chapter is till now limited to the data set considered for the tests. Now that a set of statistical instruments is identified and checked, it would be useful to extend the analysis to other data set, both from VLTI and from other interferometric instruments, to separate peculiar from general features.
\\

\noindent The analysis performed till now suffered from the lack of theoretical information on the handled signals, in particular on their noise statistics. As explained in the introductive paragraph, a stochastic model would solve many uncertainties based on the direct estimation of important features from data.
As an example, we know that our data have measurement errors:
\begin{eqnarray}
\nonumber Y_i = \tilde{Y}_i + \varepsilon_i\\
X_i = \tilde{X}_i + \delta_i
\end{eqnarray}
where $Y$ and $X$ are the measured dependent variables and regressors, $\tilde{Y}$ and $\tilde{X}$ are the true values, and $\varepsilon$ and $\delta$ are the errors of the measure, and finally $i$ ranges over the number of data samples.\\
Hence we solve the regression model:
\begin{equation}
Y_i = \beta X_i + \epsilon_i
\end{equation}
instead of:
\begin{equation}
\tilde{Y}_i = \beta \tilde{X}_i + \mu_i
\end{equation}
where $\mu$ is the vector of the errors of the hidden regression model. We have mentioned that, following Draper \& Smith, the joint moments of the random variable $\tilde{X}_i$ and $\delta_i$ can be used to correct the estimators of the coefficients.

\noindent This is an important issue especially with respect to the determination of the coefficients of normalization for signals using the photometric information (as we have seen for FINITO fringe tracker, see chap. \ref{chap:FINITO}).
\\

\begin{figure*}[!htb]
    \begin{center}
        \epsfig{figure=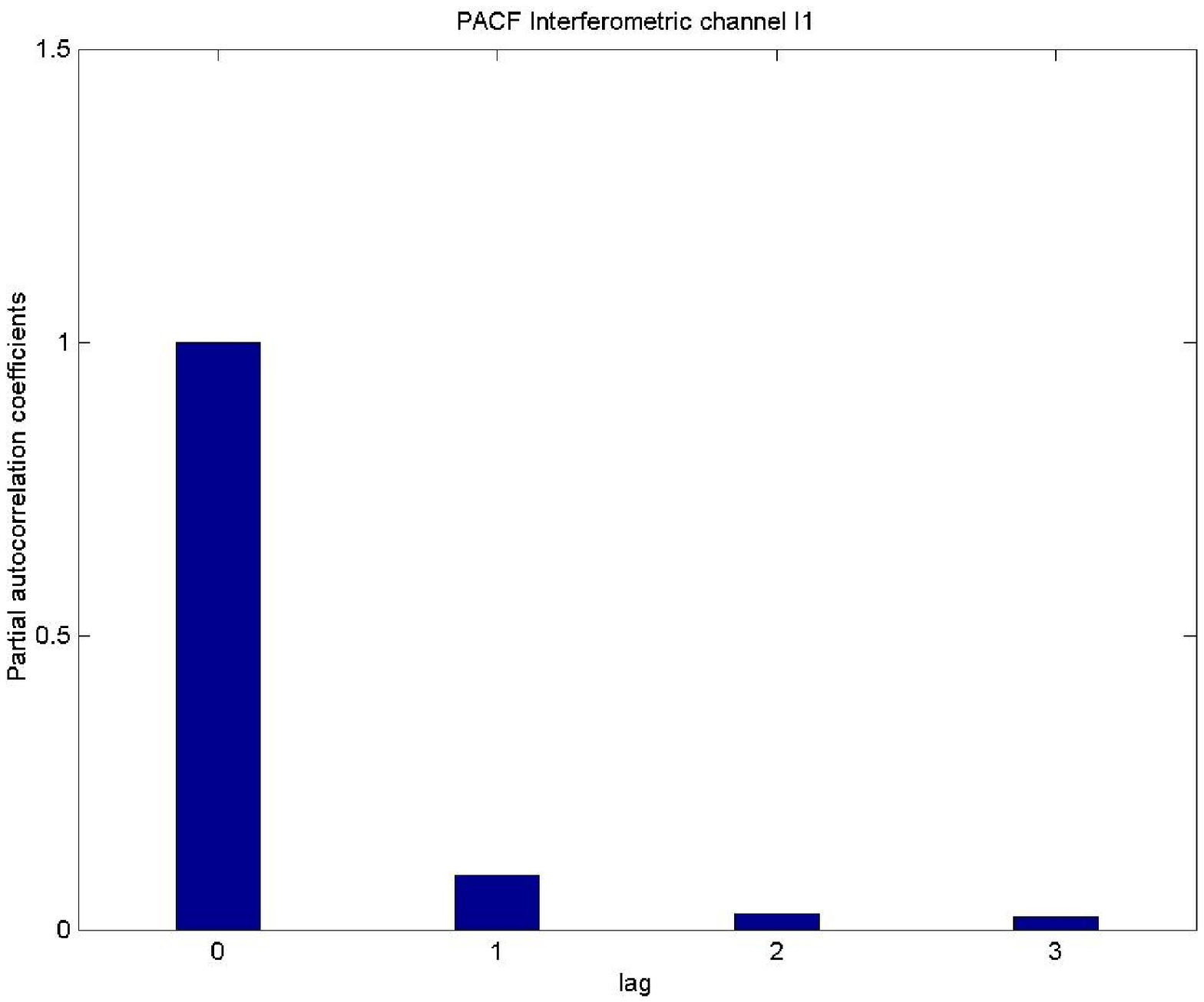,width=6.5cm}
        \epsfig{figure=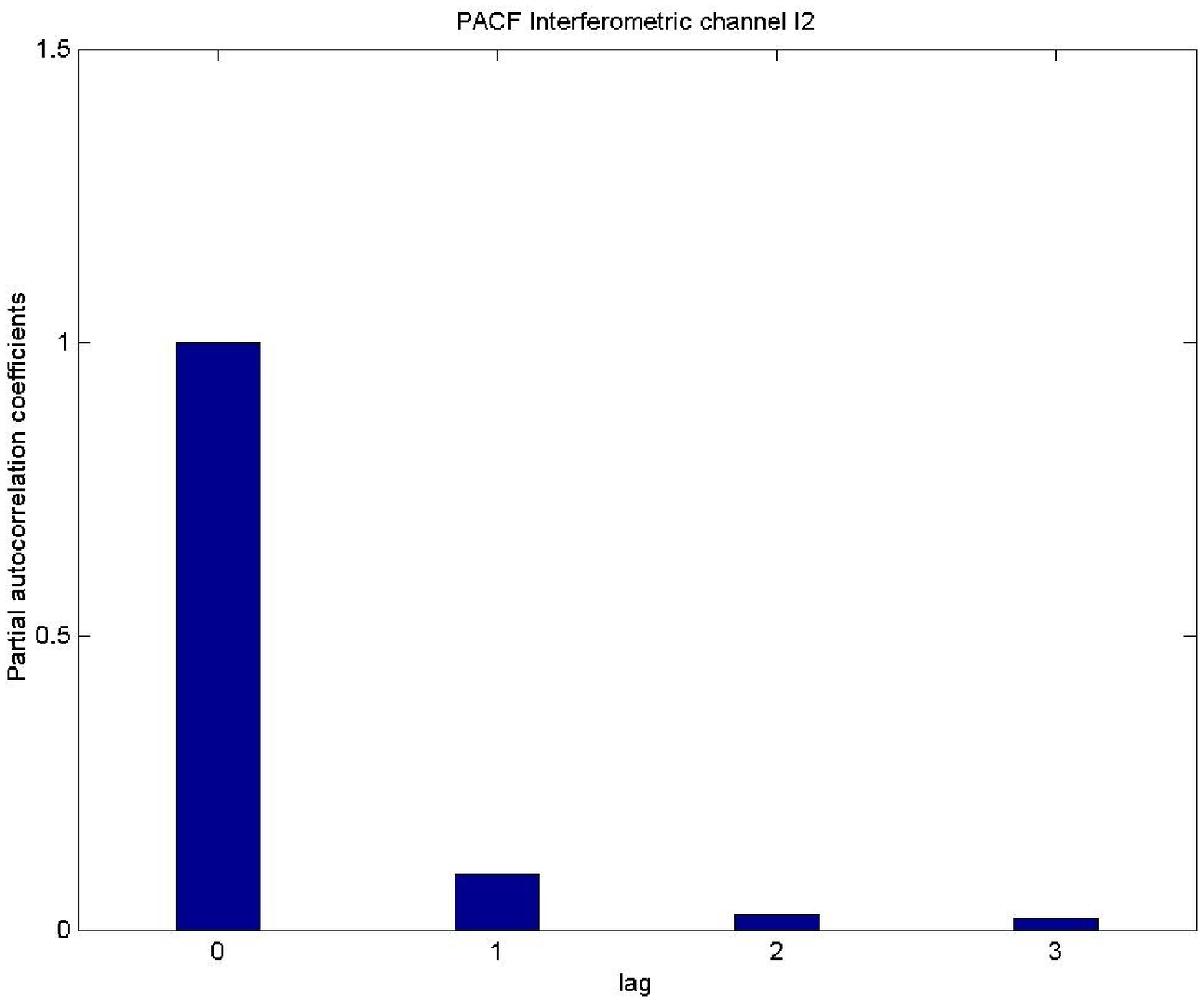,width=6.5cm}
        \epsfig{figure=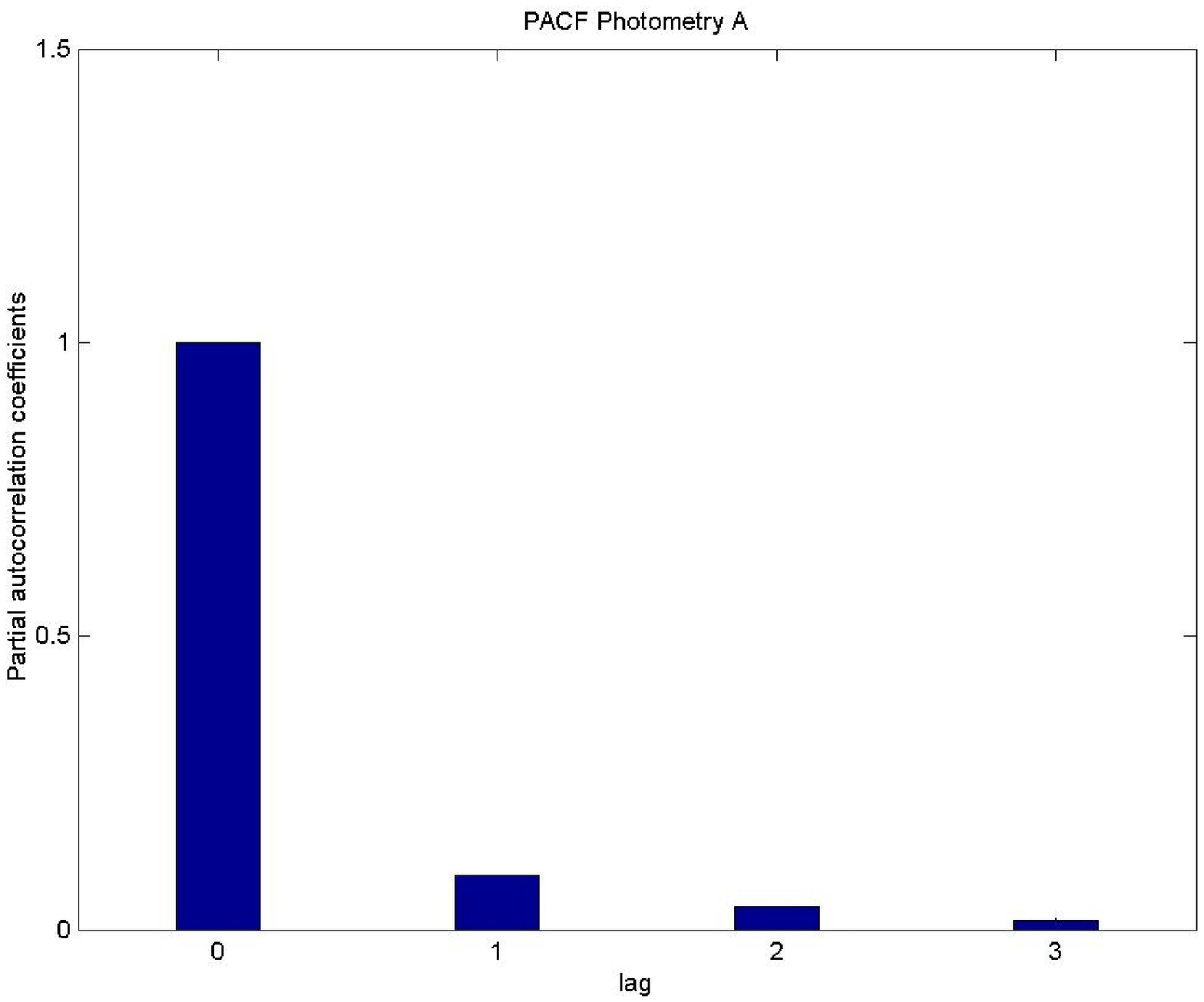,width=6.5cm}
        \epsfig{figure=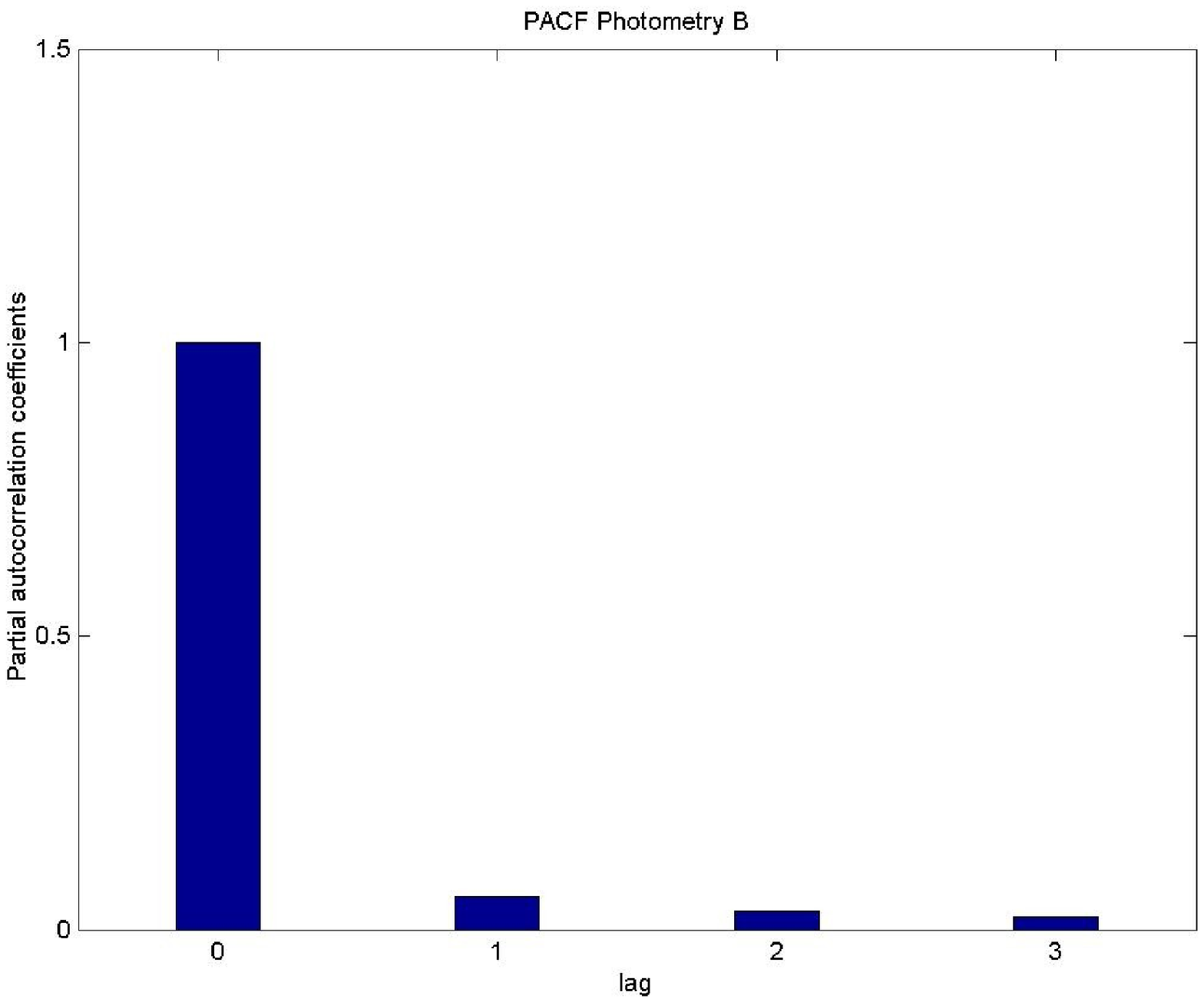,width=6.5cm}
        \caption{Partial autocorrelation points for interferometric channels (first row) and photometric channels (second row) when both arms are fed with flux. It can be seen that the PACF decreases exponentially to zero.}
        \label{fig:partial_autocorr}
    \end{center}
\end{figure*}

\noindent Preliminary works in this direction showed that the data are not easy to understand. The presence of the trend, the form of the autocorrelation of photometric data (see par. \ref{sec:time-stat}) seems to suggest a process with a memory. To refine the field of this kind of process, we have considered the partial autocorrelation function: the correlation at each lag is purified from the contribution of precedent lags. Figure \ref{fig:partial_autocorr} shows the first lags \cite{Dagum} for the observational case 4. It is clear that the partial autocorrelation points decreases exponentially to zero. It could be a moving average model, as well as an autoregressive one, or the composition of both. It is an intermediate situation, that needs a careful analysis.\\
\\

\noindent Statistical tools and software, such as Statistica, can be of help in this research.

\appendix
\chapter{Bias in the power spectral density function due to a trend removal}
\lhead[\fancyplain{}{\bfseries\thepage}]%
      {\fancyplain{}{\bfseries Bias in the spectral density function}}
\rhead[\fancyplain{}{\bfseries Bias in the spectral density function}]%
      {\fancyplain{}{\bfseries\thepage}}
\label{appendixA}
The scope of this section is the evaluation of the bias introduced on the power spectral density function (PSD) by the subtraction of an {\it estimated} linear trend. We first introduce the statistical estimators for the sample moments used in the next paragraphs, with their properties. Then we proceed to the estimation on the bias. Following \cite{Manolakis}, we will limit our analysis to wide sense stationary processes, in order to take advantage of the relationship between the autocorrelation function and the PSD. We first recall the properties of the spectral density function of a signal with zero mean, as proposed by Manolakis\cite{Manolakis} (par. \ref{subsec:const_mean}), then we consider the case of a signal with a trend. First we suppose the presence of a non-zero mean (\ref{subsec:nonzero_mean}), then of a general trend (par. \ref{subsec:bias_regression}).

\section{Statistical estimators}
\label{sec:estimator_properties}

We choose, as estimator of the mean, the variance and the autocorrelation function the correspondent time-sample estimators. We use the notation {\it time}-sample to highlight the fact that we estimate them directly from a set of subsequent samples in time, instead of a set of realizations.
Their properties are known in literature, see, e.g., Priestley\cite{Priestley}. We report here definition and properties, following the notation of \cite{Priestley}.
In the following, we will refer to signals at least stationary in the wide sense, i.e. such that the moments up to order two are not dependent on time.

\subsubsection{Time-sample mean}
\label{subsubec:mean_prop}

The {\it time-samples} mean, evaluated over the set of samples $\{X(k), \; k = 1, \ldots, N\}$, is an estimator of the mean, supposed to be constant:
\begin{equation}
\hat{\mu}_{X,N} = \frac {1} {N} \sum_{k=1}^{N} X(k)
\end{equation}

\noindent This estimator is unbiased, since \(E[\hat{\mu}_{X,N}] = \frac{1}{N} \: N \: E[X(k)] = \mu_X \). If the samples $X(k)$ are uncorrelated, it is also asymptotically consistent:
\begin{equation}\label{eq:uncorr_varOfmean}
\mbox{var}(\hat{\mu}_{X,N}) = \sigma^2_{\hat{\mu}_{X,N}} = \frac{\sigma^2_{X_s}} {N} \rightarrow_{N \rightarrow \infty} 0.
\end{equation}
But if the samples are correlated, the variance becomes:
\begin{eqnarray}
\nonumber \sigma^2_{\hat{\mu}_{X,N}} = E \left[ \frac{1}{N} \sum_{k=1}^{N} (X(k) - \mu_X) \frac{1}{N} \sum_{h=1}^{N} (X(h) - \mu_X)  \right] = \\
= \frac{1}{N^2} \sum_{k=1}^{N} \sum_{k=1}^{N} E [(X(k) - \mu_X)(X(h) - \mu_X)] = \frac{1}{N^2} \sum_{k=1}^{N} \sum_{h=1}^{N} \gamma_X(h -k)
\end{eqnarray}
where $\gamma_X(l) = \mbox{cov}(X(h),X(h+l))$; setting $r = h - k$, we obtain:
\begin{eqnarray}
\nonumber \sigma^2_{\hat{\mu}_{X,N}} = \frac{1}{N^2} \sum_{r= -(N-1)}^{N-1} (N- |r|) \gamma_X(r) = \\
= \frac{\sigma^2_X}{N} \sum_{r= -(N-1)}^{N-1} \left( 1 - \frac{|r|}{N} \right) \rho_X(r) .
\end{eqnarray}
We have, if N goes to infinity:
\begin{equation}\label{eq:samplemean_var}
\sigma^2_{\hat{\mu}_{X,N}} \rightarrow_{N \rightarrow \infty} \frac{\sigma^2_X}{N} \sum_{r= -\infty}^{+\infty} \rho_X(r)
\end{equation}
for the limiting form of the function \(g(r) = 1 - \frac{|r|}{N}\). If X is such that its autocovariance function $\gamma_X$ possesses a Fourier Transform $f(w)$, we find that
\begin{equation}
\frac{\sigma^2_X}{N} \sum_{r= -\infty}^{+\infty} \rho_X(r)  \rightarrow_{N \rightarrow \infty} \frac{\sigma^2_X 2 \pi}{N}  f(0) \rightarrow_{N \rightarrow \infty}
\end{equation}

\noindent So this estimator for the time mean is unbiased and asymptotically consistent.

\subsubsection{Time-sample variance}
\label{subsubec:var_prop}

For the estimation of the variance of $X(k)$, we use the time-samples variance with unknown mean $\hat{\mu}_{X,N}$:
\begin{equation}\label{eq:sample_var}
\hat{\sigma}^2_{X,N} = \frac {1} {N-1} \sum_{k=0}^{N-1} [X(k)-\hat{\mu}_{X,N}]^2
\end{equation}

\noindent If the samples are uncorrelated, this estimator is unbiased, while
\begin{equation}
\tilde{\sigma}^2_{X,N} = \frac {1} {N}  \sum_{k=0}^{N-1} [X(k)-\hat{\mu}_{X,N}]^2
\end{equation}
is biased. Moreover, it is asymptotically consistent.\\
If the samples are correlated, we obtain:
\begin{eqnarray}
\nonumber \hat{\sigma}^2_{X,N} &=& \frac{1}{N} \sum_{k=1}^{N} (X(k) - \hat{\mu}_{X,N})^2 = \\
\nonumber = \frac{1}{N} \sum_{k=1}^{N} (X(k) - \mu_X)^2 &-& \frac{2}{N} \sum_{k=1}^{N} (X(k) - \mu_X)(\hat{\mu}_{X,N} - \mu_X) + \\
&+& \frac{1}{N} \sum_{k=1}^{N}(\hat{\mu}_{X,N} - \mu_X)^2
\end{eqnarray}
having added and subtracted the quantity $\mu_X$. So the expectation becomes:
\begin{eqnarray}
\nonumber E[\hat{\sigma}^2_{X,N}] = \frac{1}{N} E \left[ \sum_{k=1}^{N} (X(k) - \mu_X)^2 \right] &+& \\
\nonumber - \frac{2}{N} E \left[ \sum_{k=1}^{N} (X(k) - \mu_X)(\hat{\mu}_{X,N} - \mu_X) \right] &+& \frac{1}{N}E \left[\sum_{k=1}^{N}(\hat{\mu}_{X,N} - \mu_X)^2 \right] =
\end{eqnarray}
\begin{eqnarray}\label{eq:var_exp}
\nonumber = \frac{1}{N} N \sigma^2_X &-& 2 E \left[ (\hat{\mu}_{X,N} - \mu_X) \sum_{k=1}^{N} (X(k) - \mu_X) \right] + E \left[\sum_{k=1}^{N}(\hat{\mu}_{X,N} - \mu_X)^2 \right] = \\
\nonumber &=& \sigma^2_X - E \left[ (\hat{\mu}_{X,N} - \mu_X)^2 \right] = \sigma^2_X - \sigma^2_{\hat{\mu}_{X,N}} = \\
&=& \sigma^2_X - \frac{\sigma_X^2}{N} \sum_{r=-(N-1)}^{N-1} \left(1- \frac{|r|}{N}\right) \rho_X(r)
\end{eqnarray}
As the variance of the time mean is asymptotically consistent, the expectation of the time variance is biased, but asymptotically unbiased.
\\

\noindent The variance of this estimator will be given in next paragraph as a particular case of the covariance estimator.

\subsubsection{Time-sample autocovariance function}
\label{subsubsec:autocov_prop}

For the autocovariance function, we use the time-samples autocovariance function $\hat{\gamma}_{X} (l)$, with $X$ real. As we will use it later, we distinguish the cases in which the mean is known or unknown.
\begin{enumerate}
\item{known mean $\mu$:

\begin{equation}
\nonumber \hat{\gamma}_{X} (l) = \left \{ \begin{array} {ll}
                    \frac{1}{N-l} \sum_{n = 1}^{N-l}[X(n+l)-\mu] [X(n)-\mu]    & \mbox{if $0 \leq l \leq N-1$} \\
                    \hat{\gamma}_X (-l)  & \mbox{if $-(N-1) \leq l \leq 1$}\\
                    0   & \mbox{otherwise}
                                        \end{array}
                             \right.
\end{equation}
\begin{equation}
= \left \{ \begin{array} {ll}
                    \frac{1}{N-|l|} \sum_{n = 1}^{N-|l|}[X(n+l)-\mu] [X(n)-\mu]    & \mbox{if $0 \leq |l| \leq N-1$} \\
                    0   & \mbox{otherwise}
                                        \end{array}
                             \right.
\end{equation}
If $1 \leq |l| \leq (N-1)$, the expectation becomes:
\begin{eqnarray}\label{eq:autocov_exp_unbiased}
\nonumber E[\hat{\gamma}_{X} (l)] &=& \frac{1}{N-|l|} \sum_{n = 1}^{N-|l|}E\left[\{X(n+l)-\mu\} \{X(n)-\mu \} \right] = \\
&=& \frac{1}{N-|l|} \sum_{n = 1}^{N-|l|} \gamma_X (l) = \frac{N-|l|}{N-|l|} \gamma_X (l)
\end{eqnarray}
and it is null otherwise, so this estimator is unbiased.

}

\item{ unknown mean $\mu$, estimated with $\hat{\mu}$
\begin{equation}\label{eq:sample_autocov}
\hat{\gamma}_X (l) = \left\{ \begin{array} {ll}
                    \frac{1}{N-l} \sum_{n = 1}^{N-l}[X(n+l)-\hat{\mu}] [X(n)-\hat{\mu}]    & \mbox{if $0 \leq l \leq N-1$} \\
                    \hat{\gamma}_X (-l)  & \mbox{if $-(N-1) \leq l \leq 0$} \\
                    0   & \mbox{otherwise}
                                        \end{array}
                             \right.
\end{equation}
Adding and subtracting $\mu$,for $1 \leq |l| \leq (N-1)$ we have: 
\begin{eqnarray}
\nonumber E[\hat{\gamma}_X (l)] &=& \frac{1}{N-|l|} E \left[ \sum_{n = 1}^{N-|l|}[(X(n+l) - \mu)(X(n) - \mu)] \right] + \\
\nonumber &-& \frac{1}{N-|l|} E \left[(\hat{\mu}- \mu) \sum_{n = 1}^{N-|l|}(X(n+l) - \mu) + (X(n) - \mu) \right] + \\
&+& \frac{1}{N-|l|} E \left[ \sum_{n = 1}^{N-|l|}(\hat{\mu}- \mu)^2 \right]
\end{eqnarray}
If we approximate the sum till $N-l$ with the analogous till $N$ we get:
\begin{equation}
\nonumber \sum_{n = 1}^{N-|l|}(X(n) - \mu) \approx \sum_{k = 1}^{N}(X(k) - \mu) =
N \left(\frac{\sum_{k = 1}^{N}X(k)}{N} \right) - N \mu = N (\hat{\mu} - \mu)
\end{equation}
So, the expectation can be approximated with:
\begin{eqnarray}\label{eq:autocov_exp}
\nonumber E[\hat{\gamma}_X (l)] \approx \frac{N-|l|}{N-|l|} \gamma_X (l) &-& 2 \frac{N}{N-|l|} E [(\hat{\mu}- \mu)^2] + \frac{N-|l|}{N-|l|} E [(\hat{\mu}- \mu)^2] = \\
&=& \gamma_X (l) - \frac{N + |l|}{N-|l|} \sigma_{\hat{\mu}}^2.
\end{eqnarray}

If $N$ tends to $\infty$, the fraction \(- \frac{N + |l|}{N-|l|}\) tends to $-1$. Moreover, from the properties of the time mean, we know that $\sigma_{\hat{\mu}}^2$ tends to zero if $N$ tends to $\infty$, so the expectation of $\hat{\gamma}_X (l)$ tends to $\gamma_X (l)$ if $N$ tends to $\infty$.\\
Note that the result is underestimated for all lags.\\

\noindent When $l=0$ this expression is exact and reduces to:
\begin{equation}
\nonumber E[\hat{\gamma}_X (0)] =  \gamma (0) - \sigma_{\hat{\mu}}^2.
\end{equation}
in accordance with eq. \ref{eq:var_exp}.
\\

Exact expression for the covariance of this estimate have been found by Bartlett \cite[p. 326]{Priestley} if the random process is stationary up to order four, but he also gives an approximated formula:
\begin{equation}\label{eq:autocov_cov}
\mbox{cov}\{\hat{\gamma}_X (l),\hat{\gamma}_X (l+h)\} \approx \frac{1}{N} \sum_{m=-\infty}^{+\infty} \{\gamma_X(m) \gamma_X(m + h) + \gamma_X(m + l + h) \gamma_X(m -l)\}
\end{equation}
from which:
\begin{equation}
\mbox{var}\{\hat{\gamma}_X (l)\} \approx \frac{1}{N} \sum_{m=-\infty}^{+\infty} \{\gamma_X^2(m) + \gamma_X(m + l) \gamma_X(m -l)\}
\end{equation}

From the last equation we can also deduce the variance of the time-sample variance:
\begin{equation}\label{eq:var_var}
\mbox{var}\{\hat{\sigma}_{X,N}\} \approx \frac{2}{N} \sum_{m=-\infty}^{+\infty} \gamma_X^2(m) \rightarrow_{N \rightarrow \infty} \frac{4 \pi \sigma^2_X}{N} f(0)
\end{equation}
The variance tends to zero asymptotically.
}
\end{enumerate}

\noindent  Before proceeding, we look at the properties of the following estimator for the autocovariance, as we mention it in chapter \ref{chap:stat}:
\begin{equation}
\tilde{\gamma}_X(l) = \left \{ \begin{array} {ll}
                    \frac{1}{N-|l|-k} \sum_{n = 1}^{N-|l|}[X(n+l)-\mu] [X(n)-\mu]    & \mbox{if $0 \leq |l| \leq N-1$} \\
                    0   & \mbox{otherwise}
                                        \end{array}
                             \right.
\end{equation}
with $k\leq 1$. Following the same procedure of eq. \ref{eq:autocov_exp}, we can see that this estimator is biased even if the mean is known:
\begin{equation}
\nonumber E[\tilde{\gamma}_{X} (l)] = \frac{N-|l|}{N-|l|-k} \gamma_X (l)
\end{equation}
However, it is asymptotically unbiased. We notice that it overestimates the autocovariance of $X$. If the mean is unknown, we can obtain the same result as eq. \ref{eq:autocov_exp}, which is an approximation.

\subsubsection{Time-sample autocorrelation function}
\label{subsubec:autocorr_prop}

The autocorrelation function can be estimated by the time sample autocorrelation function:
\begin{equation}
\hat{\rho}_{X} (l) = \frac{\hat{\gamma}_{X} (l)}{\sigma^2_X}
\end{equation}
If we assume the variance as known, the properties of $\hat{\rho}_{X} (l)$ can be found in a straightforward way from those of the time-sample autocovariance function, simply dividing for $\sigma^2_X$.\\

\noindent But if the variance is to be estimated, the computation is complicated by the expectation of the ratio of two random variables. However, Kendall\cite{Kendall54} has provided an approximation to order $n^{-1}$ for the expectation of the ratio between random variables in quadratic form. More precisely, if {\it A}, {\it B}, and {\it C} are the r.v., {\it a}, {\it b} and {\it c} the deviation of {\it A}, {\it B}, and {\it C} from their respective means, and r is the ratio $r = \frac{A}{\sqrt{BC}}$, we have:
\begin{eqnarray}
\nonumber E[r] = \frac{E[A]}{\sqrt{E[B]E[C]}} \left\{1 - \frac{{\frac{1}{2}}E[ab]} {E[A]E[B]} - \frac{\frac{1}{2}E[ac]} {E[A]E[C]} + \frac{{\frac{1}{4}}E[bc]} {E[B]E[C]} + \right. \\
\left. + \frac{\frac{3}{8}E[b^2]} {E^2[B]} +  \frac{{\frac{3}{8}}E[c^2]} {E^2[C]} \right\}
\end{eqnarray}
which reduces, if $B=C$, to:
\begin{equation}\label{eq:kendall}
E[r]=\frac{E[A]}{E[B]} \left\{1 - \frac{{\frac{1}{2}}E[ab]} {E[A]E[B]} + \frac{E[b^2]} {E^2[B]} \right\}
\end{equation}
Now,
\begin{equation}
E[ab] = E[(A-\mu_A)(B-\mu_B)] = \mbox{cov}(A,B)
\end{equation}
and
\begin{equation}
E[b^2] = E[(B- \mu_B)^2] = \mbox{var}(B).
\end{equation}  \\

\noindent We apply this result in our case. Then, $A_l$ is the estimator of the autocovariance function $\hat{\gamma}(l)$ defined in eq. \ref{eq:sample_autocov}, while {\it B} is the estimator of the variance (see eq. \ref{eq:sample_var}). Both are quadratic in the $X_R(n)$ variables.\\
We have already evaluated $E[A_l] = E[\hat{\gamma}_X (l)]$ (eq. \ref{eq:autocov_exp}), $E[B] = E[\hat{\sigma}^2_{X,N}]$ \linebreak (eq. \ref{eq:var_exp}), $E[b^2] = \mbox{var}[\hat{\sigma}^2_X]$ (eq. \ref{eq:var_var}); we still need the term \linebreak $\mbox{cov}(A_l,B) = \mbox{cov}(\hat{\gamma}_X (l),\hat{\gamma}_X (0))$. We can substitute in eq. \ref{eq:autocov_cov} with $h = -l$ to find:
\begin{equation}
\mbox{cov}(\hat{\gamma}_X (l),\hat{\gamma}_X (0)) \approx \frac{1}{N} \sum_{m=-\infty}^{+\infty}[\gamma_X(m) \gamma_X(m-l)]^2
\end{equation}
Substituting all these equations in eq. \ref{eq:kendall}, we obtain:
\begin{equation}
E[\hat{\rho}_{X} (l)] \approx \left(\rho_X(l)-\frac{N+|l|}{N-|l|}\frac{\sigma^2_{\hat{\mu}}}{\sigma^2_X}\right) \frac{N-\alpha}{N} \left\{1 - \frac{\delta}{4\alpha \beta} + \frac{2 \beta}{N \sigma^2_X (1-\frac{1}{N} \alpha)} \right\}
\end{equation}
where we have set:
\begin{eqnarray}
\nonumber \alpha &=& \sum_{r=-(N-1)}^{N-1}\left(1- \frac{|r|}{N} \rho_X(r) \right) \\
\nonumber \beta &=& \sum_{m=-\infty}^{+\infty} \rho^2_X(m) \\
\delta &=& \sum_{m=-\infty}^{+\infty} [\rho_X(m) \rho_X(m-l)]^2
\end{eqnarray}
The expectation is a quadratic form in the $\rho_X(l)$ function.

\section{Bias in the estimation of the power spectral density of stationary processes}
\label{subsec:PSD_manola}

A random process is said to be stationary if its moments do not depend from the time: for example, the mean and variance are constant, the covariance depends just from the lag and so on.\\
\noindent For these processes, a crucial relationship holds, linking the spectral density function and the autocorrelation function under appropriate conditions: if $\{X(t)\}$ is a zero-mean continuous parameter stationary process with (non normalized) power spectral density function $h(\omega)$ existing for all $w$, and autocovariance function $R(\tau)$, then
\begin{equation}
h(\omega) = \frac {1}{2\pi} \int_{-\infty}^{+\infty} R(\tau) e^{i \omega \tau} d\tau.
\end{equation}
A proof can be found, e.g., in Priestley\cite[p. 211]{Priestley}. A similar relation exists for the normalized power spectral density $f(\omega) = \frac{h(\omega)}{\sigma^2_X}$ and the autocorrelation function $\rho(\tau)$.
The function $f(\omega)$ is important because it has the properties of a probability density function, so it makes a connection between probability distribution of the process $X$ and its spectral density.
\\

\subsection{Bias on the PSD in presence of a zero mean}
\label{subsec:const_mean}

We can use, following e.g. Manolakis\cite[p. 210]{Manolakis}, this relationship to estimate the bias on the PSD for a stationary zero-mean signal $\{X(n)\}_{n \geq 0}$.\\
We begin estimating the autocovariance through the sample autocovariance function:

\begin{equation}
\hat{r}_X (l) = \left \{ \begin{array} {ll}
                    \frac{1}{N} \sum_{n = 0}^{N-l-1}X(n+l)X^*(n)    & \mbox{if $0 \leq l \leq N-1$} \\
                    \hat{r}^*_X (-l)  & \mbox{if $-(n-1) \leq l \leq 0$} \\
                    0   & \mbox{otherwise}
                           \end{array}
                  \right.
\end{equation}

\noindent The estimation over a finite number of samples is equivalent to the multiplication of the original samples sequence with the rectangular window
\begin{equation}\label{eq:rect_window}
w_N ( k) = \left \{ \begin{array} {ll}
                                        1   & \mbox{if $0 \leq k \leq N-1$} \\
                                        0   & \mbox{otherwise}
                                        \end{array}
                             \right.
\end{equation}
The introduction of the window induces a bias on the estimation, and we can estimate it.
\begin{equation}
E[\hat{r}_X (l)] = E\left[\frac{1}{N} \sum_{n = 0}^{N-l-1}X(n+l)X^*(n)\right], \; |l| \geq 0
\end{equation}

\noindent If $l \geq 0$, we have:
\begin{equation}
= E\left[\frac{1}{N} \sum_{n = -\infty}^{\infty}X(n+l)w(n+l)X^*(n)w(n)\right]
\end{equation}
while, if $l<0$, \(E[\hat{r}_X (l)] = E[\hat{r}^*_X (-l)]\).
We then obtain:
\begin{eqnarray}
\nonumber E[\hat{r}_X (l)] &=& \frac{1}{N} \sum_{n = -\infty}^{\infty}E[X(n+l)X^*(n)]w(n+l)w(n) = \\
&=& \frac{1}{N} r(l) \sum_{n = -\infty}^{\infty}w(n+l)w(n) = \frac{N - |l|}{N} r(l)
\end{eqnarray}
because \(\sum_{n = -\infty}^{\infty}w(n+l)w(n) = \left \{ \begin{array} {ll}
                                        N - |l|   & \mbox{if $|l| \leq N - 1$} \\
                                        0   & \mbox{otherwise}
                                        \end{array}
                             \right. \)\\

\noindent So, if the window is rectangular, the estimation is asymptotically unbiased.
We recall the estimation of the covariance of $\hat{r}_X (l)$ of eq. \ref{eq:autocov_cov}:
\begin{equation}
\mbox{cov}\{\hat{r}_X (l),\hat{r}_X (l+h)\} \approx \frac{1}{N} \sum_{m=-\infty}^{+\infty} \{r_X(m) r_X(m + h) + r_X(m + l + h) r_X(m -l)\}
\end{equation}

\noindent The covariance is small just if the lag {\it l} is small compared to N, and successive values of $\hat{r}(l)$ could be correlated.\\

\noindent The power spectrum of this kind of processes can be evaluated as the Fourier transform of the autocorrelation function:
\begin{equation}
R_X(e^{iw})\doteq \sum_{l = -\infty}^{\infty} r_X(l) e^{-iwl}
\end{equation}

\noindent We estimate it with the periodogram and the sample autocorrelation function:
\begin{equation}\label{eq:manola_PSD}
R_X(e^{iw})\doteq \sum_{l = -(N-1)}^{N-1} \hat{r}_X(l) e^{-iwl}
\end{equation}
so the mean and the variance of this estimator depends from those of the autocorrelation functions:
\begin{equation}
E[R_X(e^{iw})] = \sum_{l = -(N-1)}^{N-1} E[\hat{r}_X(l)] e^{-iwl} =
   \sum_{l = -(N-1)}^{N-1} \frac{N - |l|}{N} r_X(l) e^{-iwl}
\end{equation}

\noindent Hence, if the window is rectangular (no weights are added to the samples), the periodogram is an asymptotically unbiased estimation of the power spectrum. The bias of the estimator depends on the window chosen: if the window is not rectangular, we will have the correlation function of the window instead of the term \((N - |l|)/N\).

\noindent The variance does not tend to zero as the window increases. An approximate expression for the covariance $cov \left\{ \hat{R}_X(e^{iw_1}), \hat{R}_X(e^{iw_2}) \right\}$ has been found by Jenkins \& Watts, and it is function of both $\hat{R}_X(e^{iw_1})$ and $\hat{R}_X(e^{iw_2})$, so the variance is of order of $\hat{R}_X^2(e^{iw})$.\\

\subsection{Bias on the PSD in presence of a non-zero mean}
\label{subsec:nonzero_mean}

We now consider a $X(k)$ signal with a trend:
\begin{equation}\label{eq:staz+cost}
X(k) = X_s (k) + a(k), \;\; k  \geq 0
\end{equation}
We relax the hypothesis of sec. \ref{subsec:PSD_manola}, and we ask $X_s$ to be wide sense stationary, i.e., stationary up to order 2:
\begin{enumerate}
\item{\(\mu_{X_s} = E[X_s(k)] = \mu, \: \forall k \). We assume hereafter that \(\mu = 0\)}
\item{\(E[X_s(k+n)X_s^*(k)] = r_{X_s} (n) \: \forall k \), i.e. it depends just on the separation $n$ between the samples}
\item{\(\mbox{var}(X_s(k)) = \sigma^2_{X_s} \)}
\end{enumerate}
because for the bias on the PSD we use moments up to order 2. We ask, however, that $X_s$ has spectral density given by the Fourier transform of the autocovariance function:
\begin{equation}
f_{X_s}(\omega) \approx \frac{\sigma^2}{2 \pi} \sum_{j=-\infty}^{+\infty} \rho_{X_s}(j) e^{2\pi i j \omega}
\end{equation}
so that:
\begin{equation}
f_{X_s}(0) \approx \frac{\sigma^2}{2 \pi} \sum_{j=-\infty}^{+\infty} \rho_{X_s}(j).
\end{equation}

\noindent The simplest case is the presence of a non-zero mean: $a (k) = a \; k \geq 0$, with \(a\) constant:
\begin{equation}
X(k) = X_s (k) + a, \;\; k  \geq 0
\end{equation}
The properties of the signal \( X(k) \) depend from those of $X_s(k)$:
\begin{enumerate}
\item{\(\mu_X = E[X(k)] = \mu + a = a\)}
\item{\(\sigma_X^2 = \mbox{var}(X_s(k)+a) = \sigma^2_{X_s} \) }
\item{\(\gamma_X(l) = E[(X(k)-\mu_X)(X(k+l)-\mu_X)] = E[X_s(k)X_s(k+l)] = r_{X_s}(l), \: \forall k \) }
\item{\(\rho_X(l) = \frac{\gamma_X(l)}{\sigma_X^2} = \frac{r_{X_s}(l)} {\sigma_{X_s}^2}, \: \forall k \)}
\end{enumerate}

\subsubsection{Autocorrelation properties}
\noindent We apply the rectangular window \(w_N ( k) = \left \{ \begin{array} {ll}
                                        1   & \mbox{if $0 \leq k \leq N-1$} \\
                                        0   & \mbox{otherwise}
                                        \end{array}
                             \right. \) to the signal $X(k)$:
\[X_R (k) = X(k) w_N ( k) \]

\noindent Although the new signal is different from $\{X(k)\}$ because it is zero outside the window domain, applying such a window does not change the expectation and the variance of the set of random variables $ \{X_R(k) \}_{1 \leq k \leq N}$. Of course, if we evaluate the covariance function to the signal $X_R$ we get a different function:
\begin{eqnarray}
\nonumber \gamma_{X_R}(k,l) &=& E[X_R(k)X_R(k+l)] = E[X(k)w_R(k)X_R(k+l)w_R(l+l)]= \\
&=& E[X(k)X_R(k+l)] w_R(k)w_R(l+l)
\end{eqnarray}

\noindent because
\begin{eqnarray}
\nonumber &\left.\right.& E[X(k)w_R(k)X_R(k+l)w_R(l+l)] =\\
\nonumber &=& \sum_{t=-\infty}^{\infty}[X(k)](t)w_R(k) [X(k+l)](t)w_R(k+l)p_{\{X(k),X(k+l)\}}(t) = \\
&=&  w_R(k) w_R(k+l) \sum_{t=-\infty}^{\infty}[X(k)](t)[X(k+l)](t)p_{\{X(k),X(k+l)\}}(t)
\end{eqnarray}
where \(p_{\{X(k),X(k+l)\}}\) is the joint probability density function of $X(k)$ and \linebreak $X(k+l)$. So the autocovariance function changes its value depending on the window function $w_R$. This signal is no longer w.s.s. Moreover, we have that $\gamma_{X_R}(l) \neq \gamma_{X_R}(-l)$.\\
This effect of the application of the window is avoided by the sample autocovariance $\hat{\gamma}_{X_R} (l)$:
\begin{equation}
\hat{\gamma}_{X_R} (l) = \left \{ \begin{array} {ll}
                    \frac{1}{N-l} \sum_{n = 1}^{N-l}[X_R(n+l)-\hat{\mu}_{X,N}] [X_R^*(n)-\hat{\mu}^*_{X,N}]    & \mbox{if $0 \leq l \leq N-1$} \\
                    \hat{\gamma}^*_{X_R} (-l)  & \mbox{if $-(N-1) \leq l \leq 0$} \\
                    0   & \mbox{otherwise}
                                        \end{array}
                             \right.
\end{equation}
We are dealing with real valued signals, so we have:
\begin{equation}
\hat{\gamma}_{X_R} (l) = \left \{ \begin{array} {ll}
                    \frac{1}{N-|l|} \sum_{n = 1}^{N-|l|}[X_R(n+l)-\hat{\mu}_{X,N}] [X_R(n)-\hat{\mu}_{X,N}]    & \mbox{if $0 \leq |l| \leq N-1$} \\
                    0   & \mbox{otherwise}
                                        \end{array}
                             \right.
\end{equation}

\noindent In this definition, we have required explicitly the symmetry, and we have forced the autocovariance to be null outside the lag interval $[-(N-1), N-1]$.\\

\noindent With this estimator for the covariance, we first assume that the mean is known. As seen in sec. \ref{subsubsec:autocov_prop}, we have that the estimator is unbiased. This result can be obtained in a slightly different way remembering the particular form of the $X_R$ signal, using the window function:
\begin{eqnarray}
\nonumber E [\hat{\gamma}_{X_R} (l)] = \frac {1}{N-|l|} E \left[\sum_{n = 1}^{N-|l|}[X_R(n+l)-\mu_{X,N}] [X_R(n)- \mu_{X,N}]  \right] = \\
= \frac {1}{N-|l|} E \left[\sum_{n = -\infty}^{+\infty}[X(n+l)-\mu_{X,N}] [X(n)- \mu_{X,N}]w_N(n)w_N(n+l)  \right] =
\end{eqnarray}
There are not convergence problems transforming this finite sum into an infinite one, because we add only null terms.
\begin{eqnarray}
\nonumber = \frac {1}{N-|l|} \sum_{n = -\infty}^{+\infty}E [(X(n+l)&-&\mu_{X,N}) (X(n)- \mu_{X,N})] w_N(n)w_N(n+l) = \\
&=& \frac {1}{N-|l|} \gamma_X(l) \sum_{n = -\infty}^{+\infty} w_R(n)w_R(n+l)
\end{eqnarray}
Now, for the rectangular window $w_N$, the infinite sum results:
\begin{equation}
\sum_{n = -\infty}^{\infty}w_N(n+l)w_N(n) =
            \left \{ \begin{array} {ll}
                    N-|l|    & \mbox{if $1 \leq |l| \leq N$} \\
                    0   & \mbox{otherwise}
                      \end{array}
            \right.
\end{equation}
So the expectation becomes:
\begin{equation}
= \frac {N-|l|}{N-|l|} \gamma_X(l) = \gamma_X(l).
\end{equation}
and we find that in this case the estimator is unbiased, in agreement with eq. \ref{eq:autocov_exp_unbiased}.
\\

\noindent If the mean is unknown, we have from eq. \ref{eq:autocov_exp} that the estimator of the autocovariance function is biased, but asymptotically unbiased. We can find again this results, even with the truncated signal, using the properties of the window function $w_N$.
\begin{equation}
E[\hat{\gamma}_{X_R} (l)] = E[\frac{1}{N-|l|} \sum_{n = 1}^{N-|l|} [X_R(n+l)-\hat{\mu}_{X_R,N}] [X_R(n)-\hat{\mu}_{X_R,N}]] =
\end{equation}
\begin{eqnarray}
\nonumber &=& \frac{1}{N-|l|} \cdot \\
\nonumber &\cdot& E\left[ \sum_{n =1}^{N-|l|} X_R(n+l) - \mu_{X_R} -(\hat{\mu}_{X_R,N} - \mu_{X_R})] [X_R(n)-\mu_{X_R}-(\hat{\mu}_{X_R,N}-\mu_{X_R})]\right] =
\end{eqnarray}

\noindent Recalling that $X_R(k) = X(k) w_N(k)$ and decomposing the summation in all its terms:
\begin{eqnarray}
\nonumber &=& \frac{1}{N-|l|}E \left[ \sum_{n =1}^{N-|l|} [X_R(n+l)- \mu_{X_R}] [X_R(n)-\mu_{X_R}] + \right. \\
\nonumber - \sum_{n =1}^{N-|l|} \left[X_R(n+l)\right.&-&\left. \mu_{X_R}\right] [\hat{\mu}_{X_R,N}-\mu_{X_R}] - \sum_{n =1}^{N-|l|} [\hat{\mu}_{X_R,N} - \mu_{X_R}] [X_R(n)-\mu_{X_R}] + \\
&+& \left. \sum_{n =1}^{N-|l|} [\hat{\mu}_{X_R,N} - \mu_{X_R}] [\hat{\mu}_{X_R,N}-\mu_{X_R}] \right]
\end{eqnarray}

\noindent The expectation of this sum is the sum of the expectations. The first terms becomes:
\begin{eqnarray}
\nonumber &\left.\right.& \frac{1}{N-|l|} E \left[ \sum_{n =1}^{N-|l|} [X_R(n+l)- \mu_{X_R}] [X_R(n)-\mu_{X_R}] \right] = \\
\nonumber &=& \frac{1}{N-|l|} E \left[ \sum_{n = -\infty}^{+\infty} [X(n+l)- \mu_{X_R}] [X(n)-\mu_{X_R}]
w_N(n+l)w_N(n) \right] = \\
\nonumber &=& \frac{1}{N-|l|} \sum_{n = -\infty}^{+\infty} E \left[ (X(n+l)- \mu_{X_R}) (X(n)-\mu_{X_R}) \right] w_N(n+l)w_N(n) = \\
&=& \frac{1}{N-|l|} \gamma_X(l) \sum_{n = -\infty}^{+\infty} w_N(n+l)w_N(n) =
\end{eqnarray}

\noindent where we have used the fact that $\mu_{X_R} = \mu_X = a$. We know that the infinite sum of the rectangular window sums up to $N-|l|$, so we obtain:
\begin{equation}
= \frac{N-|l|}{N-|l|} \gamma_X(l) = \gamma_X(l).
\end{equation}

\noindent The last term is the variance of the random variable $\hat{\mu}_{X_R,N}$ :
\begin{eqnarray}
\frac{1}{N-|l|} E \left[\sum_{n =1}^{N-|l|} [\hat{\mu}_{X_R,N} - \mu_{X_R}] [\hat{\mu}_{X_R,N}-\mu_{X_R}] \right] &=& \\
\nonumber \frac{1}{N-|l|} \sum_{n =1}^{N-|l|}E \left[ (\hat{\mu}_{X_R,N} - \mu_{X_R})^2 \right] = \frac{N-|l|}{N-|l|} E \left[ (\hat{\mu}_{X_R,N} - \mu_{X_R})^2 \right] &=& \sigma_{\hat{\mu}_{X_R,N}}^2
\end{eqnarray}

\noindent and we know from eq. \ref{eq:samplemean_var} that it is equal to:
\begin{equation}
\sigma_{\hat{\mu}_{X_R,N}}^2 = \frac{\sigma^2_{X_R}}{N} \sum_{r= -(N-1)}^{N-1} \left( 1 - \frac{|r|}{N} \right) \rho_{X_R}(r)
\end{equation}
because $X_R$ is such that it possesses a Fourier Transform (it has only a finite number of non-zero values, so the coefficient integrals exist and are finite).
\\

\noindent The mid term gives:
\begin{eqnarray}
\nonumber \frac{1}{N-|l|} E \left[ \sum_{n =1}^{N-|l|} (\hat{\mu}_{X_R,N} - \mu_{X_R}) (X_R(n) - \mu_{X_R}) \right] = \\
= \frac{1}{N-|l|} E \left[ (\hat{\mu}_{X_R,N} - \mu_{X_R}) \sum_{n =1}^{N-|l|}(X_R(n) - \mu_{X_R}) \right] =
\end{eqnarray}
\noindent because $\hat{\mu}_{X_R,N} - \mu_{X_R}$ does not depend from the summation index $n$.

\noindent If we do not consider the summation up to $N-|l|$, but up to $N$, we can conclude:
\begin{equation}\label{eq:approx_cov}
=\frac{N}{N-|l|} E \left[ (\hat{\mu}_{X_R,N} - \mu_{X_R})^2 \right] = \frac{N}{N-|l|} \sigma^2_{\hat{\mu}_{X_R,N}}.
\end{equation}
So the total contribution of the mid terms is \( \approx \frac{-2N}{N-|l|} \sigma^2_{\hat{\mu}_{X_R,N}} \)
\\
\noindent Adding up all these terms, we find that the expectation of the estimator of the autocovariance coefficient $\hat{\gamma}_X(l)$ is given by the corresponding "true" autocovariance coefficient $\gamma_X(l)$ with the contribution of the error in the estimation of the media, and a third factor catching the correlation between the random variables $\hat{\mu}_{X_R}$ and the set of, in general, correlated  $\{X(k)\}$:
\begin{eqnarray}\label{eq:approx_cov_final}
\nonumber E[\hat{\gamma}_{X_R}(l)] \approx \gamma_X(l) - \frac{2N}{N-|l|} \sigma^2_{\hat{\mu}_{X_R,N}} + \sigma_{\hat{\mu}_{X_R,N}}^2 = \\
= \gamma_X(l) - \frac{N+|l|}{N-|l|} \sigma^2_{\hat{\mu}_{X_R,N}}.
\end{eqnarray}
If $N$ tends to $\infty$, $\sigma_{\hat{\mu}}^2$ tends to zero if $N$ tends to $\infty$, so the expectation of $\hat{\gamma}_X (l)$ tends to $\gamma_X (l)$ if $N$ tends to $\infty$, giving an asymptotically unbiased estimate.
We notice that the denominator of this expectation is not a problem: the lag $l$ is fixed, while $N$ grows to $\infty$: $N\gg l$.

\subsubsection{Examples}
\label{subsubsec:autocov_examples}

To illustrate the behavior of the expectation of the function $\hat{\gamma}_X (l)$ when N is varying, we consider two random processes that have stationarity properties: the moving average process of order $l$ {\it MA(l)} and the harmonic process {\it H}.
\\
\noindent The moving average process of order $l$ is defined as:
\begin{equation}
X_t = b_0 \varepsilon_t + b_1 \varepsilon_{t-1} + ... + b_l \varepsilon_{t-l}
\end{equation}
where the weights $\{b_i\}_{0 \leq i \leq l}$ are constant and the $\{\varepsilon_i\}_{0 \leq i \leq l}$ are normal distributed random variables: $\varepsilon_i \sim N(\mu_\varepsilon,\sigma_\varepsilon^2)$. We can suppose, without loss of generality, that $\varepsilon_i \sim N(0,1)$, and we obtain:

\begin{itemize}
\item{$E[X_t] = \mu_\varepsilon \sum_{i=1}^l b_i = 0$}
\item{\(\gamma_X(r) = E[X_t, X_{t+r}]=
    \left \{ \begin{array} {ll}
            \sigma^2_\varepsilon (b_0 b_r + b_1 b_{r+1} + ... + b_{l-r} b_l)    & \mbox{$0 \leq r \leq l$} \\
            0   & \mbox{$r > l$}
             \end{array}
    \right. \) \\
    \(\gamma_X(-r) = \gamma_X(r)\),\\ and so it does not depend on $t$}
\item{$\sigma^2_X = \gamma_X(0) = \sigma^2_\varepsilon \sum_{i = 0}^l b_i^2$}
\end{itemize}

\noindent In the particular case when all the weights are equal, $b_i = \frac{1}{l+1}$, $\forall i$, the previous functions simplify:
\begin{itemize}
\item{$E[X_t] = 0$}
\item{\(\gamma_X(|r|) =
    \left \{ \begin{array} {ll}
            \sigma^2_\varepsilon \frac{l-|r|+1}{(l+1)^2}    & \mbox{$0 \leq |r| \leq l$} \\
            0   & \mbox{$|r| > l$}
             \end{array}
    \right. \) }
\item{$\sigma^2_X = \gamma_X(0) = \sigma^2_\varepsilon \frac{1}{l+1}$}
\end{itemize}

\noindent In picture \ref{fig:E[gamma_X,N]_MA}, the autocovariance function and the expectation of its estimator for different N are represented, for a MA(5) with equal weights (left), and for a general MA(4) process, with weights $b = [0.9 \; 0.85 \; 0.8 \; 0.5 \; 0.1]$. In both cases the underlying random variables have normal distribution $N(0,1)$.

\begin{figure*}[ht]
\epsfig{figure=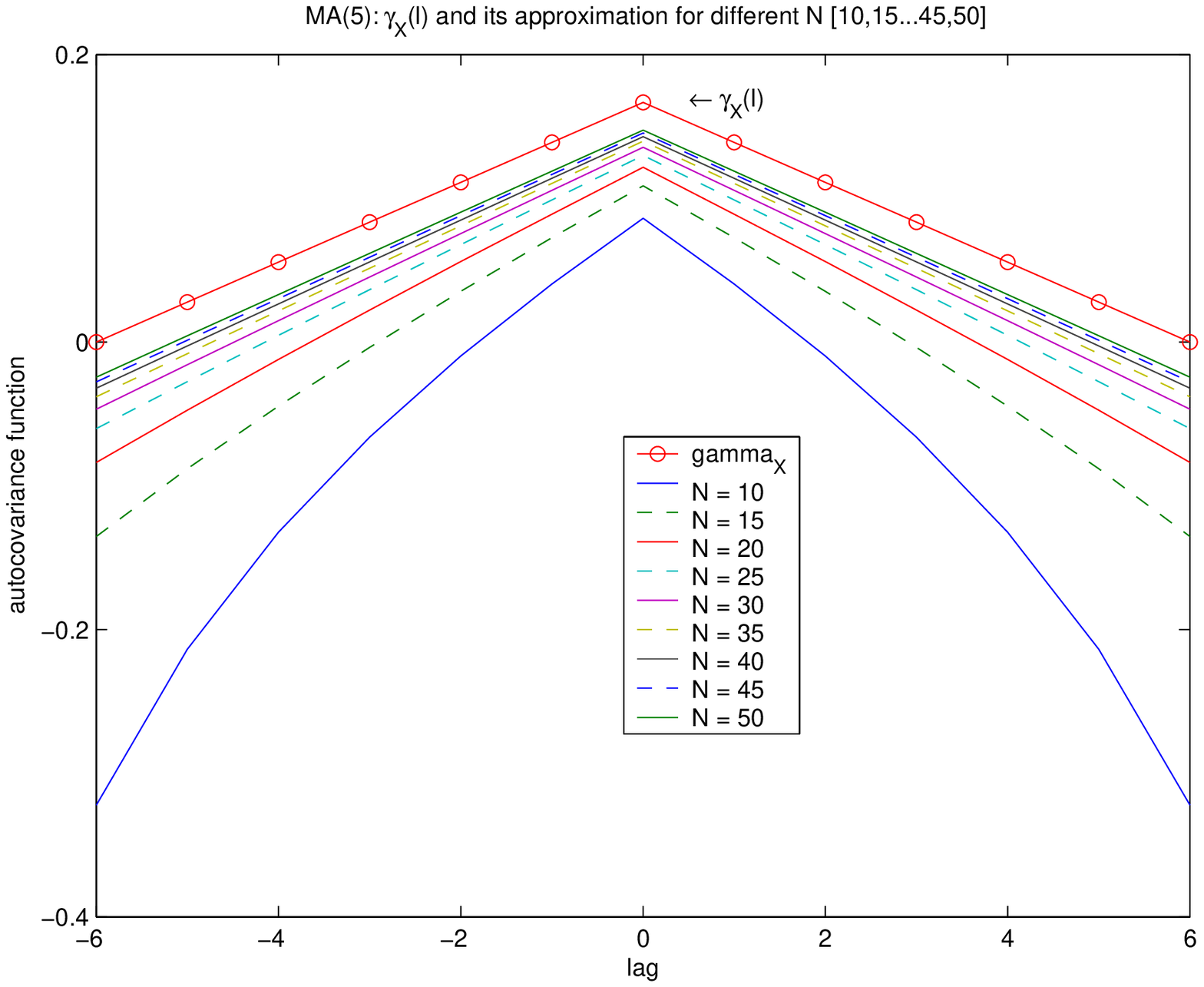,width=7cm}
\epsfig{figure=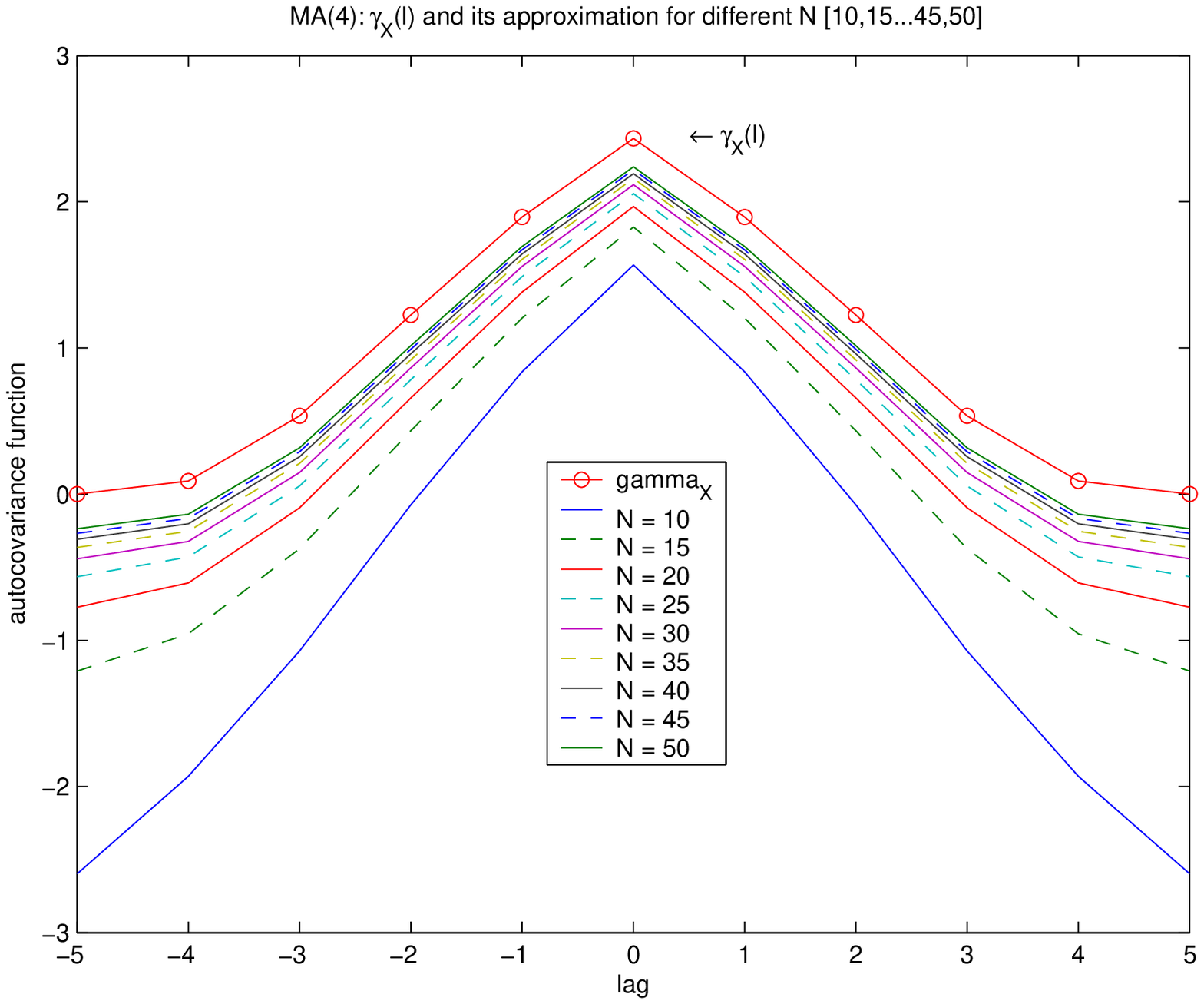,width=7cm}
\caption{Expectation of $\hat{\gamma}_X(l)$ at different N for MA(5) with equal weights (left), and in a general case (right)}
\label{fig:E[gamma_X,N]_MA}
\end{figure*}

\noindent The harmonic process is defined as:
\begin{equation}
X_t = \sum_{i = 1}^K A_i cos(w_i t + \phi_i)
\end{equation}
with $\{A_i, w_i\}_{1 \le i \le K}$ and $K$ constant, and $\{\phi_i\}_{1 \le i \le K}$ a family of random variables i.i.d. with rectangular distribution over the interval $[-\pi, \pi]$. Thanks to the properties of the $\{\phi_i\}_{1 \le i \le K}$ family, it is a stationary process $\forall \{A_i, w_i\}_{1 \le i \le K}$, $\forall K$ and $\forall t$:

\begin{itemize}
\item{$E[X_t] = 0$}
\item{$\gamma_X(r) = E[X_t, X_{t+r}]= \sum_{i = 1}^K \frac{1}{2} A_i^2 cos(w_i r)$, and so it does not depend on $t$, but it never dies out}
\item{$\sigma^2_X = \gamma_X(0) = \sum_{i = 1}^K \frac{1}{2} A_i^2$}
\end{itemize}

\noindent In figure \ref{fig:E[gamma_X,N]_harmonic}, we show the autocovariance function and the expectation of its estimator at different N, for the harmonic process:
\begin{equation}
X_t = \sum_{i = 1}^{10} 0.1 \cos(0.5 t + \phi_i)
\end{equation}
where $\phi_i, \; i=1,\ldots, 10$ is a family of uncorrelated gaussian random variables $N(\mu, \sigma^2)$.

\begin{figure}[ht]
\centering
\epsfig{figure=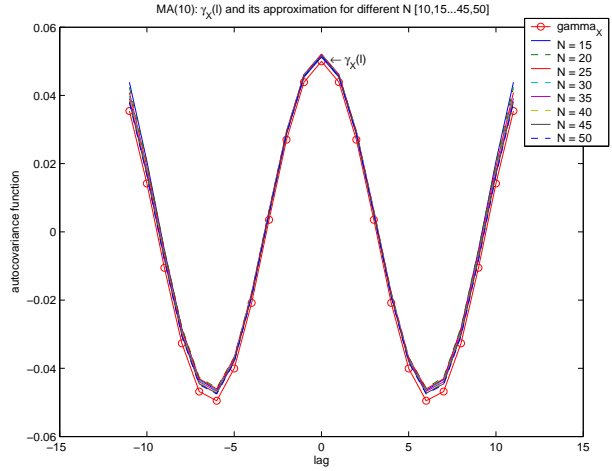,width=8cm}
\caption{Expectation of $\hat{\gamma}_X(l)$ for harmonic process at different N}
\label{fig:E[gamma_X,N]_harmonic}
\end{figure}

\subsubsection{Error of the approximation}
The error in the approximation of eq. \ref{eq:approx_cov} is larger for larger lags because we are adding more extra values in the summation. For $l=0$, in particular, the approximating formula \ref{eq:approx_cov_final} is exact, and it becomes:
\begin{eqnarray}
\nonumber E[\hat{\gamma}_{X_R}(l)] &=& \gamma_X(l) - \frac{2N}{N} \sigma^2_{\hat{\mu}_{X_R,N}} + \sigma_{\hat{\mu}_{X_R,N}}^2 = \\
&=& \gamma_X(l) - \sigma^2_{\hat{\mu}_{X_R,N}}.
\end{eqnarray}
If $l \neq 0$, we have:
\begin{eqnarray}
\nonumber \frac{1}{N-|l|} E \left[ \sum_{n =1}^{N} (\hat{\mu}_{X_R,N} - \mu_{X_R}) (X_R(n) - \mu_{X_R}) \right] = \\
\nonumber = \frac{1}{N-|l|} E \left[ \sum_{n =1}^{N-|l|} (\hat{\mu}_{X_R,N} - \mu_{X_R}) (X_R(n) - \mu_{X_R}) \right] + \\
+ \frac{1}{N-|l|} E \left[ \sum_{n =N-|l|+1}^{N} (\hat{\mu}_{X_R,N} - \mu_{X_R}) (X_R(n) - \mu_{X_R}) \right]
\end{eqnarray}

\noindent The last term is the error function:
\begin{equation}
err(l) = \frac{1}{N-|l|} E \left[ \sum_{n =N-|l|+1}^{N} (\hat{\mu}_{X_R,N} - \mu_{X_R}) (X_R(n) - \mu_{X_R}) \right]
\end{equation}

\subsubsection{Estimation properties of the power spectral density }
\noindent We now estimate the (non normalized) power spectrum density using the estimate of the autocovariance function in eq. \ref{eq:manola_PSD}:
\begin{equation}\label{eq:PSD_Vs_ACF}
\hat{R}_X(e^{iw}) = \sum_{l=-(N-1)}^{N-1} \hat{\gamma}_{X_R}(l)e^{-iwl}
\end{equation}

\noindent The expectation of this estimate depends from the expectation of the autocovariance function:
\begin{eqnarray}\label{eq:PSD_bias}
E\left[\hat{R}_X(e^{iw})\right] &=& \sum_{l=-(N-1)}^{N-1} E \left[ \hat{\gamma}_{X_R}(l) \right] e^{-iwl} = \\
\nonumber = \sum_{l=-(N-1)}^{N-1} \left\{ \gamma_{X_R}(l) \right. &-& \left. \frac{N+|l|}{N-|l|} \frac{1}{N} \sum_{r=-(N-1)}^{N-1} \left( 1 - \frac{|r|}{N} \right) \gamma_{X_R}(r) \right\} e^{-iwl} = \\
\nonumber &=& \sum_{l=-(N-1)}^{N-1} \gamma_{X_R}(l) e^{-iwl} + \\
\nonumber &-& \sum_{l=-(N-1)}^{N-1} \frac{N+|l|}{N-|l|} \frac{1}{N} \sum_{r=-(N-1)}^{N-1} \left( 1 - \frac{|r|}{N} \right) \gamma_{X_R}(r) e^{-iwl}
\end{eqnarray}
\\

\noindent The behavior of this sum is different depending on whether the underlying random process has a non-periodic or periodic autocorrelation function.\\ In the first case, assuming that the values $\gamma(l)$ are negligible for large $l$, than the first factor of the sum tends to the true periodogram value for $N \rightarrow +\infty$:
\begin{equation}
\sum_{l=-(N-1)}^{N-1} \gamma_{X_R}(l) e^{-iwl} \approx \sum_{l=-\infty}^{+\infty} \gamma_{X_R}(l) e^{-iwl} = R_{X_R}(e^{iw}).
\end{equation}
\\
\noindent If the autocorrelation function is periodic, this is no longer true. \\
\noindent The last term  of equation \ref{eq:PSD_bias} does not converge to zero if N tends to $+\infty$.
The inner sum $\sum_{r=-(N-1)}^{N-1} \left( 1 - \frac{|r|}{N}\right) \rho_{X_R}(r)$ tends to $\sum_{r=-\infty}^{+\infty} \rho_{X_R}(r) = 2 \pi f(0)$ if N tends to $+\infty$, as we have seen before.
\\

\begin{figure*}[htbp]
\epsfig{figure=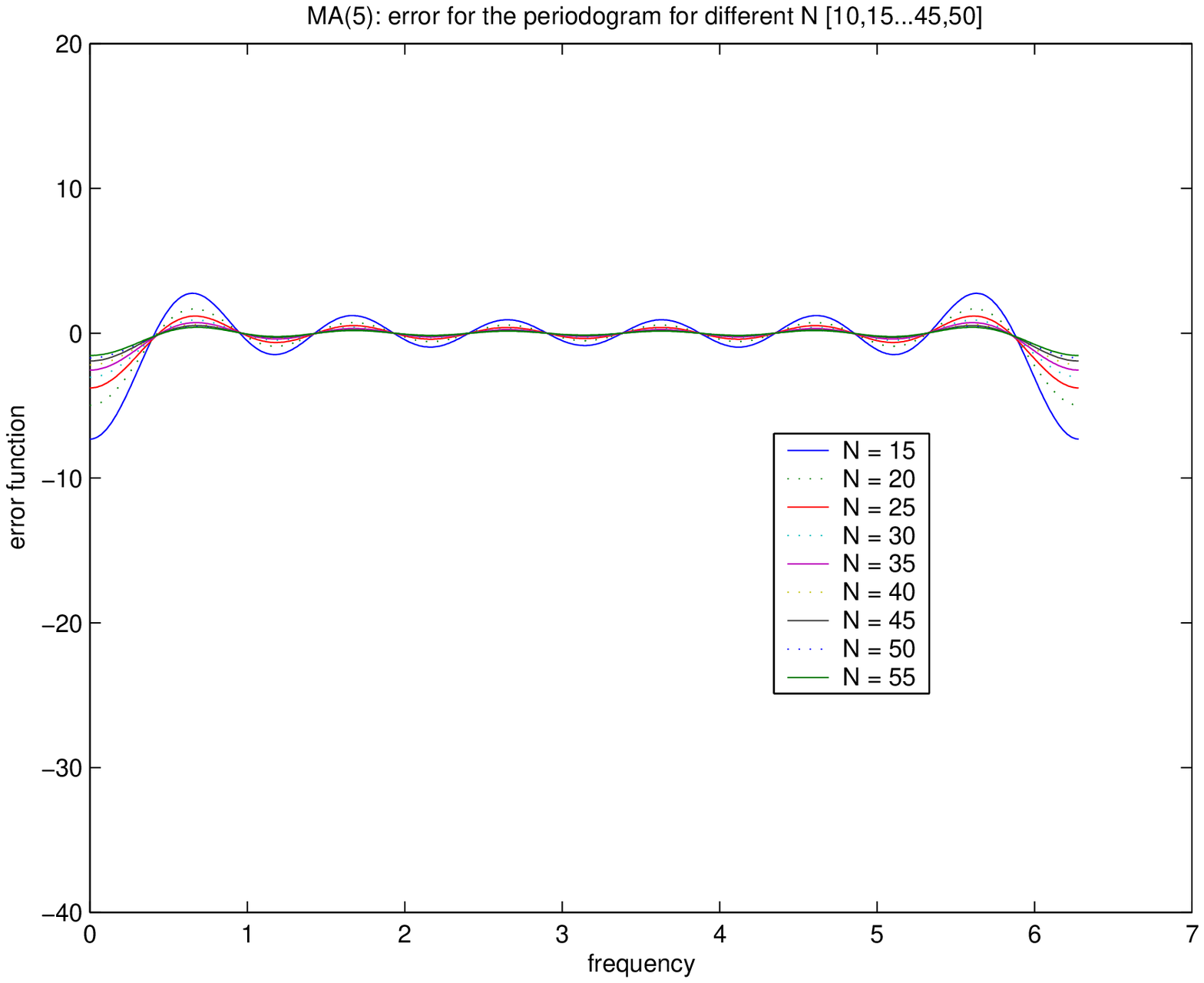,width=7cm}
\epsfig{figure=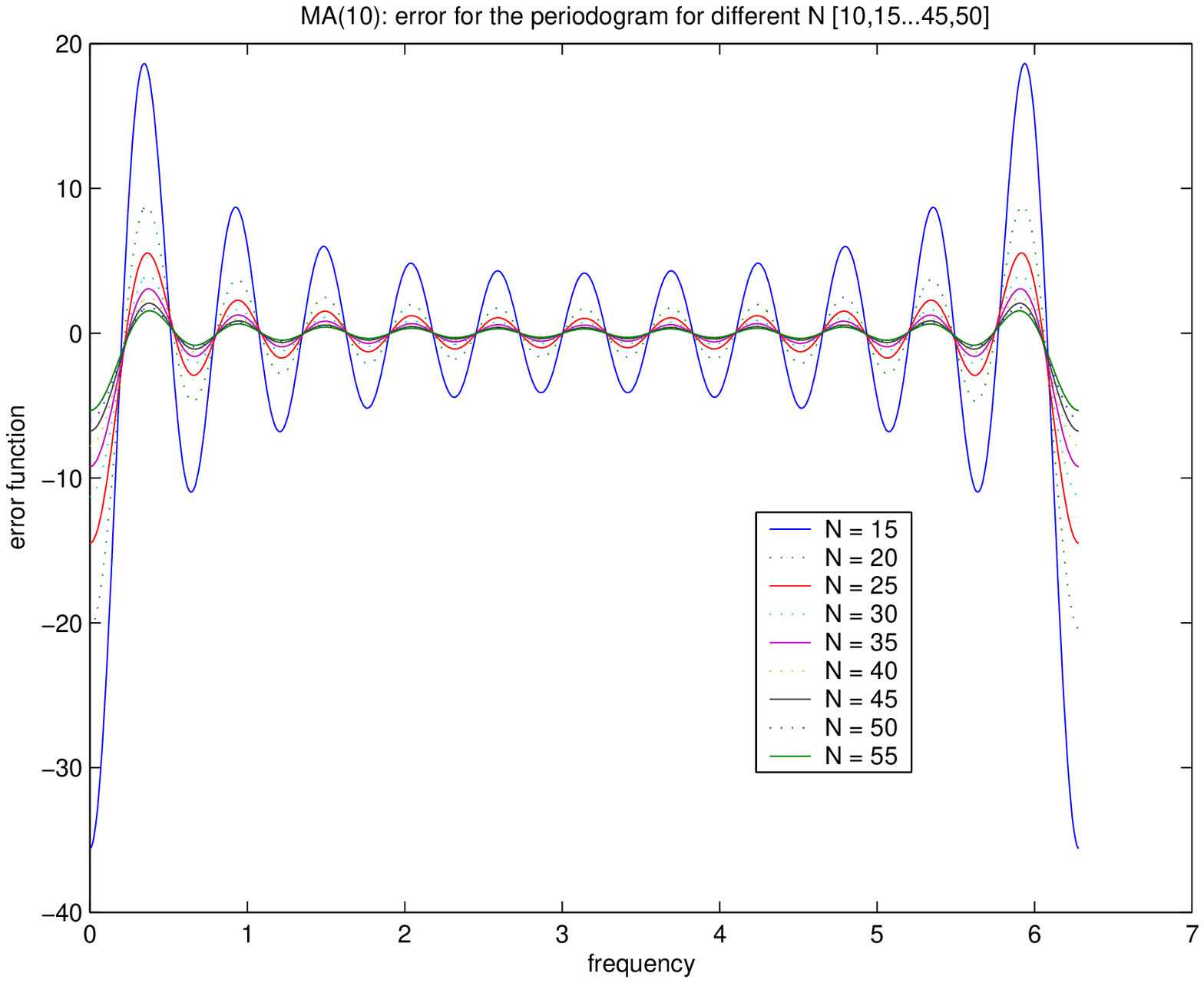,width=7cm}
\caption{Error for the expectation of periodogram at different N for MA(5) with equal weights (left), and for MA(10) with equal weights (right)}
\label{fig:errorPer_MA}
\end{figure*}

\begin{figure*}[htbp]
\epsfig{figure=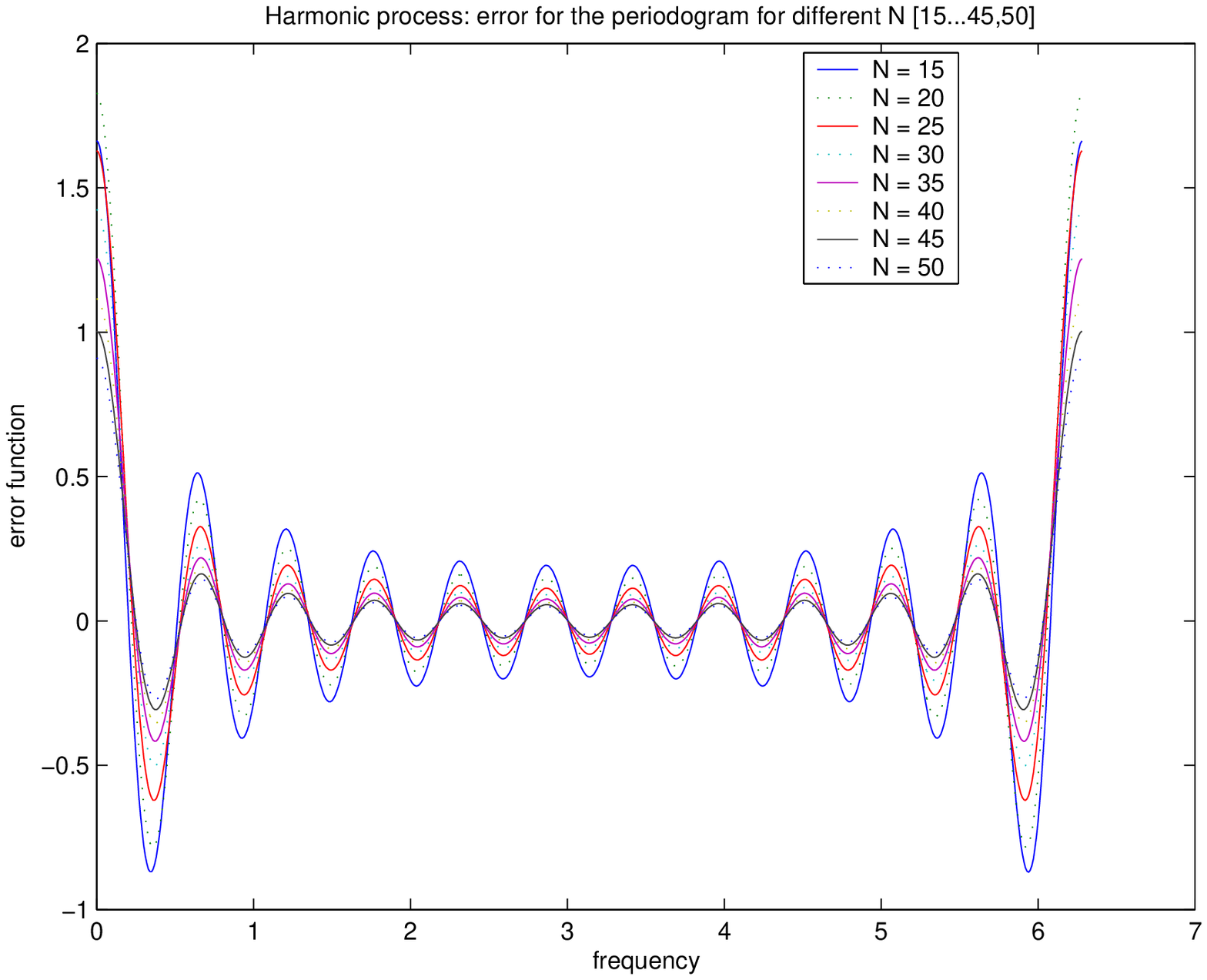,width=7cm}
\epsfig{figure=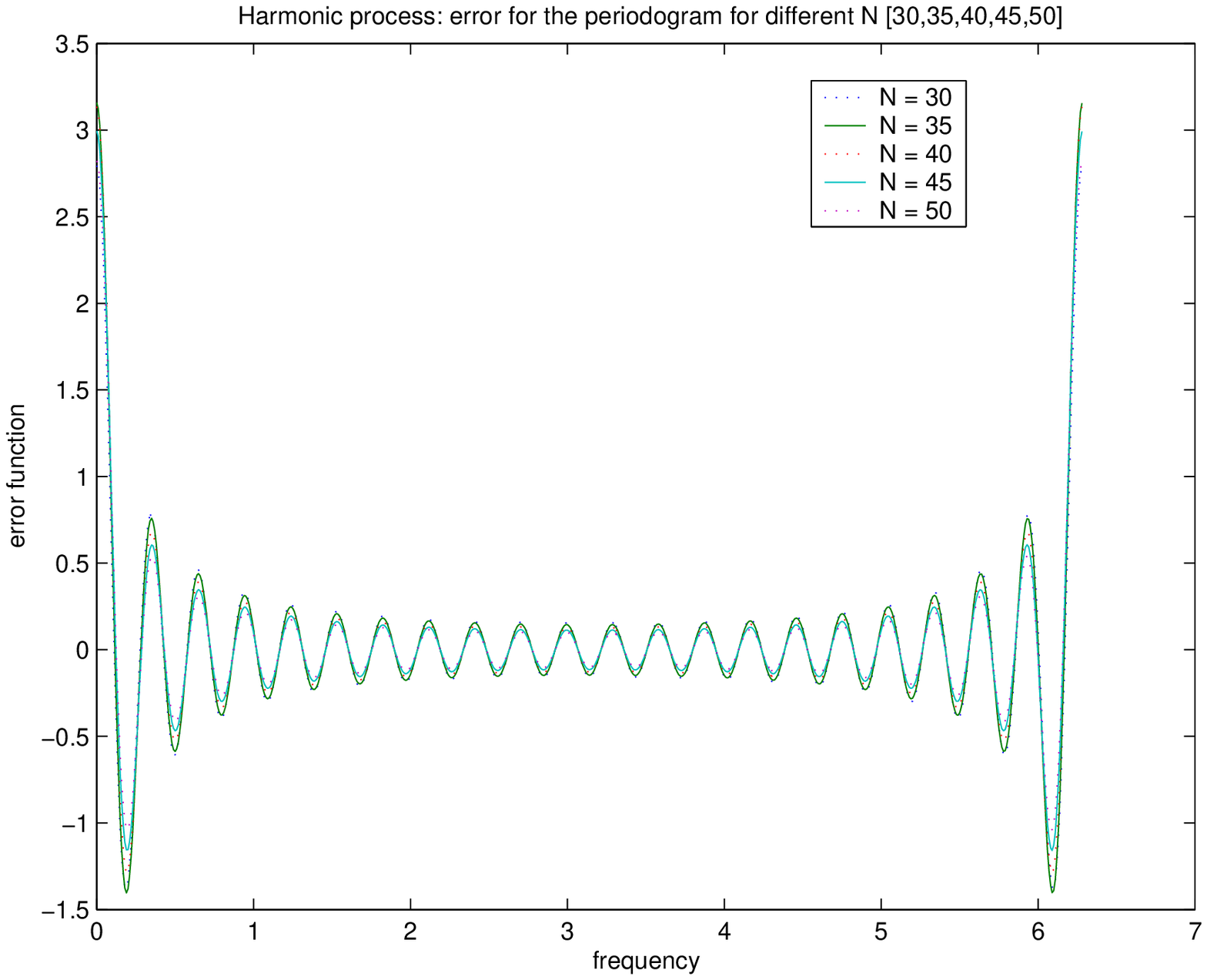,width=7cm}
\caption{Error for the expectation of periodogram at different N for harmonic process with equal frequencies; number of lags considered for autocovariance function: 21 (left), and 41 (right)}
\label{fig:errorPer_harm}
\end{figure*}

\noindent We illustrate the behavior of this error term for the processes we considered before in figures \ref{fig:errorPer_MA} and \ref{fig:errorPer_harm}. The error shape changes with the number of non-zero autocorrelation coefficients. In the MA(l) case, this depends from the order of the process, whereas for harmonic process, where the autocorrelation functions never dies out, the truncation of the autocorrelation function is a computational needs.

\subsubsection{Application to uncorrelated samples}

If the underlying process is a temporal sequence of uncorrelated random variables, the estimation of the bias simplify. However, no changes of its properties arise, since the variance of the time-sample mean of an uncorrelated process $X_u(k), \; k \geq 0$ is biased, but asymptotically unbiased (see eq. \ref{eq:uncorr_varOfmean}). The $X_u(k)$ autocovariance function is zero for all lags $l\neq0$:
\begin{equation}
\gamma_{X_u}(l) = \left \{ \begin{array} {ll}
                    \sigma^2_{X_u}    & \mbox{if $l = 0$} \\
                    0   & \mbox{otherwise}
                                        \end{array}
                             \right.
\end{equation}
so the bias of its estimator will be:
\begin{equation}
\hat{\gamma}_{X_u}(l) = \left \{ \begin{array} {ll}
                    \frac{N-1}{N^2}\sigma^2_{X_u}    & \mbox{if $l = 0$} \\
                    -\frac{N+|l|}{N(N-|l|)}\sigma^2_{X_u}   & \mbox{otherwise}
                                        \end{array}
                             \right.
\end{equation}

\noindent Substituting into eq. \ref{eq:PSD_Vs_ACF}, we finally obtain:
\begin{eqnarray}
\nonumber E\left[\hat{R}_X(e^{iw})\right] = \sum_{l=-(N-1)}^{N-1} E \left[ \hat{\gamma}_{X_R}(l) \right] e^{-iwl} = \\
= \sum_{l=-(N-1)}^{N-1} \gamma_{X_R}(l) e^{-iwl} - \sum_{l=-(N-1)}^{N-1} \frac{N+|l|}{N-|l|} \frac{1}{N} \sigma^2_X e^{-iwl}
\end{eqnarray}

\begin{figure}[htbp]
    \begin{center}
        \epsfig{figure=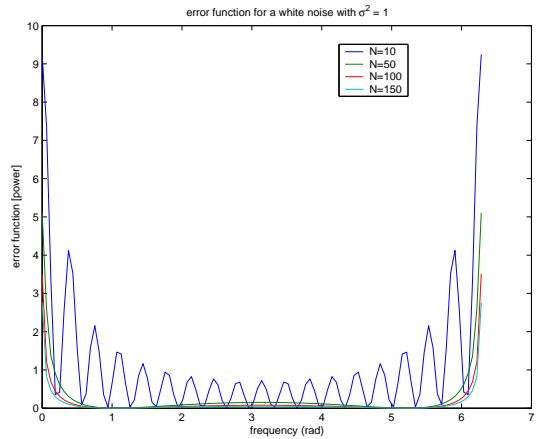,width=7cm}
        \caption{Error for the expectation of periodogram at different N for un uncorrelated process}
        \label{fig:errorBias_whiteNoise}
    \end{center}
\end{figure}

Figure \ref{fig:errorBias_whiteNoise} shows the behaviour of the error on the estimated PSD of an uncorrelated process with unitary variance for different N. We can notice that for small N ($\sim 10$) the error is relevant, while after a certain window width ($\sim 50$ samples) there are not important improvements.

\subsection{Subtraction of a trend estimated with regression}
\label{subsec:bias_regression}

If the trend is not a constant, but has some kind of a functional form, we first need to estimate it, and one technique is the regression method.
In this section we will follow the procedure of E. J. Hannan\cite{Hannan58}, and we will use his results. Under some conditions on the regressors, it can be shown (Grenander-1954) that the least squares estimate of the regression parameters has, asymptotically, the same covariance matrix as the best linear unbiased estimate.\\
Let us consider a regression:
\begin{equation}\label{eq:hannan_regr}
\mathbf{y} = \mathbf{X}\mathbf{\beta} + \mathbf{\epsilon}
\end{equation}
where $\mathbf{y}$ is a $n$ elements vector of dependent variables, $\mathbf{X}$ is a $(n \; \mbox{x} \; p)$ matrix of regressors, \textbf{$\beta$} is the $p$-dimension vector of regression coefficients and $\mathbf{\epsilon}$ is the $n$ elements vector of residuals. We limit the analysis to the case where $\mathbf{X}$ is fixed, or at least where $\mathbf{\epsilon}$ and $\mathbf{X}$ are independent. We further assume that $\mathbf{\epsilon}$ is a real valued stationary random process with continuous spectral function, with spectral density $f_{\epsilon}(\omega)$.
We want the $\mathbf{X}$ matrix with properties assuring a smart behaviour:
\begin{itemize}
\item[(i)]{$\lim_{n\rightarrow +\infty} \sum_{t=1}^{N} x_j(t)^2 = \infty, \;\; j = 1 \ldots p$}
\item[(ii)]{$\lim_{n\rightarrow +\infty} \sum_{t=1}^{N+1} x_j(t)^2 / \sum_{t=1}^{N} x_j(t)^2= 1, \;\; j = 1 \ldots p$}
\item[(iii)]{$\exists \lim_{N \rightarrow +\infty} r_{j,k}(h) = \rho_{j,k}(h)\;\; j,k = 1 \ldots p$, where
\[r_{j,k}(h) = \frac{\sum_{t=1}^N x_j(t)x_k(t+h)}{\sqrt{\sum_{t=1}^N x_j(t)^2\sum_{t=1}^N x_k(t)^2}} \]
is the `sample autocorrelation value'}
\item[(iv)]{If we extend the definition of $\rho_{j,k}(h)$, setting $x(t)=0, \; \forall t \not\in [1,N]$ we obtain a matrix $(p \; \mbox{x} \; p)$ $\mathbf{R}_{jk}(h), \; \forall h \in |\!\mbox{N}, \; j,k = 1 \ldots p$. We ask $\mathbf{R}(0)$ to be non singular.}
\end{itemize}
\noindent These requirements assure that the regressors can increase without upper bound (i) but with a slow rate (ii), that they are not linear dependent (iv) and that it is possible to define a correlation function as limit of the samples time averages (iii).
Under these assumptions, let $\mathbf{\hat{\epsilon}}$ be the residuals of the estimation of the $\mathbf{\beta}$ coefficients through least square regression. The PSD of the detrended signal will be the PSD of the $\mathbf{\hat{\epsilon}}$ residuals:
\begin{equation}
R(e^{i\omega}) = \frac{1}{N}\left|\sum_{t=1}^{N} \hat{\epsilon} e^{it\omega} \right|^2.
\end{equation}
Hannan showed that the bias on this estimate is asymptotically given by:
\begin{equation}\label{eq:hannan_bias}
\lim_{N\rightarrow + \infty} E[R(e^{i\omega})] = 2\pi f_{\epsilon}(\omega) \cdot \lim_{N\rightarrow + \infty} \left( 1 - \frac{1}{N} \sum_{\mu=1}^{p} R_{\mu} (e^{it\omega}) \right)
\end{equation}
where $R_{\mu}$ is the PSD of the function $\phi_{\mu}(\omega)$:
\begin{equation}
R_{\mu} (e^{it\omega}) = \left|\sum_{t=1}^{N} \phi_{\mu}(t) e^{it\omega} \right|^2
\end{equation}
defined, in matrix form, by: $\Phi =  \mathbf{XP}$, where $\mathbf{P}$ is such that $\mathbf{X'P'PX} = \mathbf{I}$.\\

\noindent This result is very useful because it gives an expression for the bias on the residuals that is a function of the properties of the regressors and of the process $\epsilon$.\\

\subsection{Application to stochastic processes with a linear trend}
\label{subsec:linear_trend_general}

Our aim is the subtraction of a linear trend.\\
A polynomial function agrees the assumption (i)-(iv): as $t$ increases to $\infty$, the sum of the square values of $x(t)$ goes to $\infty$; for the linearity, the limit in (ii) becomes:
\begin{equation}
\lim_{N\rightarrow +\infty} \frac{\sum_{t=1}^{N+1} x_j(t)^2} {\sum_{t=1}^{N} x_j(t)^2} = 1 - \lim_{N\rightarrow +\infty} \frac{x_j(N+1)^2} {\sum_{t=1}^{N} x_j(t)^2} = 1, \;\; j = 1 \ldots p
\end{equation}
For the same reason, and for the finite sum at the numerator against the increasing sum at the denominator in assumption (iii), the $\lim_{n\rightarrow +\infty} r_{j,k}(h)$ exist, and it is zero for all $h>N$. Note that for $h=0$, $r_{j,k}(h)$ does not depend on $n$, since we have assumed $x_j(t)=0, \; \forall \; t>N, \; j = 1 \ldots p$. We can conclude that also assumption (iv) is satisfied.\\

\noindent Let us consider the composed signal of eq. \ref{eq:staz+cost}:
\begin{equation}
X(k) = X_s (k) + \sum_{i=0}^p \alpha_i x^i, \;\; k  \geq 0
\end{equation}
where $X_s$ has the same features as in Sec. \ref{subsec:nonzero_mean}. Then $X_s$ is the $\epsilon$ factor of eq. \ref{eq:hannan_regr}. We subtract an estimate of the trend $\sum_{i=0}^p \alpha_i x^i$ through least square regression analysis. The PSD of the resulting signal is the PSD of the regression residuals.\\
Let assume linear regressor of order $p=1$:
\begin{equation}
\mathbf{X}\mathbf{\beta} = \alpha_0 + \alpha_1 x 
\end{equation}
To apply the result of Hannan, we have to find a matrix $\mathbf{P}$ such that $\mathbf{X'P'PX} = \mathbf{I}$. We limit our analysis at the case $x(t) \in [-1,1]$. In this situation, we can find several orthonormal basis, otherwise, we should search for approximate solutions. We choose the Legendre polynomials (see, e.g., \cite{Gradsh}):
\begin{eqnarray}
\nonumber \phi_0 &=& \sqrt{\frac{1}{2}}\\
\phi_1 &=& \sqrt{\frac{3}{2}} x
\end{eqnarray}
The $\mathbf{P}$ matrix is:
\begin{equation}
\mathbf{P} = \left( \begin{array}{cc}
                      \frac{1}{\sqrt{2} \alpha_0} & 0 \\
                      0 & \frac{\sqrt{3}}{\sqrt{2} \alpha_1}
                    \end{array}
 \right)
\end{equation}

\noindent The bias on the PSD of the $X_s$ signal will be, according to eq. \ref{eq:hannan_bias}:
\begin{eqnarray}
\nonumber E[R(e^{i\omega})] = 2\pi f_{\epsilon}(\omega) \cdot \left( 1 - \lim_{N\rightarrow + \infty} \frac{1}{N} \sum_{\mu=0}^{1} R_{\mu} (e^{it\omega}) \right) = \\
= 2\pi f_{\epsilon}(\omega) \cdot \left[ 1 - \lim_{N\rightarrow + \infty} \frac{1}{N} \left(\left|\sum_{t=1}^{N} \phi_0(t) e^{it\omega} \right|^2 + \left|\sum_{t=1}^{N} \phi_1(t) e^{it\omega} \right|^2 \right)\right)
\end{eqnarray}

\noindent The finite summations can be written as:

\begin{equation}\frac{1}{N}\left|\sum_{t=1}^{N} \phi_0(t) e^{it\omega} \right|^2 = \frac{1}{2N} \left( \frac{\sin (\omega N/2)}{\sin (\omega /2)} \right)^2 \rightarrow_{N\rightarrow +\infty} 0, \; \omega \neq k\pi, \; k = 0, \pm 1, \pm 2, \ldots
\end{equation}

\begin{eqnarray}
\nonumber \frac{1}{N}\left|\sum_{t=1}^{N} \phi_1(t) e^{it\omega} \right|^2 = \frac{3}{2N} \left|\sum_{t=1}^{N} t \cos(\omega t) + i \sum_{t=1}^{N} t \sin(\omega t) \right|^2 = \\
\nonumber = \frac{3}{2N}\left(\left[N\sin(N\omega) + \frac{\sin(N\omega)}{4\sin^2(\omega/2)} -  \frac{\cos(\frac{2N-1}{2}\omega)} {2\sin(\omega/2)} \right]^2 +  \right.\\
\left. + \left[N\cos(N\omega) - \frac{1 - \cos(N\omega)}{4\sin^2(\omega/2)}+  \frac{N\sin(\frac{2N-1}{2}\omega)}{2\sin(\omega/2)} \right]^2\right)
\end{eqnarray}
where we have used the relations\cite[p. 38]{Gradsh}:
\begin{eqnarray}
\nonumber \sum_{t=1}^{N-1} t \sin(\omega t) = \frac{\sin(N\omega)}{4\sin^2(\omega/2)} -  \frac{N \cos(\frac{2N-1}{2}\omega)} {2\sin(\omega/2)}  \\
\sum_{t=1}^{N-1} t \cos(\omega t) = - \frac{1 - \cos(N\omega)}{4\sin^2(\omega/2)} +  \frac{N\sin(\frac{2N-1}{2}\omega)}{2\sin(\omega/2)}
\end{eqnarray}
There are not convergence problems if $\omega \neq k\pi, \; k = 0, \pm 1, \pm 2, \ldots$. The latter are discontinuity points for which the limit goes to $\infty$ independently from $N$. For all other $\omega$, however, for $N\rightarrow \infty$ the limit does not converge, as we obtain:
\begin{equation}
\nonumber \frac{3}{2N}\left|\sum_{t=1}^{N} \phi_1(t) e^{it\omega} \right|^2 = \frac{3}{2N} N^2 \left( 1 + \frac{1} {4\sin^2(\omega/2)} - \frac{\sin(2N\omega - \omega/2)}{2 \sin(\omega/2)} \right)
\end{equation}

\noindent From the last equation arises the need of an appropriate windowing of the detrended data for the bias compensation.

\subsubsection{Application to uncorrelated samples}

Again, we apply the result to samples whose residuals are uncorrelated, since this seems to be the case of the residuals of the data we analyze in chap. \ref{chap:stat}, sec. \ref{sec:time-stat}, after the detrending operation. In this case, the spectral density function is
\begin{equation}
f_{\epsilon}(\omega) = \frac{\sigma^2_{\epsilon}}{2 \pi}, \;\; \forall \omega
\end{equation}
Let $\sigma^2_{\epsilon} = 1$.

\begin{figure}[htb]
    \begin{center}
        \epsfig{figure=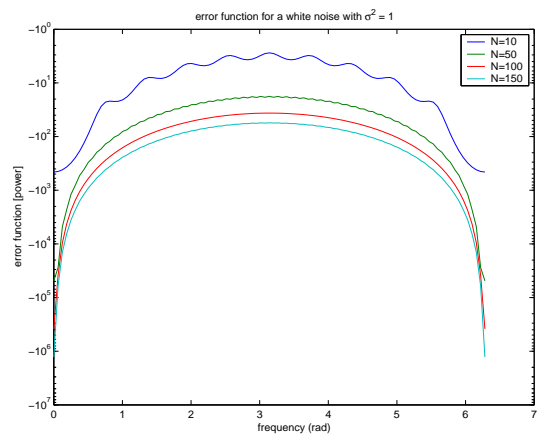,width=7cm}
        \caption{Error for the expectation of periodogram at different N for un uncorrelated process}
        \label{fig:bias_whiteNoise_trend}
    \end{center}
\end{figure}

\noindent We apply eq. \ref{eq:hannan_bias} at different window sizes. Figure \ref{fig:bias_whiteNoise_trend} shows the behaviour of the expectation of the estimation through the periodogram for increasing $N$. The vertical axis is in logarithmic scale. The estimation goes to $\infty$ at the extreme values of the frequencies interval: these are the points for which $\sin(\omega/2) \rightarrow 0$.

\noindent Note that if we subtract just a constant, the PSD is asymptotically unbiased, since the correction limit tends to zero as $N$ tends to $\infty$. This completes the results of Section \ref{subsec:nonzero_mean}.

\chapter{Graphics omitted in chapter 4}
\lhead[\fancyplain{}{\bfseries\thepage}]%
      {\fancyplain{}{\bfseries Graphics omitted in chapter 4}}
\rhead[\fancyplain{}{\bfseries Graphics omitted in chapter 4}]%
      {\fancyplain{}{\bfseries\thepage}}
\label{appendixB}
In this appendix are reported the graphics that we did not insert in chapter \ref{chap:stat} for ease of reading.

\section{Statistical analysis in the time domain}

\subsection{Void channels}

\begin{figure*}[htbp]
   \begin{center}
    \epsfig{figure=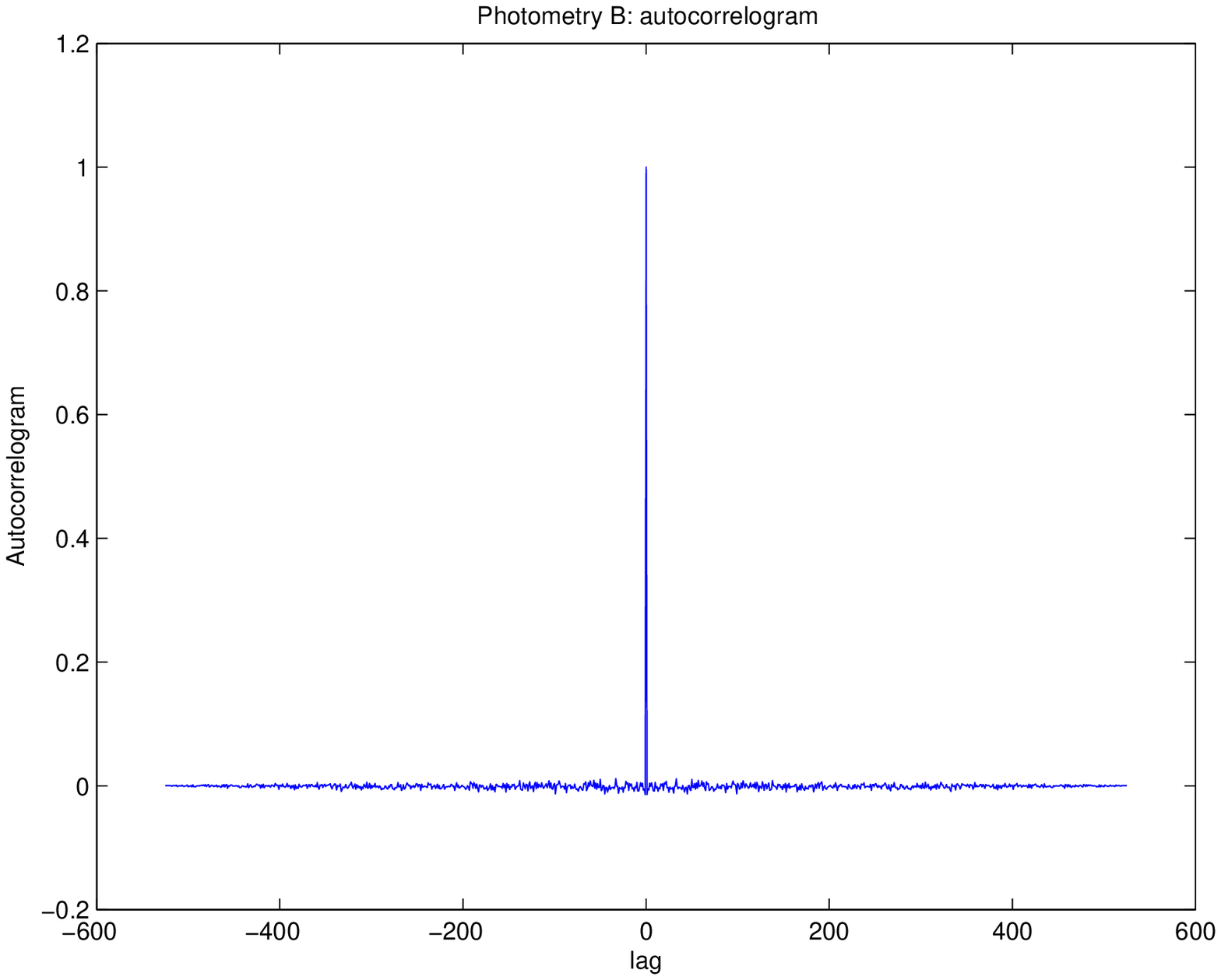,width=6.5cm}
    \epsfig{figure=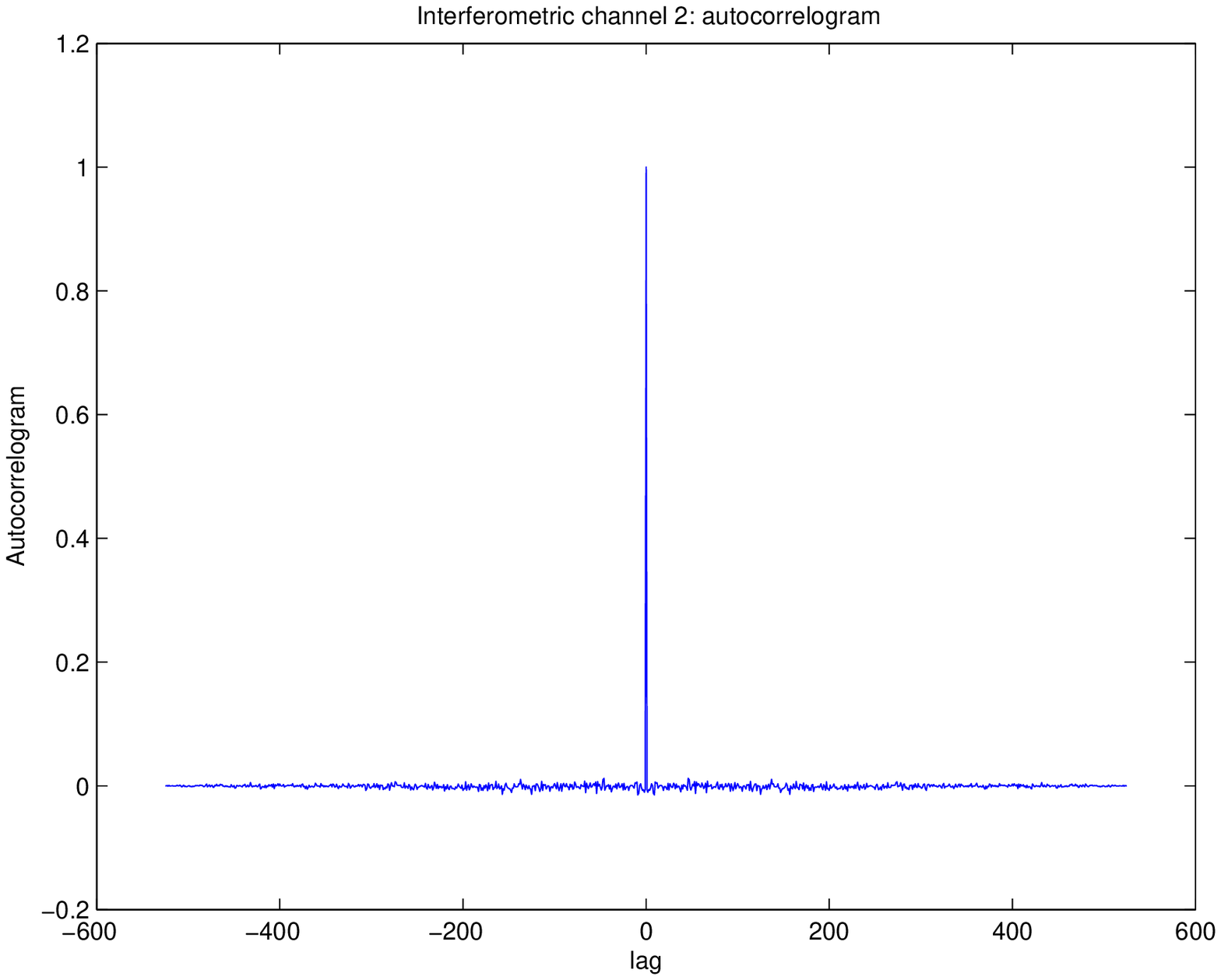,width=6.5cm}
      \caption{Case 1: Autocorrelation function estimate for photometric channel $PB$ (left) and interferometric channel $I2$ (right). }
   \end{center}
\end{figure*}

\begin{figure*}[htbp]
    \begin{center}
    \epsfig{figure=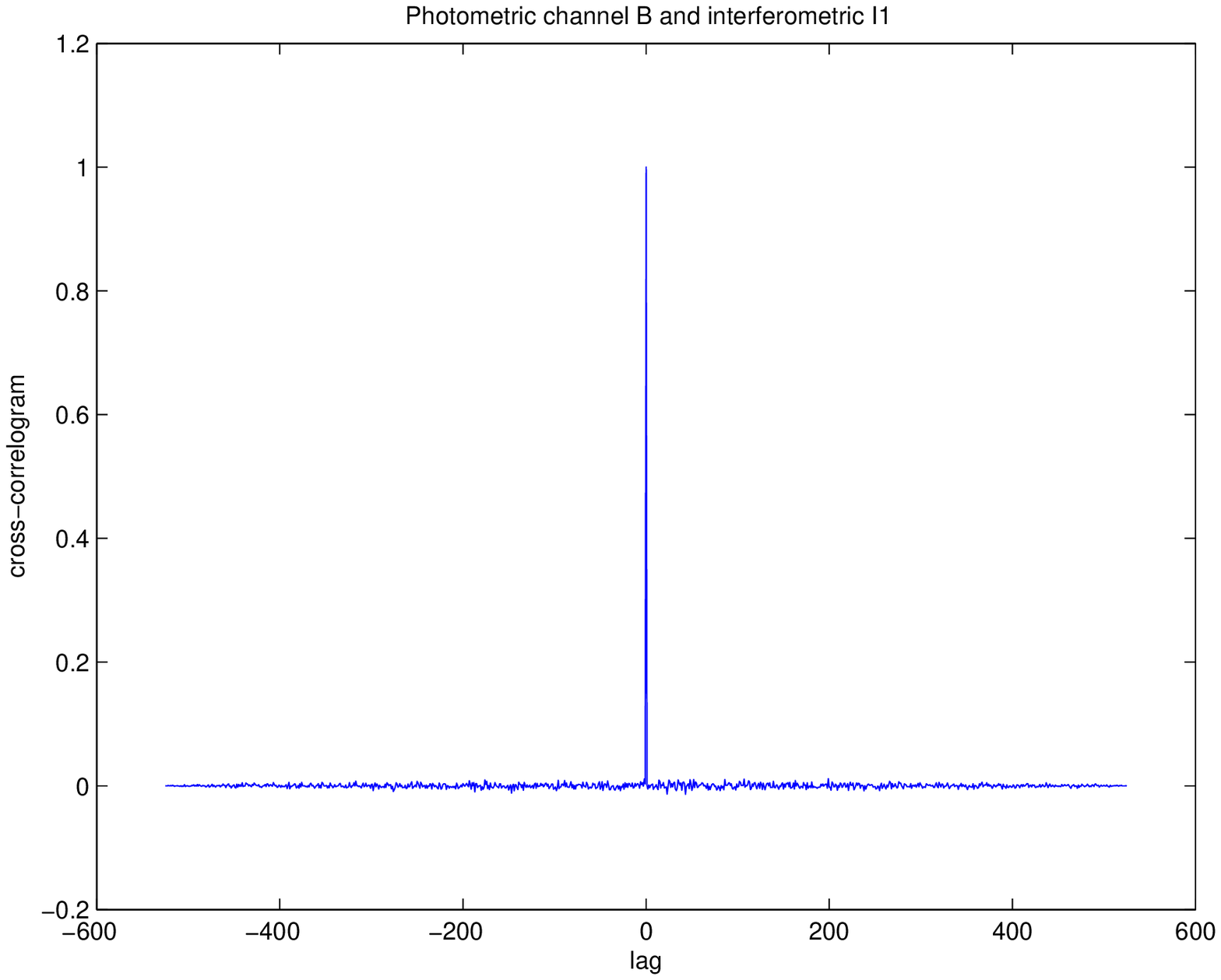,width=6.5cm}
    \epsfig{figure=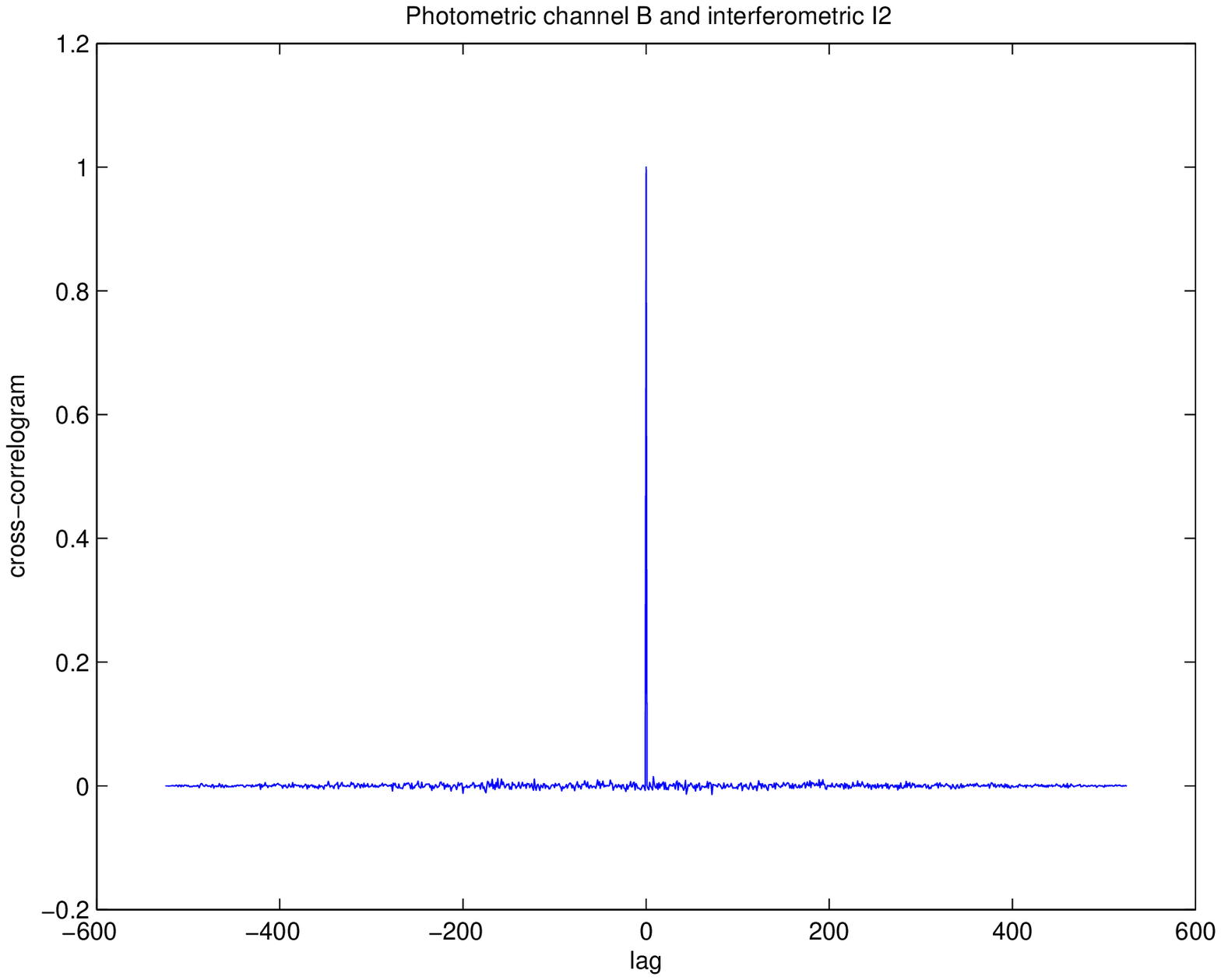,width=6.5cm}
    \caption{Case 1: cross-correlogram between inputs and outputs - $PB$ and $I1$ (left) and $PB$ and $I2$ (right)}
    \end{center}
\end{figure*}
\pagebreak

\subsection{Input: photometric signals}

\begin{figure*}[htbp]
    \epsfig{figure=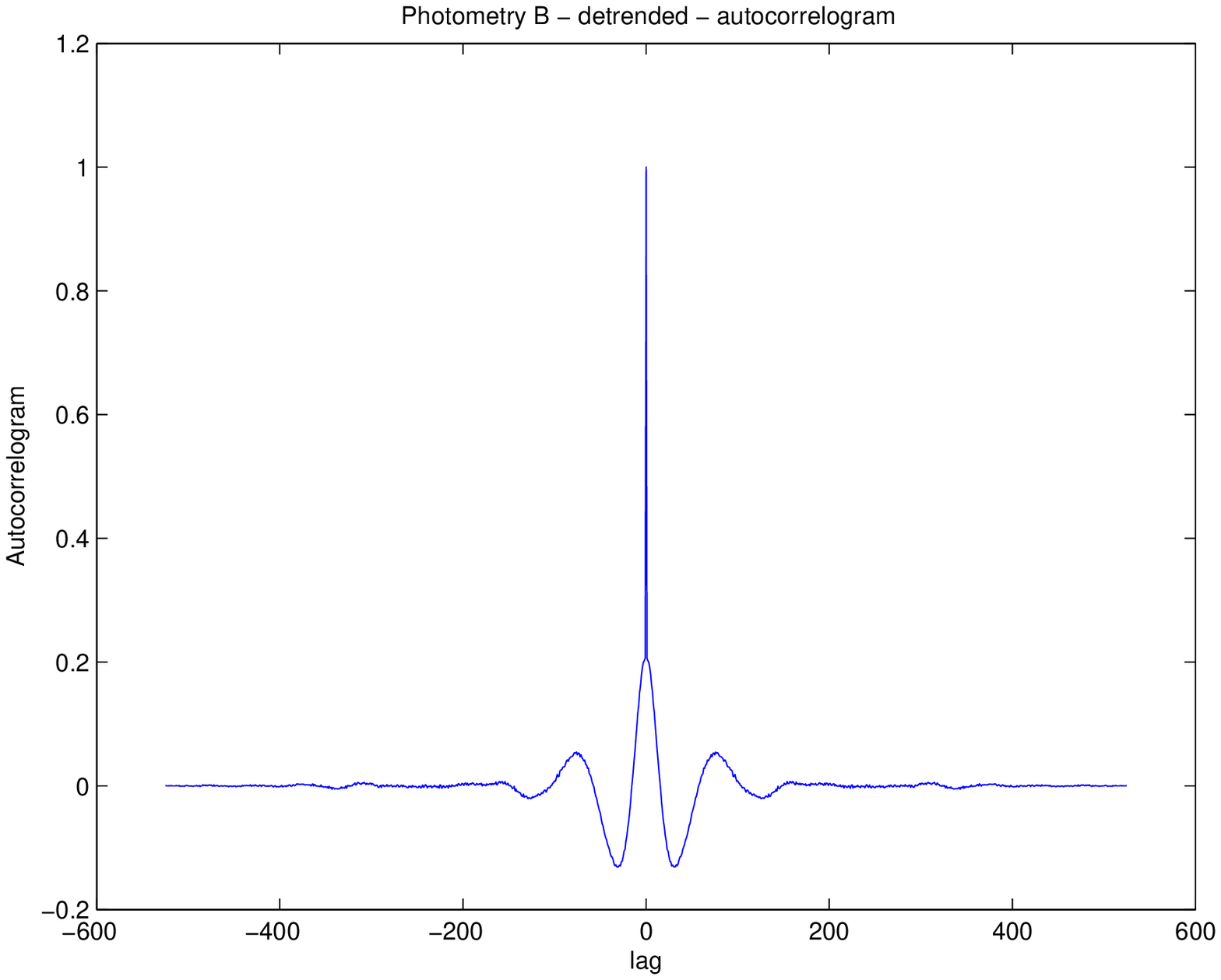,width=6.5cm}
    \epsfig{figure=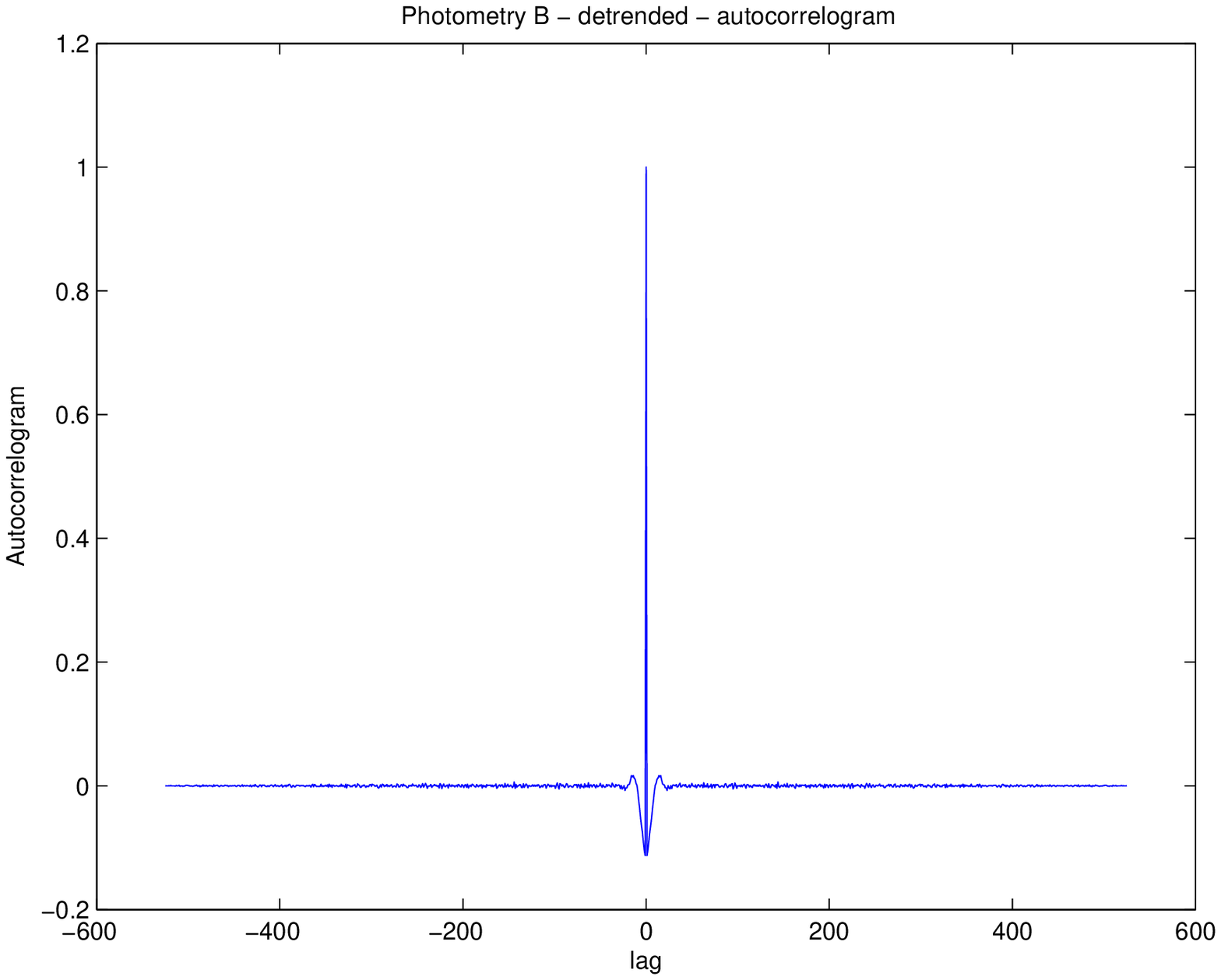,width=6.5cm}
    \begin{center}
        \caption{Case 4: raw data, autocorrelation functions for photometric channel B. Left, the linear trend to subtract is evaluated as a piecewise polynomial with breakpoints every $50$ samples; right, breakpoints are every $10$ samples.}
    \end{center}
\end{figure*}

\begin{figure*}[htbp]
    \epsfig{figure=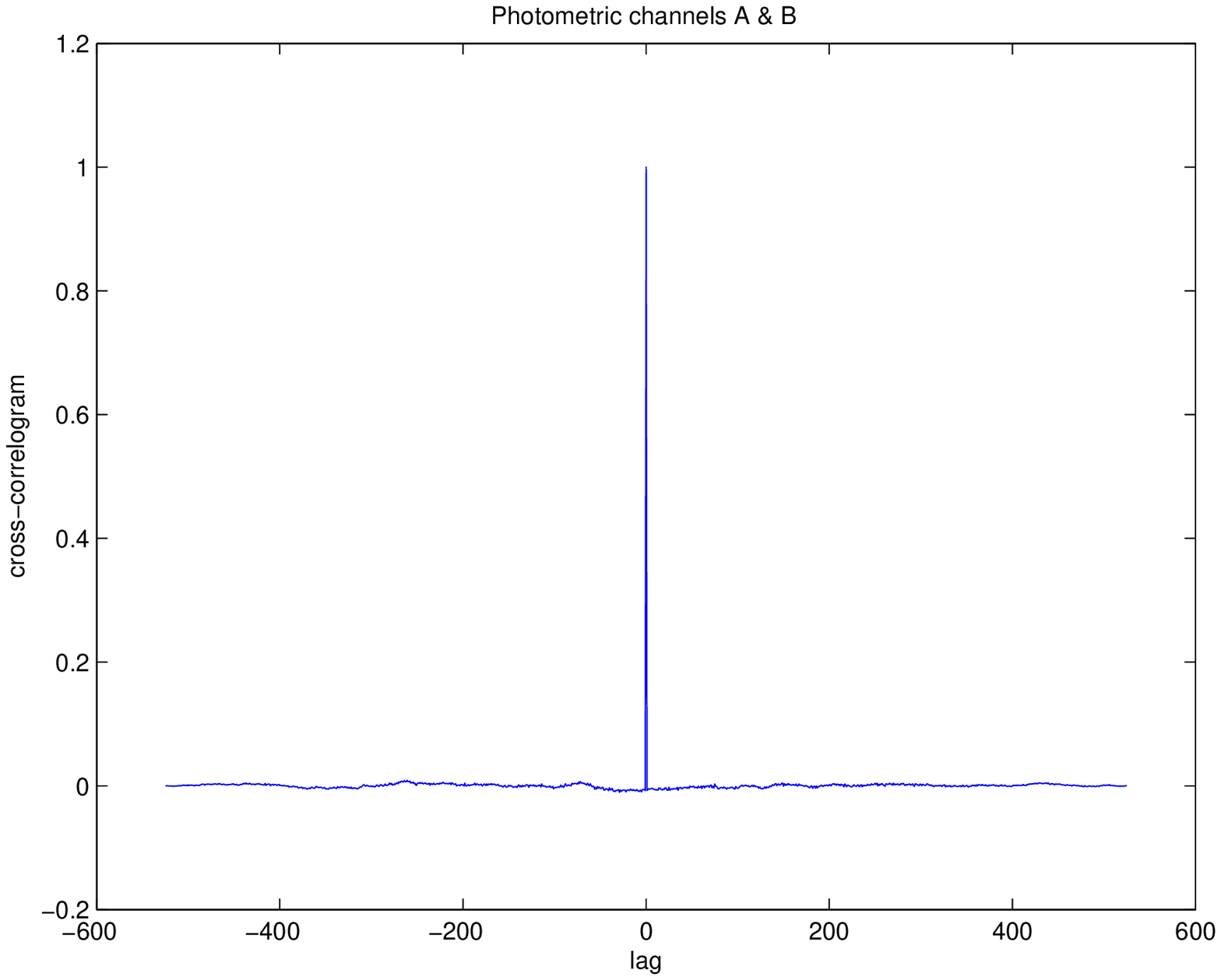,width=6.5cm}
    \epsfig{figure=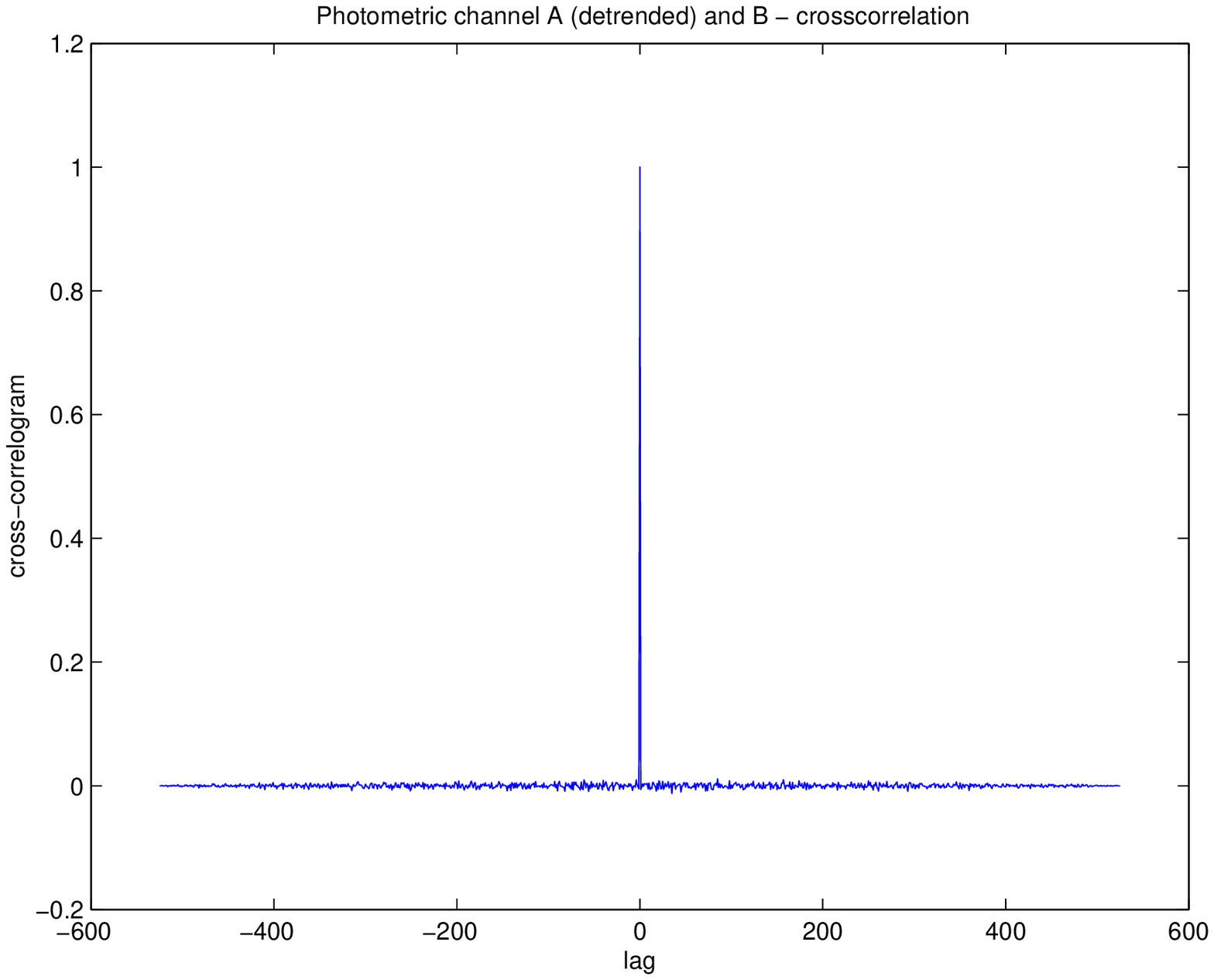,width=6.5cm}
    \begin{center}
        \caption{Case 2: Cross-correlation functions for photometric channel A and B. Left, raw data; right, linear trend subtracted.}
    \end{center}
\end{figure*}
\pagebreak

\subsection{Output in calibration mode}

\begin{figure*}[htbp]
    \begin{center}
    \epsfig{figure=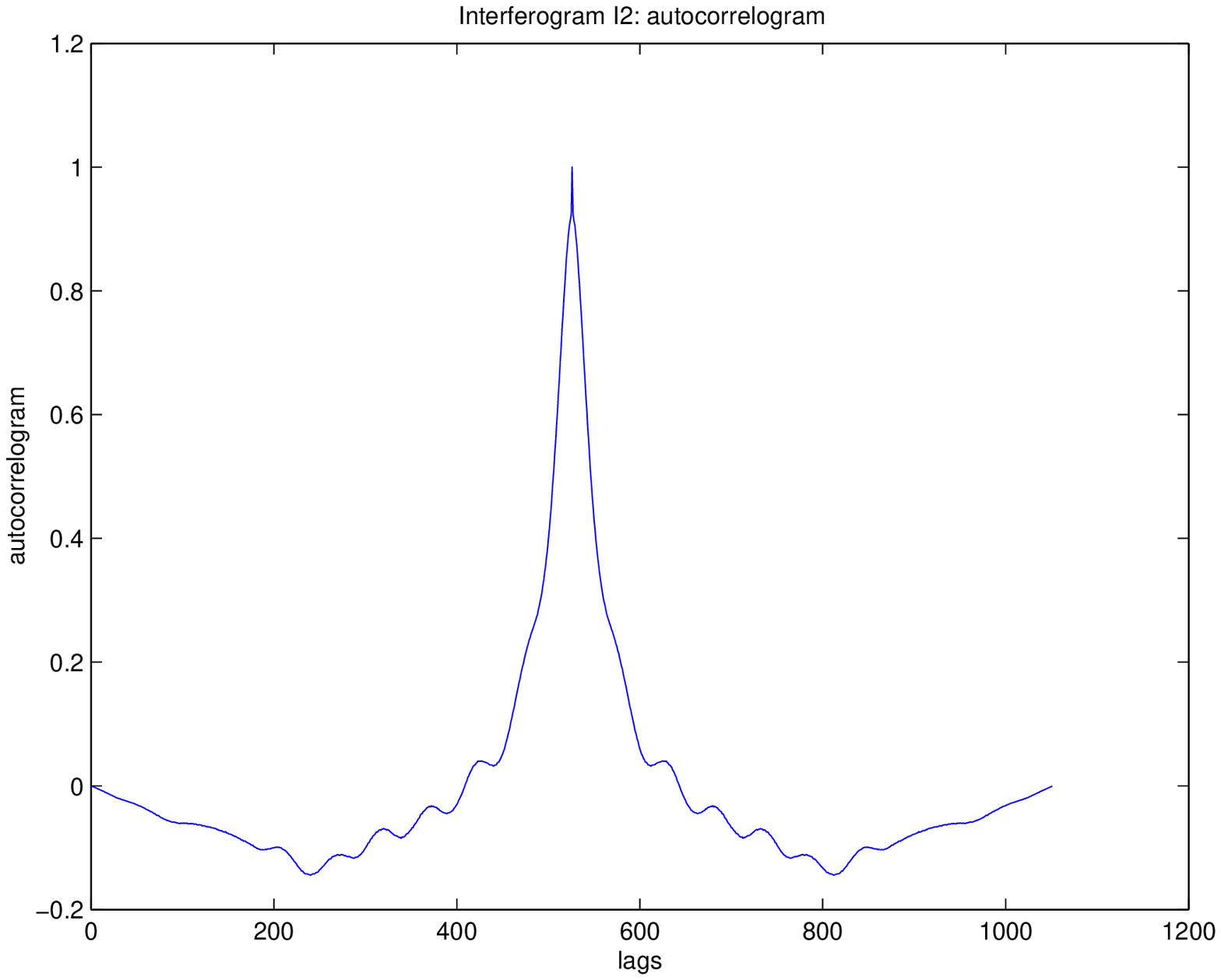,width=6.5cm}
    \epsfig{figure=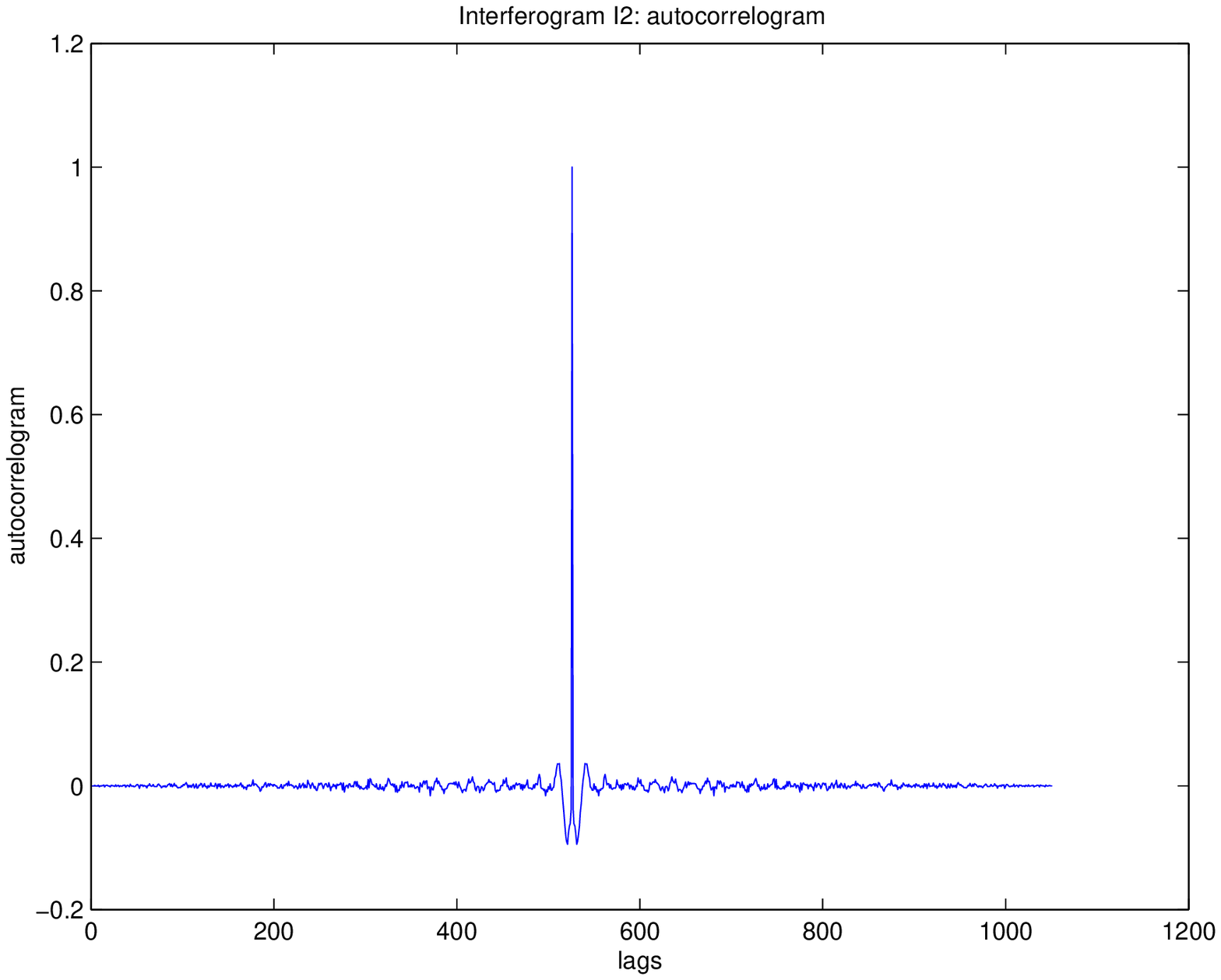,width=6.5cm}
   \caption{Case 2: Autocorrelation functions for interferometric channel $I2$: raw data (left) and detrended (right).}
    \end{center}
\end{figure*}

\begin{figure*}[htbp]
    \begin{center}
    \epsfig{figure=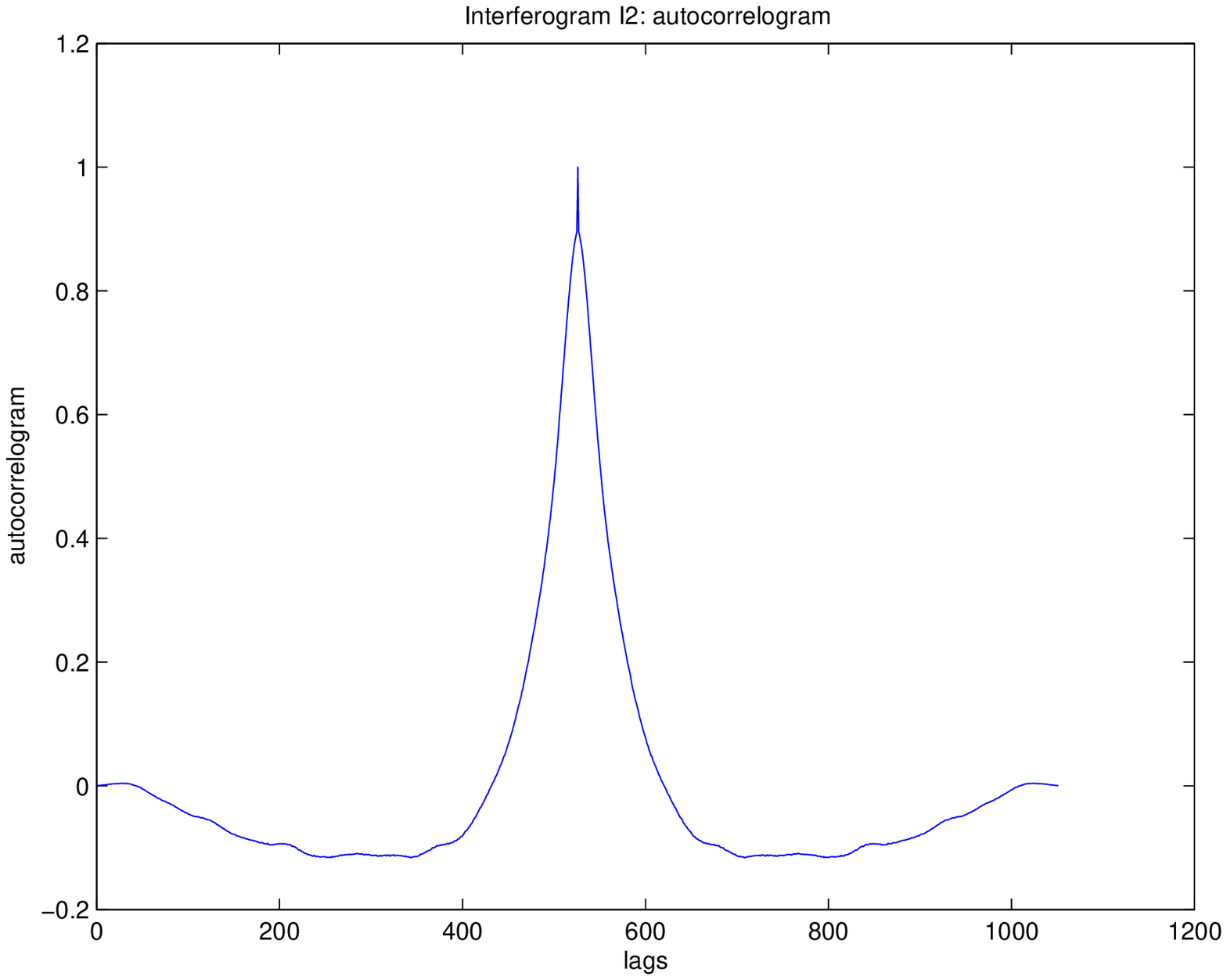,width=6.5cm}
    \epsfig{figure=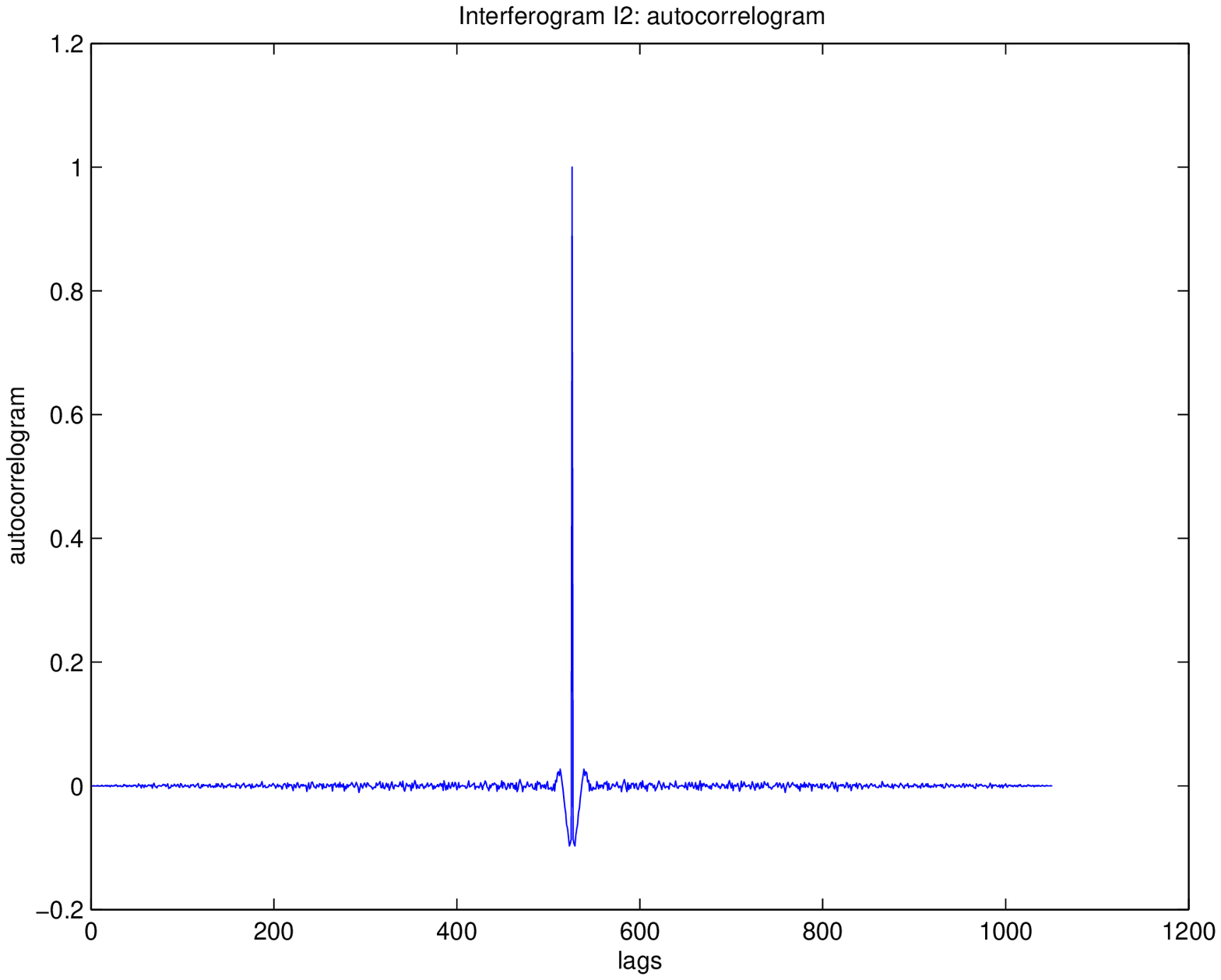,width=6.5cm}
    \caption{Case 3: Autocorrelation functions for interferometric channel $I2$: raw data (left) and detrended (right).}
    \end{center}
\end{figure*}

\begin{figure*}[htbp]
    \begin{center}
    \epsfig{figure=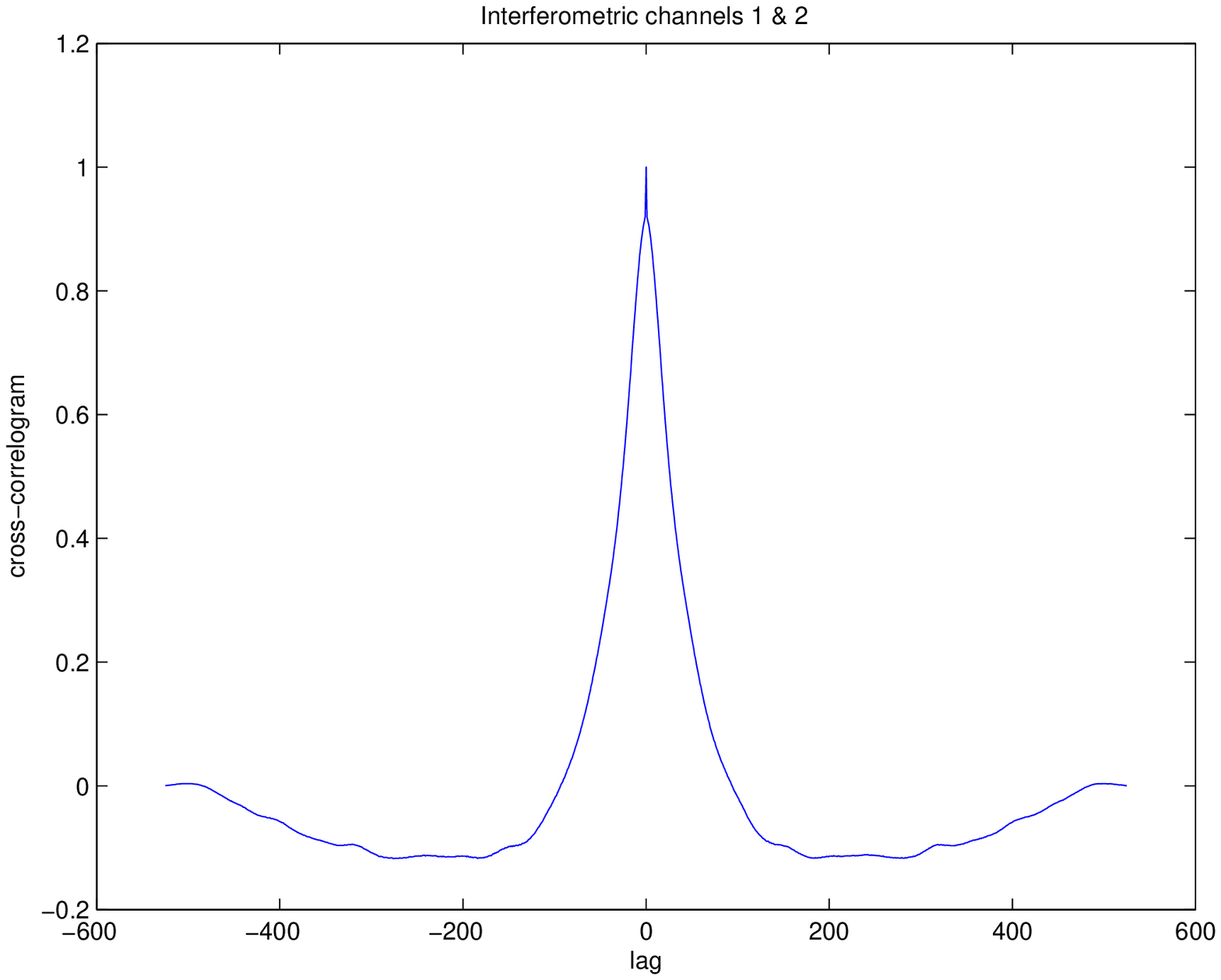,width=6.5cm}
    \epsfig{figure=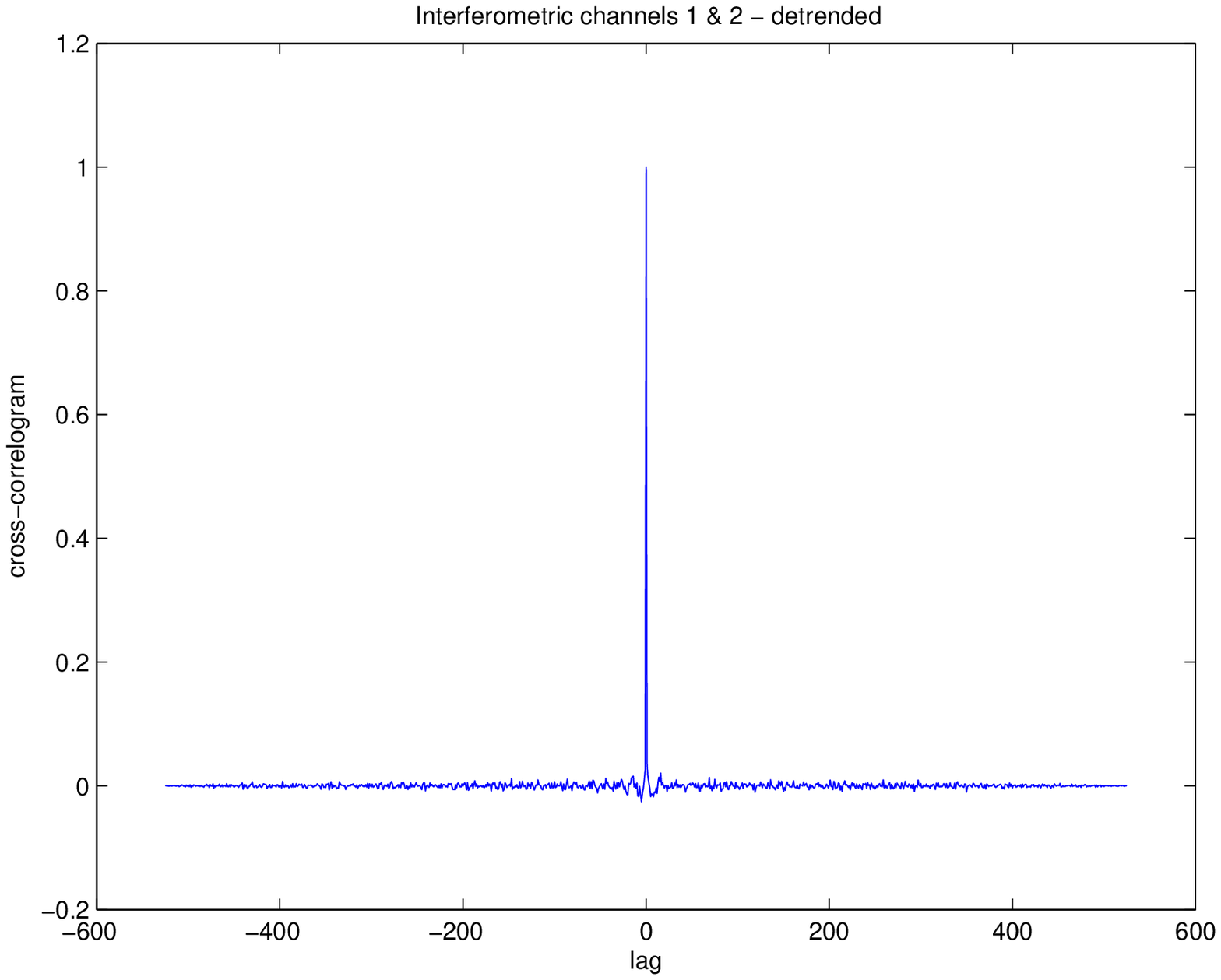,width=6.5cm}
    \caption{Case 3: Cross-correlation functions for interferometric channel $I1$ and $I2$: raw data (left) and detrended (right).}
    \end{center}
\end{figure*}
\pagebreak

\subsection{Output in observational mode: Interferometric signals }
\begin{figure*}[htbp]
    \begin{center}
    \epsfig{figure=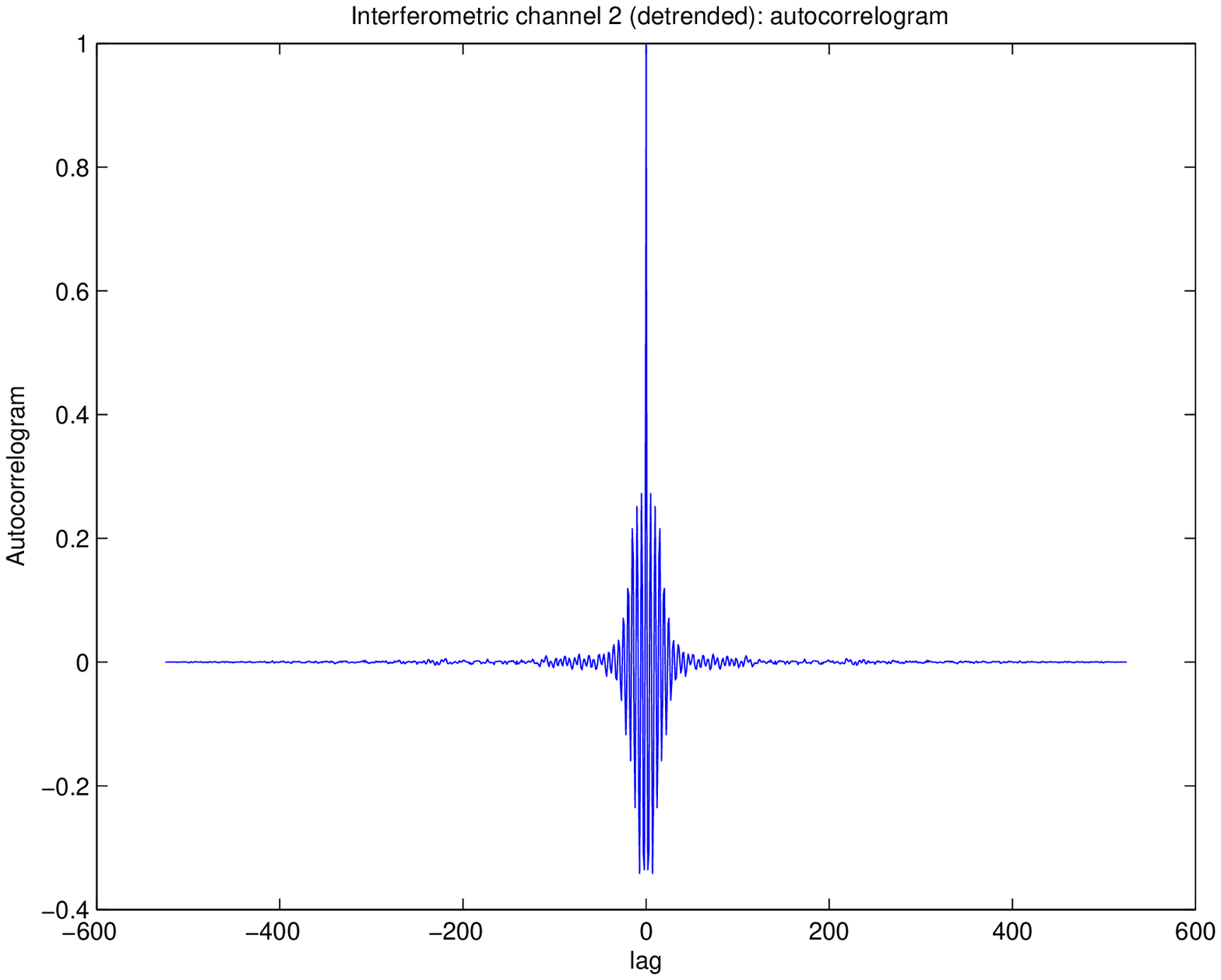,width=6.5cm}
    \epsfig{figure=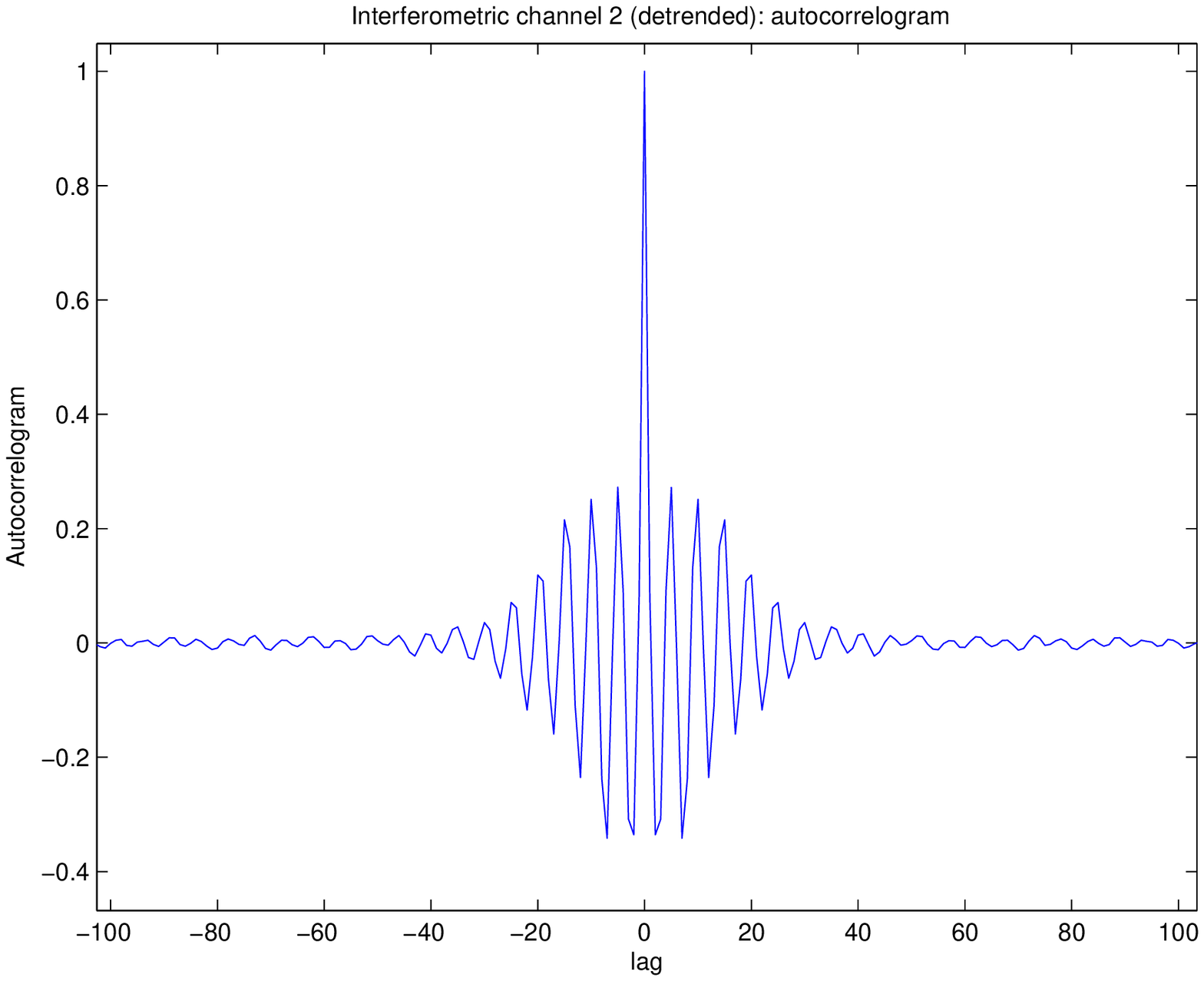,width=6.5cm}
    \caption{Case 4. autocorrelation function after a detrend for interferometric channel $I2$ (left) and a zoom in the central lags area (right).}
    \end{center}
\end{figure*}

\subsection{Cross-correlation between photometric inputs and interferometric outputs}

\begin{figure*}[htbp]
    \epsfig{figure=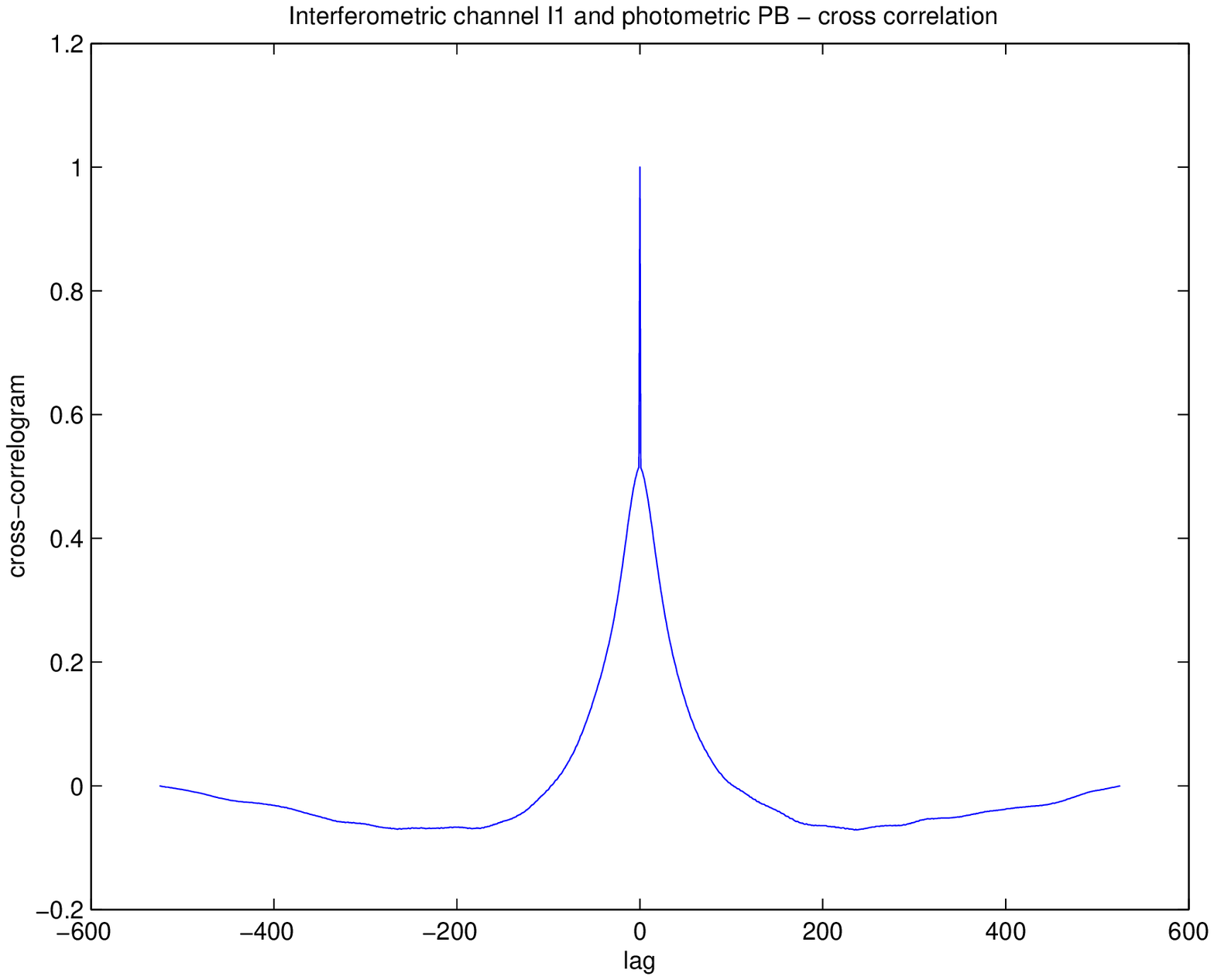,width=6.5cm}
    \epsfig{figure=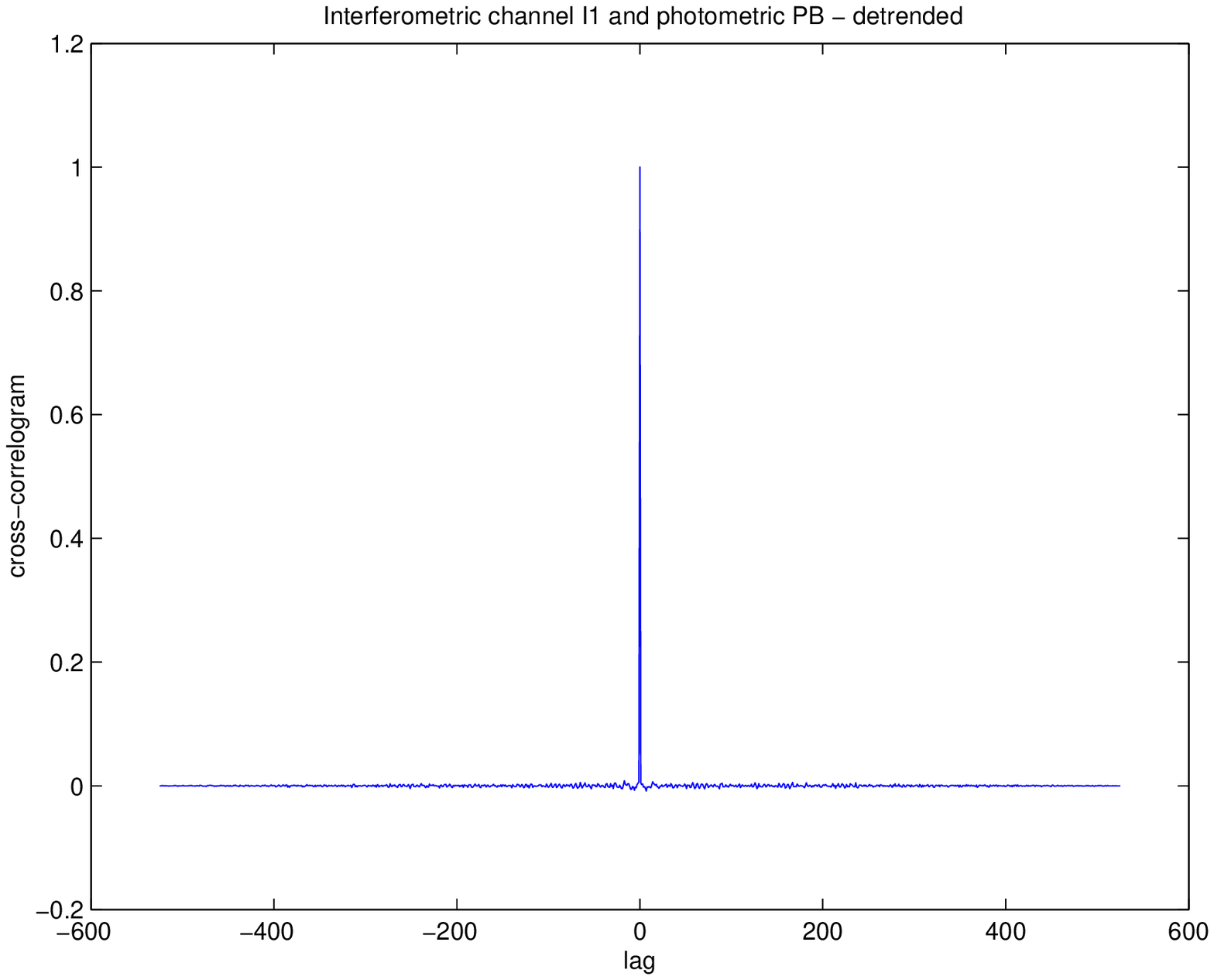,width=6.5cm}
    \begin{center}
        \caption{Case 4. Cross-correlation functions for interferometric $I1$ and photometric $PB$ channels. Left, raw data; right, linear trend subtracted.}
    \end{center}
\end{figure*}

\begin{figure*}[htbp]
    \epsfig{figure=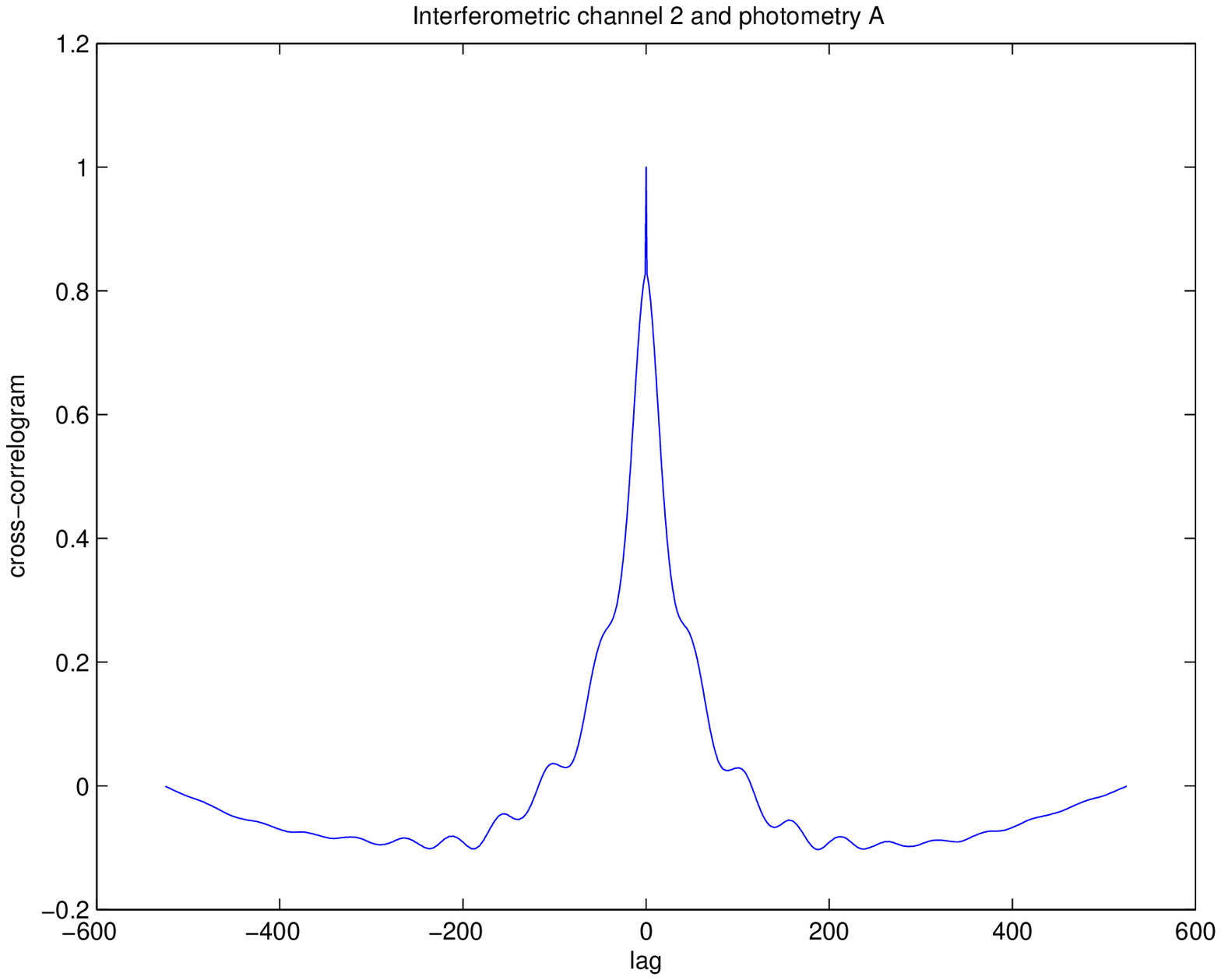,width=6.5cm}
    \epsfig{figure=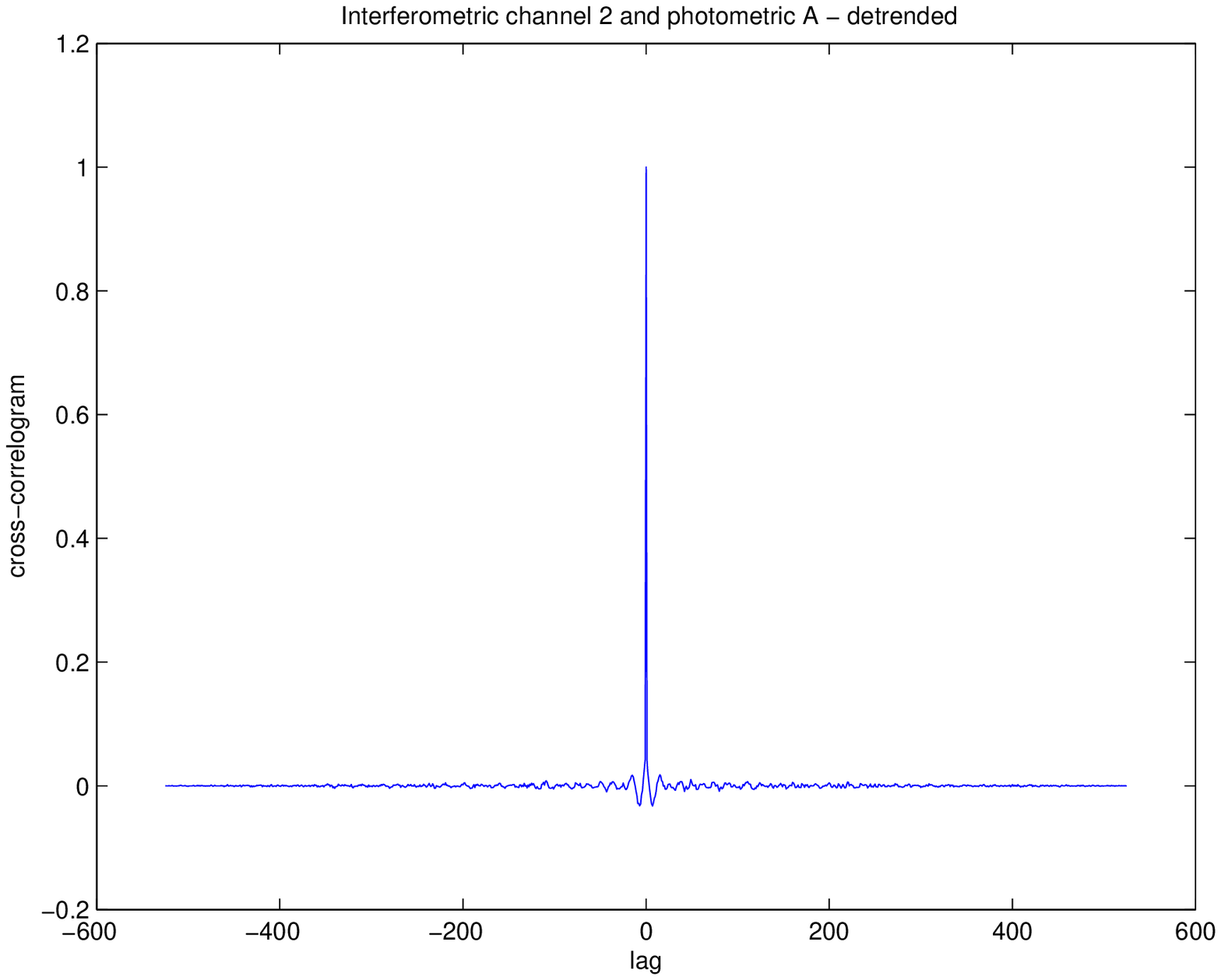,width=6.5cm}
    \begin{center}
        \caption{Case 4. Cross-correlation functions for interferometric $I2$ and photometric $PA$ channels. Left, raw data; right, linear trend subtracted.}
    \end{center}
\end{figure*}

\begin{figure*}[htbp]
    \epsfig{figure=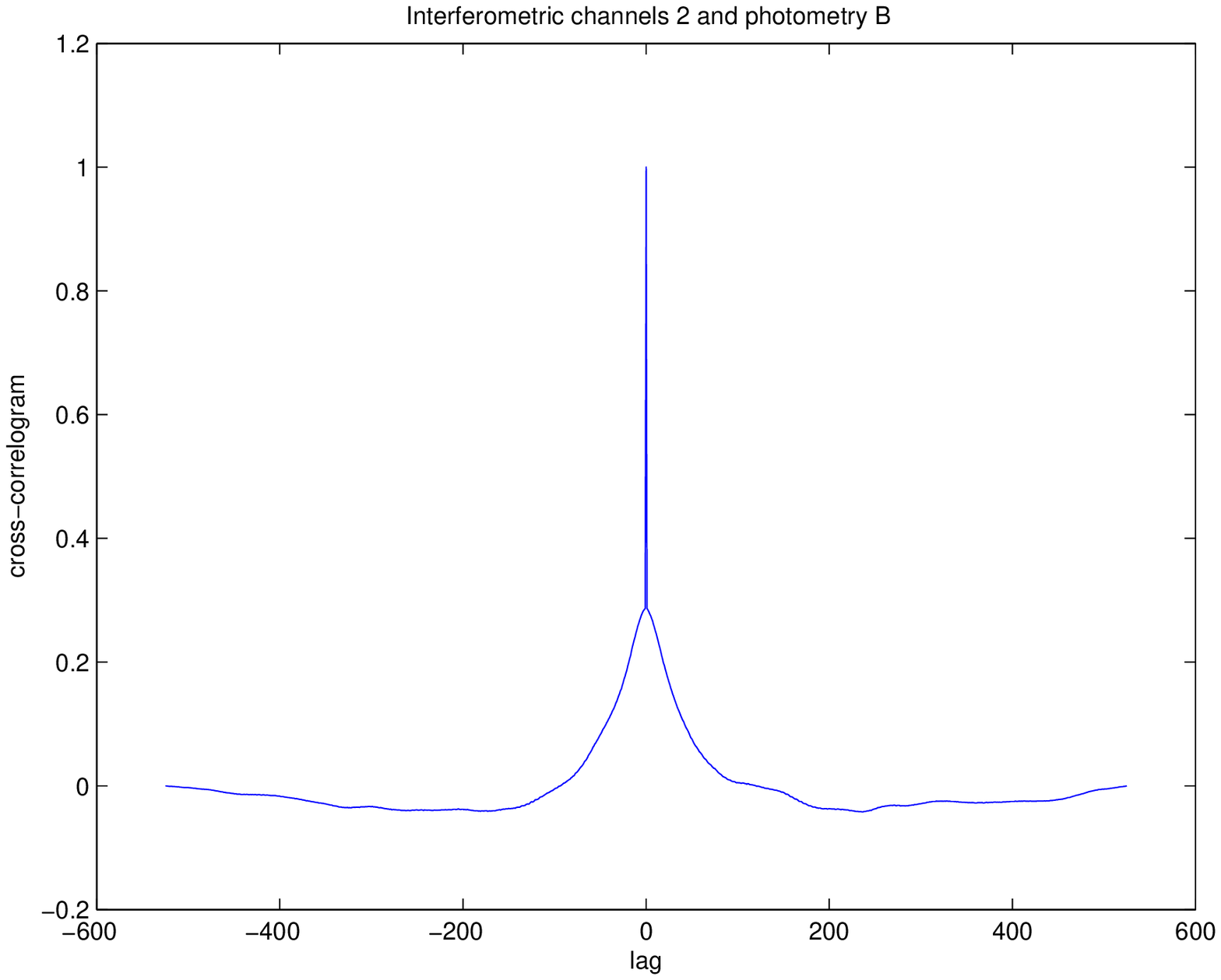,width=6.5cm}
    \epsfig{figure=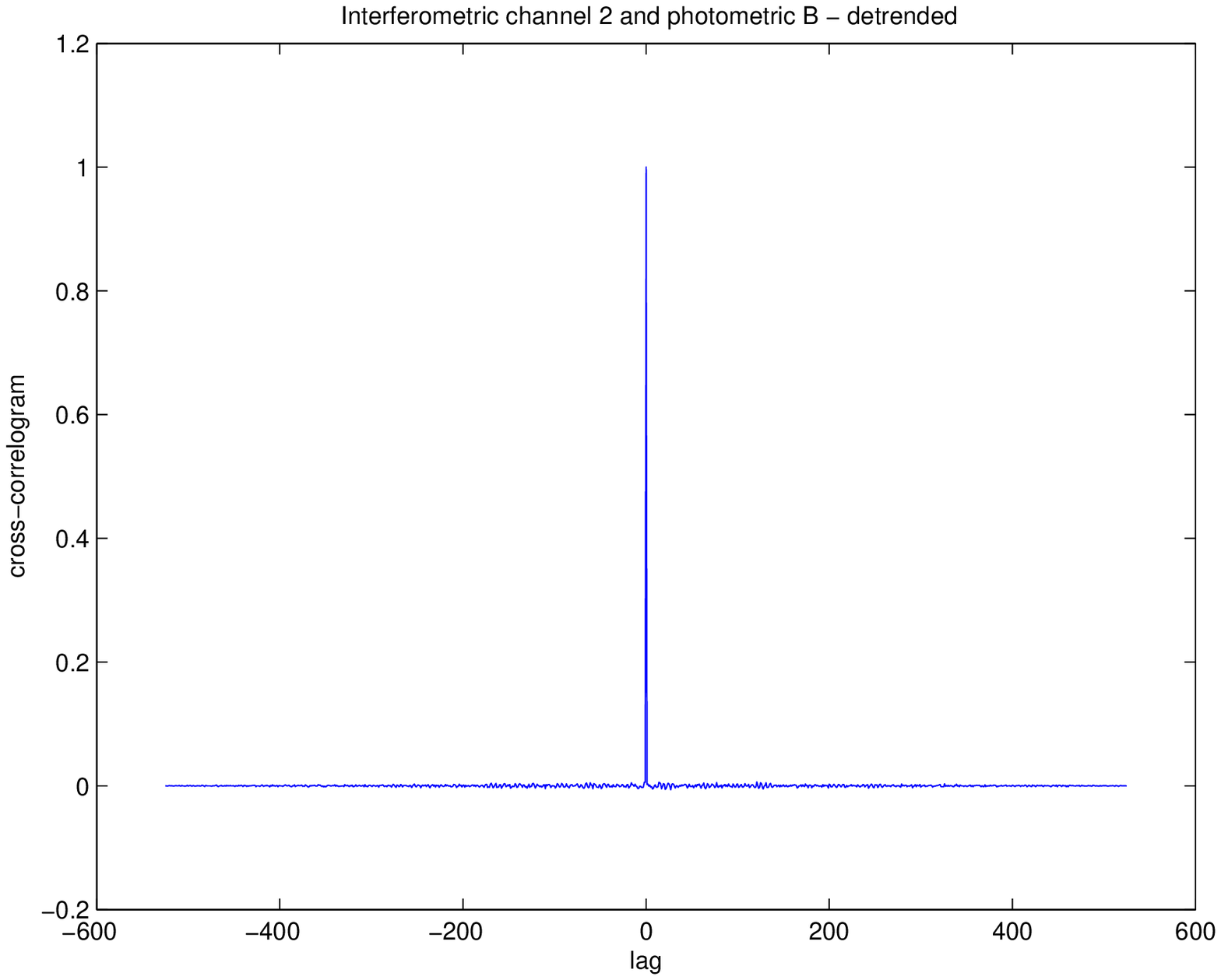,width=6.5cm}
    \begin{center}
        \caption{Case 4. Cross-correlation functions for interferometric $I2$ and photometric $PB$ channels. Left, raw data; right, linear trend subtracted.}
    \end{center}
\end{figure*}


\section{Statistical analysis in the frequency domain}

\begin{figure*}[htbp]
    \begin{center}
    \epsfig{figure=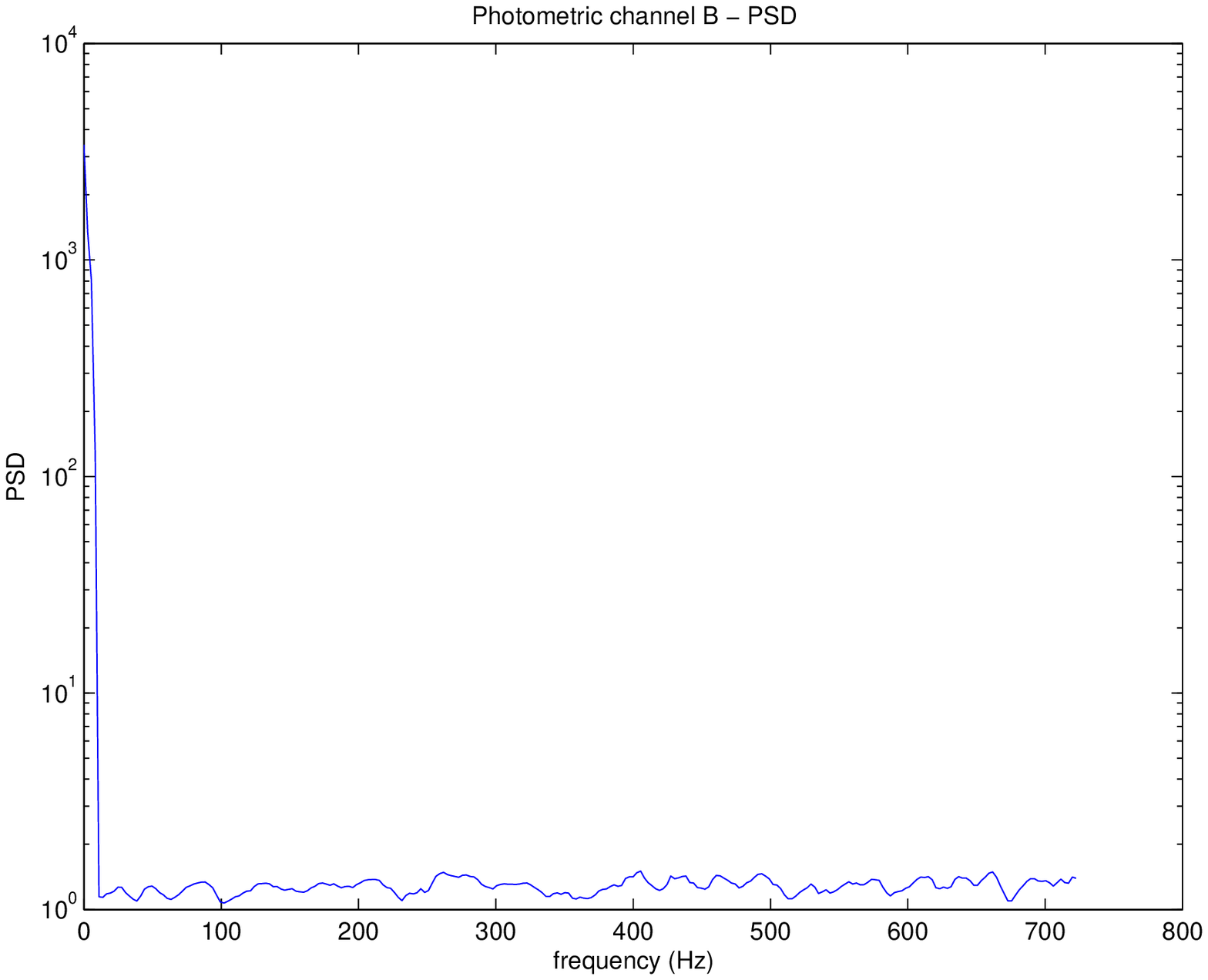,width=6.5cm}
    \epsfig{figure=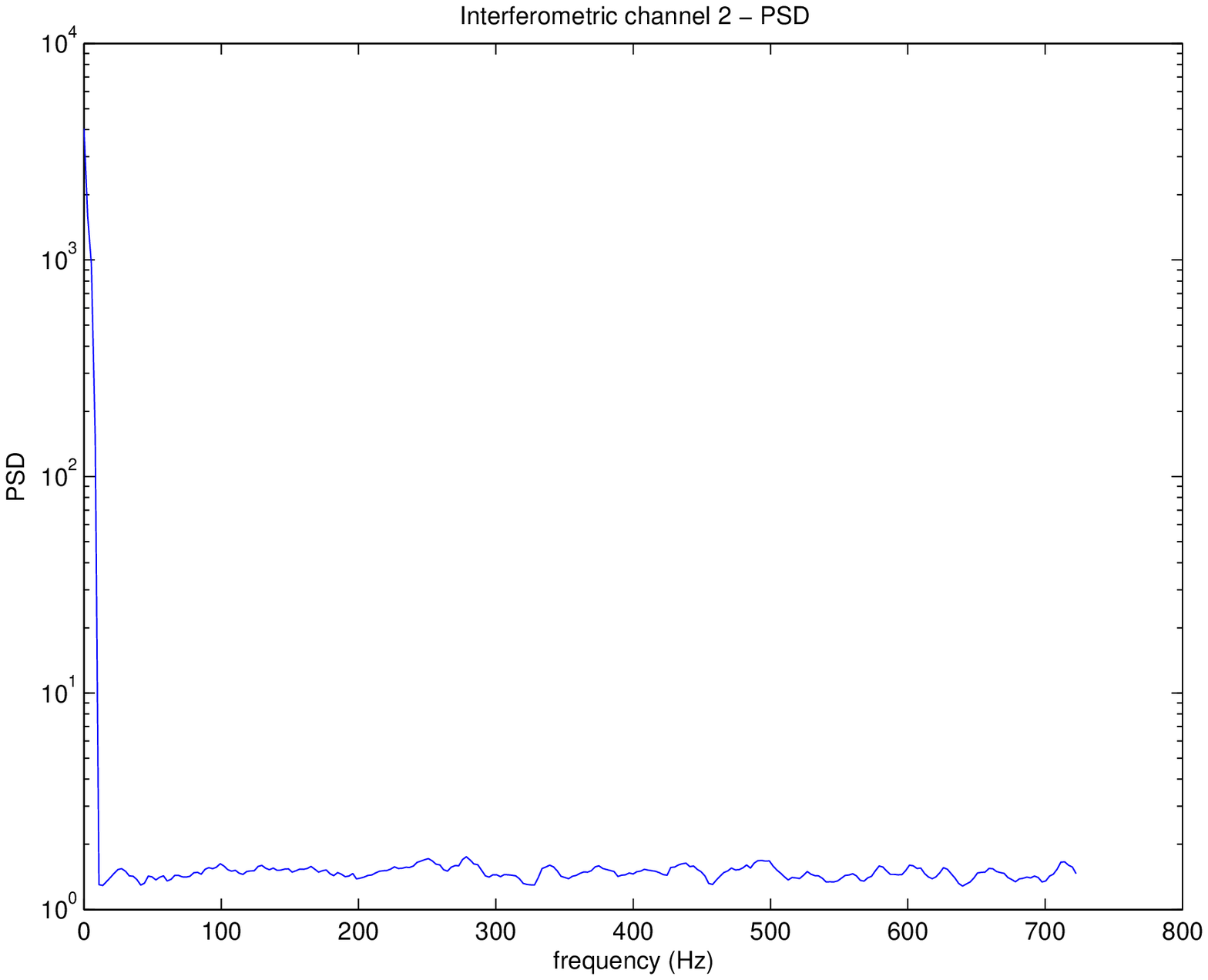,width=6.5cm}
    \caption{Case 1: Power Spectral Density functions for photometric input $PB$ (left) and interferometric output $I2$ (right).}
   \end{center}
\end{figure*}

\begin{figure*}[htbp]
    \begin{center}
    \epsfig{figure=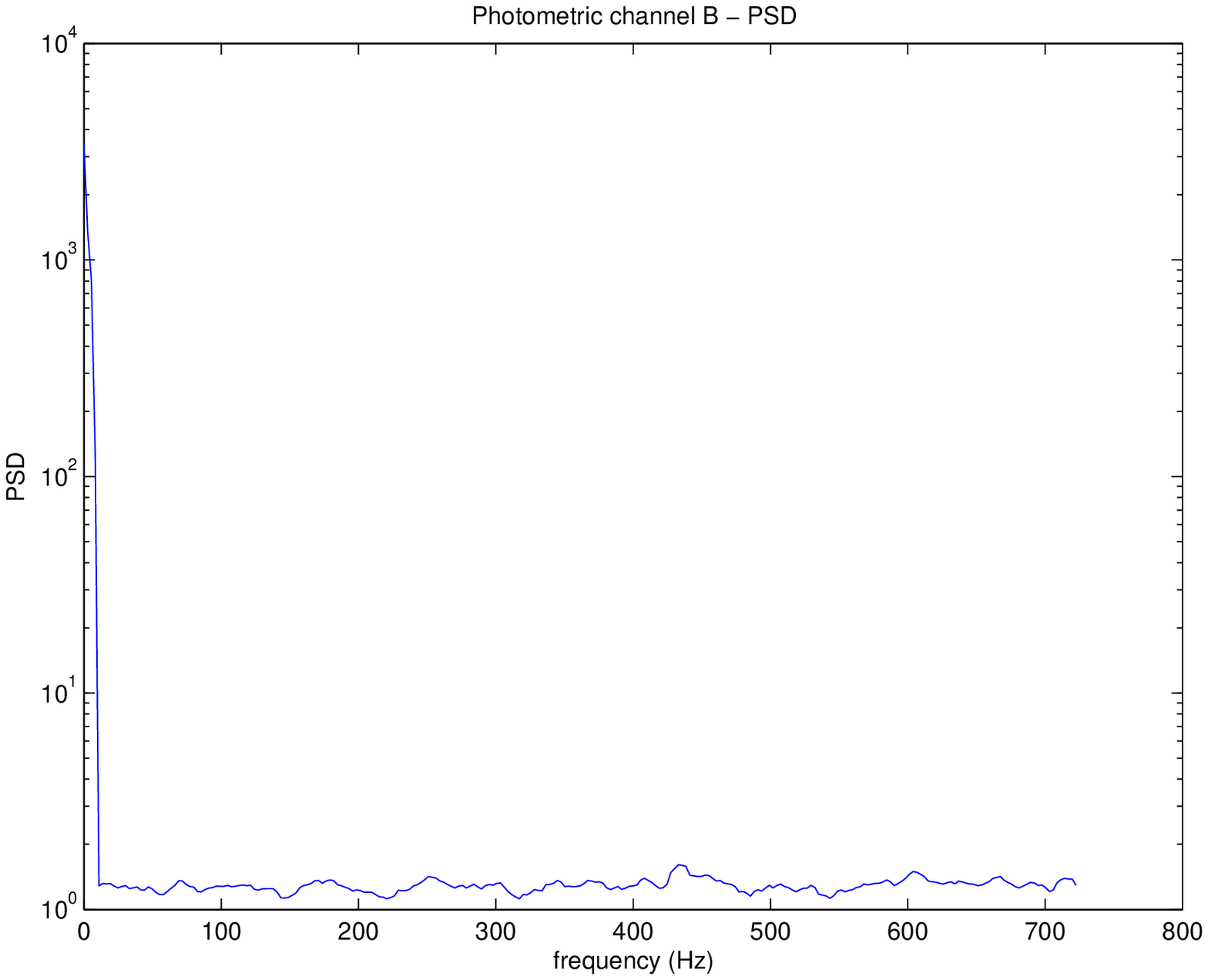,width=6.5cm}
    \epsfig{figure=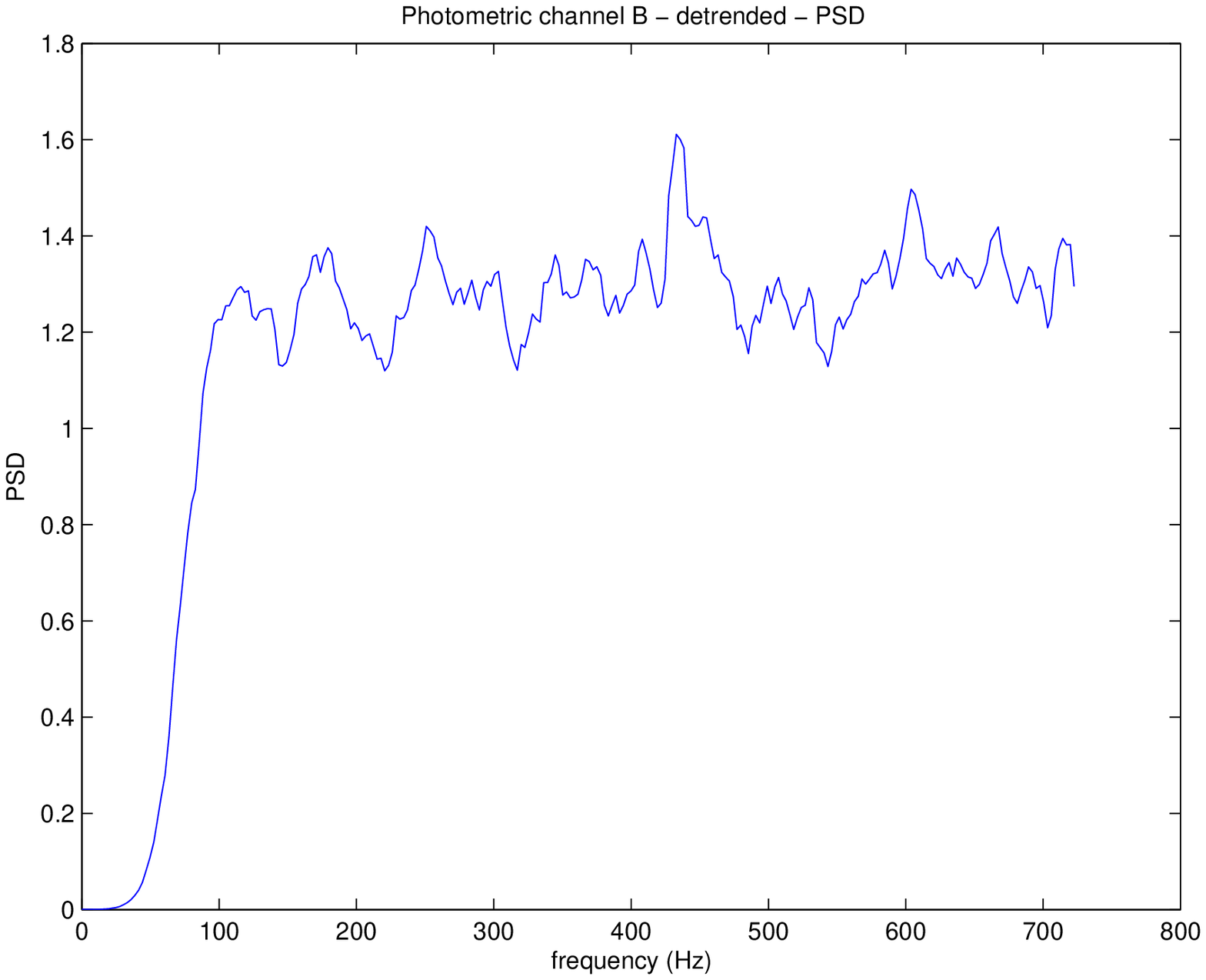,width=6.5cm}
    \caption{Case 2: Power Spectral Density functions for photometric input $PB$: raw data (left) and detrended (right).}
 
  \end{center}
\end{figure*}

\begin{figure*}[htbp]
    \begin{center}
    \epsfig{figure=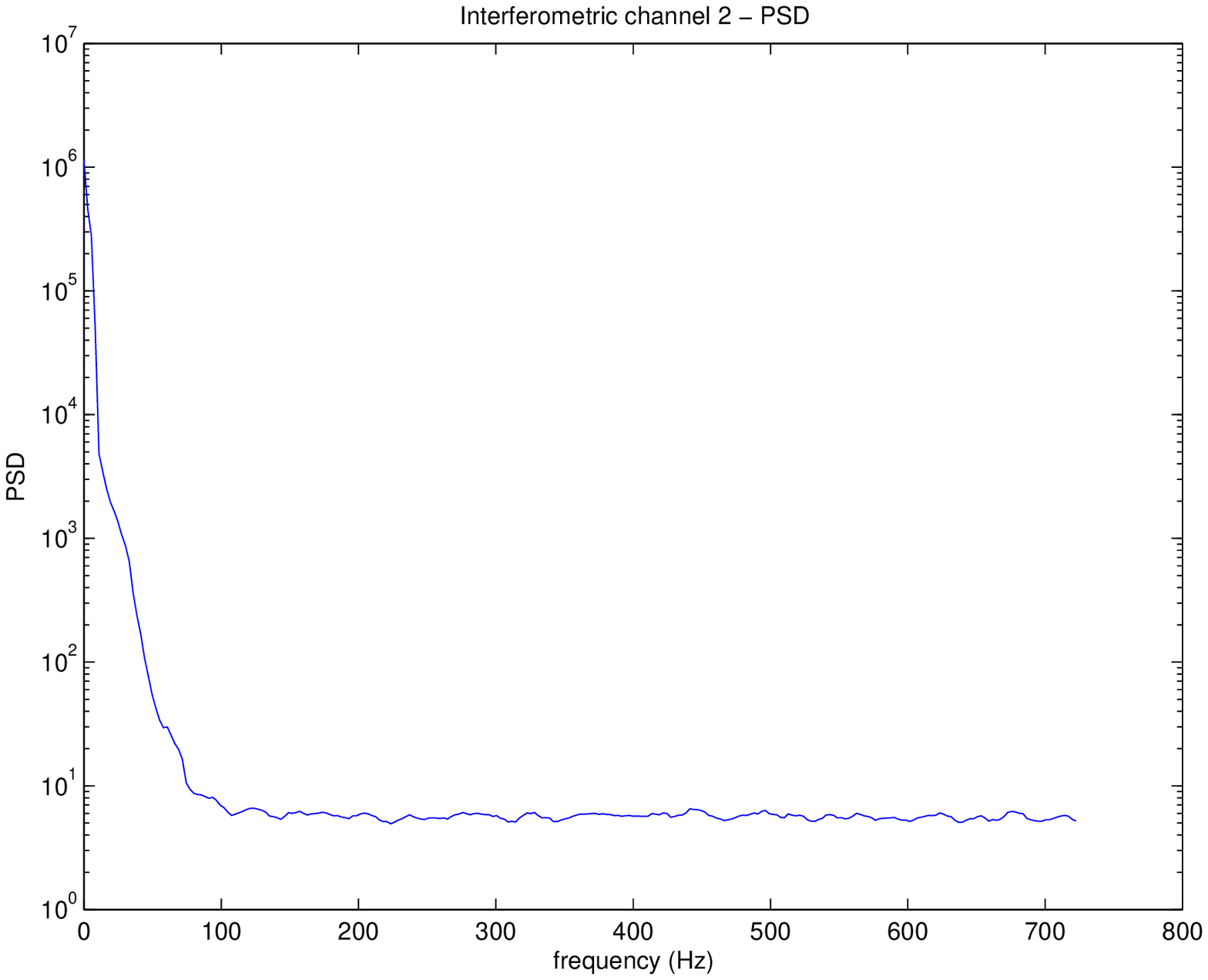,width=6.5cm}
    \epsfig{figure=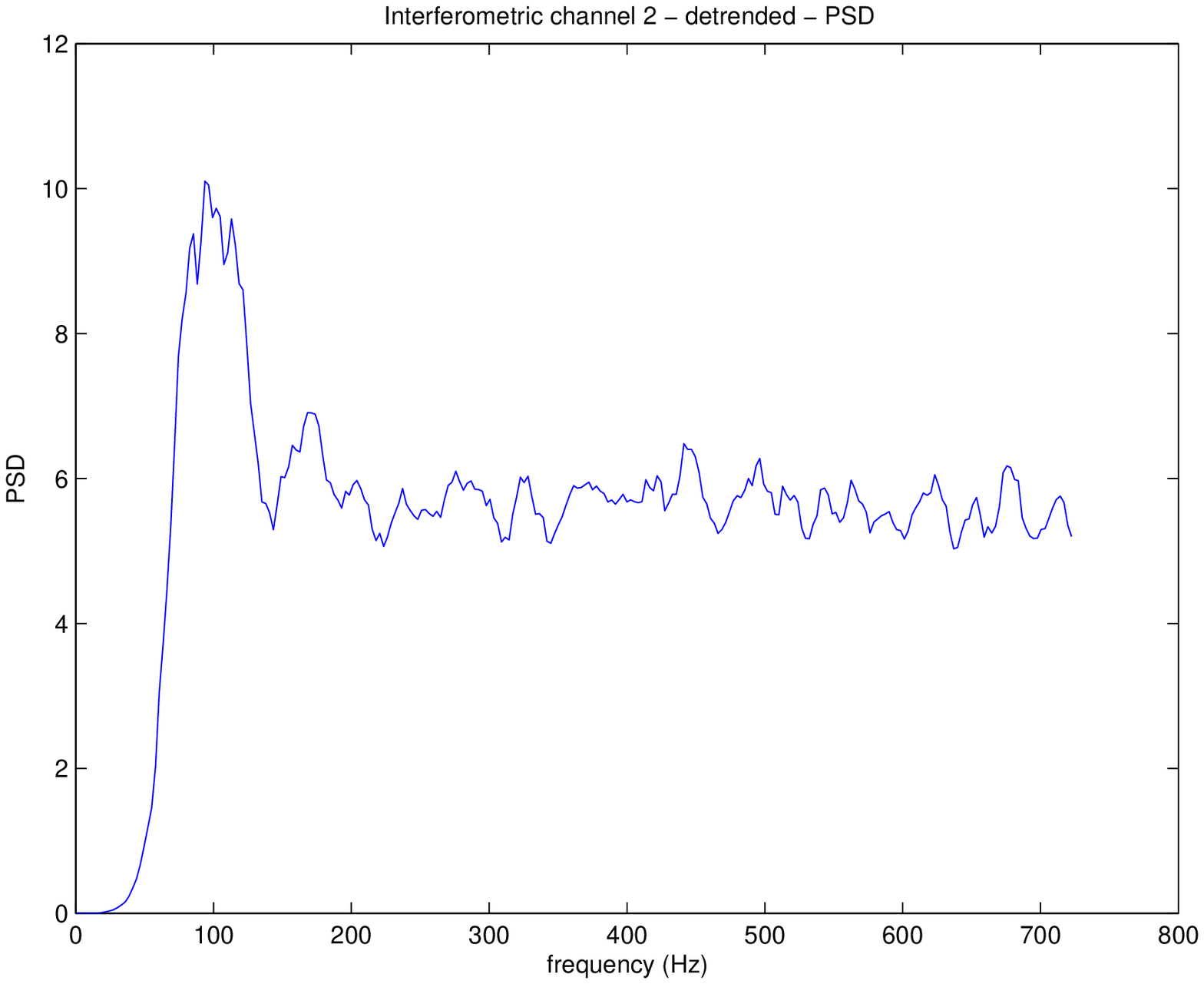,width=6.5cm}
    \caption{Case 2: Power Spectral Density functions for interferometric output $I2$: raw data (left) and detrended (right).}
   \end{center}
\end{figure*}
\begin{figure*}[htbp]
    \begin{center}
    \epsfig{figure=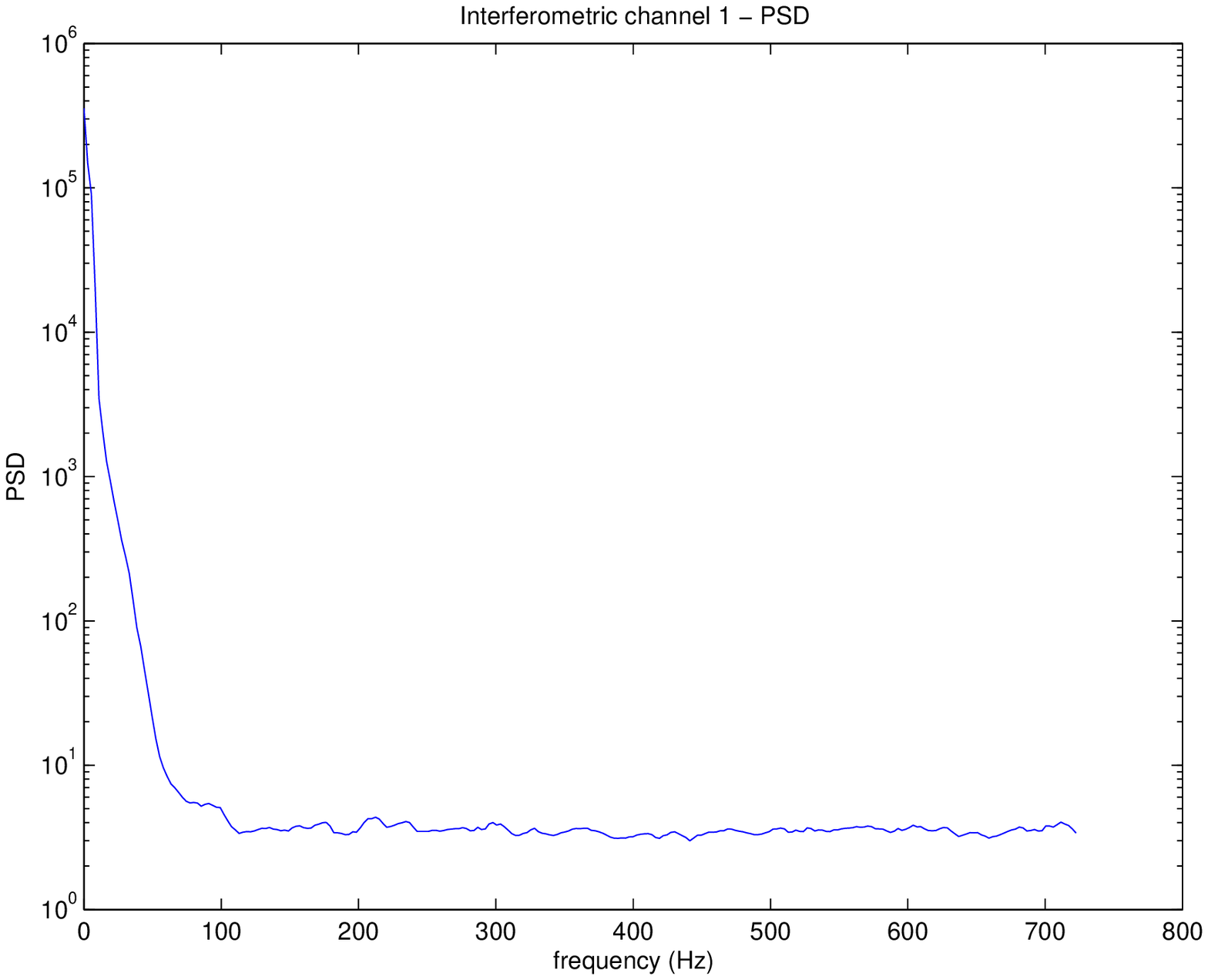,width=6.5cm}
    \epsfig{figure=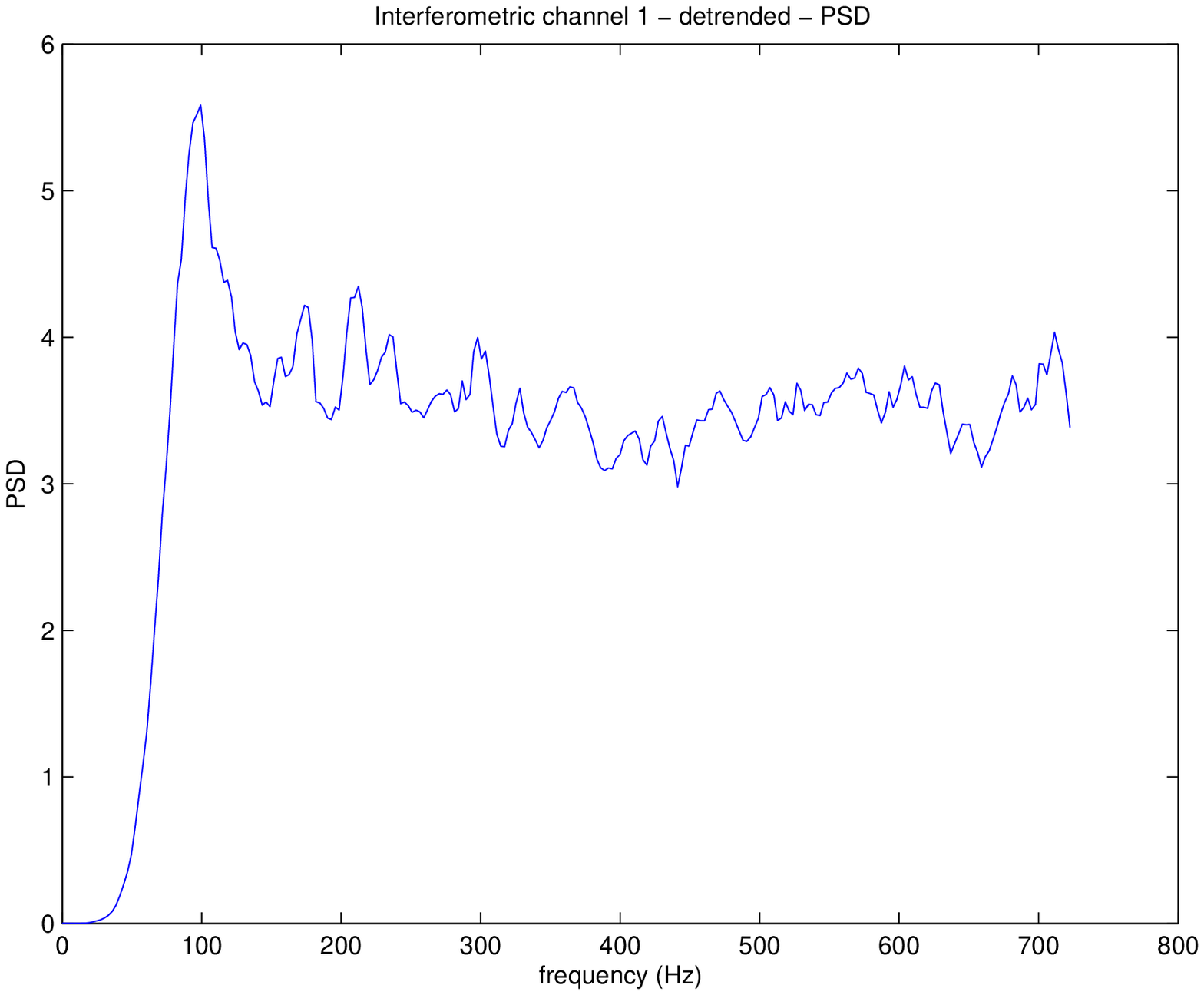,width=6.5cm}
    \epsfig{figure=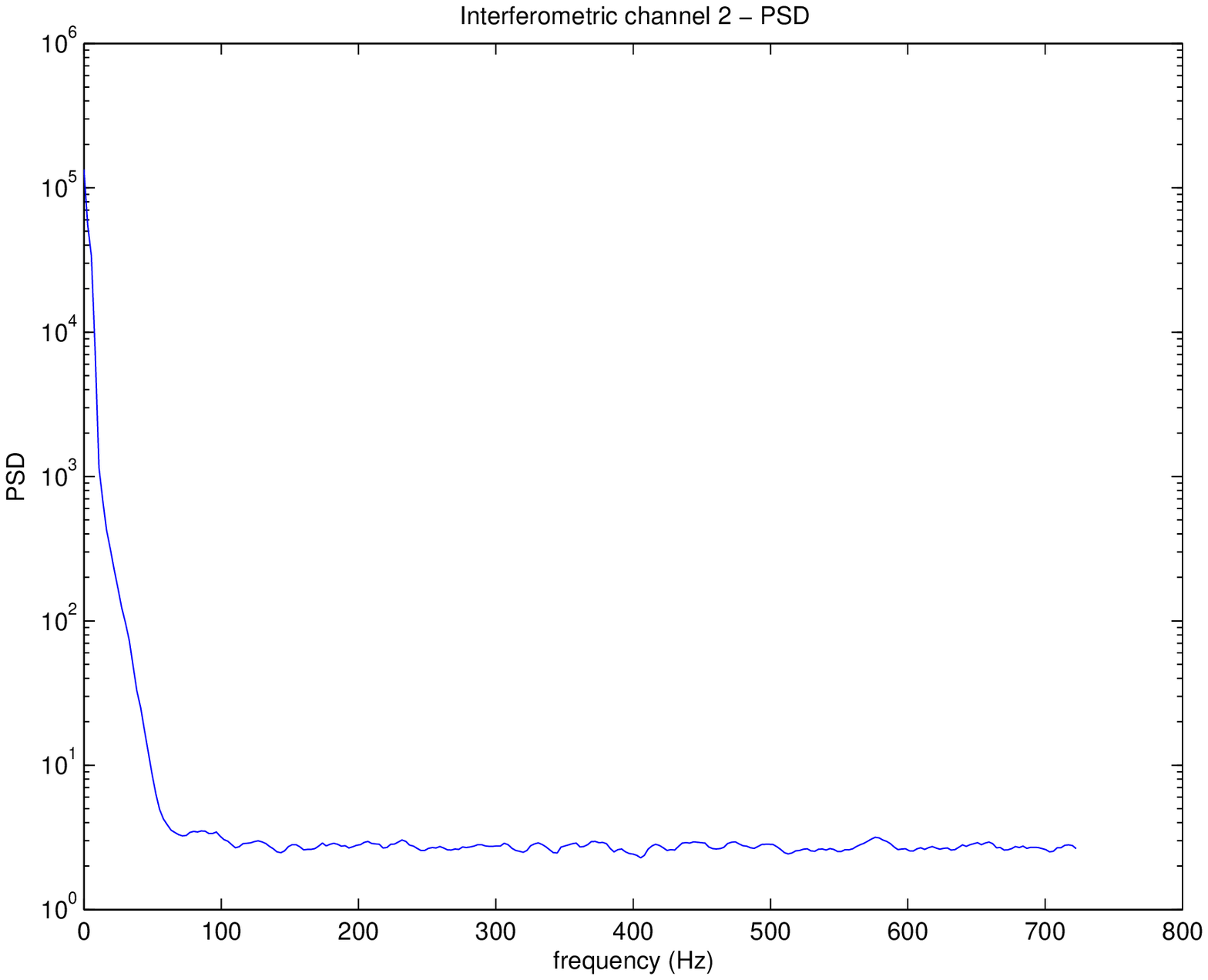,width=6.5cm}
    \epsfig{figure=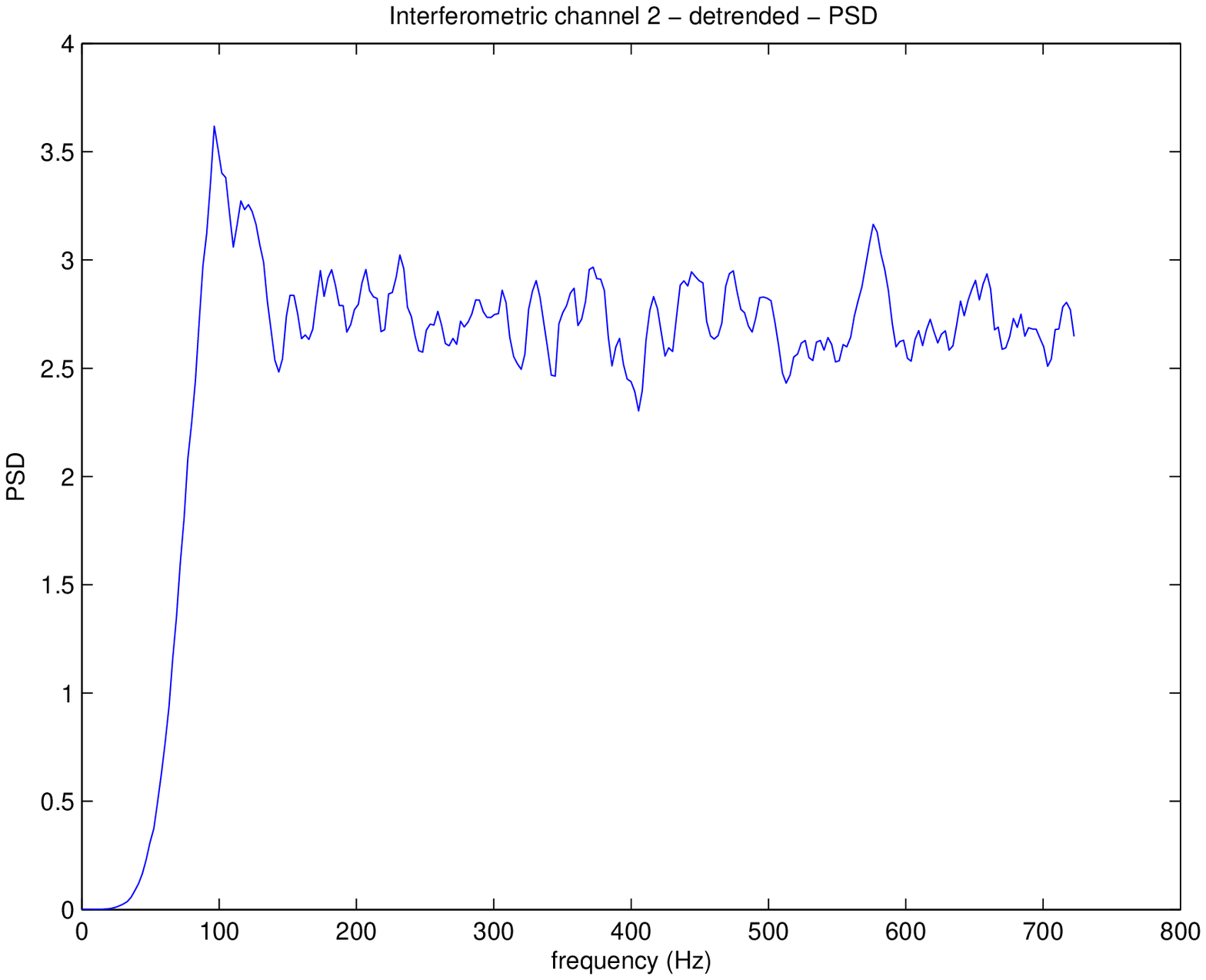,width=6.5cm}
    \caption{Case 3: Power Spectral Density functions for interferometric input $I1$ (first row) and $I2$ (second row): raw data (left) and detrended (right).}
    \end{center}
\end{figure*}

\begin{figure*}[htbp]
    \begin{center}
    \epsfig{figure=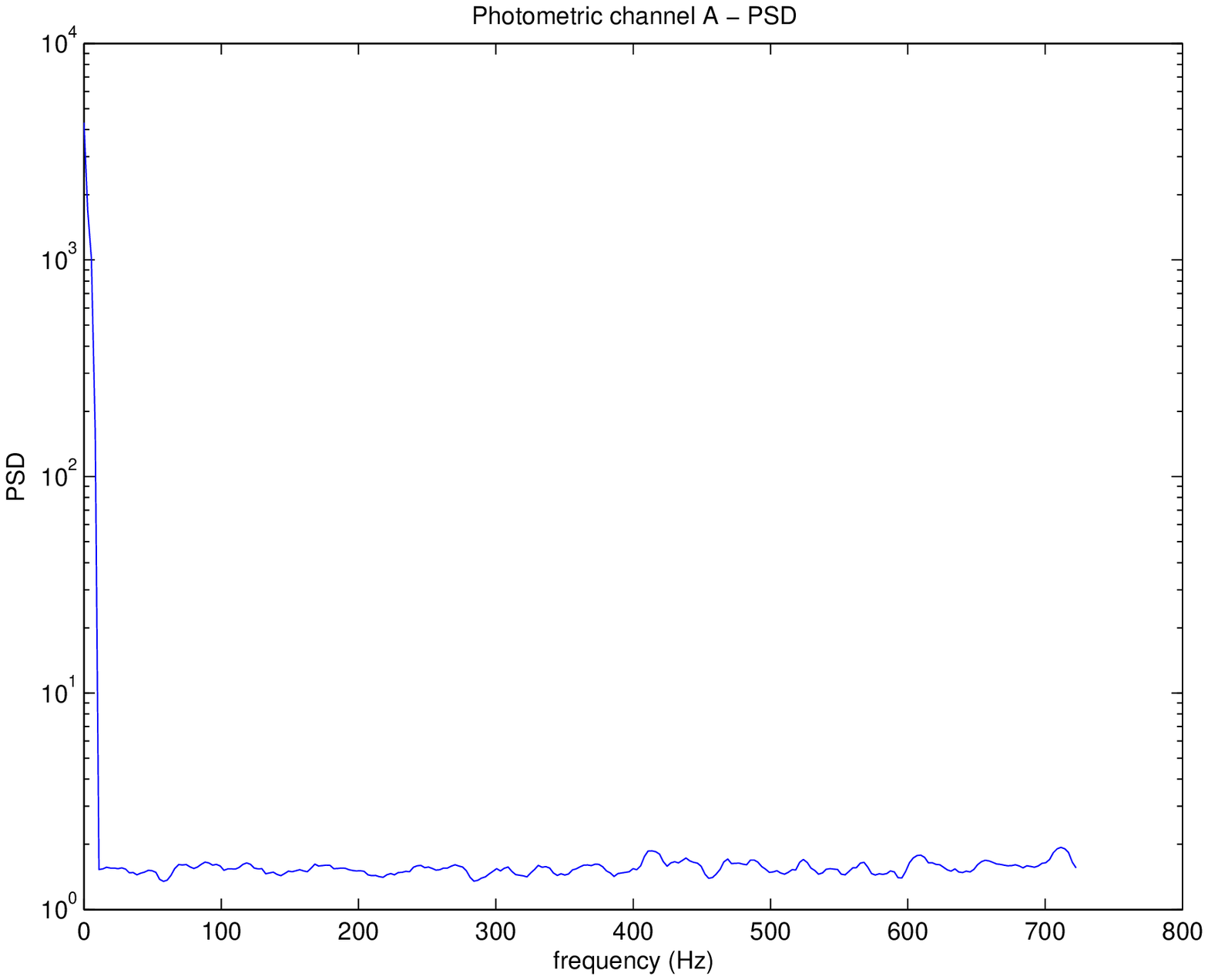,width=6.5cm}
    \epsfig{figure=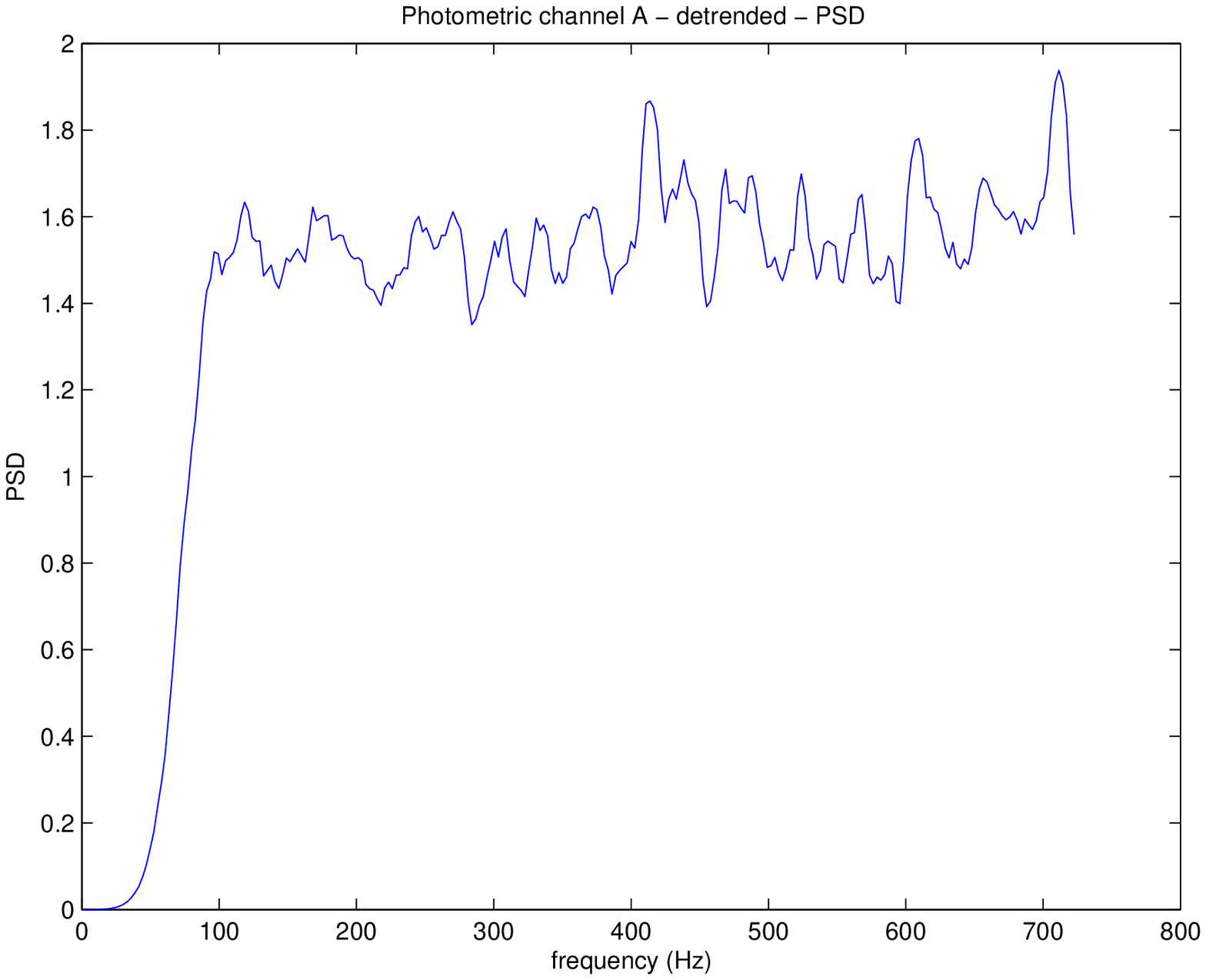,width=6.5cm}
    \epsfig{figure=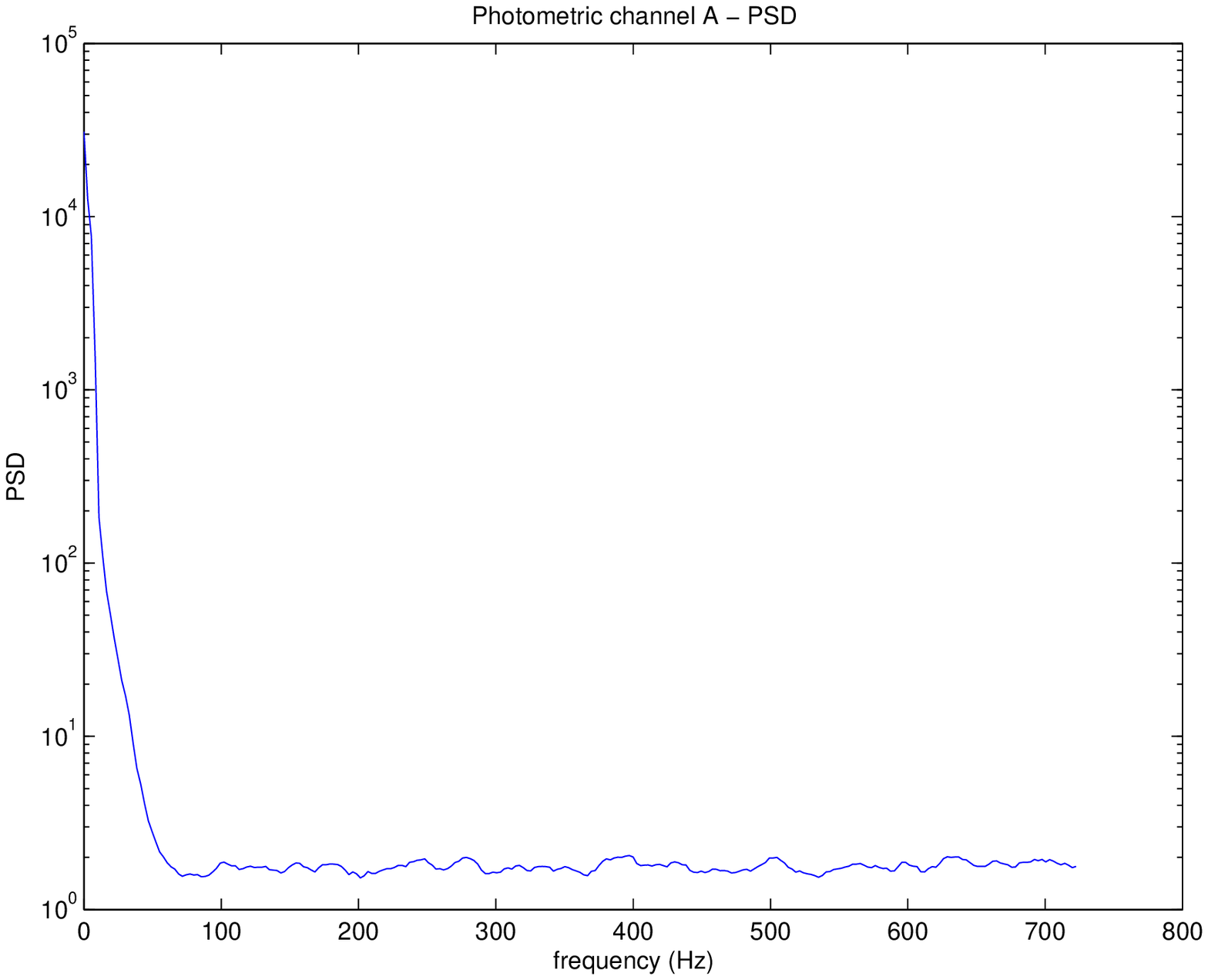,width=6.5cm}
    \epsfig{figure=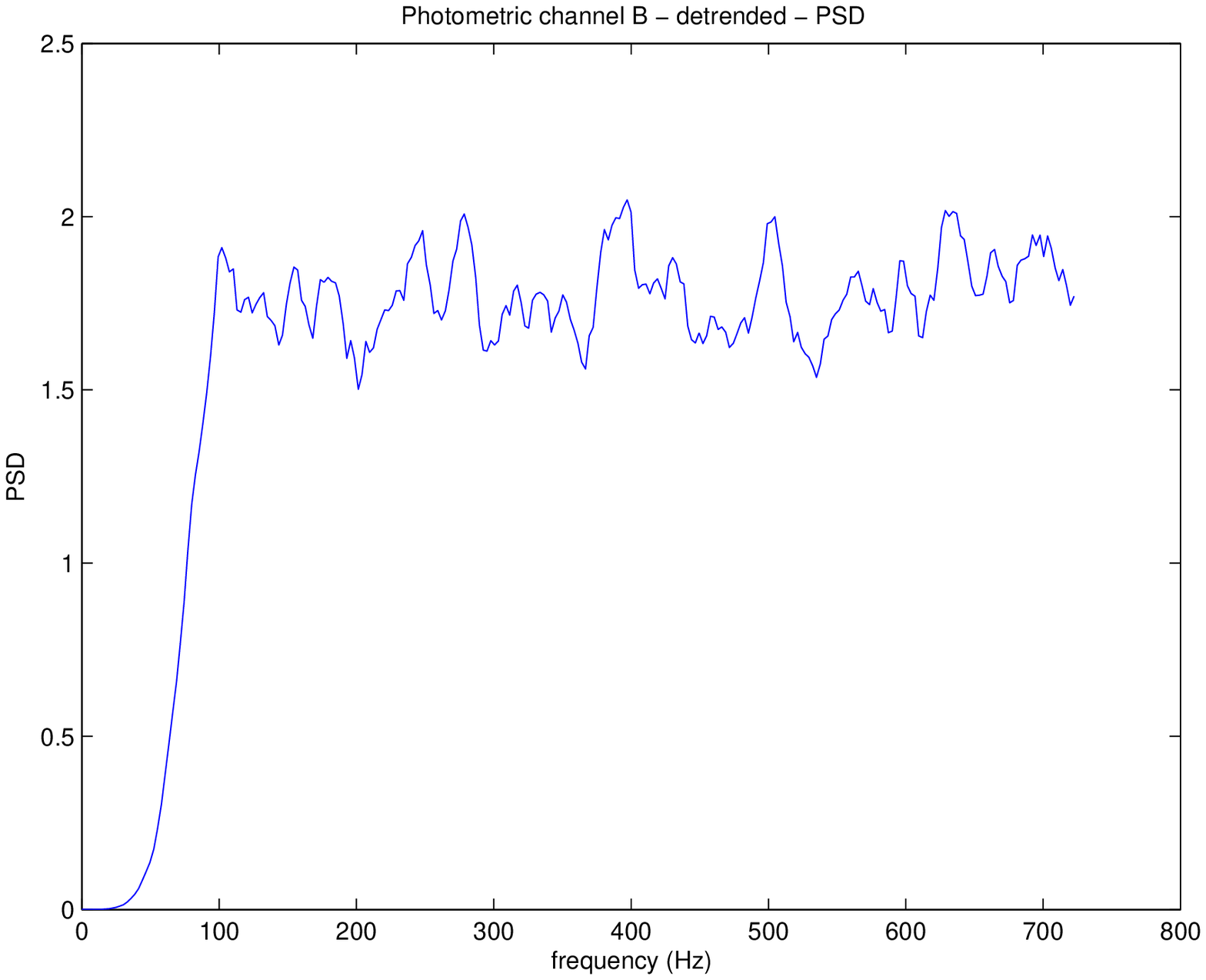,width=6.5cm}
    \caption{Case 3: Power Spectral Density functions for photometric input $PA$ (first row) and $PB$ (second row): raw data (left) and detrended (right).}
    \end{center}
\end{figure*}

\begin{figure*}[htbp]
    \begin{center}
    \epsfig{figure=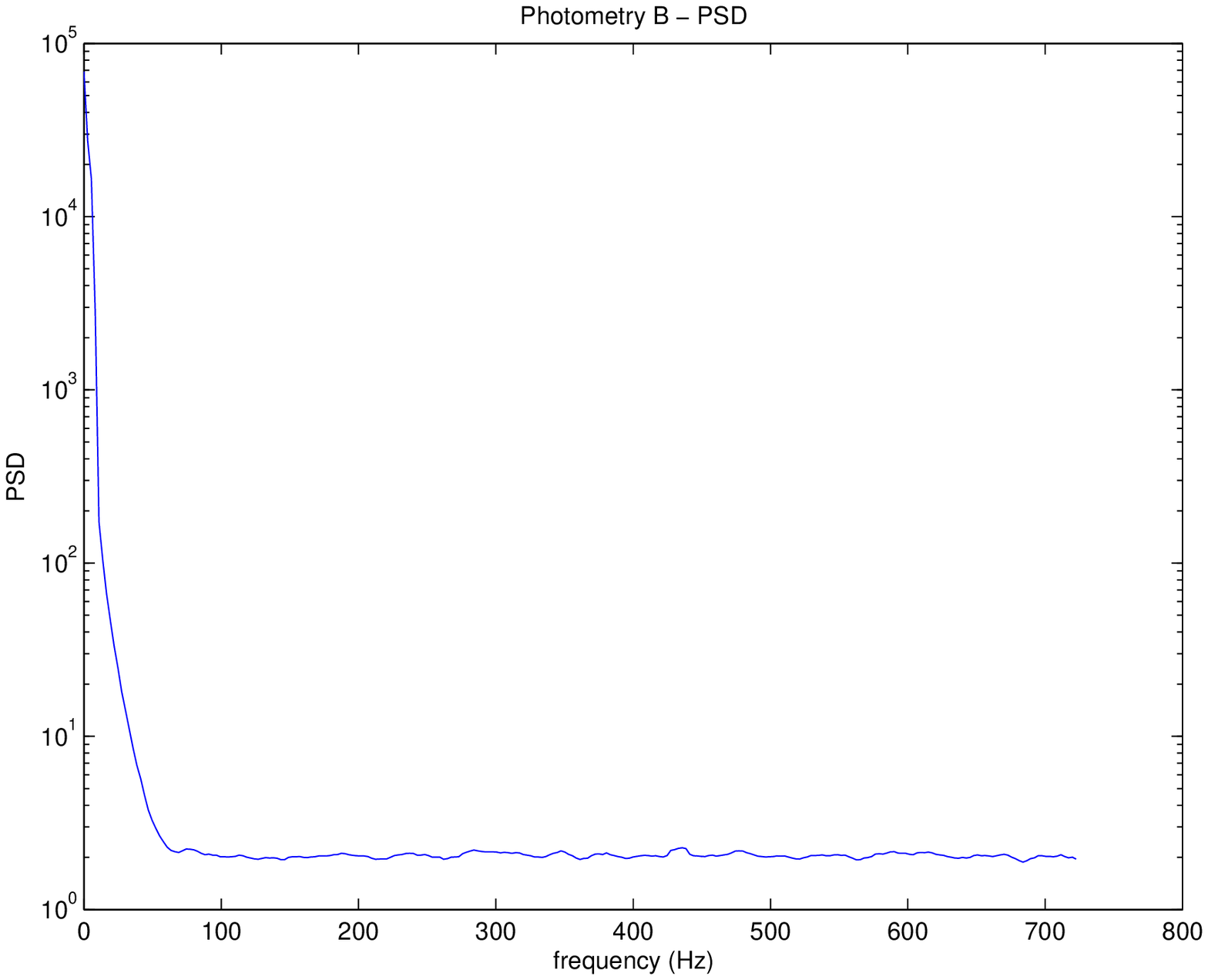,width=6.5cm}
    \epsfig{figure=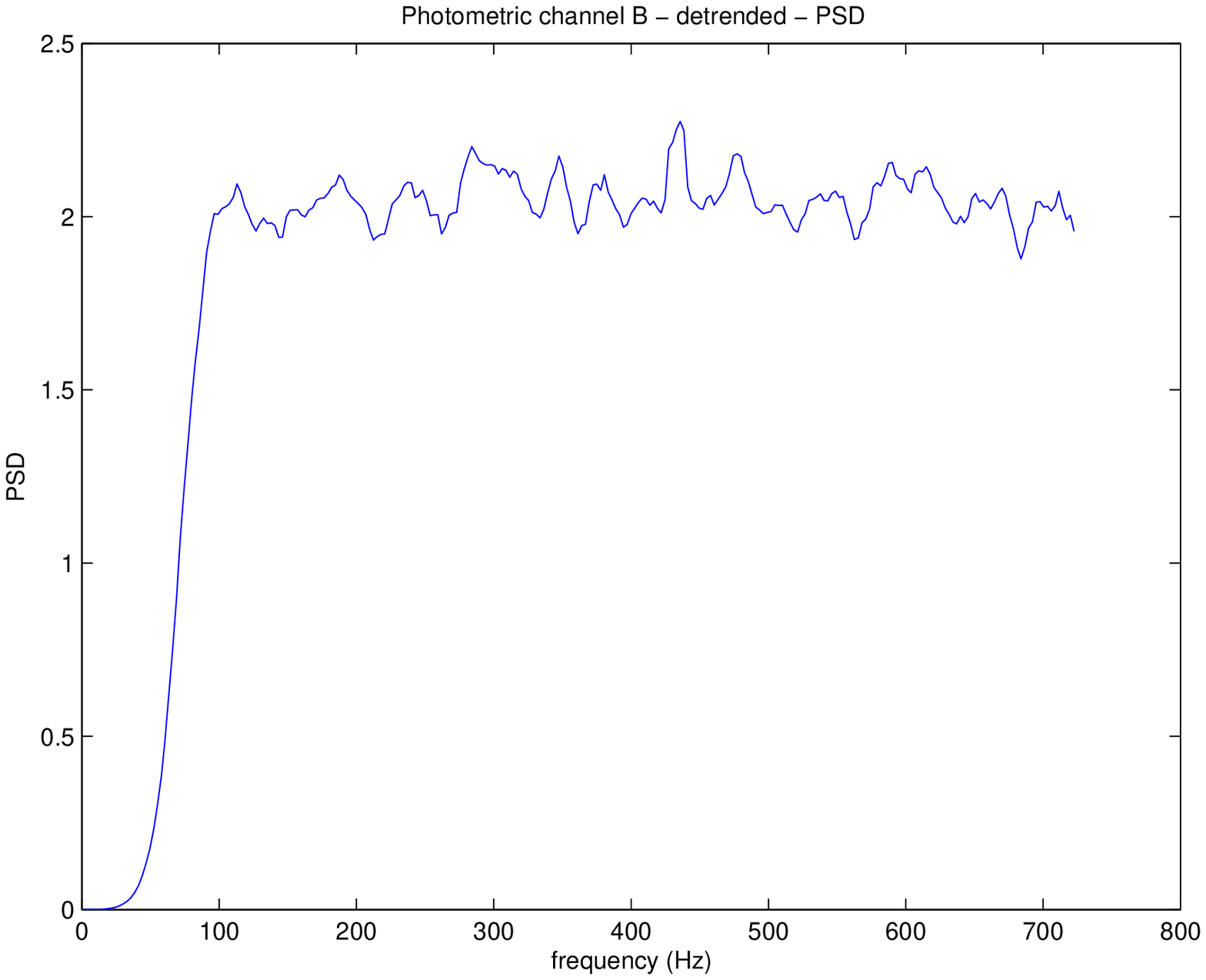,width=6.5cm}
    \epsfig{figure=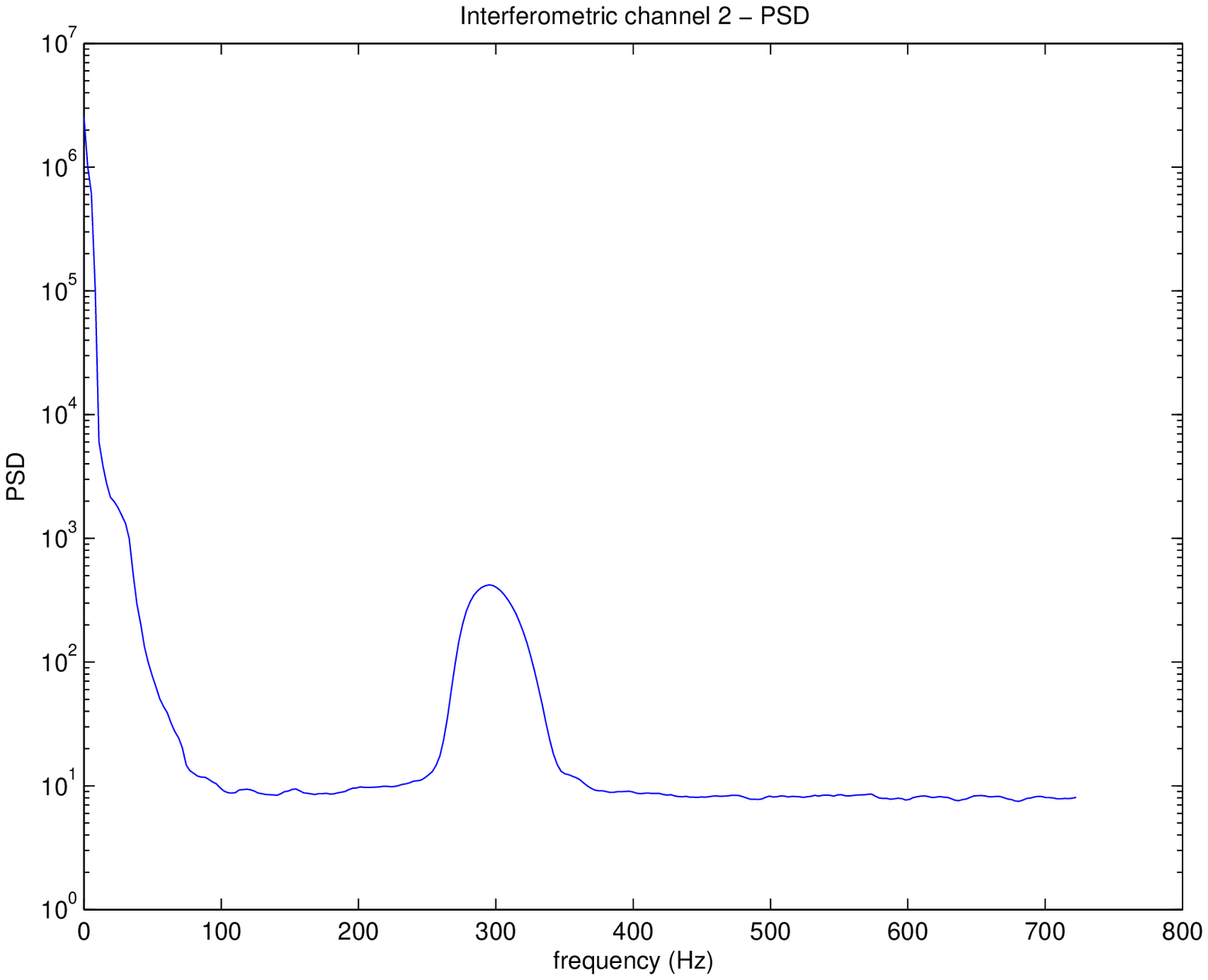,width=6.5cm}
    \epsfig{figure=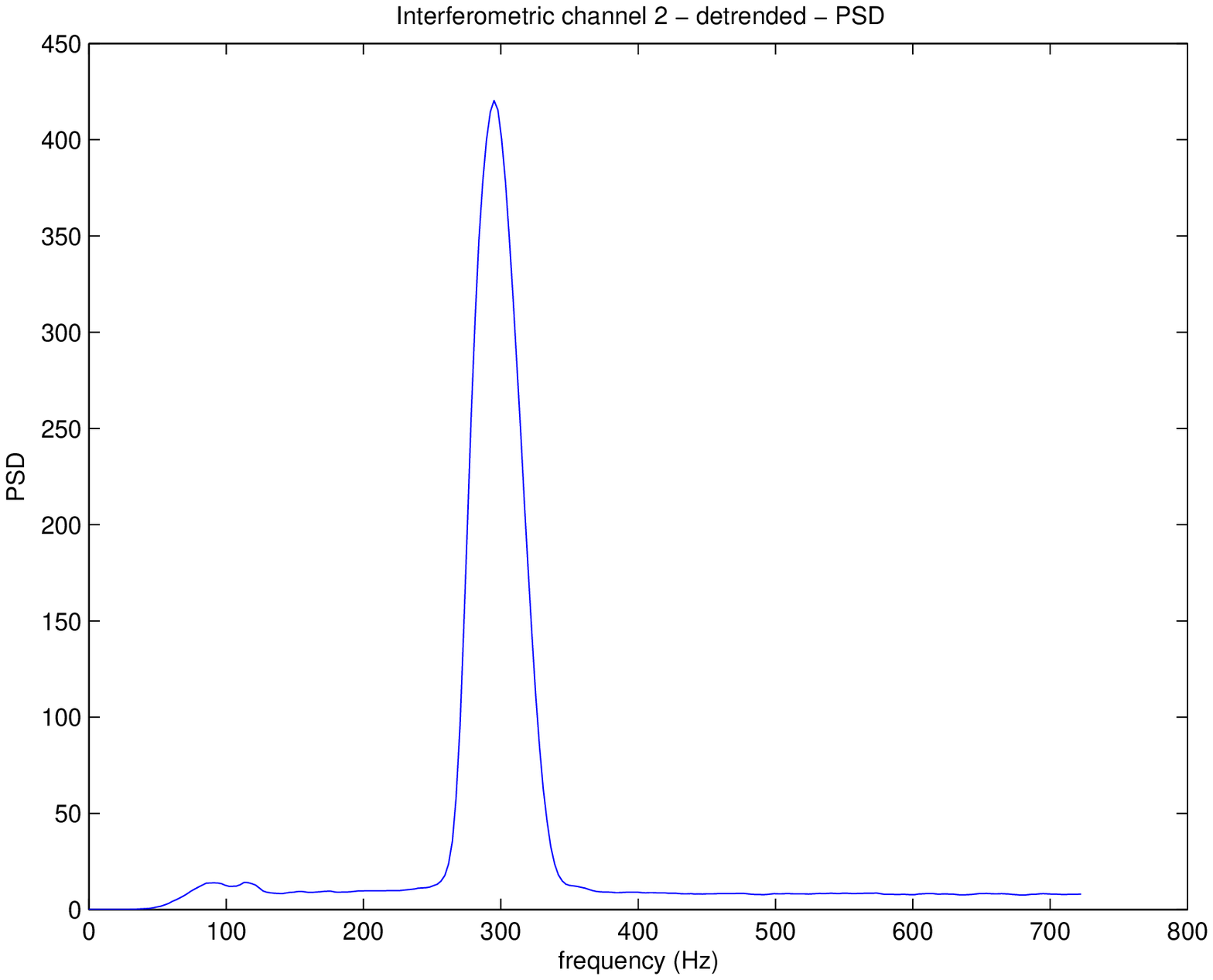,width=6.5cm}
    \caption{Case 4: Power Spectral Density functions for photometric input $PB$ (first row) and for interferometric output $I2$ (second row): raw data (left) and detrended (right).}
    \end{center}
\end{figure*}

\pagebreak

\section{Allan variance}
\begin{figure*}[htbp]
    \begin{center}
    \epsfig{figure=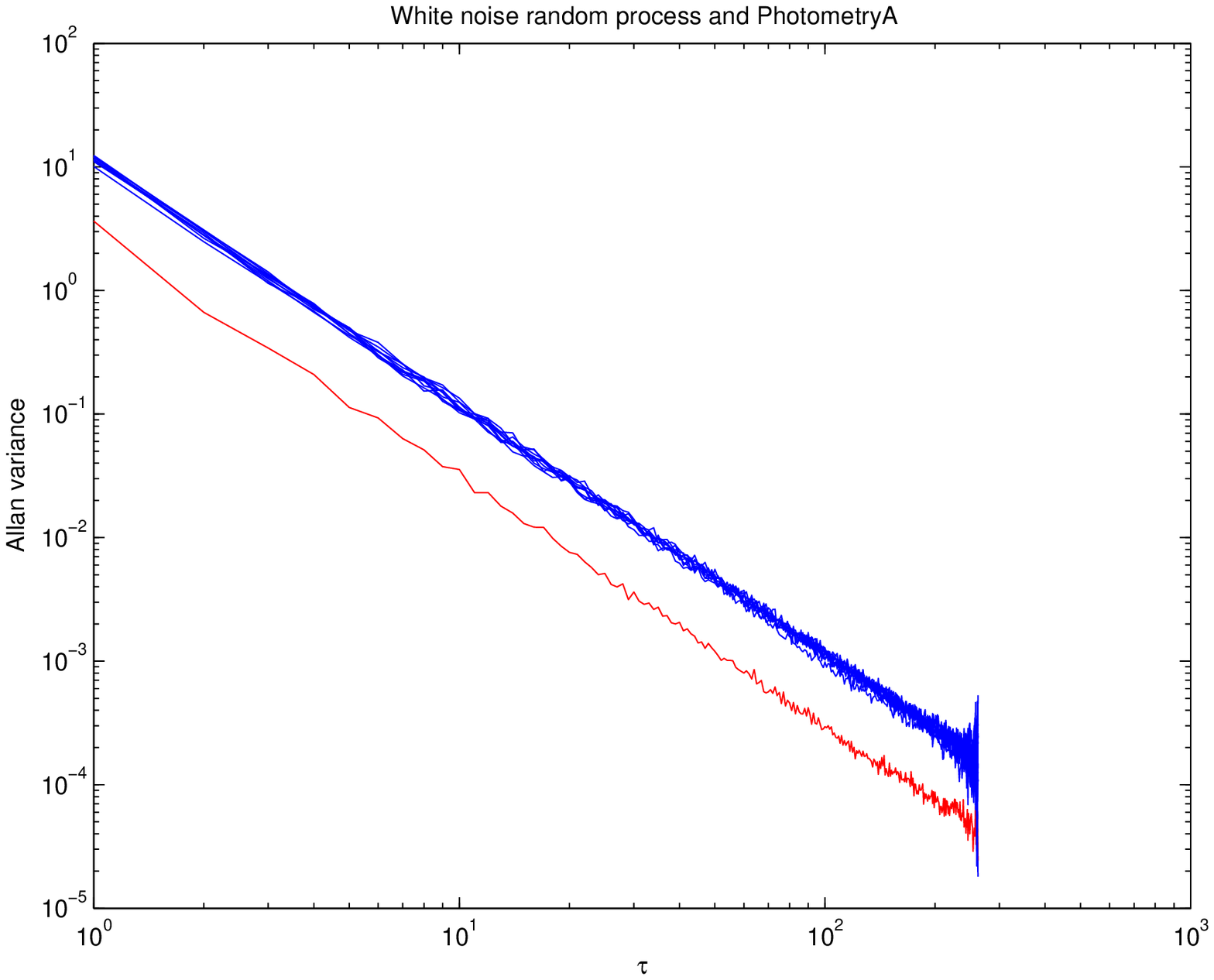,width=6.5cm}
    \epsfig{figure=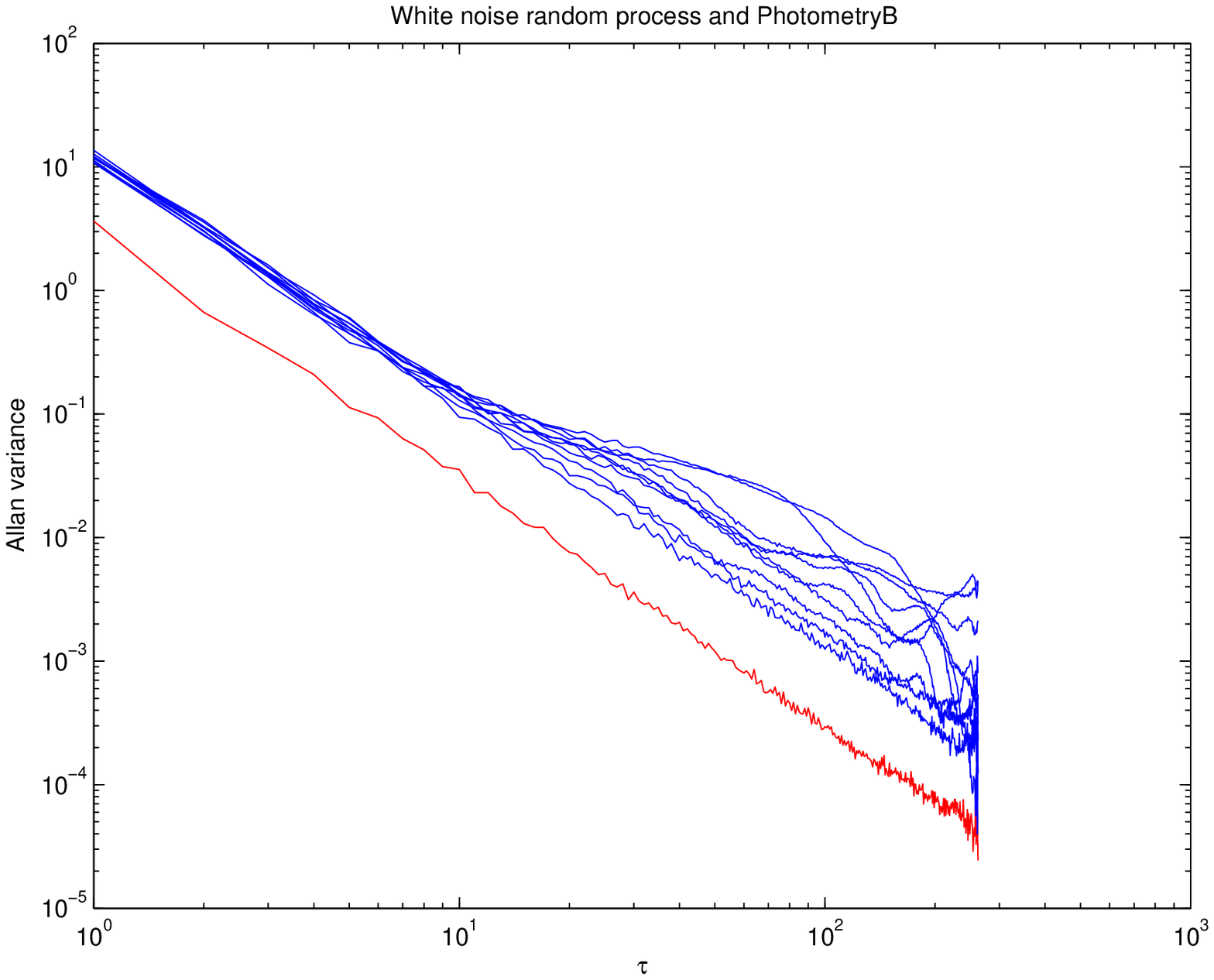,width=6.5cm}
    \epsfig{figure=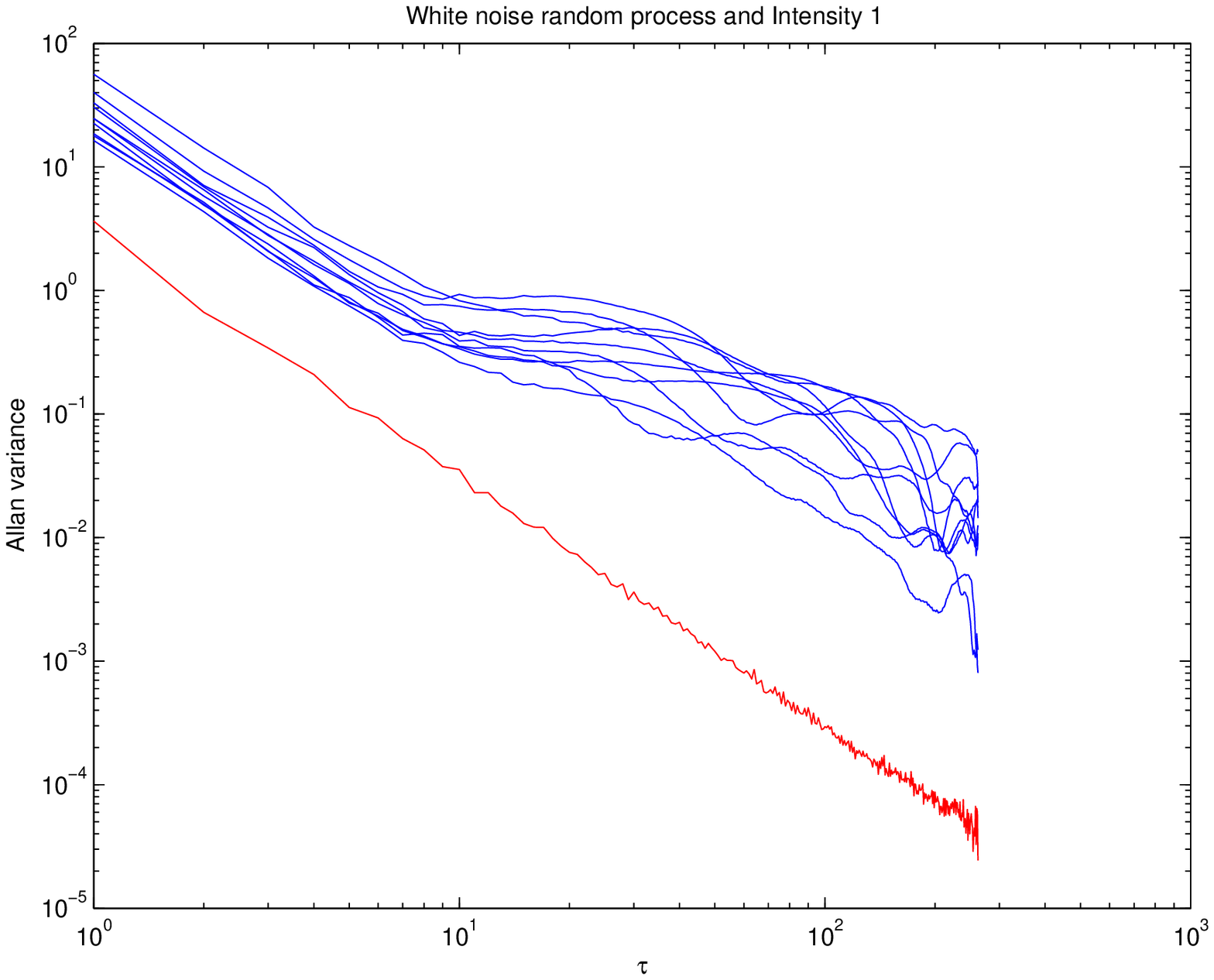,width=6.5cm}
    \epsfig{figure=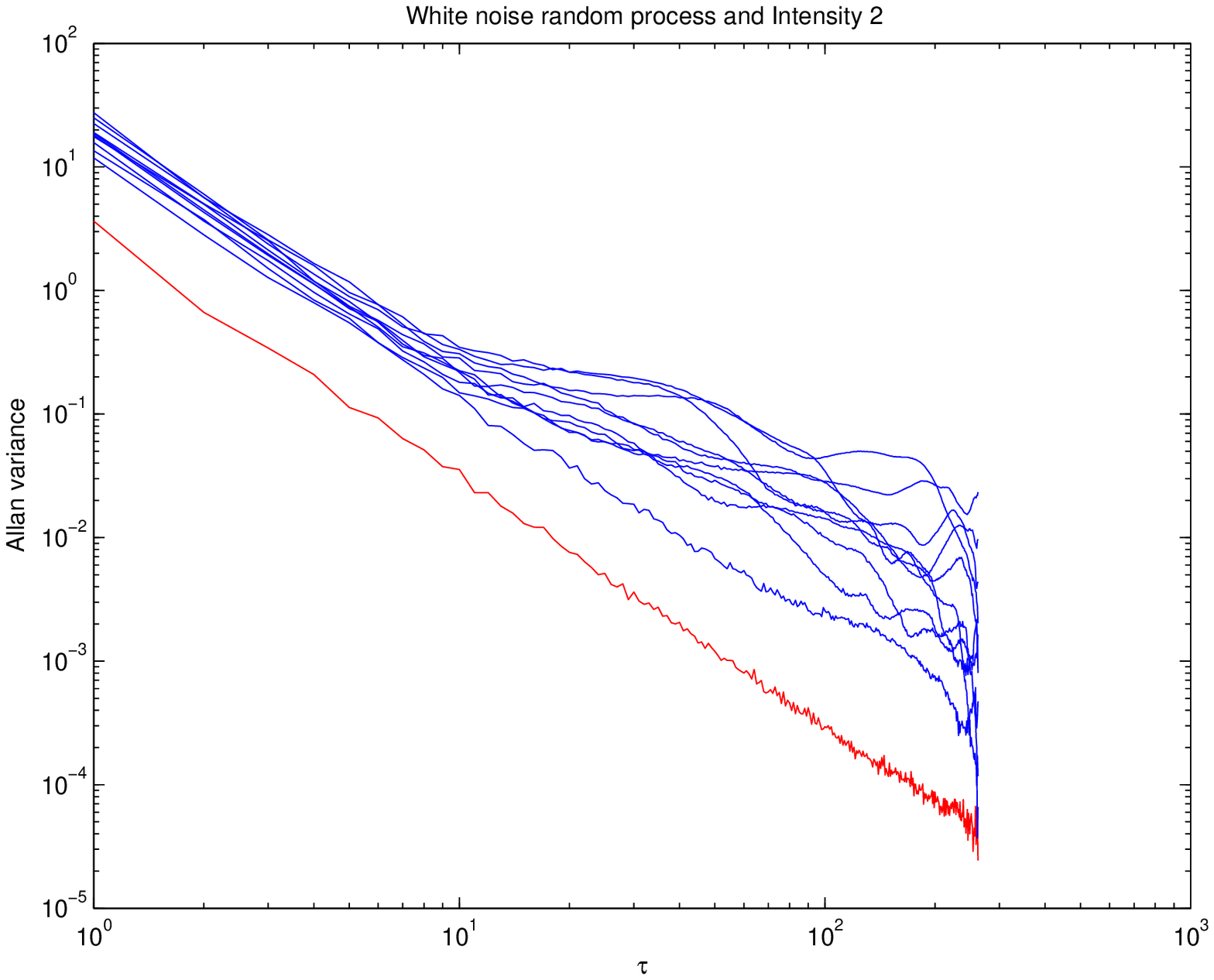,width=6.5cm}
    \caption{Case 2. Allan variance comparison between a realization of a gaussian white noise (red) and ten records (blue) of the photometric channels (first row) and the interferometric ones (second row) for case 3. Flux is injected in channel $PB$ (first row, left), while $PA$ is void (first row, right). }
    \end{center}
\end{figure*}



\newpage{\pagestyle{empty}\cleardoublepage}
\chapter*{Bibliography}
\lhead[\fancyplain{}{\bfseries\thepage}]%
      {\fancyplain{}{\bfseries Bibliography}}
\rhead[\fancyplain{}{\bfseries Bibliography}]%
      {\fancyplain{}{\bfseries\thepage}}
\addcontentsline{toc}{chapter}{Bibliography}
\bibliographystyle{unsrt}
\bibliography{tesi_Bonino}

\begin{thebibliography}{10}

\bibitem{Michelson}
Peter Lawson, editor.
\newblock {\em Course Note of the 1999 Michelson Summer School, Pasadena,
  August 1999}.
\newblock JPL Publications, 2000.

\bibitem{Herschel}
William Herschel.
\newblock {\em Phil. Trans. Royal Society London}, 95:31--64, 1804.

\bibitem{Fizeau}
Hippolyte Fizeau.
\newblock Prix bordin: Rapport sur le concours de l'ann\'{e}e 1867.
\newblock {\em C. R. Acad. Sci. Paris}, 66:932--934, 1868.

\bibitem{stephane1874}
E.~St\'{e}phane.
\newblock Sur l'extr\^{e}me petitesse du diam\`{e}tre apparent des \'{e}toiles
  fixes.
\newblock {\em C. R. Acad. Sci. Paris}, 78:1008--1012, 1874.

\bibitem{hamy1893}
M.~Hamy.
\newblock Sur la mesure des faibles diam\`{e}tres.
\newblock {\em Bulletin Astronomique}, 10:489--504, 1893.

\bibitem{Michelson1891}
A.~A. Michelson.
\newblock Measurement of jupiter's satellites by interference.
\newblock {\em Nature}, 45:160--161, 1891.

\bibitem{hambury56}
R.~Hambury Brown and R.~Q. Twiss.
\newblock Correlation between photons in two coherent beams of light.
\newblock {\em Nature}, 177:27--32, 1956.

\bibitem{Thompson}
Richard Thompson, James Moran, and George Swenson.
\newblock {\em Interferometry and Synthesis in Radio Astronomy}.
\newblock John Wiley \& sons, 1986.

\bibitem{colavita00}
M.~M. Colavita and P.~L. Wizinowich.
\newblock {Keck} {I}nterferometer: progress report.
\newblock {\em SPIE}, 4006:310--320, 2000.

\bibitem{glindemann00}
A.~Glindemann, R.~Abuter, F.~Carbognani, F.~Delplancke, F.~Derie, A.~Gennai,
  P.~B. Gitton, P.~Kervella, B.~Koehler, S.~A. L\'ev\^eque, S.~Menardi,
  A.~Michel, F.~Paresce, T.~P. Duc, A.~Richichi, M.~Sch\"oller, M.~Tarenghi,
  A.~Wallander, and R.~Wilhelm.
\newblock {\em SPIE}, 4006, 2000.

\bibitem{Monnier03}
John~D Monnier.
\newblock Optical interferometry in astronomy.
\newblock {\em Reports on Progress in Physics}, 66:789--857, 2003.

\bibitem{Kervella04}
Kervella Pierre, S\'{e}gransan D, and Coud\'{e} du~Foresto~V.
\newblock Data reduction methods for single-mode optical interferometry.
\newblock {\em A \& A}, 425:1164--1174, 2004.

\bibitem{Vasisht03}
G.~Vasisht, A.~J. Booth, M.~M. Colavita, R.~L. Johnson, E.~R. Ligon, J.~D.
  Moore, and D.~L. Palmer.
\newblock Performance and verification of the {K}eck {I}nterferometer fringe
  detection and tracking system.
\newblock {\em SPIE}, 4838:824--834, 2003.

\bibitem{coudedeforesto97}
V.~Coud\'e du~Foresto, S.~Ridgway, and J.~M. Mariotti.
\newblock Deriving object visibilities from interferograms obtained with a
  fiber stellar interferometer.
\newblock {\em A\&AS}, 121:379, 1997.

\bibitem{Kervella00}
P.~Kervella, V.~Coud\'{e} du~Foresto, A.~Glindemann, and R.~Hofmann.
\newblock {VINCI}: the {VLT} {I}nterferometer commissioning instrument.
\newblock {\em SPIE}, 4006:31--42, 2000.

\bibitem{petrov07}
R.~G. Petrov et~al.
\newblock {AMBER}, the near-infrared spectro-interferometric three-telescope
  {VLTI} instrument.
\newblock {\em A\&A}, 464:1--12, 2007.

\bibitem{leinert98}
Christopher Leinert and Uwe Graser.
\newblock {MIDI}: a mid-infrared interferometric instrument for the {VLTI}.
\newblock {\em SPIE}, 3350:389--393, 1998.

\bibitem{GayRabbia94}
Jean Gay and Yves Rabbia.
\newblock {L.A.M.P.} : a concept for the {ESO-VLTI} fringe sensor.
\newblock {\em SPIE}, 2200:195--203, 1994.

\bibitem{Bonnet06}
H.~Bonnet et~al.
\newblock Enabling fringe tracking at the {VLTI}.
\newblock {\em The Messenger ESO}, (126), 2006.

\bibitem{DelPlancke00}
Fran\c{c}oise Delplancke et~al.
\newblock Phase-referenced imaging and micro-arcsecond astrometry with the
  vlti.
\newblock {\em SPIE}, 4006:365, 2000.

\bibitem{Gai-SPIE04}
Mario Gai, Serge Menardi, Stefano Cesare, Bertrand Bauvir, Donata Bonino,
  Leonardo Corcione, Martin Dimmler, Giuseppe Massone, F.~Reynaud, and Anders
  Wallander.
\newblock The {VLTI} fringe sensors: {FINITO} and {PRIMA FSU}.
\newblock {\em SPIE}, 5491:528, 2004.

\bibitem{Gennai01}
Serge Menardi and Alberto Gennai.
\newblock Technical specifications for the {PRIMA} {F}ringe {S}ensor {U}nit.
\newblock Technical Report VLT-SPE-ESO-15740-2210, ESO, 2001.

\bibitem{Goodman}
J.~W. Goodman.
\newblock {\em Statistical Optics}.
\newblock Wiley Classics Library, 1985.

\bibitem{Rabbia94}
Yves Rabbia, Serge M\'{e}nardi, Jean Gay, et~al.
\newblock Prototype for the {ESO-VLTI} fringe sensor.
\newblock {\em SPIE}, 2200:204--215, 1994.

\bibitem{WalkupGoodman73}
J.~F. Walkup and J.~W. Goodman.
\newblock Limitations of fringe-parameter estimation at low light levels.
\newblock {\em Journal of Optical Society of America}, 63:399--407, 1973.

\bibitem{OCA93}
Group~Project {VLTI}.
\newblock {VLTI} {F}ringe {S}ensor – {P}hase {A}2 additional report.
\newblock Technical report, Observatoire de la {C}\^{o}te d'{A}zur, 1993.

\bibitem{Daigne99}
G.~Daigne and J.~F. Lestrade.
\newblock Astrometric optical interferometry with non-evacuated delay lines.
\newblock {\em A\&A Supplement Series}, 138:355--363, 1999.

\bibitem{Gardiol}
D.~Gardiol.
\newblock Fringe {T}racking simulations.
\newblock {VLTI-FINITO OAT}o internal memo, {OAT}o, 2000.

\bibitem{FINITOTestParanal}
A.~Wallander, B.~Bauvir, P.~Gitton, and S.~Menardi.
\newblock Technical report on fringe tracking with {UT}s - results from paranal
  tests {A}ugust 2004.
\newblock Technical Report VLT-TRE-ESO-15430-3391, ESO, 2004.

\bibitem{VLTScience}
Oscar von~der L\"{u}he.
\newblock An introduction to interferometry with the {ESO} {V}ery {L}arge
  {T}elescope.
\newblock In F.~Paresce, editor, {\em Science with the {VLT} {I}nterferometer}.
  Springer, 1996.

\bibitem{Bonino04}
Bonino D., Gai M., Corcione L., and Massone G.
\newblock Models for {VLTI} {Fringe} {Sensors}: {FINITO} and {PRIMA FSU}.
\newblock {\em SPIE}, 5491:1463, 2004.

\bibitem{errorbudget}
Alenia and {OAT}o {PRIMA FSU}~working group.
\newblock Final design review - report.
\newblock Technical Report ALS-FSU-BDG-0001, Alenia Spazio - {INAF OAT}o, 2004.

\bibitem{Gai-PASP1998}
Gai M., Casertano S., Carollo D., and Lattanzi M.G.
\newblock Location estimators for interferometric fringes.
\newblock {\em PASP}, 110:848--862, 1998.

\bibitem{Kervella03}
Kervella Pierre, Th\'{e}venin F, S\'{e}gransan D, et~al.
\newblock The diameters of $\alpha$ centauri {A} and {B}.
\newblock {\em A \& A}, 404:1087--1097, 2003.

\bibitem{Priestley}
M.~B. Priestley.
\newblock {\em Spectral analysis and time series}.
\newblock Academic Press, Probability and Mathematical Statistics Series, 1981.

\bibitem{Allan}
David~W. Allan.
\newblock Statistics of atomic frequency standards.
\newblock {\em IEEE}, 54--2:221--230, 1966.

\bibitem{Manolakis}
D.~G. Manolakis, V.~K. Ingle, and S.~M. Kogon.
\newblock {\em Statistical and Adaptive Signal Processing}.
\newblock Mc Graw Hill, 2005.

\bibitem{Allan87}
David~W. Allan.
\newblock Time and frequency (time-domain) characterization, estimation and
  prediction of precision clocks and oscillators.
\newblock {\em IEEE}, UFFC-34:647--654, 1987.

\bibitem{Schieder}
R.~Schieder and C.~Kramer.
\newblock Optimization of heterodyne observations using allan variance
  measurements.
\newblock {\em A \& A}, 373:746--756, 2001.

\bibitem{Colavita}
M.~M. Colavita.
\newblock Measurements of the atmospheric limit to narrow-angle interferometric
  astrometry using the mark iii stellar interferometer.
\newblock {\em A \& A}, 283:1027--1036, 1994.

\bibitem{Galleani}
L.~Galleani and P.~Tavella.
\newblock The characterization of clock behaviour with the dynamic allan
  variance.
\newblock In {\em Proc. Joint Meeting European Frequency and Time Forum and
  {IEEE} Frequency Control Symposium}, 2003.

\bibitem{hopkins}
G.~V. Glass and K.~D. Hopkins.
\newblock {\em Statistical methods in education and Psychology}.
\newblock Allyn and Bacon, New York, 1996.

\bibitem{rawlings}
J.~O. Rawlings, S.~G. Pantula, and D.~A. Dickey.
\newblock {\em Applied Regression Analysis. A research tool}.
\newblock Springer Text in Statistics, 2001.

\bibitem{Ryan}
T.~P. Ryan.
\newblock {\em Modern Regression Methods}.
\newblock Wiley Series in Probability and Statistics, 1997.

\bibitem{DraperSmith}
N.~Draper and H.~Smith.
\newblock {\em Applied Regression Analysis, second edition}.
\newblock Wiley Series in Probability and Statistics, 1966.

\bibitem{durbin-watson51}
J.~Durbin and G.~S. Watson.
\newblock Testing for serial correlation in least squares regression. ii.
\newblock {\em Biometrika}, 38:159--178, 1951.

\bibitem{Dagum}
E.~B. Dagum.
\newblock {\em Analisi delle serie storiche: modellistica, previsione e
  scomposizione}.
\newblock Springer, 2002.

\bibitem{Kendall54}
M.~G. Kendall.
\newblock Note on bias in the estimation of autocorrelation.
\newblock {\em Biometrika}, 41:403--404, 1954.

\bibitem{Hannan58}
E.~J. Hannan.
\newblock The estimation of the spectral density after trend removal.
\newblock {\em Journal Royal Statistical Society B}, 20:322--333, 1958.

\bibitem{Gradsh}
I.~S. Gradshteyn and I.M. Ryzhik.
\newblock {\em Table of Integrals, Series and Products}.
\newblock Academic Press, 1963.

\end{thebibliography}
\end{document}